\documentclass[twocolumn]{aastex62}

\usepackage{xspace}
\usepackage{graphicx}
\usepackage{natbib}
\usepackage{amsmath}
\usepackage{afterpage}
\usepackage{longtable}

\newcommand{\hide}[1]{}

\newcommand{\lsim}{\ensuremath{\,\lesssim\,}\xspace}
\newcommand{\kms}{\ensuremath{\,{\rm km\,s^{-1}}}\xspace}

\newcommand{\K}{\ensuremath{\,{\rm K}}\xspace}

\newcommand{\hi}{{\rm H\,{\footnotesize I}}\xspace}
\newcommand{\hii}{{\rm H\,{\footnotesize II}}\xspace}

\newcommand{\heh}{\ensuremath{N(^4\textnormal{He}^+)/N(\textnormal{H}^+)}\xspace}

\shorttitle{Ionization Profiles of Galactic H\,ii Regions}
\shortauthors{Luisi et al.}

\begin{document}

\title{Ionization Profiles of Galactic H{\footnotesize{\,II}} Regions}

\author[0000-0001-8061-216X]{Matteo~Luisi}
\affiliation{Department of Physics and Astronomy, West Virginia University, Morgantown WV 26506, USA}
\affiliation{Center for Gravitational Waves and Cosmology, West Virginia University, Chestnut Ridge Research Building, Morgantown WV 26505, USA}

\author[0000-0001-8800-1793]{L.~D.~Anderson}
\affiliation{Department of Physics and Astronomy, West Virginia University, Morgantown WV 26506, USA}
\affiliation{Center for Gravitational Waves and Cosmology, West Virginia University, Chestnut Ridge Research Building, Morgantown WV 26505, USA}
\affiliation{Adjunct Astronomer at the Green Bank Observatory, P.O. Box 2, Green Bank WV 24944, USA}

\author[0000-0002-1311-8839]{Bin~Liu}
\affiliation{CAS Key Laboratory of FAST, National Astronomical Observatories, Chinese Academy of Sciences, Beijing 100101, People's Republic of China}

\author[0000-0002-1732-5990]{D.~Anish~Roshi}
\affiliation{National Radio Astronomy Observatory, 520 Edgemont Road, Charlottesville VA 22903-2475, USA}

\author{Ed~Churchwell}
\affiliation{Department of Astronomy, University of Wisconsin-Madison, Madison WI 53706-1507, USA}

\begin{abstract}
Using Green Bank Telescope radio recombination line (RRL) data, we analyze the role of leaking radiation from \hii\ regions in maintaining the ionization of the interstellar medium. We observed a sample of eight Galactic \hii\ regions of various sizes, morphologies, and luminosities. For each region the hydrogen RRL intensity decreases roughly as a power-law with distance from the center of the region. This suggests that radiation leaking from the \hii\ region is responsible for the majority of surrounding ionized gas producing RRL emission. Our results further indicate that the hydrogen RRL intensity appears to be fundamentally related to the \hii\ region sizes traced by their photodissociation regions, such that physically smaller \hii\ regions show a steeper decrease in intensity with increasing distance from the region centers. As a result, giant \hii\ regions may have a much larger effect in maintaining the ionization of the interstellar medium. For six of the eight observed \hii\ regions we find a decrease in the $^4\textnormal{He}^+/\textnormal{H}^+$ abundance ratio with increasing distance, indicating that He-ionizing photons are being absorbed within the ionization front of the \hii\ region. There is enhanced carbon RRL emission toward directions with strong continuum background, suggesting that the carbon emission is amplified by stimulated emission.
\end{abstract}


\section{Introduction}\label{sec:intro}

\hii\ regions, first described by \citet{Stroemgren1939}, are regions of ionized gas surrounding O and B-type stars. Only these short-lived and massive stars emit a sufficient number of high-energy photons ($>13.6$\,eV) to fully ionize their surroundings.  \hii\ regions are among the most luminous objects in our Galaxy at radio wavelengths and can be studied using radio recombination line (RRL) and radio free-free continuum emission. Compared to observations at near- to mid-infrared (IR) wavelengths, these radio observations are faint, but have the benefit of being essentially free from extinction. 

Between the \hii\ region and the ambient diffuse interstellar medium (ISM) lies the \hi\ front, the boundary between fully ionized hydrogen within the region and neutral hydrogen outside, followed by a photodissociation region (PDR). In PDRs, hydrogen is predominantly neutral, but carbon and other species with ionization potentials lower than that of hydrogen are mostly ionized. PDRs can be studied using numerous atomic or molecular transitions, e.g.~IR emission from polycyclic aromatic hydrocarbons (PAHs) or C$^+$ emission at 158\,$\mu$m.

Low-density diffuse gas known as the ``Diffuse Ionized Gas" or ``Warm Ionized Medium" (WIM) is a major component of the ISM of our Galaxy. The WIM, with electron temperatures ranging from 6000\,K to 10,000\,K, accounts for over 90\% of all ionized hydrogen in the ISM \citep{Haffner2009}. It has a scale height of $\sim$1\,kpc and was found to be in a lower ionization state than gas in \hii\ regions \citep{Haffner1999,Madsen2006}. The ``Extended Low-Density Medium" (ELDM), another diffuse component of the ISM \citep[see][]{Gottesman1970,Mezger1978}, has a smaller scale height of only $\sim$100\,pc and has been observed to be spatially correlated with the locations of discrete \hii\ regions \citep{Alves2012}.

While it is still not fully known how the WIM maintains its ionization \citep{Haffner2009}, it is believed that O-stars are the most likely source of ionizing photons since all other possible ionization mechanisms (e.g., supernova explosions) cannot fulfill the energy requirements \citep{Domgoergen1994,Hoopes2003}. Given the distribution of the WIM, however, it remains unclear precisely how the radiation from O-stars within \hii\ regions is able to propagate through their surrounding PDRs and across kiloparsec size-scales into the ISM. While \citet{Wood2010} argue that a supernova-driven turbulent ISM has low-density paths that would allow ionizing photons to reach and ionize gas several kiloparsecs above the midplane, it is unknown whether this scenario could explain the WIM distribution in the Galactic plane.

Observations have shown that a significant amount of ionizing radiation is leaking from individual \hii\ regions. While most of these analyses focus on \hii\ regions in external galaxies \citep[e.g.][]{Oey1997,Zurita2002,Giammanco2005,Pellegrini2012}, a few studies were performed on Milky Way \hii\ regions. \citet{Anderson2015} showed using H-alpha emission data that the Galactic \hii\ region RCW\,120 is leaking $\sim$25\% of its ionizing radiation into the ISM. They further showed that the PDR is clumpy at 8.0\,$\mu$m and that photons preferentially escape through low-density pathways into the ISM. We performed a similar analysis on the compact Galactic \hii\ region NCG\,7538 \citep[][hereafter L16]{Luisi2016} using Green Bank Telescope (GBT) RRL and radio continuum data to better understand how a single \hii\ region may contribute to the ionization of the WIM. We computed an ionizing leaking fraction of $15 \pm 5$\% and found that, unlike giant \hii\ region complexes, the radiation leaking from NCG\,7538 seems to only affect the local ambient medium. 

It is not well-understood how emission from the WIM is affected by the presence of \hii\ regions, their sizes, and morphologies. We showed in L16 that the hydrogen RRL intensity around the compact \hii\ region NGC\,7538 decreases rapidly with distance from the central region, whereas the RRL emission decrease in the giant \hii\ region complex W43 is much less steep. This result implies that giant \hii\ regions may have a much larger effect on maintaining the ionization of the WIM compared to compact \hii\ regions. In fact, \citet{Zurita2002} suggest that essentially all ionizing radiation escapes from \hii\ regions with the highest luminosities. Using a model from \citet{Beckman2000}, \citet{Zurita2000} suggest that radiation leaking from luminous clusters of \hii\ regions may be sufficient to ionize the diffuse gas. As a result, a large escape fraction of less luminous regions may not necessarily be required to maintain the ionization of the WIM. 

The spectrum of the radiation field provides additional information on the physical processes within \hii\ regions and the effect of leaking radiation on the WIM. Within an \hii\ region, the radiation field depends on the temperature of the ionizing star(s). The ratio of emitted helium ($E > 24.6$\,eV) to hydrogen ($E > 13.6$\,eV) ionizing photons, $Q_1/Q_0$, can be estimated indirectly by measuring the \heh ionic abundance ratio. While $Q_1/Q_0$ is determined by the temperature of the ionizing star(s), its value is only $\sim 0.25$ even for the hottest O3V stars \citep[see][]{Martins2005,Draine2011}. For an O9V star, $Q_1/Q_0$ is reduced to 0.015. As the radiation travels through the stellar atmosphere, metals will selectively absorb more energetic photons in a process known as line blanketing. While dependent on the metallicity of the star, this process will generally result in a decrease in the ratio of He-to-H ionizing photons \citep[e.g.,][]{Pankonin1980,Afflerbach1997}. The He-to-H ionizing photon ratio is also affected by dust, which causes selective attenuation of ultraviolet (UV) photons.

As radiation propagates through an \hii\ region, its spectrum changes further due to absorption and re-emission processes. \citet{Wood2004} suggested that these interaction processes preferentially result in a hardening of the H-ionizing continuum and a suppression of He-ionizing photons. Recently, \citet{Weber2018} showed that for O-stars with low effective temperatures ($< 35000$\,K) nearly the entire He-ionizing radiation is absorbed within the \hii\ regions. The radiation hardening has been demonstrated by \citet{Osterbrock1989} based on the dependence of the absorption cross-section on frequency: ionizing photons with energies E$\gtrsim$13.6\,eV are preferentially absorbed by hydrogen compared to photons with much higher energies. Since the ionization cross-section of He is much greater than that of H, a large fraction of He-ionizing photons is absorbed well within the ionization front of the \hii region, resulting in a depletion of He-ionizing photons outside the PDR. 

\begin{deluxetable*}{lcccccc}
\tabletypesize{\footnotesize}
\tablecaption{\hii\ Region Properties}
\tablehead{Source  & \colhead{$\ell$} & \colhead{$b$} & \colhead{Radius} & \colhead{$R_{\rm gal}$} & \colhead{Distance} & \colhead{Spectral Type}\\
  & \colhead{(deg.)} & \colhead{(deg.)} & \colhead{(pc)} & \colhead{(kpc)} & \colhead{(kpc)} & }
\startdata
  M17 (S45)                 & \phn 15.098 & \phn     $-0.729$ &     11.3 & \phn 6.6 & 2.0 & O4-O4$^{[1]}$\\
  M16 (S49)                 & \phn 16.993 & \phn \phs $0.874$ &     14.2 & \phn 6.1 & 2.6 & O5$^{[2]}$\\
  N49                 & \phn 28.823 & \phn     $-0.226$ & \phn 3.6 & \phn 4.5 & 5.5 & O5$^{[3]}$\\
  G45.45+0.06 (G45)   & \phn 45.453 & \phn \phs $0.055$ & \phn 9.9 & \phn 6.5 & 8.4 & O6$^{[4]}$\\
  S104	              & \phn 74.769 & \phn \phs $0.622$ & \phn 2.4 & \phn 8.2 & 1.5 & O5$^{[5]}$\\
  S206                &     150.596 & \phn     $-0.955$ & \phn 7.9 &     11.6 & 3.4 & O6$^{[6]}$\\
  Orion (S281)                &     209.107 &         $-19.509$ & \phn 3.4 & \phn 8.9 & 0.4 & O6$^{[7]}$\\
  G29.96$-$0.02 (G29) & \phn 29.956 & \phn     $-0.020$ & \phn 2.4 & \phn 4.7 & 5.3 & O5.5$^{[8]}$\\  
\enddata
\tablenotetext{\,}{\,\textbf{References} --- [1] \citet{Broos2007}, [2] \citet{Sota2011}, [3] \citet{Watson2008}, [4] \citet{Moises2011}, [5] \citet{Lahulla1985}, [6] \citet{Georgelin1973}, [7] \citet{ODell2017}, [8] \citet{Cesaroni1994}.}
\label{tab:hii}
\end{deluxetable*}

The suppression of He-ionizing photons implies a reduced \heh ionic abundance ratio by number outside the \hii\ region as there will be a fewer number of photons with sufficient energy to ionize He compared to H. Such \hii\ regions may be density bounded beyond the He$^+$ zone, but within the H$^+$ zone \citep[see][]{Reynolds1995a}. This effect was observed by \citet{Pankonin1980}, who show that the ionized helium abundance in the Orion Nebula decreases with distance from the exciting star. We also indirectly confirmed these results in L16 for the compact \hii\ region NCG\,7538 by observing a decrease in the \heh ratio with increasing distance from the region's central position. It is, however, unknown whether these findings are applicable to Galactic \hii\ regions in general and their relation to age or geometry of the region.

Previous observational work and simulations confirm that the ratio of He-to-H ionizing photons in the WIM is lower than that found in \hii\ regions. In a study of optical emission lines toward faint H$\alpha$-emitting regions in the Milky Way, \citet{Madsen2006} show that the He\,{\footnotesize I}/H$\alpha$ line ratio is suppressed compared to that of \hii\ regions, indicating a softer radiation field. Using Monte Carlo photoionization simulations, \citet{Wood2004} suggest that this line ratio depends strongly on the \hii\ region leaking fraction. They find that He\,{\footnotesize I}/H$\alpha$ is significantly reduced only for low escape fractions ($\sim 15$\%). This result is in disagreement with \citet{Roshi2012}, who observed a \heh\ upper limit of only 0.024 in the diffuse gas near the \hii\ region G49, despite an apparent escape fraction of $\sim 63$\% and \heh\ ratios of $>0.066$ within the \hii\ region \citep{Churchwell1974,Lichten1979,Thum1980,McGee1981,Mehringer1994,Bell2011}. Clearly, a larger sample size is required for further study.

The goal of this study is to observe a variety of \hii\ regions to determine the role of leaking radiation from \hii\ regions in maintaining the ionization of the WIM. Our observed sample includes \hii\ regions of different sizes and morphologies for which the PDR boundary can be identified and for which the spectral class of its central star(s) is known. The properties of our observed \hii\ regions are summarized in Table~\ref{tab:hii}, which lists the source name, the Galactic longitude and latitude, the radius of the \hii\ region, its Galactocentric radius, the distance to the Sun, and the spectral type of the ionizing source(s). 

The GBT RRL observations are described in \S \ref{sec:obs} of this paper and we outline the process of defining PDR boundaries in \S \ref{sec:pdr}. In \S \ref{sec:h} and \S \ref{sec:yplus} we analyze the hydrogen RRL emission around the observed \hii\ regions and derive \heh\ ionic abundance ratios for the observed directions, respectively. Physical properties of the ionized gas, including electron temperatures, emission measures, and electron densities are derived in \S \ref{sec:properties}. In \S \ref{sec:lineprofile} a line profile analysis is performed as an alternative method to derive electron temperatures. We analyze carbon and doubly-ionized helium emission in \S \ref{sec:c} and \S \ref{sec:hei}, respectively, and conclude in \S \ref{sec:conclusions}.

\section{Observations and Data Reduction}\label{sec:obs}

We observed 8 \hii\ regions with the Green Bank Observatory GBT from 2017 February to 2017 May. The properties of the observed \hii\ regions are given in Table~\ref{tab:hii}, which lists the source, the Galactic longitude and latitude, the radius of the \hii\ regions as defined in \S \ref{sec:pdr}, the Galactocentric radius and heliocentric distance given by the WISE Catalog of Galactic \hii\ Regions \citep{Anderson2014}, and the spectral type of the ionizing source(s). For each targeted \hii\ region, several positions within and outside the region's PDR were observed (see \S \ref{sec:pdr} for a description of how the PDR boundaries were determined). We employed total-power position-switching observations with variable off-source and on-source integration times, ranging from 3 to 36 minutes, depending on the expected brightness of the position. For integration times exceeding 6 minutes, the observation was split up into blocks of 6 minutes each to reduce the overall impact of radio-frequency interference (RFI).

The individual pointings for each observed \hii\ region lie along an imaginary line on the plane of the sky intersecting the region center. The angle of the line was chosen such that there is as little confusion by other radio sources (e.g.,~nearby \hii\ regions) as possible. For each targeted \hii\ region, between 7 and 17 positions were observed. The goal was to include as many positions as possible within the regions' PDR boundaries, separated by as little as the average GBT half-power beam width (HPBW) of 123\arcsec. Outside the PDR boundaries, our pointings are spaced apart by one to four GBT HPBWs, depending on the spatial extent of the source and our total available observing time. For the largest regions, we sample up to a maximum distance of 47\arcmin\ from the center. The locations of the observed positions for each source are shown in Figure~\ref{fig:pdr}. The locations are labeled according to their predominant direction from the region center in Galactic coordinates (N\,...\,north, S\,...\,south, E\,...\,east, W\,...\,west) and their distance from the center of the regions in multiples of 2$\,\times\,$HPBW. The coordinates of all observed directions are given in Table~\ref{tab:overview}, which lists the source, the source coordinates, and the location of the observed direction with respect to the PDR boundary (see \S \ref{sec:pdr}).

The GBT C--band instantaneous bandpass in combination with the VEGAS backend includes 22 Hn$\alpha$ lines, 19 Hn$\beta$ lines, and 8 Hn$\gamma$ lines between 4.054 and 7.793\,GHz. In our setup, the $\alpha$ lines range from $n = 95$ to $n = 117$, the $\beta$ lines range from $n = 120$ to $n = 146$, and the $\gamma$ lines range from $n = 138$ to $n = 147$ \citep[see][]{Luisi2018}. In addition, we tuned to 7 HeIn$\alpha$ lines and 8 molecular lines, including formaldehyde and methanol. To achieve a higher signal-to-noise ratio, we average together all Hn$\alpha$, Hn$\beta$, and Hn$\gamma$ lines, respectively, using TMBIDL\footnote{V7.1, see https://github.com/tvwenger/tmbidl.git.} after first re-gridding and shifting the spectra so that they are aligned in velocity \citep[see][]{Balser2006}. Spectra affected by RFI were discarded before averaging. The averaged spectra were smoothed to a resolution of 1.86\,\kms and a fourth-order polynomial baseline was subtracted.

Gaussian models were fit to the H and He profiles for which the signal-to-noise ratio (S/N) is at least 5 as defined by the method given by \citet{Lenz1992},
\begin{equation}
{\rm S/N} = 0.7 \left( \frac{T_{\rm L}}{\rm rms} \right) \left( \frac{\Delta V}{\Delta V_{\rm k}} \right) ^{0.5},
\end{equation}
where $T_{\rm L}$ is the peak line intensity, rms is the root-mean-squared spectral noise, $\Delta V$ is the full width at half maximum (FWHM) of the line, and $\Delta V_{\rm k} = 1.86$\kms is the FWHM of the Gaussian smoothing kernel. We also fit carbon lines with a S/N of at least 3. Here, the narrow width of the carbon lines allows for a lower S/N threshold as they are less likely to be confused with baseline fluctuations. In the few cases where more than one hydrogen line velocity component is detected, we assume that the brightest component is due to the \hii\ region. 

The continuum antenna temperature, $T_{\rm C}$, was derived for all observed positions from the continuum background of our averaged spectral line data by removing all lines above the 3$\sigma$ level. We also remove 20\% of all channels towards each end of the bandpass, since the bandpass edges are prone to instabilities, and average the antenna temperature over the remaining baseline to find $T_{\rm C}$. Since the baseline stability is the dominant uncertainty contribution, the error in $T_{\rm C}$ is estimated by averaging over individual segments of each baseline, 1000 channels in width, and by computing the standard deviation between these segments. All continuum antenna temperatures are given in Table~\ref{tab:temperatures}, which lists the source, $T_{\rm C}$, the LTE electron temperatures ($T_{\rm e}^*$; see \S \ref{sec:te}), and the emission measures (EM; see \S \ref{sec:ne}), including all 1$\sigma$ uncertainties.

The averaged Hn$\alpha$ spectra are shown in Figure~\ref{fig:spectra} and the derived Hn$\alpha$, Hn$\beta$, and Hn$\gamma$ RRL parameters are given in Table~\ref{tab:alpha} which lists the source, the transition, the element, the line intensity, the FWHM line width, the LSR velocity, the rms noise in the spectrum, and the total on-source integration time for each position, including the corresponding 1$\sigma$ uncertainties of the Gaussian fits. The averaged Hn$\beta$ and Hn$\gamma$ spectra are shown in Figures~\ref{fig:spectrabeta} and \ref{fig:spectragamma}, respectively.

\section{The PDR Boundaries}\label{sec:pdr}
Defining the PDR boundaries of the observed \hii\ regions is a crucial step in determining the region properties. Since PAHs are abundant in PDRs where they emit strongly in the 8\,$\mu$m, 12\,$\mu$m, and 24\,$\mu$m bands \citep[e.g.][]{Hollenbach1997}, enhancements in the 12\,$\mu$m \emph{WISE} emission can be used to trace the PDR itself. We define the PDR boundary for all observed regions by following the enhanced 12\,$\mu$m emission surrounding the \hii\ region by hand. We estimate the inner and outer PDR boundary for each region such that the width of the PDR is the FWHM of the enhanced emission. Although this characterization of the PDR structure is by no means unique, results from L16 show that the above method appears to reliably trace the PDR around compact \hii\ regions.

\begin{figure*}[htp]
\centering
\begin{tabular}{cc}
\includegraphics[width=.49\textwidth]{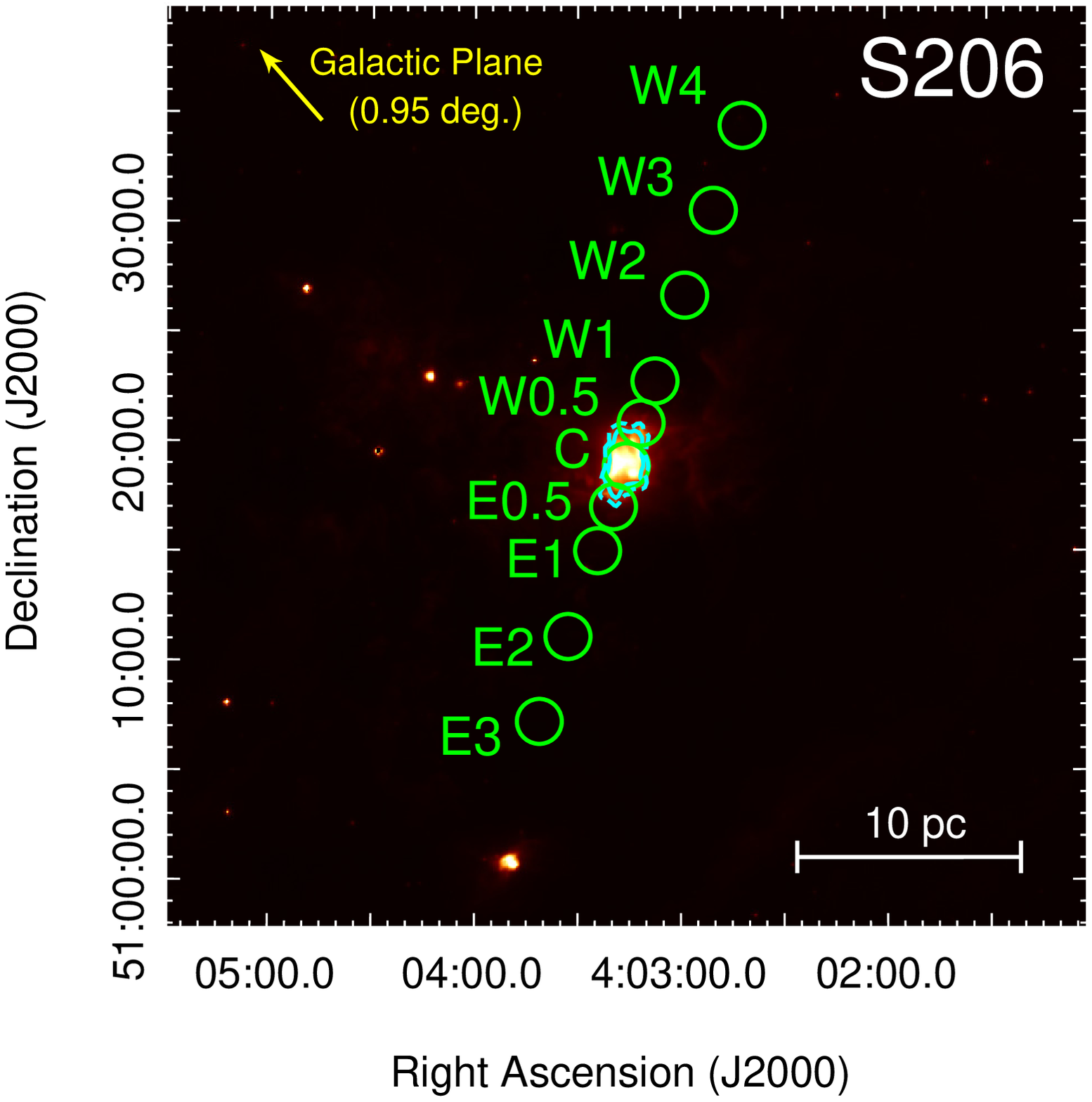} &
\includegraphics[width=.49\textwidth]{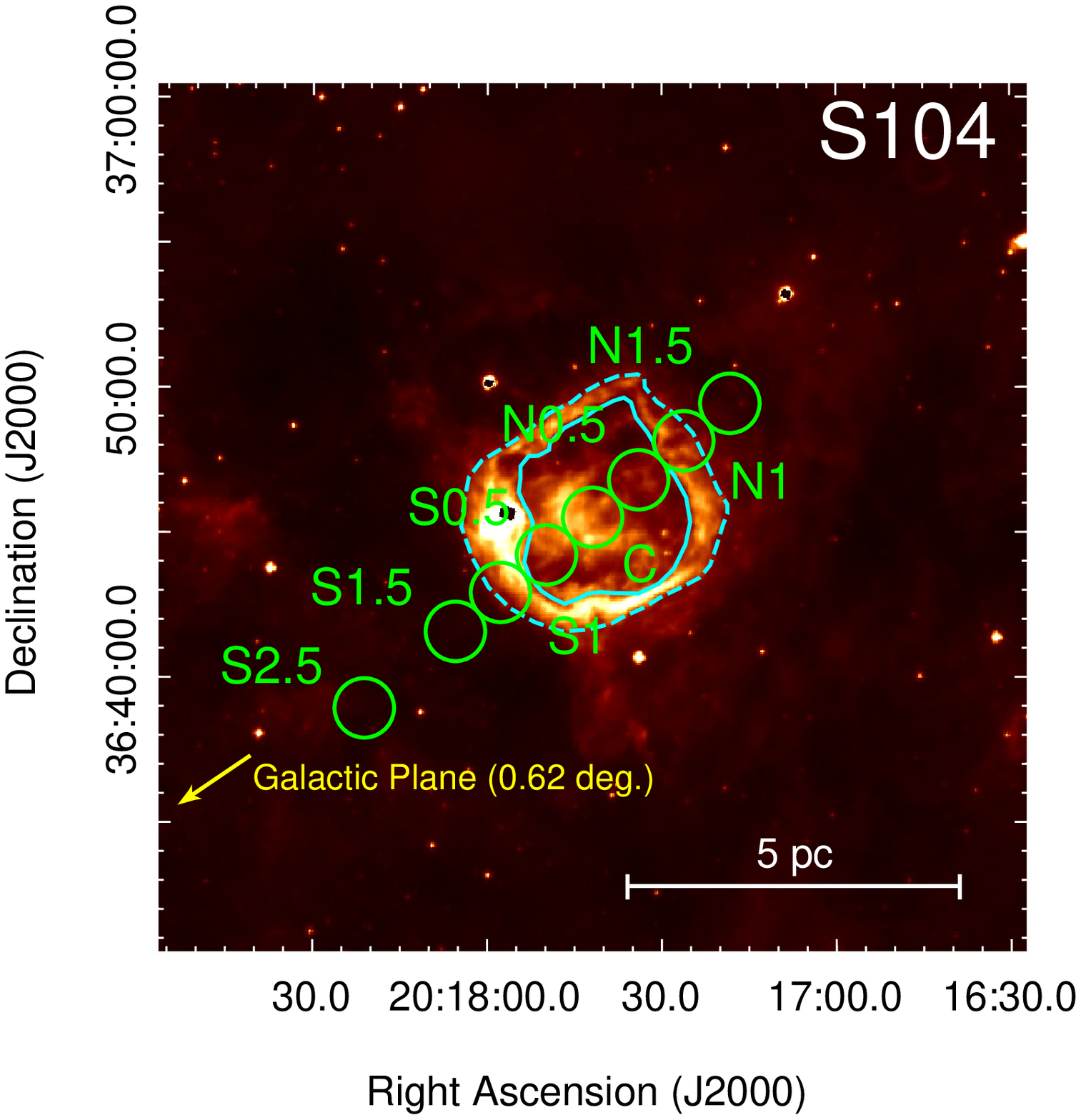} \\
\includegraphics[width=.49\textwidth]{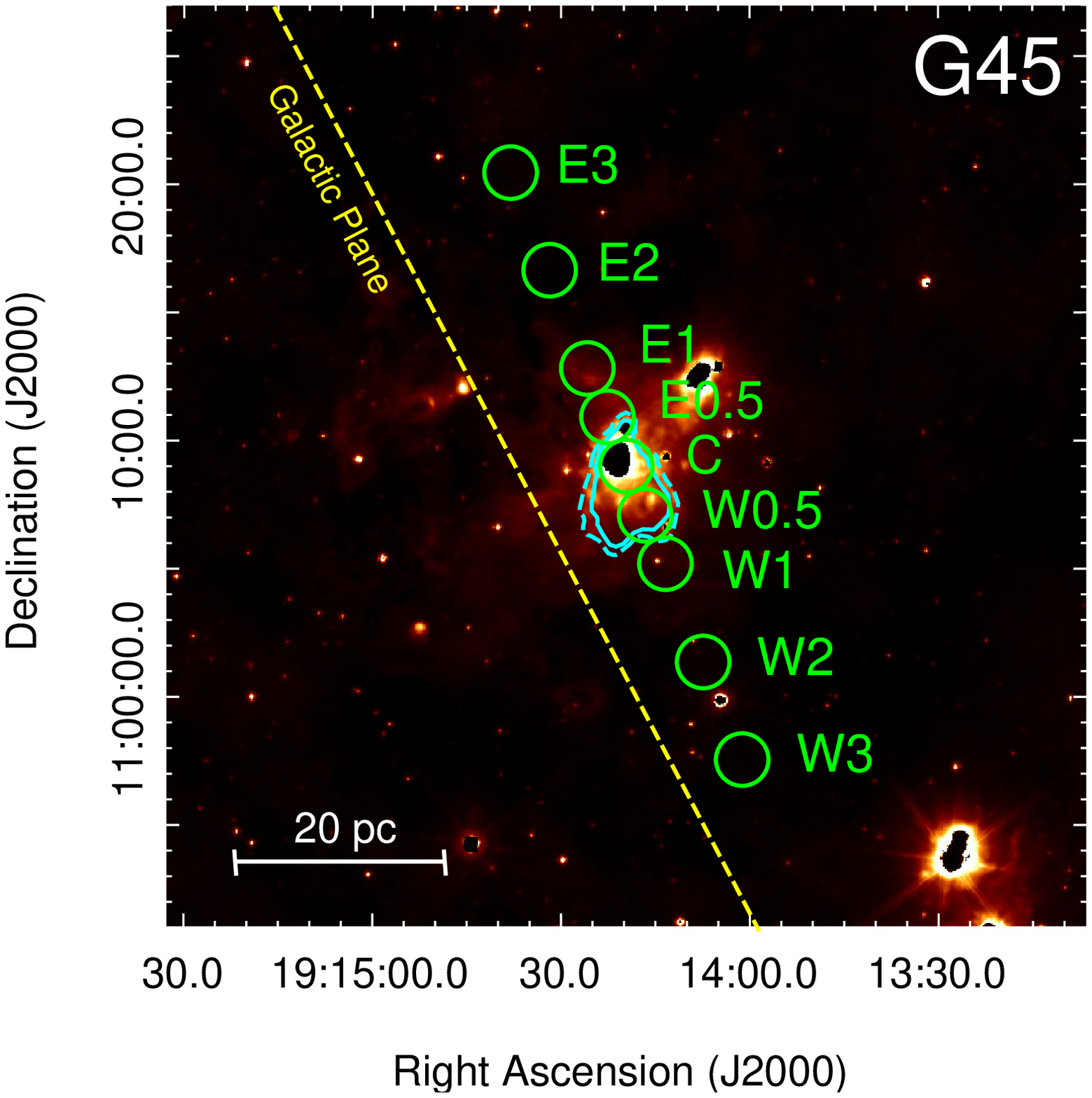} &
\includegraphics[width=.49\textwidth]{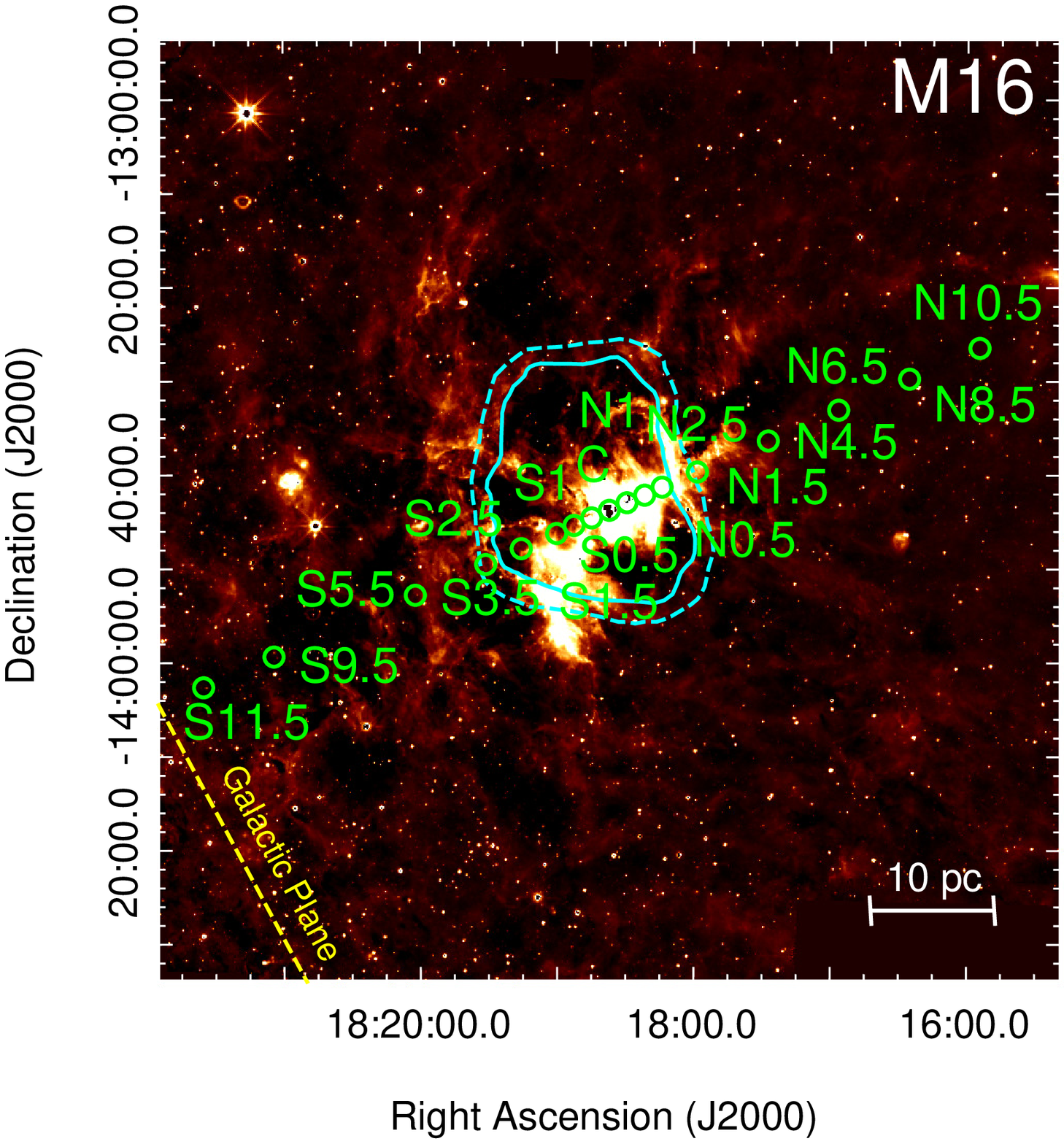} \\
\end{tabular}
\caption{The observed \hii\ regions in 12\,$\mu$m \emph{WISE} emission. This page: S206 (top left), S104 (top right), G45 (bottom left), and M16 (bottom right). The green circles show the positions that we observed with the GBT; the size of the circles corresponds to the average GBT beam width at the observed frequencies. The solid and dashed light blue regions indicate the inner and outer PDR boundaries for each \hii\ region as defined in \S \ref{sec:pdr} and the scale bars are derived from the distances given in Table~\ref{tab:hii}. The yellow arrow or yellow dashed line indicates the location of the Galactic plane. The angle next to the arrow gives the distance between the central position of the \hii\ region and the Galactic plane. \label{fig:pdr}}
\end{figure*}
\renewcommand{\thefigure}{\arabic{figure}}

\renewcommand{\thefigure}{\arabic{figure} (cont.)}
\addtocounter{figure}{-1}
\begin{figure*}[htp]
\centering
\begin{tabular}{cc}
\includegraphics[width=.49\textwidth]{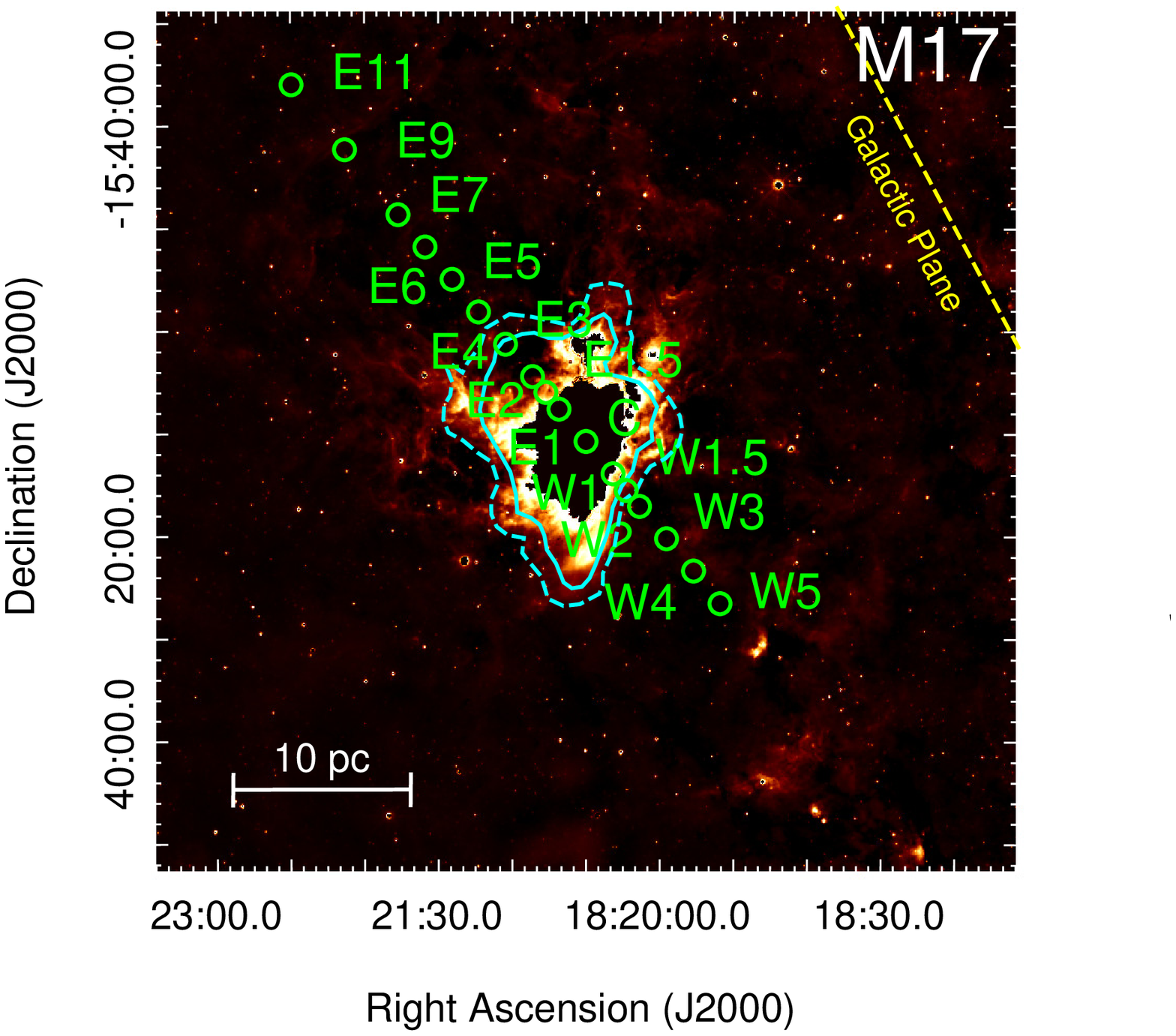} &
\includegraphics[width=.49\textwidth]{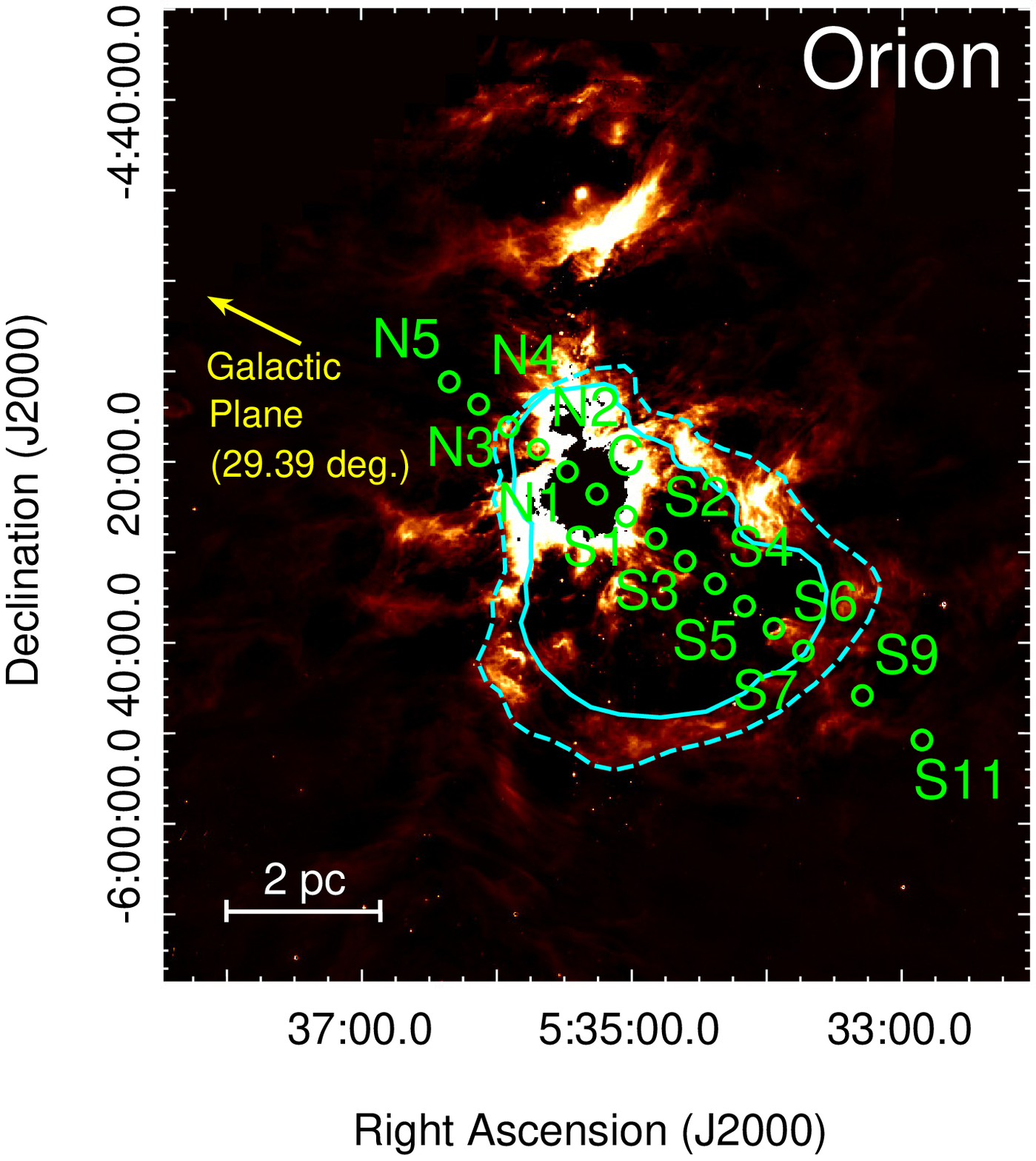} \\
\includegraphics[width=.49\textwidth]{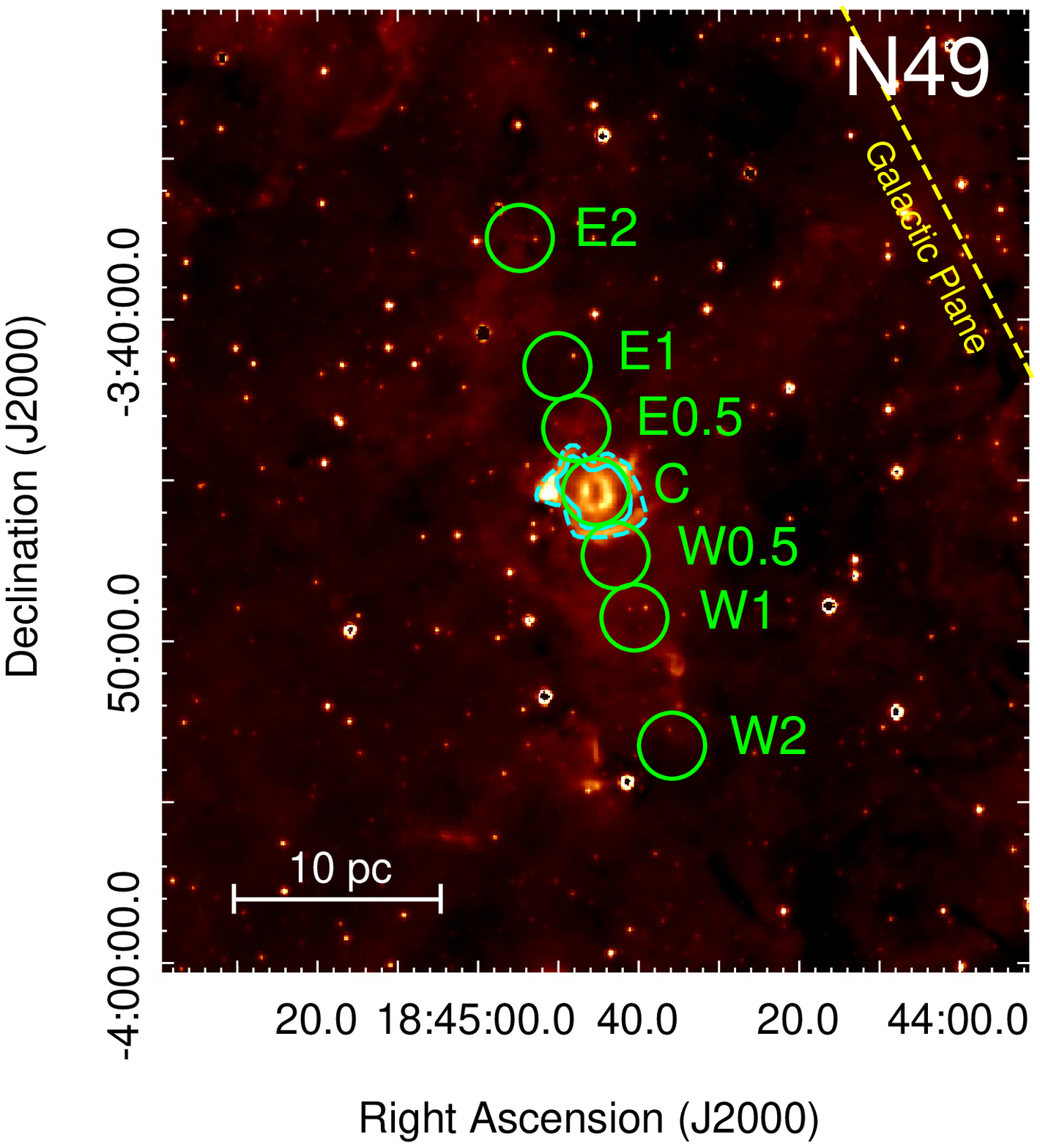} &
\includegraphics[width=.49\textwidth]{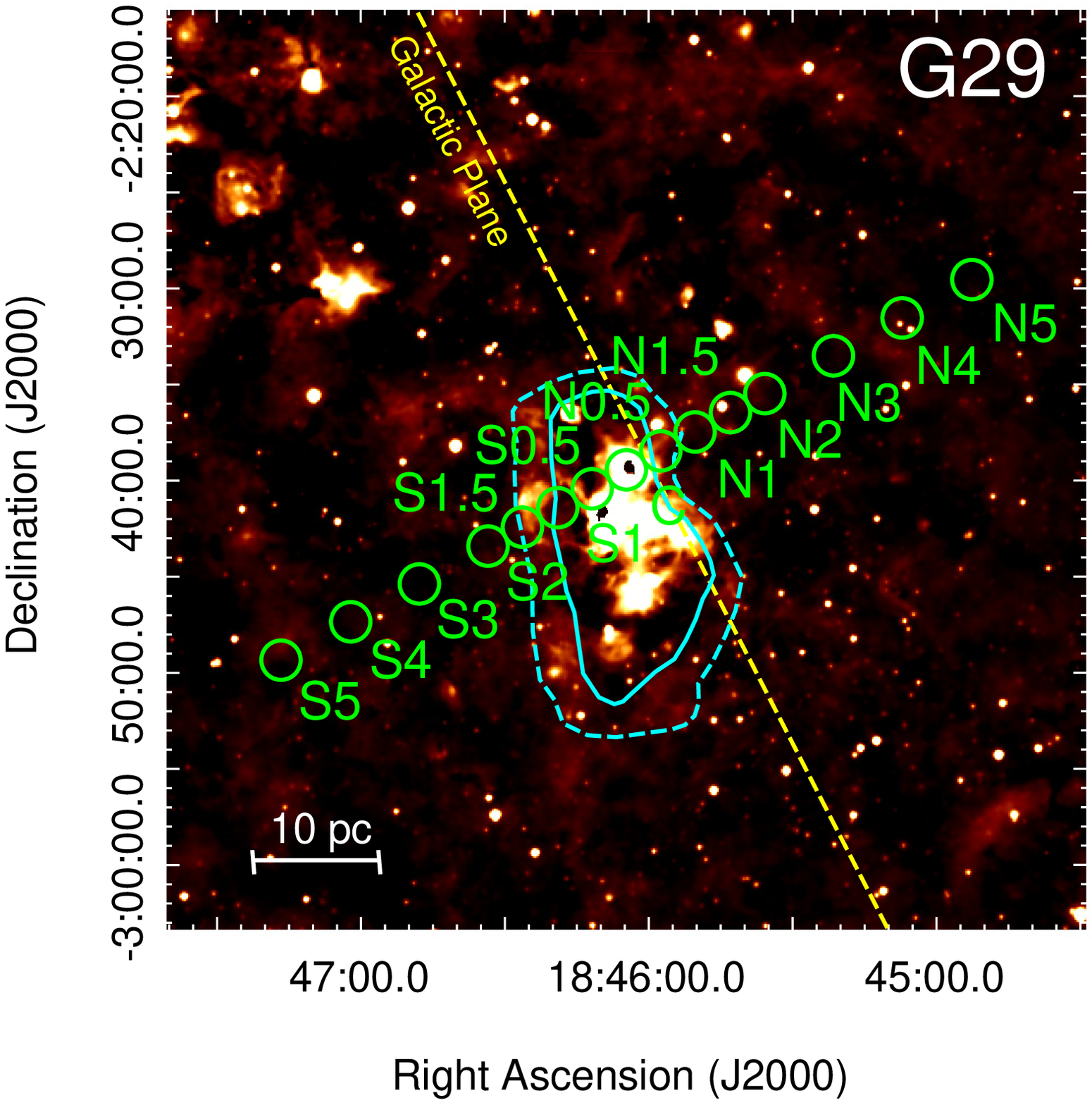} \\
\end{tabular}
\caption{The observed \hii\ regions in 12\,$\mu$m \emph{WISE} emission. This page: M17 (top left), Orion (top right), N49 (bottom left), and G29 (bottom right).}
\end{figure*}
\renewcommand{\thefigure}{\arabic{figure}}

\begin{figure*}[htp]
\centering
\begin{tabular}{cc}
\includegraphics[width=.5\textwidth]{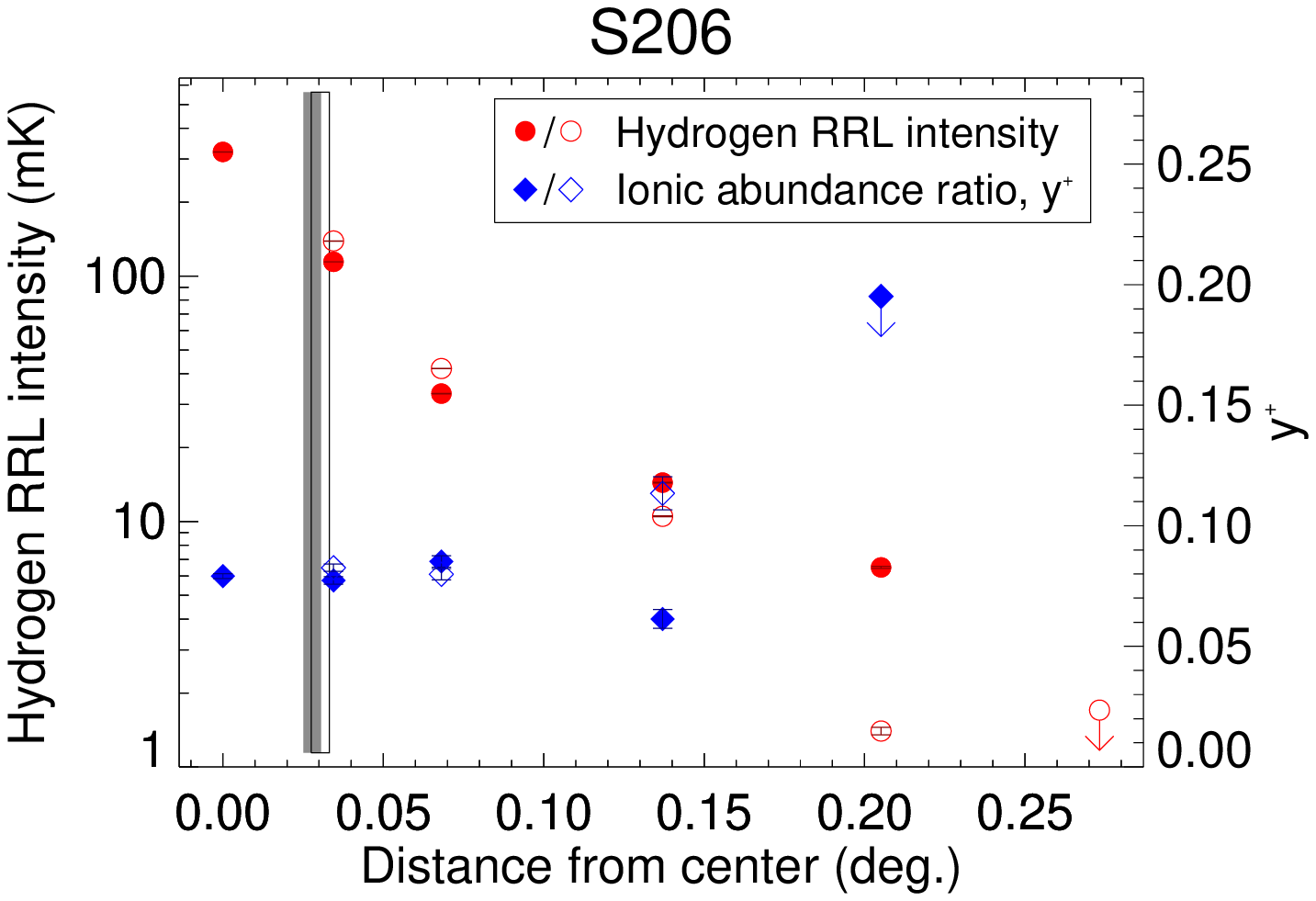} &
\includegraphics[width=.5\textwidth]{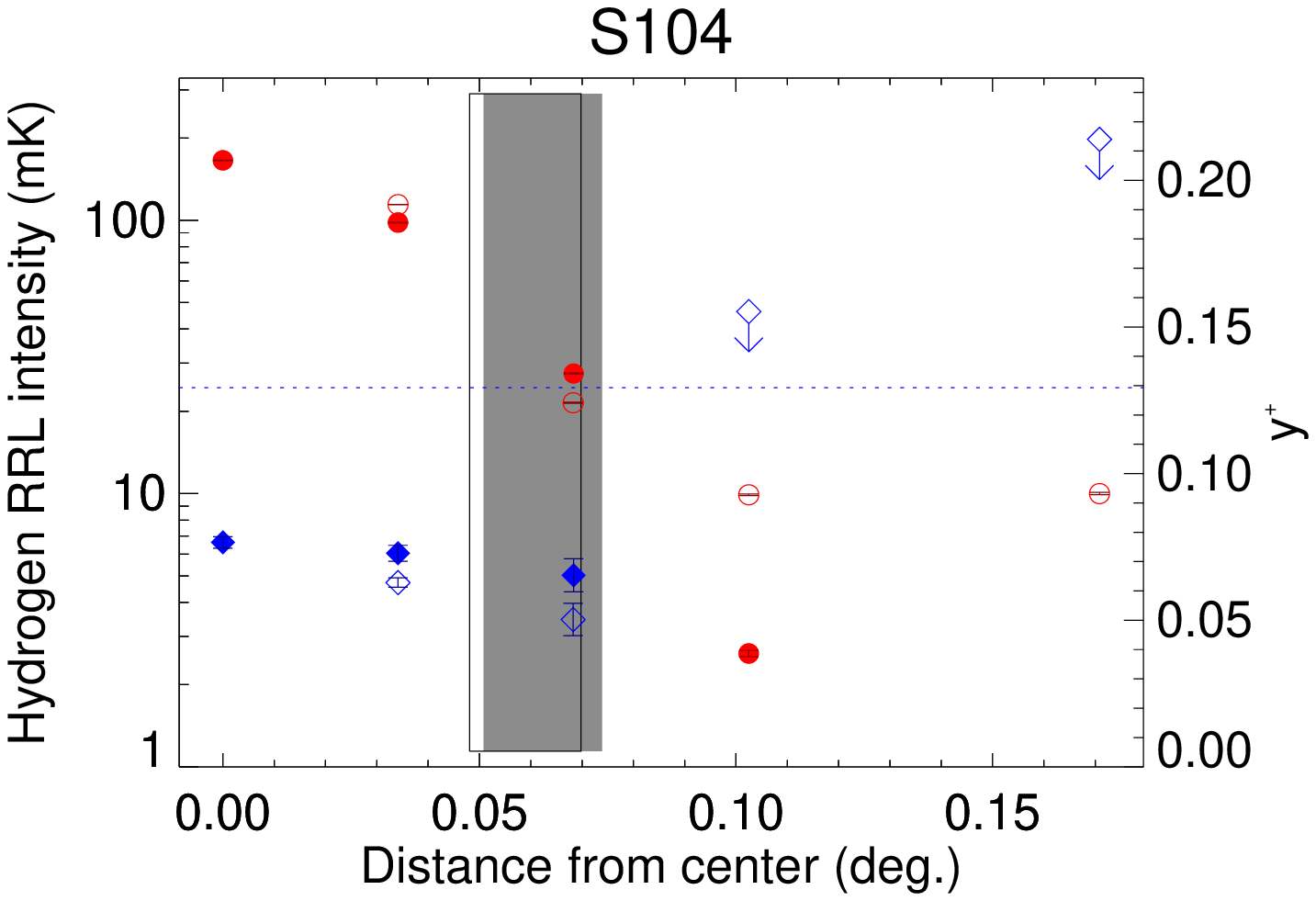} \\
\includegraphics[width=.5\textwidth]{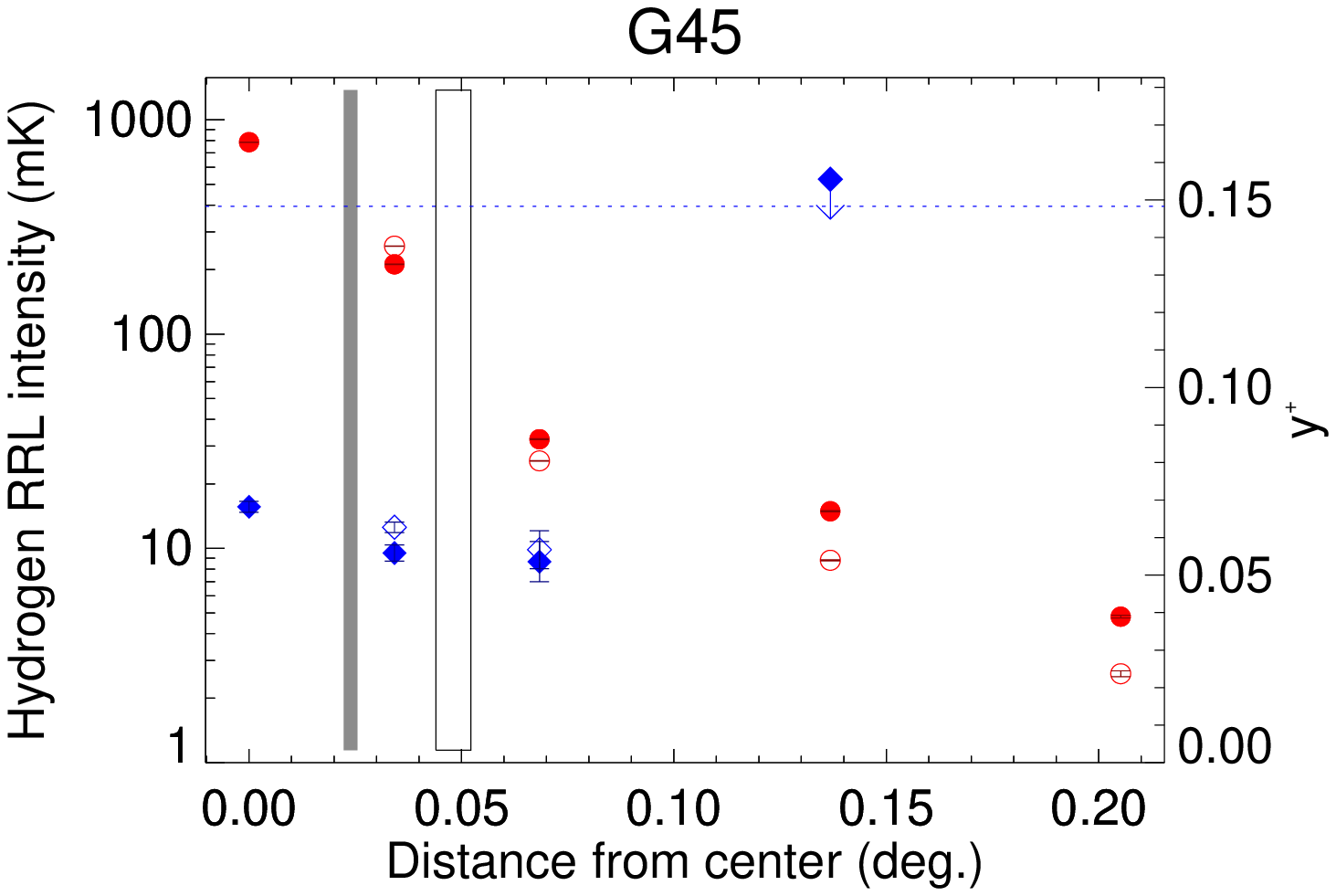} &
\includegraphics[width=.5\textwidth]{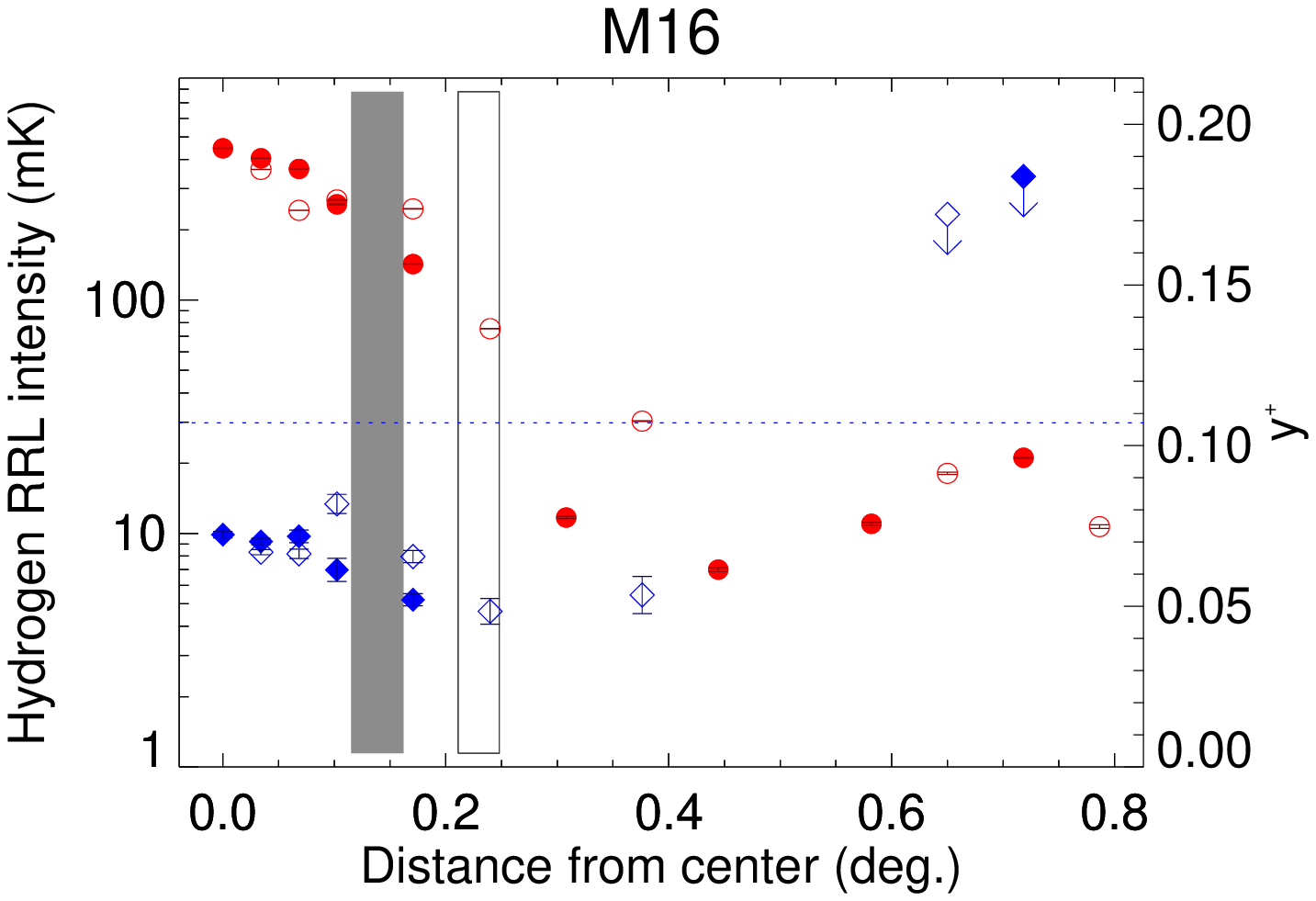} \\
\end{tabular}
\caption{The hydrogen RRL intensity (red circles) and ionic abundance ratio, $y^+$ (blue diamonds), as a function of distance from the region center for all observed \hii\ regions. On this page are shown: S206 (top left), S104 (top right), G45 (bottom left), and M16 (bottom right). The shaded and unshaded symbols indicate pointings towards the two different observed directions from the center (shaded: north/east, unshaded: south/west). Upper limits are denoted as symbols with downward arrows. The large unshaded and shaded regions in the background show the extent of the PDR boundary towards the two directions. The combined upper limit of $y^+$ for all helium non-detections is marked as the dotted blue horizontal line. \label{fig:heh}}
\end{figure*}
\renewcommand{\thefigure}{\arabic{figure}}

\renewcommand{\thefigure}{\arabic{figure} (cont.)}
\addtocounter{figure}{-1}
\begin{figure*}[htp]
\centering
\begin{tabular}{cc}
\includegraphics[width=.5\textwidth]{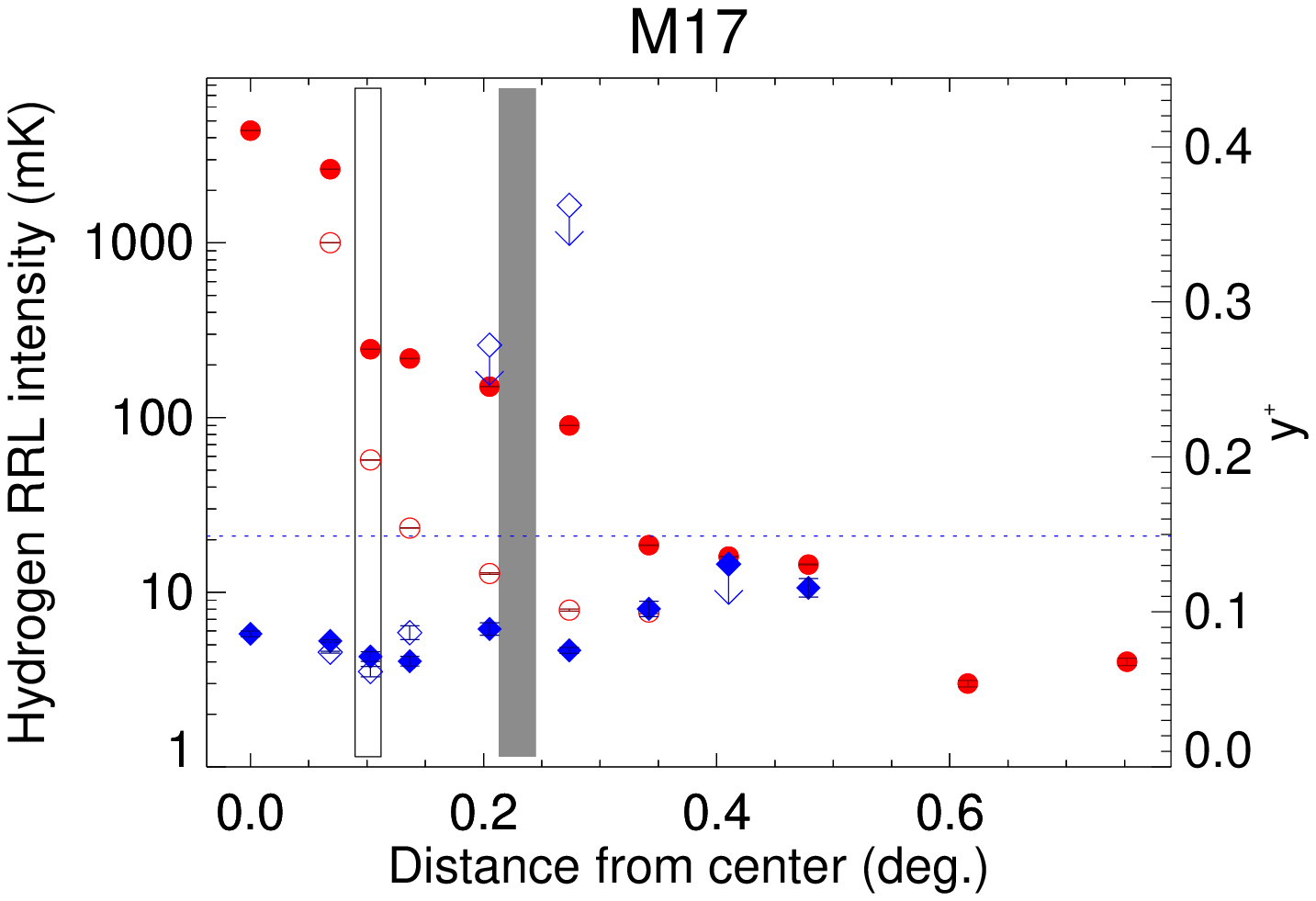} &
\includegraphics[width=.5\textwidth]{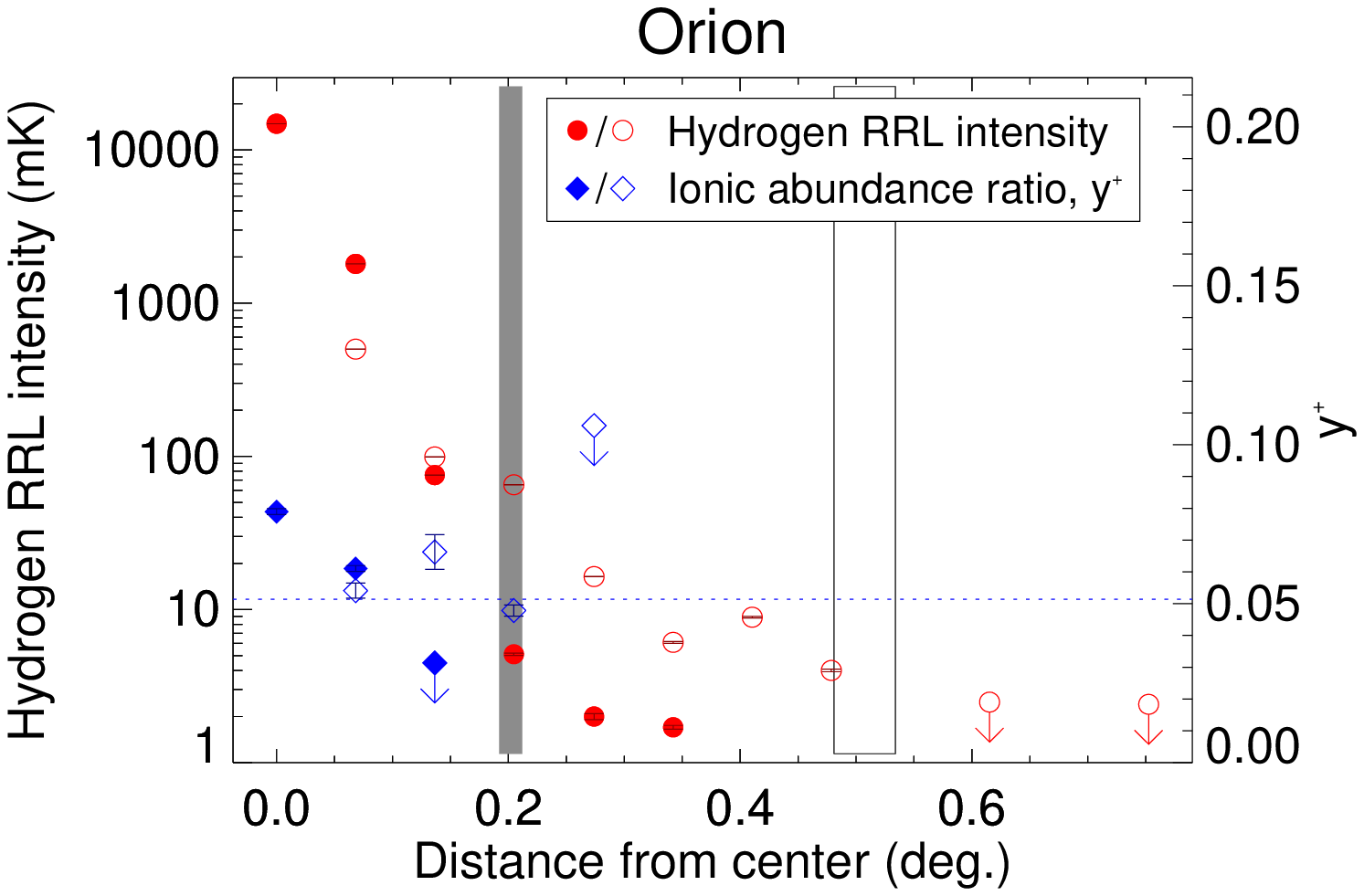} \\
\includegraphics[width=.5\textwidth]{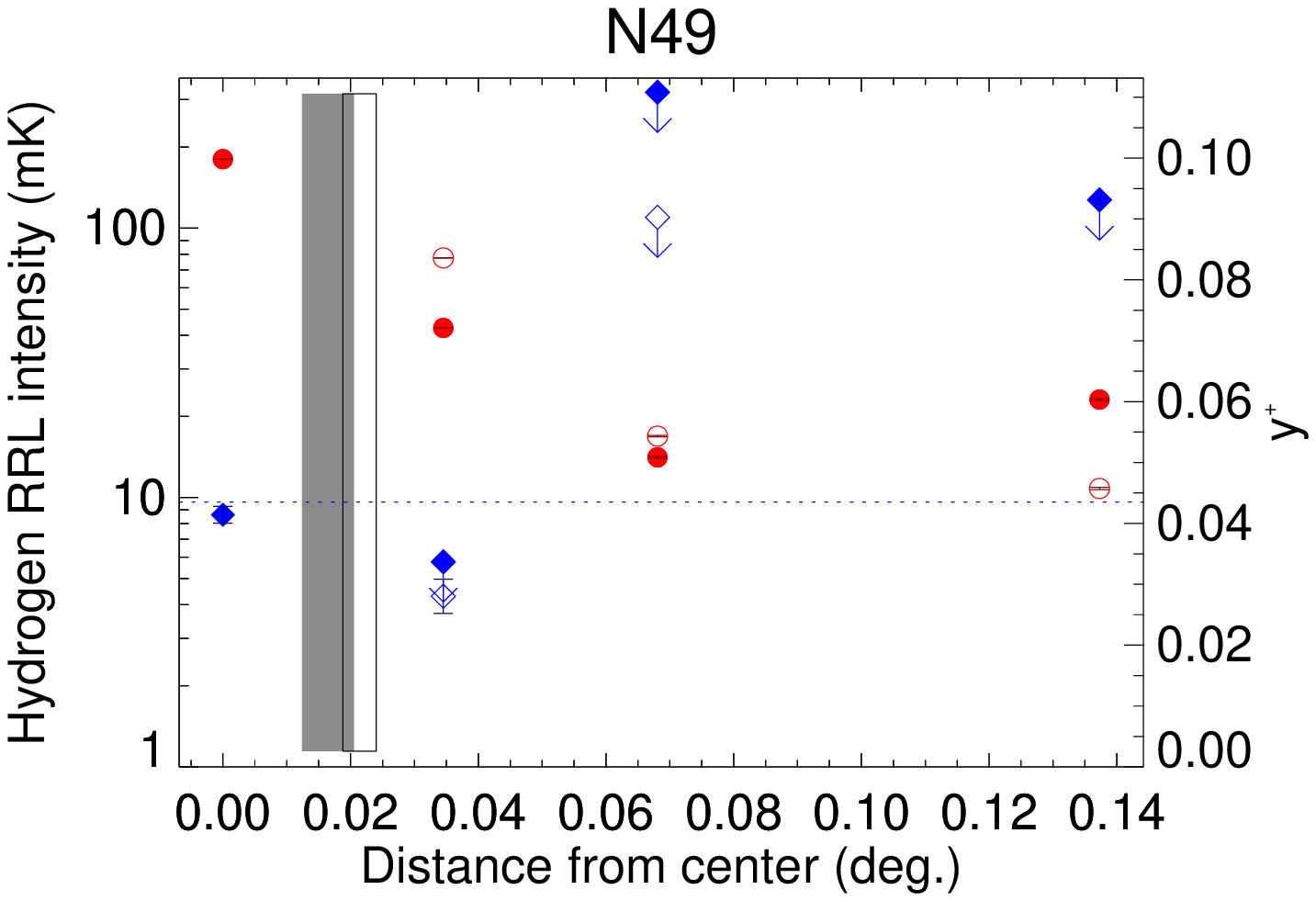} &
\includegraphics[width=.5\textwidth]{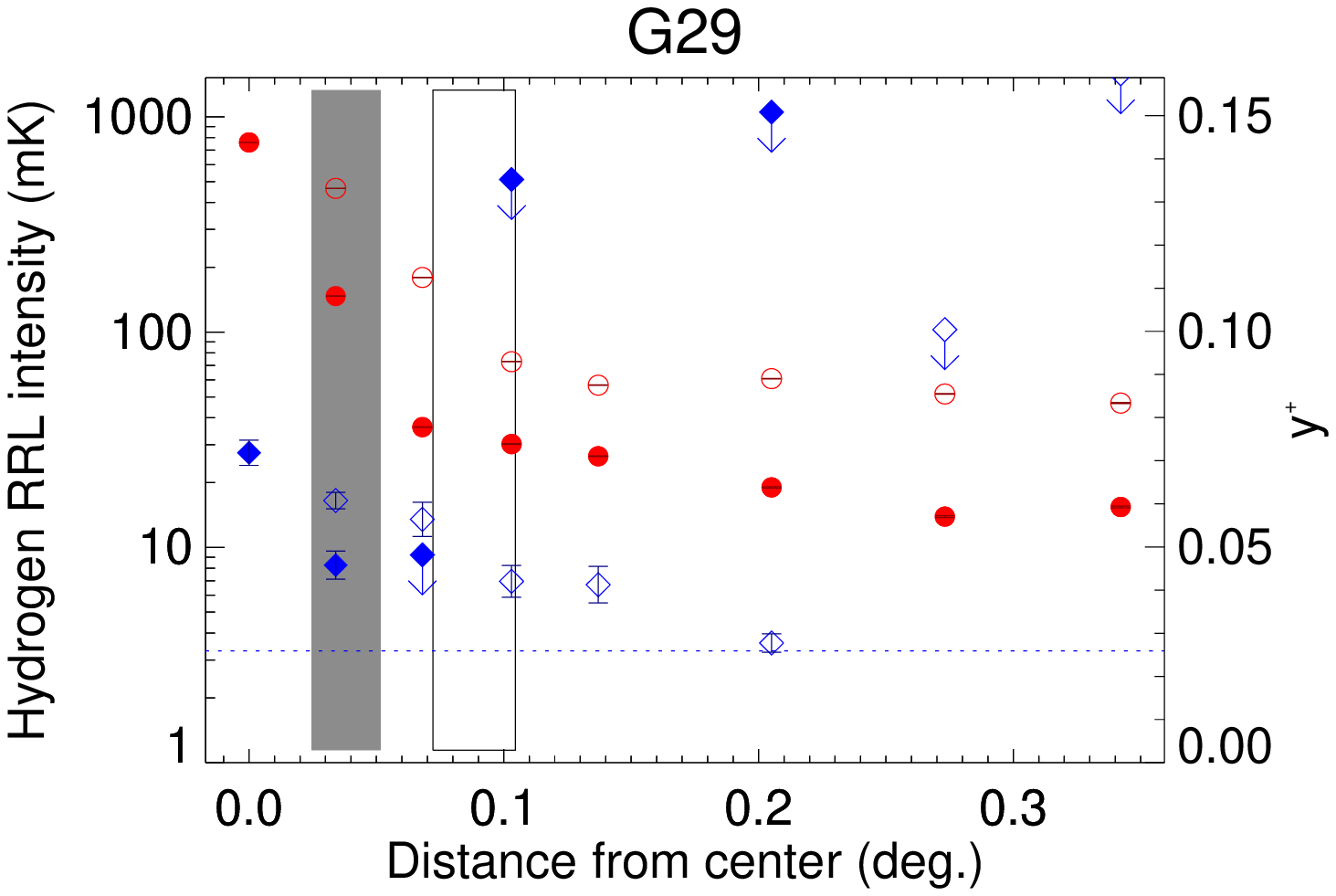} \\
\end{tabular}
\caption{The hydrogen RRL intensity (red circles) and ionic abundance ratio, $y^+$ (blue diamonds), as a function of distance from the region center for all observed \hii\ regions. On this page are shown: M17 (top left), Orion (top right), N49 (bottom left), and G29 (bottom right).}
\end{figure*}
\renewcommand{\thefigure}{\arabic{figure}}

The PDR boundaries are determined using the above method for each of the 8 observed \hii\ regions and the results are shown in Figure~\ref{fig:pdr}. The term ``strong" refers to a PDR boundary for which the enhanced 12\,$\mu$m emission shows the largest contrast to the surrounding medium, whereas a ``weak" PDR is barely distinguishable from the 12\,$\mu$m background. Several PDR boundaries are asymmetrical or incomplete in the associated 12\,$\mu$m emission. For example, in the case of G45, the western PDR boundary is much weaker and further away from the region center than the eastern boundary. For M16, the PDR shows discontinuities to the east and west, but is strong toward the north (the sampled directions). Similarly, the PDR boundary of Orion appears to be strongest toward the east and west and less so towards the south where it is at the largest distance from the center. Defining the PDR boundary of G29 is perhaps the most challenging of all observed regions due to its very weak 12\,$\mu$m emission, many discontinuities, and proximity to several nearby continuum sources.

For M16 and Orion, several PDR components are visible in a given direction. In these cases, we define the strongest PDR boundary as the ``main" PDR. For a number of regions, the sharp decrease in 12\,$\mu$m intensity makes it difficult to unambiguously locate the PDR boundary toward some directions.

\section{Hydrogen RRL Emission}\label{sec:h}
\begin{figure}
\centering
\includegraphics[width=0.49\textwidth]{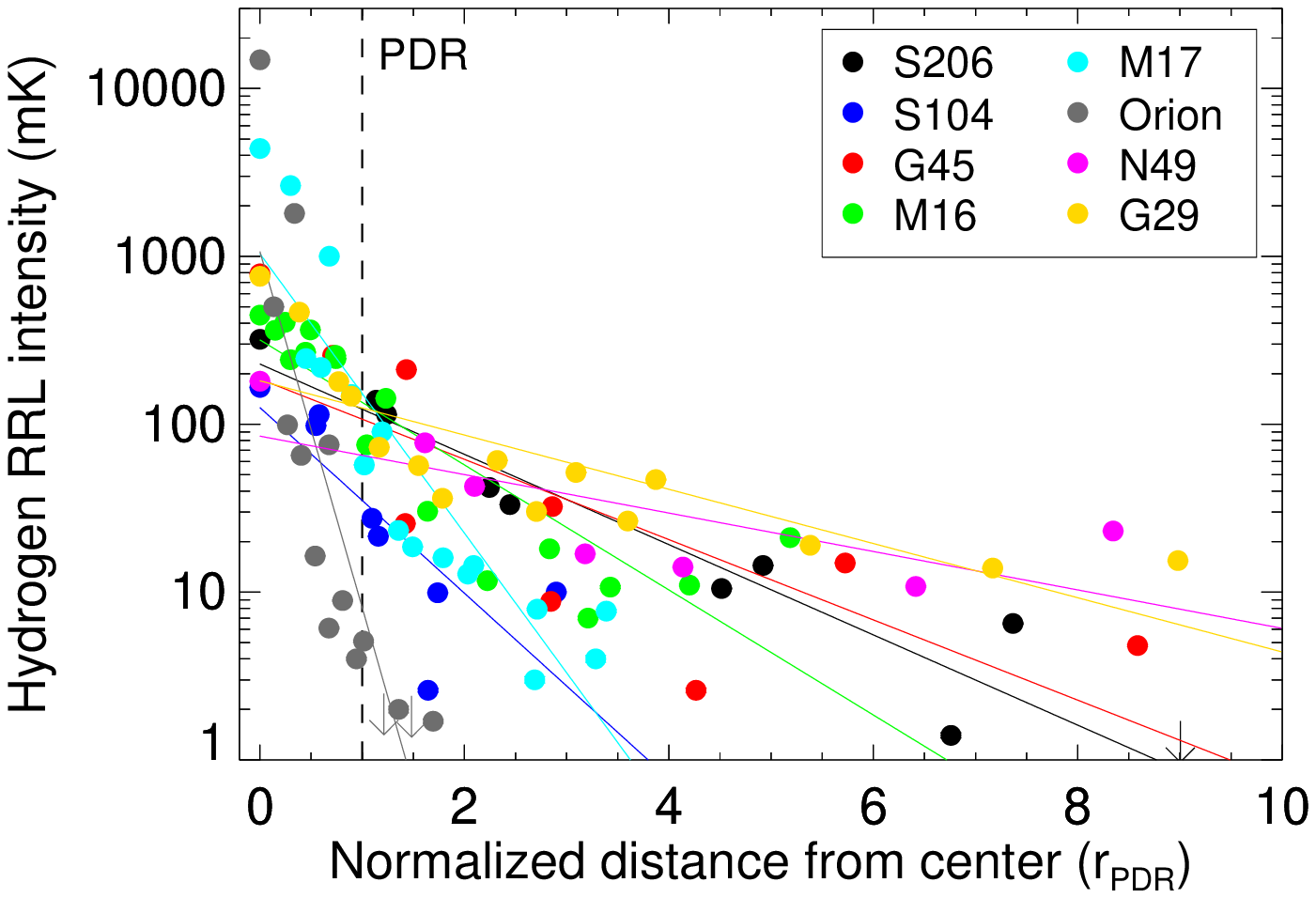}
\caption{The hydrogen RRL intensity for all observed positions as a function of angular offset from the \hii\ regions. The distance is normalized by the radius of the PDR boundary along the observed direction. Downward arrows show upper limits. The solid lines are exponential fits to the data and the vertical dashed line indicates the PDR boundary. This figure shows that, with the exception of Orion and perhaps M17, the hydrogen line intensity decreases with normalized distance at roughly the same rate for all observed \hii\ regions. \label{fig:h_all}}
\end{figure}

Using our GBT data, we test the hypothesis that high luminosity \hii\ regions have a greater effect on maintaining the ionization of the WIM compared to compact \hii\ regions, possibly because they allow a larger fraction of ionizing radiation to escape into the ISM. In Figure~\ref{fig:heh} the hydrogen RRL intensity is shown, averaged over the Hn$\alpha$ transitions given in \S \ref{sec:obs}, and the ionic abundance ratio (see \S \ref{sec:yplus}) as a function of distance from the region center for all observed \hii\ regions. As expected, the RRL intensity decreases with increasing distance for all regions. Regions with more extended PDR boundaries (e.g., M16, M17, and Orion) have more extended emission than the observed compact regions, possibly because they are in a later evolutionary stage. In Figure~\ref{fig:h_all} the hydrogen RRL emission is shown for all observed regions with distance, normalized by the radius of the PDR boundary along the given direction. It is striking that all observed \hii\ regions except for Orion and perhaps M17 exhibit roughly the same hydrogen RRL emission gradient. We also show exponential fits of the form $I = a \times exp(b\,r_{\rm PDR})$ (where $I$ is the hydrogen RRL intensity, $r_{\rm PDR}$ is the normalized distance from the region center, and $a$ and $b$ are the fit parameters) to the data to highlight the similarity between the gradients.

The inverse correlation between \hii\ region size and the slope of the hydrogen RRL intensity may indicate that larger regions allow more ionizing photons to escape through their PDRs which in turn maintain the ionization of the surrounding WIM. Our results suggest that the total amount of escaping ionizing radiation is fundamentally correlated with the radius of the PDR boundary of that region. It is possible that the PDR boundaries surrounding large and luminous \hii\ regions are generally weaker or more inhomogeneous than those surrounding more compact regions. In \S \ref{sec:pdr} we show that the PDRs around M16 and Orion are not as well-defined as the PDR boundaries around the more compact regions in our sample. For M16, this result could be related to the fact that large, high-luminosity \hii\ regions are more likely to be density-bounded rather than radiation-bounded \citep{Beckman1998}, however, the PDR boundaries may be less well-defined for regions with multiple sources of ionization.

For a number of sources (M16, G29, N49), the intensity does not continue to decrease with distance from the center but rather flattens out beyond a certain radius. We hypothesize that this emission far from the region center is not due to the \hii\ region itself but from the WIM. The observed hydrogen line intensities of $\sim$10--50\,mK are not uncommon for WIM emission \citep[see][]{Anderson2015a,Luisi2017}. It is noteworthy that all these regions are at relatively low Galactic latitudes where presumably emission from the WIM is the strongest \citep{Alves2012}.

Several regions have more than one hydrogen line component. The southern positions of G29 exhibit a second component approximately --40\kms offset from the velocity of the \hii\ region itself. There is also a second hydrogen line visible in N49 and in the southernmost positions of M16. Due to their low RRL intensities and since the existence of an additional \hii\ region along the line of sight is unlikely, these additional line components are probably due to emission from the WIM \citep[see][]{Anderson2015a}. The second hydrogen component in the central position of M17 is too strong and spatially constrained to be caused by the WIM. It may instead be due to expansion processes within the \hii\ region itself.

\section{Ionic Abundance Ratios}\label{sec:yplus}
It has been suggested that He-ionizing photons are suppressed as UV photons escape from \hii\ regions \citep{Hoopes2003,Wood2004}. The complex absorption and re-emission processes in the surrounding gas, however, have never been observed in detail and it is unclear whether this result is applicable to the Galactic \hii\ region population as a whole. The hardness of the interstellar radiation field for our region sample can be constrained by deriving the $y^+ = N(^4\textnormal{He}^+)/N(\textnormal{H}^+)$ ionic abundance ratio using our GBT data. Since helium (with an ionization potential of $\sim$24.6\,eV) is ionized by harder radiation compared to hydrogen ($\sim$13.6\,eV), a larger value of $y^+$ indicates a more energetic radiation field. 

We calculate $y^+$ using
\begin{equation} y^+ = \frac{T_{\rm L}(^4\textnormal{He}^+)\Delta V (^4\textnormal{He}^+)}{T_{\rm L}(\textnormal{H}^+)\Delta V (\textnormal{H}^+)},\label{eq:heh} \end{equation}
where $T_{\rm L}(^4\textnormal{He}^+)$ and $T_{\rm L}(\textnormal{H}^+)$ are the line temperatures of helium and hydrogen, respectively, and $\Delta V (^4\textnormal{He}^+)$ and $\Delta V (\textnormal{H}^+)$ are the corresponding FWHM line widths \citep{Peimbert1992}. For positions with hydrogen but no helium detections, we use upper limits of ${T_{\rm L}(^4\textnormal{He}^+) = 3 \times \textnormal{rms}}$ and $\Delta V (^4\textnormal{He}^+) = \bar{x} \, \Delta V (\textnormal{H}^+)$, where $\bar{x}$ is the average line width ratio $\Delta V(^4\textnormal{He}^+) / \Delta V (\textnormal{H}^+)$ for the observed region. Our values for $\bar{x}$ range from 0.59 to 0.87, which is consistent with previous studies by L16 ($\bar{x} = 0.84$) and \citet[][$\bar{x} = 0.77 \pm 0.25$]{Wenger2013}. Since the atomic mass of hydrogen is approximately one-fourth that of helium, $\bar{x}$ should be equal to $\sim 0.5$ in the absence of turbulence. The above values for $\bar{x}$ therefore indicate that turbulence plays a significant role in broadening the observed line widths of our positions (see \S \ref{sec:lineprofile}).

\begin{figure}
\centering
\includegraphics[width=.49\textwidth]{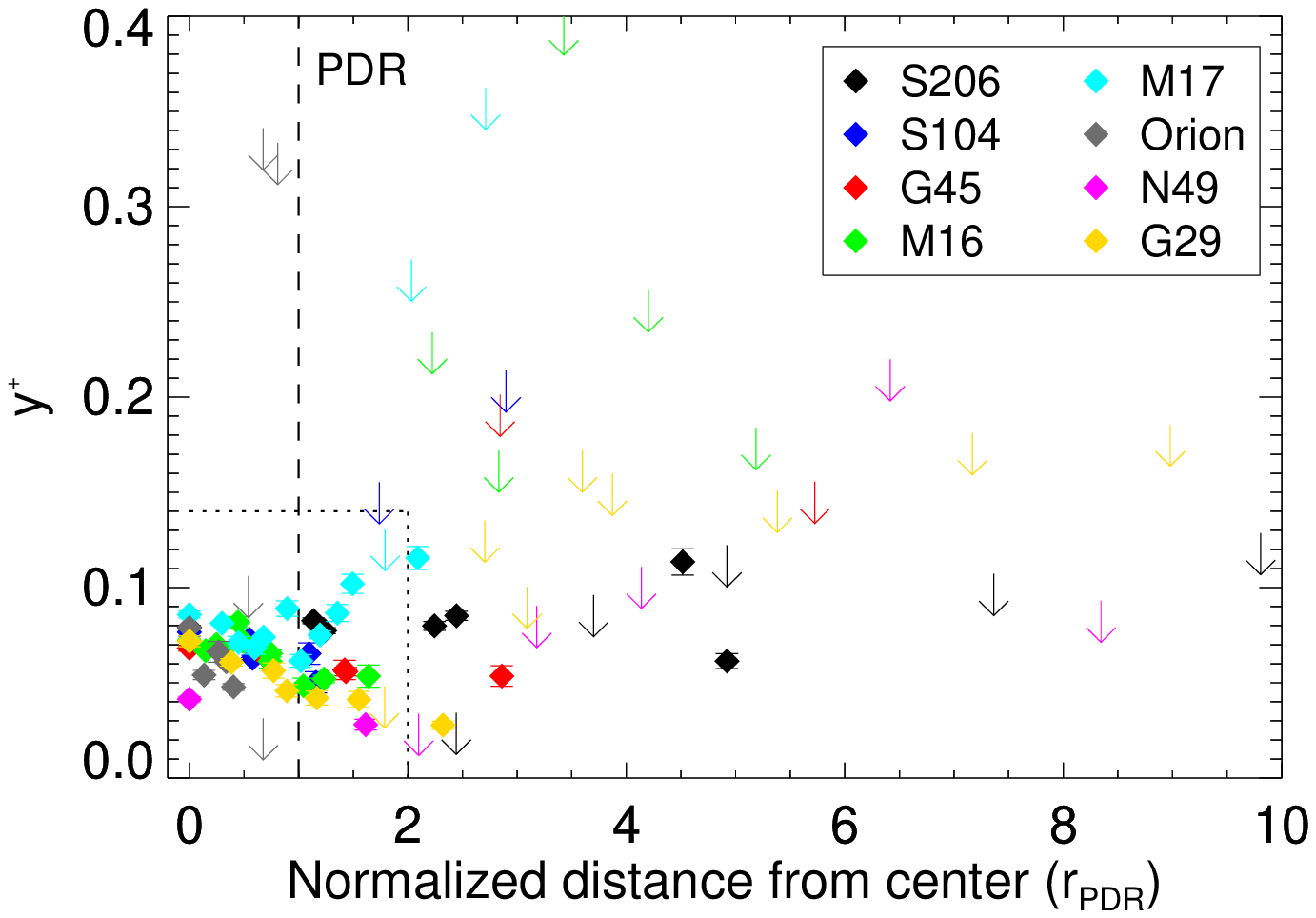} \\
\includegraphics[width=.49\textwidth]{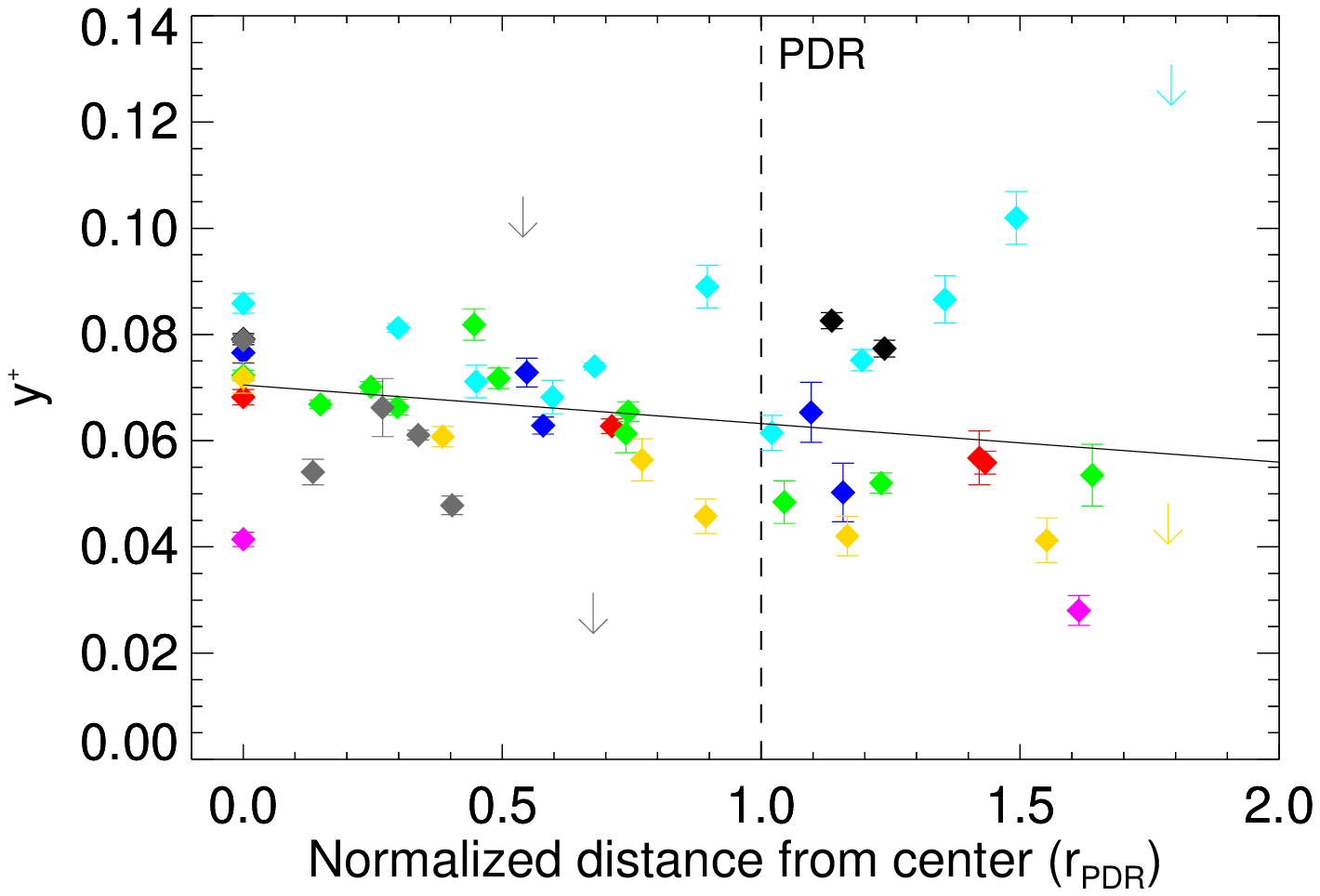}
\caption{Top: The $y^+ = N(^4\textnormal{He}^+)/N(\textnormal{H}^+)$ ionic abundance ratio as a function of angular offset from the \hii\ region. The distance is normalized by the radius of the PDR boundary along the observed direction. Downward arrows show upper limits and the vertical dashed line indicates the PDR boundary. Except for S206 and M17, $y^+$ decreases for all regions with distance. Bottom: Same, zoomed into the dotted region shown in the top panel. The solid black line is a linear fit, $y = a + bx$, to the data shown. The fit parameters are $a = 0.0705 \pm 0.0036$ and $b = -0.0073 \pm 0.0041$. \label{fig:heh_all}}\vspace{14pt}
\end{figure}

$y^+$ is shown for each individual \hii\ region in Figure~\ref{fig:heh} and for all observed regions in Figure~\ref{fig:heh_all}. We observe a decrease in $y^+$ with distance from the center for most regions. While previous results indicated an approximately constant value of $y^+$ within the region, and a decrease outside the PDR boundary \citep[L16; see also][]{Balser2001}, our sample shows a relatively steady decrease with angular offset regardless of the sampled location with respect to the PDR. We fit a linear profile, $y^+ = a + b \times r$ (where $r$ is the distance to the center of the \hii\ region in degrees, and $a$ and $b$ are the fitting parameters), to the ionic abundance ratios of each region, separately for each direction from the region center. We also calculate Spearman's rank correlation coefficient, $\rho$, both for each direction separately and for each \hii\ region as a whole. We give the fitting parameters and Spearman's $\rho$ in Table~2. Our results support the hypothesis that a large fraction of He-ionizing photons is being absorbed well within the \hii\ region boundary. We note that the measured $y^+$ gradient may also partly be due to the geometries of the observed \hii\ regions since we implicitly assume in the derivation of $y^+$ that both H$^+$ and He$^+$ fill the beam. We expect to find environments near ionization fronts where hydrogen exists in predominantly ionized form, but where helium remains mostly neutral \citep[see][]{Pankonin1980}.  Depending on the geometry of the region, the telescope beam may intersect several ionization fronts, and for these lines of sight the value of $y^+$ would be lower than expected.

\begin{deluxetable*}{lccccccc}
\tabletypesize{\footnotesize}
\tablecaption{Fitted $y^+$ Gradients and Spearman Coefficients}
\label{tab:heh}
\tablehead{\colhead{Source}  & \colhead{Direction} & \colhead{$a$} & \colhead{$\delta a$} & \colhead{$b$} & \colhead{$\delta b$} & \colhead{Spearman's $\rho$} & \colhead{Spearman's $\rho$}\\
  & & &  & \colhead{(deg$^{-1}$)} & \colhead{(deg$^{-1}$)} & (along direction) & (total)}
\startdata
  S206  & East & $0.0798$ & $0.0019$ & $-0.0533$ & $0.0238$ & $-0.400$ & $+0.202$\\
        & West & $0.0788$ & $0.0022$ & \phs $0.0963$ & $0.0288$ & $+0.800$ & \\
  S104  & North & $0.0768$ & $0.0032$ & $-0.1408$ & $0.0741$ & $-1.000$ & $-0.616$\\  
        & South & $0.0765$ & $0.0026$ & $-0.3947$ & $0.0631$ & $-1.000$ & \\
  G45   & East & $0.0678$ & $0.0027$ & $-0.2931$ & $0.0605$ & $-1.000$ & $-0.738$\\
        & West & $0.0682$ & $0.0022$ & $-0.1619$ & $0.0521$ & $-1.000$ & \\  
  M16   & North & $0.0735$ & $0.0012$ & $-0.1116$ & $0.0120$ & $-0.900$ & $-0.743$\\    
        & South & $0.0705$ & $0.0016$ & $-0.0448$ & $0.0088$ & $-0.750$ & \\ 
  M17   & East & $0.0802$ & $0.0017$ & \phs $0.0087$ & $0.0072$ & $+0.310$ & $+0.522$\\   
        & West & $0.0835$ & $0.0020$ & $-0.1330$ & $0.0246$ & $+0.200$ & \\   
  Orion & North & $0.0790$ & $0.0009$ & $-0.2622$ & $0.0192$ & $-1.000$ & $-0.667$\\
        & South & $0.0777$ & $0.0013$ & $-0.1541$ & $0.0095$ & $-0.800$ & \\ 
  N49   & East & \nodata & \nodata & \nodata & \nodata & \nodata & $-1.000$\\
        & West & $0.0414$ & $0.0022$ & $-0.3877$ & $0.0920$ & $-1.000$ & \\  
  G29   & North & $0.0718$ & $0.0031$ & $-0.7658$ & $0.1294$ & $-1.000$ & $-0.937$\\  
        & South & $0.0687$ & $0.0017$ & $-0.2042$ & $0.0144$ & $-1.000$ & \\ 
\enddata
\end{deluxetable*}

Two regions do not follow the general trend of decreasing $y^+$: S206 and M17. S206 shows an increase of $y^+$ with distance and for M17 the results are inconclusive. It is unclear why these two regions exhibit such different behavior than the rest of our sample. S206 is a relatively compact \hii\ region without much extended emission, whereas M17 is one of the largest and brightest \hii\ regions in the Milky Way. This suggests that the measured increase in $y^+$ may not be related to the size or morphology of the these two regions.

\section{Physical Properties of the Ionized Gas}\label{sec:properties}
\subsection{LTE Electron Temperatures}\label{sec:te}

The electron temperature, $T_{\rm e}$, is a proxy for the metallicity of an \hii\ region \citep[e.g.,][]{Rubin1985} and can be used to study its intrinsic heating and cooling processes. Previous studies have found conflicting results regarding the relationship between $T_{\rm e}$ inside and outside the PDR boundaries of \hii\ regions. Most research shows a relatively constant electron temperature distribution within \hii\ regions \citep[e.g.,][]{Roelfsema1992,Adler1996,Rubin2003}, but there has been evidence for a decrease of $T_{\rm e}$ with increasing distance from Orion A \citep{Wilson2015}. Under the assumption of LTE, $T_{\rm e}$ can be derived by

\begin{multline} \left( \frac{T_{\rm e}^*}{\textnormal{K}} \right) = \left \{ 7103.3 \left( \frac{\nu _{\rm L}}{\textnormal{GHz}}\right)^{1.1} \left[ \frac{T_{\rm C}}{T_{\rm L}(\textnormal{H}^+)} \right] \right. \\
\left. \times \left[ \frac{\Delta V (\textnormal{H}^+)}{\textnormal{km\,s}^{-1}} \right]^{-1} \times \left[ 1+y^+ \right]^{-1} \right \}^{0.87},\label{eq:etemp}
\end{multline}

where $\nu_{\rm L} = 6$\,GHz is the average frequency of our
Hn$\alpha$ recombination lines, $T_{\rm C}$ is the continuum antenna
temperature, $T_{\rm L}$ is the H line antenna temperature, $\Delta V
(\textnormal{H}^+)$ is the FWHM line width, and $y^+$ is the ionic
abundance ratio from Eq.\,\ref{eq:heh} \citep{Mezger1968,Quireza2006}. 

\begin{figure}
\centering
\includegraphics[width=.49\textwidth]{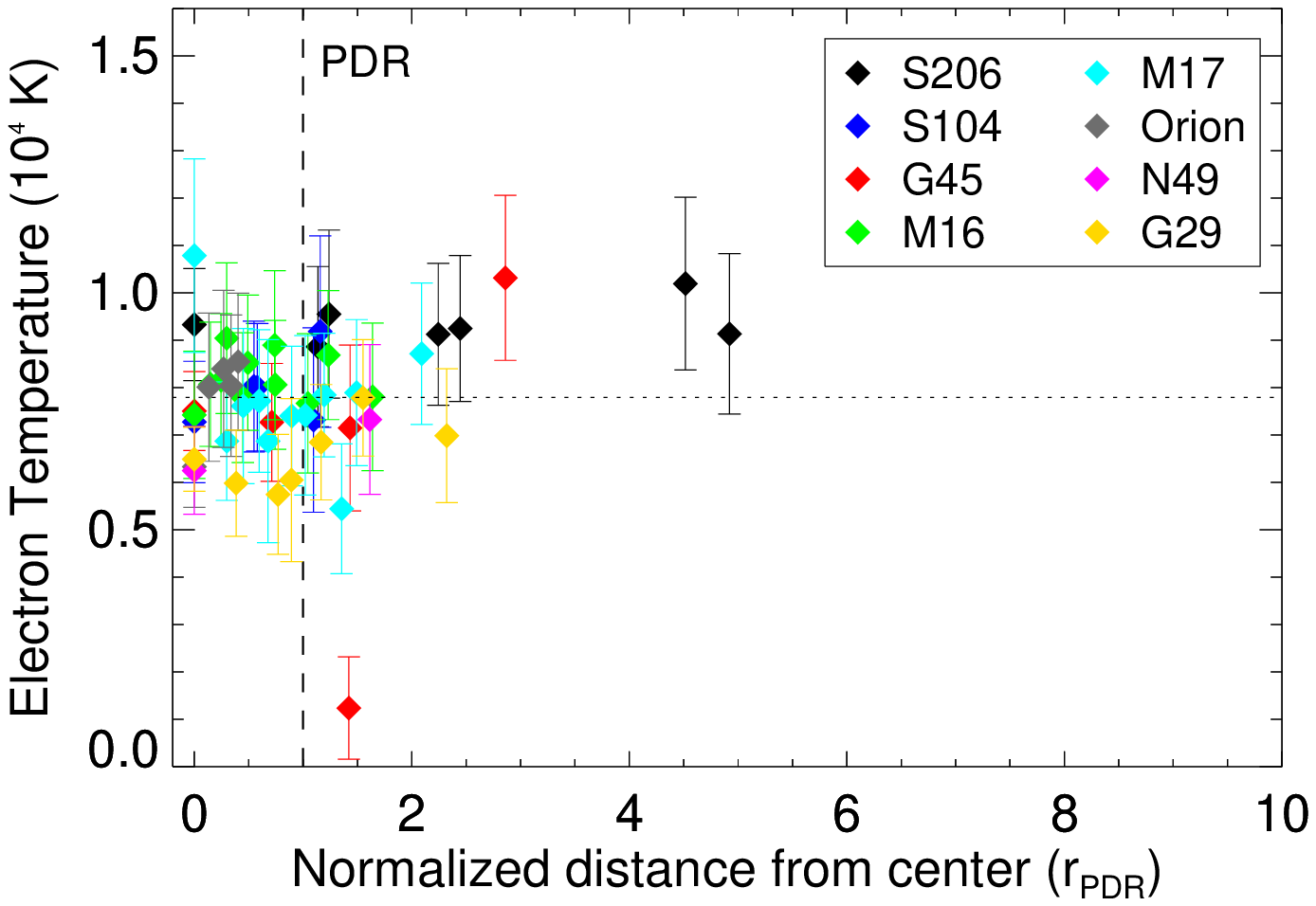} \\
\includegraphics[width=.49\textwidth]{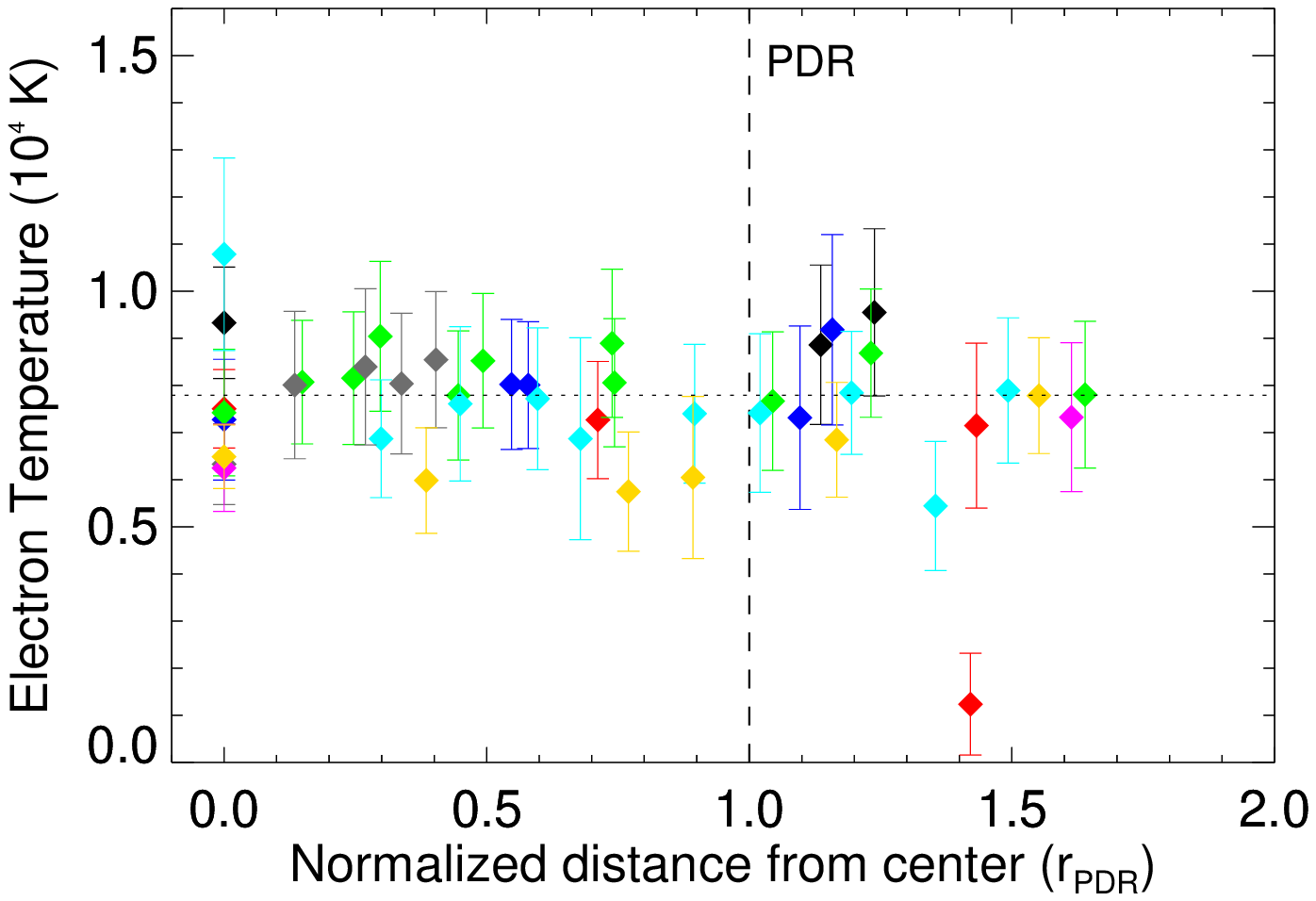} 
\caption{Top: The LTE electron temperature, $T_{\rm e}^*$, as a function of angular offset from the \hii\ region. The distance is normalized by the radius of the PDR boundary along the observed direction. The vertical dashed line indicates the PDR boundary and the horizontal dotted line shows the average derived electron temperature, $T_{\rm e}^* = 7790 \pm 1480$\K. Bottom: Same, zoomed in. \label{fig:etemp_all}}
\end{figure}

The derived $T_{\rm e}^*$ are shown in Figure~\ref{fig:etemp_all} for all observed positions where the He line could be detected. While the average electron temperature for each \hii\ region is slightly different, there are no large variations of $T_{\rm e}^*$ with distance for any of the observed regions. For G45, our derived electron temperature of $1240 \pm 1080$\K at a normalized distance of $r_{\rm PDR} =1.42$ is abnormally low. At this position, our calculated continuum antenna temperature used to derive $T_{\rm e}^*$ is affected by severe baseline instability in our the spectra. We therefore argue that here our value of $T_{\rm e}^*$ does not reflect the actual electron temperature. This assumption is supported by our line profile analysis (\S \ref{sec:lineprofile}), which yields an estimated electron temperature of $\sim 6000$\,K for this direction. We disregard this position for all further analysis.

\subsection{Electron Densities}\label{sec:ne}
\begin{figure*}
\centering
\includegraphics[width=1.055\textwidth]{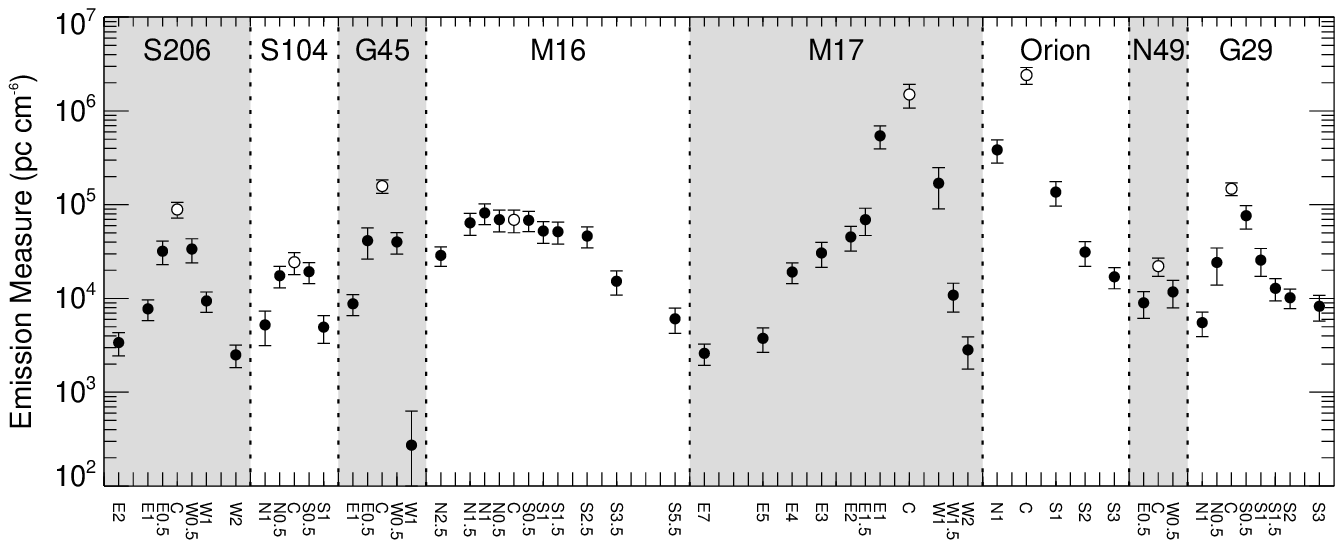} \vspace{-20pt} \\
\includegraphics[width=1.055\textwidth]{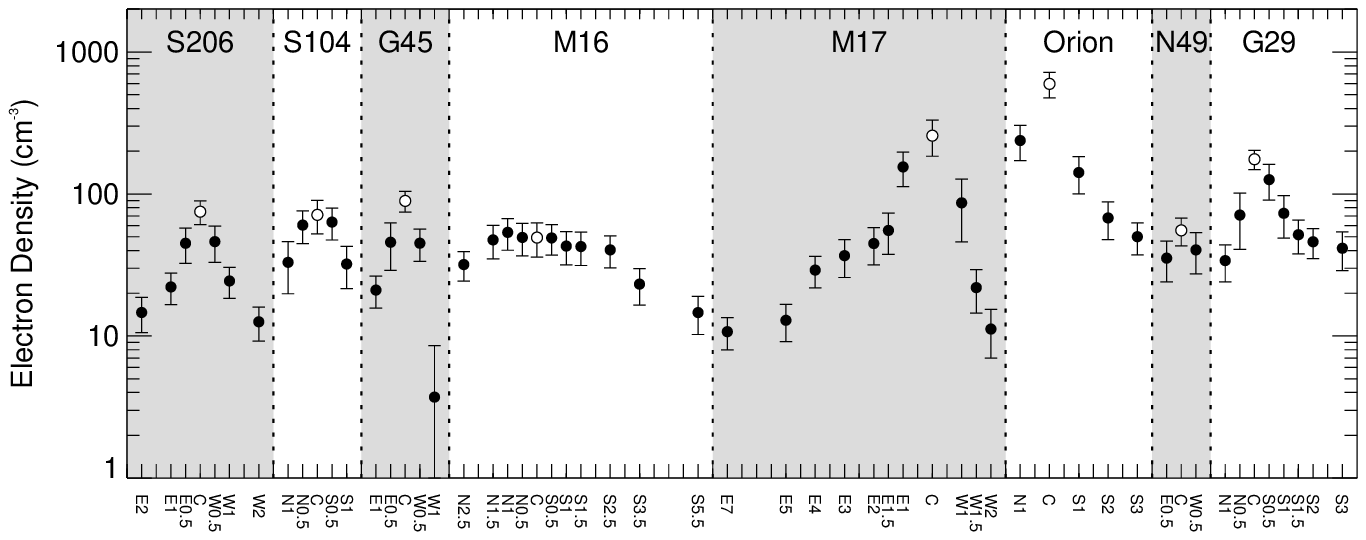} \vspace{-20pt}
\caption{Top: The emission measure, EM, for each observed direction with a He line detection. The central directions of the \hii\ regions are indicated by the unfilled circles. The largest EM values are found toward these directions. Bottom: Same, for the root mean square electron density, $\bar{n}_{\rm e}$, assuming a simple slab geometry for each region. \label{fig:EM}}
\end{figure*}

Assuming that the ionized gas is in LTE, we are able to constrain the emission measure and the root mean square electron number density for each observed direction. The emission measure, EM, is defined as the integral of the electron density squared, $n_{\rm e}^2$, along the line of sight. Because the emission measure is proportional to the optical depth at the line center, $\tau_{\rm L}$, the brightness temperature of a recombination emission line can be estimated by
\begin{multline} 
\frac{T_{\rm b}}{\rm K} \approx T_{\rm e} \tau_{\rm L} \approx 1.92 \times 10^3 \left( \frac{T_{\rm e}}{\rm K} \right) ^{-3/2} \\
\times \left( \frac{\rm EM}{\rm pc\,cm^{-6}} \right) \left( \frac{\Delta \nu}{\rm kHz} \right) ^{-1},\label{eq:em1}
\end{multline}
where $T_{\rm b}$ is the brightness temperature at the line center and $\Delta \nu$ is the line FWHM \citep{Condon2016}. Assuming that the RRL emitting region is extended evenly across the GBT beam and using a GBT main beam efficiency of 0.94 at C-band \citep{Maddalena2010, Maddalena2012}, the emission measure can be expressed as
\begin{multline} \frac{\rm EM}{\rm pc\,cm^{-6}} \approx 1.109 \times 10^{-2} \left( \frac{T_{\rm L} (\textnormal{H}^+)}{\rm K} \right) \\
\times \left( \frac{\Delta V (\textnormal{H}^+)}{\textnormal{km\,s}^{-1}} \right) \left( \frac{T_{\rm e}}{\rm K} \right) ^{3/2}.\label{eq:em2}
\end{multline}
We use our LTE electron temperatures, $T_{\rm e}^*$, to calculate the EM for all positions for which an He line was detected. The resulting EM values range from $270$\,pc\,cm$^{-6}$ to $2.4 \times 10^6$\,pc\,cm$^{-6}$, with the largest values found toward the central directions of Orion and M17 (see Figure~\ref{fig:EM}, top panel). All observed \hii\ regions show a decrease in EM with distance from the center, with the exception of M16 whose EM remains roughly constant within the PDR boundaries. The derived EM values are given in Table~\ref{tab:temperatures}.

The root mean square electron number density, $\bar{n}_{\rm e}$, can be estimated from the derived emission measures, assuming the \hii\ region geometry is approximated by a slab of constant line-of-sight path length and uniform density. We further assume that the path length for each \hii\ region is twice its radius, $R_{\rm PDR}$, as given in Table~\ref{tab:hii}. By definition,
\begin{equation} \left( \frac{\rm EM}{\rm pc\,cm^{-6}} \right) = \int_{\rm los} \left( \frac{n_{\rm e}}{\rm cm^{-3}} \right) ^2 d \left( \frac{s}{\rm pc} \right),\label{eq:em3}
\end{equation}
such that
\begin{equation} \frac{\bar{n}_{\rm e}}{\rm cm^{-3}} \approx 0.707\,\left( \frac{\rm EM}{\rm pc\,cm^{-6}} \right)^{1/2} \left( \frac{R_{\rm PDR}}{\rm pc} \right)^{-1/2}.\label{eq:n_e}
\end{equation}

We calculate $\bar{n}_{\rm e}$ for all positions with an He line detection and show the calculated $\bar{n}_{\rm e}$ values in the bottom panel of Figure~\ref{fig:EM}. Due to our assumption of constant path length for each region, $\bar{n}_{\rm e}$ follows the same trends as EM.

\subsection{Non-LTE Analysis}\label{sec:lte}
There is debate in the literature on whether \hii\ regions and their surroundings are typically close to LTE. In a study of 72 Galactic \hii\ regions, \citet{Balser2011} argue that non-LTE effects, such as stimulated emission should be small within \hii\ regions. \citet{Dupree1970} show for three Galactic \hii\ regions that the observed intensity ratios of Hn$\alpha$ and higher-order lines systematically deviate from the theoretical values; in their sample the Hn$\beta$ to Hn$\alpha$ line intensity ratio is generally 20--30\% lower than the LTE value. They suggest that this effect is due to departure from LTE and further argue that non-LTE effects are different for each level population \citep[see also][]{Zuckerman1967}. While stimulated emission typically affects $\alpha$-lines more strongly than $\beta$-lines at the same frequency and could account for the lower line ratios, \citet{Shaver1979} suggest that instead the observed line ratios are produced by pressure broadening. They argue that consequently there is no clear evidence of non-LTE effects in single-dish RRL observations of Galactic \hii\ regions \citep{Shaver1980}.

Unfortunately, our $\beta$- and $\gamma$-line data cannot be used to test for LTE since we did not detect RRLs in most individual Hn$\beta$ and Hn$\gamma$ spectra. While the hydrogen line is detected toward many directions after averaging together the $\beta$- and $\gamma$-lines, respectively, the averaged spectra are centered at different frequencies than our $\alpha$-lines. Thus, the average beam size varies and a different region of space is sampled for each level population.

The departure from LTE can, however, be directly quantified using our Hn$\alpha$ line data. The following analysis is based on our derived LTE electron temperatures (\S \ref{sec:te}) and root mean square electron densities (\S \ref{sec:ne}). The necessary calculations are from \citet{Brocklehurst1977} and \citet{Salem1979}, and include the effect of stimulated emission due to an external radiation field, as well as collisional transitions from excited atom--electron collisions. As a first-order approximation, we assume that the conditions toward our observed directions are typical for \hii\ regions but that there is no incident radiation present \citep[see][]{Salem1979}. Given our average $n=105$ for the observed $\alpha$-lines, we perform a bilinear interpolation between the electron temperature and electron density values from \citet[][their Table~1]{Salem1979} to find the departure coefficient ($b_{\rm n}$) that matches our values of $T_{\rm e}^*$ and $\bar{n}_{\rm e}$. $b_{\rm n}$ relates the number of atoms in level $n$ to the number that would be there if the system were in thermodynamic equilibrium. Using the same method, we also find the amplification factor ($\beta_{\rm n}$) which describes an enhancement of the stimulated emission due to the overpopulation of level $n$ relative to lower states.

Our values for $b_{\rm n}$ range from 0.76 to 0.96 and $1- \beta_{\rm n} = kT_{\rm e}d({\rm ln}\,b_{\rm n})/dE_{\rm n}$ ranges from 46 to 136.  With average $b_{\rm n} = 0.86 \pm 0.04$, there is only a small deviation from LTE for all observed directions. We notice a trend of decreasing $b_{\rm n}$ with increasing distance from the all \hii\ regions in our sample. This effect is particularly strong for the more luminous \hii\ regions (e.g., M17) and is likely due to the steep decrease in electron density while $T_e^*$ remains roughly constant. The largest $b_{\rm n}$ values are found toward the central positions of M17, Orion, and G29, suggesting that due to the large electron densities the collision rates dominate the level populations at these locations.

We note that our non-LTE analysis of the observed positions is only a rough approximation due to the following two reasons. First, $T_{\rm e}^*$ and $\bar{n}_{\rm e}$ are mean values averaged along the line of sight and over the beam. Given our large average beam ($\textnormal{HPBW} \approx 123$\arcsec) and the \hii\ region geometries, many different physical environments must contribute to these parameters for each observed position. This becomes especially important near the central locations and the PDRs of each \hii\ region where the spatial gradients of electron temperature and density are presumably the steepest. Second, we use $T_{\rm e}^*$ and $\bar{n}_{\rm e}$ to estimate $b_{\rm n}$ and $\beta_{\rm n}$. Since $T_{\rm e}^*$ and $\bar{n}_{\rm e}$ are based on the assumption of LTE, this approach is only valid for positions that are near LTE.

While our derived $b_{\rm n}$ and $\beta_{\rm n}$ values assume constant density and temperature, and are therefore not truly representative of the values in real \hii\ regions, we argue that $T_{\rm e}^*$ is a reasonable approximation of the true electron temperature averaged along the line of sight and over the HPBW. The line opacities are small for these sources at the observed frequencies, which decreases the impact of stimulated emission. Therefore, $b_{\rm n}$ provides a good estimate of the deviation from LTE despite typical values of $\geq 100$ for $\beta_{\rm n}$ \citep[see, e.g.,][]{Salem1979}. With $b_{\rm n}$ close to unity for all observed positions, it is unlikely that the true average electron temperatures deviate significantly from $T_{\rm e}^*$.

\section{Line Profile Analysis}\label{sec:lineprofile}
The observed line widths of RRLs within and near \hii\ regions predominantly depend on two variables: the temperature of the plasma (thermal line broadening) and the amount of turbulence in the local ISM (turbulent line broadening). Other mechanisms that affect the observed line widths, such as natural broadening or pressure broadening, are thought to be negligible given our observed frequencies and electron densities \citep{Hoang-Binh1972}.

Here we assume that the RRL widths are only affected by thermal broadening and turbulent broadening. Therefore,
\begin{equation} \Delta V_{\rm H} = 2.355 \left( \frac{kT_{\rm e}}{m_{\rm H}} + V^2_{\rm turb} \right) ^{1/2},\label{eq:broadening1}
\end{equation}
where $\Delta V_{\rm H}$ is the observed FWHM of the hydrogen RRL, $T_{\rm e}$ is the electron temperature, $m_{\rm H}$ is the mass of the hydrogen atom, and $V_{\rm turb}$ is the velocity contribution due to turbulence. The first term in the parentheses thus corresponds to the thermal contribution, while the constant of 2.355 accounts for the conversion from the one-dimensional velocity dispersion to FWHM. Since we can measure $\Delta V_{\rm H}$ directly, only two unknowns remain, $T_{\rm e}$ and $V_{\rm turb}$. These, however, can be expressed in terms of each other by observing the helium RRL toward the same direction, as long as $T_{\rm e}$ and $V_{\rm turb}$ are unchanged between the two species. After accounting for natural constants and atomic masses,
\begin{equation} \left( \frac{T_{\rm e}}{\rm K} \right) = 29.22 \left[ \left( \frac{\Delta V_{\rm H}}{\rm km\,s^{-1}} \right)^2 - \left( \frac{\Delta V_{\rm He}}{\rm km\,s^{-1}} \right)^2 \right],\label{eq:broadening2}
\end{equation}
where $\Delta V_{\rm He}$ is the observed FWHM of the helium RRL.

\begin{figure}
\centering
\includegraphics[width=.49\textwidth]{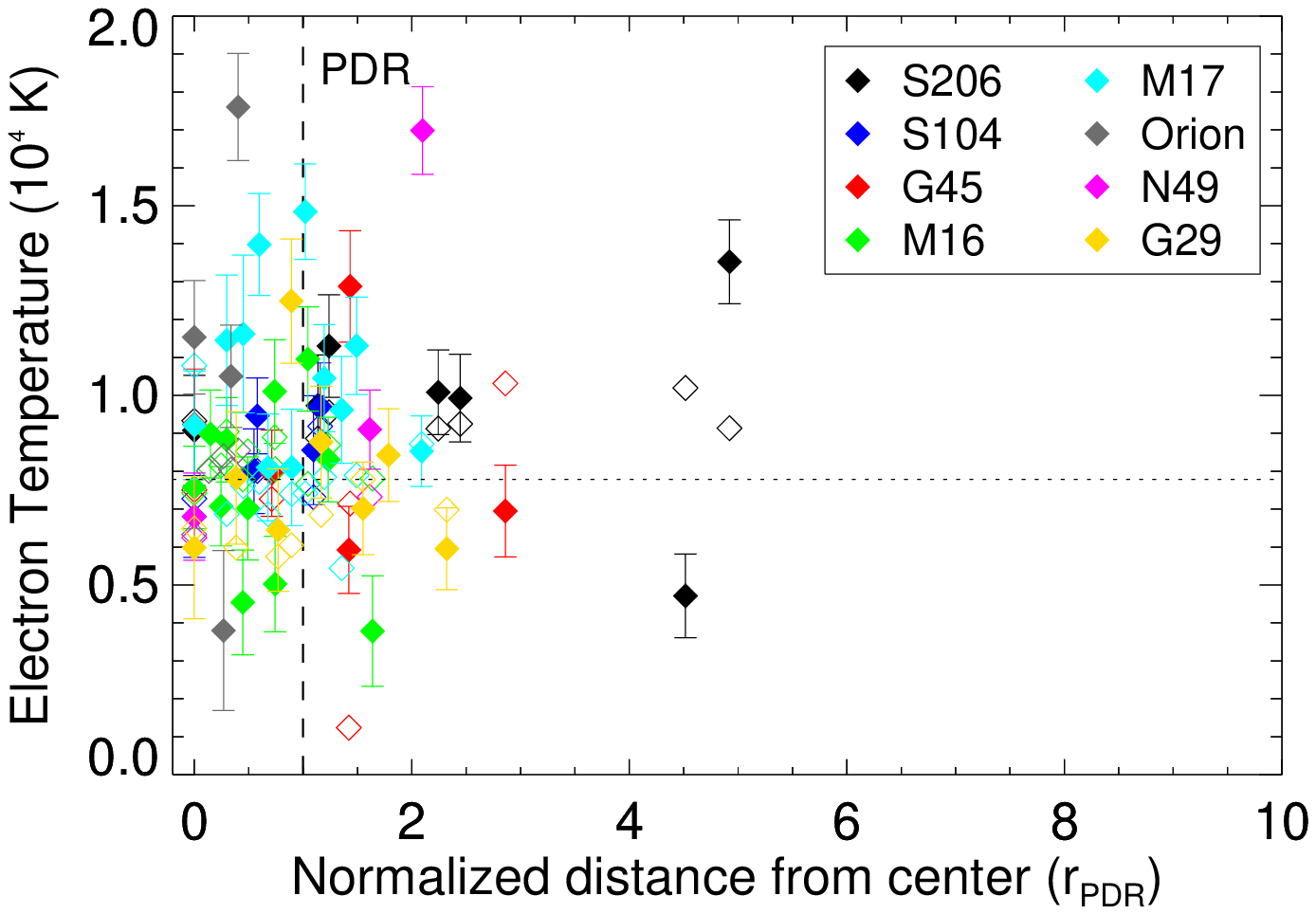}\\
\includegraphics[width=.49\textwidth]{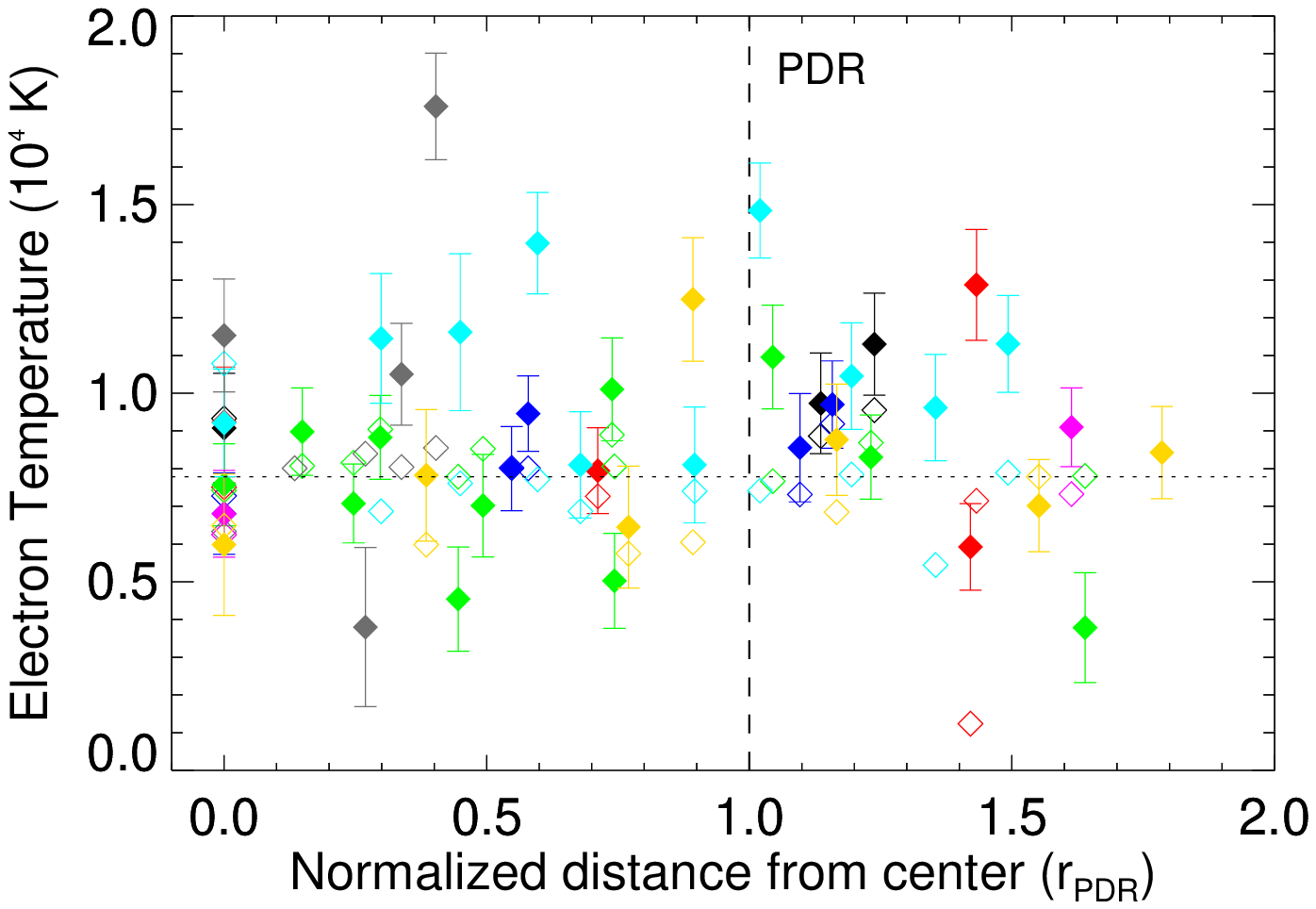}
\caption{Top: The electron temperature derived from our line profile analysis (filled symbols), $T_{\rm e}$, as a function of angular offset from the \hii\ region. Shown with the unfilled symbols is the LTE electron temperature derived in \S \ref{sec:te}. The distance is normalized by the radius of the PDR boundary along the observed direction. The vertical dashed line indicates the PDR boundary and the horizontal dotted line shows the average derived LTE electron temperature from Figure\,\ref{fig:etemp_all}. Bottom: Same, zoomed in. \label{fig:etemp_lte}}
\end{figure}

Our results for $T_{\rm e}$ are shown in  Figure~\ref{fig:etemp_lte}. As for $T^*_{\rm e}$, only directions for which the helium line was detected are included. The average electron temperature is $T_{\rm e} = 7940 \pm 4720$\,K outside the \hii\ region PDRs and $T_{\rm e} = 9180 \pm 3950$\,K within. As for $T_{\rm e}^*$, this difference is not statistically significant. The larger deviation between individual values of $T_{\rm e}$ suggests that this method of calculation is less robust than that described in \S \ref{sec:te}. This may be due to the strong dependence of $T_{\rm e}$ on the width of the observed hydrogen and helium RRLs, and the fact that $\Delta V_{\rm He}$ may have larger uncertainties than those given in Table~\ref{tab:alpha} because of its low intensity and sensitivity to baseline variations. Due to the large deviation between individual values of $T_{\rm e}$, we use $T_{\rm e}$ only as a consistency check in this work.

\section{Carbon RRLs}\label{sec:c}
Previous results suggest that carbon RRL emission is often observed from \hii\ region PDRs \citep[e.g.,][L16]{Hollenbach1999}. This is possibly due to its first ionization potential of only $\sim$11.3\,eV, which is lower than that of hydrogen or helium. The carbon in \hii\ region PDRs may therefore be ionized by soft-UV photons ($E < 13.6$\,eV) that pass through the \hii\ region essentially undisturbed, aside from being attenuated by dust.

\begin{figure}
\centering
\includegraphics[width=.49\textwidth]{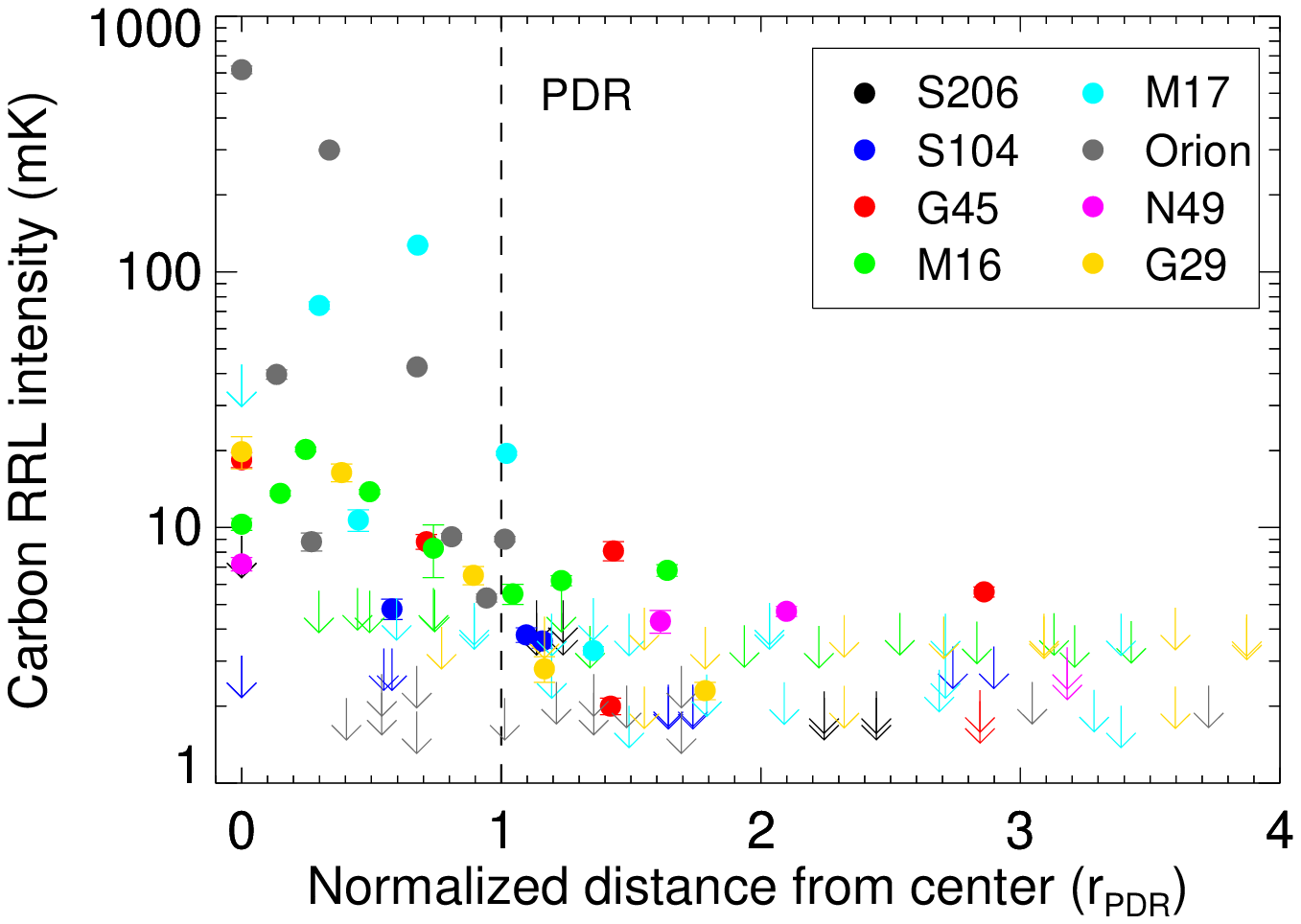} \\
\includegraphics[width=.49\textwidth]{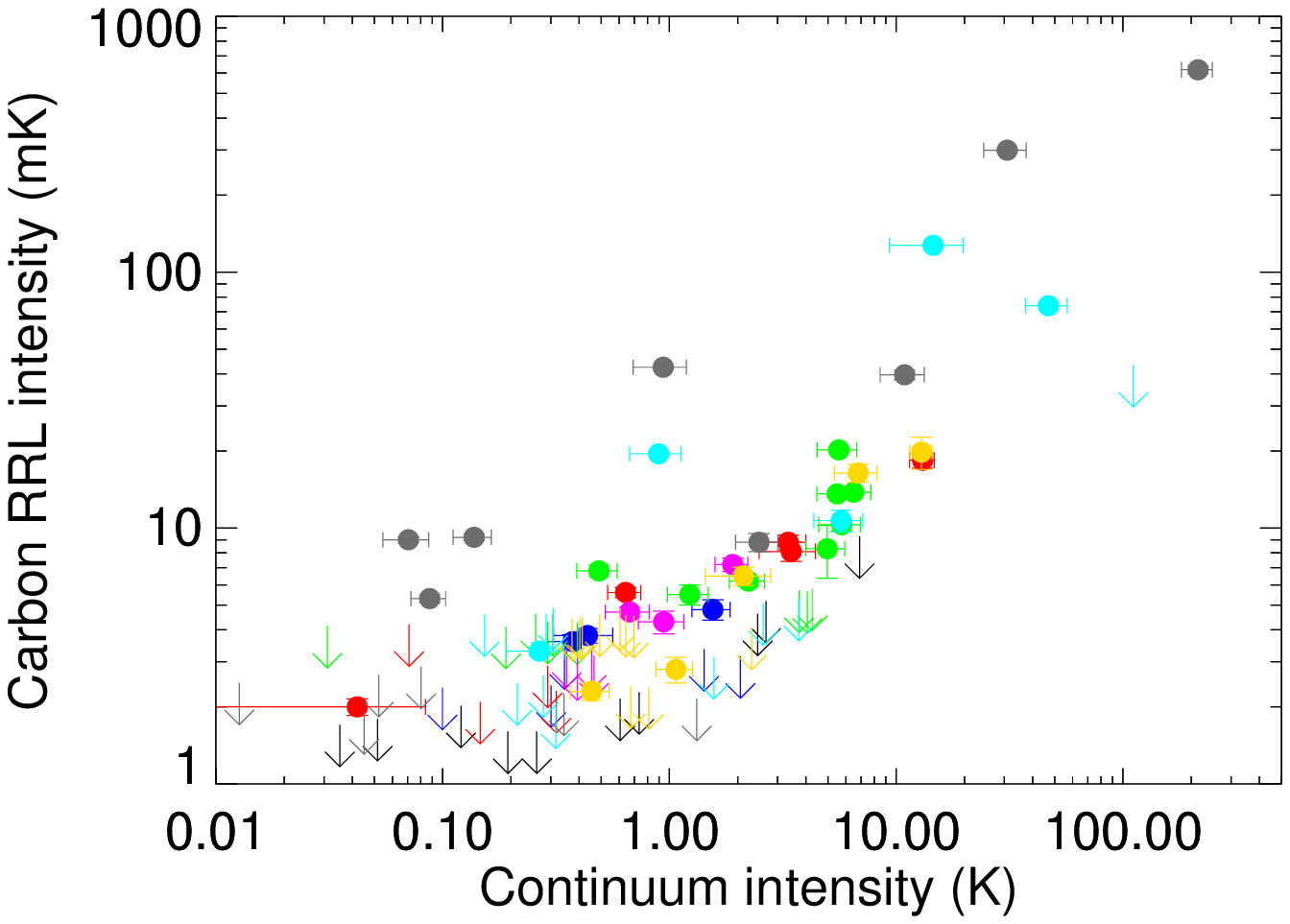} 
\caption{Top: The carbon line intensity as a function of angular offset from the \hii\ region. The distance is normalized by the radius of the PDR boundary along the observed direction. Downward arrows show upper limits and the vertical dashed line indicates the PDR boundary. There is no increase in the carbon intensity near the PDR boundaries. Bottom: Correlation between the carbon line intensity and the continuum intensity. The correlation may indicate that our carbon non-detections at large distance offsets are due to a lack of amplification by stimulated emission or that the carbon abundance in the diffuse medium is low. \label{fig:c_all}}
\end{figure}

We do not observe enhanced carbon RRL emission near the PDR boundaries for most \hii\ regions in our sample (see top panel of Figure~\ref{fig:c_all}). Instead, the emission is strongest within the \hii\ regions and decreases steadily with distance. While it is likely that a fraction of the PDR is contained within the telescope beam along the line of sight, this may also suggest that a large number of soft-UV photons are attenuated by dust present within the \hii\ regions. S104 is the only observed source for which the line emission may be increased near the PDR. The large number of non-detections for S104, however, casts doubt on the statistical significance of this interpretation.

At low frequencies, carbon can be observed in either absorption lines or amplified emission lines due to stimulated emission from inverted populations. Observations of Cas A at 26\,MHz show spectral features consistent with the detection of a carbon-$\alpha$ absorption line at $n \sim 630$ \citep{Konovalenko1981,Walmsley1982}. In an Ooty Radio Telescope study of RRLs near 327\,MHz, \citet{Roshi2002} found evidence of stimulated emission, resulting a strong correlation between carbon RRL intensity and continuum emission.

To test whether carbon RRL emission is amplified by stimulated emission at our higher average observing frequency of $\sim$6\,GHz, we show the carbon RRL intensity as a function of continuum intensity in the bottom panel of Figure~\ref{fig:c_all}. While there is a correlation between the carbon emission and the continuum intensity, the spread in our data is quite large. Most of the spread, however, is caused by only two sources, Orion and M17, which show stronger carbon RRL emission than expected given their background continuum intensity. These two sources are among the largest and most luminous \hii\ regions in our sample. This suggests that much of the carbon emission may be amplified by stimulated emission. At large distance offsets, a possible lack of amplification of the carbon lines by stimulated emission may explain our carbon non-detections where the continuum emission is weak. It is also possible that a low carbon abundance in the diffuse medium at these positions is responsible for the non-detections.

\section{He$^{++}$ Emission}\label{sec:hei}
The second ionization potential of helium at $\sim$54.4\,eV exceeds helium's first ionization potential by over a factor of two and thus He$^+$ is ionized by only the most energetic radiation fields within \hii\ regions. Previous GBT observations by \citet{Roshi2017} failed to detect emission from He$^{++}$ in the diffuse gas surrounding ultracompact \hii (UC\hii) regions. Their combined 1$\sigma$ upper limit for their setup is 4\,mK, however, they only observed 3 UC\hii regions with a total of 18 independent pointings.

Although many of our sampled positions are cospatial with luminous \hii\ regions, we did not detect the He$^{++}$ line. The rms values for our individual positions range from 3 to 20\,mK, corresponding to upper limits for the He$^{++}$ line of 9 to 60\,mK. Our best upper limit for the He$^{++}$/He$^{+}$ line ratio is 0.064, sampled at the central position of Orion. We average together all positions for which the line intensity of singly-ionized He is at least 30\,mK after shifting them in velocity, and again fail to detect the He$^{++}$ line. The rms of our combined, averaged spectrum is 0.6\,mK, corresponding to a 3$\sigma$ upper limit for the He$^{++}$ line of 1.8\,mK.

\section{Conclusions}\label{sec:conclusions}
Despite our recent studies \citep[][L16]{Anderson2015}, radiation leaking from Galactic \hii\ regions is still not well-understood. Models and observations of external galaxies suggest that 30--70\% of the emitted ionizing radiation escapes from \hii\ regions into the ISM \citep{Oey1997,Zurita2002,Giammanco2005,Pellegrini2012}. Due to observational constraints, however, these studies tend to be biased towards the largest and most luminous \hii\ regions which only make up a small fraction of the total number of \hii\ regions in a given galaxy. Here we observe a sample of Galactic \hii\ regions of various sizes, geometries, and luminosities, including several compact (radius $<5$\,pc) \hii\ regions whose extragalactic counterparts may have been missed in previous studies.

If ionizing radiation leaking from an individual \hii\ region is responsible for maintaining the ionization of the WIM around that region, we expect a decrease in RRL intensity with distance from the region, for which the slope of the decrease is believed to roughly constrain the amount of leaking photons (see L16). We observe a power-law decrease in hydrogen RRL emission intensity outside the PDR of all observed \hii\ regions (see Figure~3). We showed in \S \ref{sec:h} that the hydrogen line intensity decreases with roughly the same slope for all our targets except Orion when normalizing the angular offset from the region center by the radius of the PDR. This suggests that the slope in the decrease of hydrogen RRL intensity with distance is directly related to the size of the region. The case of Orion is further complicated by its highly asymmetrical geometry and the presence of several PDR boundaries toward the observed directions. In fact, when defining the innermost visible enhancement in 12\,$\mu$m emission as the main PDR boundary, its slope is similar to that of the other sources, suggesting that the power-law decrease of hydrogen RRL intensity with distance and region size may fundamentally be the same for all Galactic \hii\ regions. The physical implications of this result are unclear, although it may be related to a transition from ionization-bounded to density-bounded behavior \citep[e.g.,][]{Rozas1998,Zurita2002}.

Using the ionic abundance ratio, $y^+$, the hardness of the interstellar radiation field can be constrained for our region sample. We observe a general trend of decreasing $y^+$ with distance from the \hii\ regions. This is indicative of absorption of He-ionizing photons within the ionization front of the \hii\ region. It is unclear why two of the observed regions, S206 and M17, do not follow this trend. The difference in their physical properties, such as M17 being ionized by more than one O star, makes it unlikely that this is due to the region's size, luminosity, or morphology.

The electron temperature, $T_{\rm e}$, is believed to remain relatively constant within \hii\ regions \citep{Roelfsema1992,Adler1996,Rubin2003}, although recently \citet{Wilson2015} reported on a decrease in $T_{\rm e}$ with increasing distance from Orion A. We calculate the electron temperature for our positions with He detections using two independent methods. The first method assumes that LTE is satisfied for all observed directions \citep[see also][]{Mezger1968, Quireza2006}. We find no general trends in the derived electron temperatures with distance from the \hii\ regions and no statistically significant difference between $T_{\rm e}$ within and outside the \hii\ region PDRs. The second method is based on a line profile analysis and assumes that $T_{\rm e}$ and the amount of turbulent line broadening remain the same between hydrogen and helium. Although LTE is not required, this method appears to be less robust than the first, possibly due to the larger contribution of line width uncertainties on the overall uncertainty in $T_{\rm e}$. We, again, do not detect a significant difference between electron temperatures within and outside the \hii\ region PDRs.

Our \hii\ region sample spans a wide range of emission measures, EM, and root mean square electron densities, $\bar{n}_{\rm e}$. While most observed directions have $10^2 \lsim \bar{n}_{\rm e} \lsim 10^3$\,cm$^{-3}$, the electron densities at the central locations of M17 and Orion are as large as 258\,cm$^{-3}$ and 596\,cm$^{-3}$, respectively. These values are significantly lower than those found by optical line tracers \citep[e.g.,][]{Danks1970,Kitchin1987}, possibly because they are based on $T^*_{\rm e}$ averaged along the line of sight and over the HPBW. With an average departure coefficient of $b_{\rm n} = 0.86 \pm 0.04$, we do not find strong departure from LTE in our sample. We note that $b_{\rm n}$ is largest toward the high-$\bar{n}_{\rm e}$ regions at the central locations of our \hii\ regions sample where the level populations are dominated by collisions. As expected, $b_{\rm n}$ decreases with increasing distance from all observed \hii\ regions, suggesting that non-LTE effects become more significant in the lower-density envelopes surrounding Galactic \hii\ regions.

Finally, we do not find enhanced carbon RRL emission near the PDR boundaries, as has been observed previously \citep[][L16]{Hollenbach1999}. This may indicate selective attenuation of soft-UV photons by dust within the \hii\ regions. However, there exists a correlation between carbon RRL intensity and continuum intensity. This suggests that the carbon emission is amplified by stimulated emission, a relationship that has previously only been observed at frequencies below $\sim$1\,GHz \citep{Roshi2002}.

\acknowledgements
We thank the referee for insightful comments that improved the clarity of this manuscript. We thank Zachary Tallman for his help in identifying the PDR boundaries used in this work. This work is supported by NSF grant AST1516021 to LDA. This research has made use of the NASA/IPAC Infrared Science Archive, which is operated by the Jet Propulsion Laboratory, California Institute of Technology, under contract with the National Aeronautics and Space Administration.

\textit{Facility:} Green Bank Telescope, WISE.
\textit{Software:} TMBIDL \citep{Bania2014}.

\bibliographystyle{aasjournal}
\bibliography{HII}

\begin{thebibliography}{}
\expandafter\ifx\csname natexlab\endcsname\relax\def\natexlab#1{#1}\fi
\providecommand{\url}[1]{\href{#1}{#1}}
\providecommand{\dodoi}[1]{doi:~\href{http://doi.org/#1}{\nolinkurl{#1}}}
\providecommand{\doeprint}[1]{\href{http://ascl.net/#1}{\nolinkurl{http://ascl.net/#1}}}
\providecommand{\doarXiv}[1]{\href{https://arxiv.org/abs/#1}{\nolinkurl{https://arxiv.org/abs/#1}}}

\bibitem[{{Adler} {et~al.}(1996){Adler}, {Wood}, \& {Goss}}]{Adler1996}
{Adler}, D.~S., {Wood}, D.~O.~S., \& {Goss}, W.~M. 1996, ApJ, 471, 871,
  \dodoi{10.1086/178014}

\bibitem[{{Afflerbach} {et~al.}(1997){Afflerbach}, {Churchwell}, \&
  {Werner}}]{Afflerbach1997}
{Afflerbach}, A., {Churchwell}, E., \& {Werner}, M.~W. 1997, ApJ, 478, 190,
  \dodoi{10.1086/303771}

\bibitem[{{Alves} {et~al.}(2012){Alves}, {Davies}, {Dickinson}, {Calabretta},
  {Davis}, \& {Staveley-Smith}}]{Alves2012}
{Alves}, M.~I.~R., {Davies}, R.~D., {Dickinson}, C., {et~al.} 2012, MNRAS, 422,
  2429, \dodoi{10.1111/j.1365-2966.2012.20796.x}

\bibitem[{Anderson {et~al.}(2014)Anderson, Bania, Balser, V., V., Johnstone, \&
  Armentrout}]{Anderson2014}
Anderson, L.~D., Bania, T.~M., Balser, D.~S., {et~al.} 2014, ApJS, 212, 1,
  \dodoi{10.1088/0067-0049/212/1/1}

\bibitem[{{Anderson} {et~al.}(2015){Anderson}, {Hough}, {Wenger}, {Bania}, \&
  {Balser}}]{Anderson2015a}
{Anderson}, L.~D., {Hough}, L.~A., {Wenger}, T.~V., {Bania}, T.~M., \&
  {Balser}, D.~S. 2015, ApJ, 810, 42, \dodoi{10.1088/0004-637X/810/1/42}

\bibitem[{Anderson {et~al.}(2015)Anderson, Deharveng, Zavagno, Tremblin, Lowe,
  Cunningham, Jones, Mullins, \& Redman}]{Anderson2015}
Anderson, L.~D., Deharveng, L., Zavagno, A., {et~al.} 2015, ApJ, 800, 101,
  \dodoi{10.1088/0004-637X/800/2/101}

\bibitem[{Balser(2006)}]{Balser2006}
Balser, D.~S. 2006, AJ, 132, 2326, \dodoi{10.1086/508515}

\bibitem[{{Balser} {et~al.}(2001){Balser}, {Goss}, \& {De Pree}}]{Balser2001}
{Balser}, D.~S., {Goss}, W.~M., \& {De Pree}, C.~G. 2001, AJ, 121, 371,
  \dodoi{10.1086/318028}

\bibitem[{{Balser} {et~al.}(2011){Balser}, {Rood}, {Bania}, \&
  {Anderson}}]{Balser2011}
{Balser}, D.~S., {Rood}, R.~T., {Bania}, T.~M., \& {Anderson}, L.~D. 2011, ApJ,
  738, 27, \dodoi{10.1088/0004-637X/738/1/27}

\bibitem[{Bania {et~al.}(2014)Bania, Wenger, Balser, \& Anderson}]{Bania2014}
Bania, T., Wenger, T., Balser, D., \& Anderson, L. 2014, tmbidl: TMBIDL v7.1,
  Zenodo.
\newblock \url{dx.doi.org/10.5281/zenodo.32790}

\bibitem[{{Beckman} {et~al.}(1998){Beckman}, {Rozas}, \&
  {Knapen}}]{Beckman1998}
{Beckman}, J.~E., {Rozas}, M., \& {Knapen}, J.~H. 1998, PASA, 15, 83,
  \dodoi{10.1071/AS98083}

\bibitem[{{Beckman} {et~al.}(2000){Beckman}, {Rozas}, {Zurita}, {Watson}, \&
  {Knapen}}]{Beckman2000}
{Beckman}, J.~E., {Rozas}, M., {Zurita}, A., {Watson}, R.~A., \& {Knapen},
  J.~H. 2000, AJ, 119, 2728, \dodoi{{10.1086/301380}}

\bibitem[{{Bell} {et~al.}(2011){Bell}, {Avery}, {MacLeod}, \&
  {Vall{\'e}e}}]{Bell2011}
{Bell}, M.~B., {Avery}, L.~W., {MacLeod}, J.~M., \& {Vall{\'e}e}, J.~P. 2011,
  \apss, 333, 377, \dodoi{10.1007/s10509-011-0662-5}

\bibitem[{{Brocklehurst} \& {Salem}(1977)}]{Brocklehurst1977}
{Brocklehurst}, M., \& {Salem}, M. 1977, Computer Physics Communications, 13,
  39, \dodoi{10.1016/0010-4655(77)90025-X}

\bibitem[{{Broos} {et~al.}(2007){Broos}, {Feigelson}, {Townsley}, {Getman},
  {Wang}, {Garmire}, {Jiang}, \& {Tsuboi}}]{Broos2007}
{Broos}, P.~S., {Feigelson}, E.~D., {Townsley}, L.~K., {et~al.} 2007, ApJS,
  169, 353, \dodoi{10.1086/512068}

\bibitem[{{Cesaroni} {et~al.}(1994){Cesaroni}, {Churchwell}, {Hofner},
  {Walmsley}, \& {Kurtz}}]{Cesaroni1994}
{Cesaroni}, R., {Churchwell}, E., {Hofner}, P., {Walmsley}, C.~M., \& {Kurtz},
  S. 1994, A\&A, 288, 903

\bibitem[{Churchwell {et~al.}(1974)Churchwell, Mezger, \&
  Huchtmeier}]{Churchwell1974}
Churchwell, E., Mezger, P.~G., \& Huchtmeier, W. 1974, A\&A, 32, 283

\bibitem[{Condon \& Ransom(2016)}]{Condon2016}
Condon, J.~J., \& Ransom, S.~M. 2016, Essential Radio Astronomy, ed. D.~N.
  Spergel (Princeton University Press)

\bibitem[{{Danks}(1970)}]{Danks1970}
{Danks}, A.~C. 1970, A\&A, 9, 175

\bibitem[{Domg{\"o}rgen \& Mathis(1994)}]{Domgoergen1994}
Domg{\"o}rgen, H., \& Mathis, J.~S. 1994, ApJ, 428, 647, \dodoi{10.1086/174275}

\bibitem[{Draine(2011)}]{Draine2011}
Draine, B.~T. 2011, Physics of the Interstellar and Intergalactic Medium, ed.
  D.~N. Spergel (Princeton)

\bibitem[{{Dupree} \& {Goldberg}(1970)}]{Dupree1970}
{Dupree}, A.~K., \& {Goldberg}, L. 1970, ARA\&A, 8, 231,
  \dodoi{10.1146/annurev.aa.08.090170.001311}

\bibitem[{{Georgelin} {et~al.}(1973){Georgelin}, {Georgelin}, \&
  {Roux}}]{Georgelin1973}
{Georgelin}, Y.~M., {Georgelin}, Y.~P., \& {Roux}, S. 1973, A\&A, 25, 337

\bibitem[{{Giammanco} {et~al.}(2005){Giammanco}, {Beckman}, \&
  {Cedr{\'e}s}}]{Giammanco2005}
{Giammanco}, C., {Beckman}, J.~E., \& {Cedr{\'e}s}, B. 2005, A\&A, 438, 599,
  \dodoi{10.1051/0004-6361:20042268}

\bibitem[{Gottesman \& Gordon(1970)}]{Gottesman1970}
Gottesman, S.~T., \& Gordon, M.~A. 1970, ApJL, 162, L93, \dodoi{10.1086/180631}

\bibitem[{{Haffner} {et~al.}(1999){Haffner}, {Reynolds}, \&
  {Tufte}}]{Haffner1999}
{Haffner}, L.~M., {Reynolds}, R.~J., \& {Tufte}, S.~L. 1999, ApJ, 523, 223,
  \dodoi{10.1086/307734}

\bibitem[{Haffner {et~al.}(2009)Haffner, Dettmar, Beckman, Wood, Slavin,
  Giammanco, Madsen, Zurita, \& Reynolds}]{Haffner2009}
Haffner, L.~M., Dettmar, R.-J., Beckman, J.~E., {et~al.} 2009, RvMP, 81, 969,
  \dodoi{10.1103/RevModPhys.81.969}

\bibitem[{{Hoang-Binh}(1972)}]{Hoang-Binh1972}
{Hoang-Binh}, D. 1972, in Les Spectres des Astres dans l'Infrarouge et les
  Microondes, 367--370

\bibitem[{{Hollenbach} \& {Tielens}(1997)}]{Hollenbach1997}
{Hollenbach}, D.~J., \& {Tielens}, A.~G.~G.~M. 1997, ARA\&A, 35, 179,
  \dodoi{10.1146/annurev.astro.35.1.179}

\bibitem[{Hollenbach \& Tielens(1999)}]{Hollenbach1999}
Hollenbach, D.~J., \& Tielens, A. G. G.~M. 1999, RvMP, 71, 173,
  \dodoi{10.1103/RevModPhys.71.173}

\bibitem[{Hoopes \& Walterbos(2003)}]{Hoopes2003}
Hoopes, C.~G., \& Walterbos, R. A.~M. 2003, ApJ, 586, 902

\bibitem[{{Kitchin}(1987)}]{Kitchin1987}
{Kitchin}, C.~R. 1987, Stars, nebulae and the interstellar medium.
  Observational physics and astrophysics (Bristol: Hilger, 1987)

\bibitem[{{Konovalenko} \& {Sodin}(1981)}]{Konovalenko1981}
{Konovalenko}, A.~A., \& {Sodin}, L.~G. 1981, \nat, 294, 135,
  \dodoi{10.1038/294135a0}

\bibitem[{{Lahulla}(1985)}]{Lahulla1985}
{Lahulla}, J.~F. 1985, A\&AS, 61, 537

\bibitem[{{Lenz} \& {Ayres}(1992)}]{Lenz1992}
{Lenz}, D.~D., \& {Ayres}, T.~R. 1992, PASP, 104, 1104, \dodoi{10.1086/133096}

\bibitem[{{Lichten} {et~al.}(1979){Lichten}, {Rodriguez}, \&
  {Chaisson}}]{Lichten1979}
{Lichten}, S.~M., {Rodriguez}, L.~F., \& {Chaisson}, E.~J. 1979, ApJ, 229, 524,
  \dodoi{10.1086/156985}

\bibitem[{{Luisi} {et~al.}(2016){Luisi}, {Anderson}, {Balser}, {Bania}, \&
  {Wenger}}]{Luisi2016}
{Luisi}, M., {Anderson}, L.~D., {Balser}, D.~S., {Bania}, T.~M., \& {Wenger},
  T.~V. 2016, ApJ, 824, 125, \dodoi{10.3847/0004-637X/824/2/125}

\bibitem[{{Luisi} {et~al.}(2017){Luisi}, {Anderson}, {Balser}, {Wenger}, \&
  {Bania}}]{Luisi2017}
{Luisi}, M., {Anderson}, L.~D., {Balser}, D.~S., {Wenger}, T.~V., \& {Bania},
  T.~M. 2017, ApJ, 849, 117, \dodoi{10.3847/1538-4357/aa8fd2}

\bibitem[{{Luisi} {et~al.}(2018){Luisi}, {Anderson}, {Bania}, {Balser},
  {Wenger}, \& {Kepley}}]{Luisi2018}
{Luisi}, M., {Anderson}, L.~D., {Bania}, T.~M., {et~al.} 2018, PASP, 130,
  084101, \dodoi{10.1088/1538-3873/aac8e9}

\bibitem[{Maddalena(2010)}]{Maddalena2010}
Maddalena, R.~J. 2010, Theoretical Ratio of Beam Efficiency to Aperture
  Efficiency, Tech. rep., National Radio Astronomy Observatory

\bibitem[{Maddalena(2012)}]{Maddalena2012}
---. 2012, Modeling the Elevation and Frequency Dependence of Aperture
  Efficiency for the GBT's Pipeline, Tech. rep., National Radio Astronomy
  Observatory

\bibitem[{Madsen {et~al.}(2006)Madsen, Reynolds, \& Haffner}]{Madsen2006}
Madsen, G.~J., Reynolds, R.~J., \& Haffner, L.~M. 2006, ApJ, 652, 401,
  \dodoi{10.1086/508441}

\bibitem[{{Martins} {et~al.}(2005){Martins}, {Schaerer}, \&
  {Hillier}}]{Martins2005}
{Martins}, F., {Schaerer}, D., \& {Hillier}, D.~J. 2005, A\&A, 436, 1049,
  \dodoi{10.1051/0004-6361:20042386}

\bibitem[{{McGee} \& {Newton}(1981)}]{McGee1981}
{McGee}, R.~X., \& {Newton}, L.~M. 1981, MNRAS, 196, 889,
  \dodoi{10.1093/mnras/196.4.889}

\bibitem[{{Mehringer}(1994)}]{Mehringer1994}
{Mehringer}, D.~M. 1994, ApJS, 91, 713, \dodoi{10.1086/191953}

\bibitem[{Mezger(1978)}]{Mezger1978}
Mezger, P.~G. 1978, A\&A, 70, 565

\bibitem[{{Mezger} \& {Ellis}(1968)}]{Mezger1968}
{Mezger}, P.~G., \& {Ellis}, S.~A. 1968, ApJL, 1, 159

\bibitem[{{Mois{\'e}s} {et~al.}(2011){Mois{\'e}s}, {Damineli}, {Figuer{\^e}do},
  {Blum}, {Conti}, \& {Barbosa}}]{Moises2011}
{Mois{\'e}s}, A.~P., {Damineli}, A., {Figuer{\^e}do}, E., {et~al.} 2011, MNRAS,
  411, 705, \dodoi{10.1111/j.1365-2966.2010.17713.x}

\bibitem[{{O'Dell} {et~al.}(2017){O'Dell}, {Kollatschny}, \&
  {Ferland}}]{ODell2017}
{O'Dell}, C.~R., {Kollatschny}, W., \& {Ferland}, G.~J. 2017, ApJ, 837, 151,
  \dodoi{10.3847/1538-4357/aa6198}

\bibitem[{Oey \& Kennicutt(1997)}]{Oey1997}
Oey, M.~S., \& Kennicutt, R. C.~J. 1997, MNRAS, 291, 827

\bibitem[{Osterbrock(1989)}]{Osterbrock1989}
Osterbrock, D. 1989, Astrophysics of gaseous nebulae and active galactic nuclei
  (University Science Books)

\bibitem[{{Pankonin} {et~al.}(1980){Pankonin}, {Walmsley}, \&
  {Thum}}]{Pankonin1980}
{Pankonin}, V., {Walmsley}, C.~M., \& {Thum}, C. 1980, A\&A, 89, 173

\bibitem[{{Peimbert} {et~al.}(1992){Peimbert}, {Rodriguez}, {Bania}, {Rood}, \&
  {Wilson}}]{Peimbert1992}
{Peimbert}, M., {Rodriguez}, L.~F., {Bania}, T.~M., {Rood}, R.~T., \& {Wilson},
  T.~L. 1992, ApJ, 395, 484, \dodoi{10.1086/171668}

\bibitem[{{Pellegrini} {et~al.}(2012){Pellegrini}, {Oey}, {Winkler}, {Points},
  {Smith}, {Jaskot}, \& {Zastrow}}]{Pellegrini2012}
{Pellegrini}, E.~W., {Oey}, M.~S., {Winkler}, P.~F., {et~al.} 2012, ApJ, 755,
  40, \dodoi{10.1088/0004-637X/755/1/40}

\bibitem[{Quireza {et~al.}(2006)Quireza, Rood, Bania, Balser, \&
  Maciel}]{Quireza2006}
Quireza, C., Rood, R.~T., Bania, T.~M., Balser, D.~S., \& Maciel, W.~J. 2006,
  ApJ, 653, 1226, \dodoi{10.1086/508803}

\bibitem[{{Reynolds} {et~al.}(1995){Reynolds}, {Tufte}, {Kung}, {McCullough},
  \& {Heiles}}]{Reynolds1995a}
{Reynolds}, R.~J., {Tufte}, S.~L., {Kung}, D.~T., {McCullough}, P.~R., \&
  {Heiles}, C. 1995, ApJ, 448, 715, \dodoi{10.1086/175999}

\bibitem[{{Roelfsema} {et~al.}(1992){Roelfsema}, {Goss}, \&
  {Mallik}}]{Roelfsema1992}
{Roelfsema}, P.~R., {Goss}, W.~M., \& {Mallik}, D.~C.~V. 1992, ApJ, 394, 188,
  \dodoi{10.1086/171570}

\bibitem[{{Roshi} {et~al.}(2017){Roshi}, {Churchwell}, \&
  {Anderson}}]{Roshi2017}
{Roshi}, D.~A., {Churchwell}, E., \& {Anderson}, L.~D. 2017, ApJ, 838, 144,
  \dodoi{10.3847/1538-4357/aa662b}

\bibitem[{{Roshi} {et~al.}(2002){Roshi}, {Kantharia}, \&
  {Anantharamaiah}}]{Roshi2002}
{Roshi}, D.~A., {Kantharia}, N.~G., \& {Anantharamaiah}, K.~R. 2002, A\&A, 391,
  1097, \dodoi{10.1051/0004-6361:20020899}

\bibitem[{Roshi {et~al.}(2012)Roshi, Plunkett, Rosero, \& Sravani}]{Roshi2012}
Roshi, D.~A., Plunkett, A., Rosero, V., \& Sravani, V. 2012, ApJ, 749, 49,
  \dodoi{10.1088/0004-637X/749/1/49}

\bibitem[{{Rozas} {et~al.}(1998){Rozas}, {Castaneda}, \& {Beckman}}]{Rozas1998}
{Rozas}, M., {Castaneda}, H.~O., \& {Beckman}, J.~E. 1998, A\&A, 330, 873

\bibitem[{{Rubin}(1985)}]{Rubin1985}
{Rubin}, R.~H. 1985, ApJS, 57, 349, \dodoi{10.1086/191007}

\bibitem[{{Rubin} {et~al.}(2003){Rubin}, {Martin}, {Dufour}, {Ferland},
  {Blagrave}, {Liu}, {Nguyen}, \& {Baldwin}}]{Rubin2003}
{Rubin}, R.~H., {Martin}, P.~G., {Dufour}, R.~J., {et~al.} 2003, MNRAS, 340,
  362, \dodoi{10.1046/j.1365-8711.2003.06185.x}

\bibitem[{{Salem} \& {Brocklehurst}(1979)}]{Salem1979}
{Salem}, M., \& {Brocklehurst}, M. 1979, ApJS, 39, 633, \dodoi{10.1086/190588}

\bibitem[{{Shaver}(1980)}]{Shaver1980}
{Shaver}, P.~A. 1980, A\&A, 90, 34

\bibitem[{{Shaver} \& {Wilson}(1979)}]{Shaver1979}
{Shaver}, P.~A., \& {Wilson}, T.~L. 1979, A\&A, 79, 312

\bibitem[{{Sota} {et~al.}(2011){Sota}, {Ma{\'{\i}}z Apell{\'a}niz}, {Walborn},
  {Alfaro}, {Barb{\'a}}, {Morrell}, {Gamen}, \& {Arias}}]{Sota2011}
{Sota}, A., {Ma{\'{\i}}z Apell{\'a}niz}, J., {Walborn}, N.~R., {et~al.} 2011,
  ApJS, 193, 24, \dodoi{10.1088/0067-0049/193/2/24}

\bibitem[{{Str{\"o}mgren}(1939)}]{Stroemgren1939}
{Str{\"o}mgren}, B. 1939, ApJ, 89, 526, \dodoi{10.1086/144074}

\bibitem[{{Thum} {et~al.}(1980){Thum}, {Mezger}, \& {Pankonin}}]{Thum1980}
{Thum}, C., {Mezger}, P.~G., \& {Pankonin}, V. 1980, A\&A, 87, 269

\bibitem[{{Walmsley} \& {Watson}(1982)}]{Walmsley1982}
{Walmsley}, C.~M., \& {Watson}, W.~D. 1982, ApJ, 260, 317,
  \dodoi{10.1086/160256}

\bibitem[{{Watson} {et~al.}(2008){Watson}, {Povich}, {Churchwell}, {Babler},
  {Chunev}, {Hoare}, {Indebetouw}, {Meade}, {Robitaille}, \&
  {Whitney}}]{Watson2008}
{Watson}, C., {Povich}, M.~S., {Churchwell}, E.~B., {et~al.} 2008, ApJ, 681,
  1341, \dodoi{10.1086/588005}

\bibitem[{{Weber} {et~al.}(2018){Weber}, {Pauldrach}, \&
  {Hoffmann}}]{Weber2018}
{Weber}, J.~A., {Pauldrach}, A.~W.~A., \& {Hoffmann}, T.~L. 2018, ArXiv
  e-prints

\bibitem[{Wenger {et~al.}(2013)Wenger, Bania, Balser, \& Anderson}]{Wenger2013}
Wenger, T.~V., Bania, T.~M., Balser, D.~S., \& Anderson, L.~D. 2013, ApJ, 764,
  34, \dodoi{10.1088/0004-637X/764/1/34}

\bibitem[{{Wilson} {et~al.}(2015){Wilson}, {Bania}, \& {Balser}}]{Wilson2015}
{Wilson}, T.~L., {Bania}, T.~M., \& {Balser}, D.~S. 2015, ApJ, 812, 45,
  \dodoi{10.1088/0004-637X/812/1/45}

\bibitem[{{Wood} {et~al.}(2010){Wood}, {Hill}, {Joung}, {Mac Low}, {Benjamin},
  {Haffner}, {Reynolds}, \& {Madsen}}]{Wood2010}
{Wood}, K., {Hill}, A.~S., {Joung}, M.~R., {et~al.} 2010, ApJ, 721, 1397,
  \dodoi{10.1088/0004-637X/721/2/1397}

\bibitem[{Wood \& Mathis(2004)}]{Wood2004}
Wood, K., \& Mathis, J.~S. 2004, MNRAS, 353, 1126,
  \dodoi{10.1111/j.1365-2966.2004.0784}

\bibitem[{{Zuckerman} {et~al.}(1967){Zuckerman}, {Palmer}, {Penfield}, \&
  {Lilley}}]{Zuckerman1967}
{Zuckerman}, B., {Palmer}, P., {Penfield}, H., \& {Lilley}, A.~E. 1967, ApJL,
  149, L61, \dodoi{10.1086/180057}

\bibitem[{Zurita {et~al.}(2002)Zurita, Beckman, Rozas, \& Ryder}]{Zurita2002}
Zurita, A., Beckman, J.~E., Rozas, M., \& Ryder, S. 2002, A\&A, 386, 801,
  \dodoi{10.1051/0004-6361:20020212}

\bibitem[{{Zurita} {et~al.}(2000){Zurita}, {Rozas}, \& {Beckman}}]{Zurita2000}
{Zurita}, A., {Rozas}, M., \& {Beckman}, J.~E. 2000, A\&A, 363, 9

\end{thebibliography}

\appendix
\section{Observed RRL Parameters and Derived Source Properties}

\renewcommand\thetable{\thesection.\arabic{table}}
\setcounter{table}{0}

\renewcommand\thefigure{\thesection.\arabic{figure}}
\setcounter{figure}{0}

\iftrue
\startlongtable
\begin{deluxetable*}{lccccl}
\tabletypesize{\footnotesize}
\tablecaption{Observed Positions \label{tab:overview}}
\tablehead{Source  & \colhead{RA} & \colhead{Dec} & \colhead{$\ell$} & \colhead{$b$} & \colhead{PDR\tablenotemark{a}} \\
 &  \colhead{(J2000)} & \colhead{(J2000)} & \colhead{(deg.)} & \colhead{(deg.)} & }
\startdata
S206\_C    & 04:03:15.8 & 51:18:54    & 150.593 & $-$0.951   & in  \\
S206\_E0.5 & 04:03:20.1 & 51:16:57    & 150.623 & $-$0.968   & out \\
S206\_E1   & 04:03:24.2 & 51:15:02    & 150.652 & $-$0.985   & out \\
S206\_E2   & 04:03:32.8 & 51:11:07    & 150.712 & $-$1.019   & out \\
S206\_E3   & 04:03:41.1 & 51:07:15    & 150.771 & $-$1.053   & out \\
S206\_W0.5 & 04:03:11.5 & 51:20:51    & 150.563 & $-$0.934   & out \\
S206\_W1   & 04:03:07.5 & 51:22:46    & 150.534 & $-$0.917   & out \\
S206\_W2   & 04:02:58.8 & 51:26:41    & 150.474 & $-$0.883   & out \\
\enddata
\tablenotetext{a}{\,``out" = outside the PDR, ``on" = on the PDR, ``in" = inside the PDR.}

\end{deluxetable*}
\fi

\clearpage

\iftrue
\startlongtable
\begin{deluxetable*}{lclcccccccr}
\tabletypesize{\footnotesize}
\tablecaption{H, He, and C RRL Parameters \label{tab:alpha}}
 \tablehead{Source & \colhead{Transition} & \colhead{Element} & \colhead{$T_{\rm L}$} & \colhead{$\delta T_{\rm L}$} & \colhead{$\Delta V$} & \colhead{$\delta \Delta V$} & \colhead{$V_{\rm LSR}$} & \colhead{$\delta V_{\rm LSR}$} & \colhead{rms} & \colhead{t$_{\rm intg}$}\\
  & & & \colhead{(mK)} & \colhead{(mK)} & \colhead{(\kms)} & \colhead{(\kms)} & \colhead{(\kms)} & \colhead{(\kms)} & \colhead{(mK)} & \colhead{(min)}}
\startdata
S206\_C    & $\alpha$ & H  & $\phn \phn 320.5$         & $\phn   0.2$ & $27.8$     & $0.1$   & \phn $-26.0$          & $0.1$   & \phn   $3.1$ & 3  \\
           & $\alpha$ & He & $\phn \phn \phn 32.8$     & \phn $0.3$                                             & $21.5$     & $0.2$   & \phn $-26.0$          & $0.1$   &                                                        &  \\      
           & $\beta$ & H  & \phn \phn \phn 86.3               & \phn 0.3      & 27.1     & 0.1     & \phn $-$25.5       & 0.1     & \phn 3.6      &  \\
           & $\beta$ & He & \phn \phn \phn 10.4               & \phn 0.3      & 34.2     & 1.3     & \phn $-$22.0       & 0.5     &          &    \\  
           & $\gamma$ & H  & \phn \phn \phn 41.2               & \phn 0.4     & 23.3      & 0.3     & \phn $-$25.5       & 0.1     & \phn 4.4      &   \\    
S206\_E0.5 & $\alpha$ & H  & $\phn \phn 114.3$         & $\phn 0.1$                                             & $27.0$     & $0.1$   & \phn $-23.6$          & $0.1$   & \phn $1.5$                                             & 6  \\
           & $\alpha$ & He & $\phn \phn \phn 12.9$     & \phn $0.2$                                             & $18.5$     & $0.3$   & \phn $-23.5$          & $0.1$   &                                                        &    \\
           & $\beta$ & H  & \phn \phn \phn 33.0               & \phn 0.2      & 26.3     & 0.1     & \phn $-$23.1       & 0.1     & \phn 1.9      &  \\
\enddata
\tablecomments{All uncertainties are $\pm$1$\sigma$.}
\end{deluxetable*}
\fi

\iftrue 
\begin{figure*}
\centering
\begin{tabular}{cccc}
\includegraphics[width=.23\textwidth]{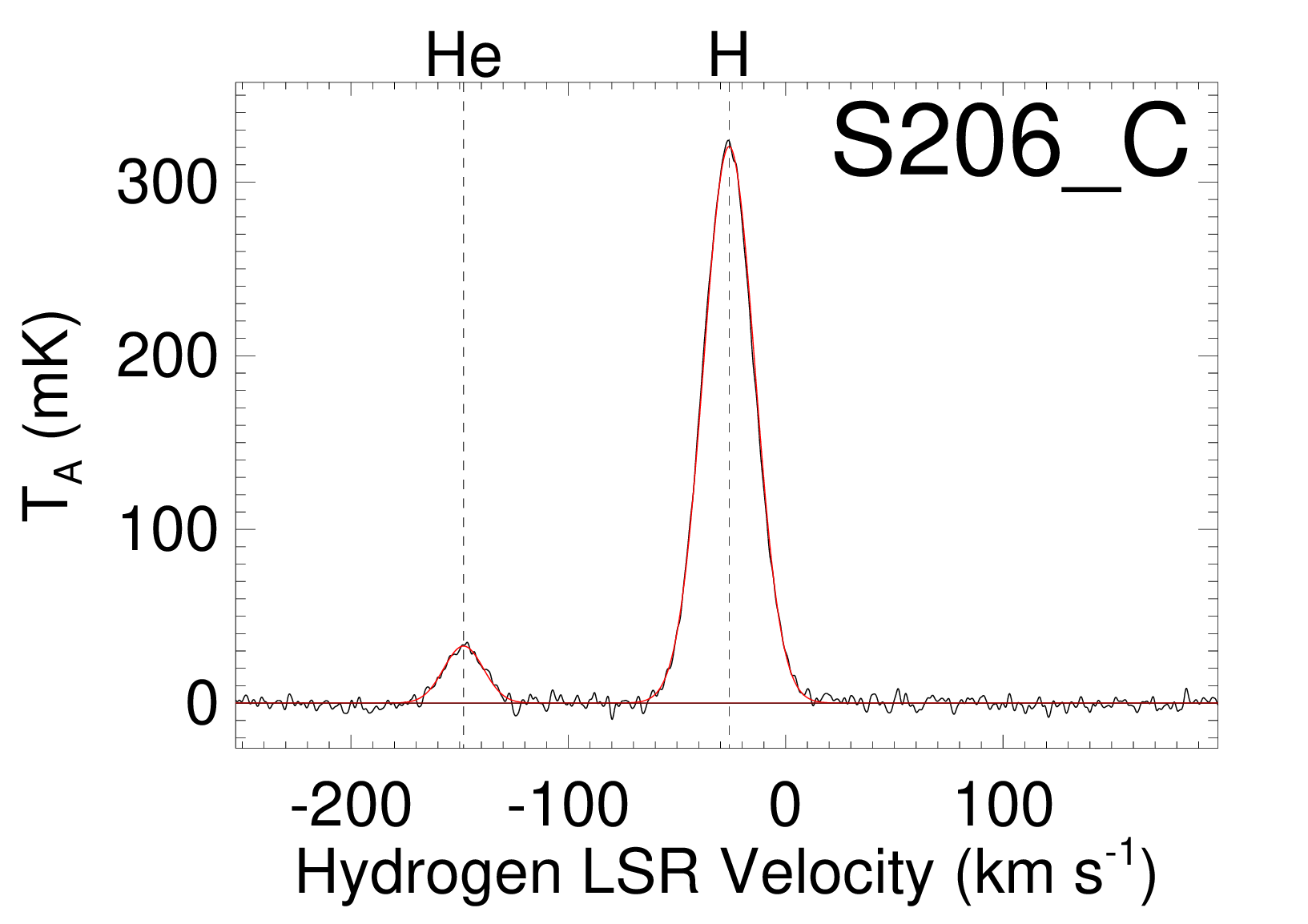} &
\includegraphics[width=.23\textwidth]{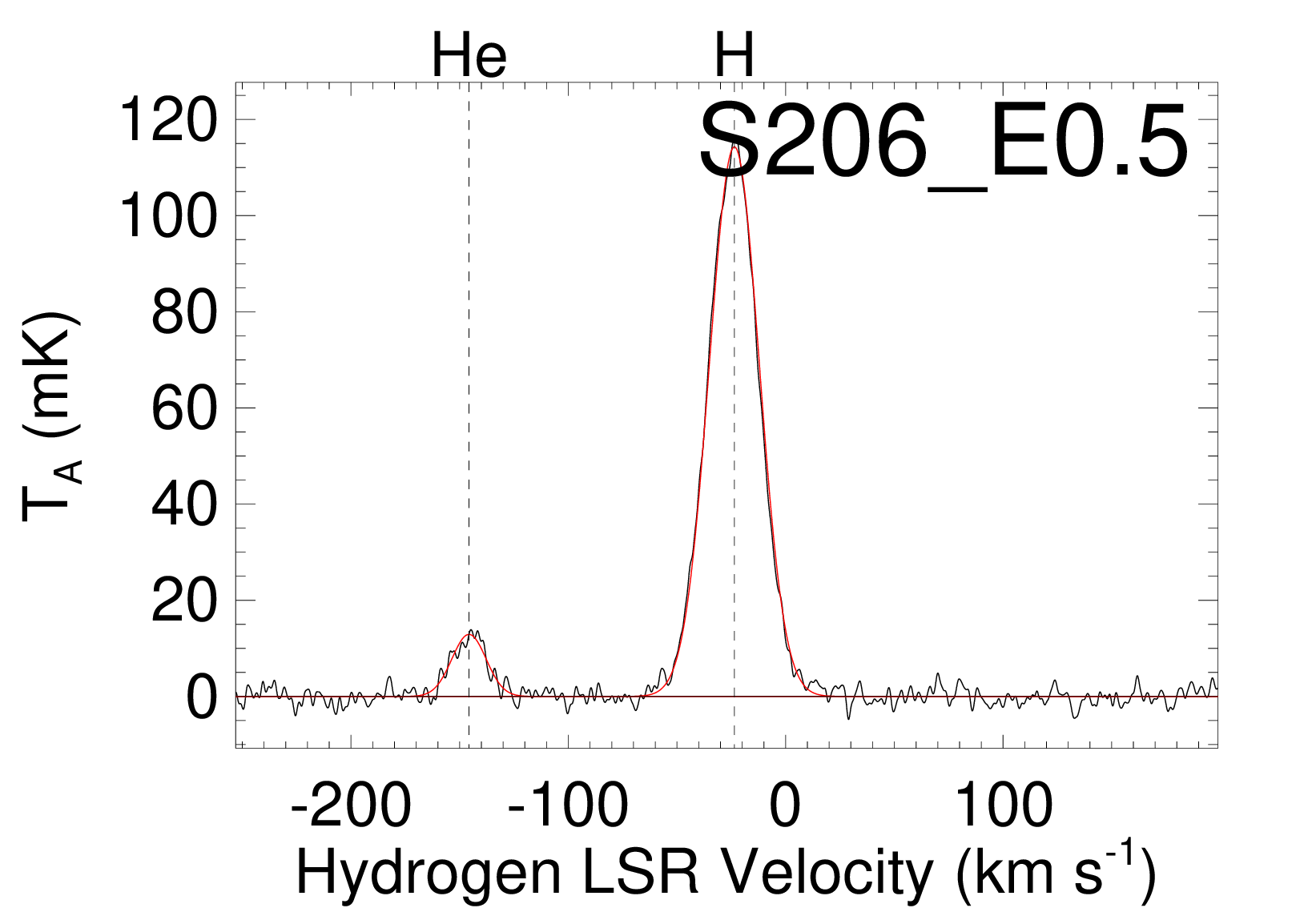} &
\includegraphics[width=.23\textwidth]{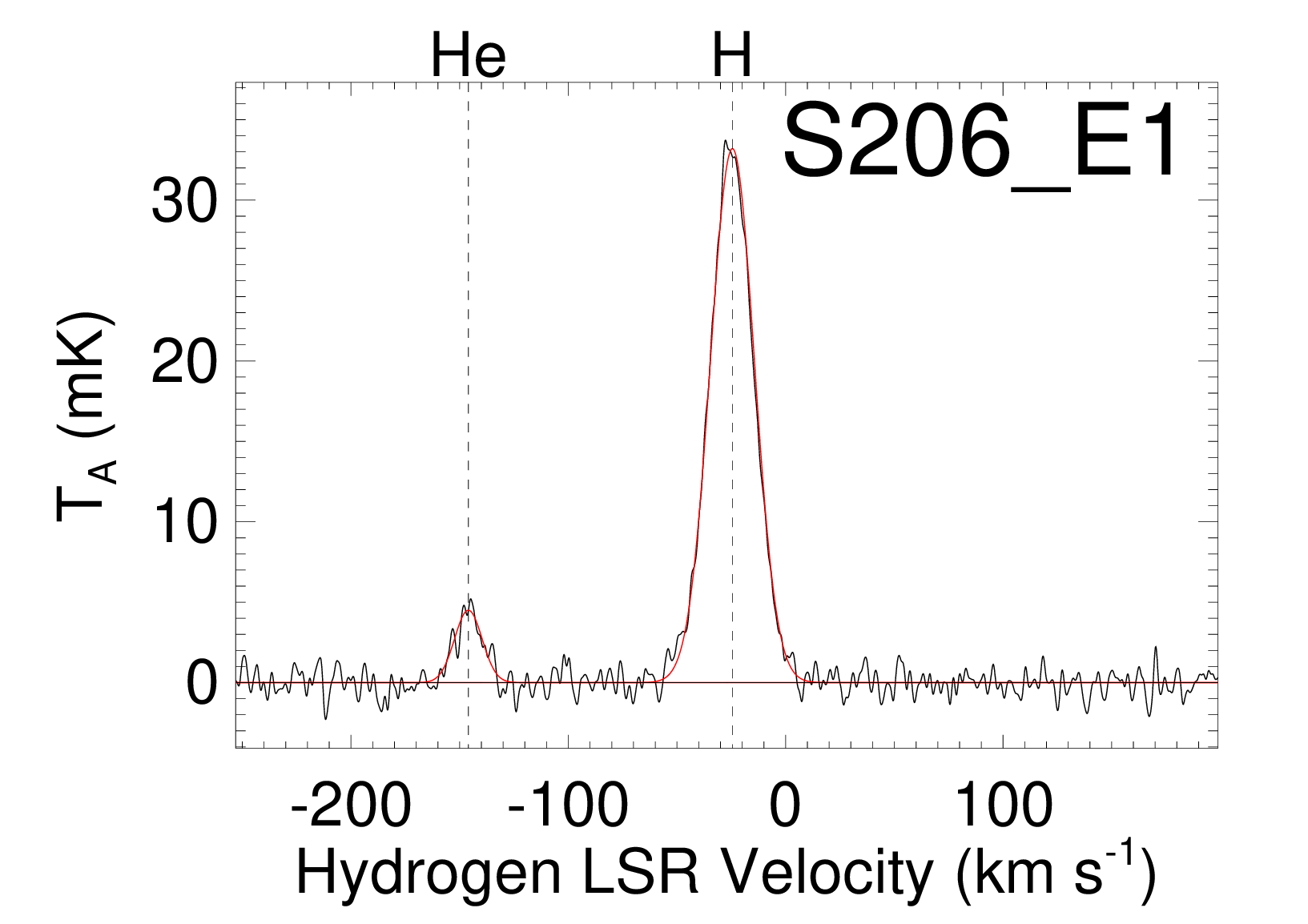} &
\includegraphics[width=.23\textwidth]{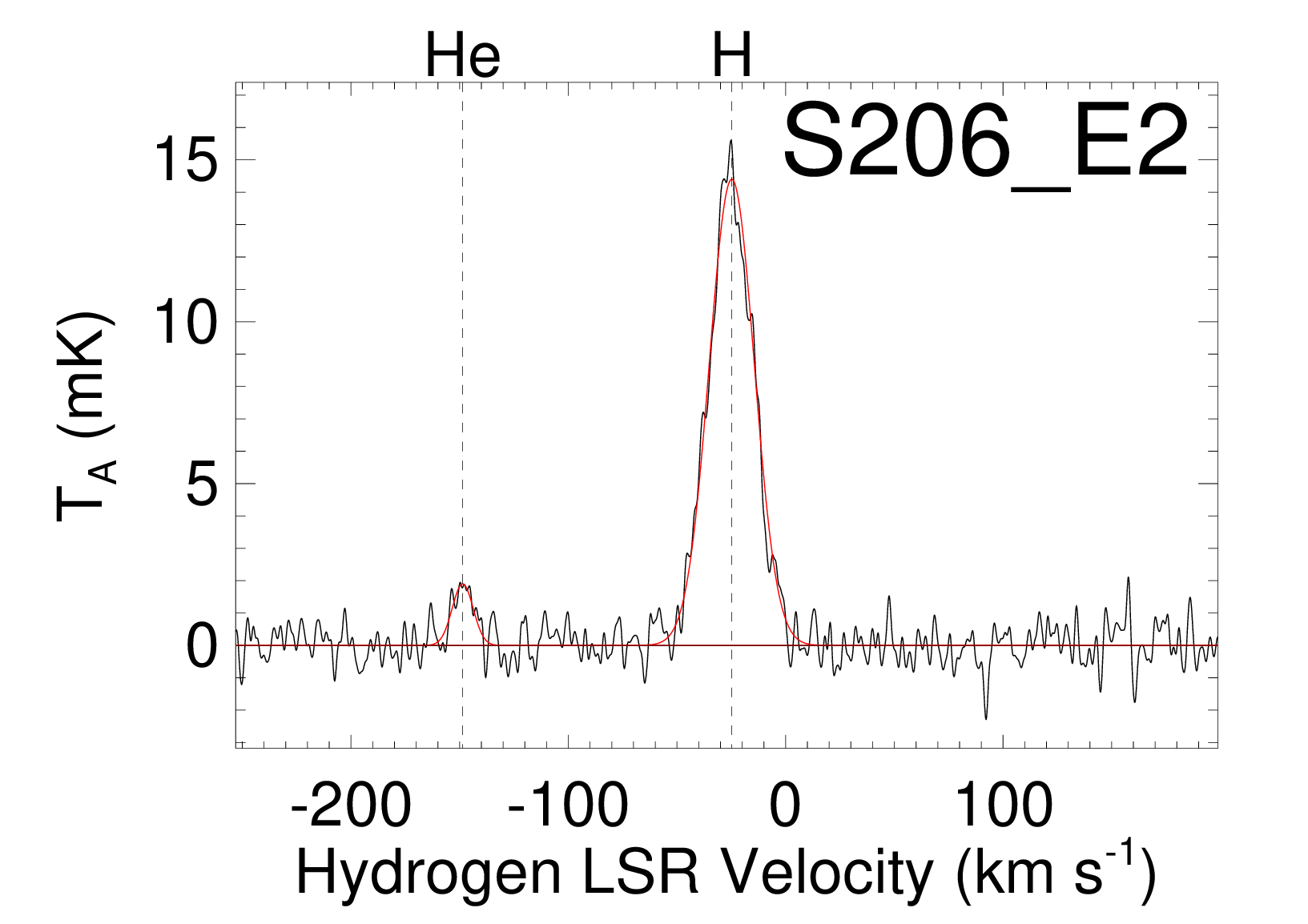} \\
\includegraphics[width=.23\textwidth]{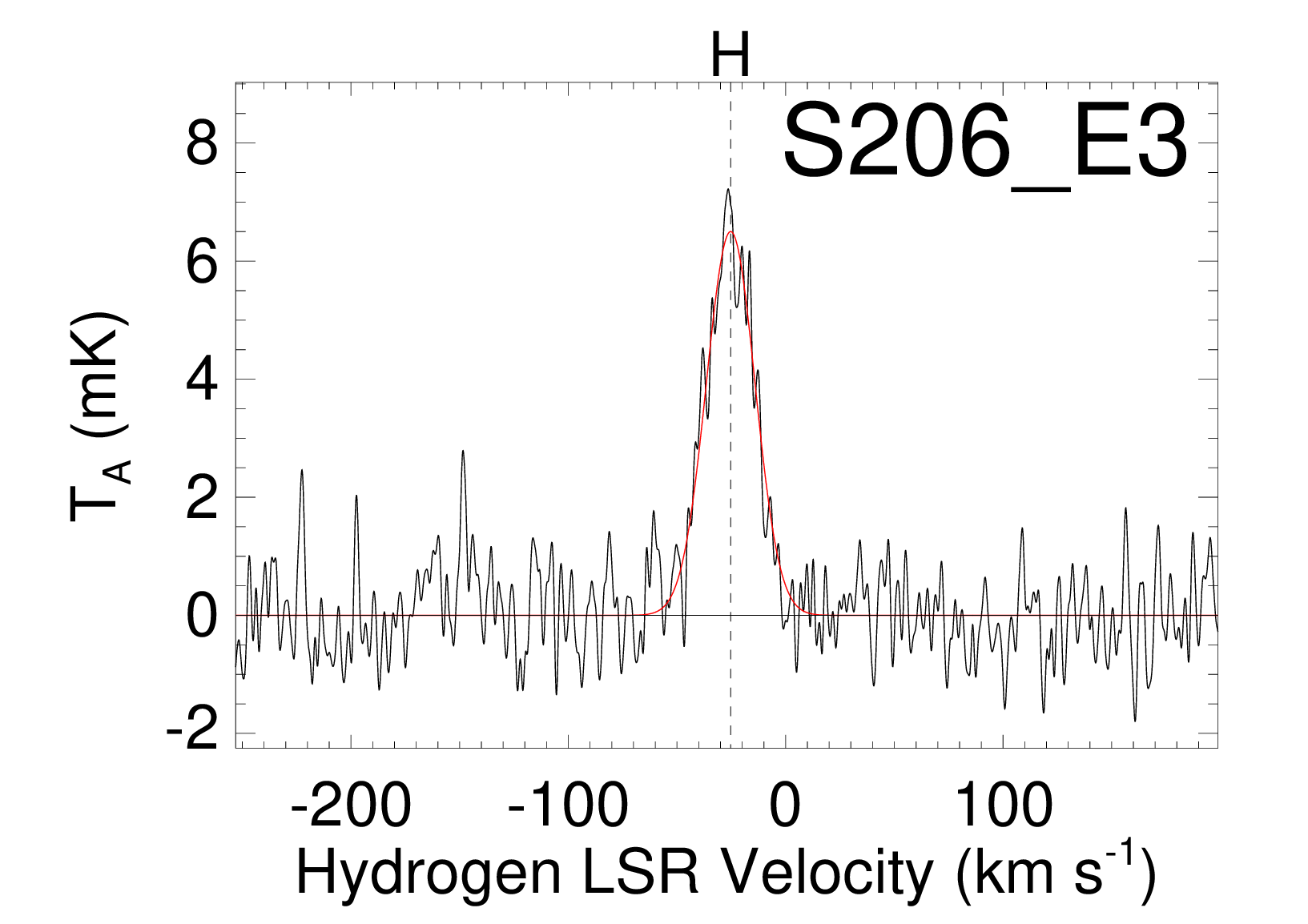} &
\includegraphics[width=.23\textwidth]{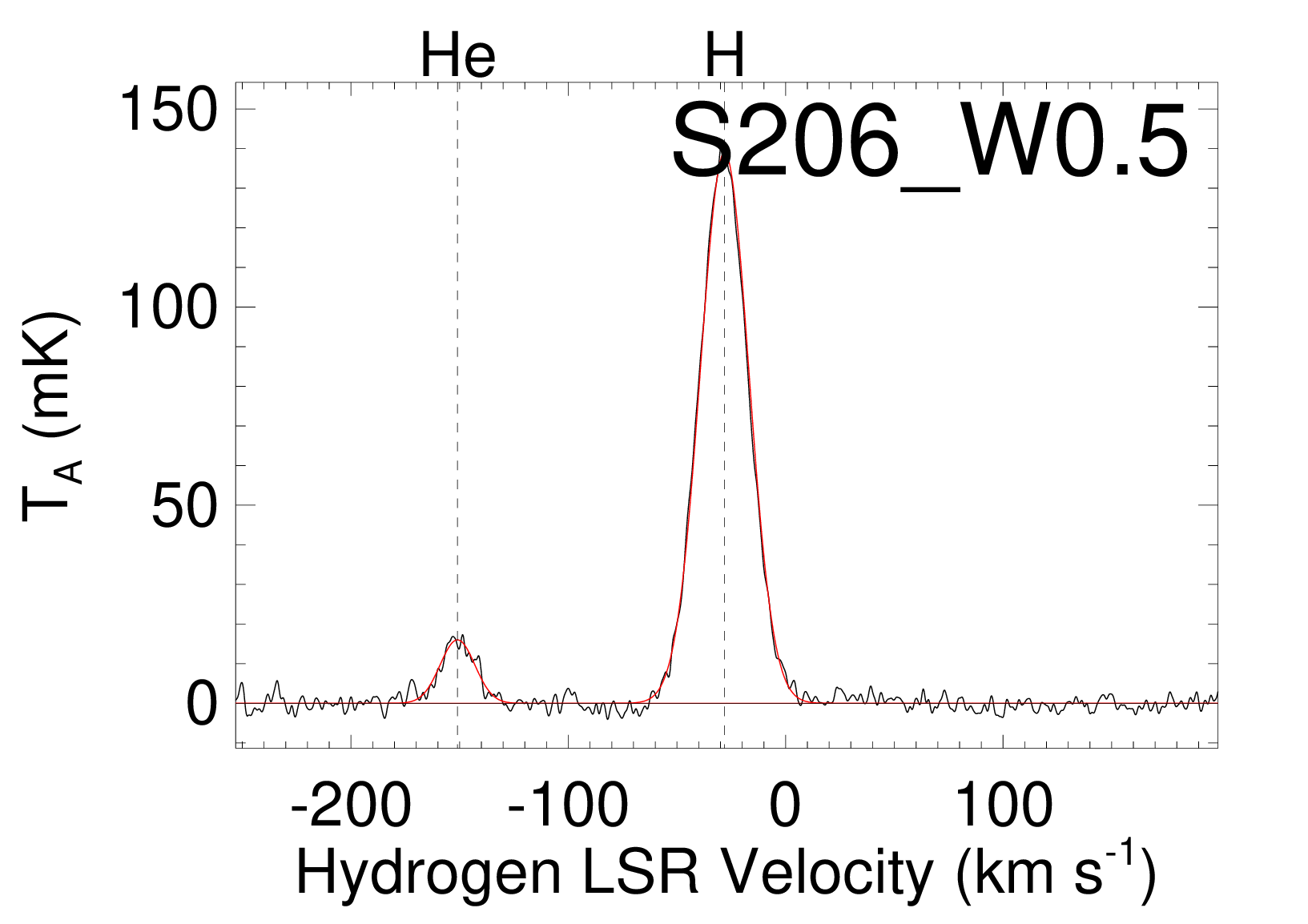} &
\includegraphics[width=.23\textwidth]{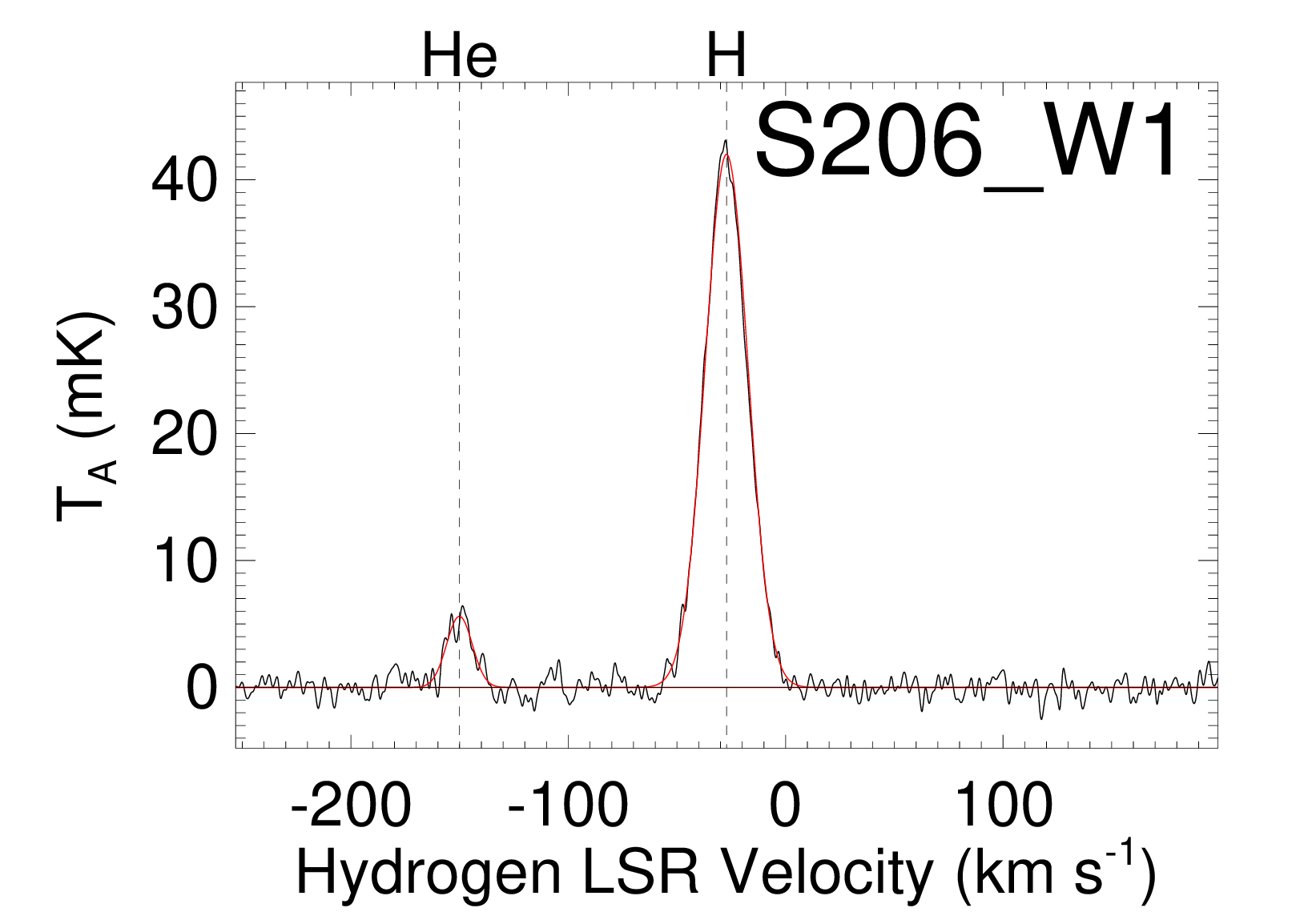} &
\includegraphics[width=.23\textwidth]{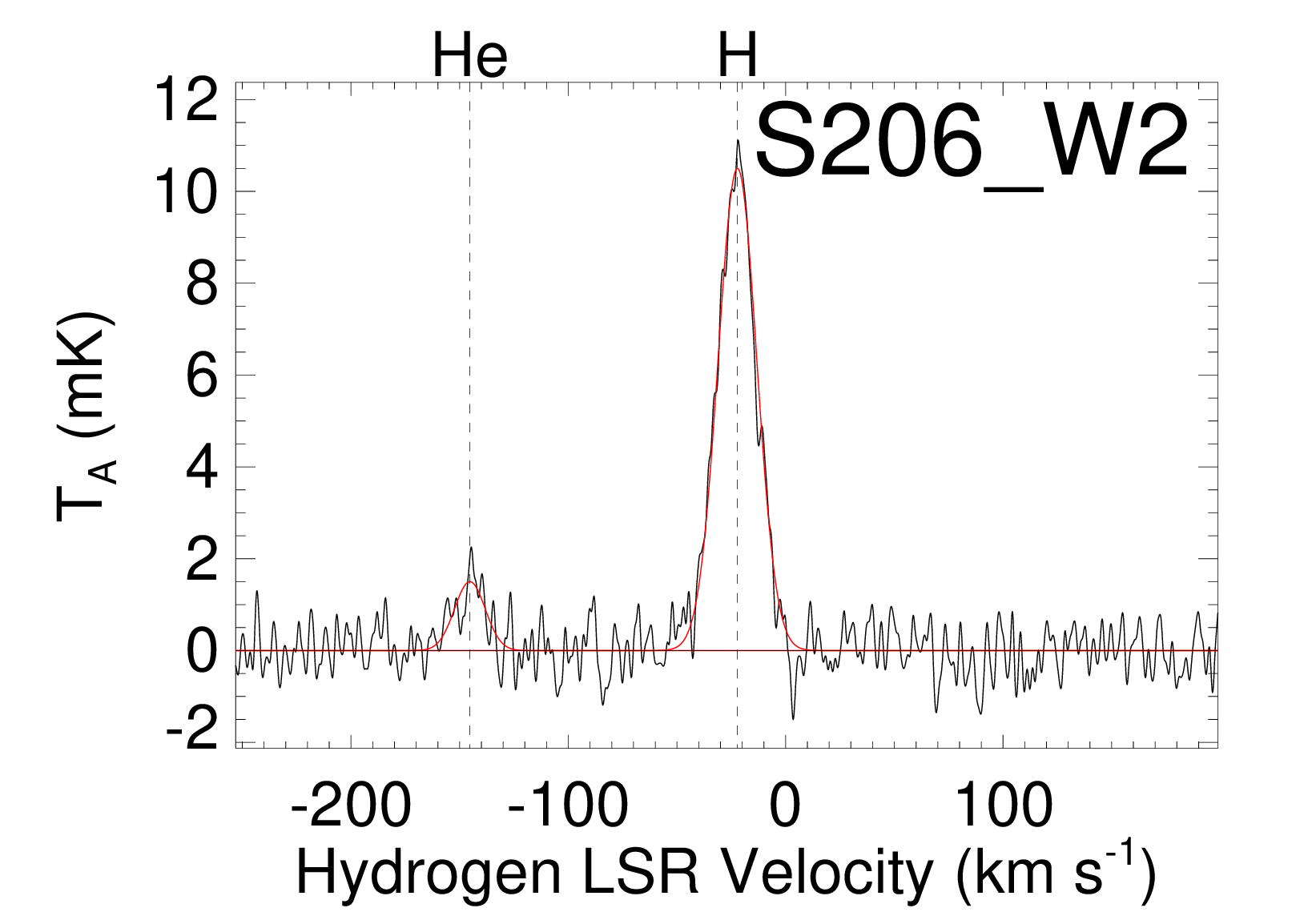} \\
\includegraphics[width=.23\textwidth]{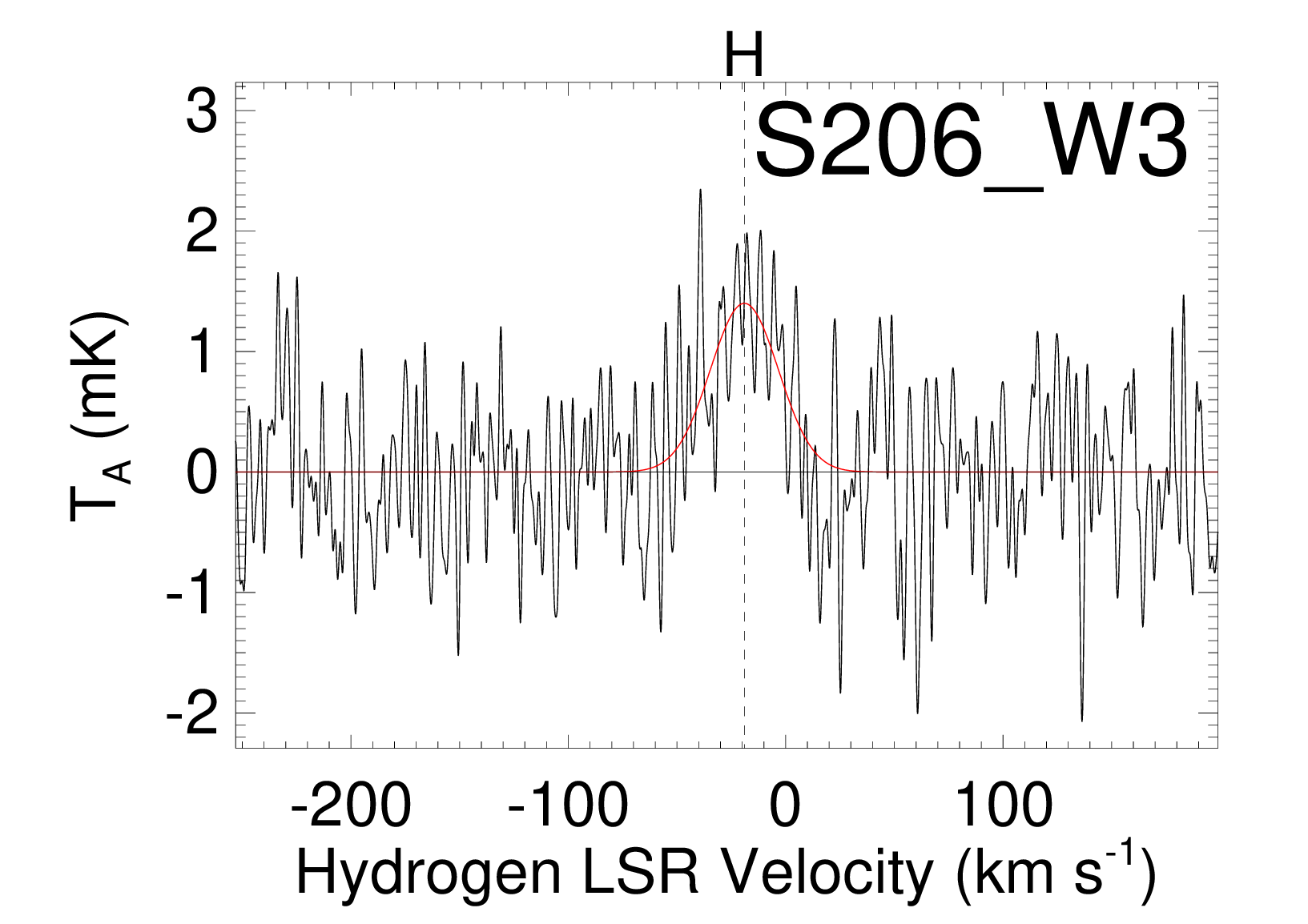} &
\includegraphics[width=.23\textwidth]{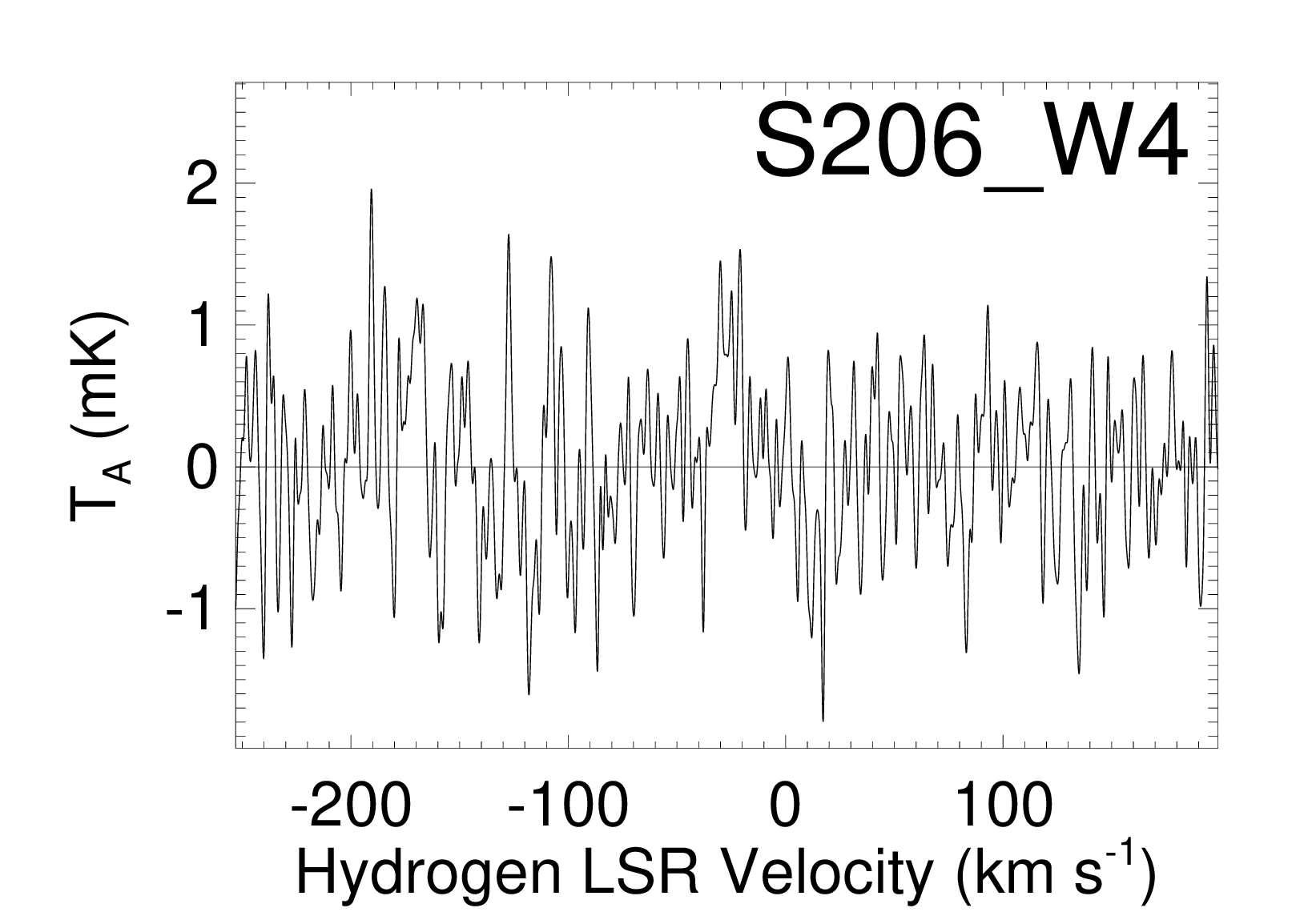} &
\includegraphics[width=.23\textwidth]{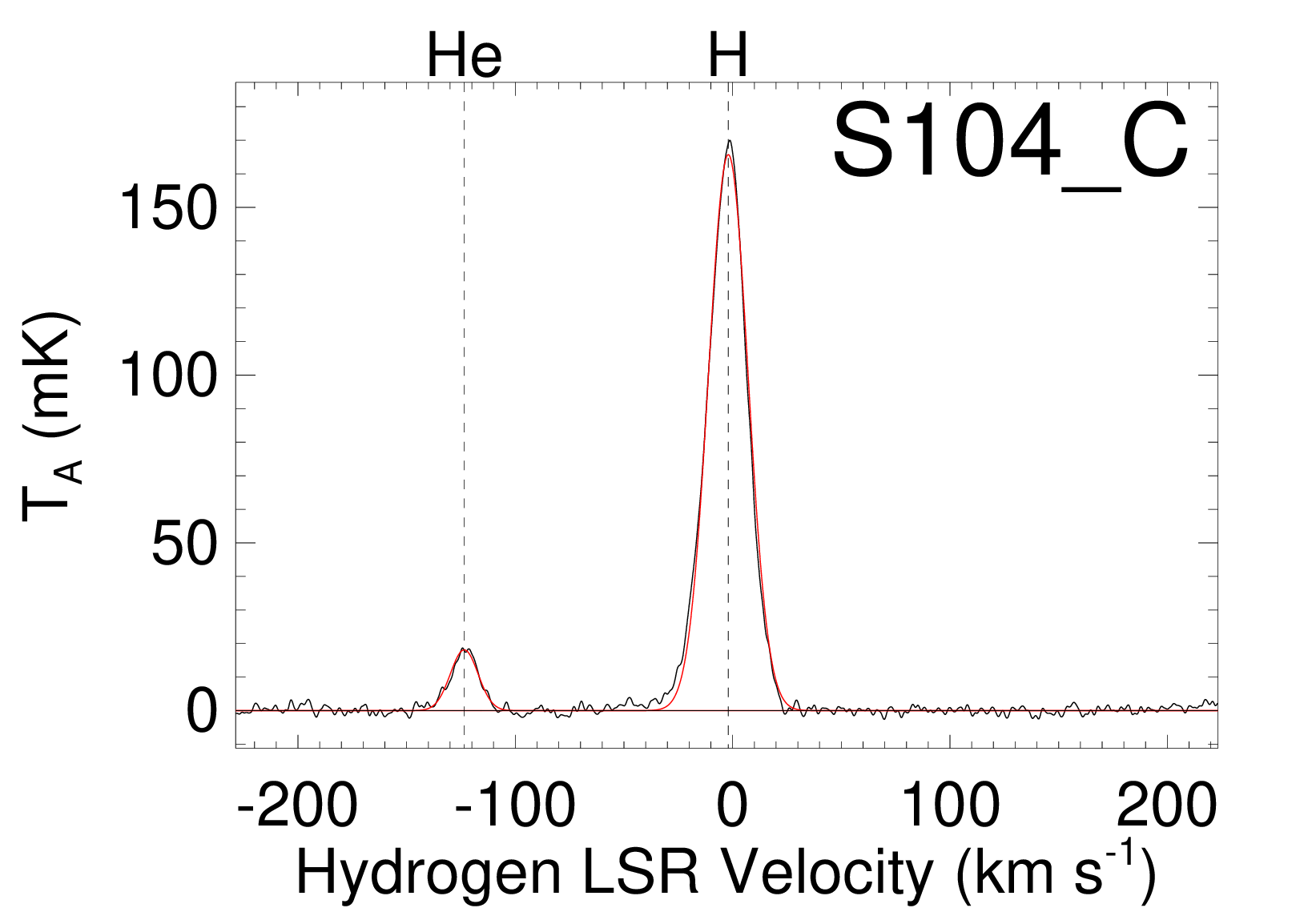} &
\includegraphics[width=.23\textwidth]{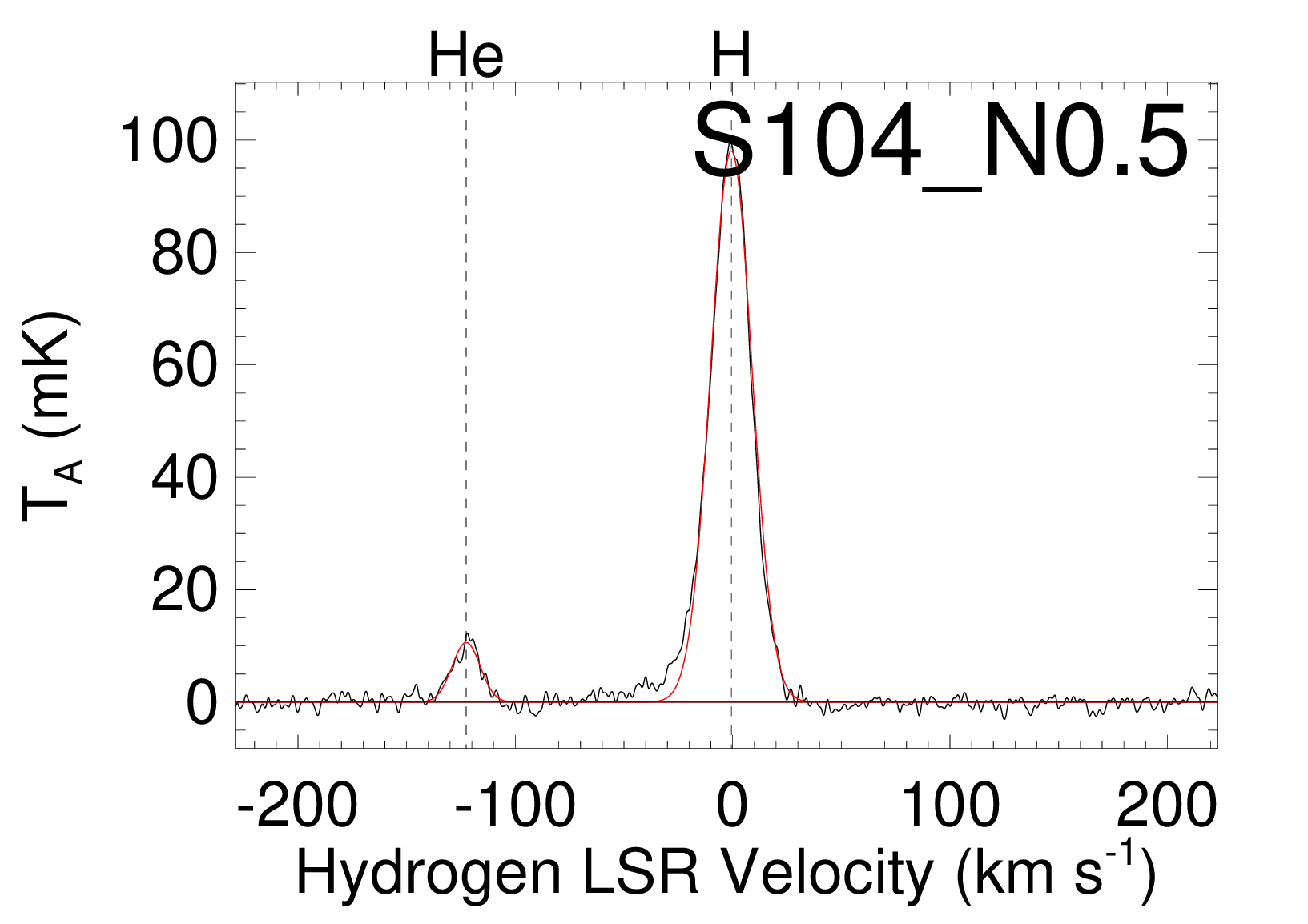} \\
\includegraphics[width=.23\textwidth]{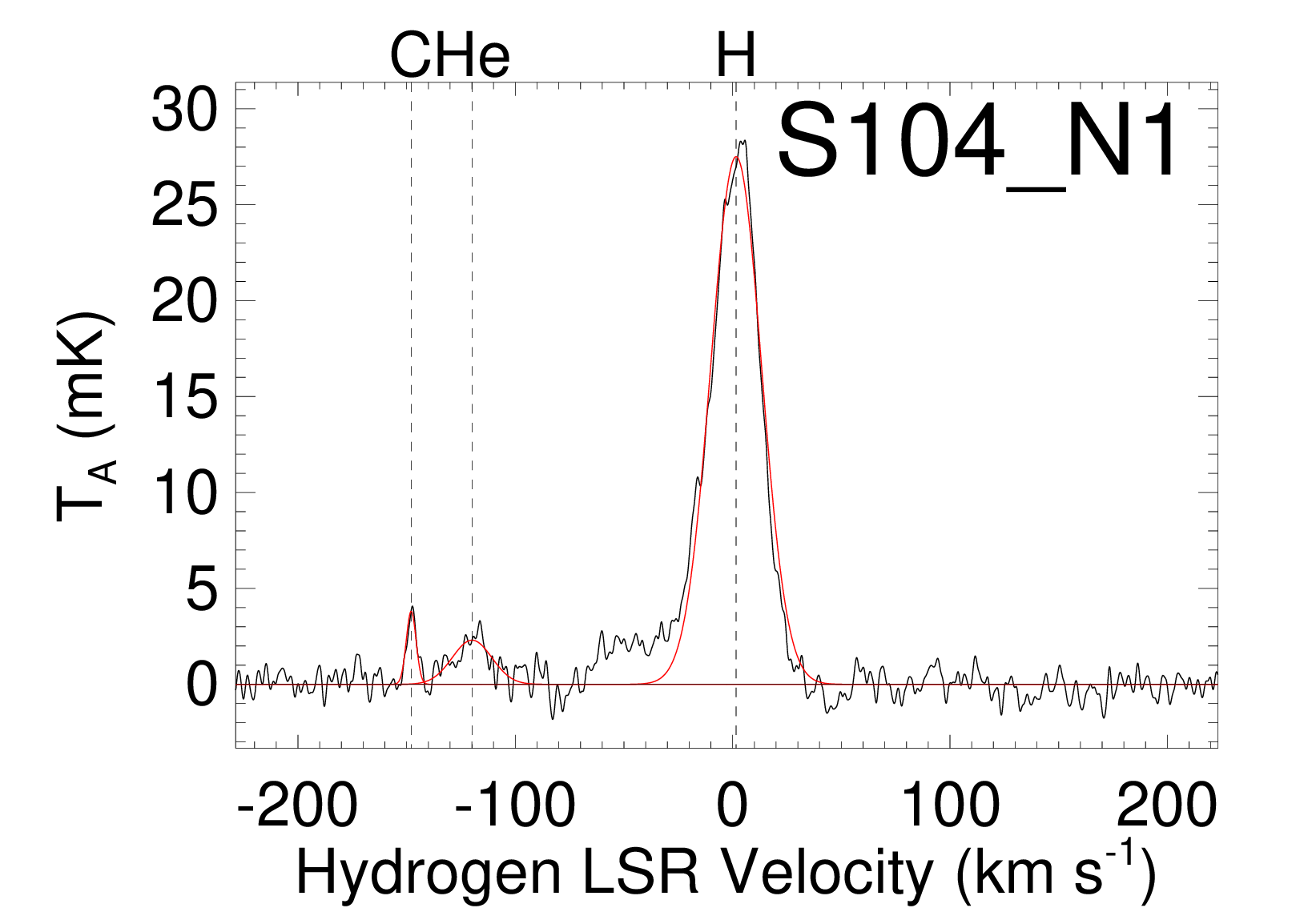} &
\includegraphics[width=.23\textwidth]{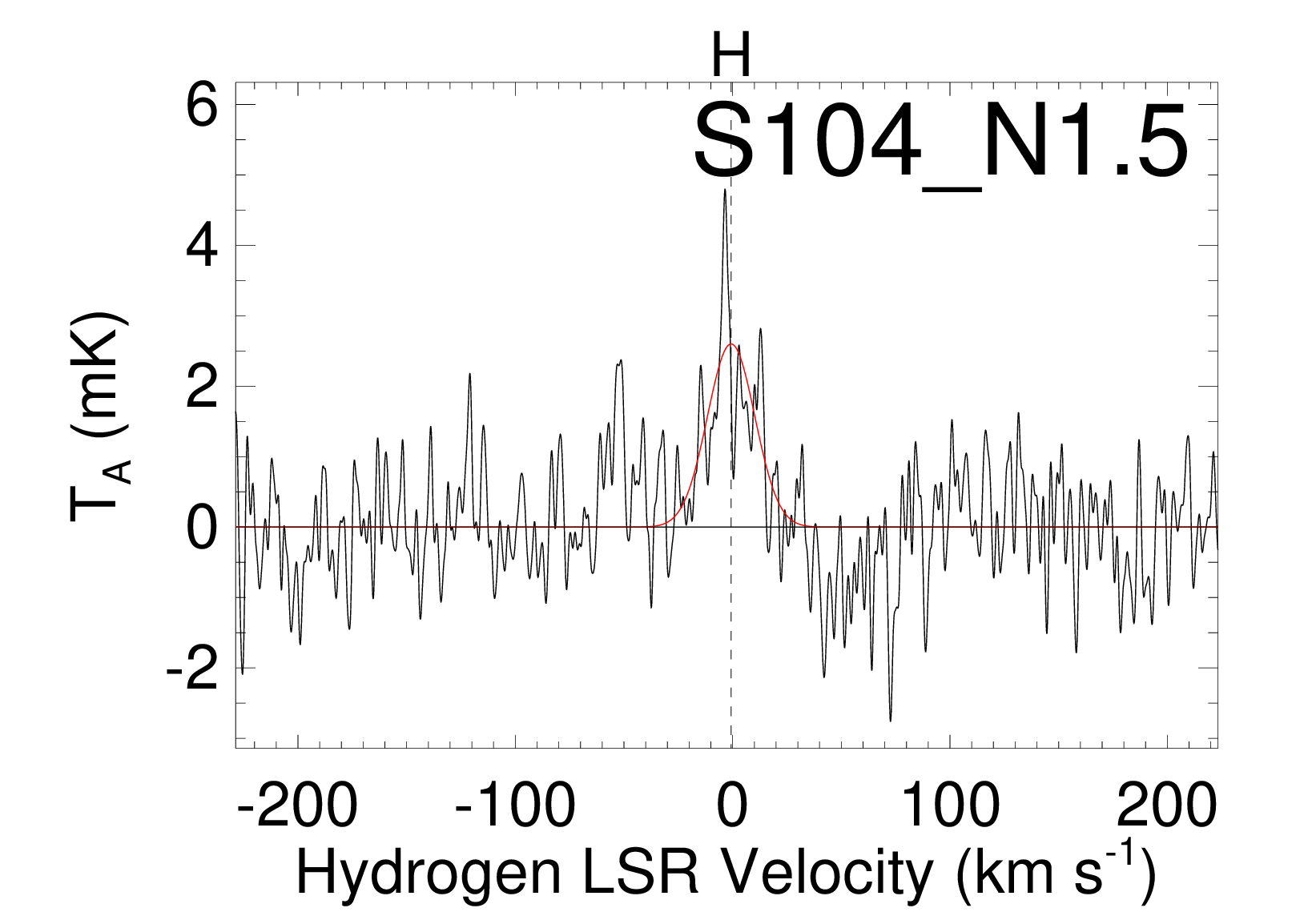} &
\includegraphics[width=.23\textwidth]{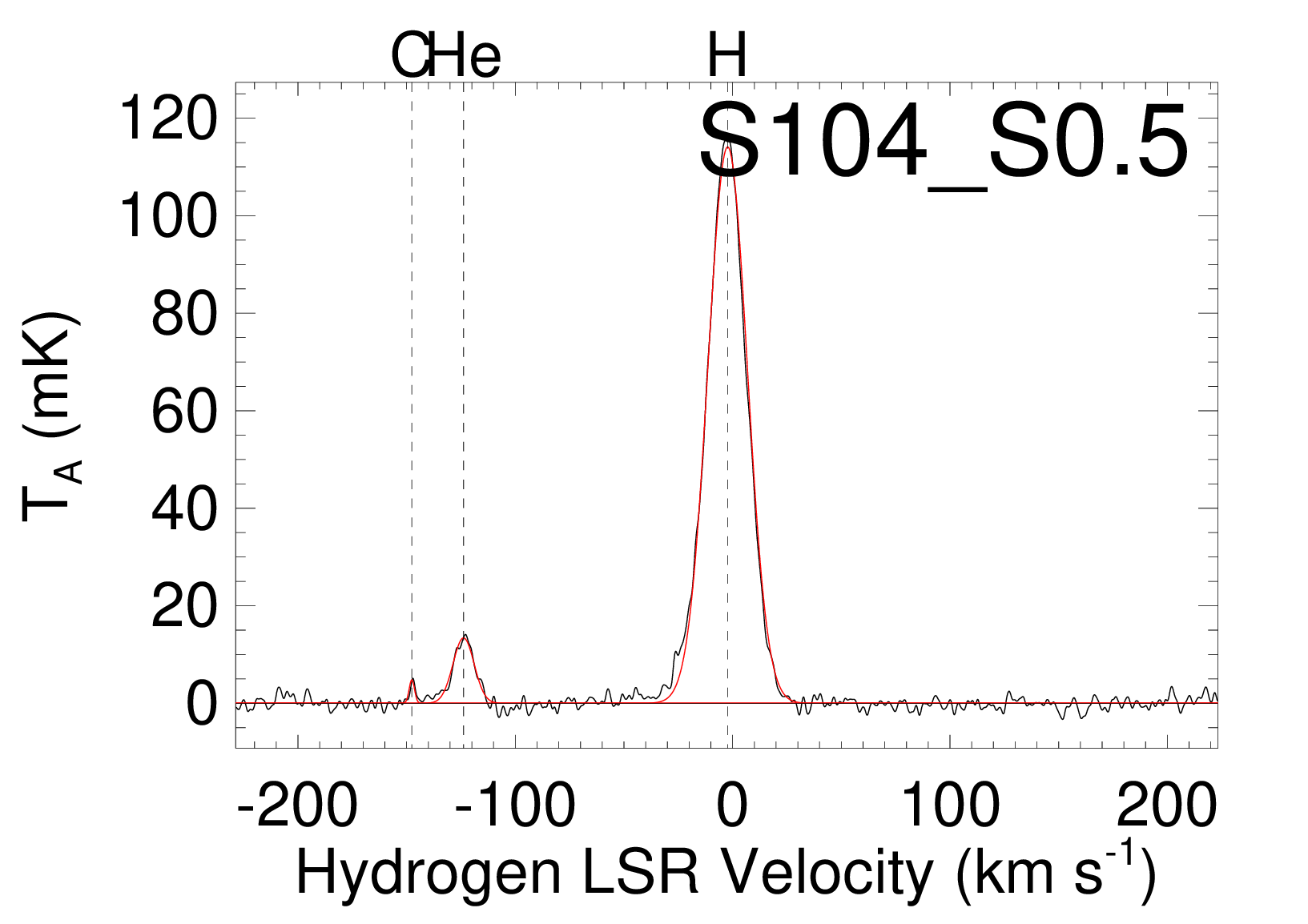} &
\includegraphics[width=.23\textwidth]{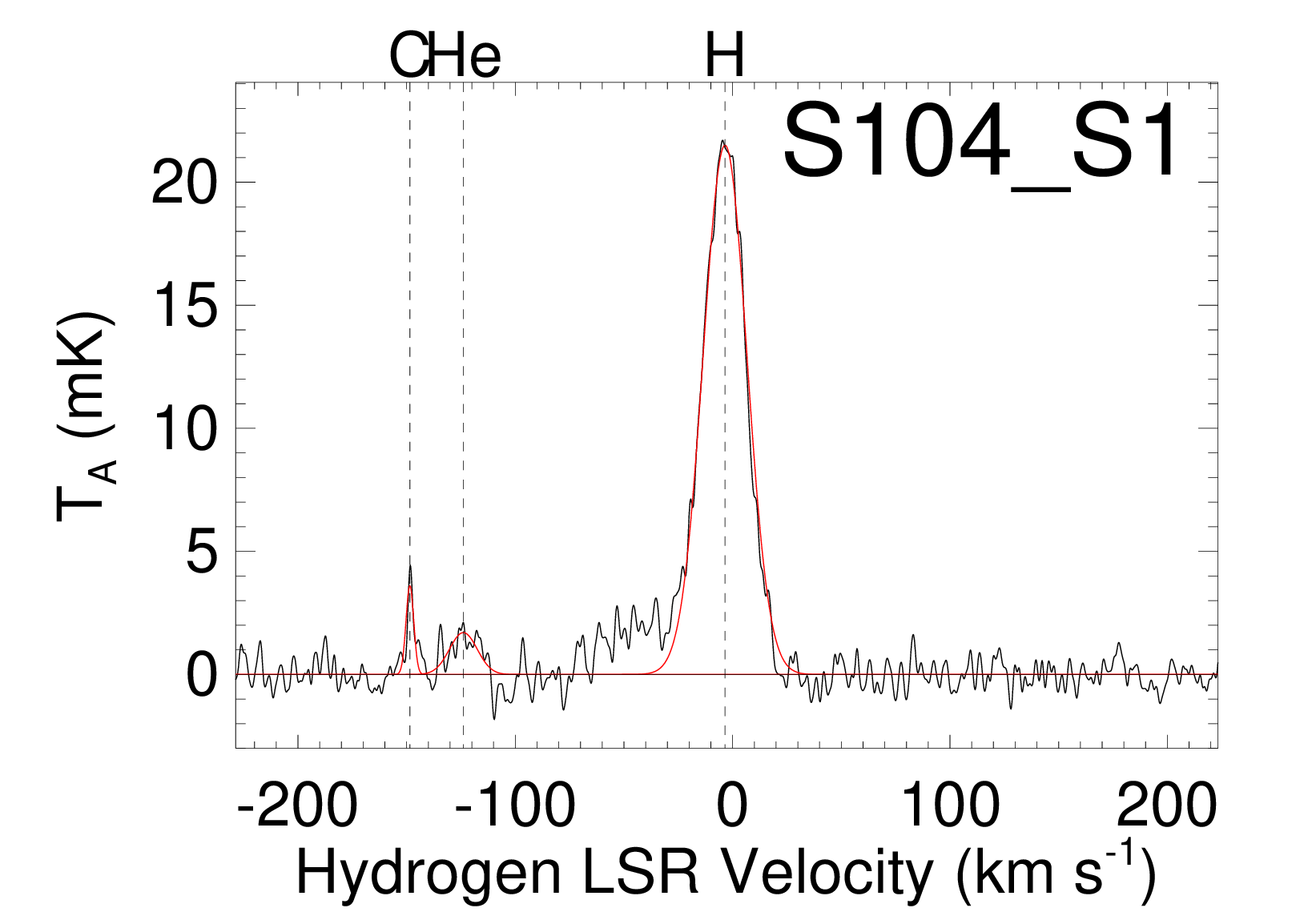} \\
\includegraphics[width=.23\textwidth]{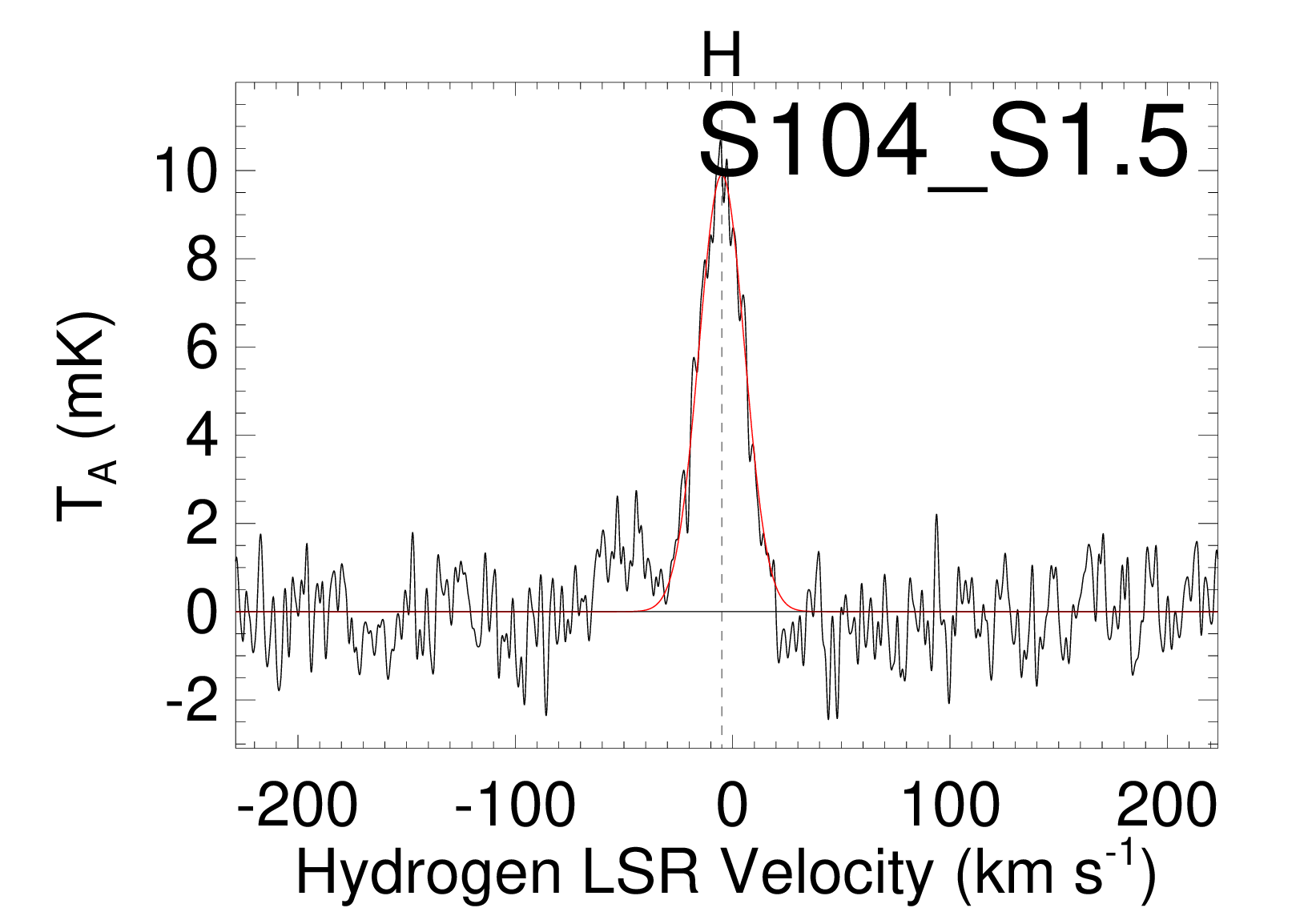} &
\includegraphics[width=.23\textwidth]{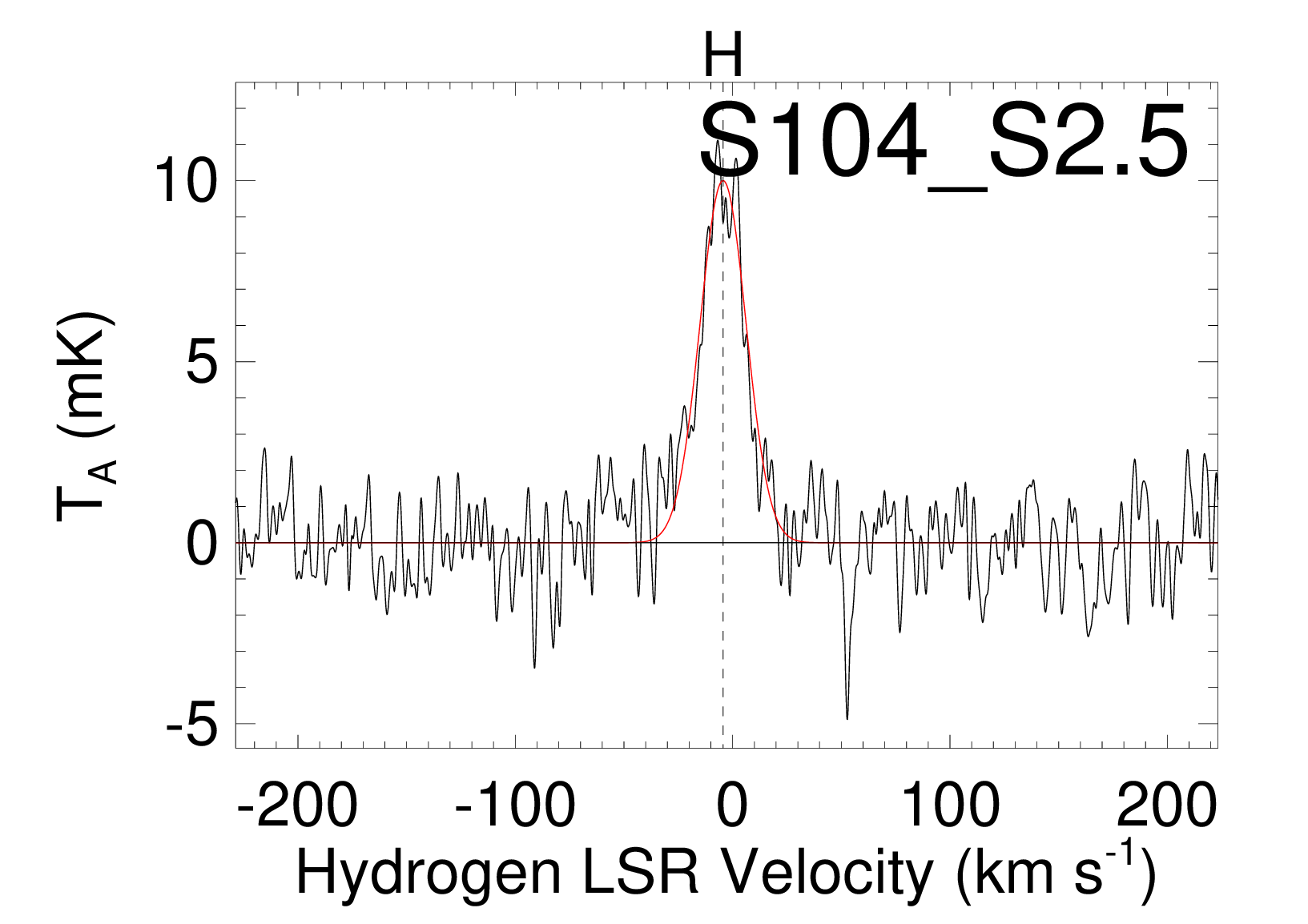} &
\includegraphics[width=.23\textwidth]{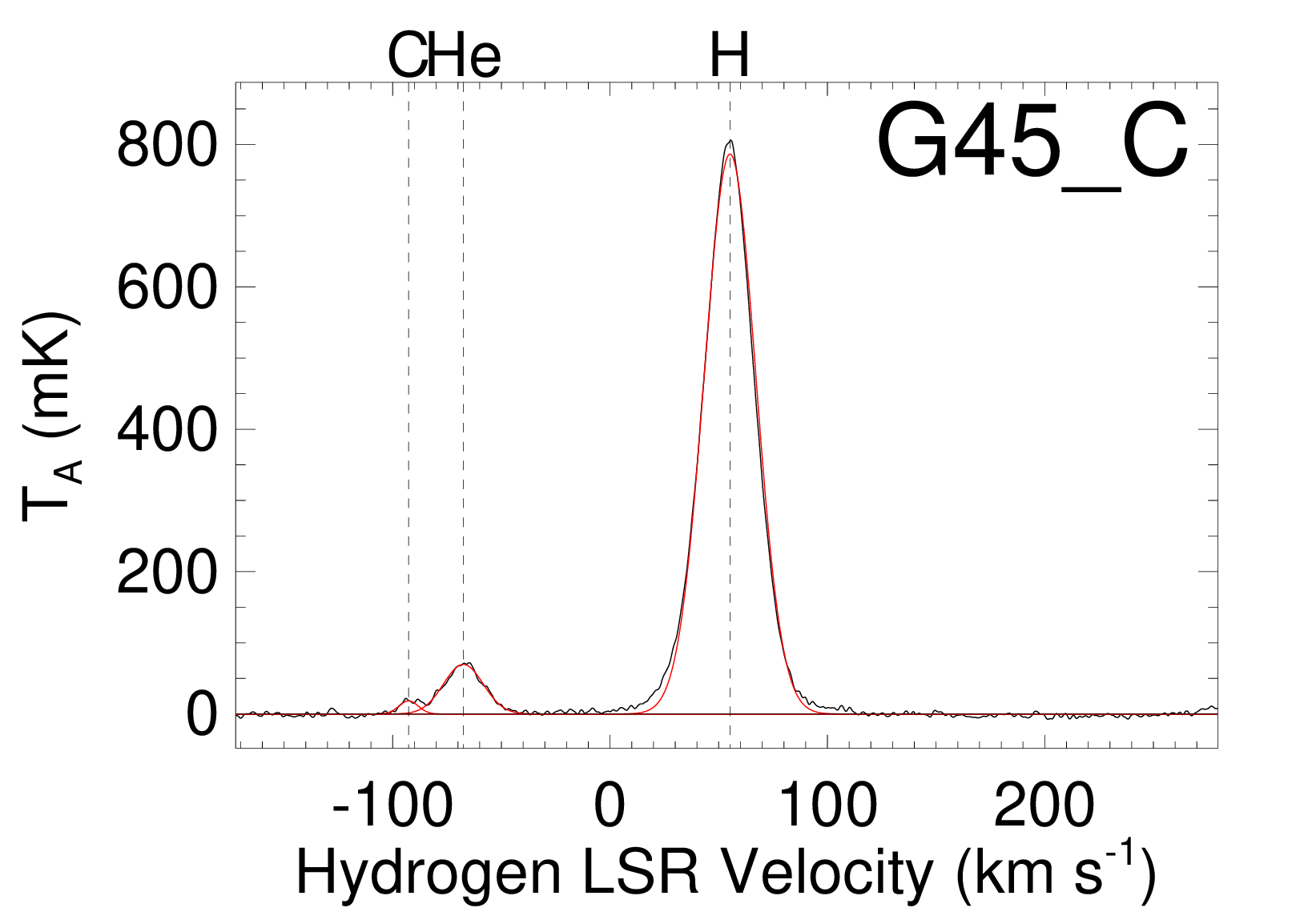} &
\includegraphics[width=.23\textwidth]{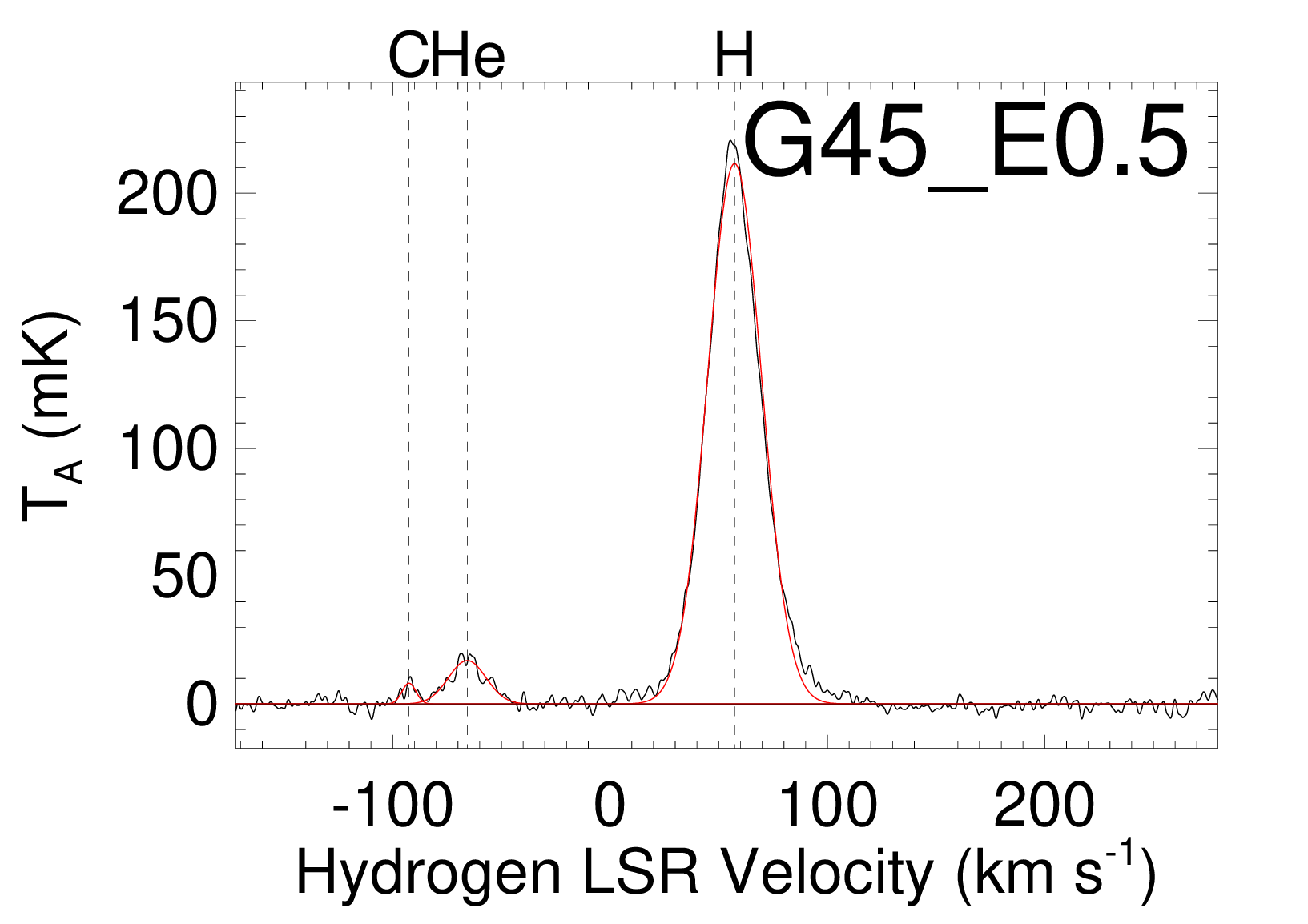} \\
\includegraphics[width=.23\textwidth]{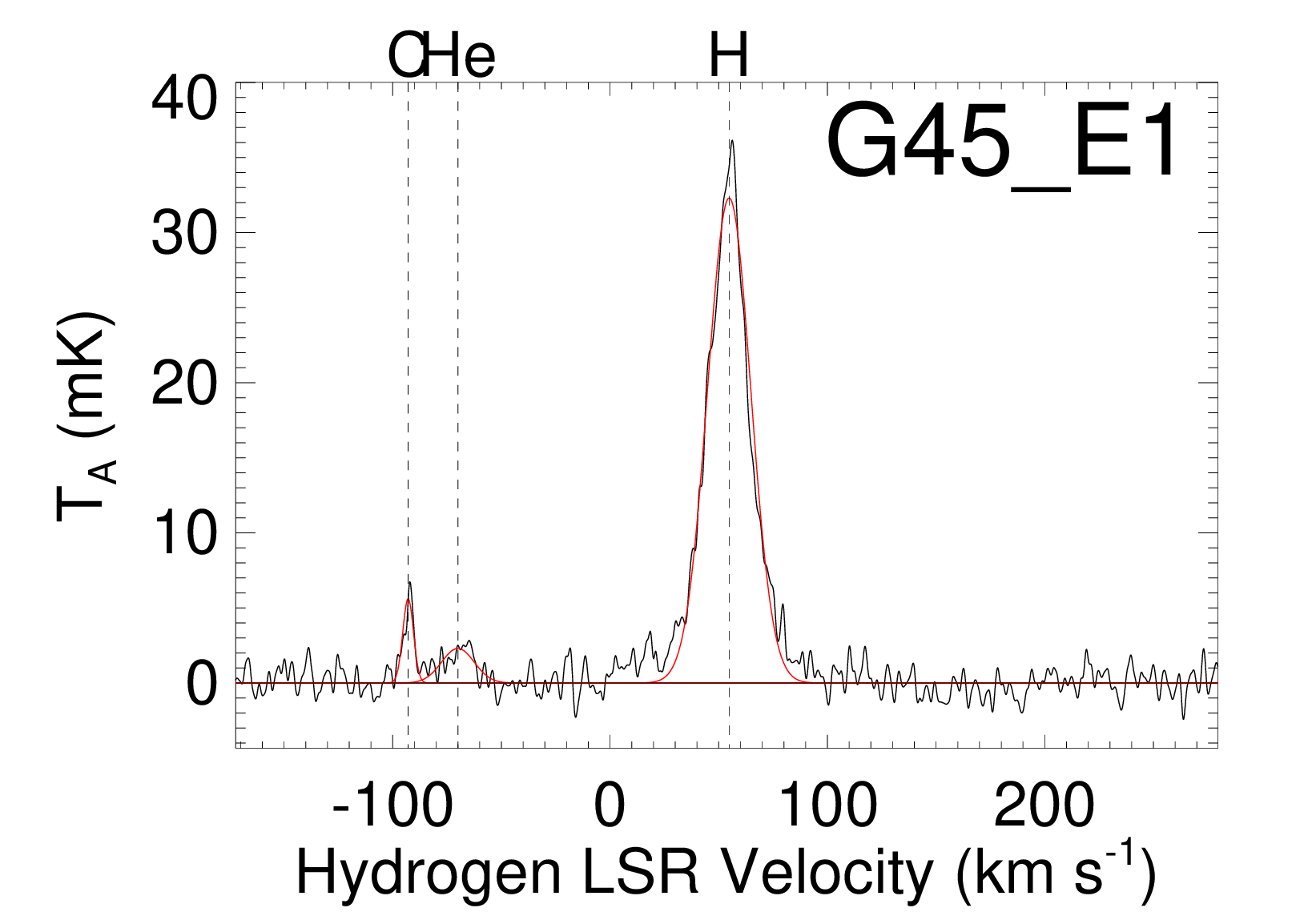} &
\includegraphics[width=.23\textwidth]{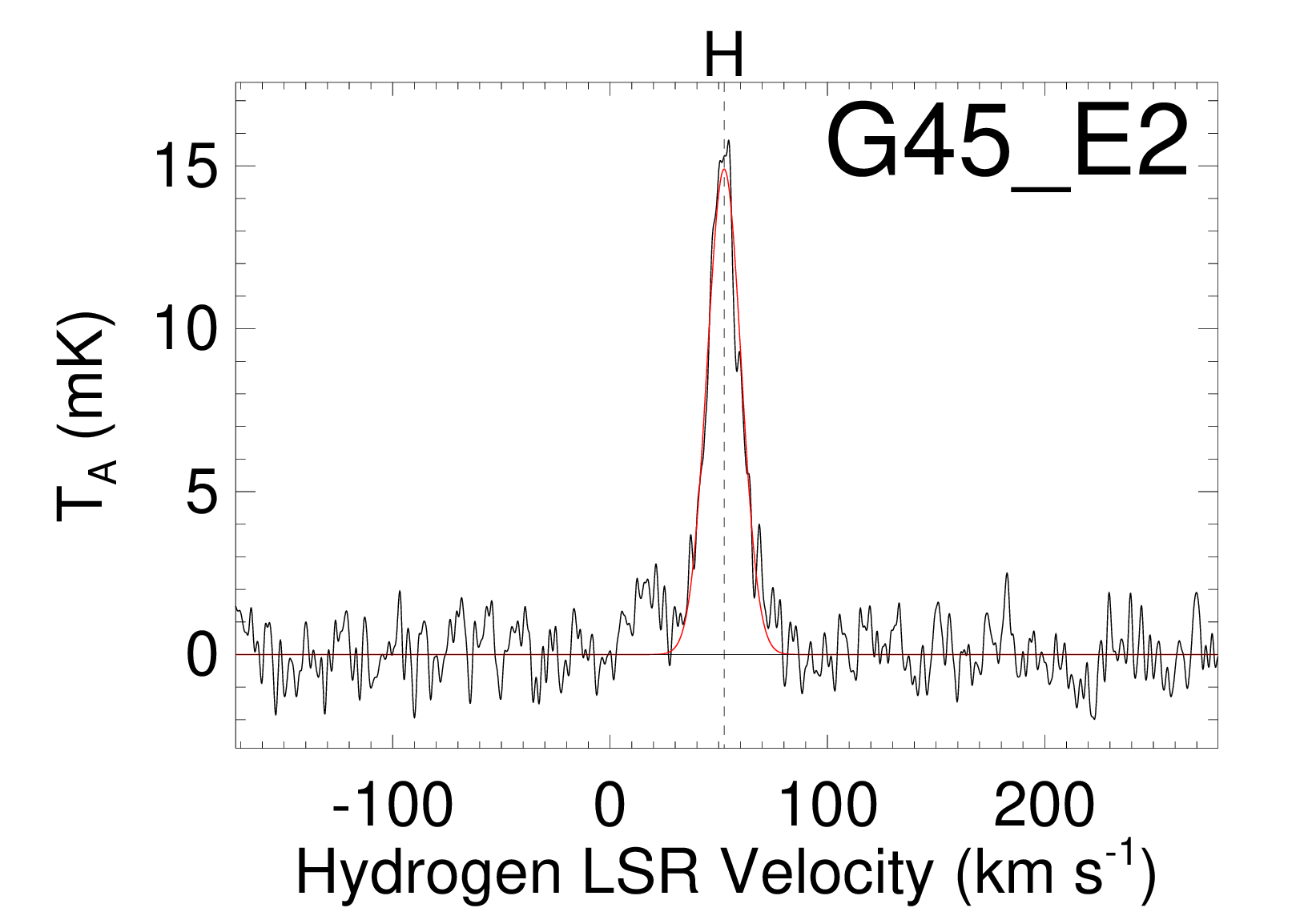} &
\includegraphics[width=.23\textwidth]{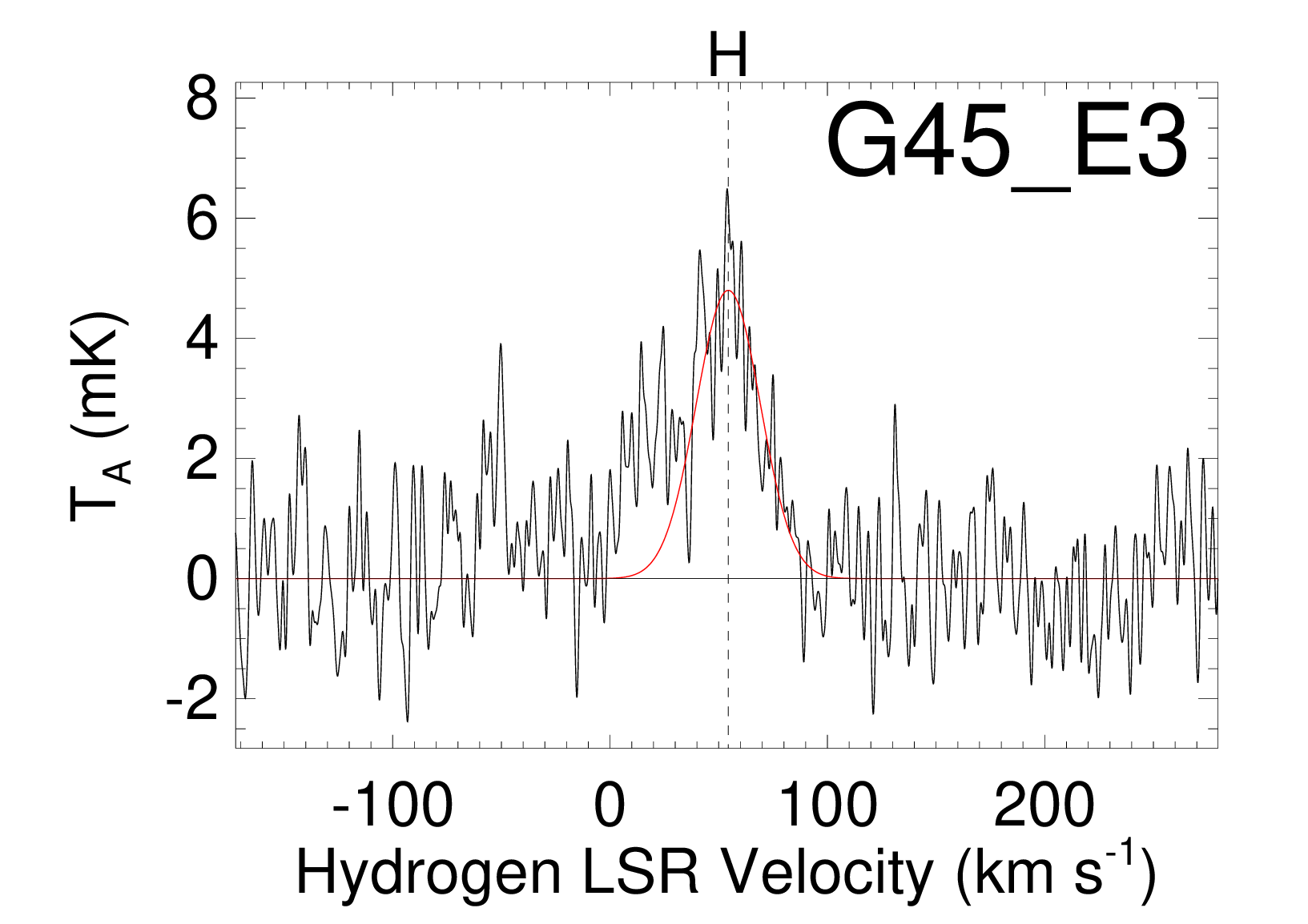} &
\includegraphics[width=.23\textwidth]{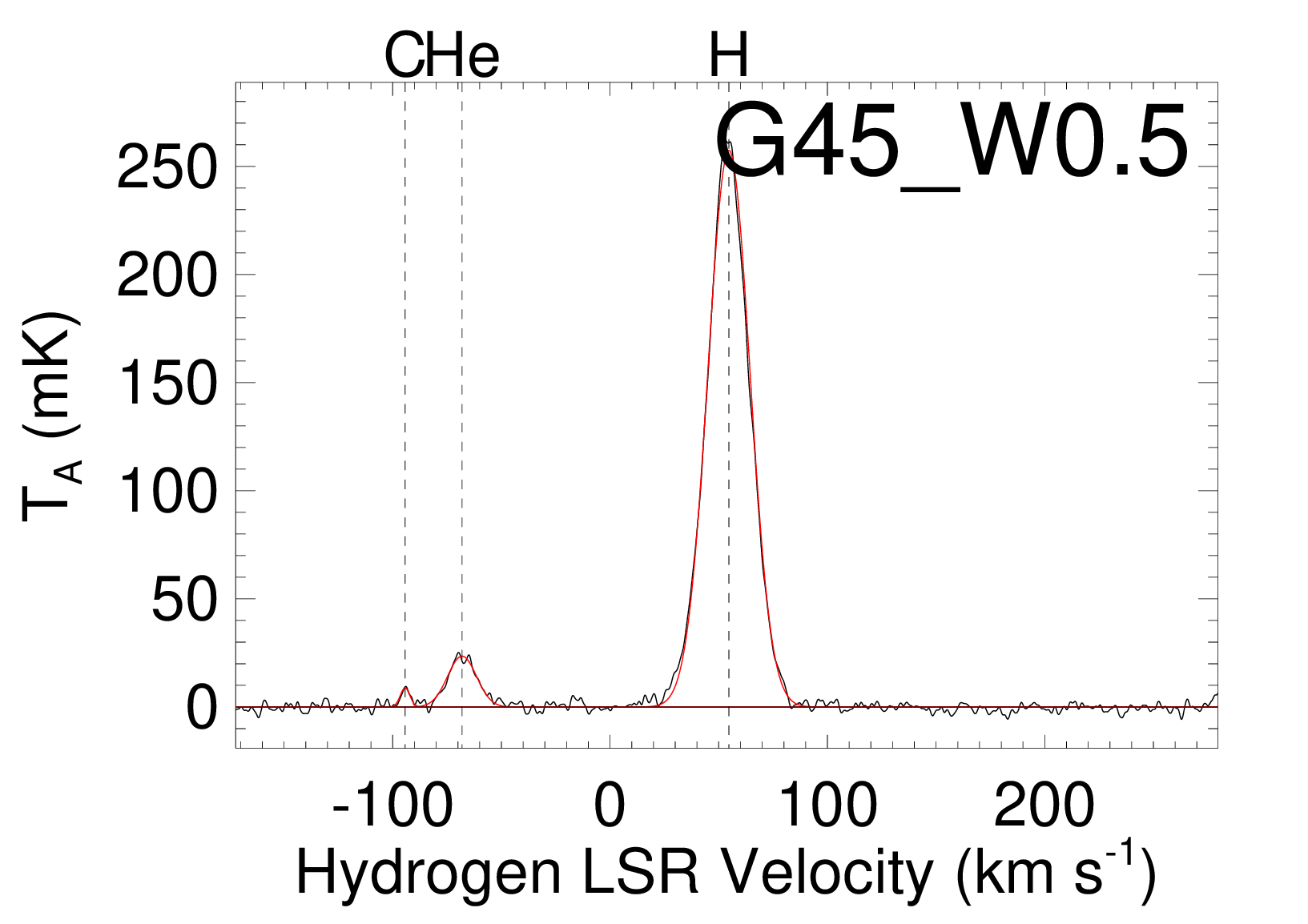} \\
\end{tabular}
\caption{$\alpha$ RRL spectra of all observed positions, smoothed to a spectral resolution of 1.86\kms. Plotted is the antenna temperature as a function of hydrogen LSR velocity. The helium and carbon lines are offset from hydrogen by $-124$\kms and $-149$\kms, respectively. We approximate hydrogen, helium, and carbon emission above the S/N threshold defined in \S \ref{sec:obs} with the Gaussian model fits shown in red. The centers of the Gaussian peaks are indicated by dashed vertical lines. The name of the observed position is given in the upper right-hand corner of each plot. \label{fig:spectra}}
\end{figure*}

\renewcommand\thefigure{\thesection.\arabic{figure} (Cont.)}
\addtocounter{figure}{-1}
\begin{figure*}
\centering
\begin{tabular}{cccc}
\includegraphics[width=.23\textwidth]{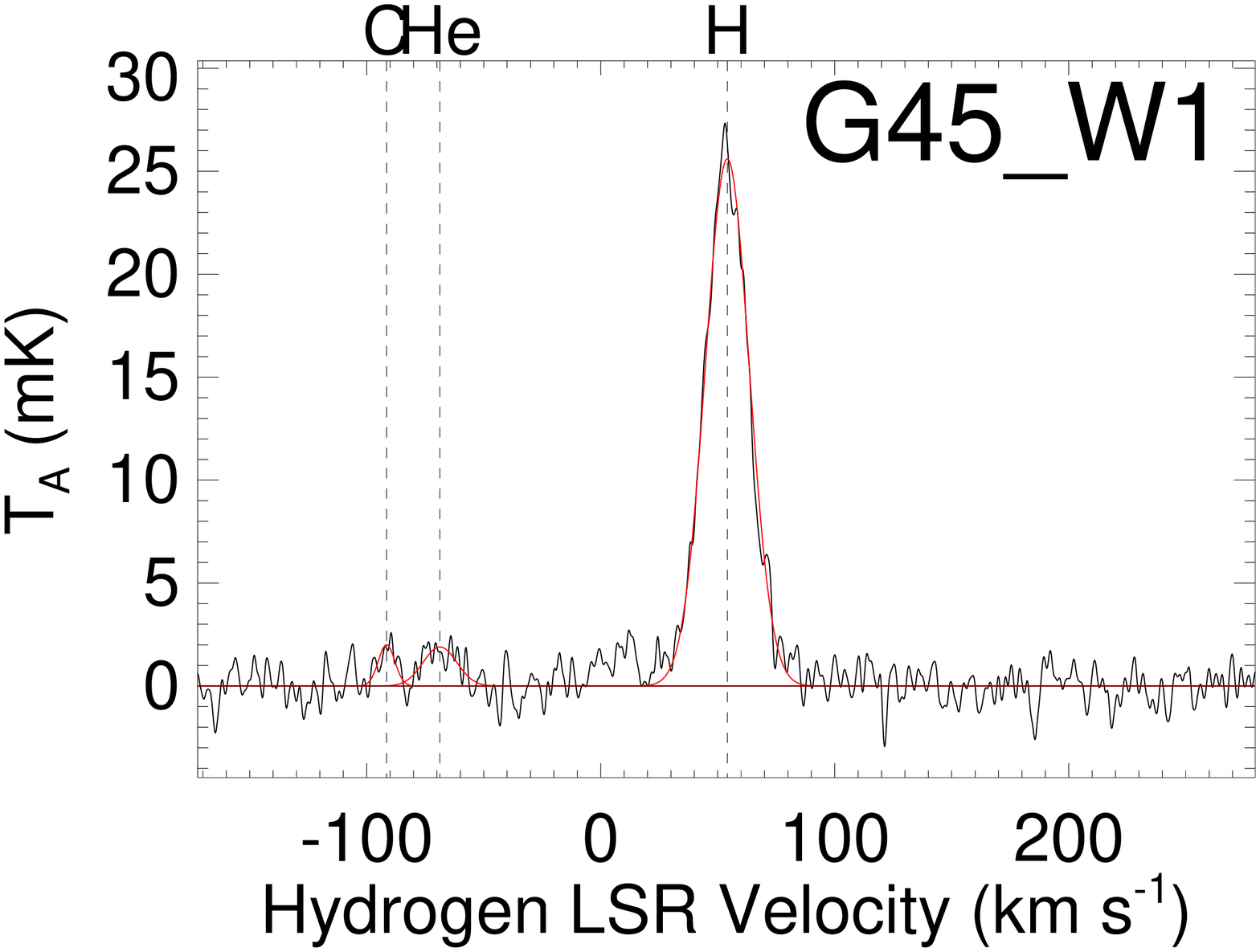} &
\includegraphics[width=.23\textwidth]{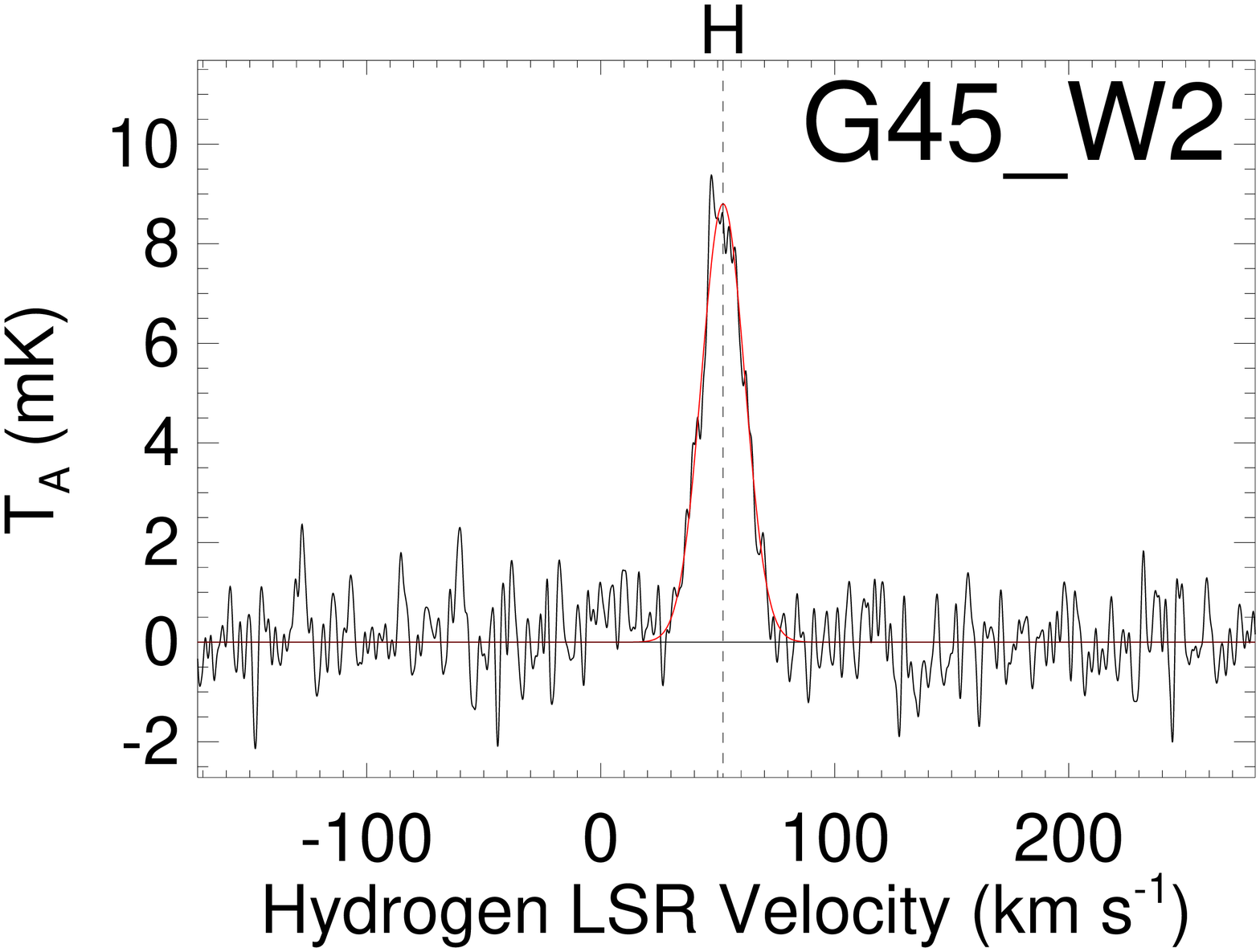} &
\includegraphics[width=.23\textwidth]{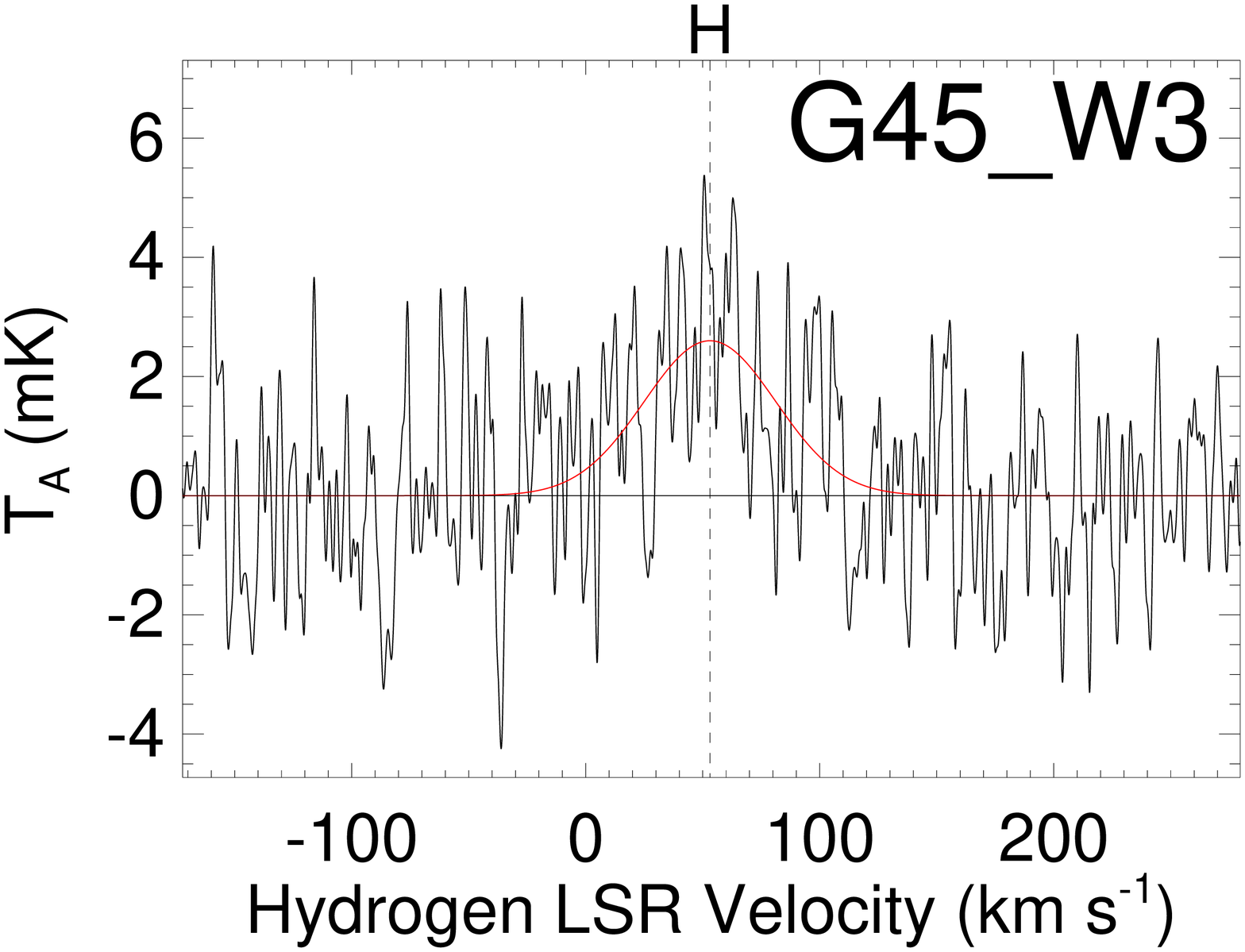} &
\includegraphics[width=.23\textwidth]{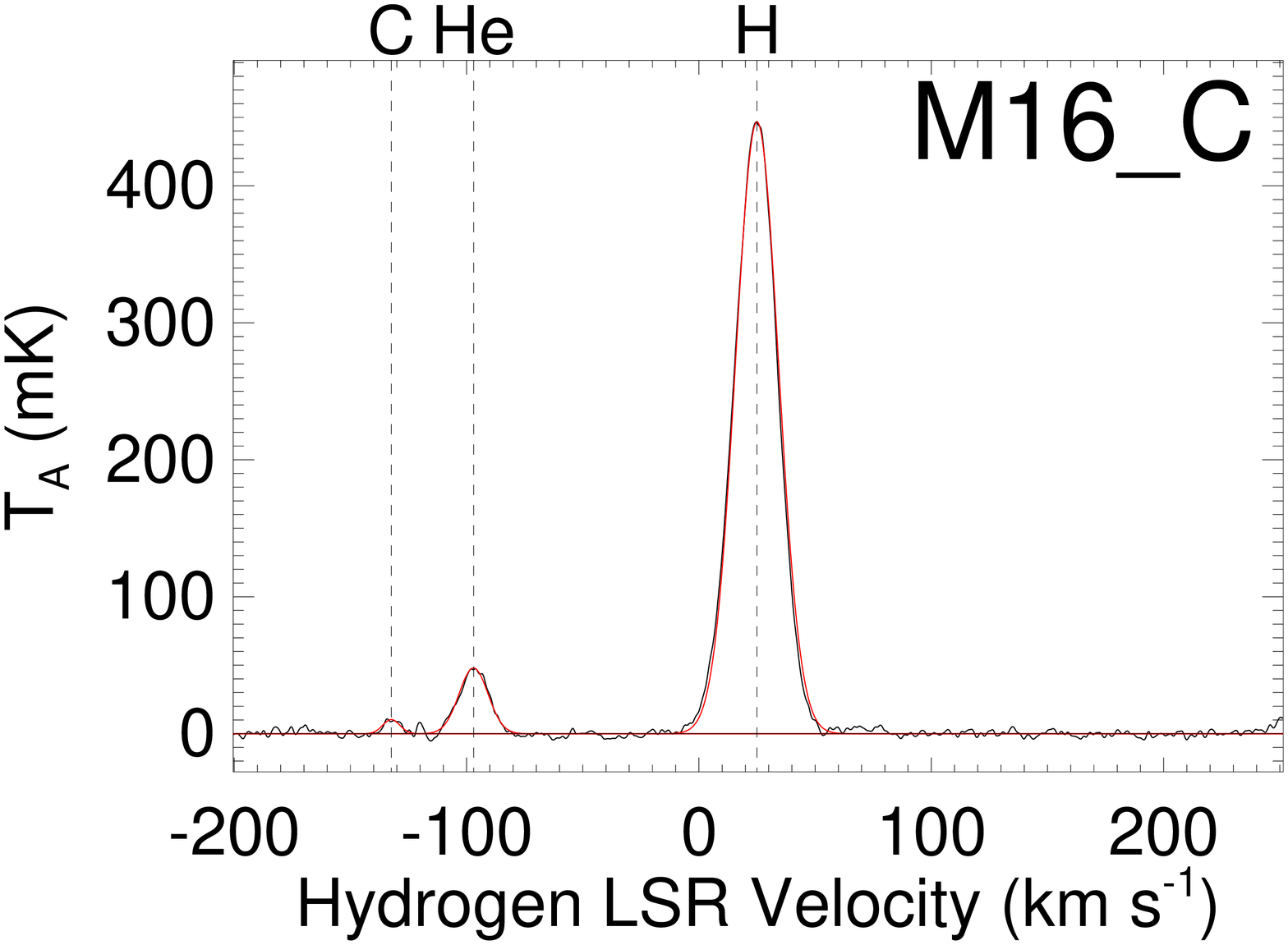} \\
\includegraphics[width=.23\textwidth]{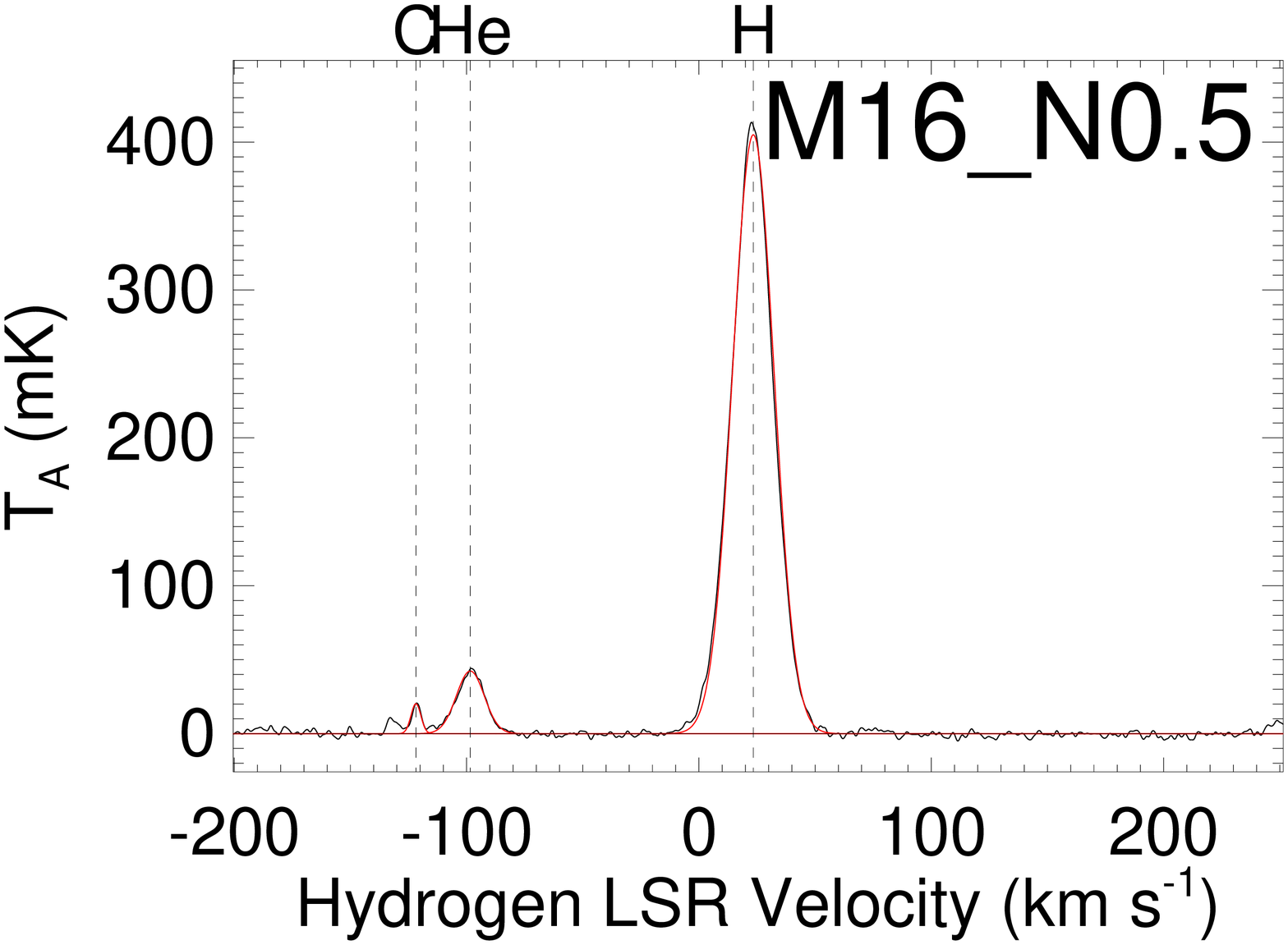} &
\includegraphics[width=.23\textwidth]{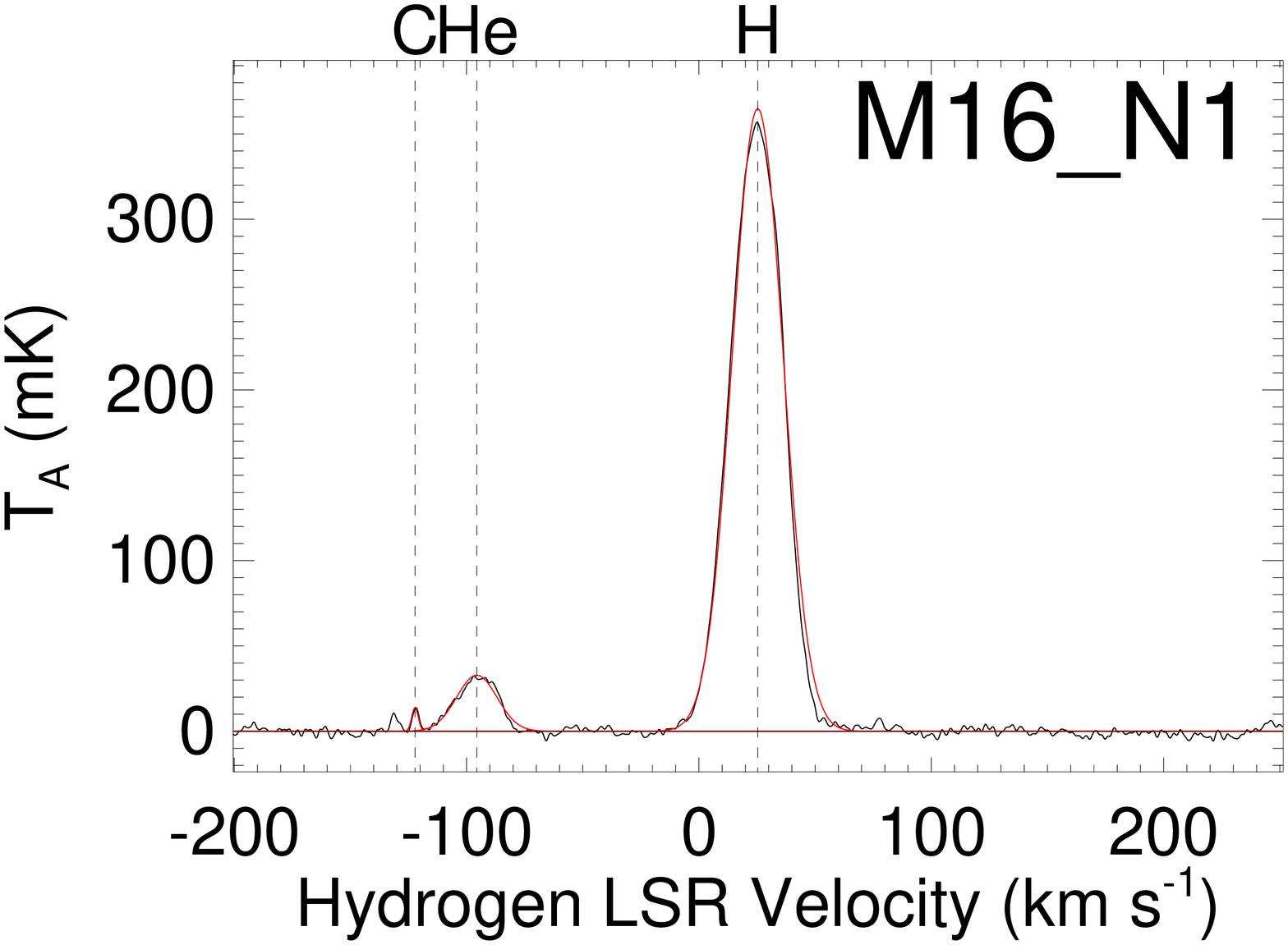} &
\includegraphics[width=.23\textwidth]{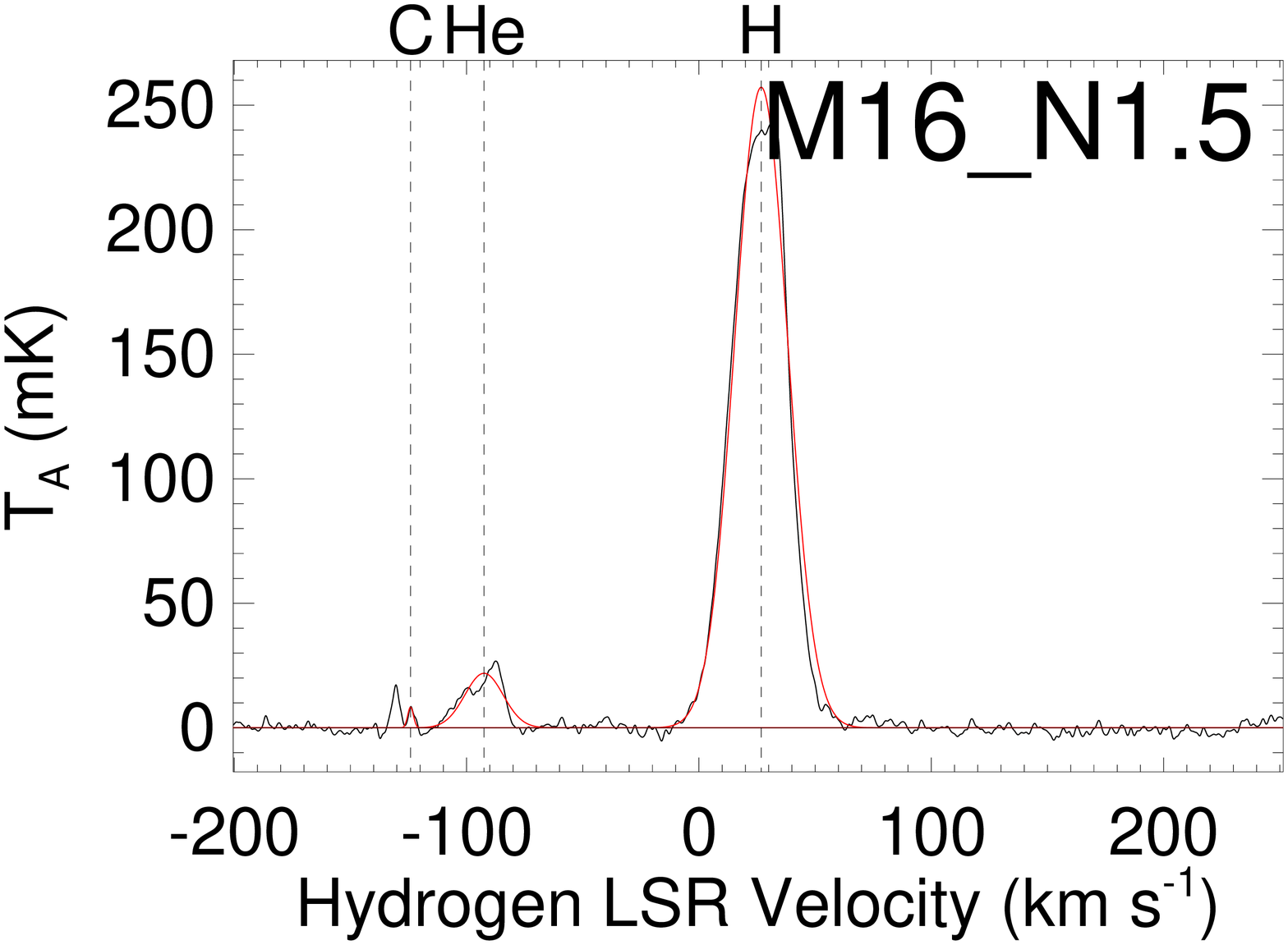} &
\includegraphics[width=.23\textwidth]{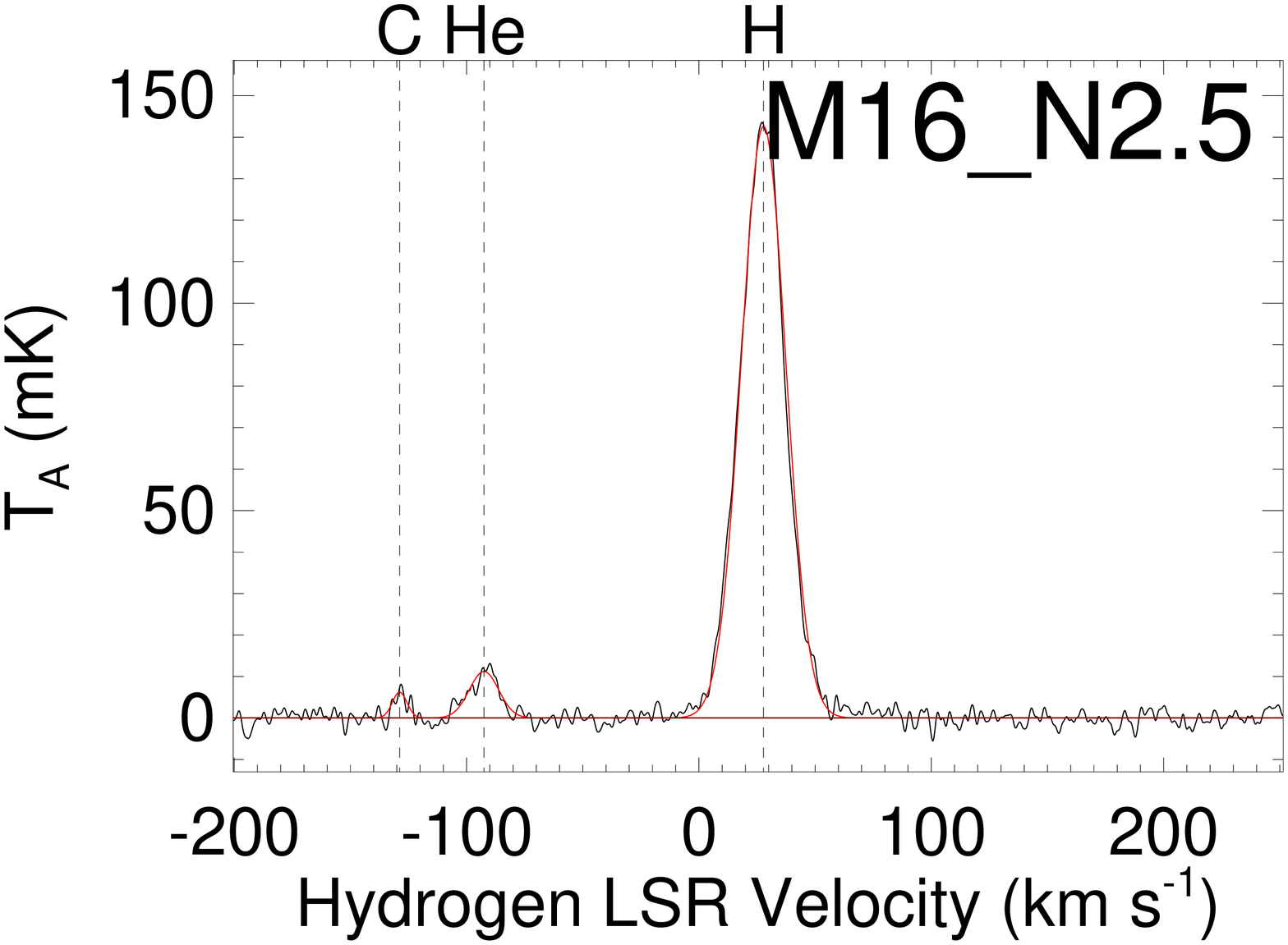} \\
\includegraphics[width=.23\textwidth]{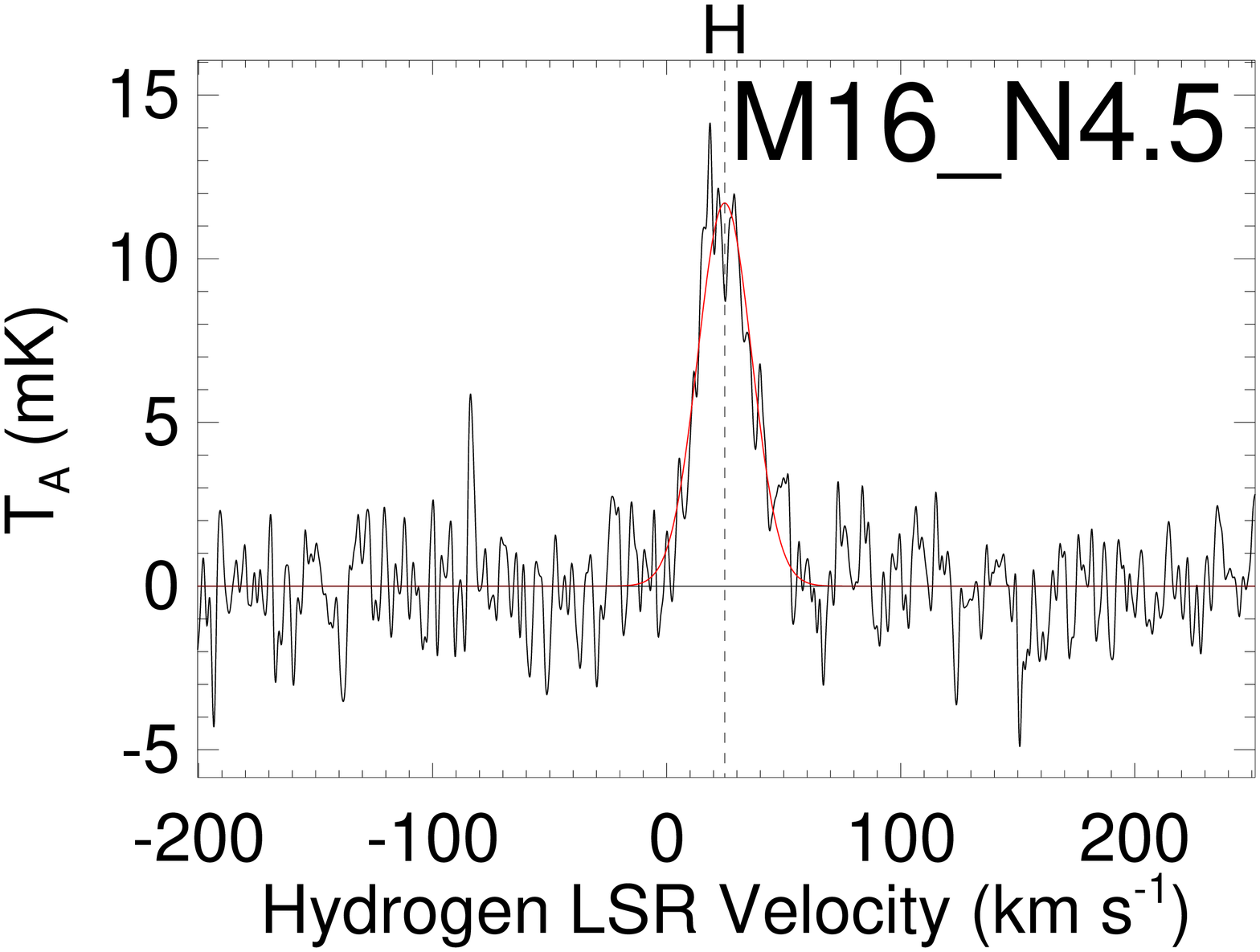} &
\includegraphics[width=.23\textwidth]{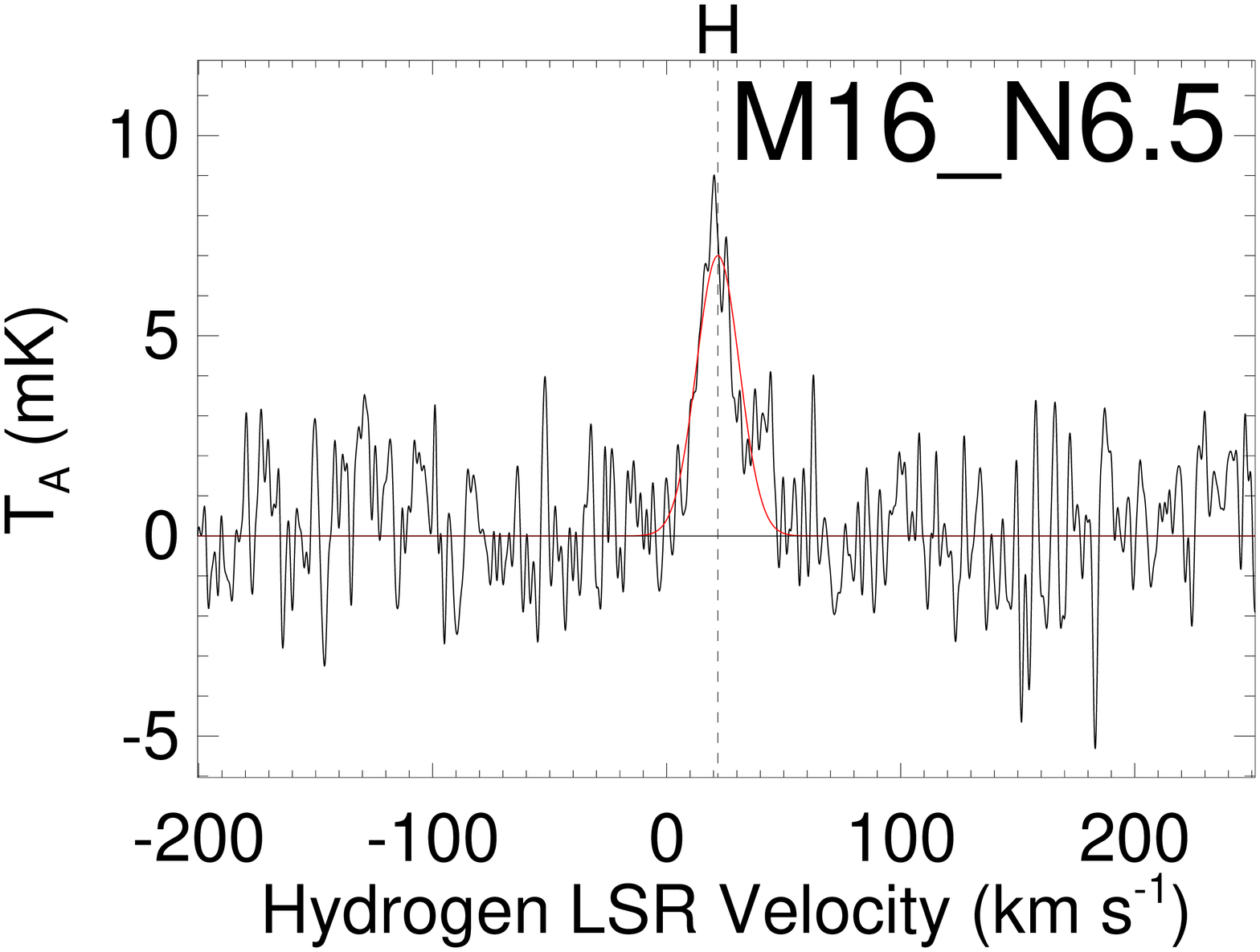} &
\includegraphics[width=.23\textwidth]{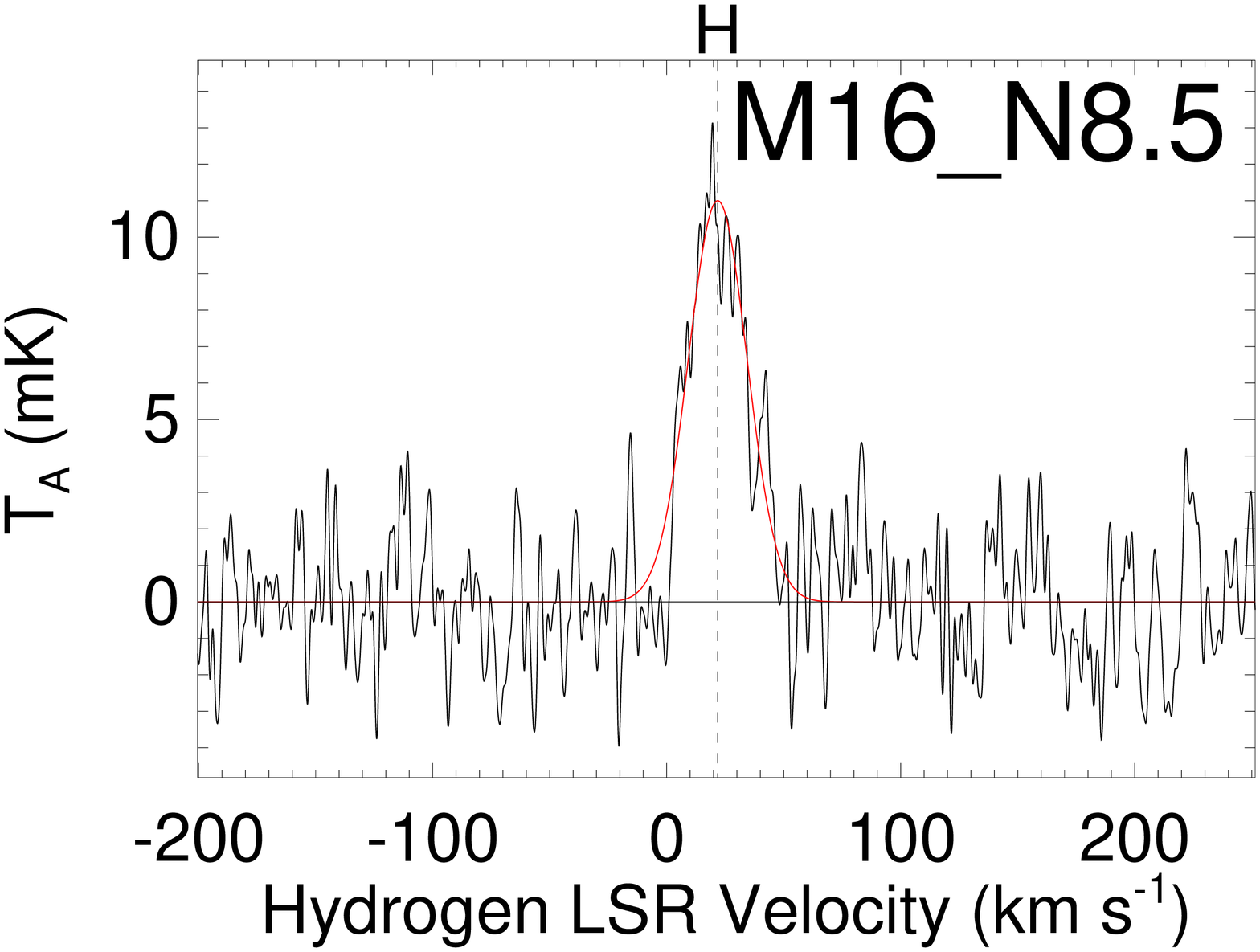} &
\includegraphics[width=.23\textwidth]{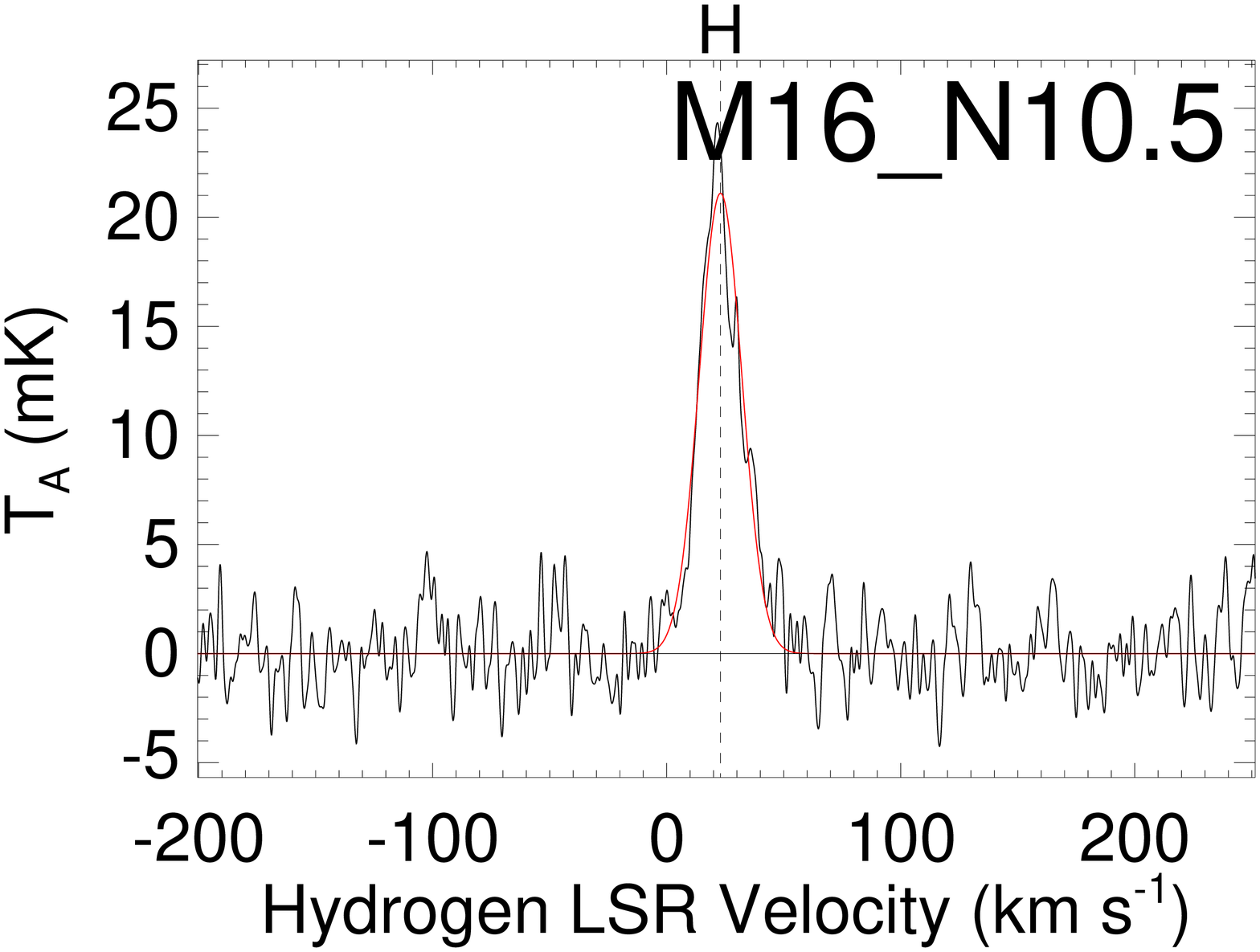} \\
\includegraphics[width=.23\textwidth]{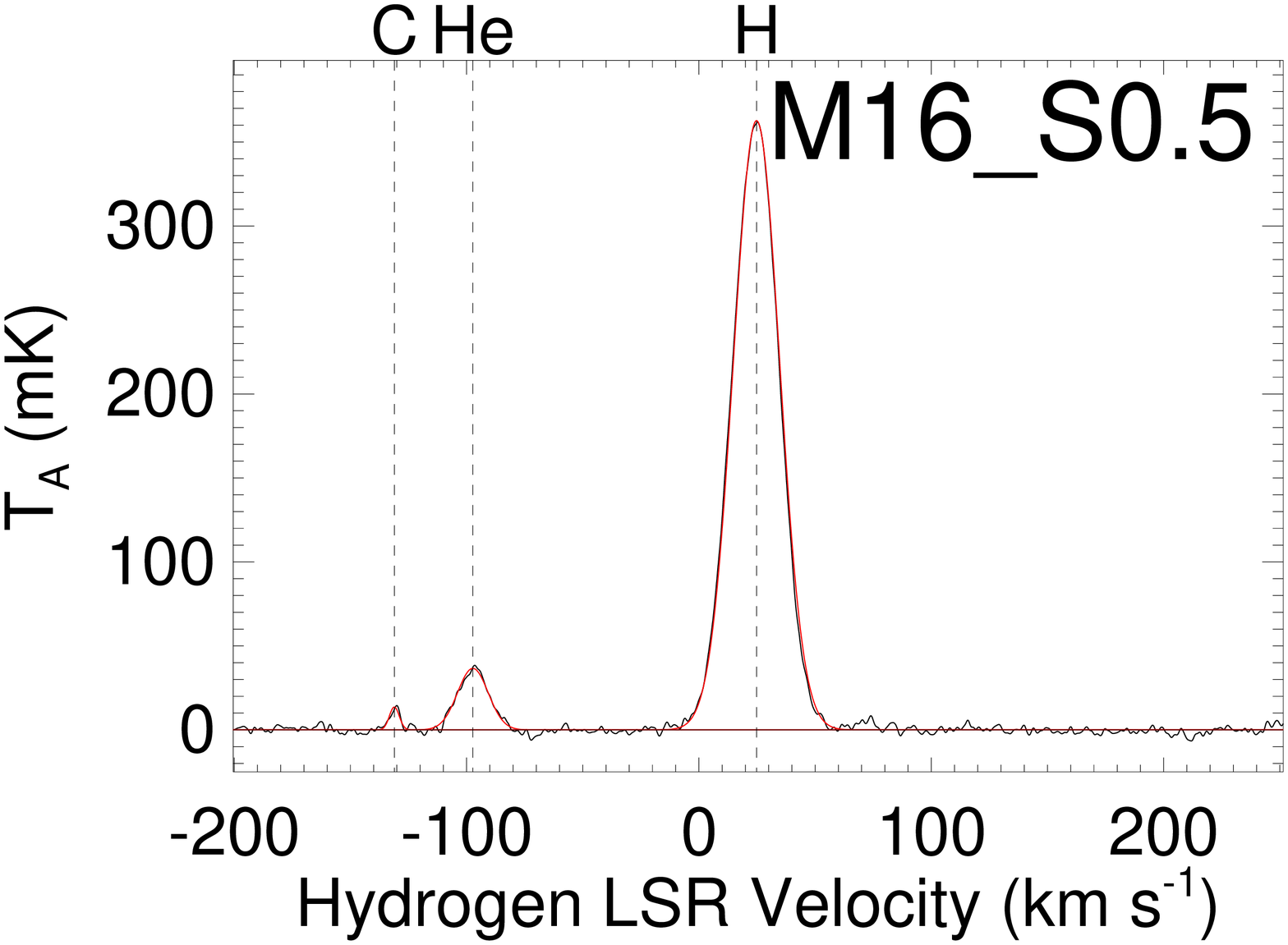} &
\includegraphics[width=.23\textwidth]{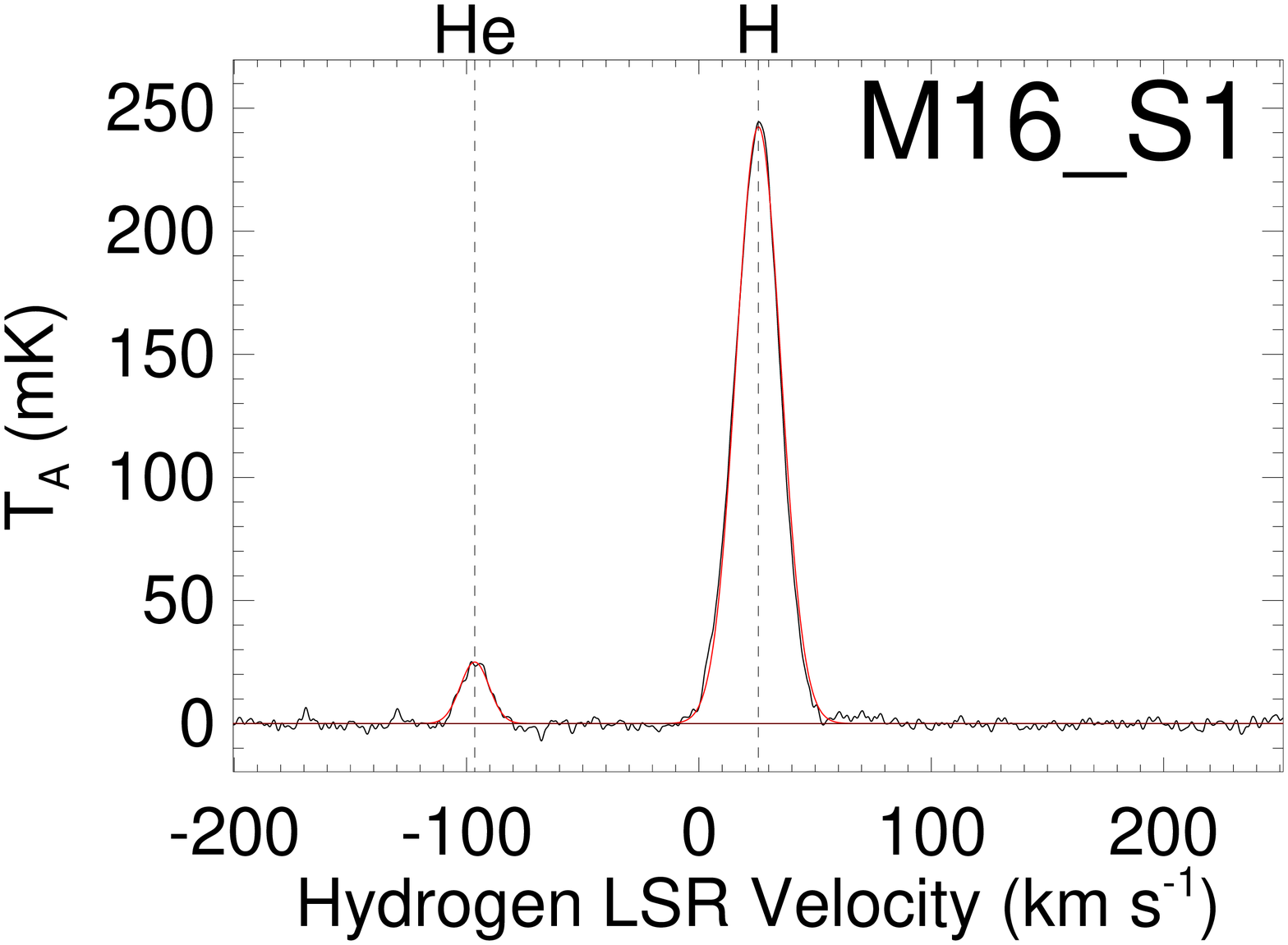} &
\includegraphics[width=.23\textwidth]{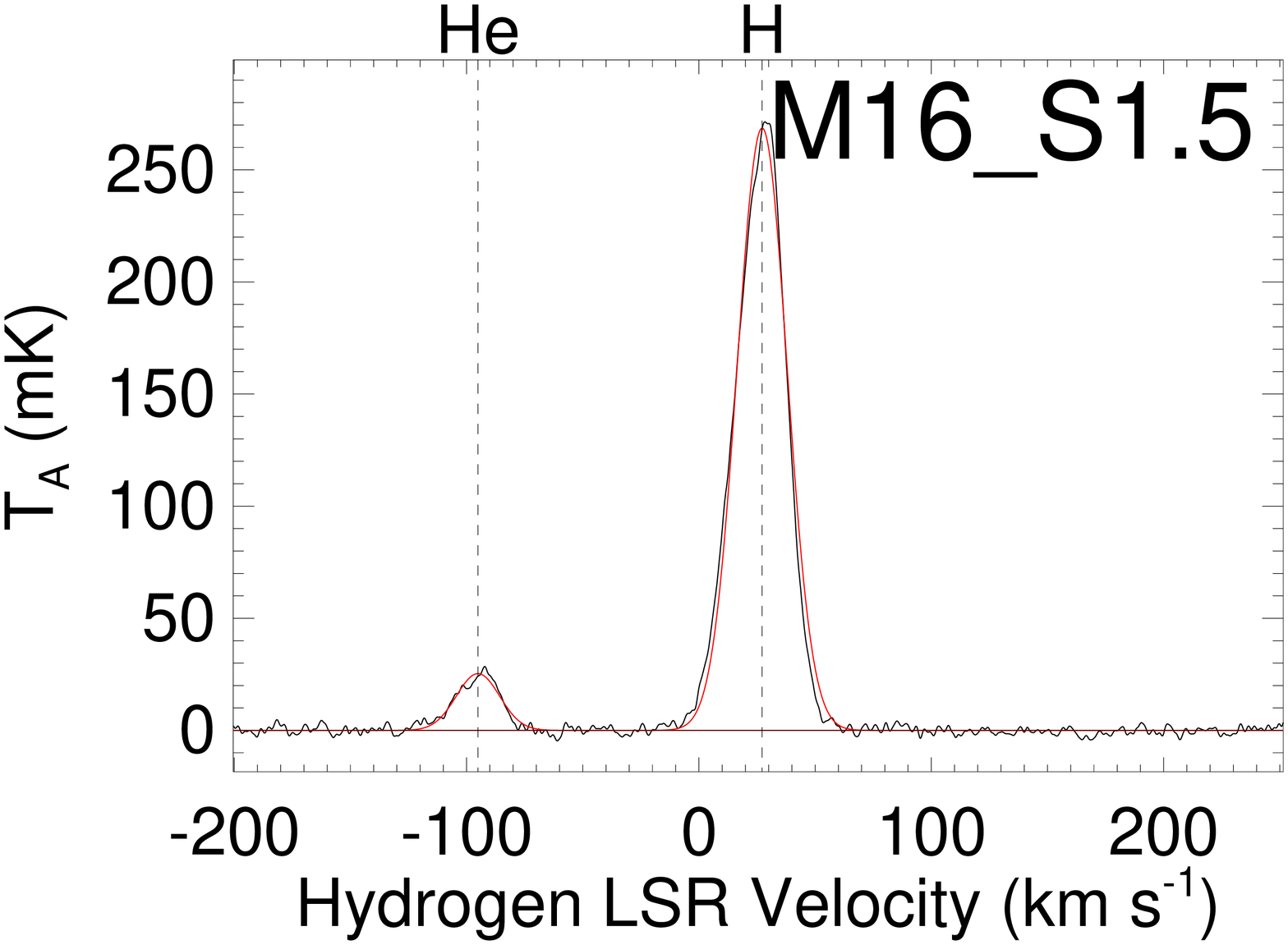} &
\includegraphics[width=.23\textwidth]{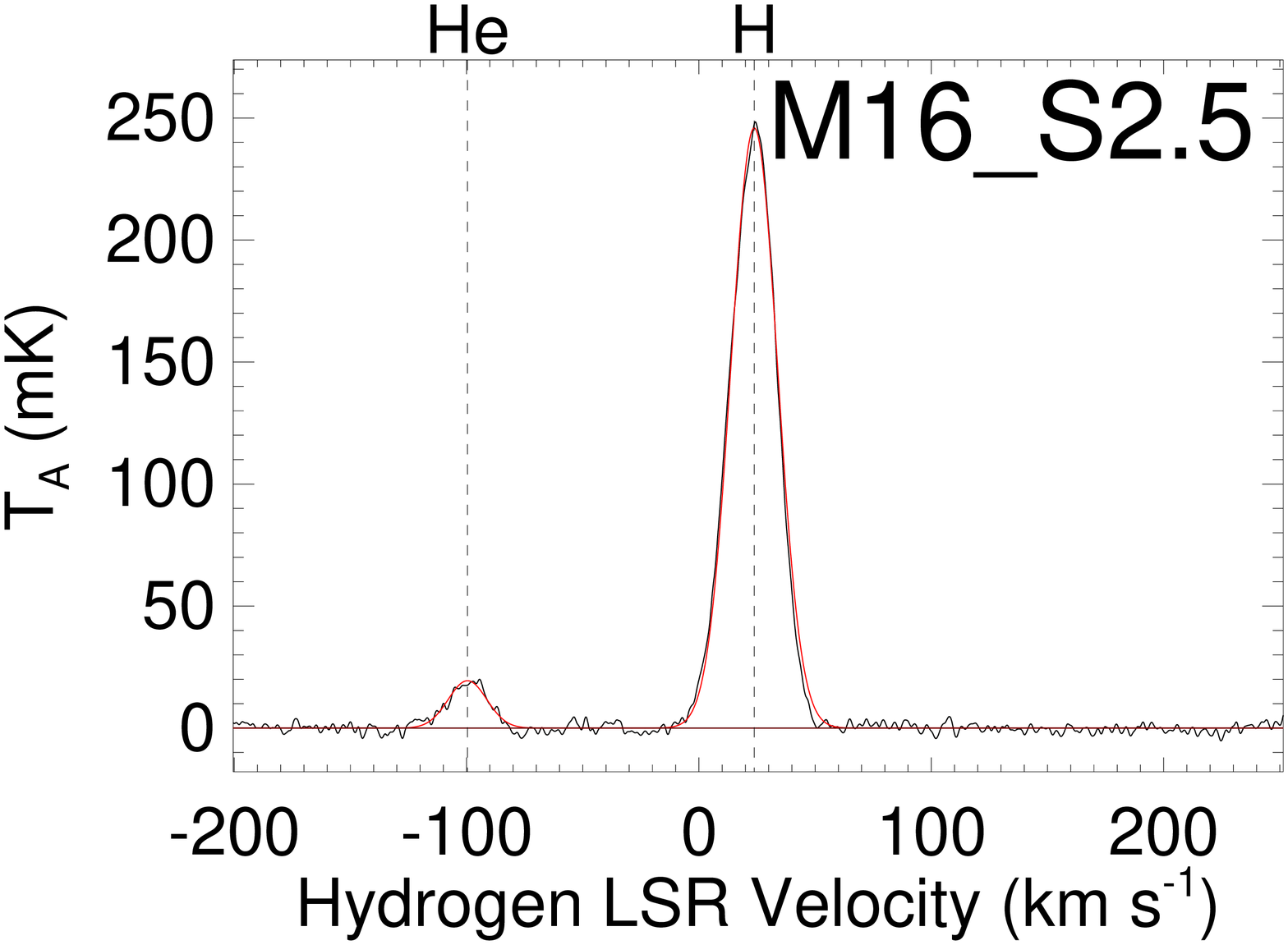} \\
\includegraphics[width=.23\textwidth]{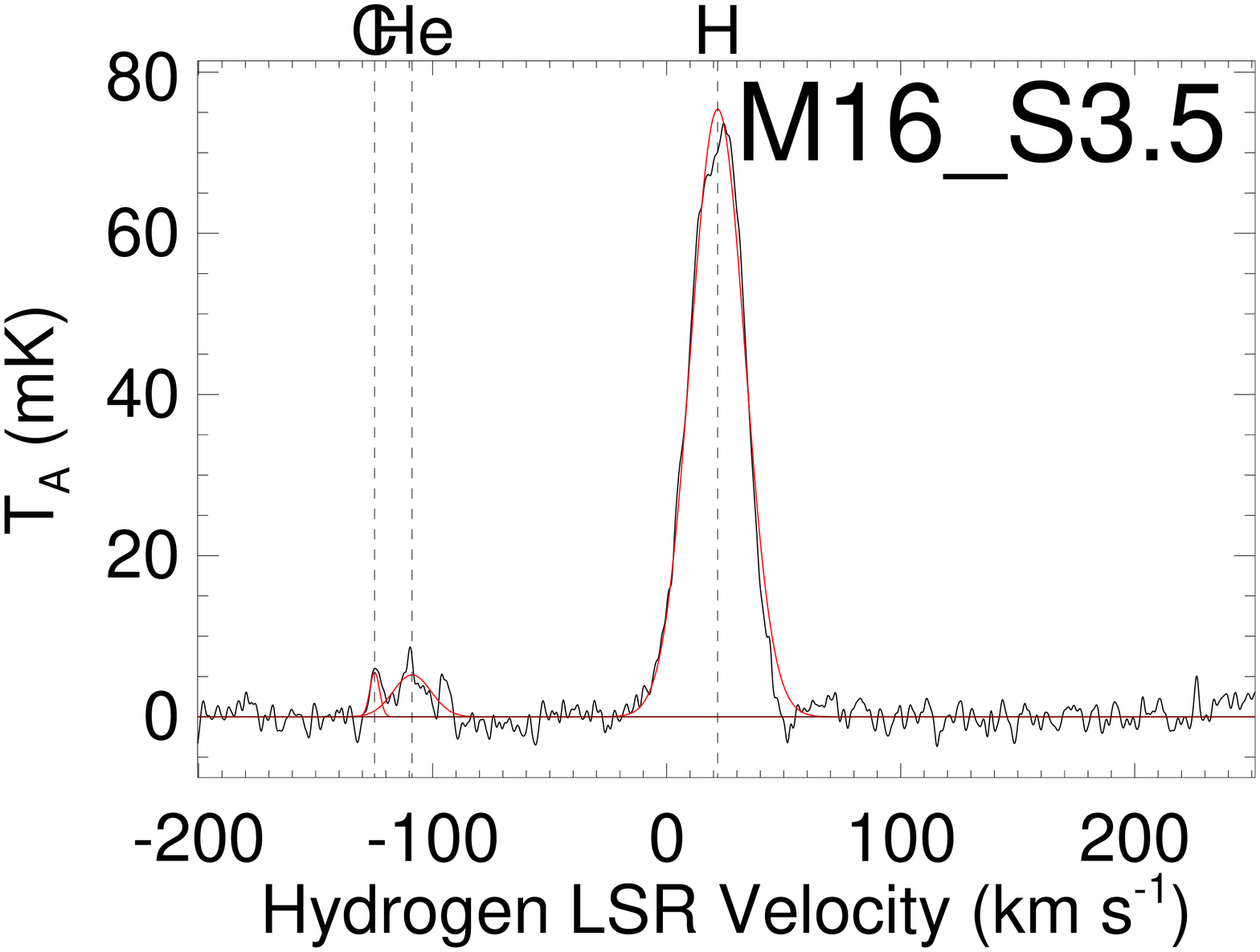} &
\includegraphics[width=.23\textwidth]{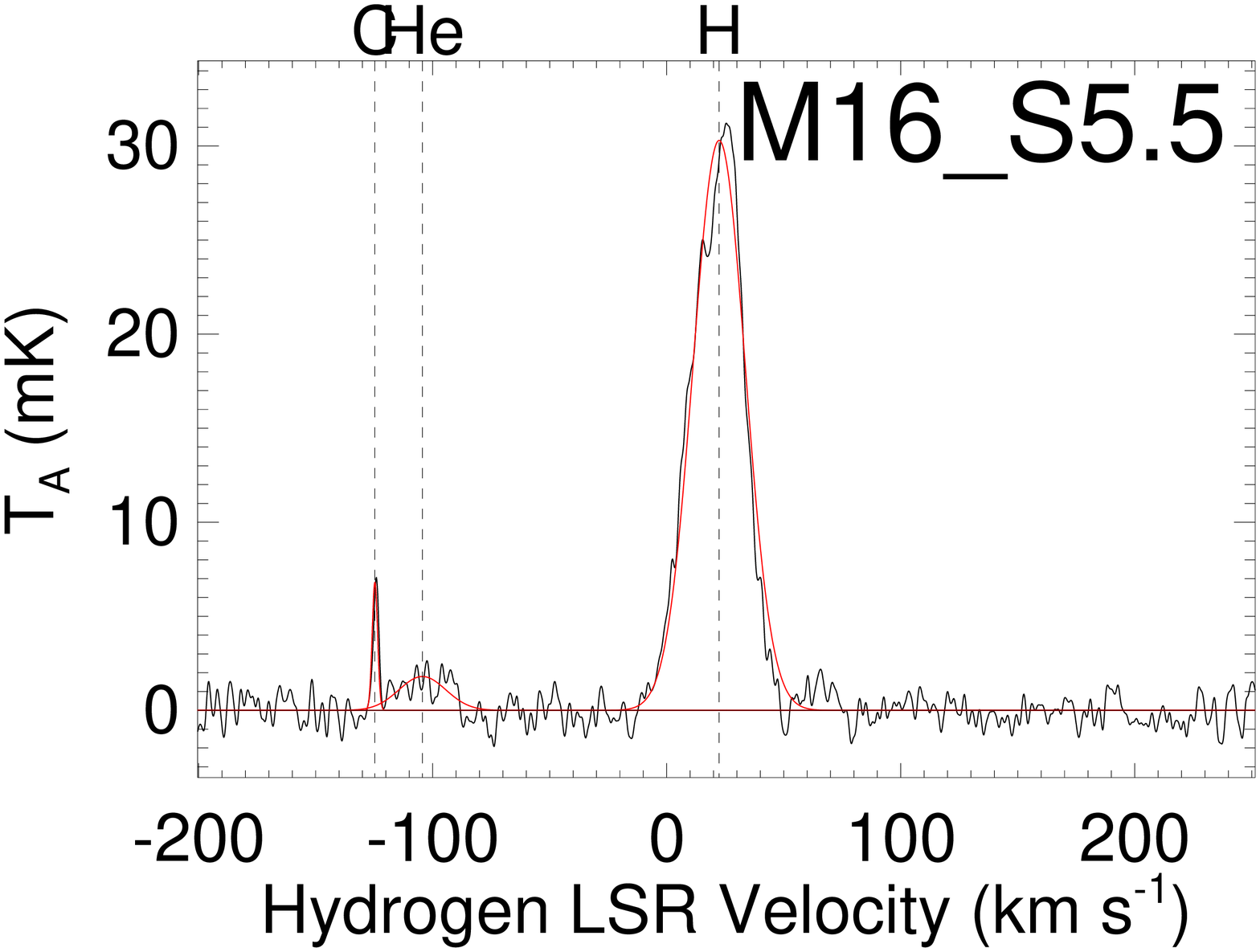} &
\includegraphics[width=.23\textwidth]{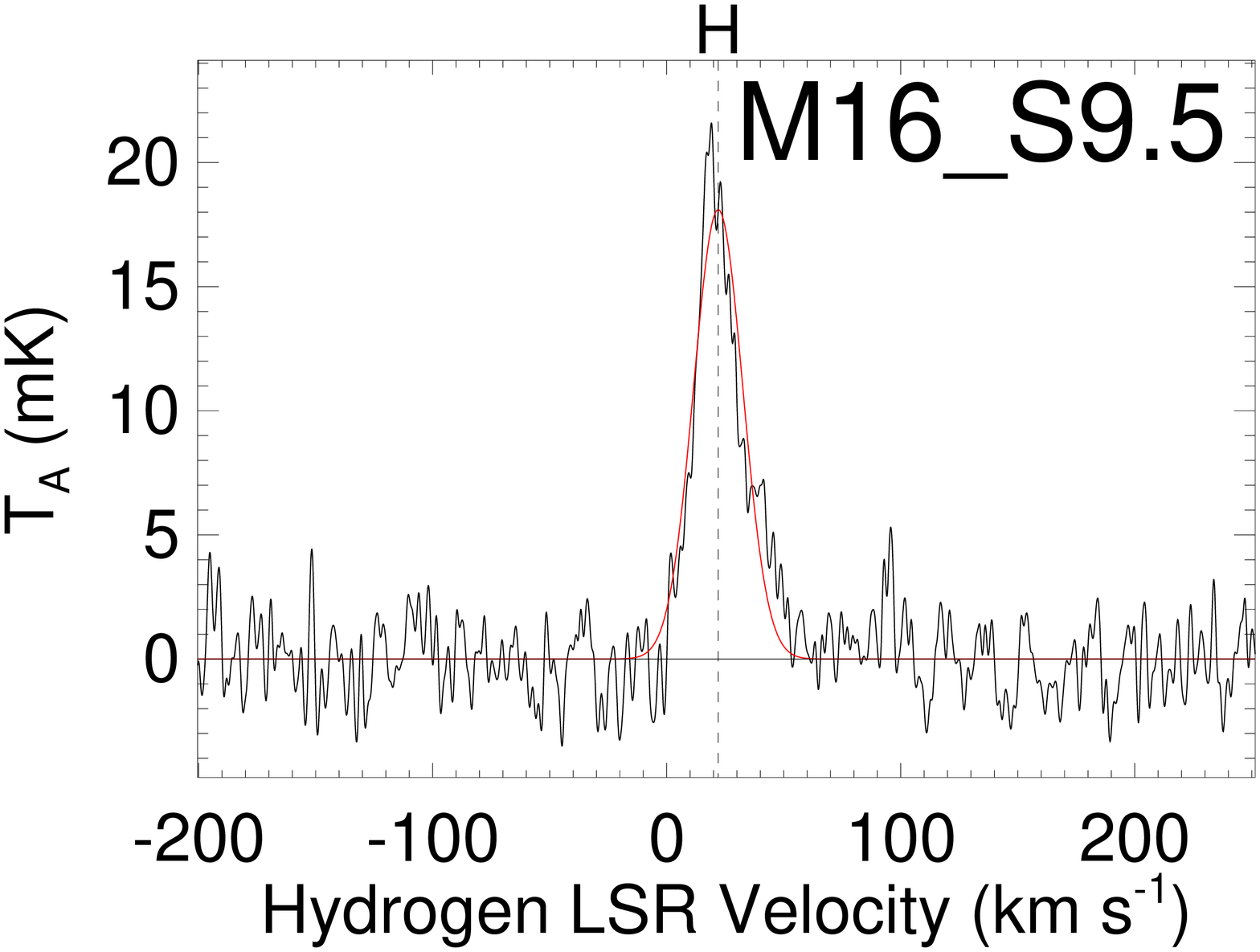} &
\includegraphics[width=.23\textwidth]{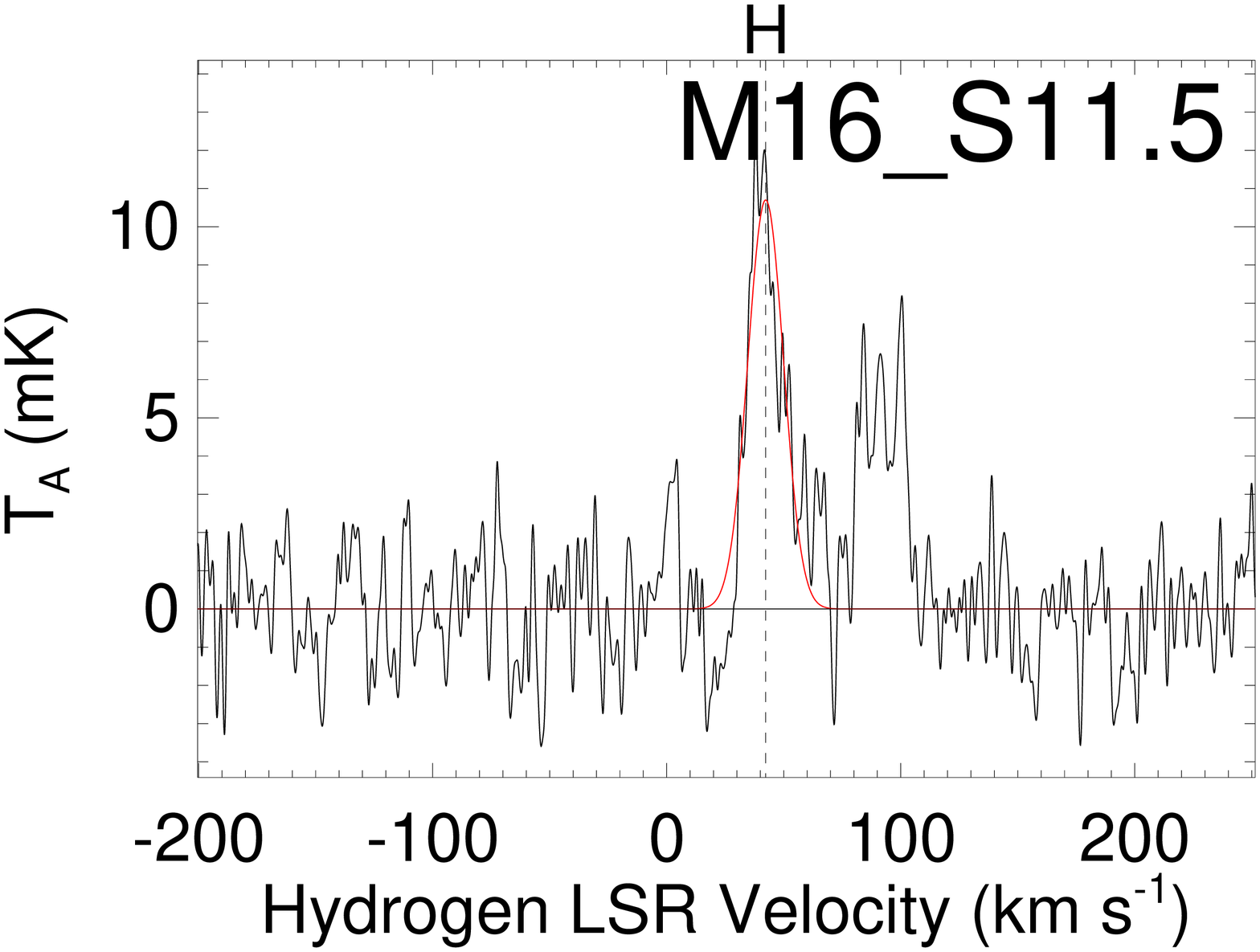} \\
\includegraphics[width=.23\textwidth]{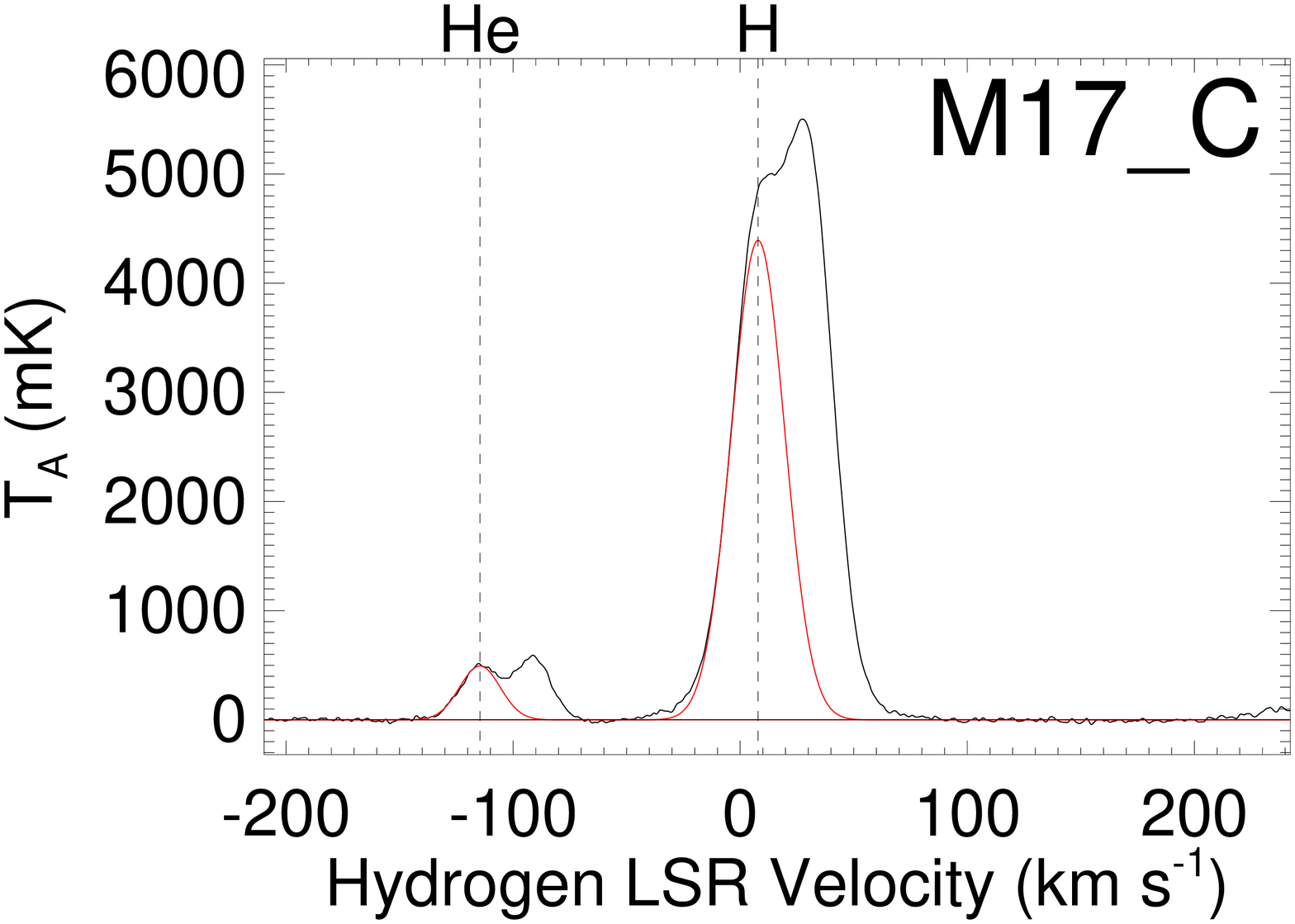} &
\includegraphics[width=.23\textwidth]{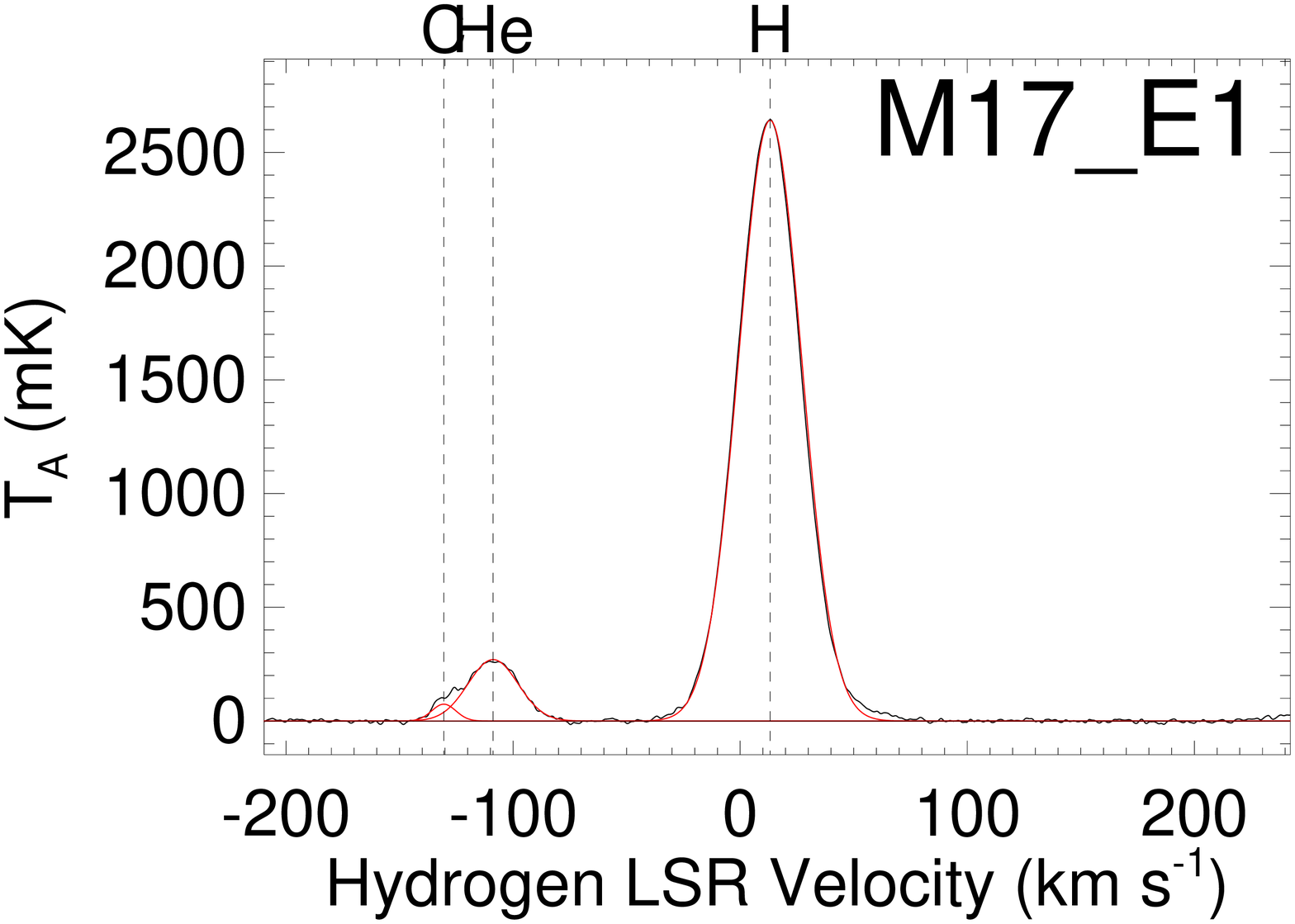} &
\includegraphics[width=.23\textwidth]{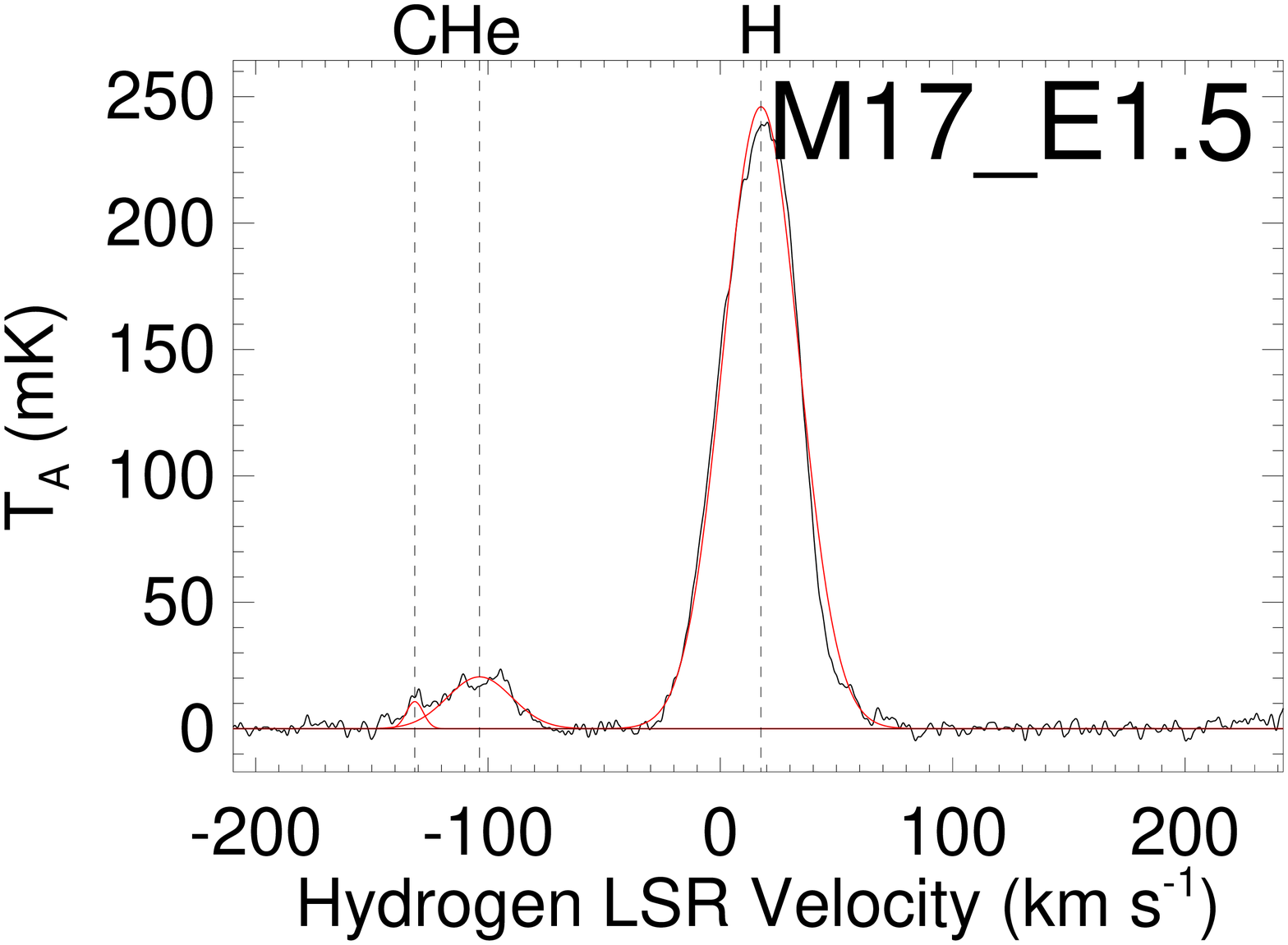} &
\includegraphics[width=.23\textwidth]{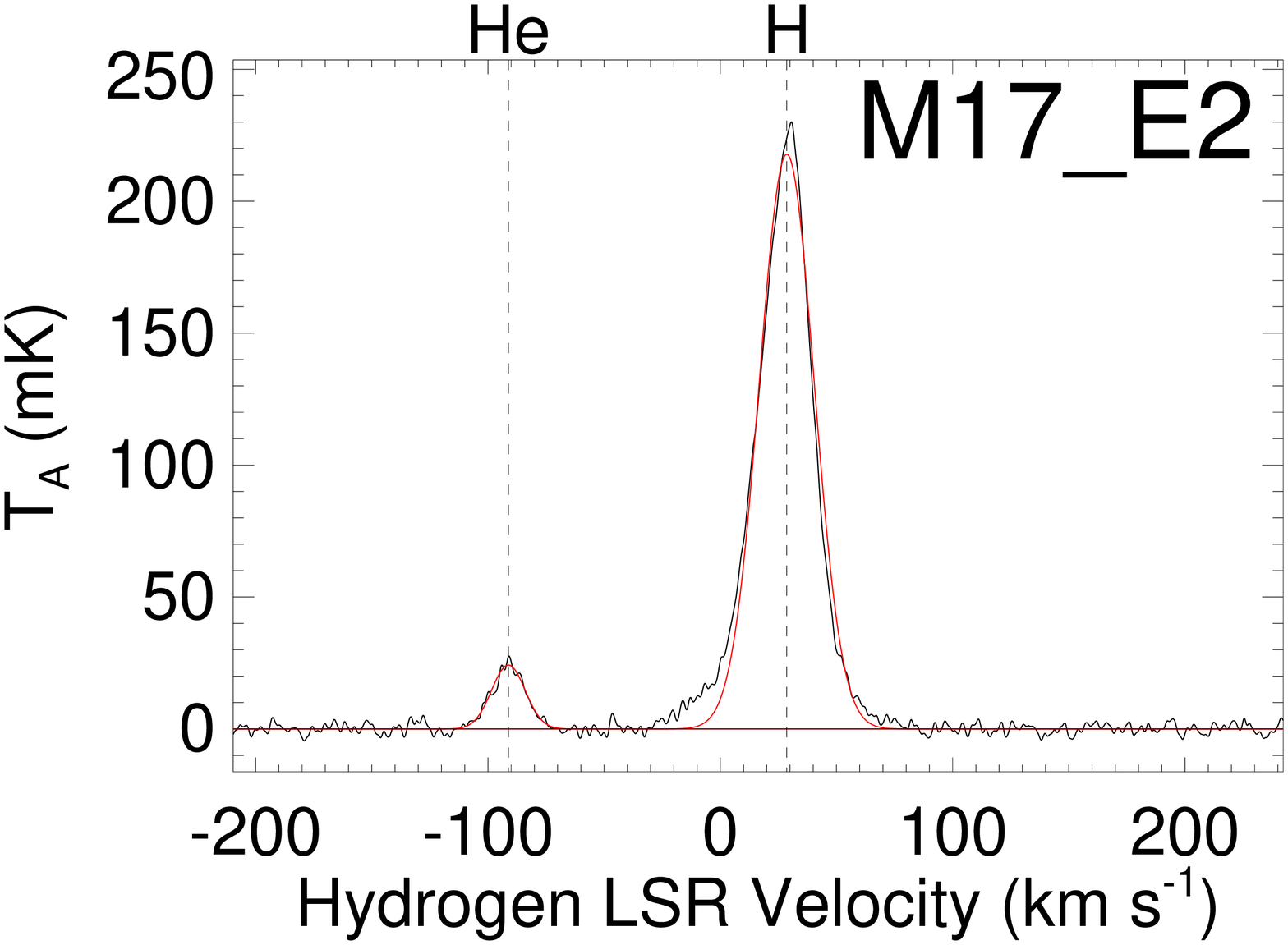} \\
\includegraphics[width=.23\textwidth]{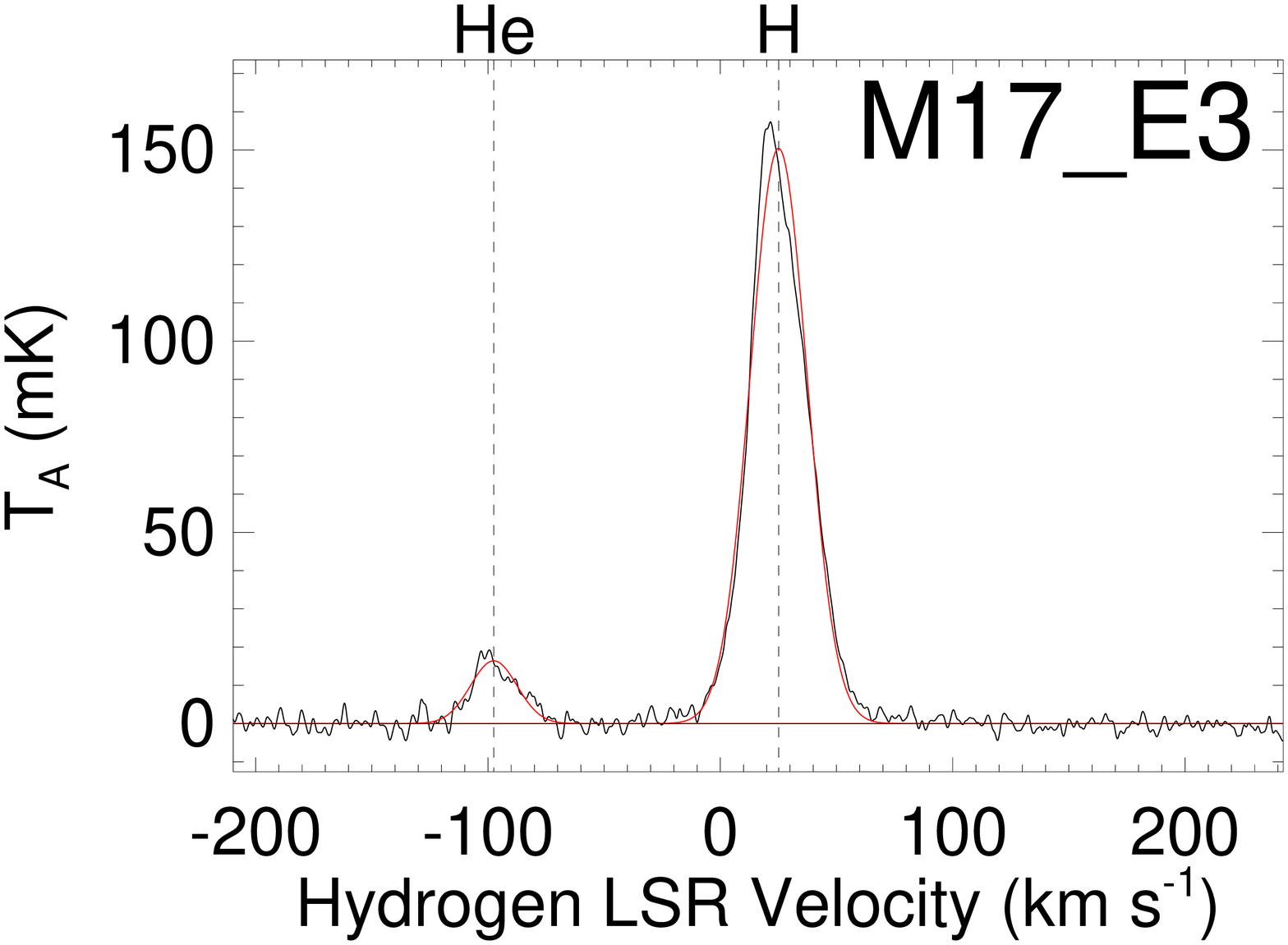} &
\includegraphics[width=.23\textwidth]{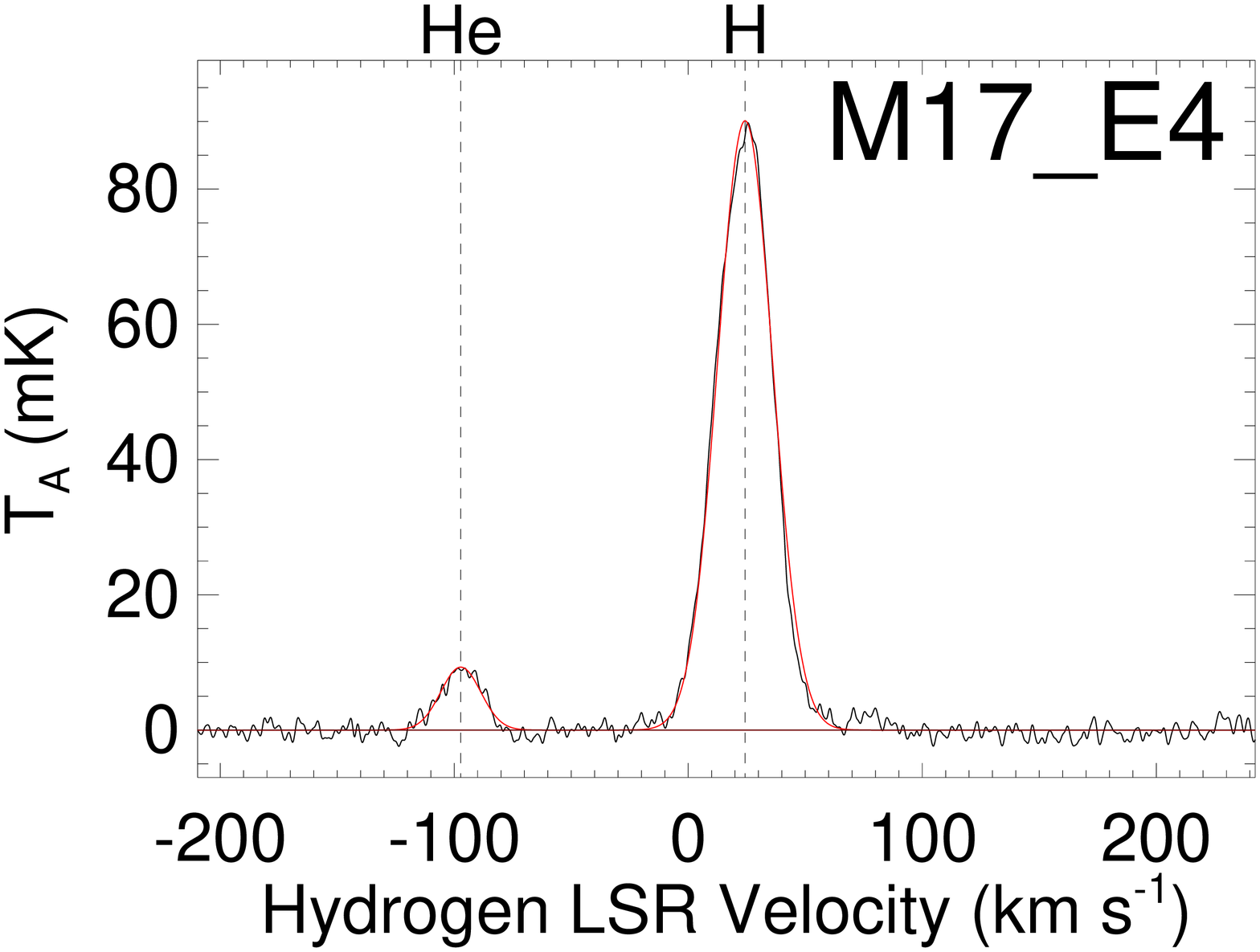} &
\includegraphics[width=.23\textwidth]{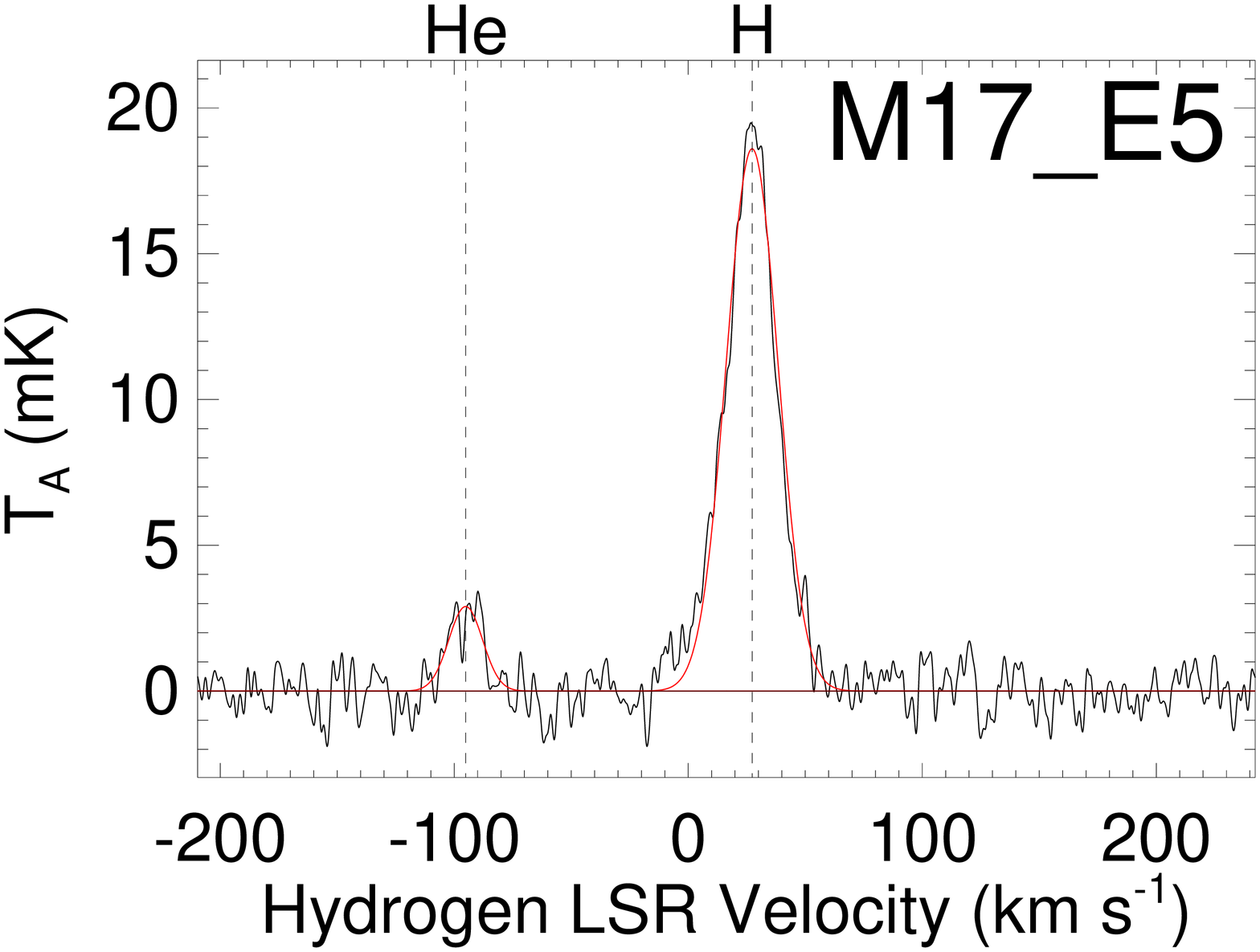} &
\includegraphics[width=.23\textwidth]{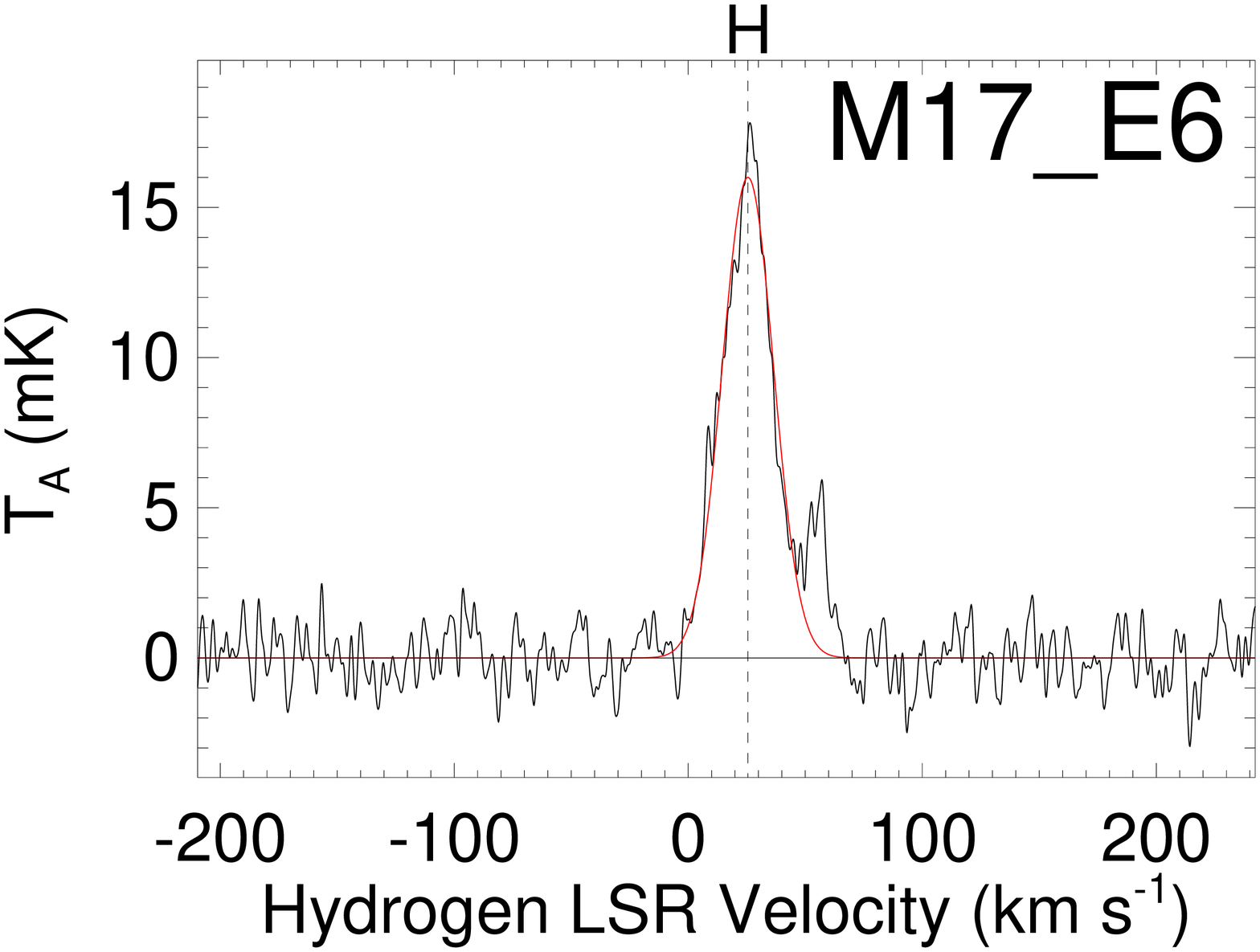} \\
\end{tabular}
\caption{}
\end{figure*}
\renewcommand{\thefigure}{\thesection.\arabic{figure}}

\renewcommand\thefigure{\thesection.\arabic{figure} (Cont.)}
\addtocounter{figure}{-1}
\begin{figure*}
\centering
\begin{tabular}{cccc}
\includegraphics[width=.23\textwidth]{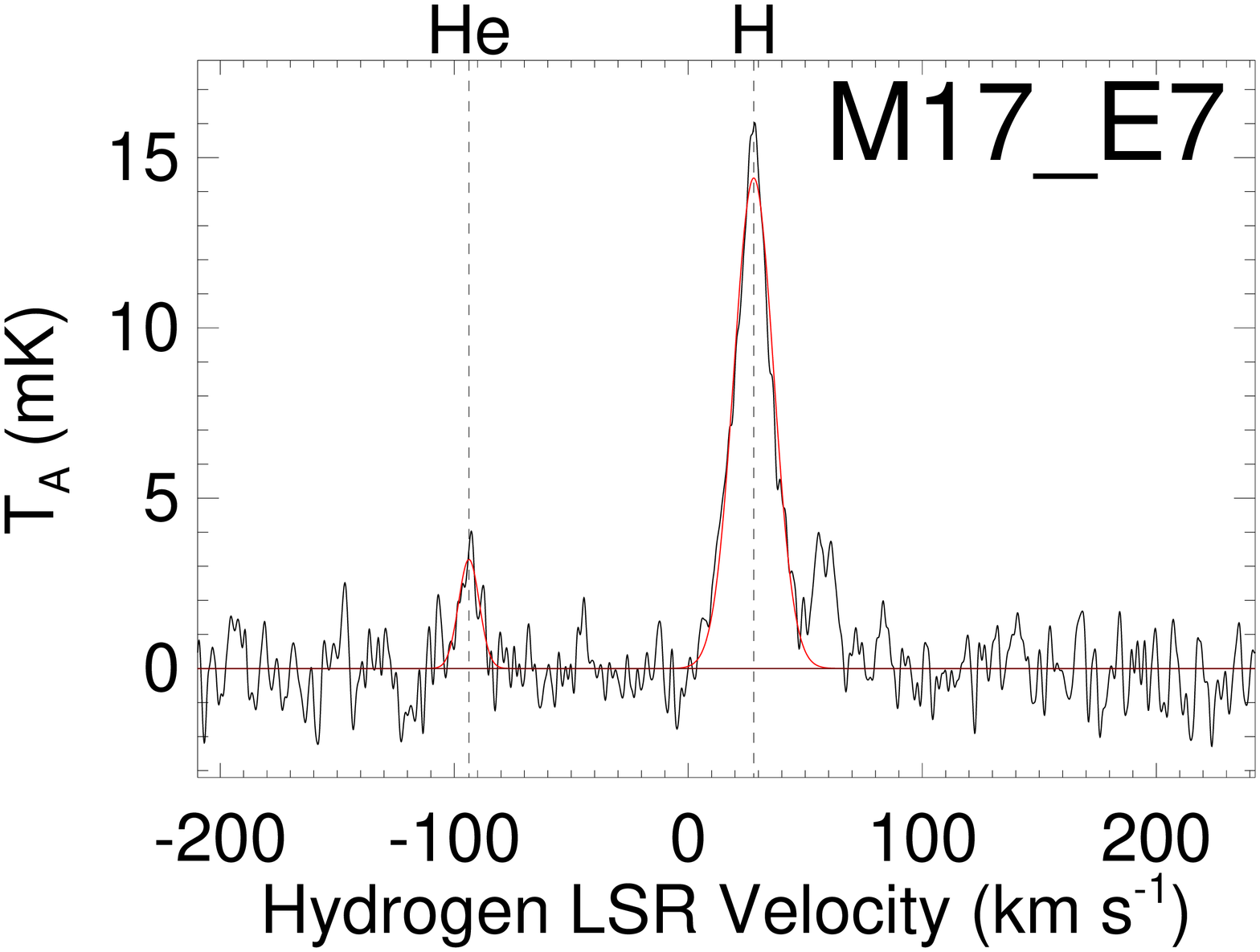} &
\includegraphics[width=.23\textwidth]{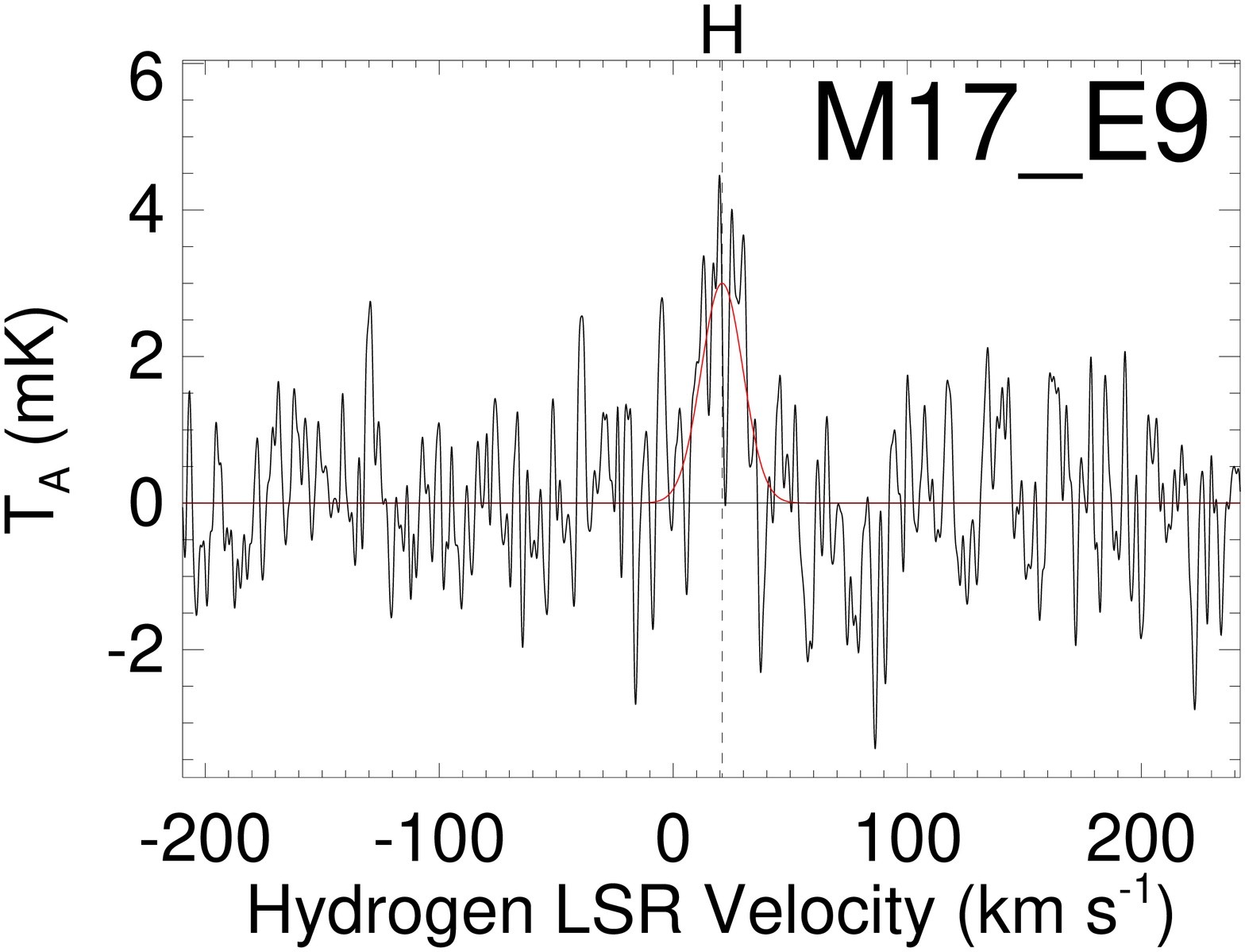} &
\includegraphics[width=.23\textwidth]{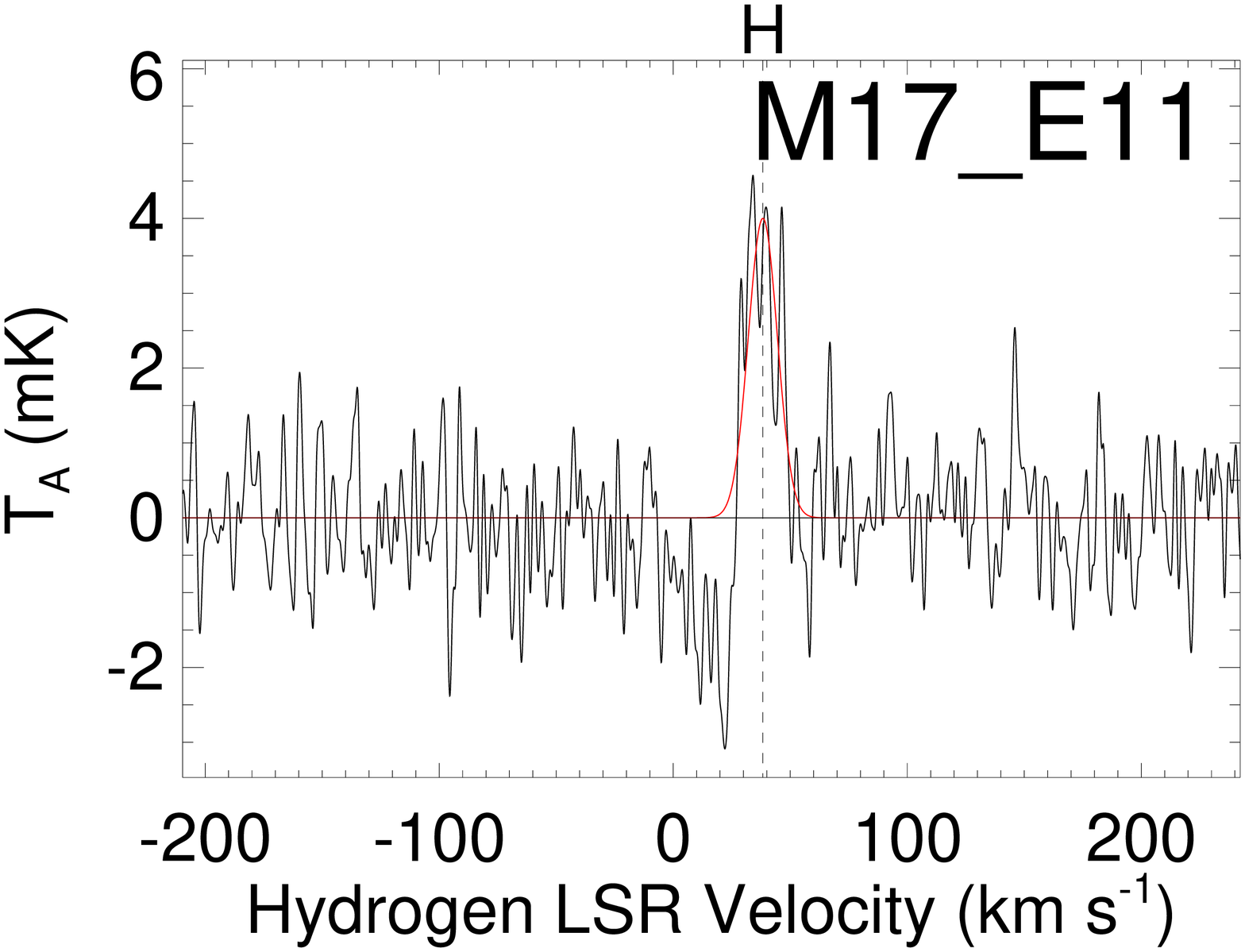} &
\includegraphics[width=.23\textwidth]{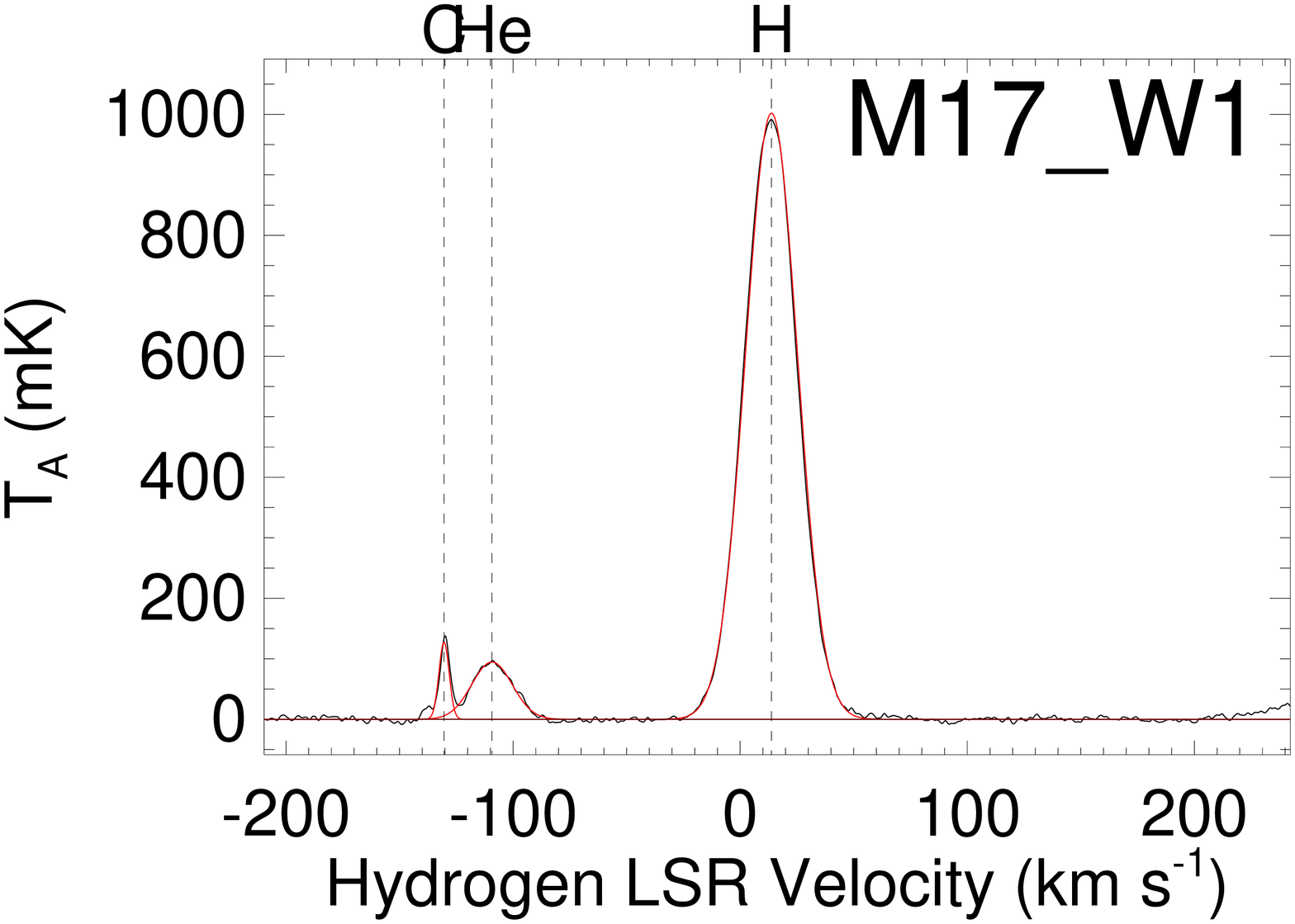} \\
\includegraphics[width=.23\textwidth]{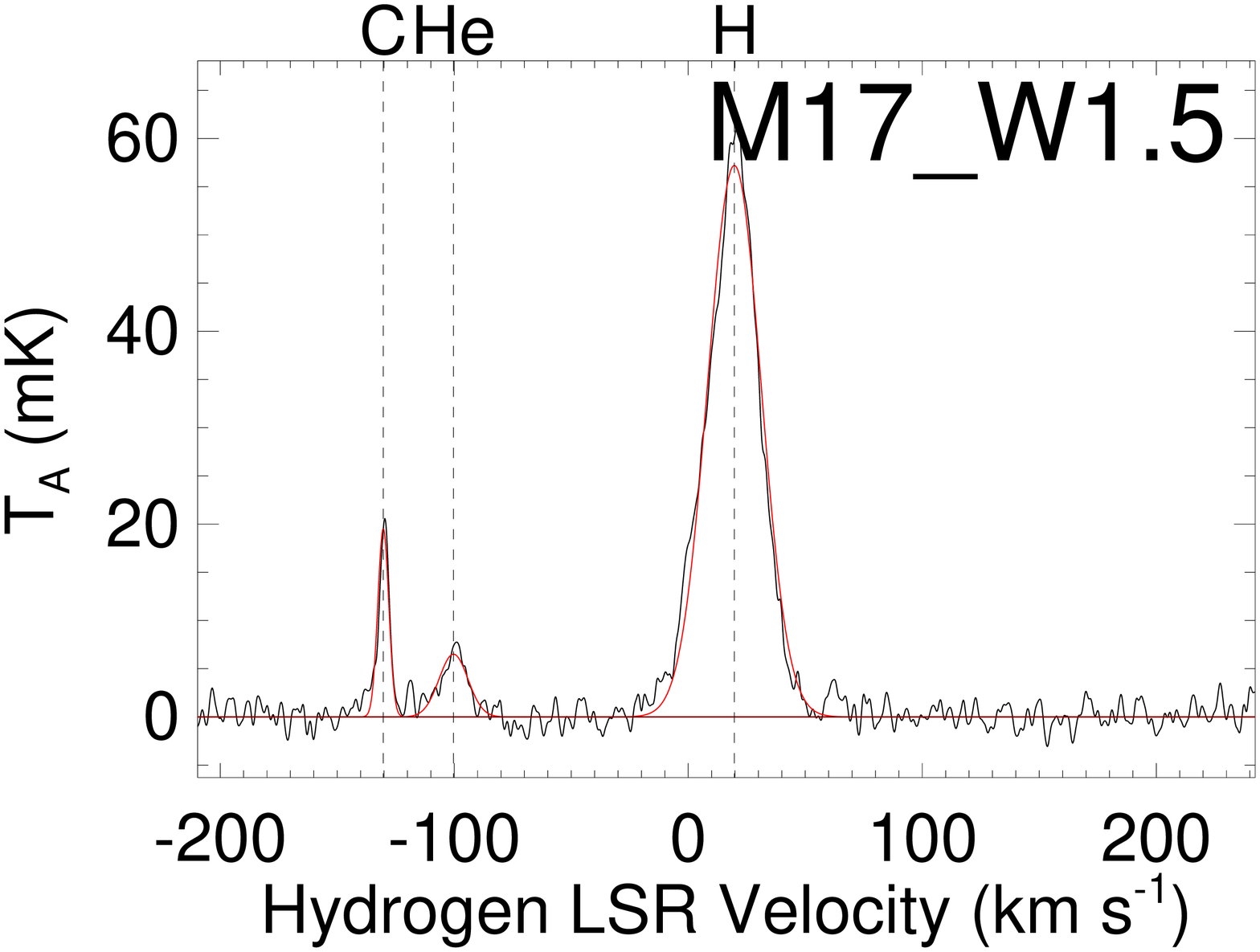} &
\includegraphics[width=.23\textwidth]{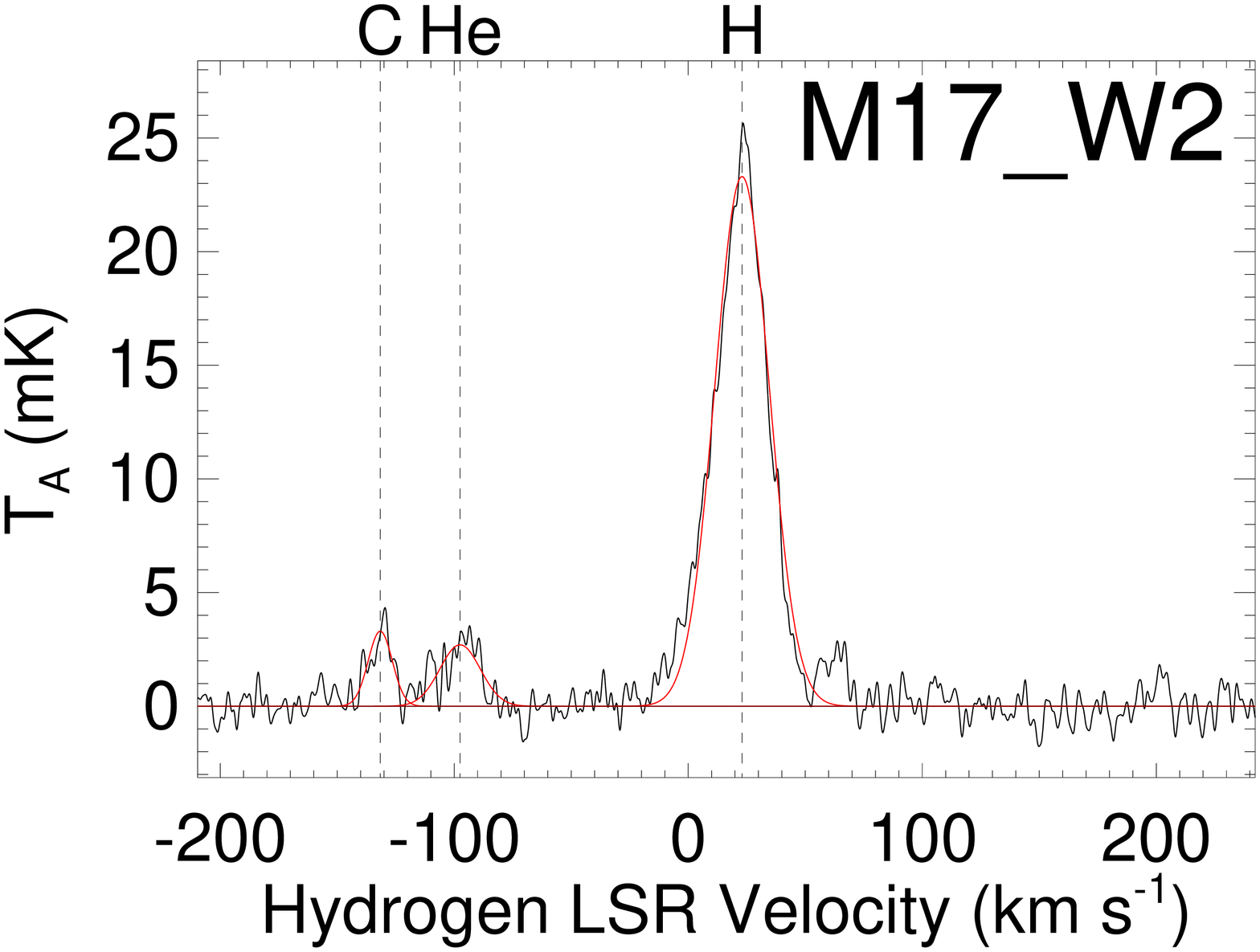} &
\includegraphics[width=.23\textwidth]{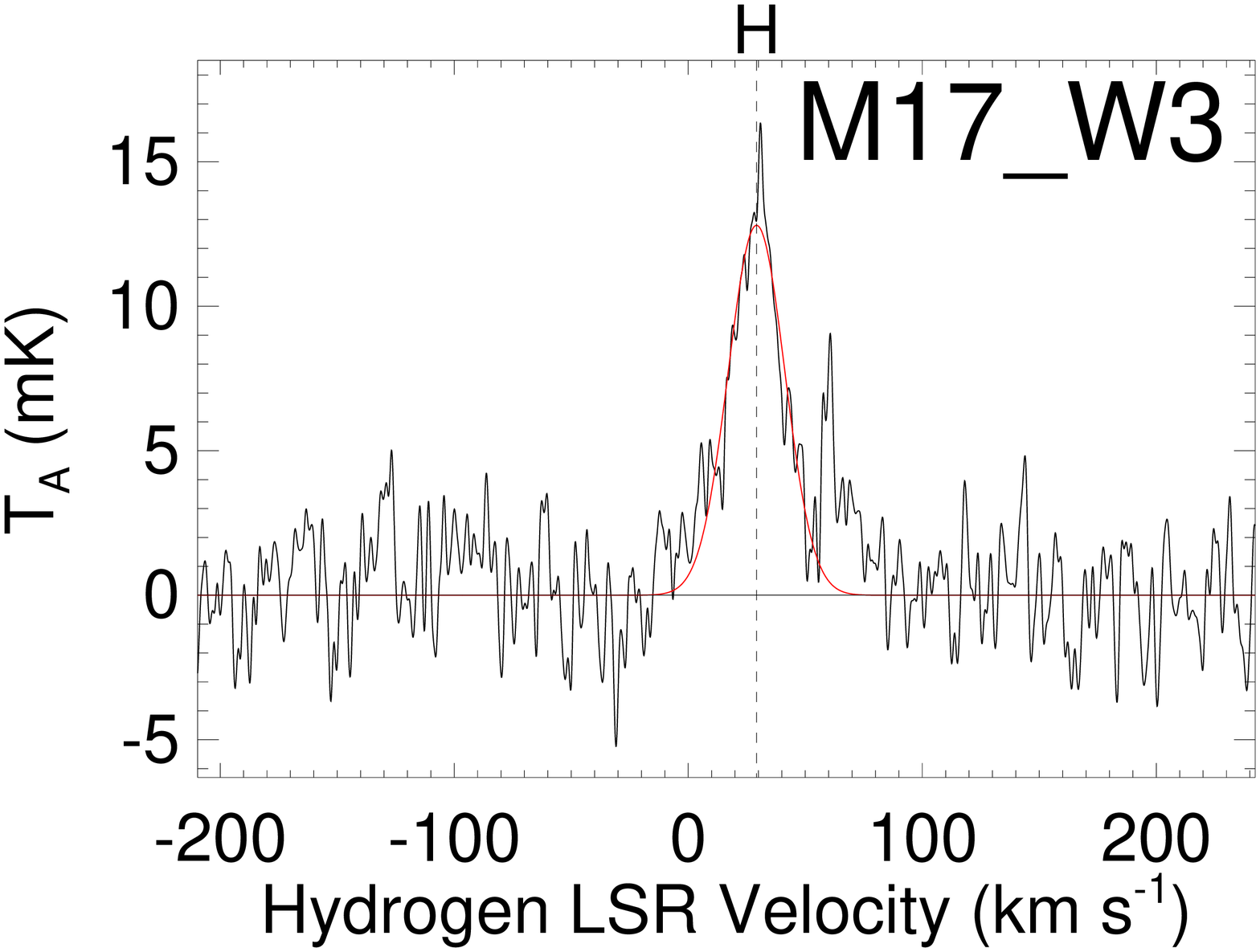} &
\includegraphics[width=.23\textwidth]{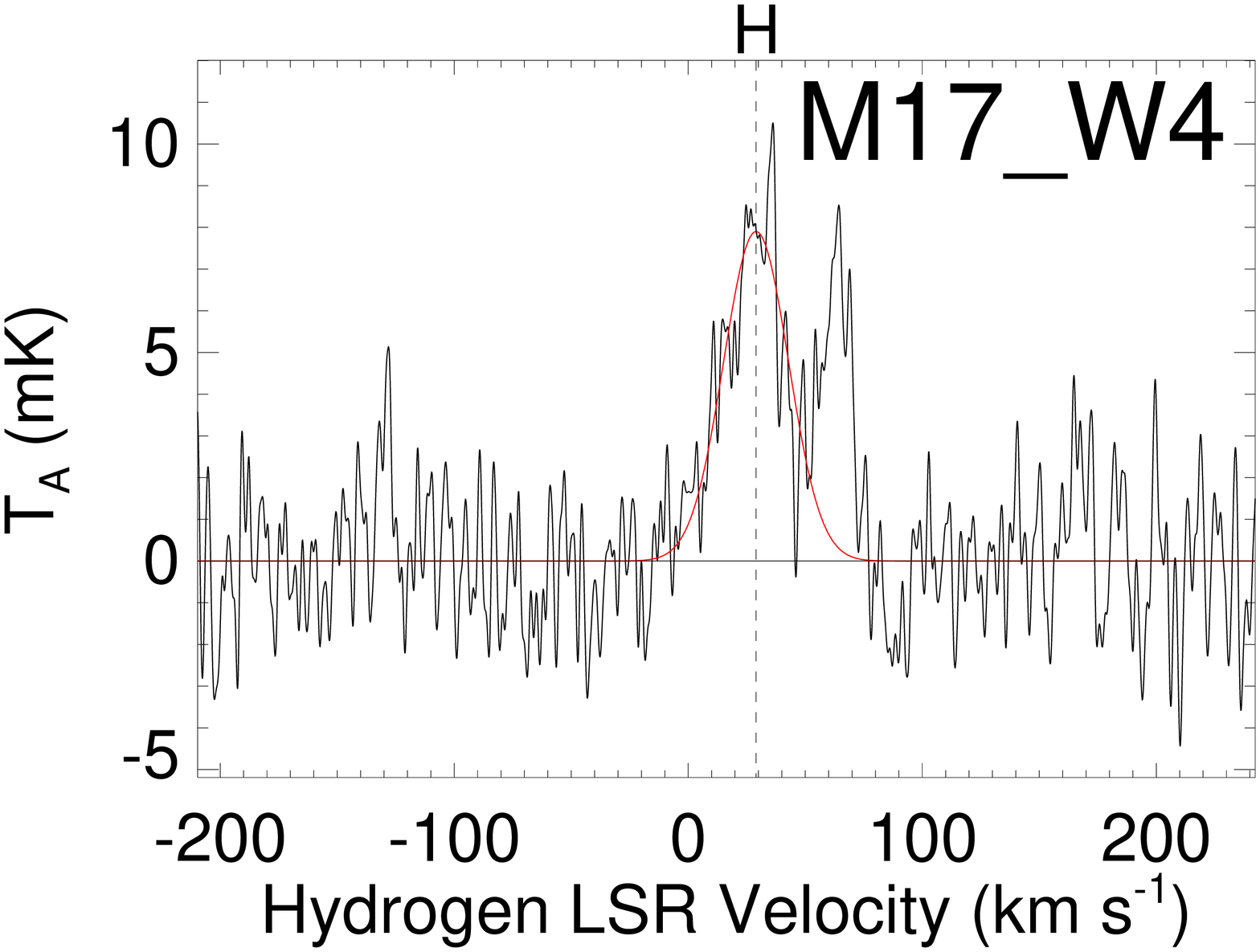} \\
\includegraphics[width=.23\textwidth]{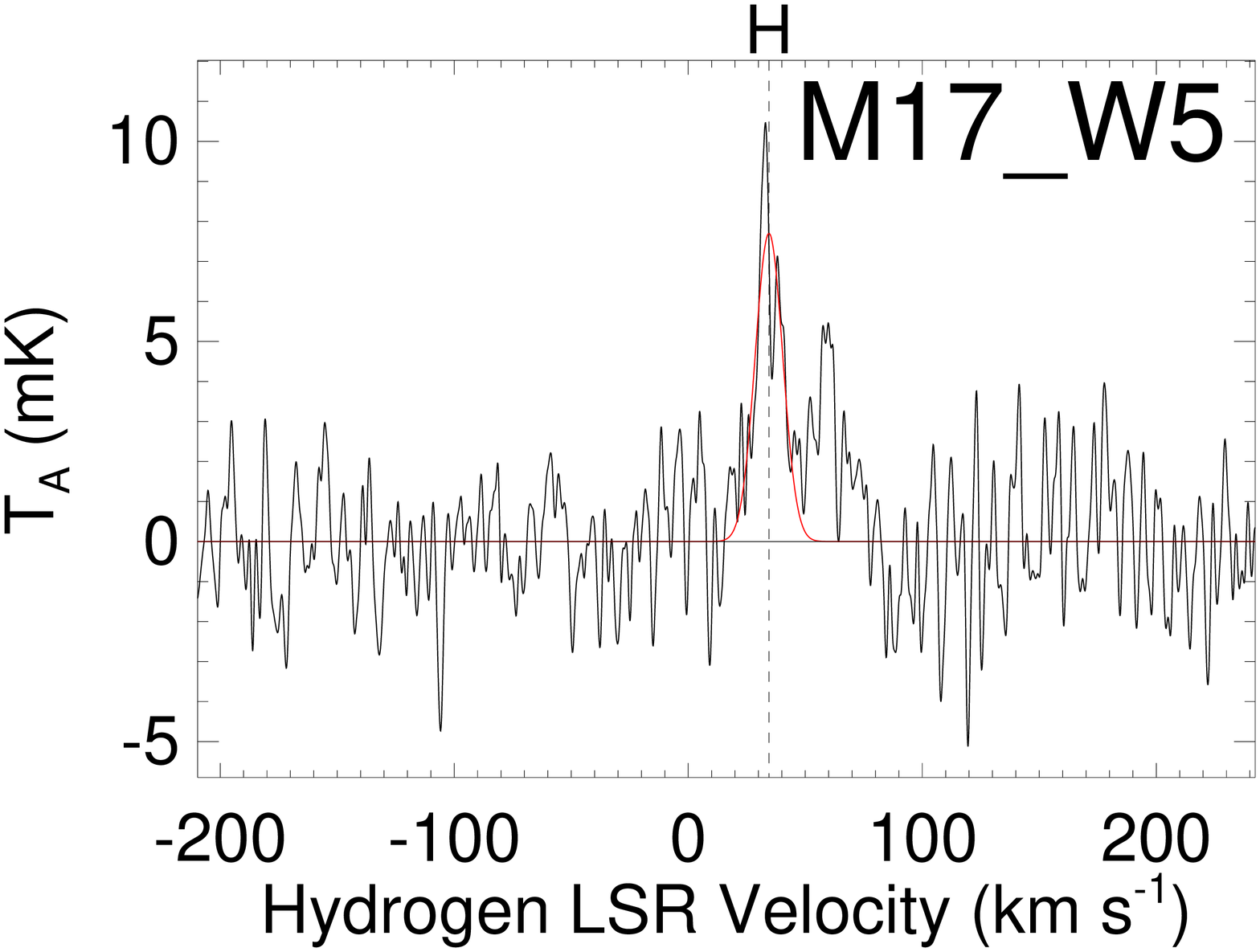} &
\includegraphics[width=.23\textwidth]{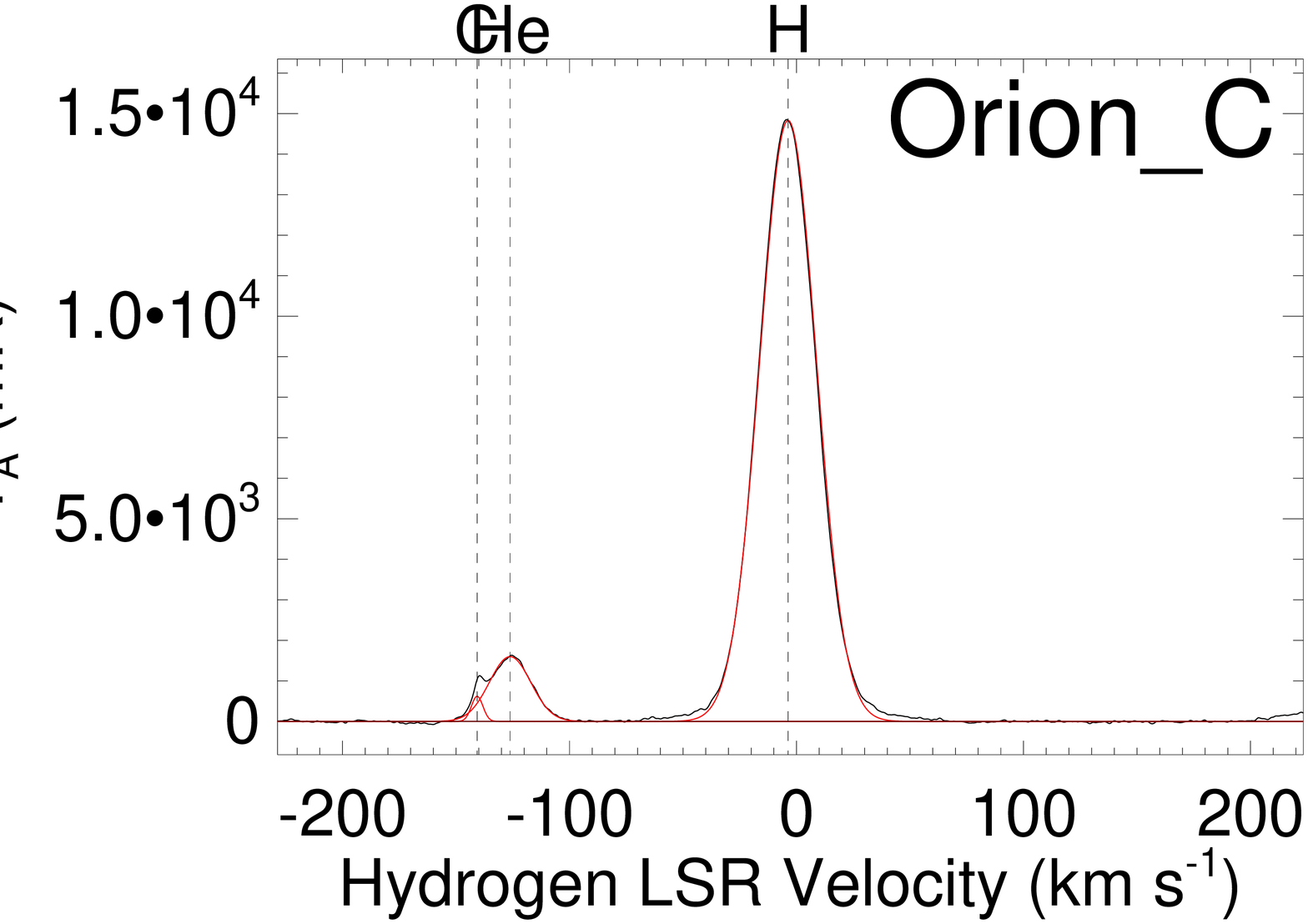} &
\includegraphics[width=.23\textwidth]{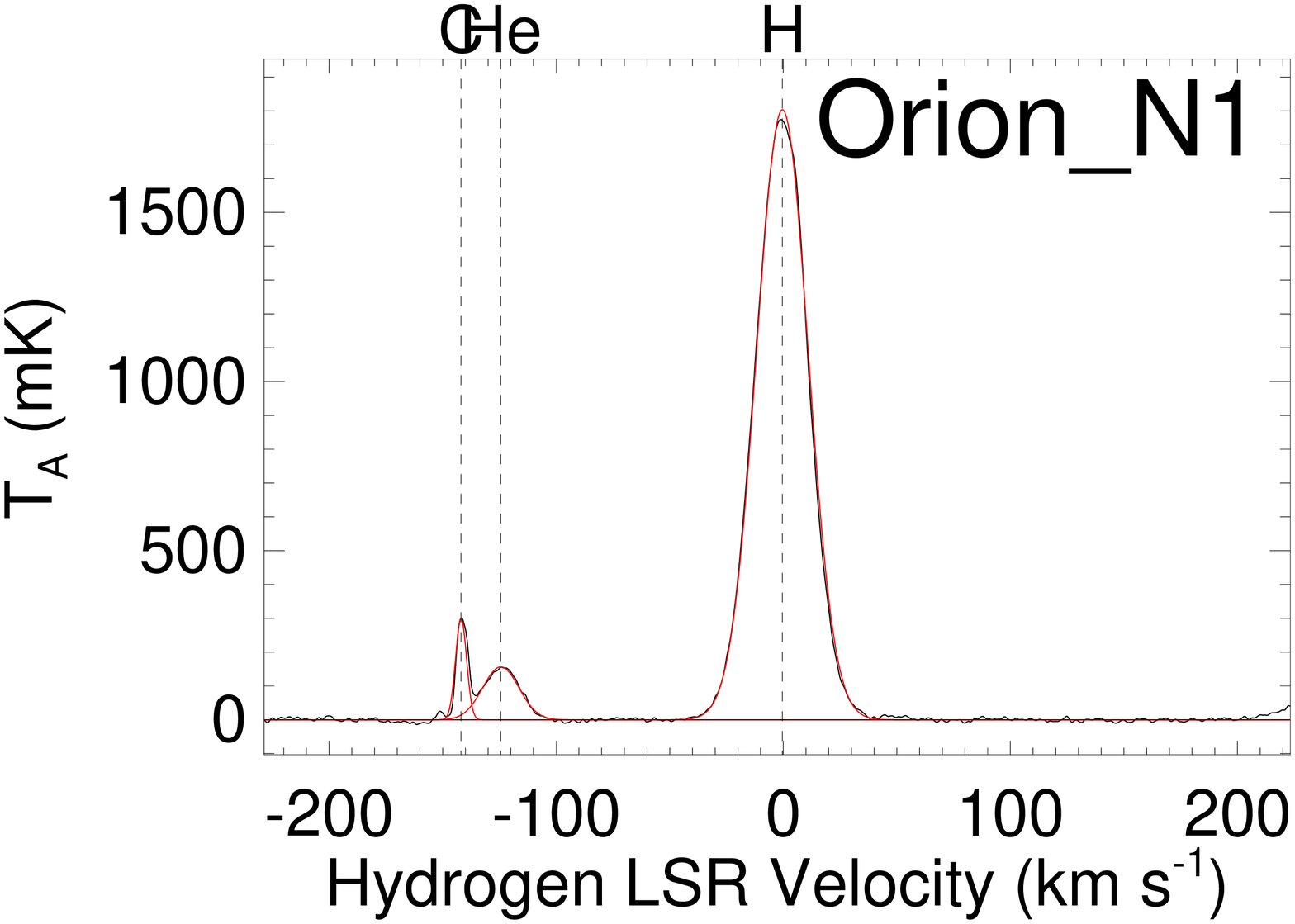} &
\includegraphics[width=.23\textwidth]{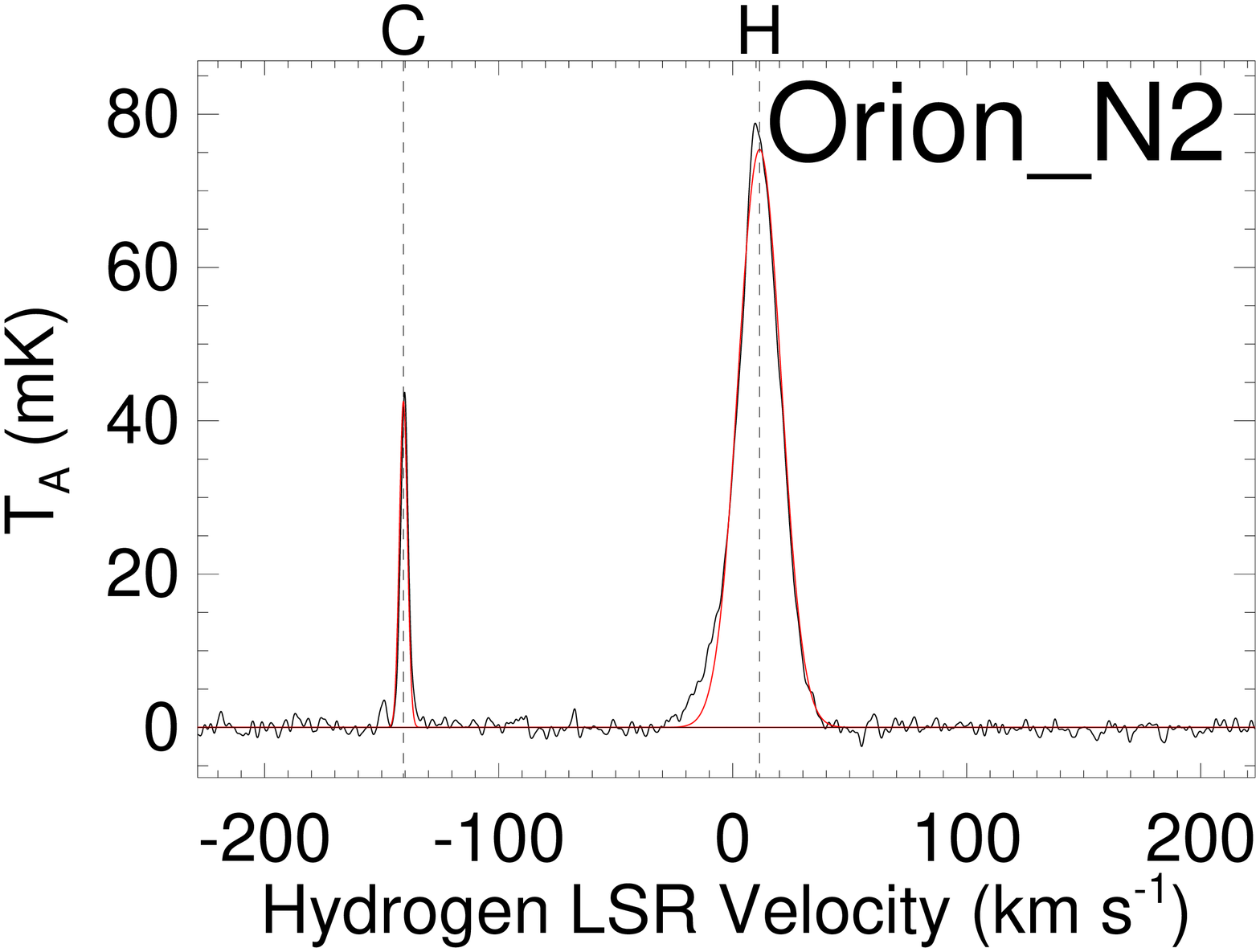} \\
\includegraphics[width=.23\textwidth]{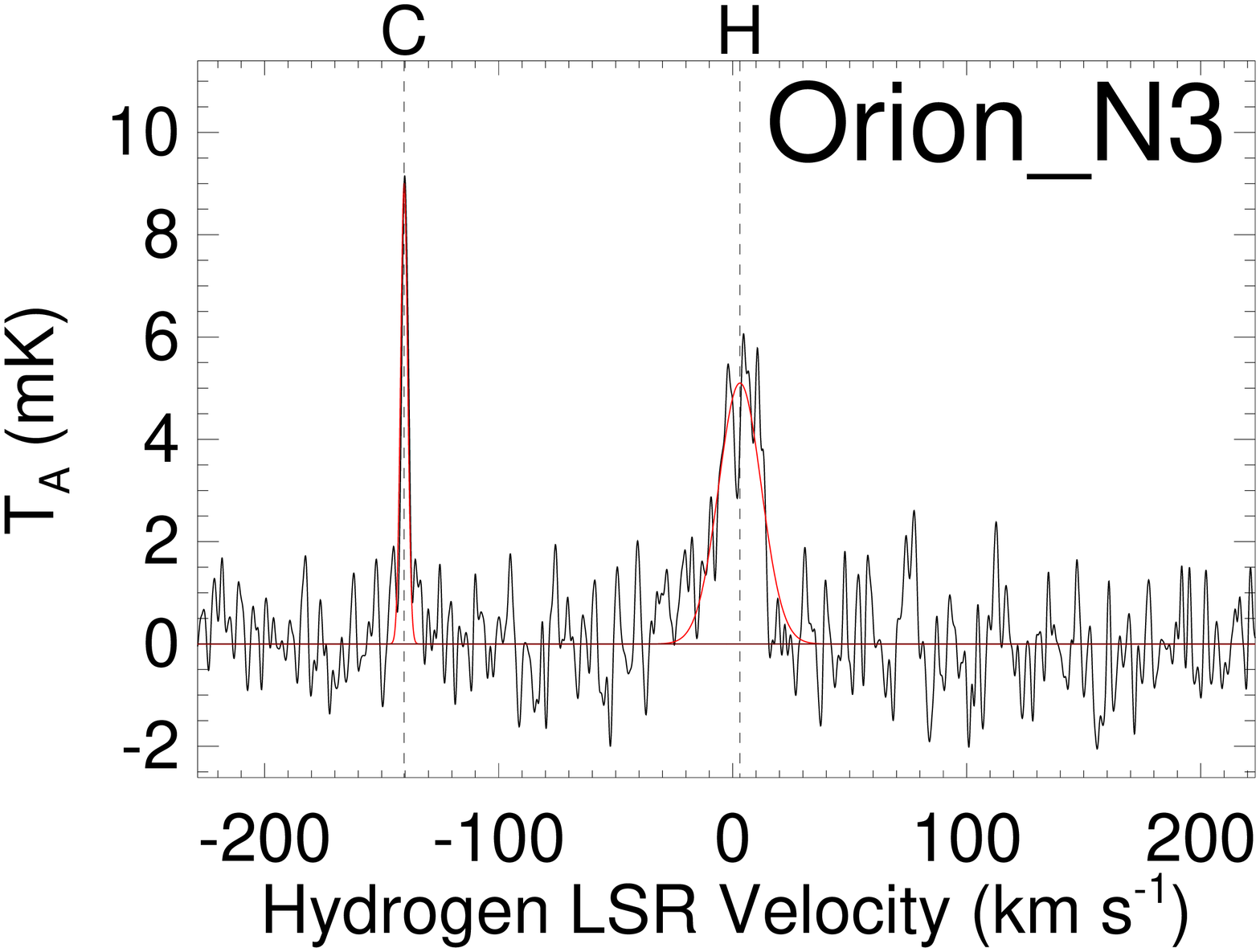} &
\includegraphics[width=.23\textwidth]{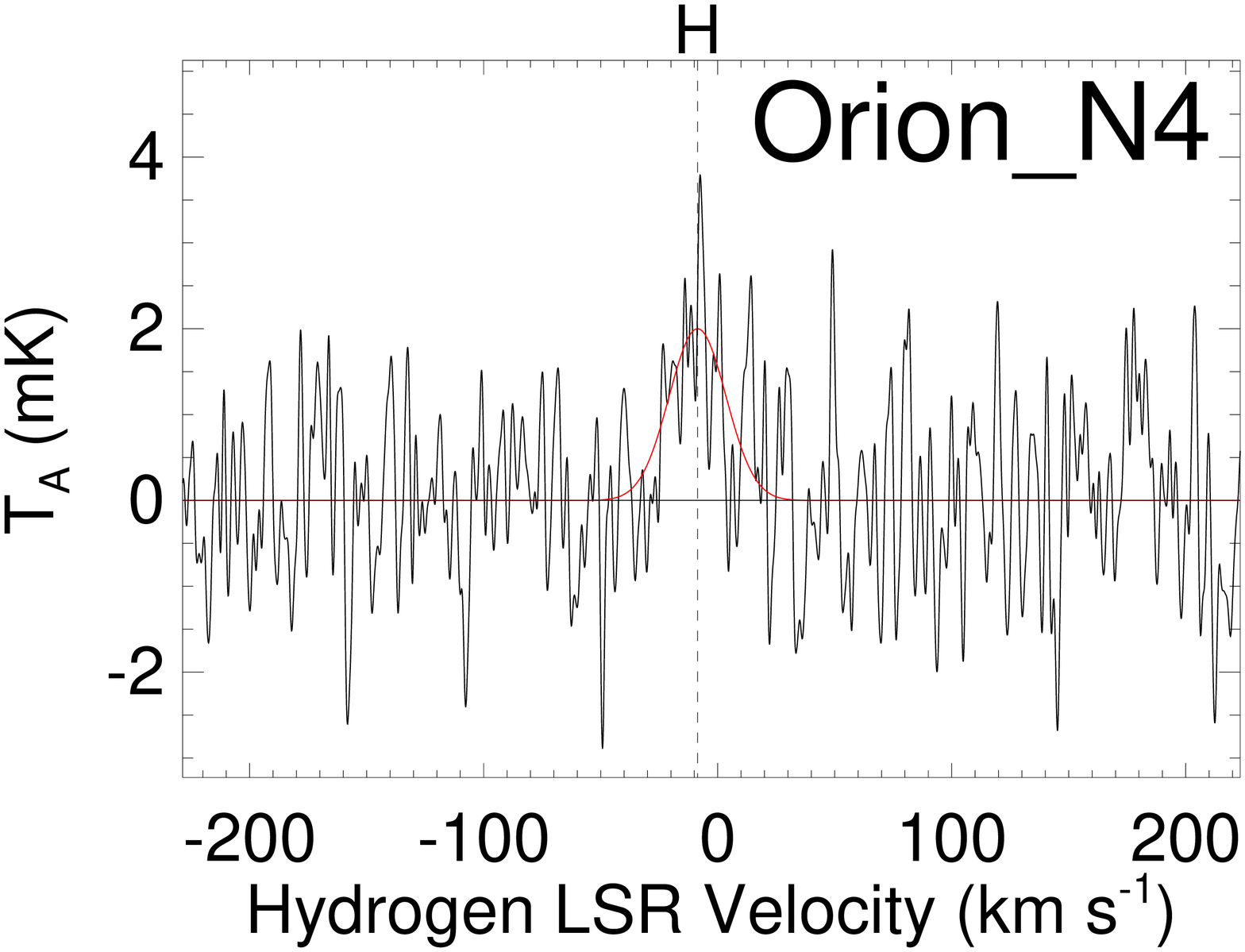} &
\includegraphics[width=.23\textwidth]{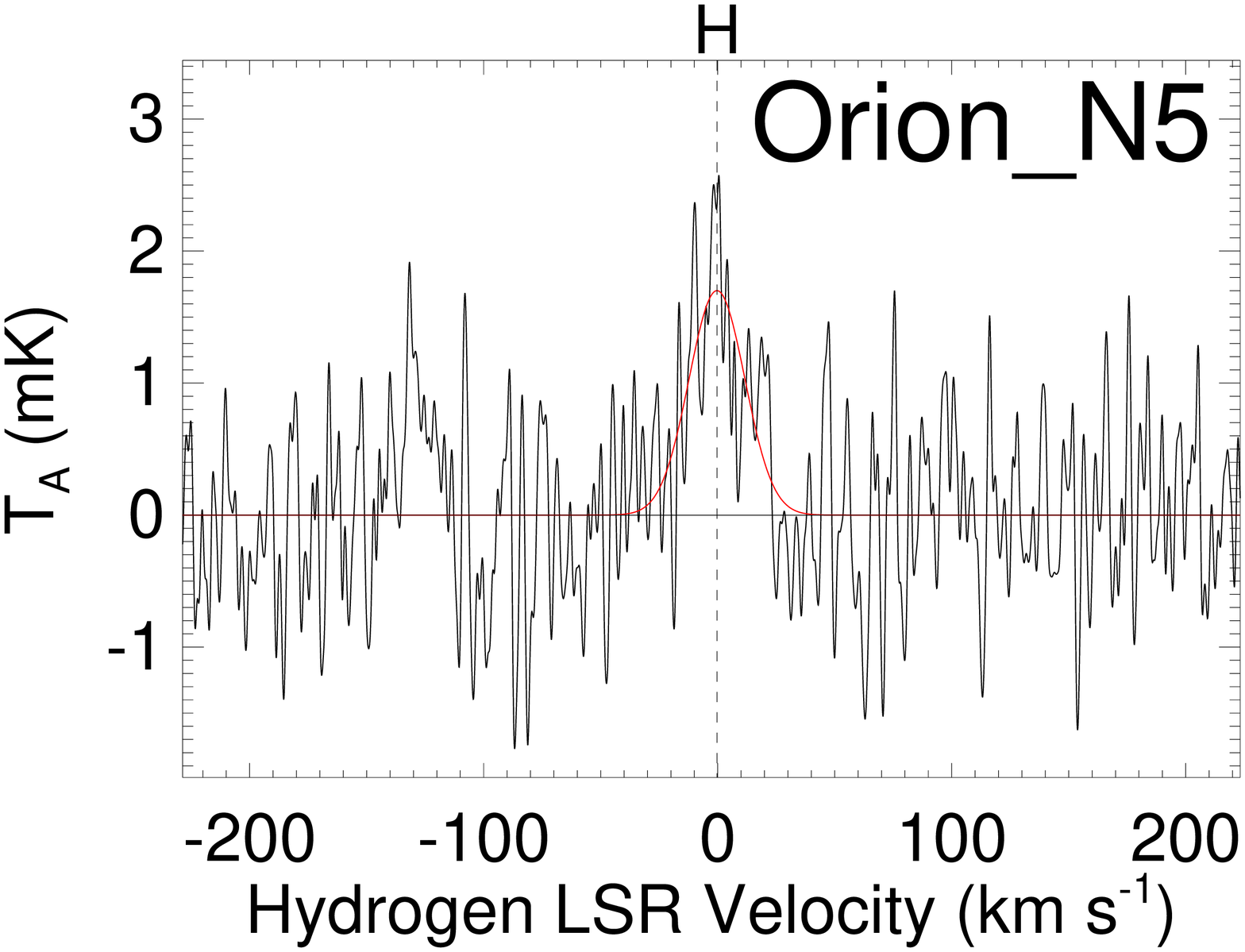} &
\includegraphics[width=.23\textwidth]{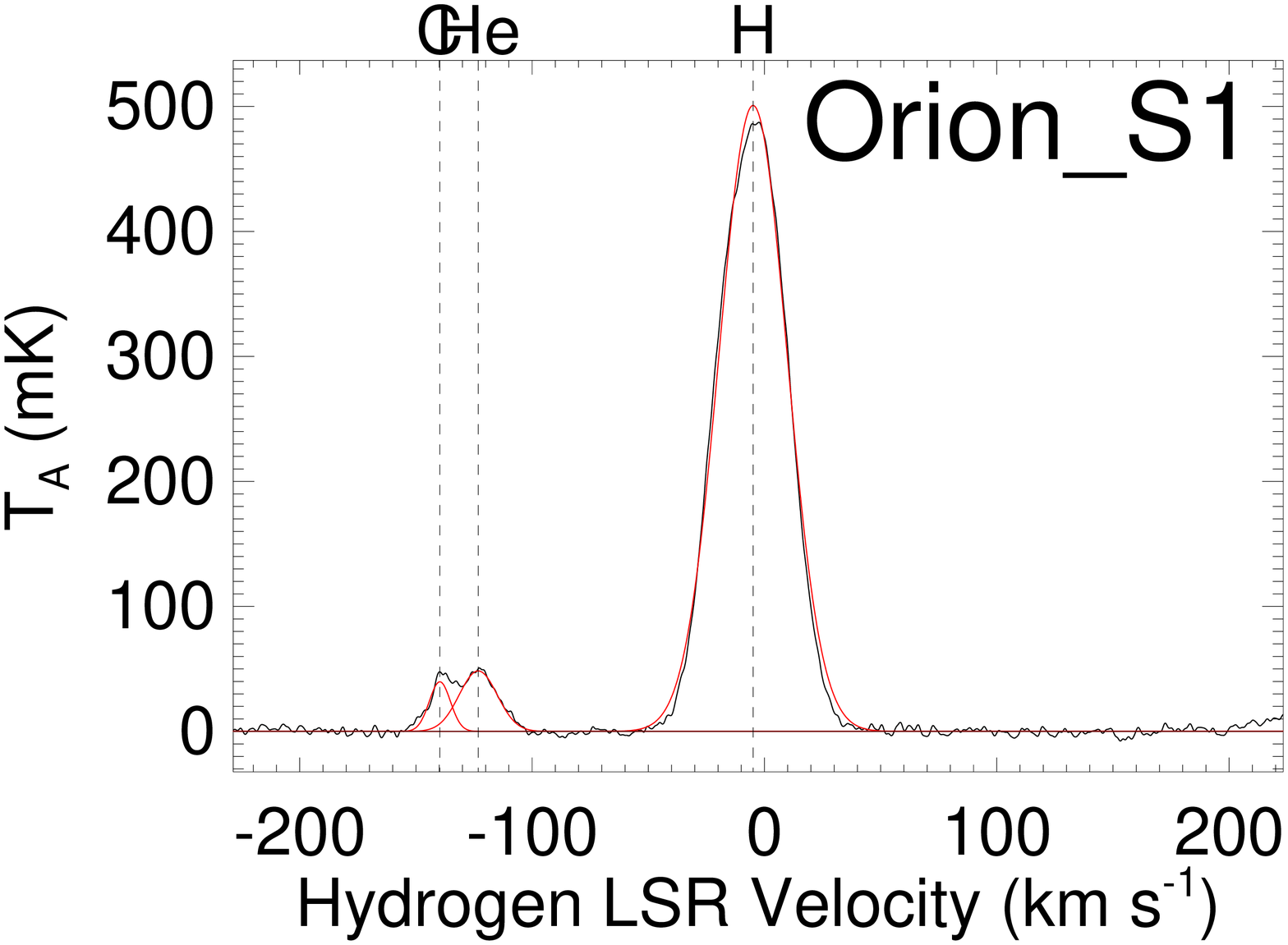} \\
\includegraphics[width=.23\textwidth]{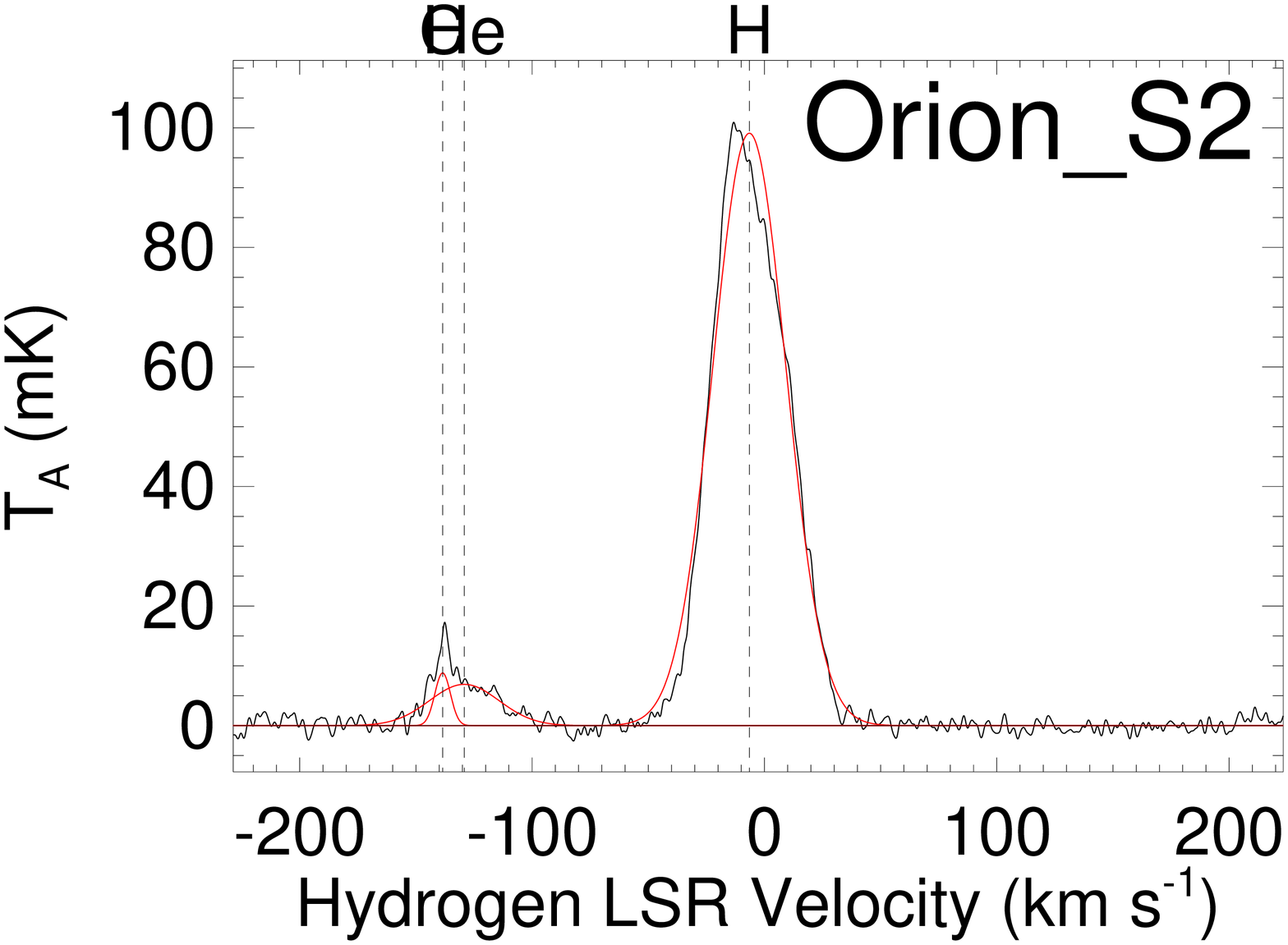} &
\includegraphics[width=.23\textwidth]{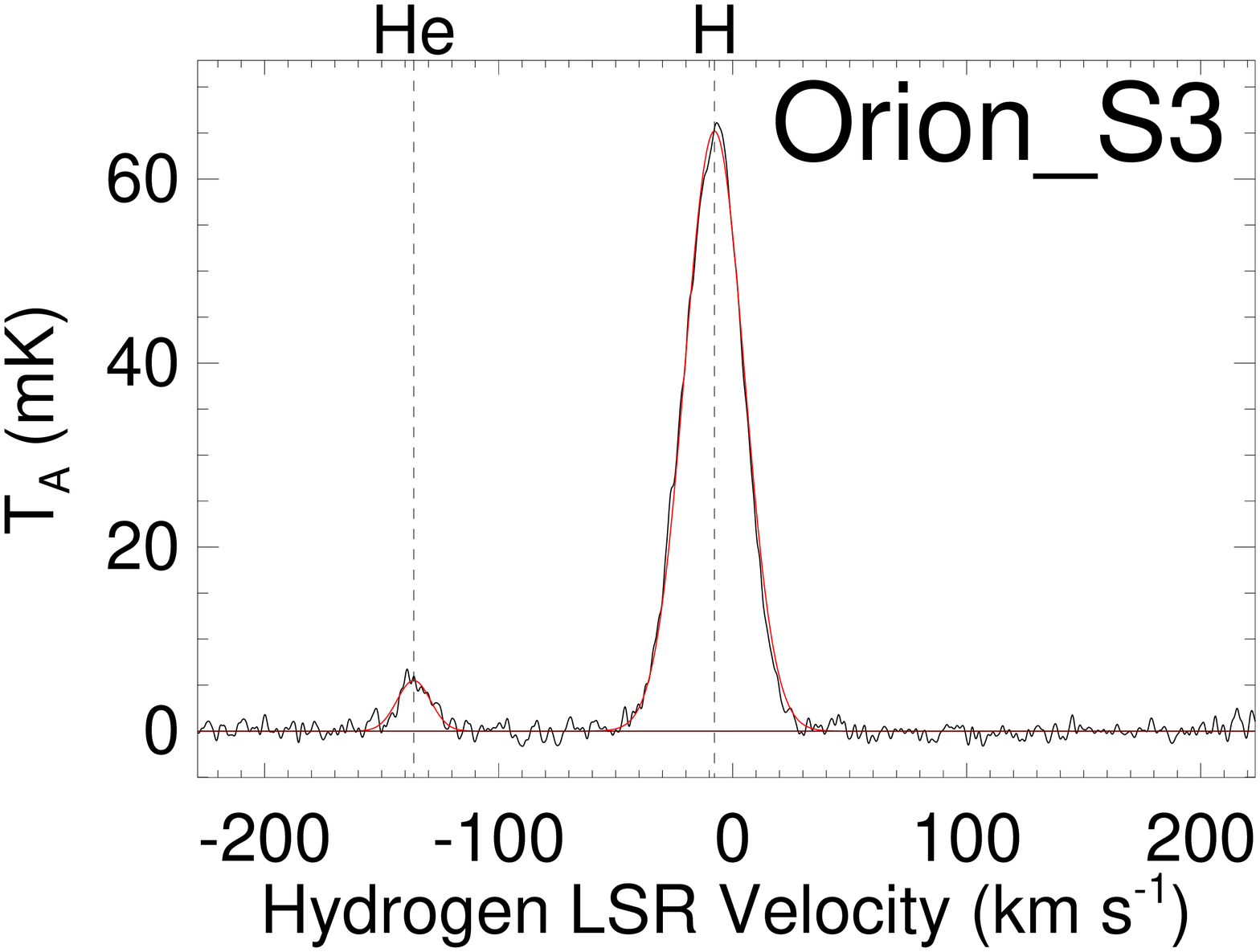} &
\includegraphics[width=.23\textwidth]{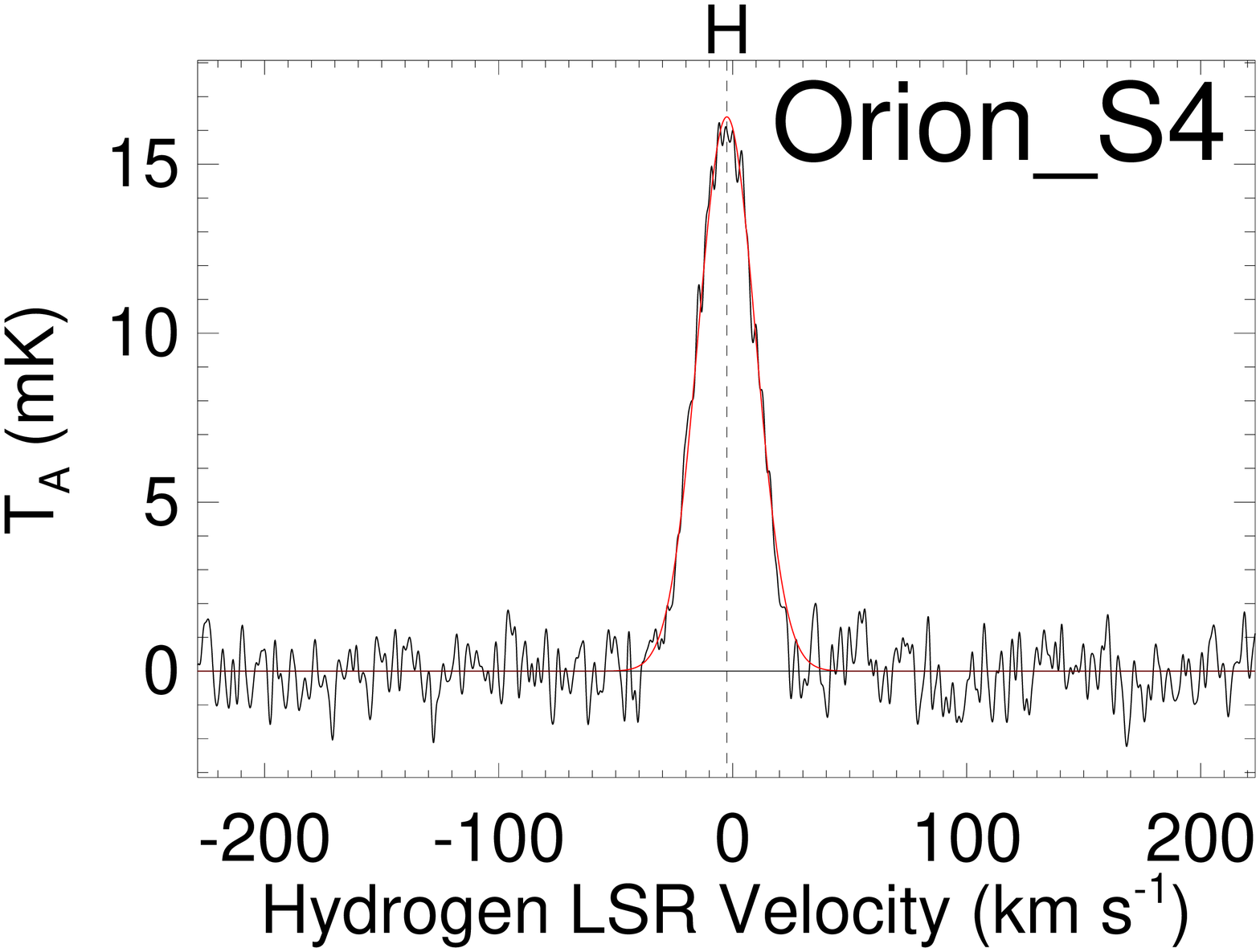} &
\includegraphics[width=.23\textwidth]{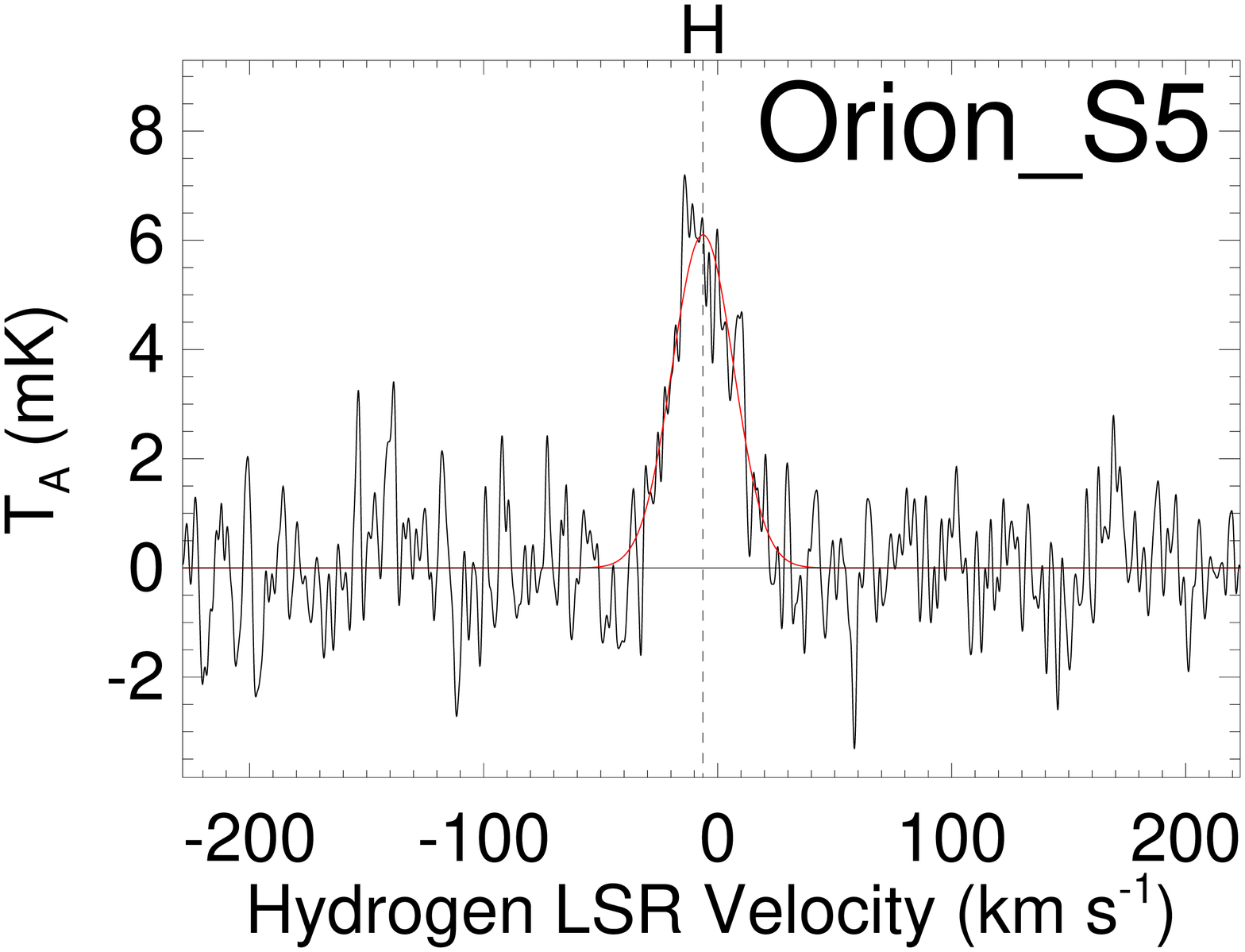}\\
\includegraphics[width=.23\textwidth]{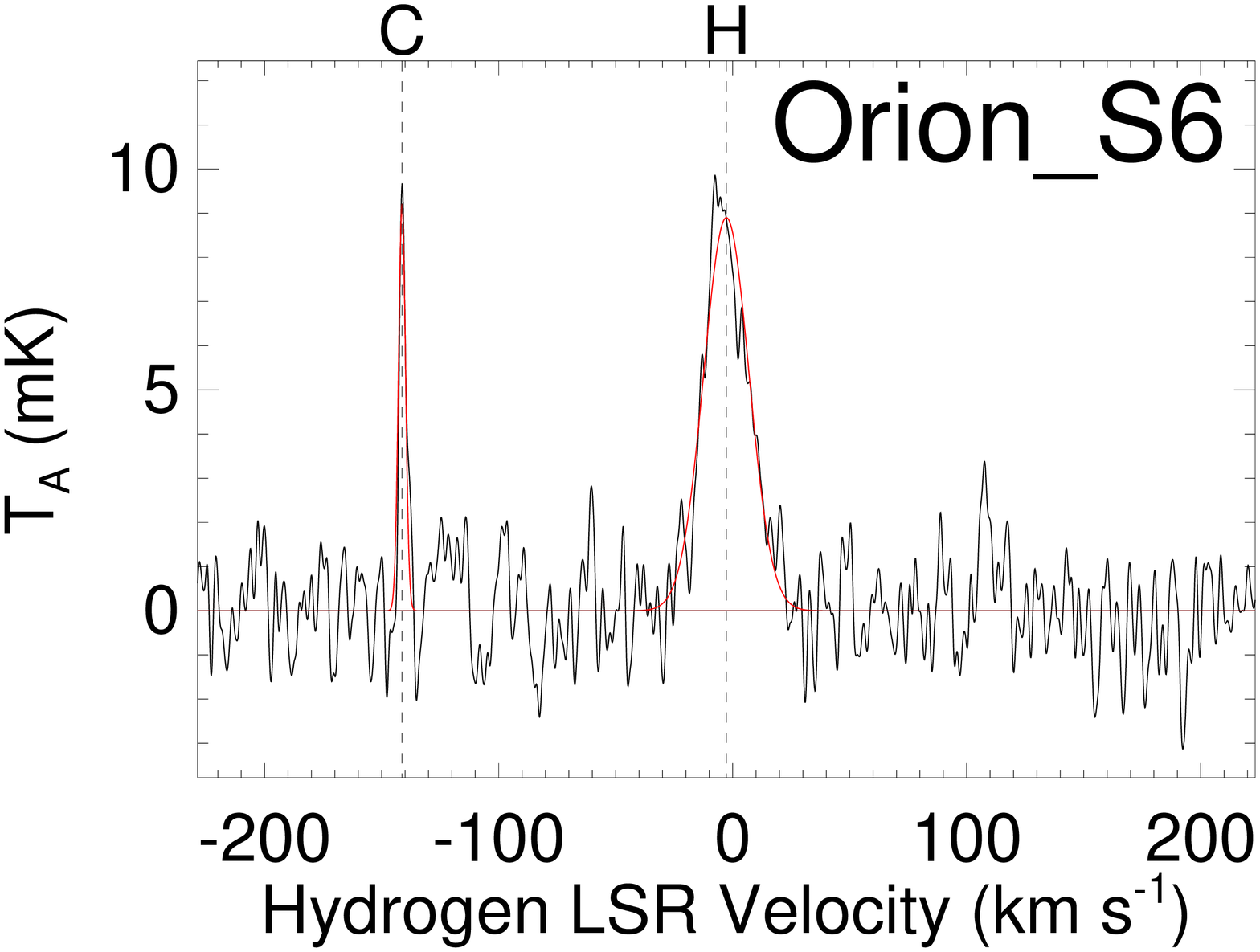} &
\includegraphics[width=.23\textwidth]{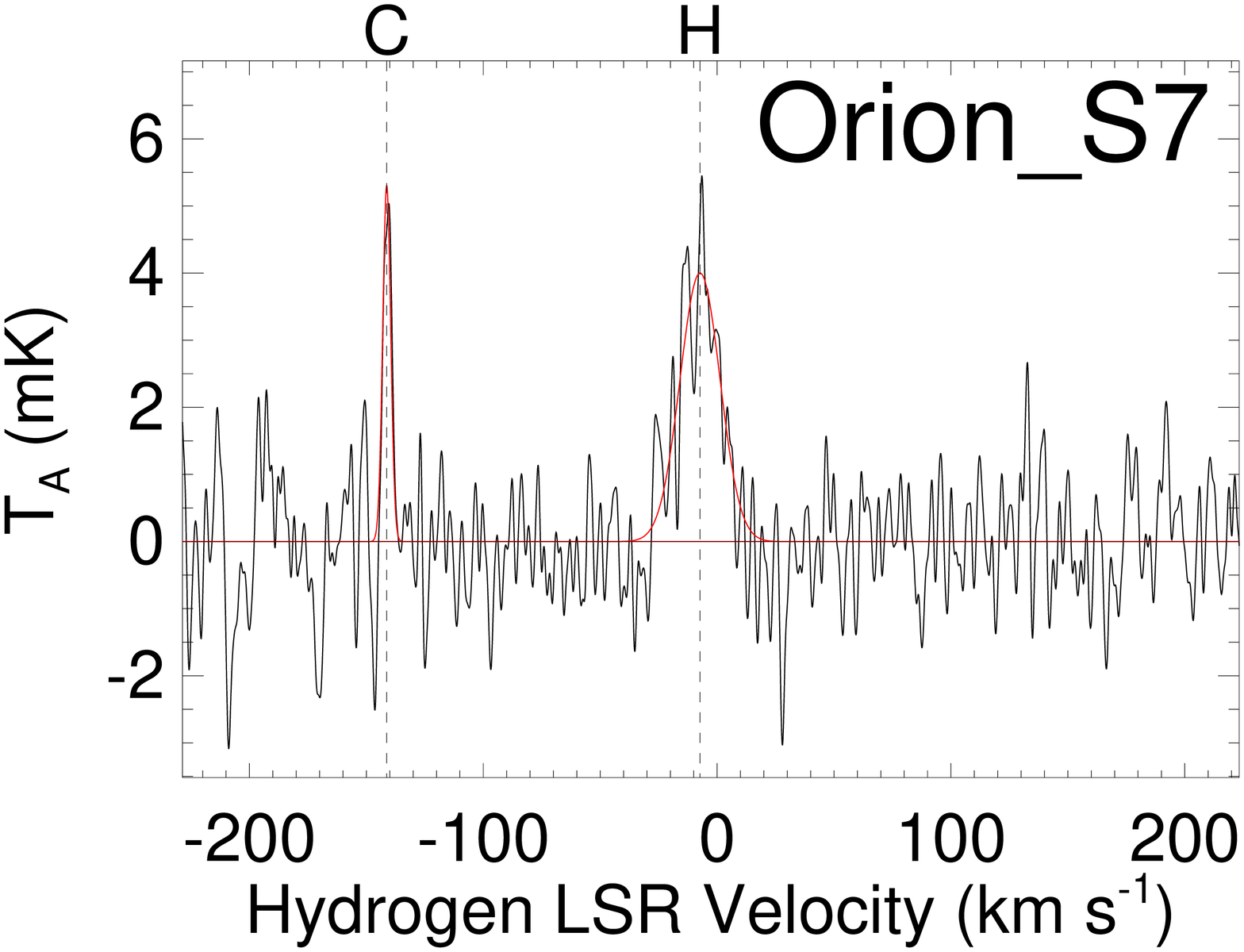} &
\includegraphics[width=.23\textwidth]{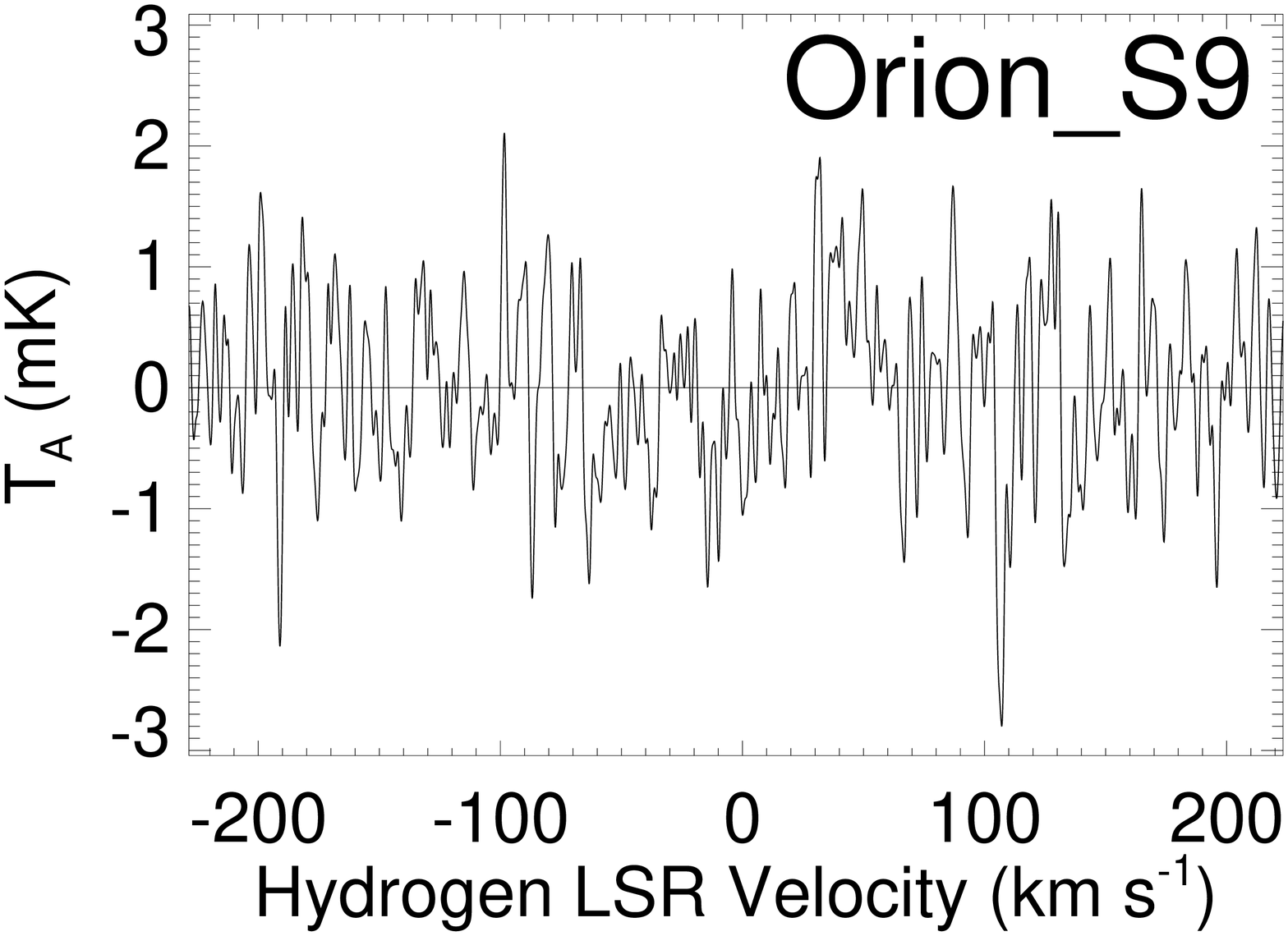} &
\includegraphics[width=.23\textwidth]{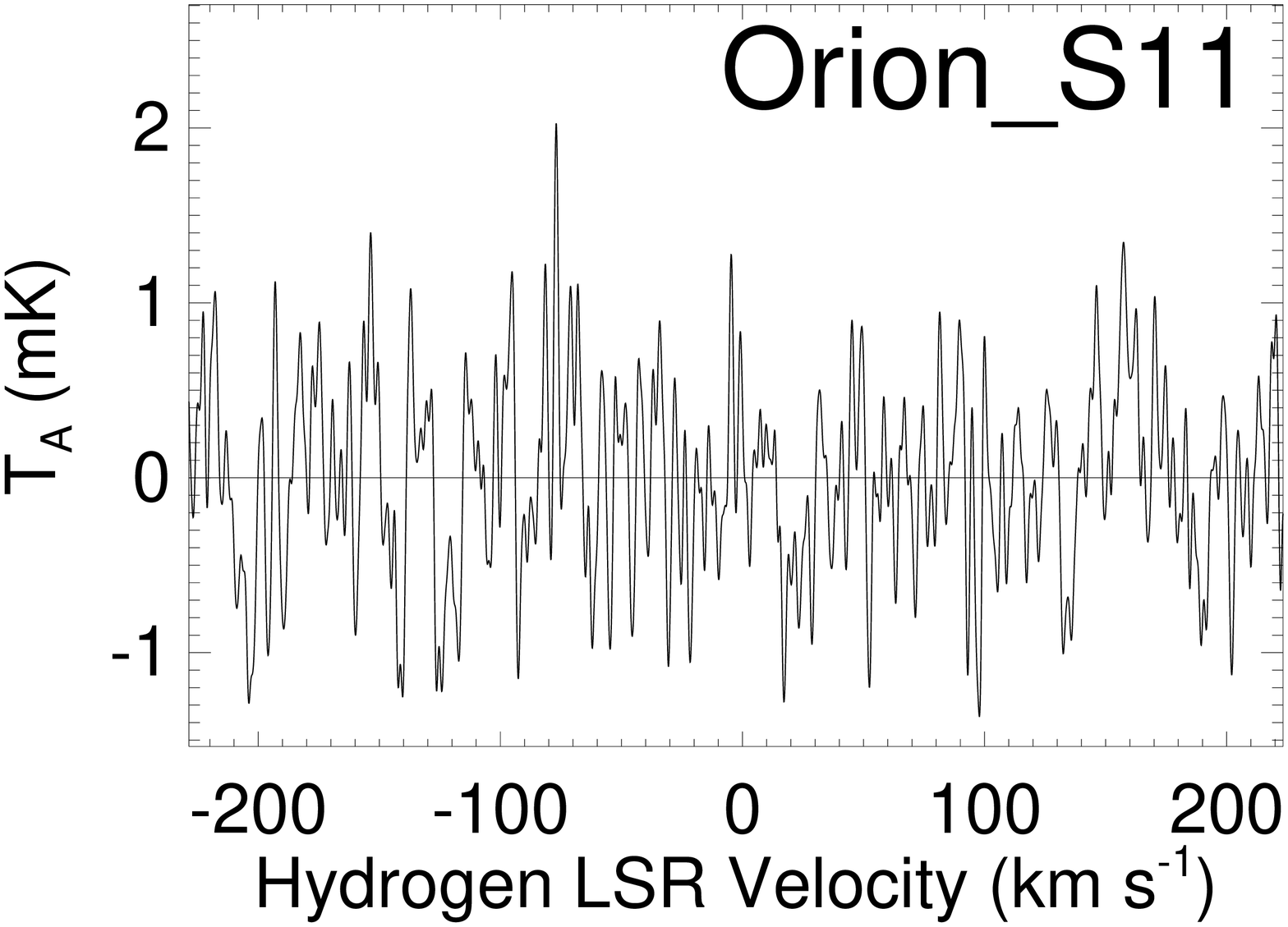} \\
\includegraphics[width=.23\textwidth]{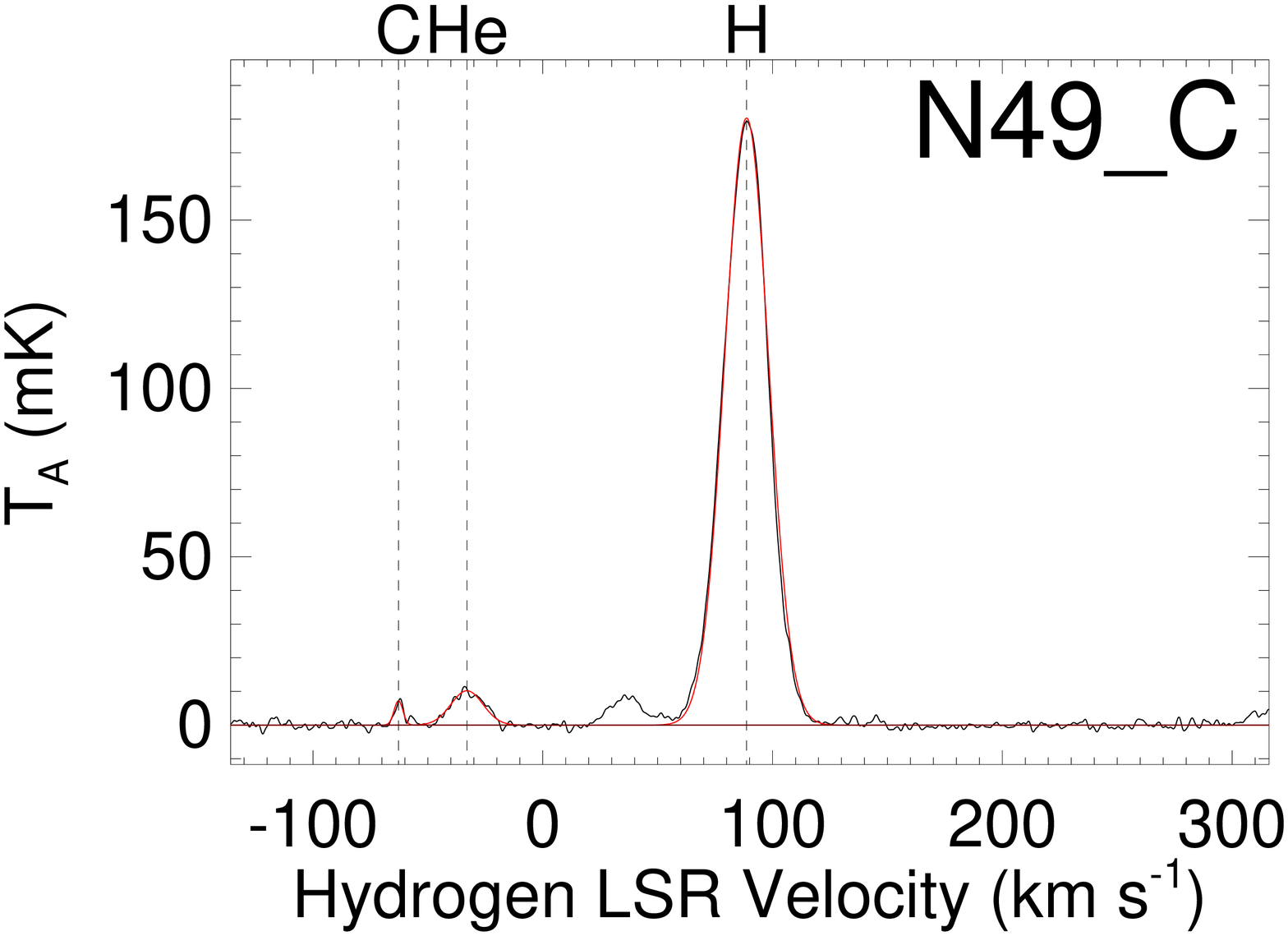} &
\includegraphics[width=.23\textwidth]{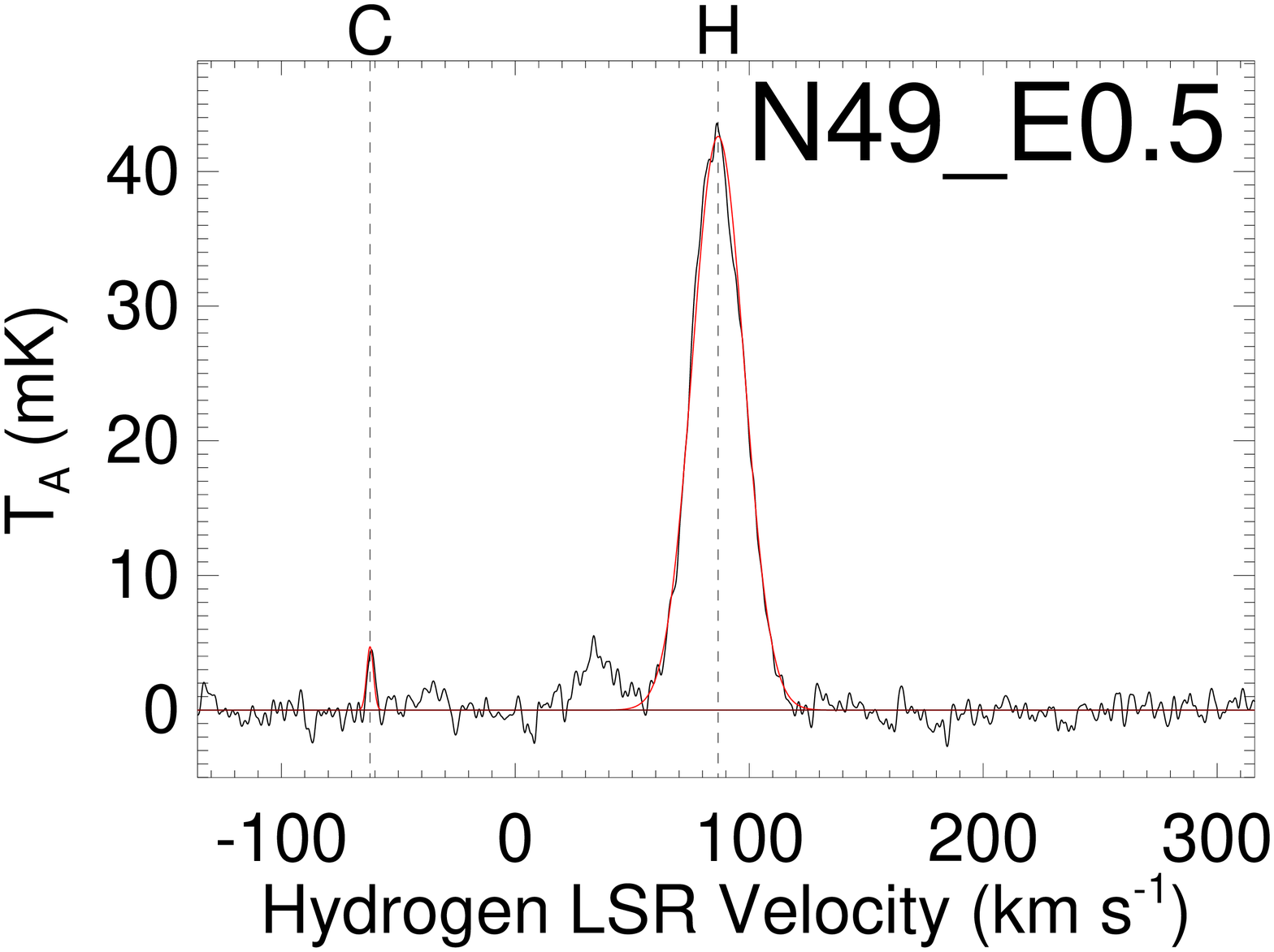} &
\includegraphics[width=.23\textwidth]{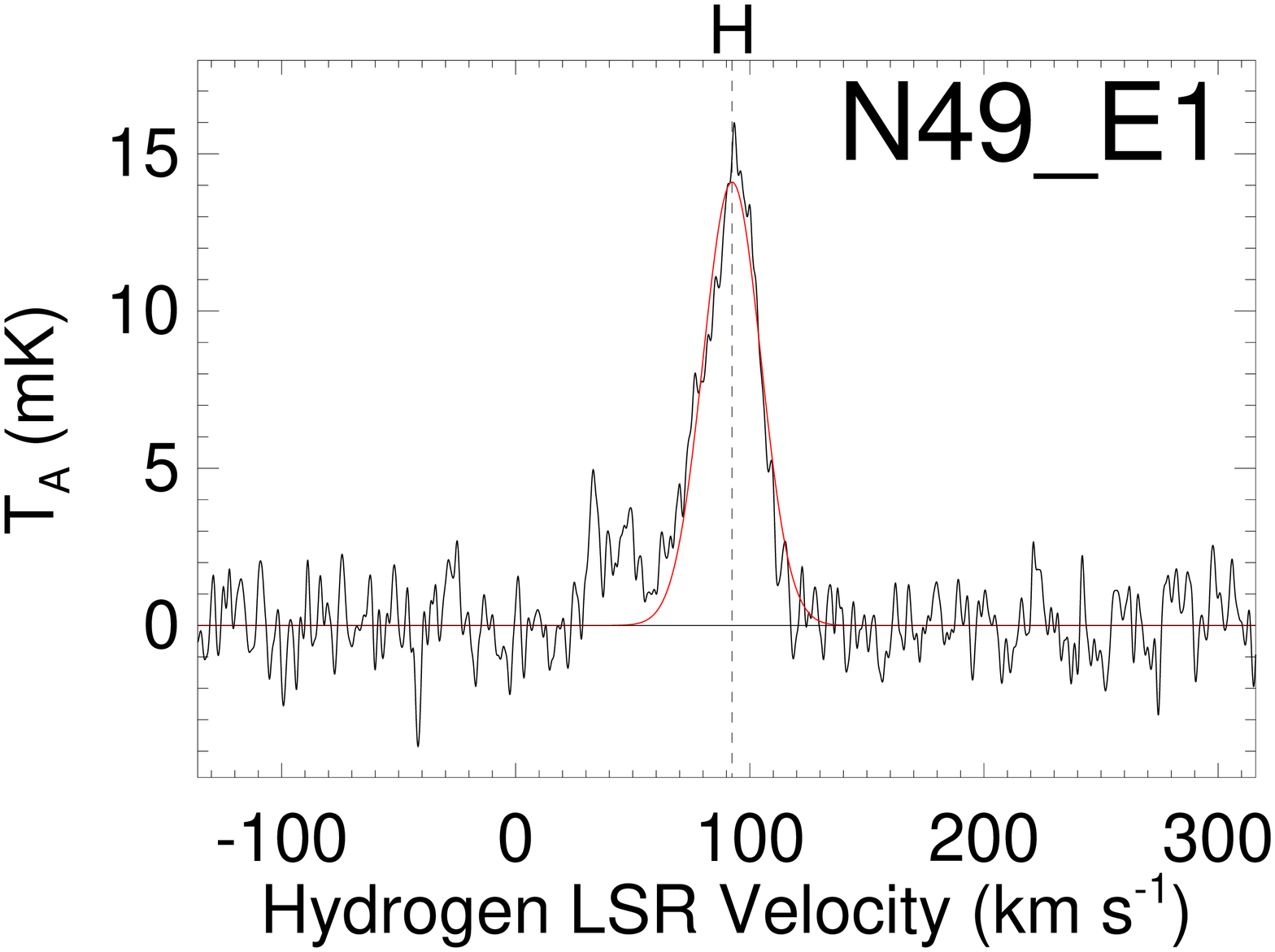} &
\includegraphics[width=.23\textwidth]{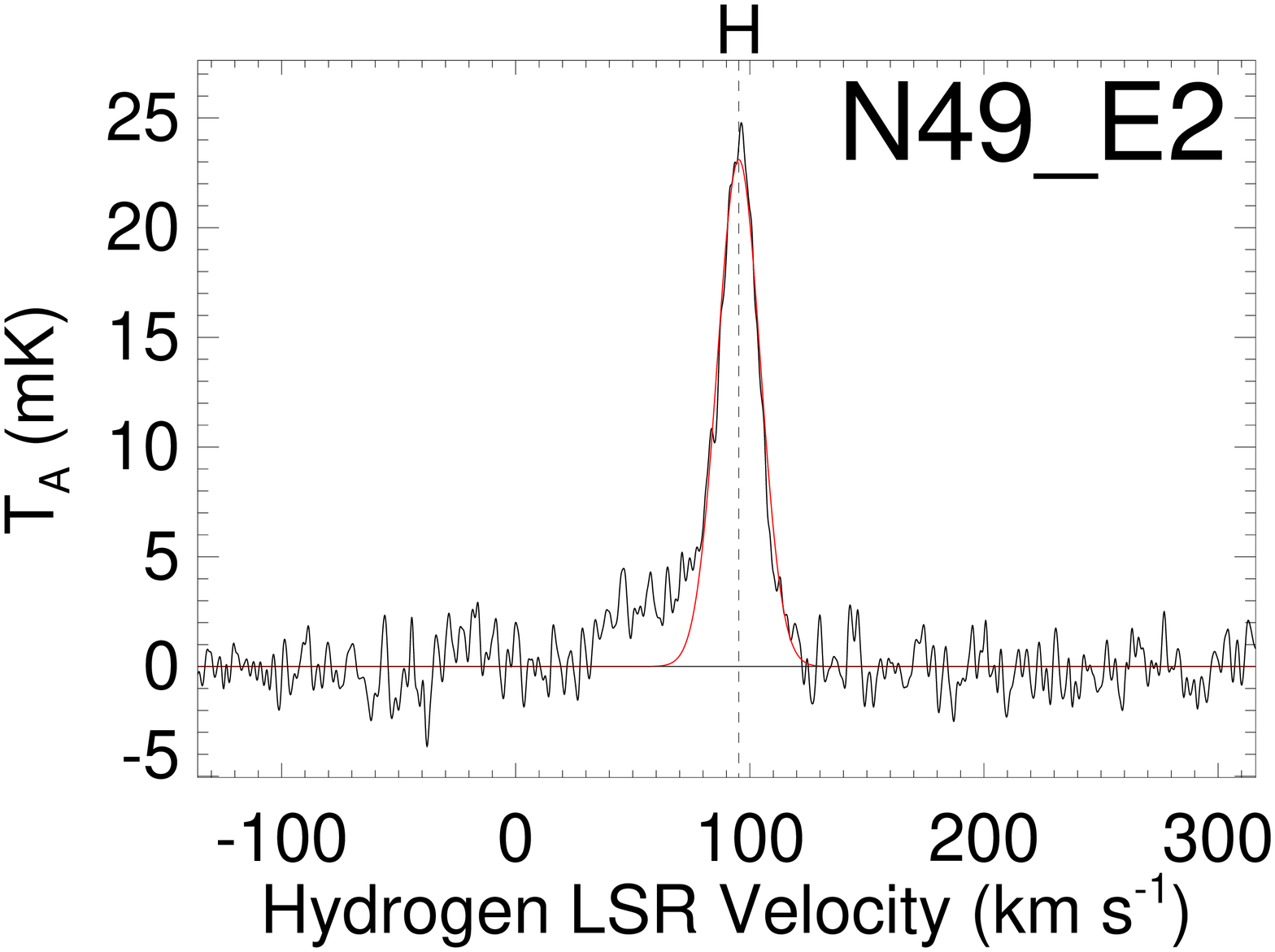} \\
\end{tabular}
\caption{}
\end{figure*}
\renewcommand{\thefigure}{\thesection.\arabic{figure}}

\renewcommand\thefigure{\thesection.\arabic{figure} (Cont.)}
\addtocounter{figure}{-1}
\begin{figure*}
\centering
\begin{tabular}{cccc}
\includegraphics[width=.23\textwidth]{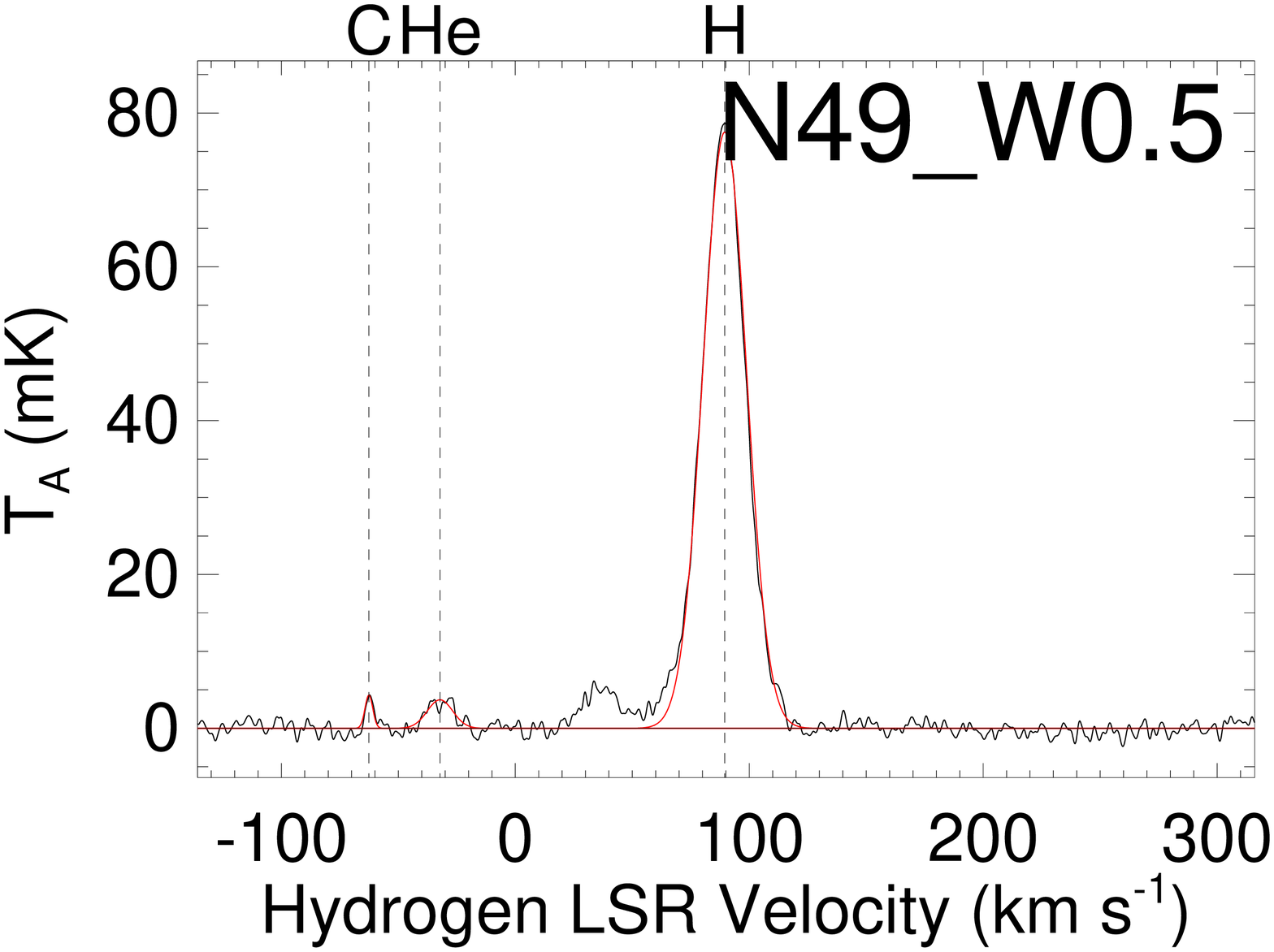} &
\includegraphics[width=.23\textwidth]{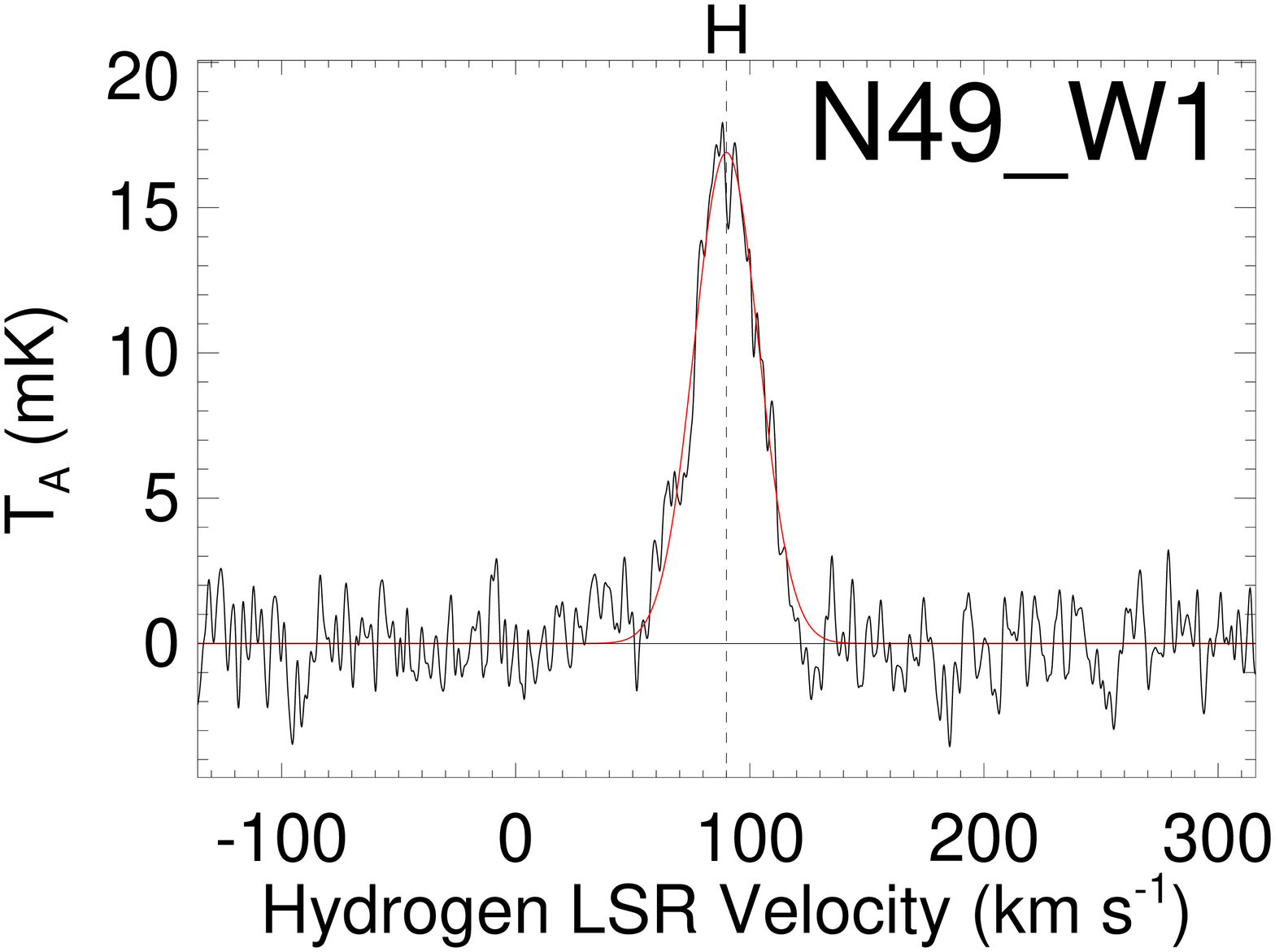} &
\includegraphics[width=.23\textwidth]{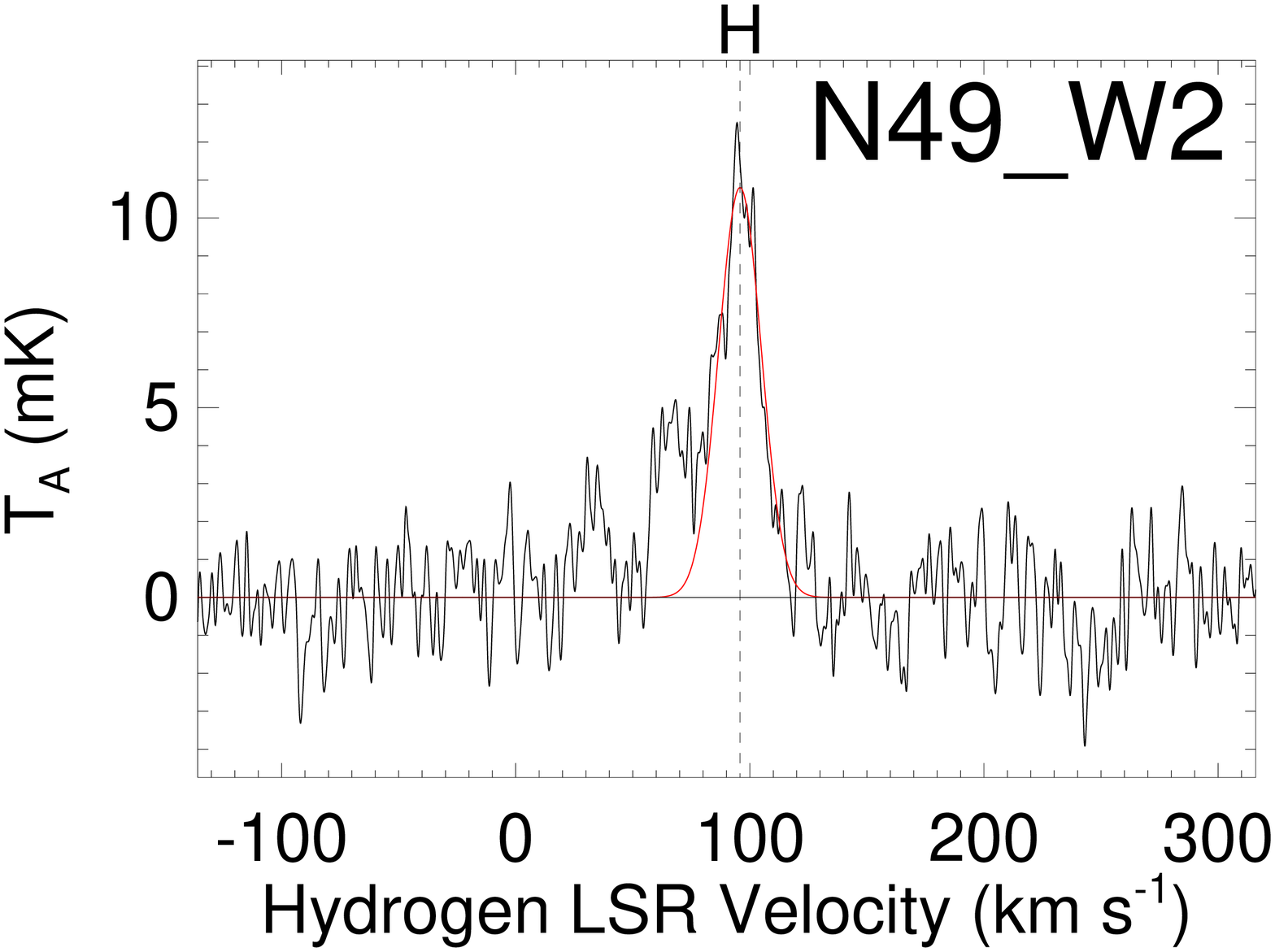} &
\includegraphics[width=.23\textwidth]{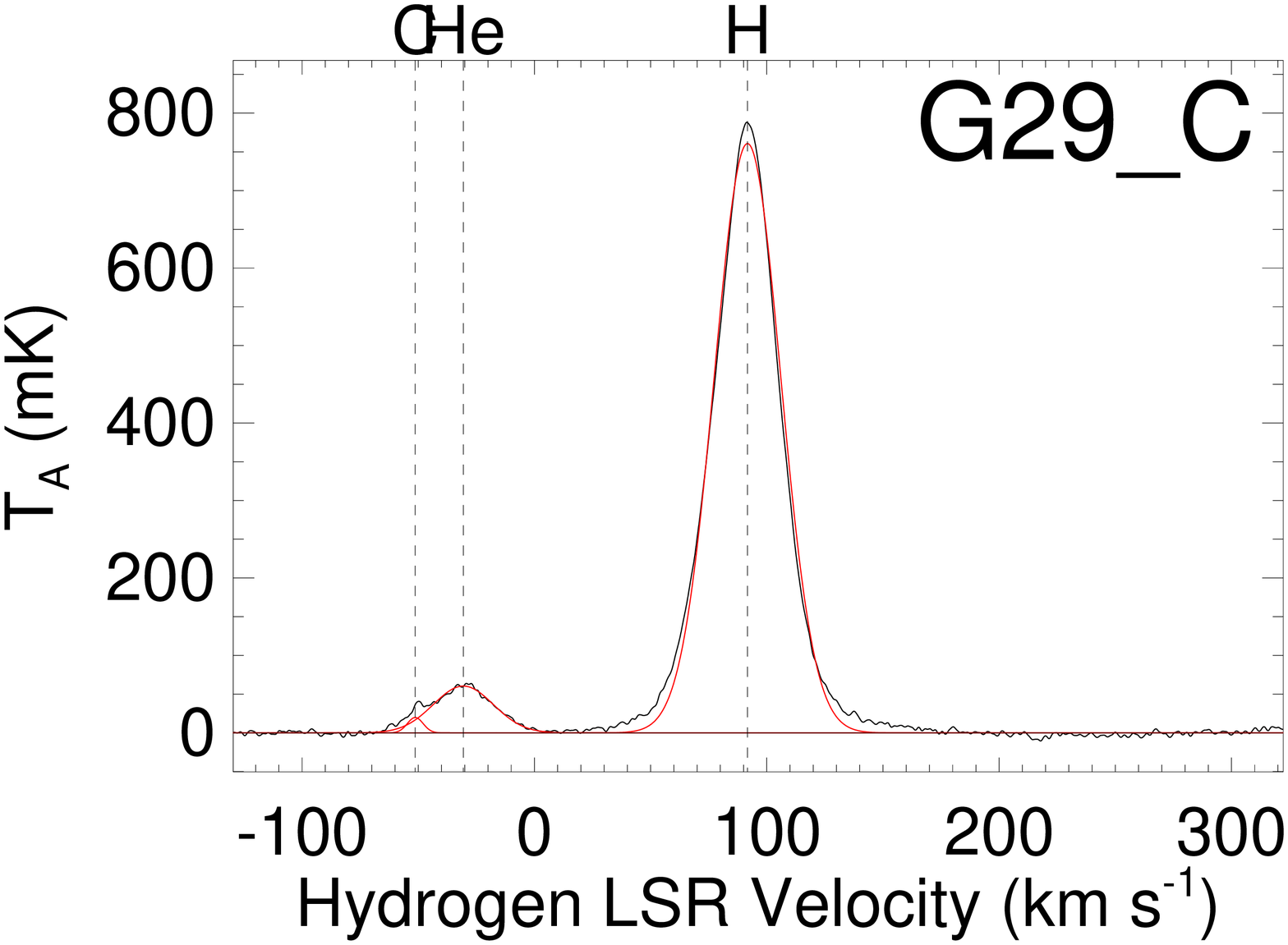} \\
\includegraphics[width=.23\textwidth]{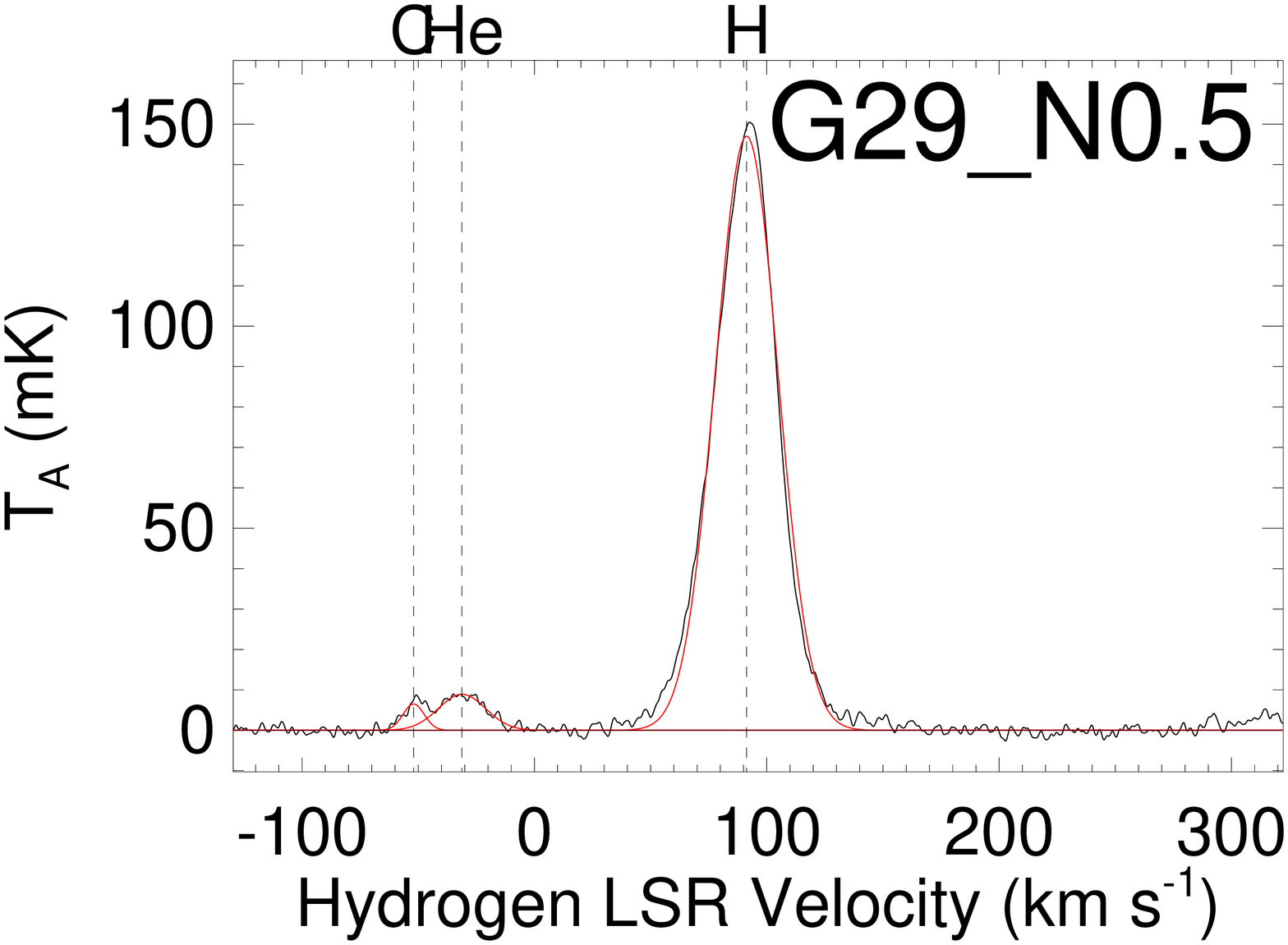} &
\includegraphics[width=.23\textwidth]{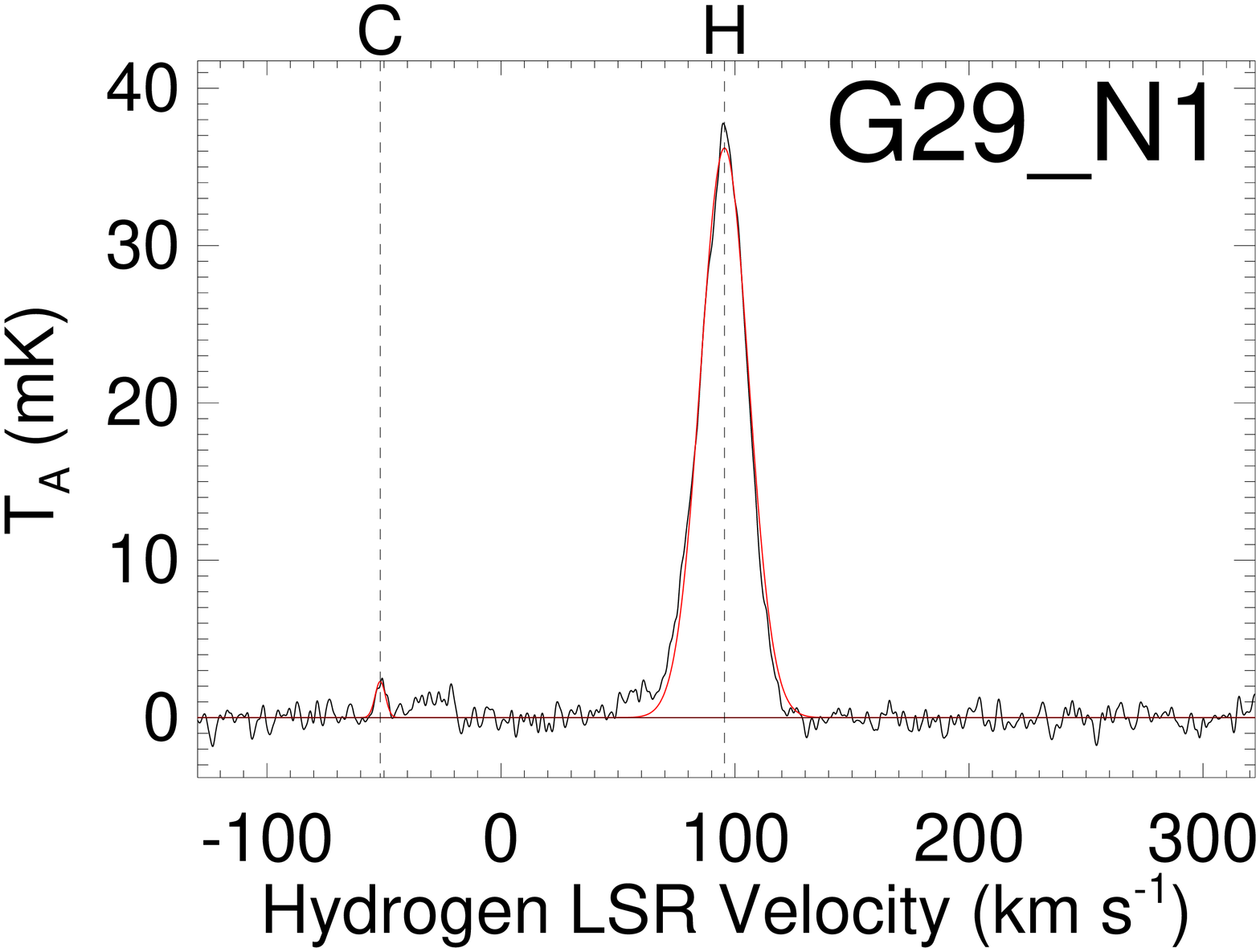} &
\includegraphics[width=.23\textwidth]{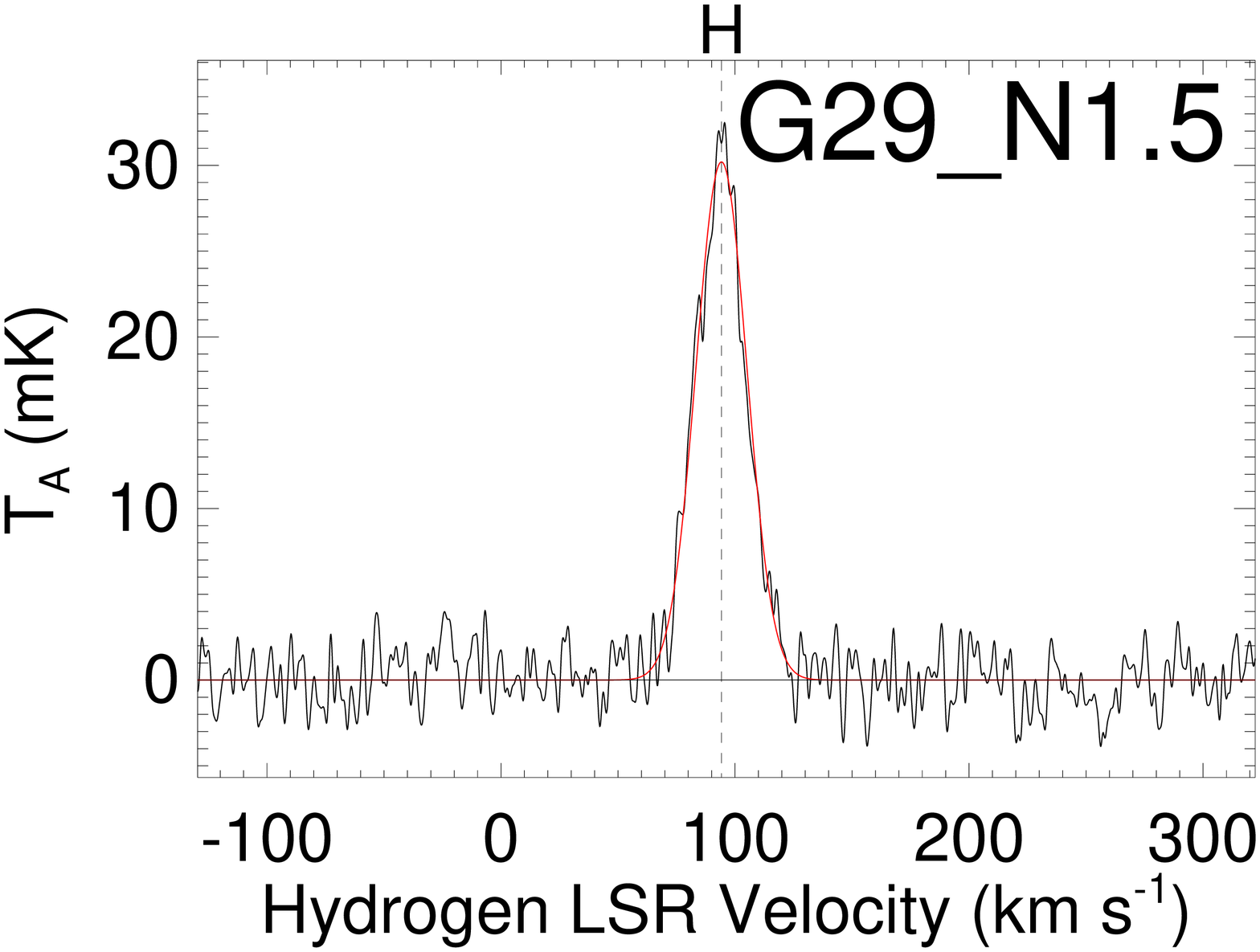} &
\includegraphics[width=.23\textwidth]{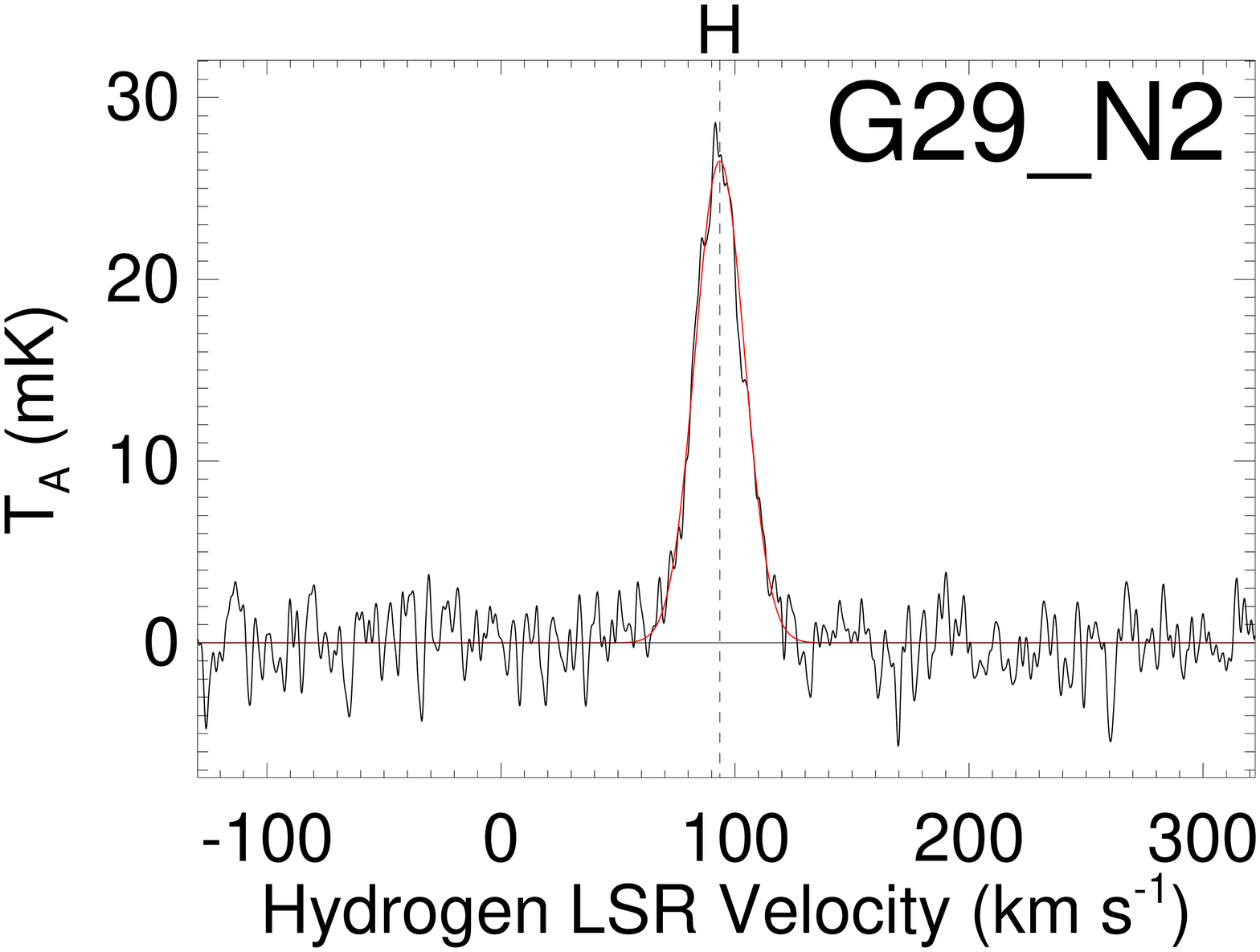} \\
\includegraphics[width=.23\textwidth]{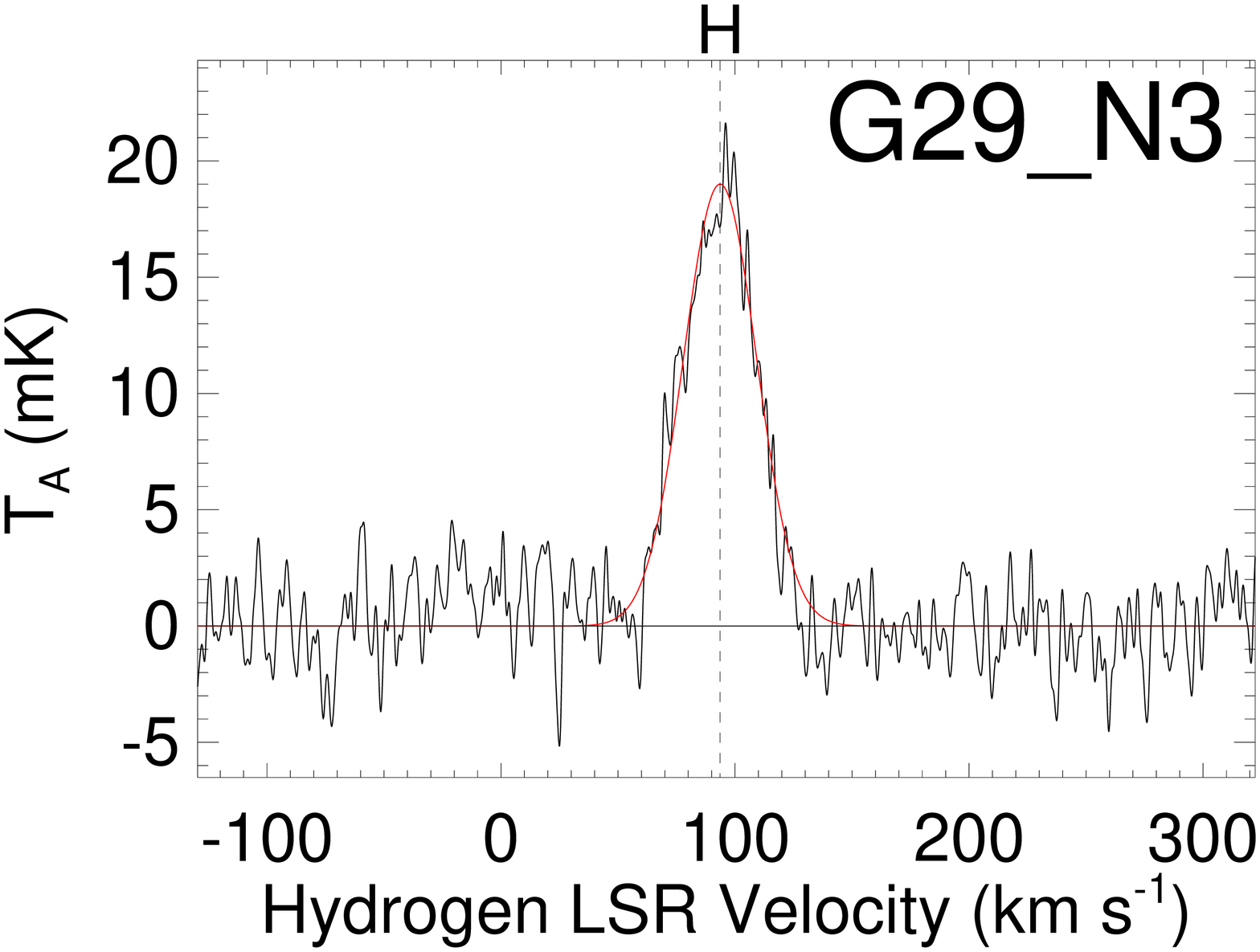} &
\includegraphics[width=.23\textwidth]{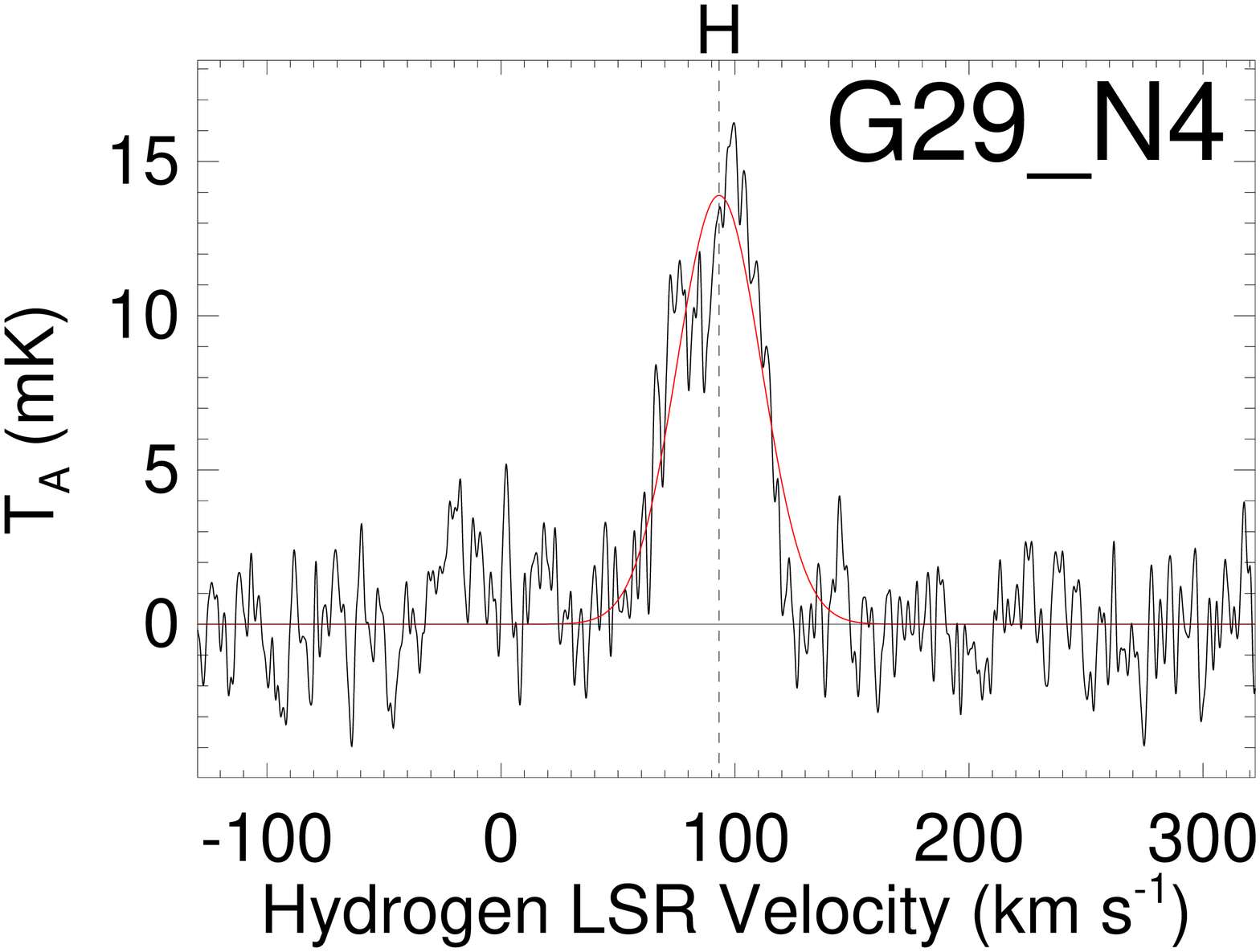} &
\includegraphics[width=.23\textwidth]{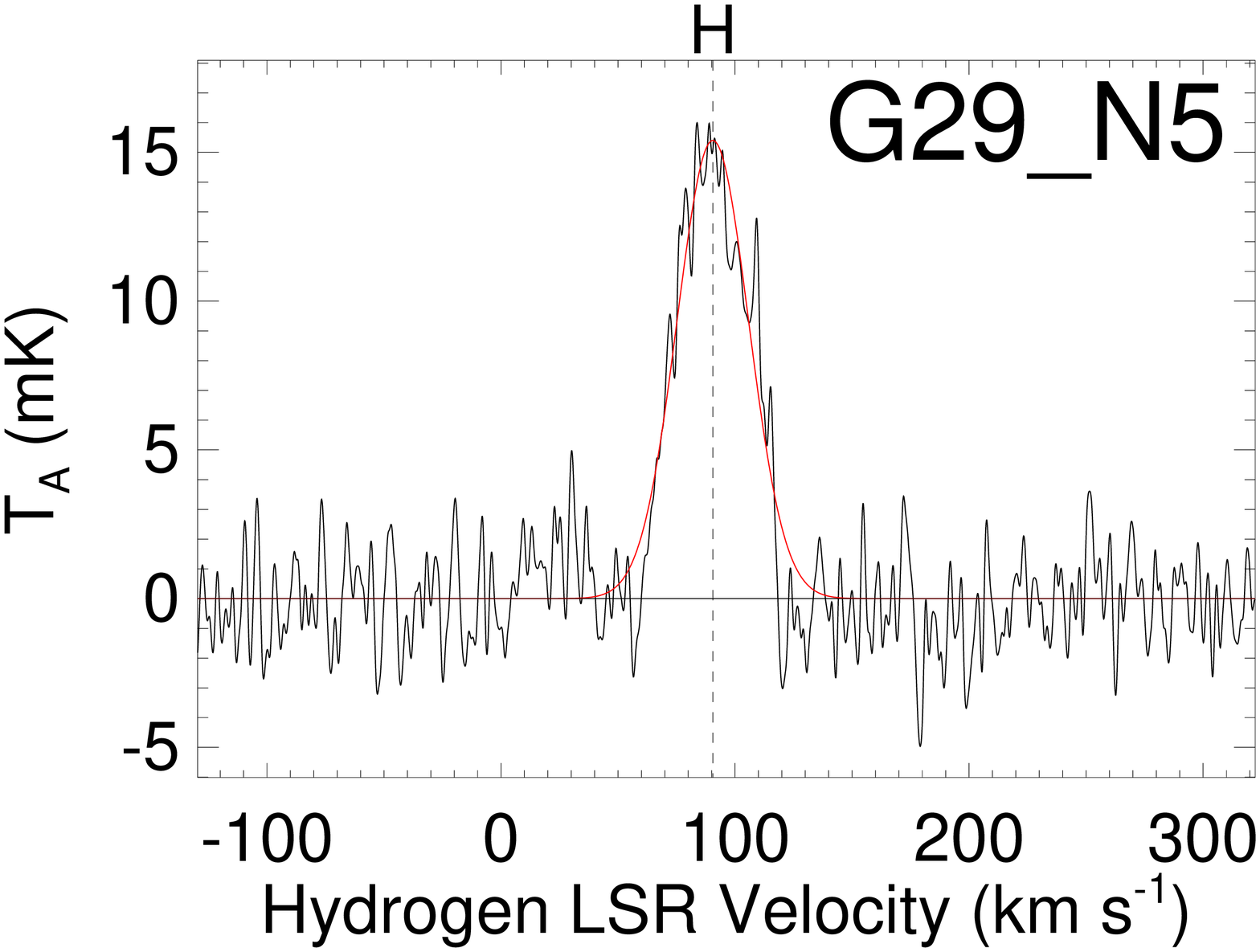} &
\includegraphics[width=.23\textwidth]{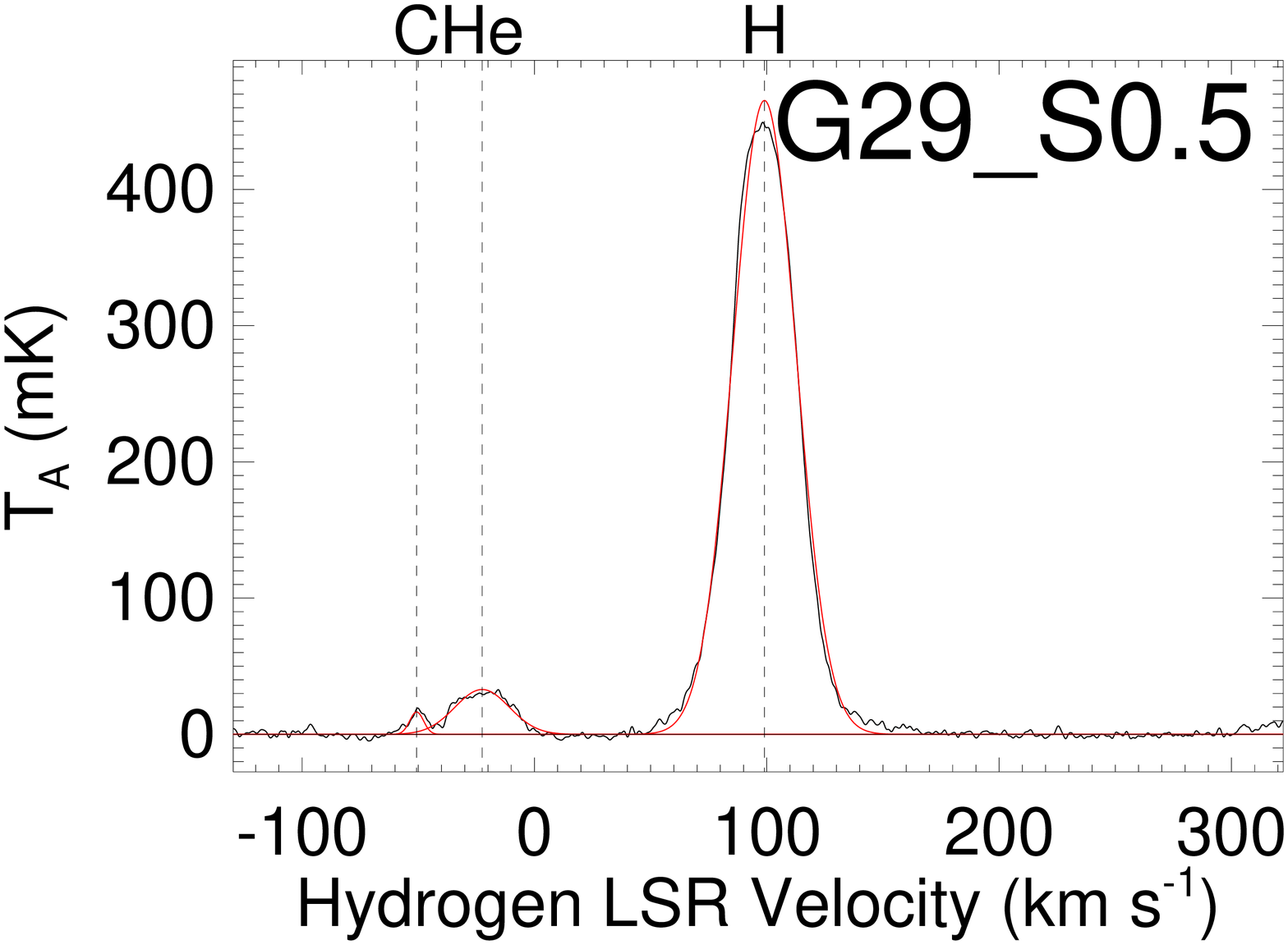} \\
\includegraphics[width=.23\textwidth]{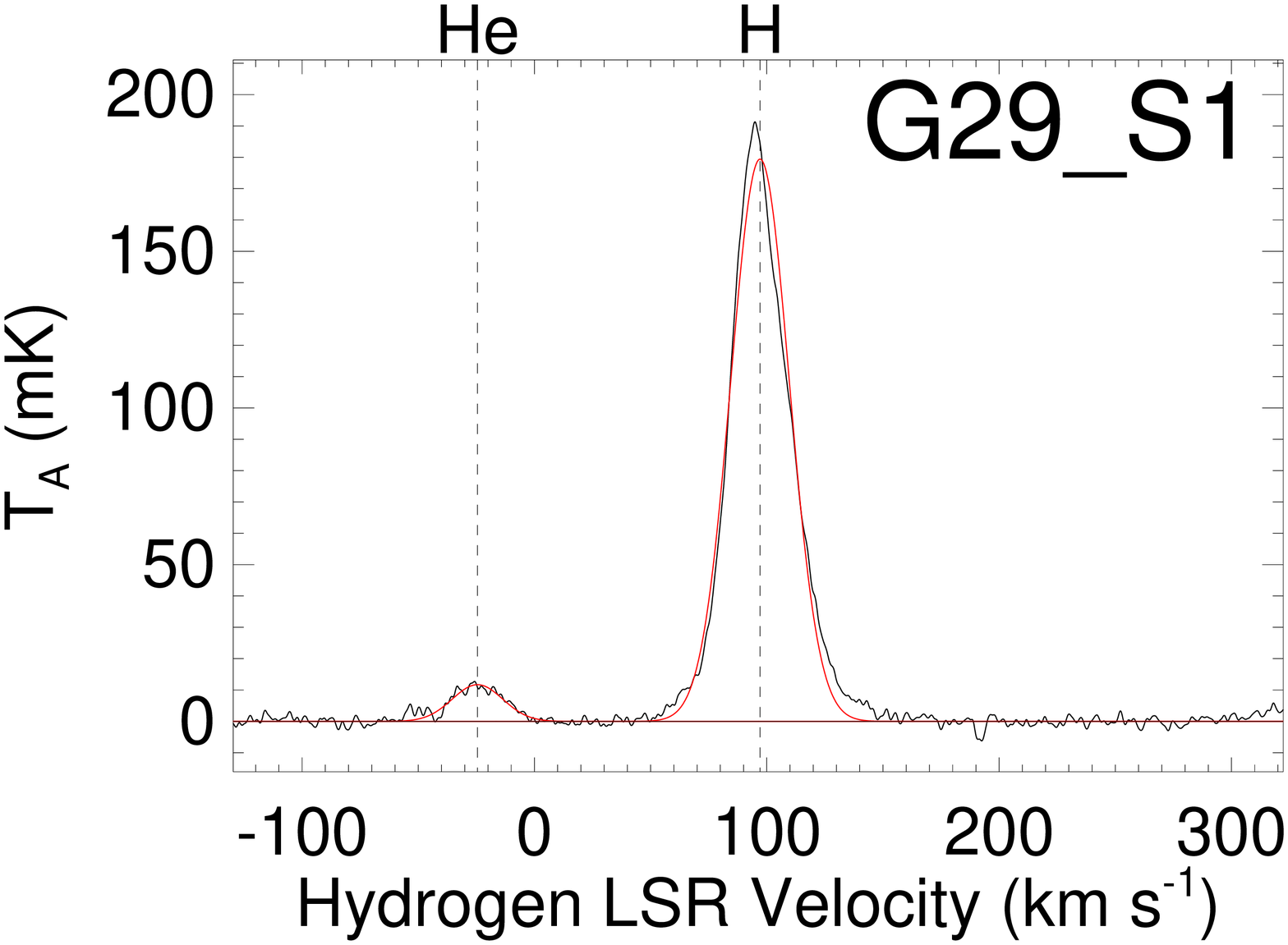} &
\includegraphics[width=.23\textwidth]{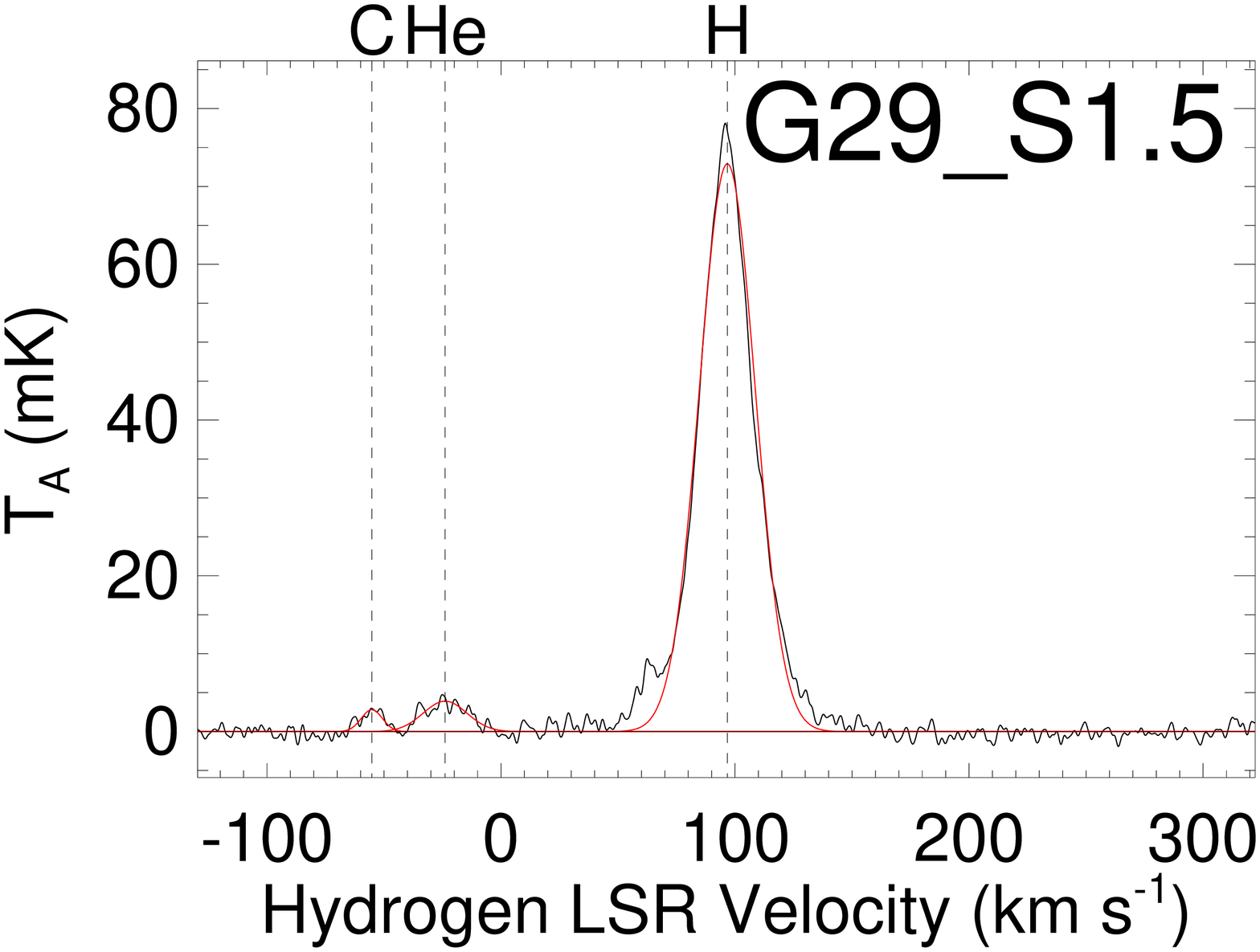} &
\includegraphics[width=.23\textwidth]{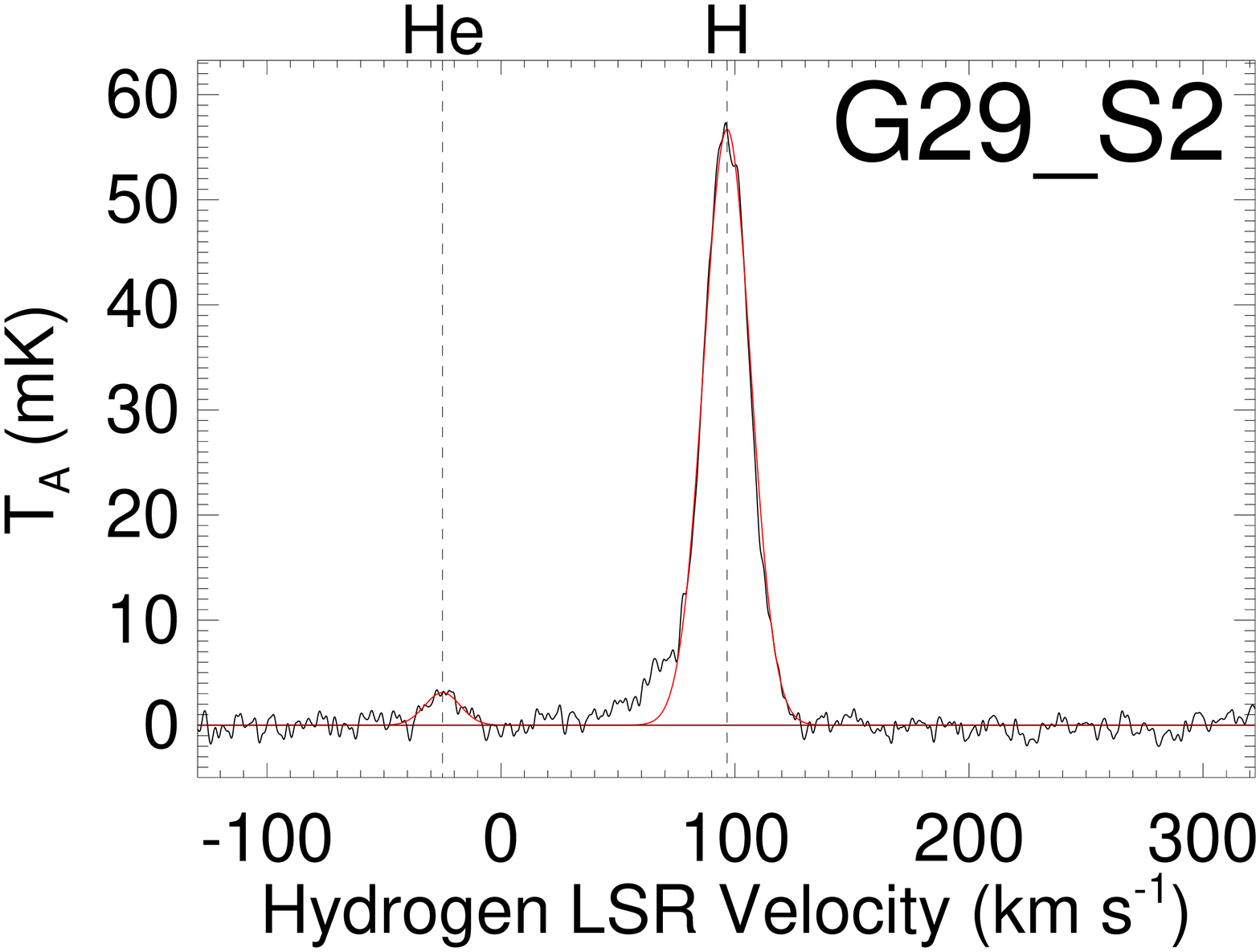} &
\includegraphics[width=.23\textwidth]{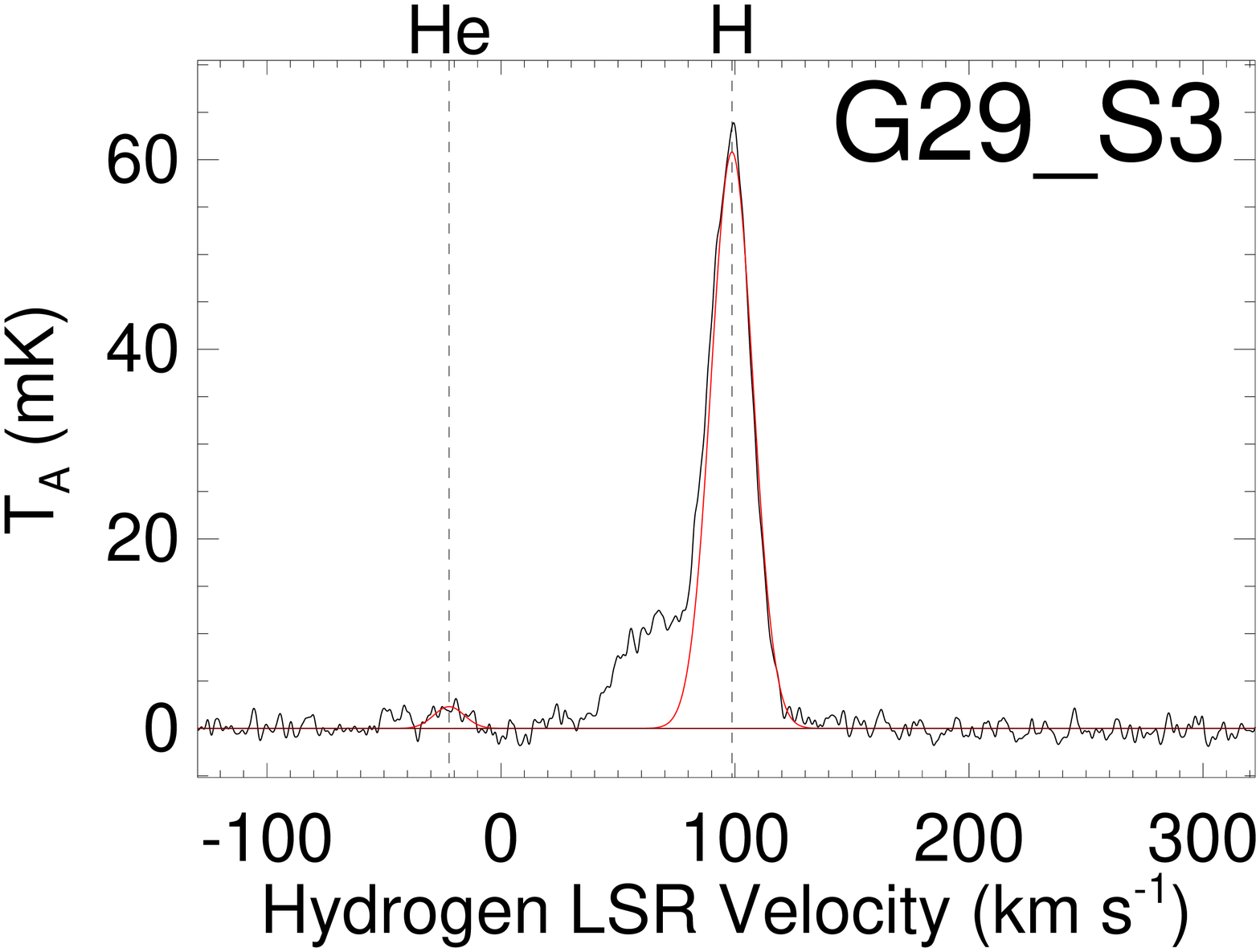} \\
\includegraphics[width=.23\textwidth]{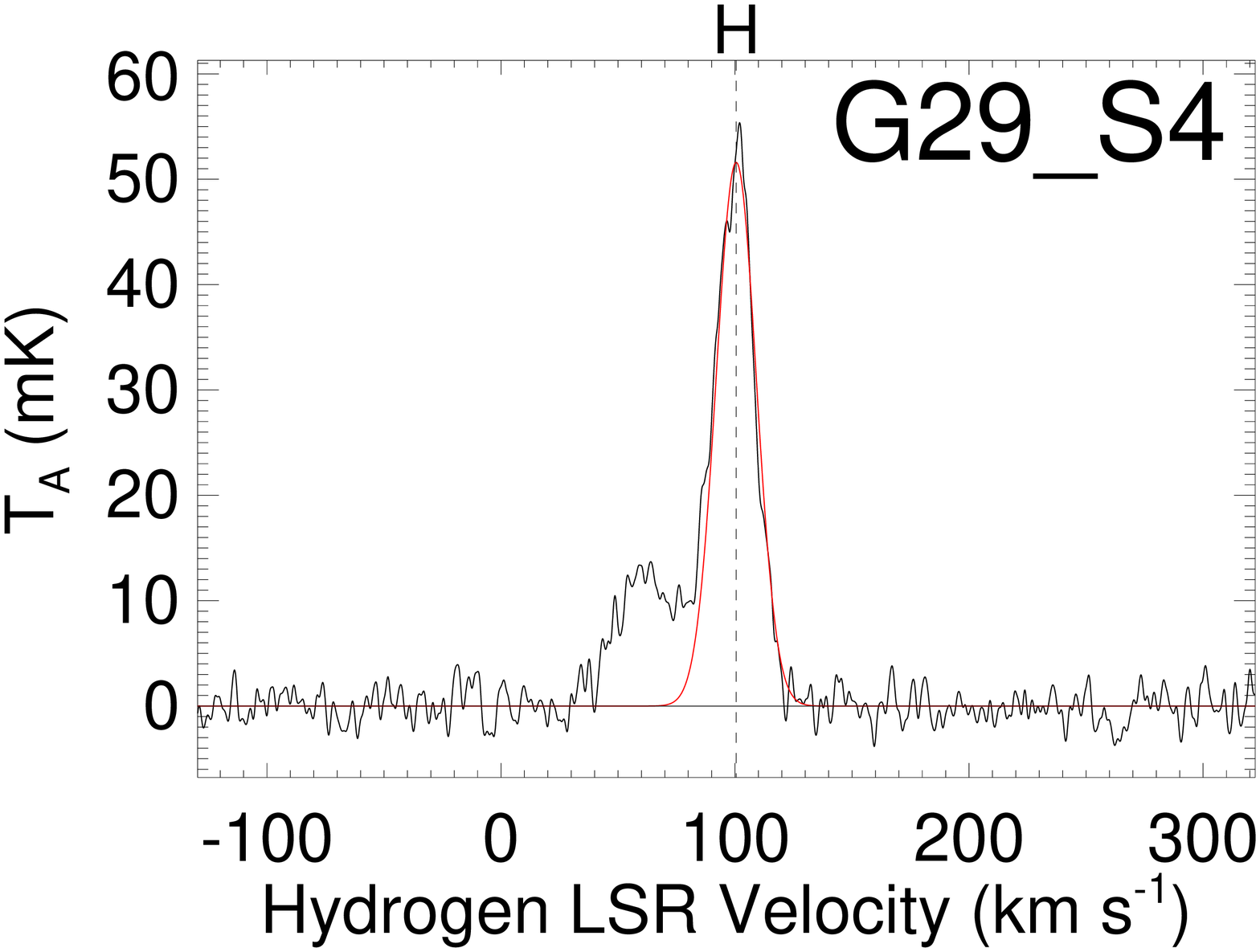} &
\includegraphics[width=.23\textwidth]{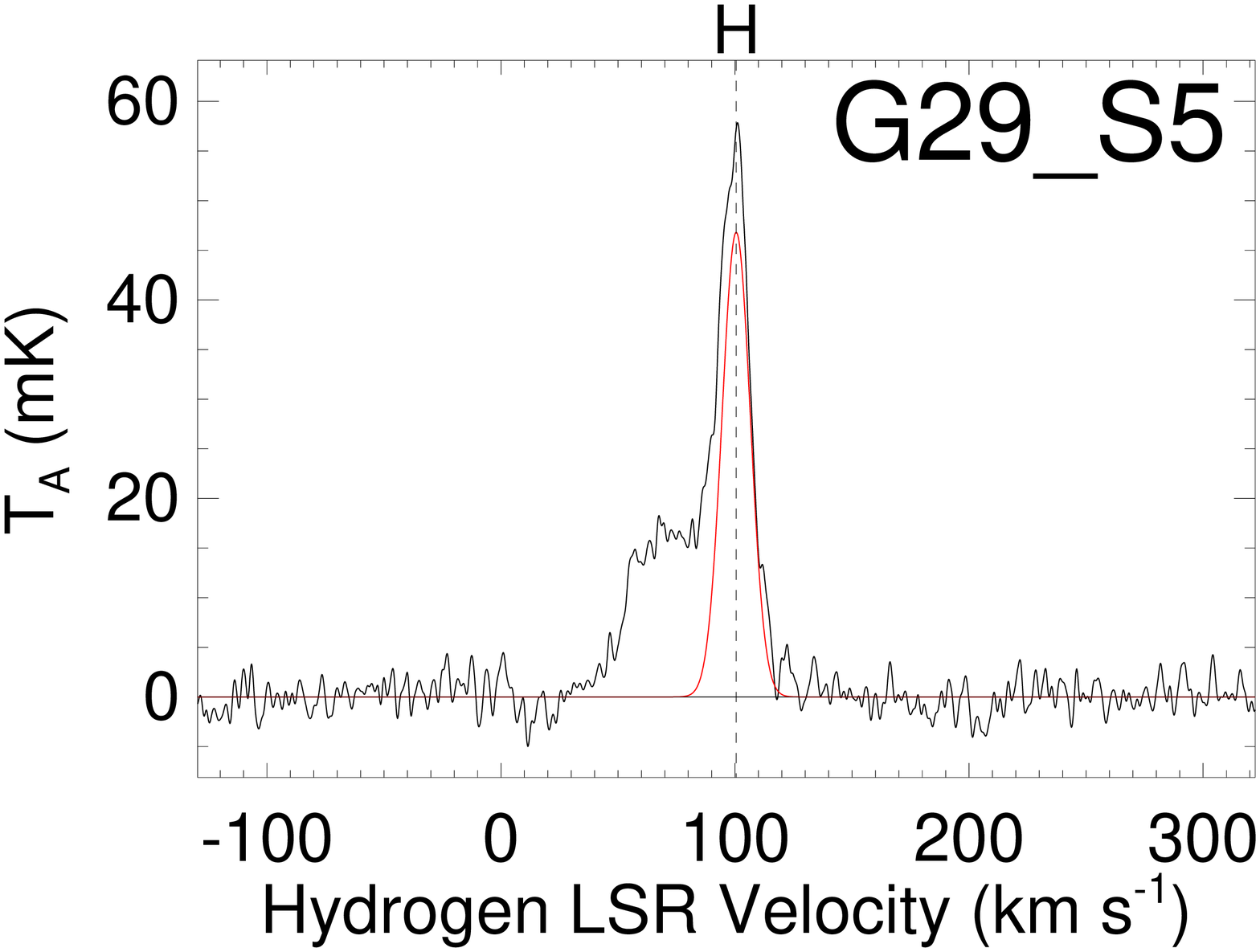} &
&
 \\
\end{tabular}
\caption{}
\end{figure*}
\renewcommand{\thefigure}{\thesection.\arabic{figure}}

\begin{figure*} 
\centering
\begin{tabular}{cccc}
\includegraphics[width=.23\textwidth]{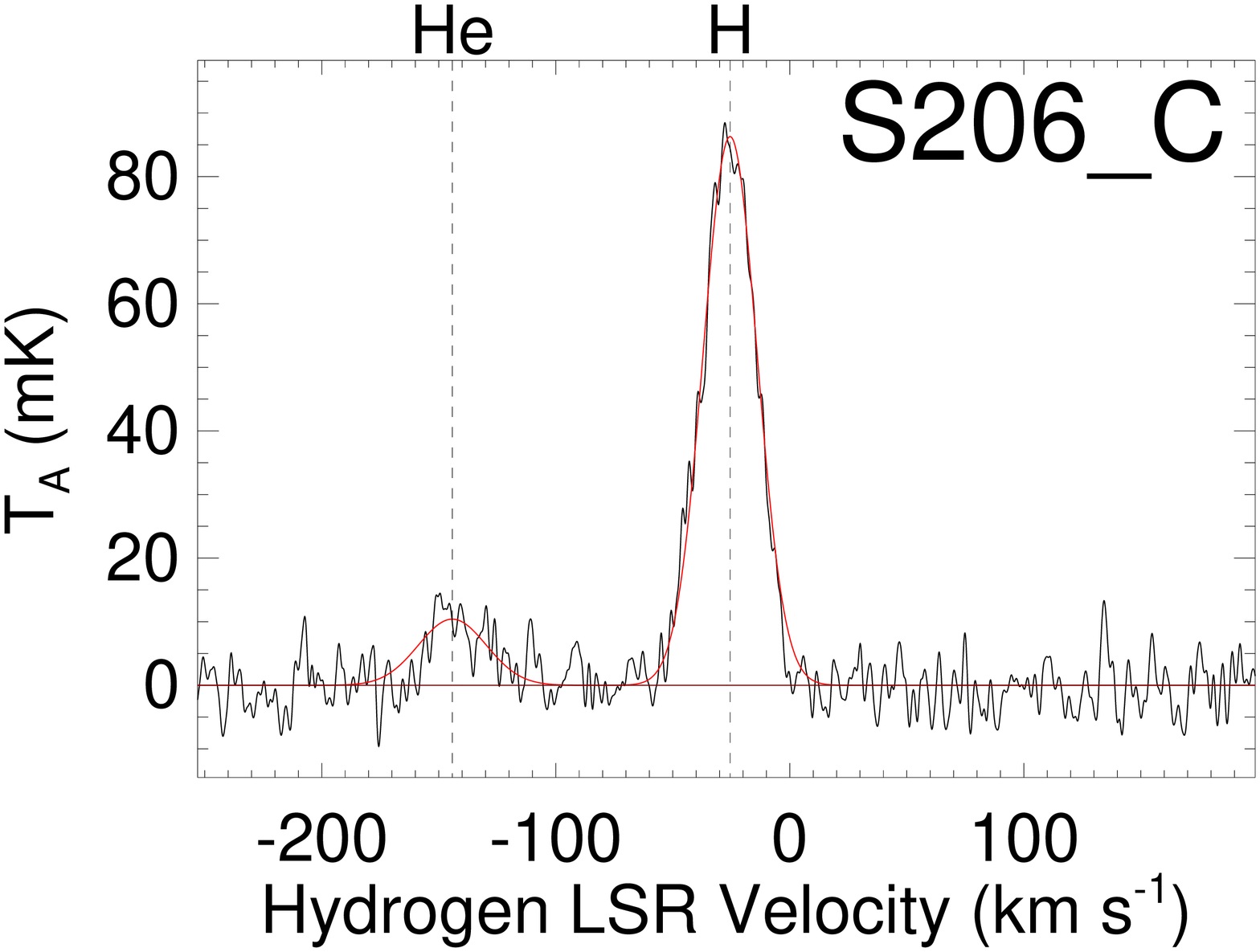} &
\includegraphics[width=.23\textwidth]{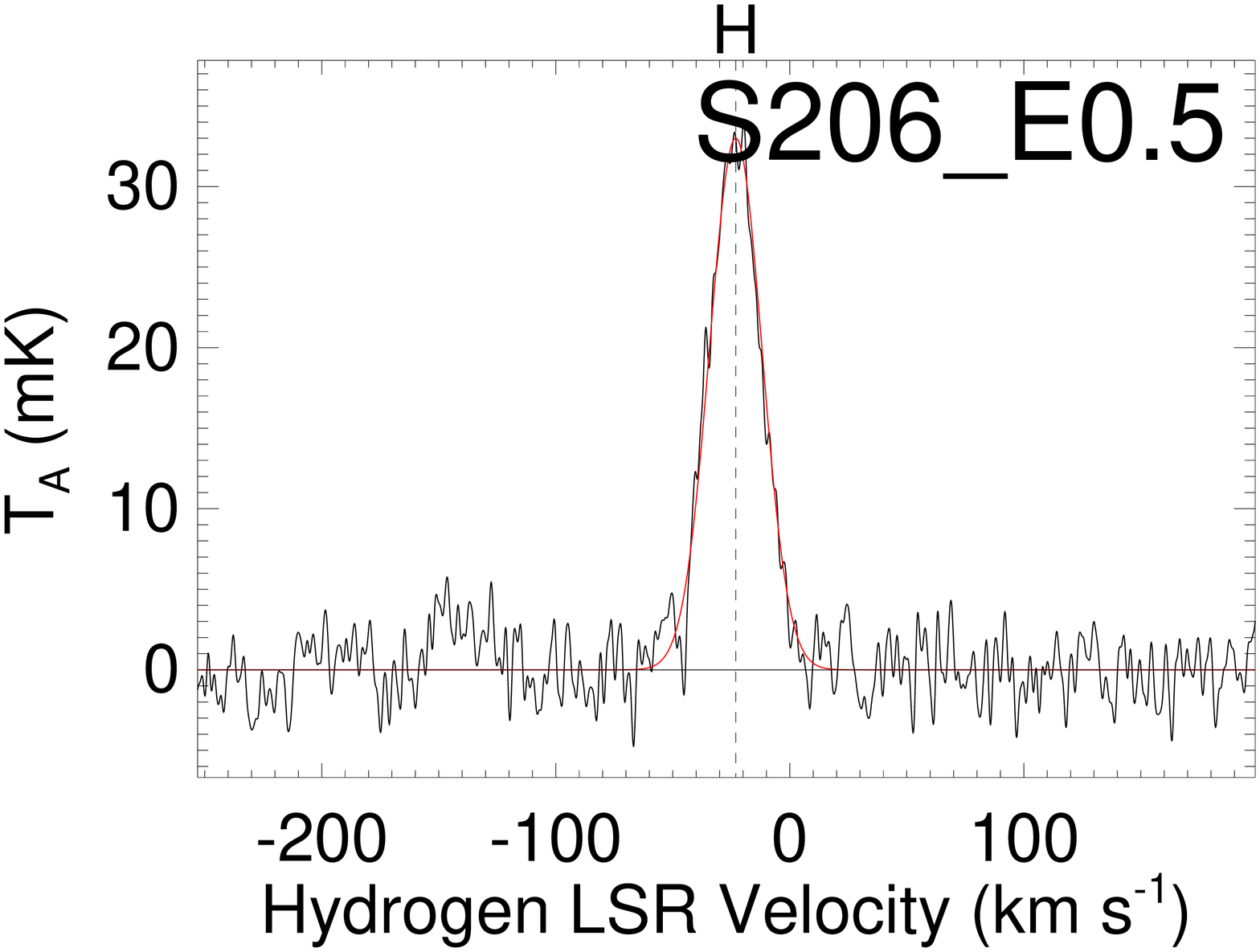} &
\includegraphics[width=.23\textwidth]{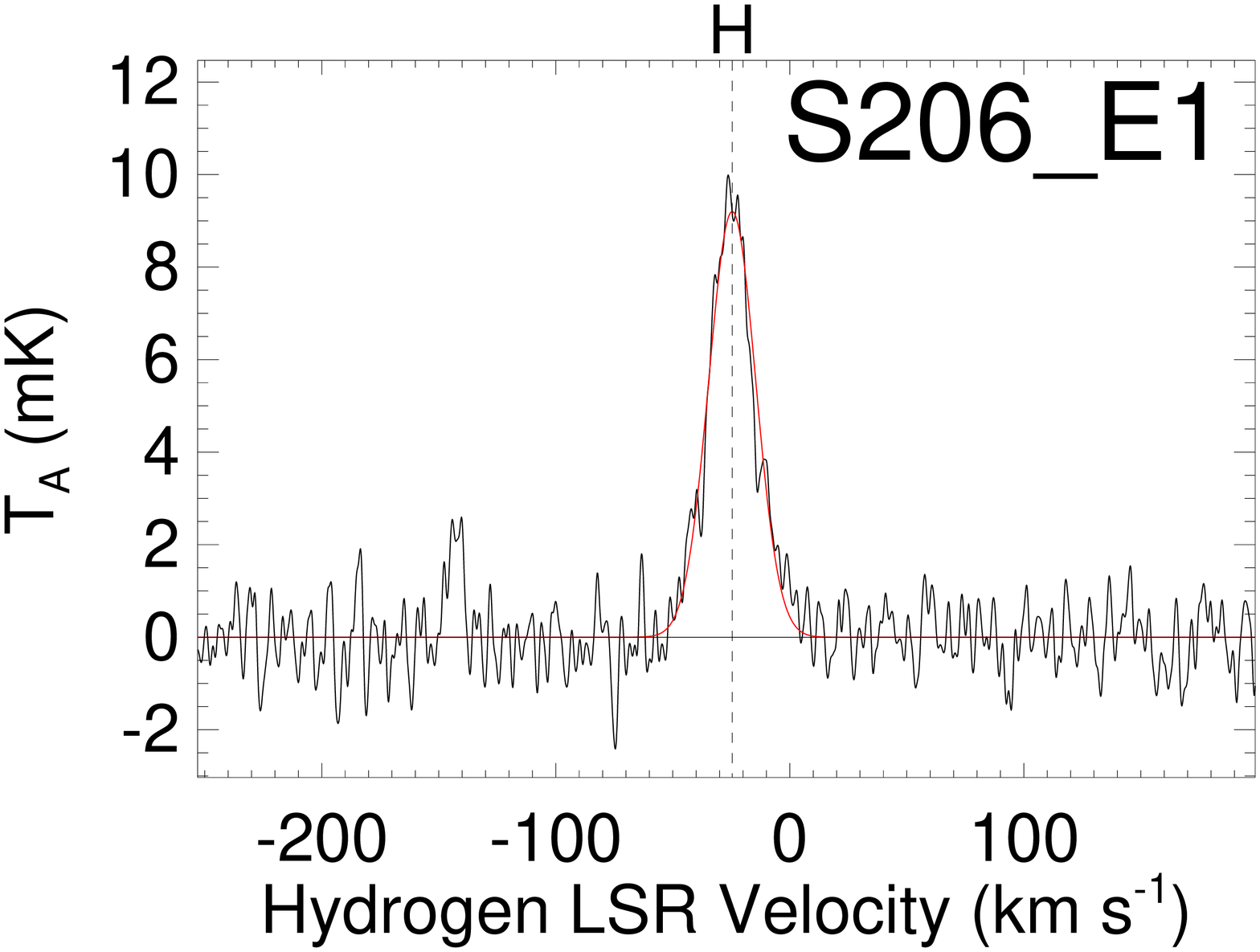} &
\includegraphics[width=.23\textwidth]{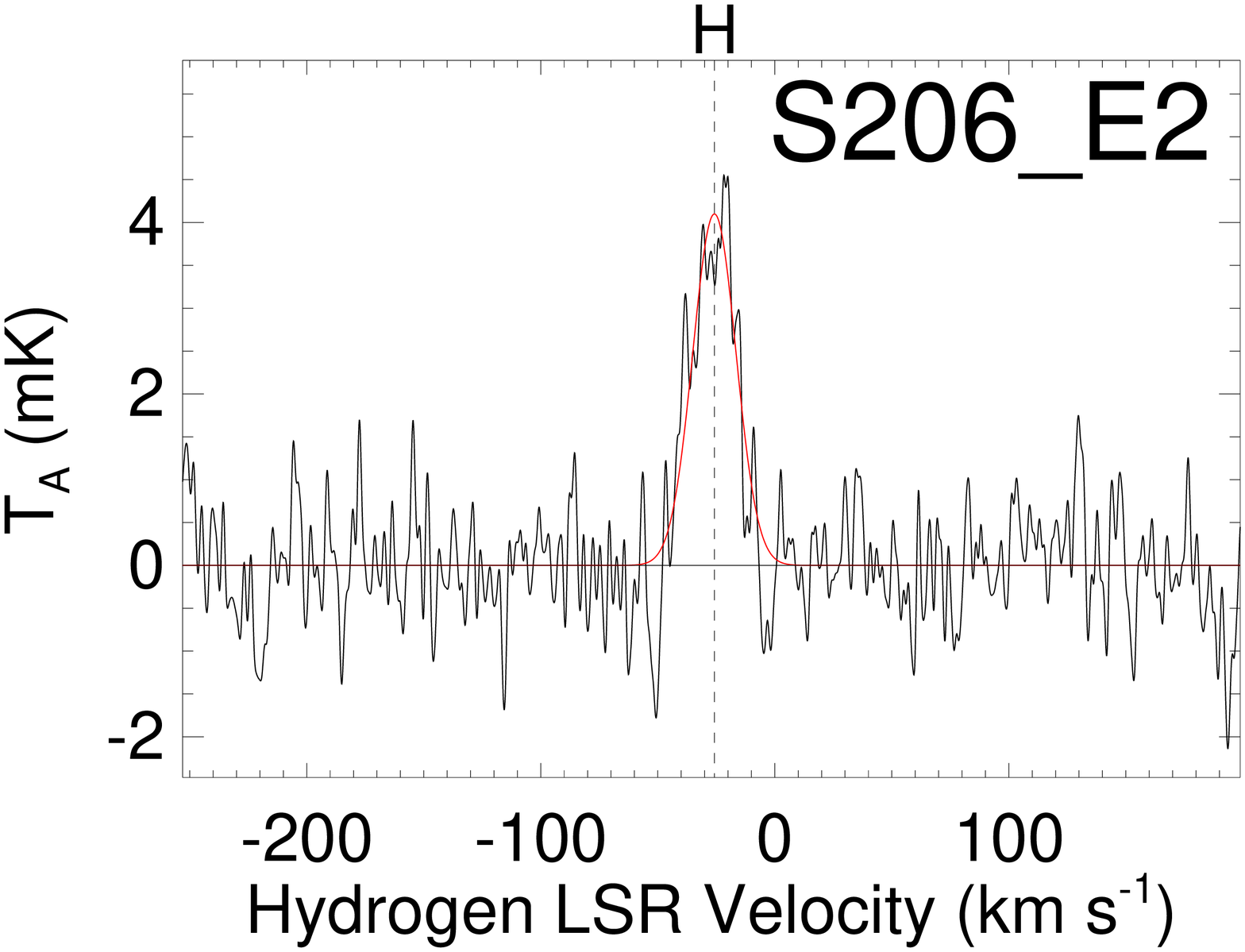} \\
\includegraphics[width=.23\textwidth]{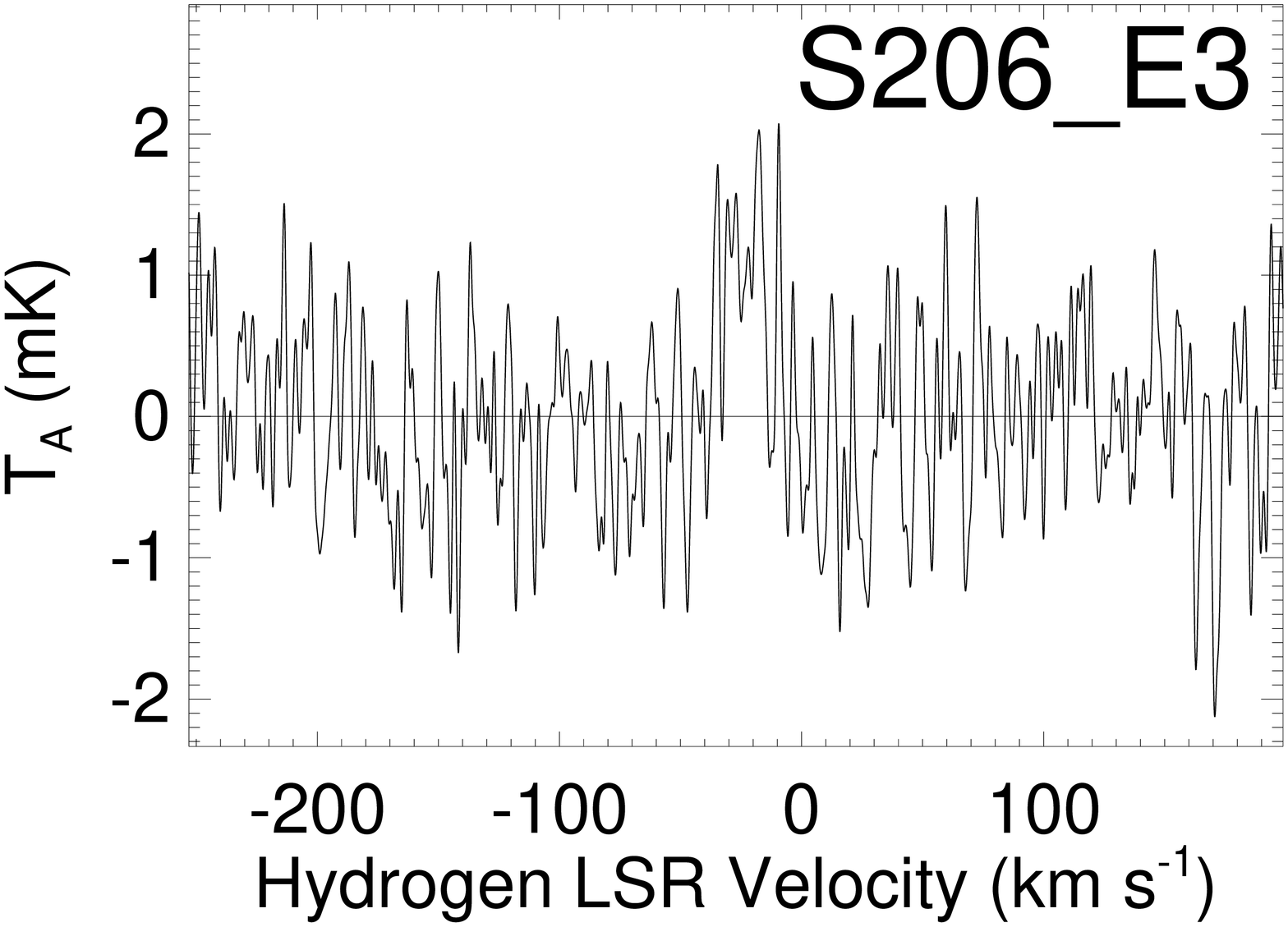} &
\includegraphics[width=.23\textwidth]{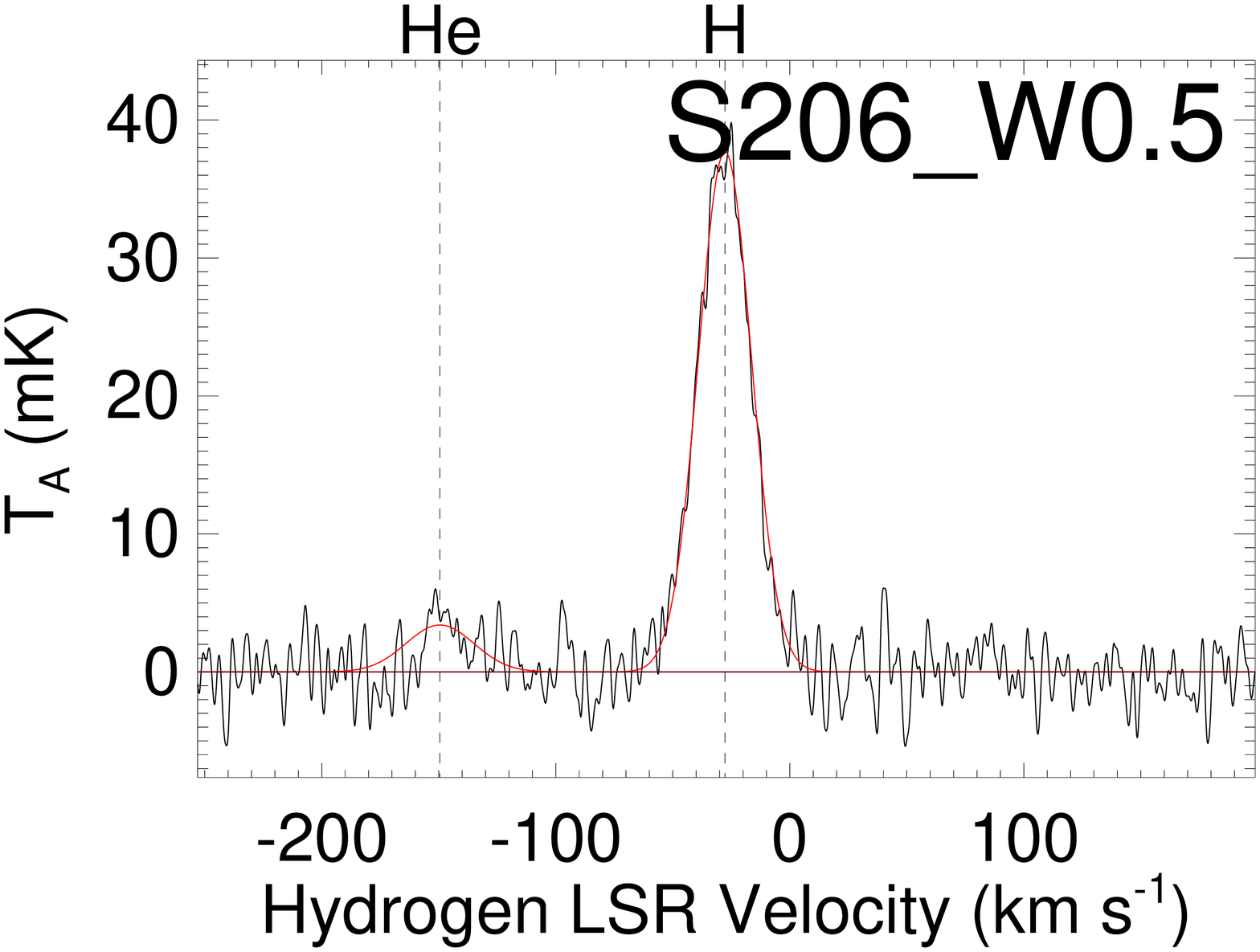} &
\includegraphics[width=.23\textwidth]{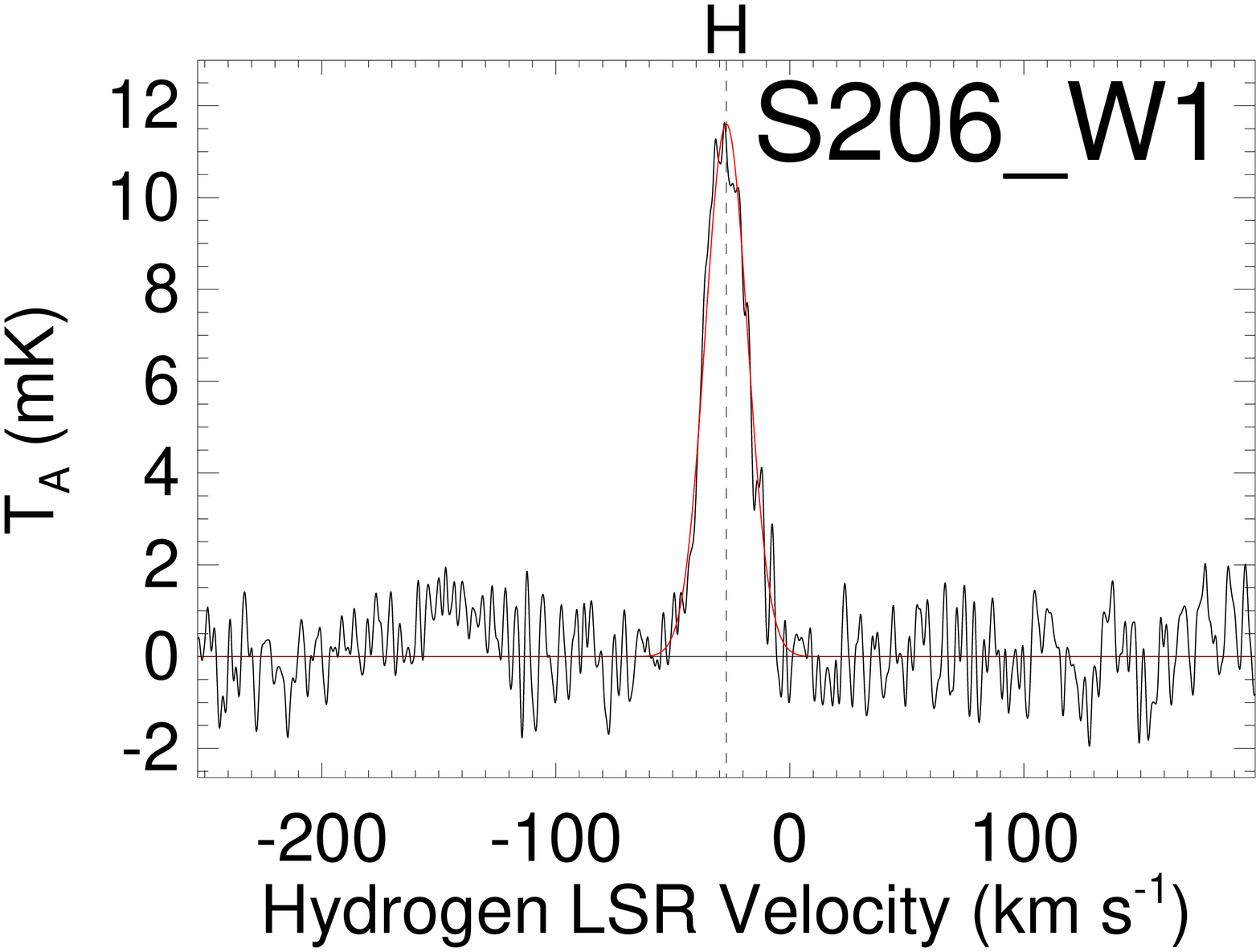} &
\includegraphics[width=.23\textwidth]{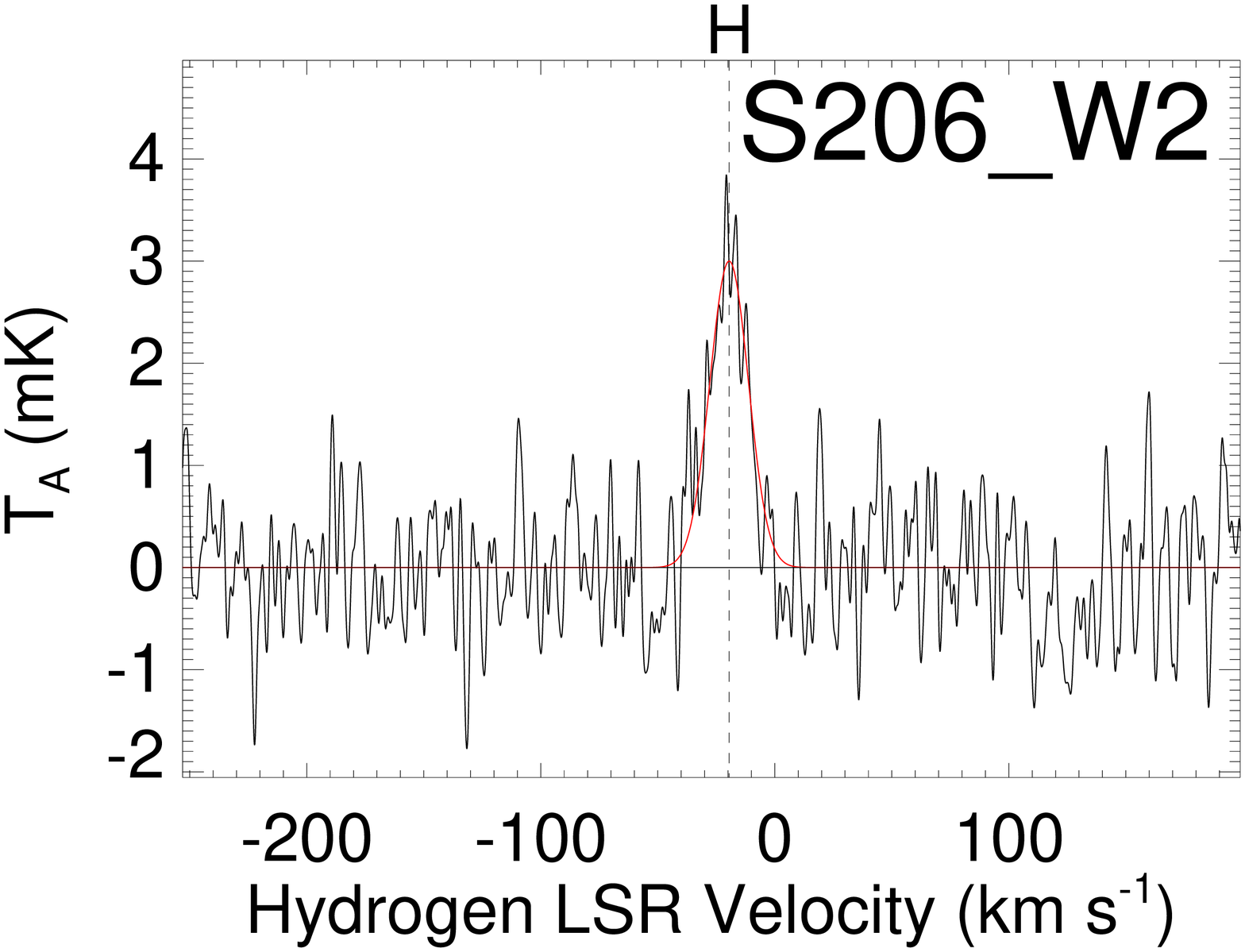} \\
\includegraphics[width=.23\textwidth]{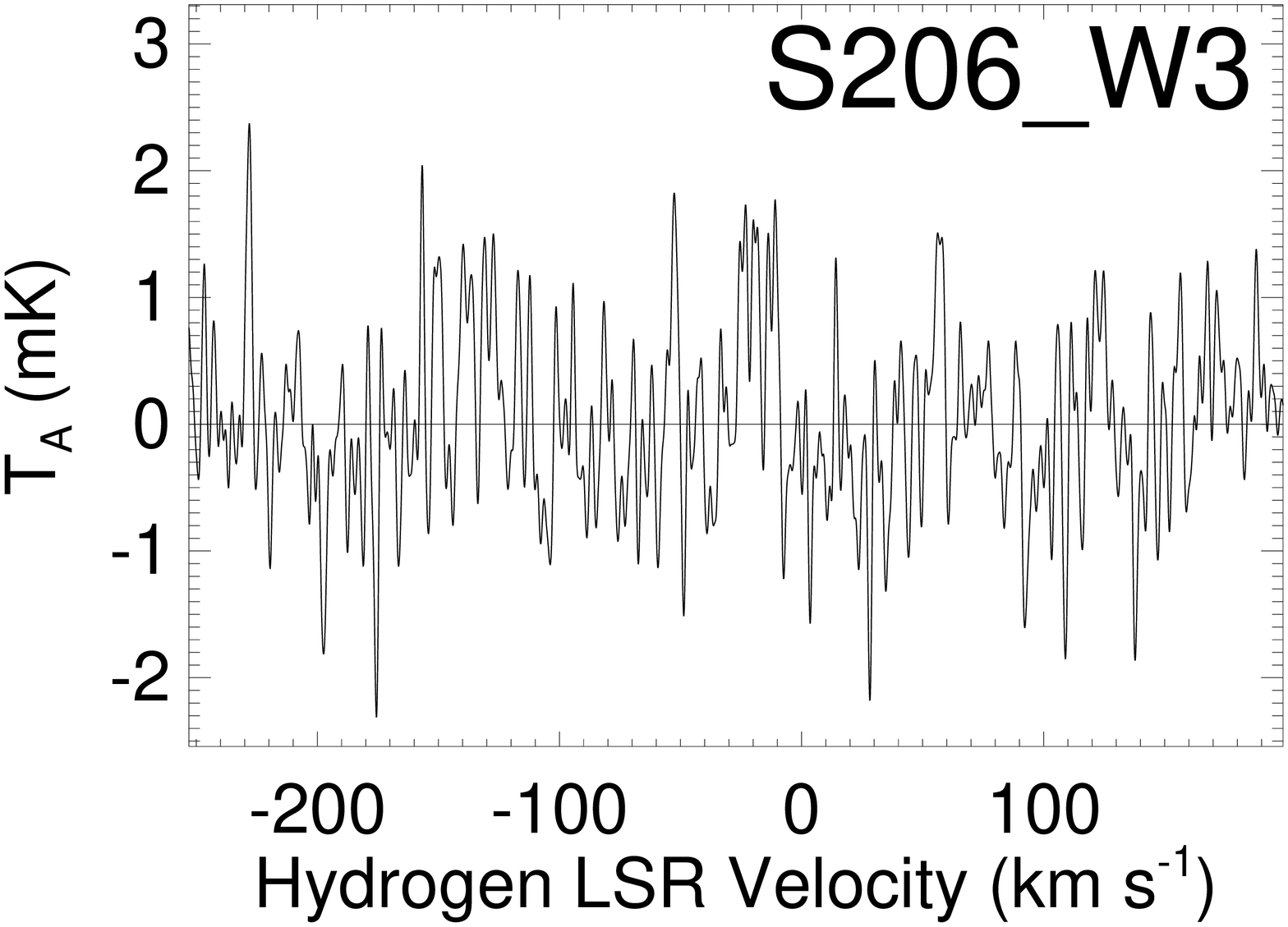} &
\includegraphics[width=.23\textwidth]{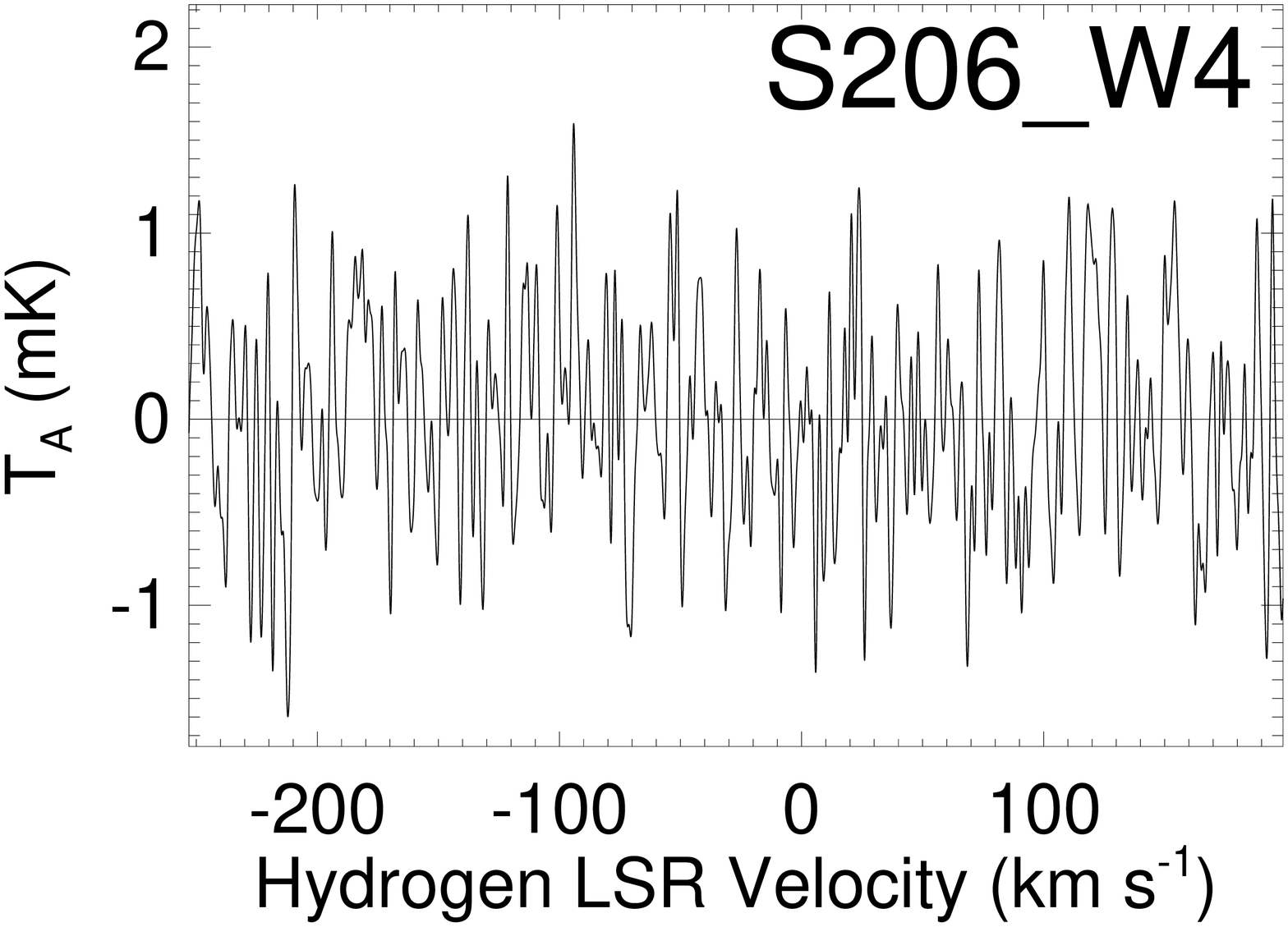} &
\includegraphics[width=.23\textwidth]{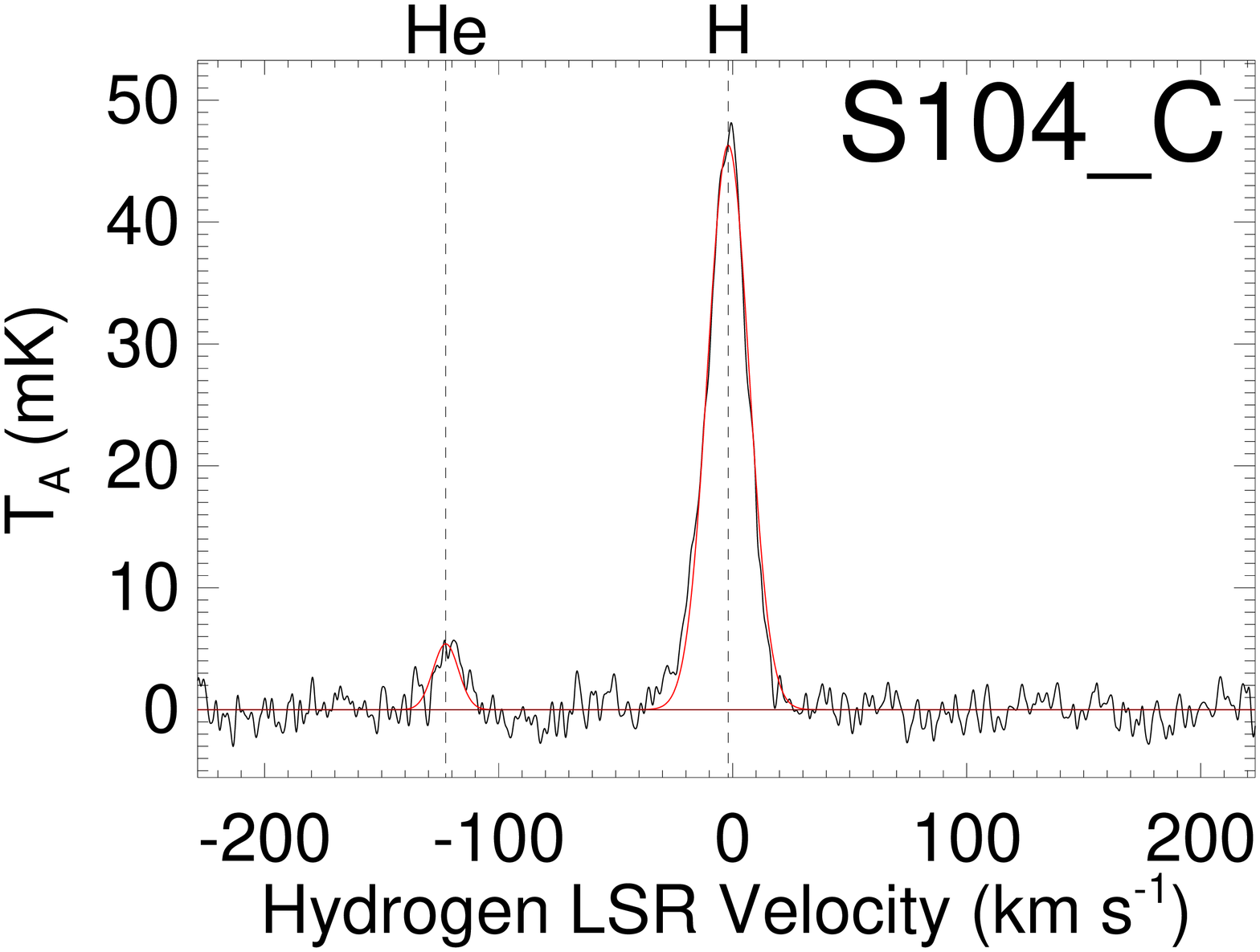} &
\includegraphics[width=.23\textwidth]{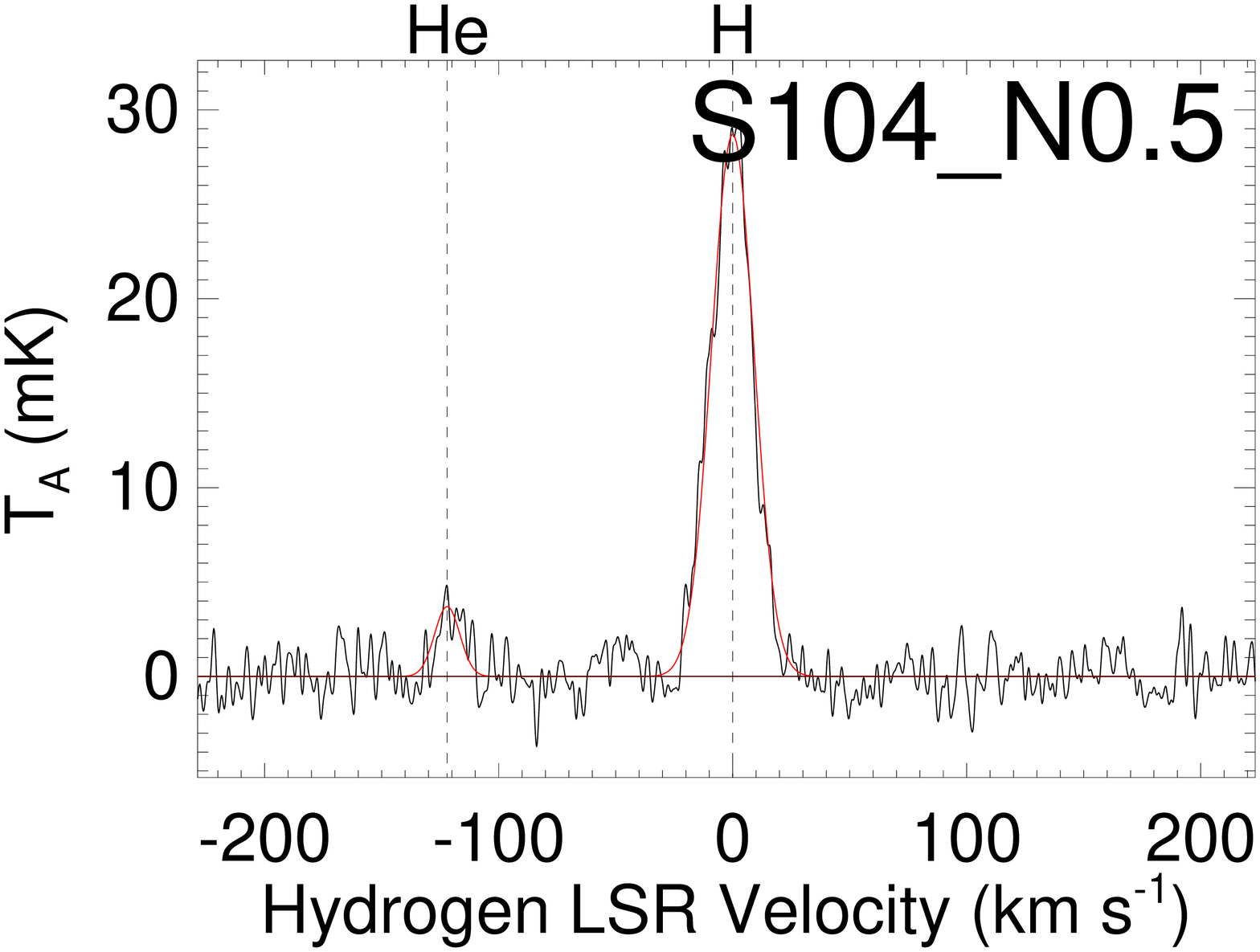} \\
\includegraphics[width=.23\textwidth]{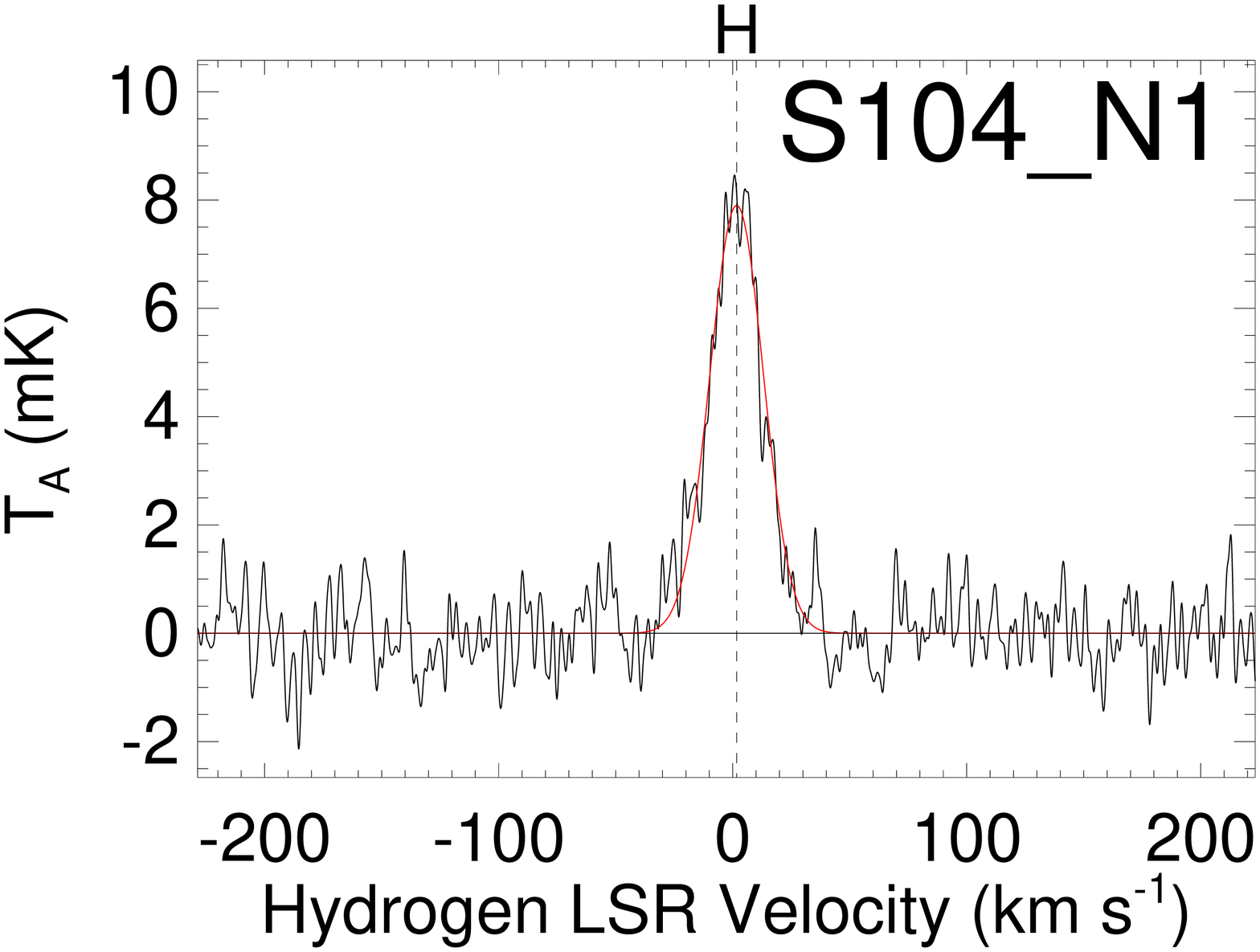} &
\includegraphics[width=.23\textwidth]{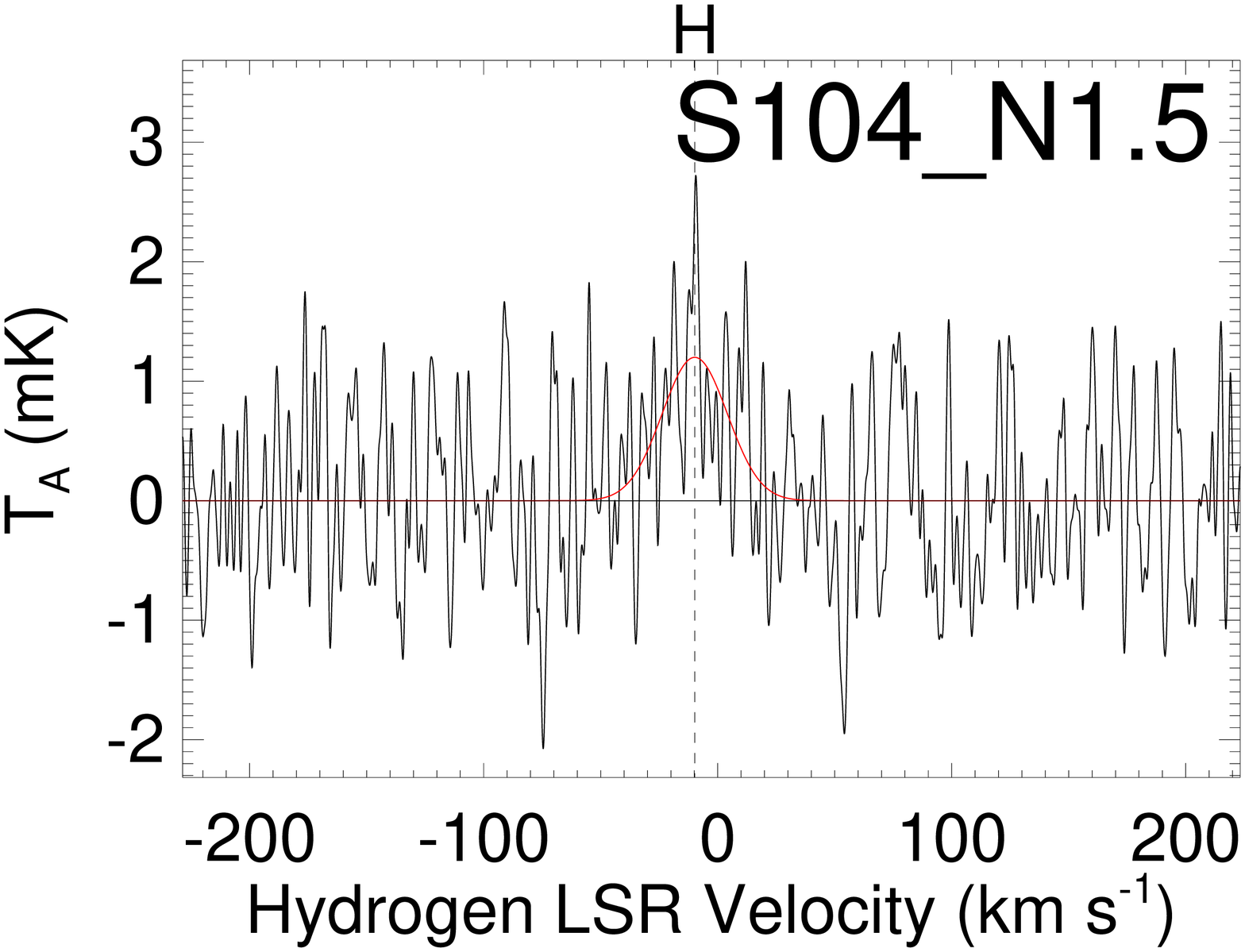} &
\includegraphics[width=.23\textwidth]{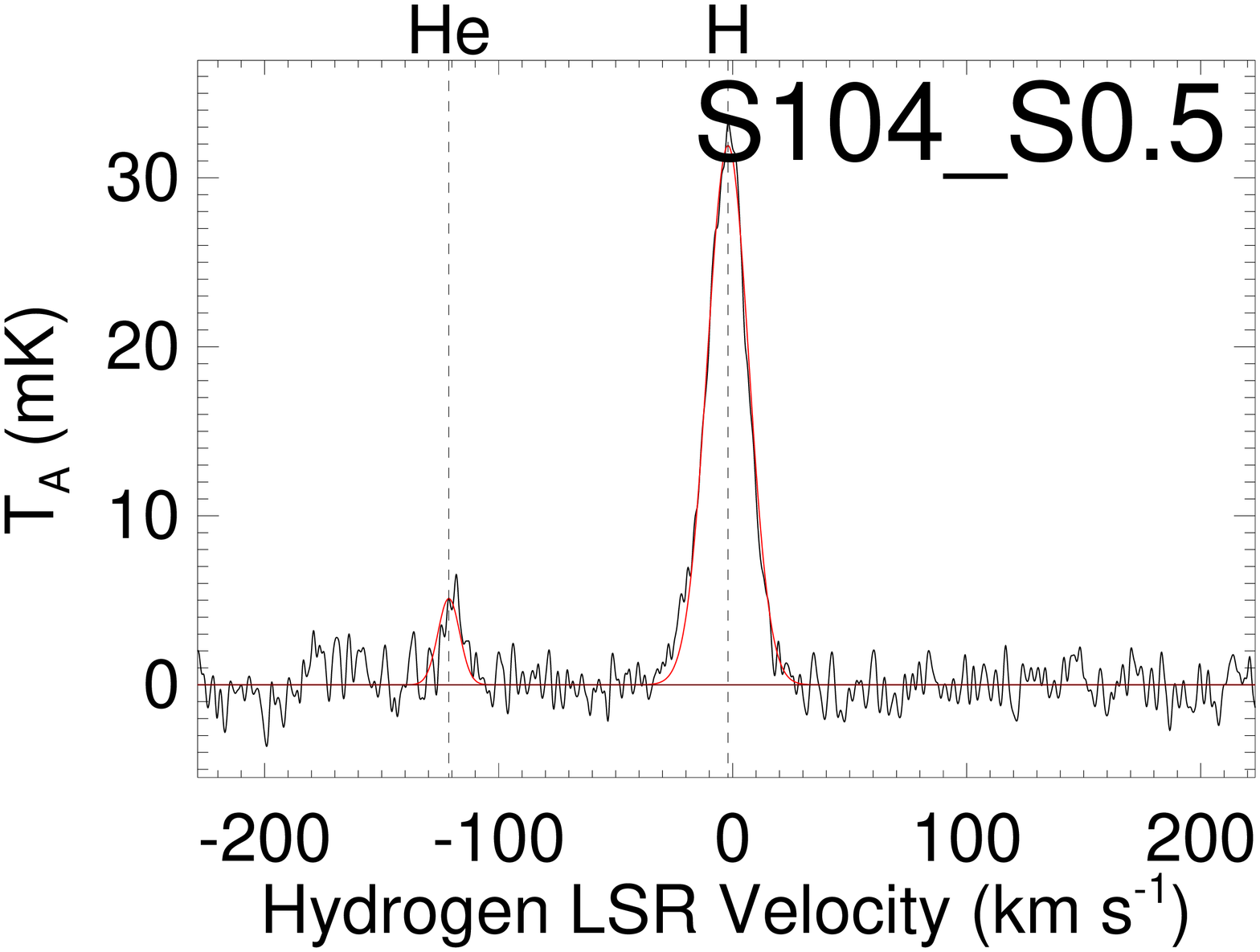} &
\includegraphics[width=.23\textwidth]{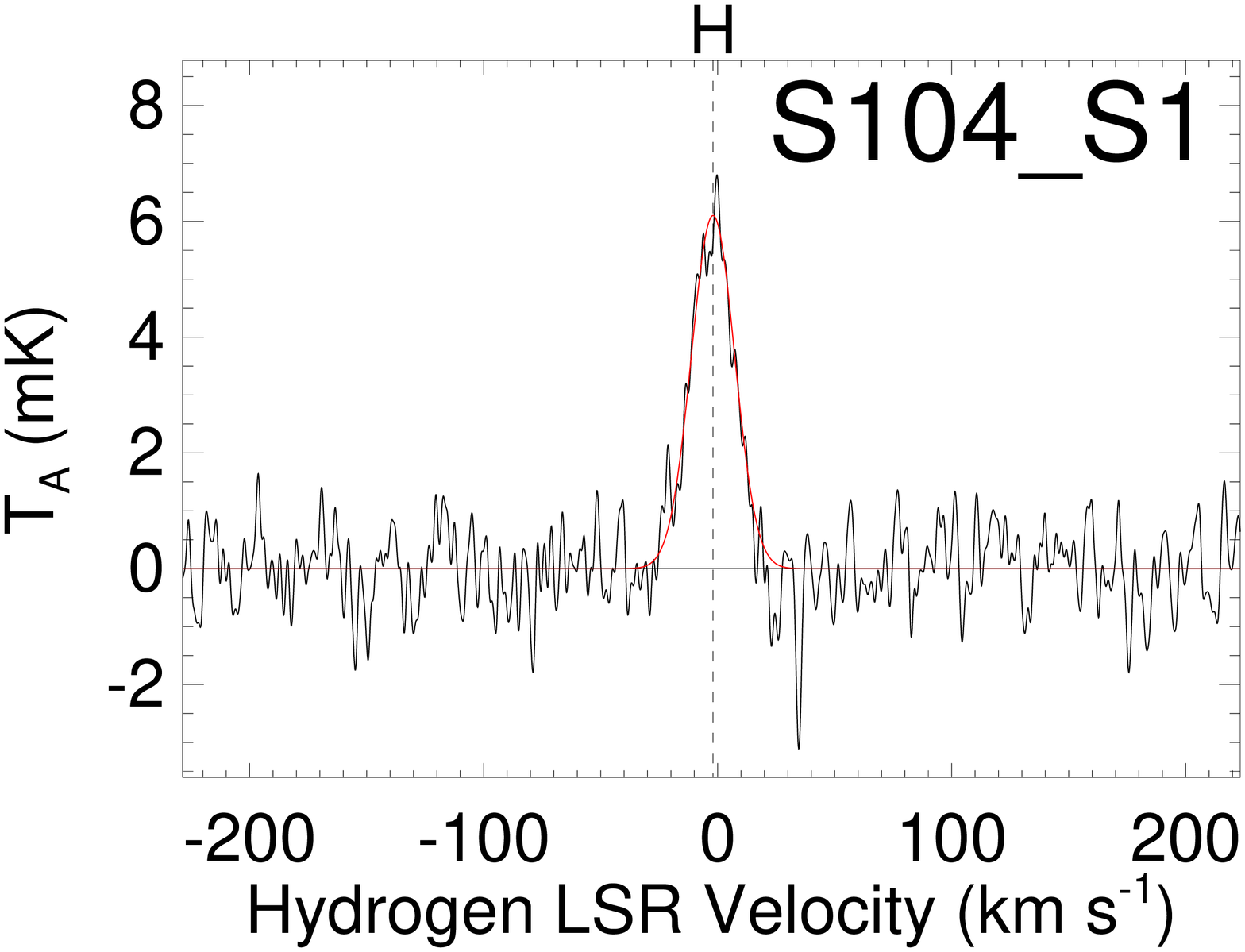} \\
\includegraphics[width=.23\textwidth]{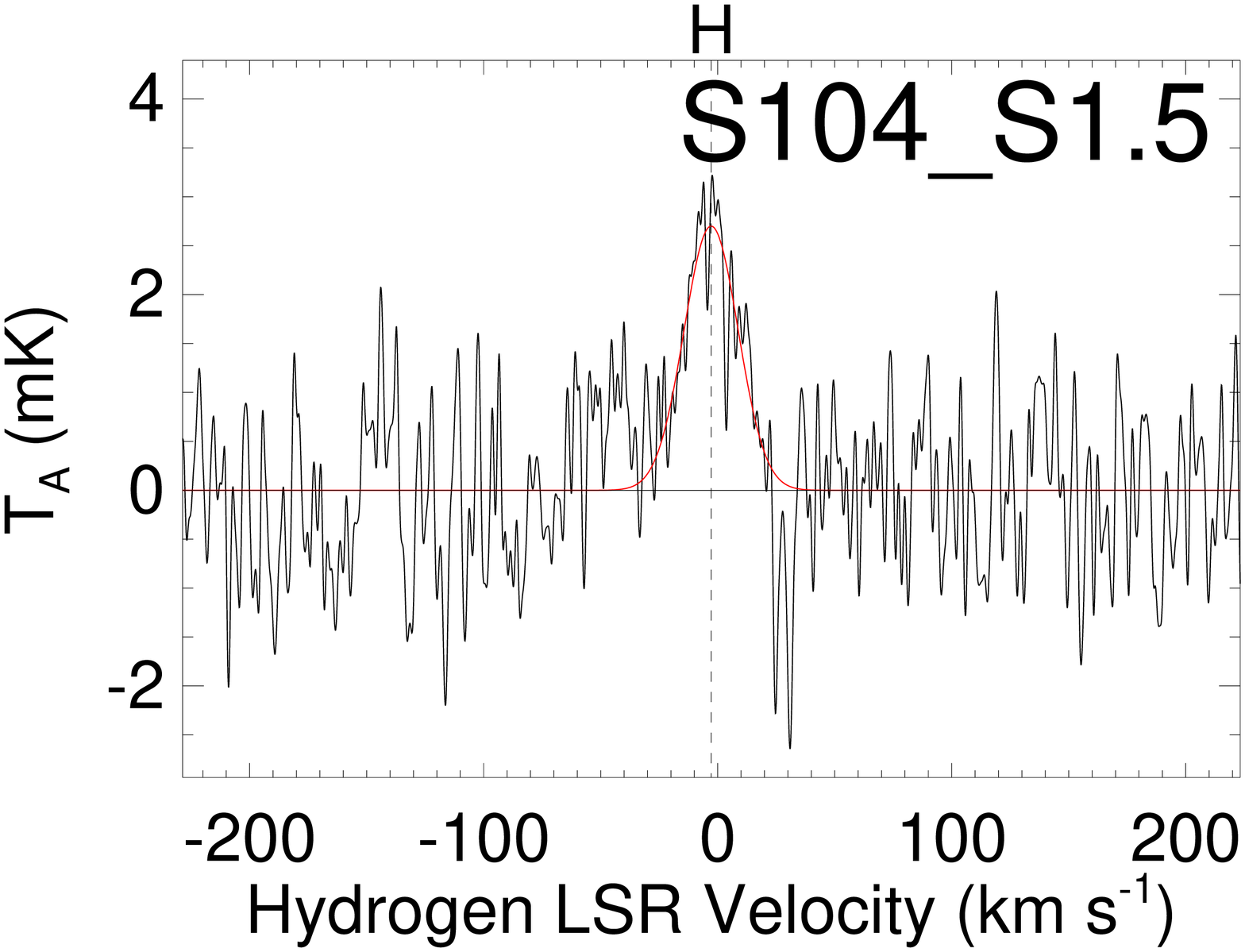} &
\includegraphics[width=.23\textwidth]{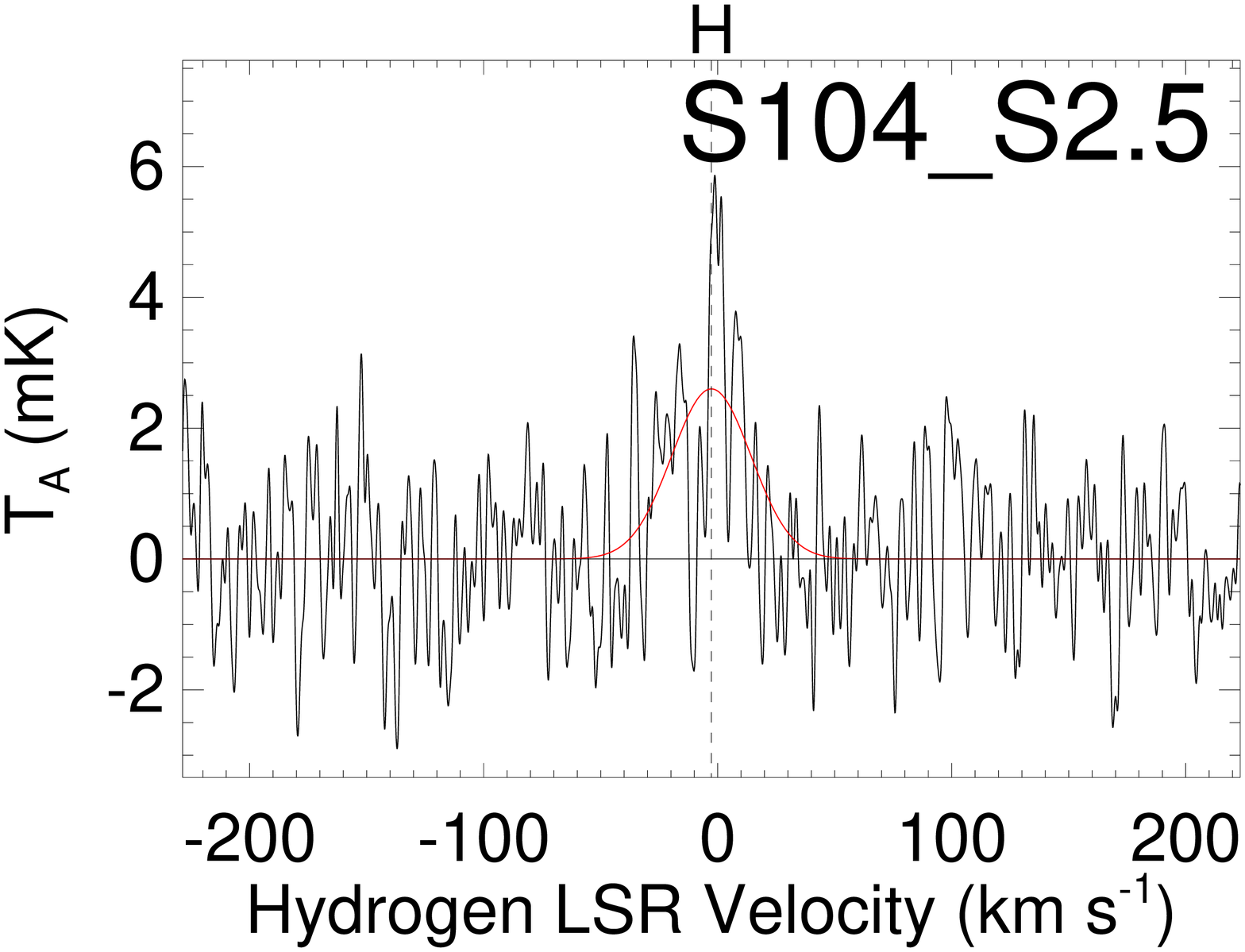} &
\includegraphics[width=.23\textwidth]{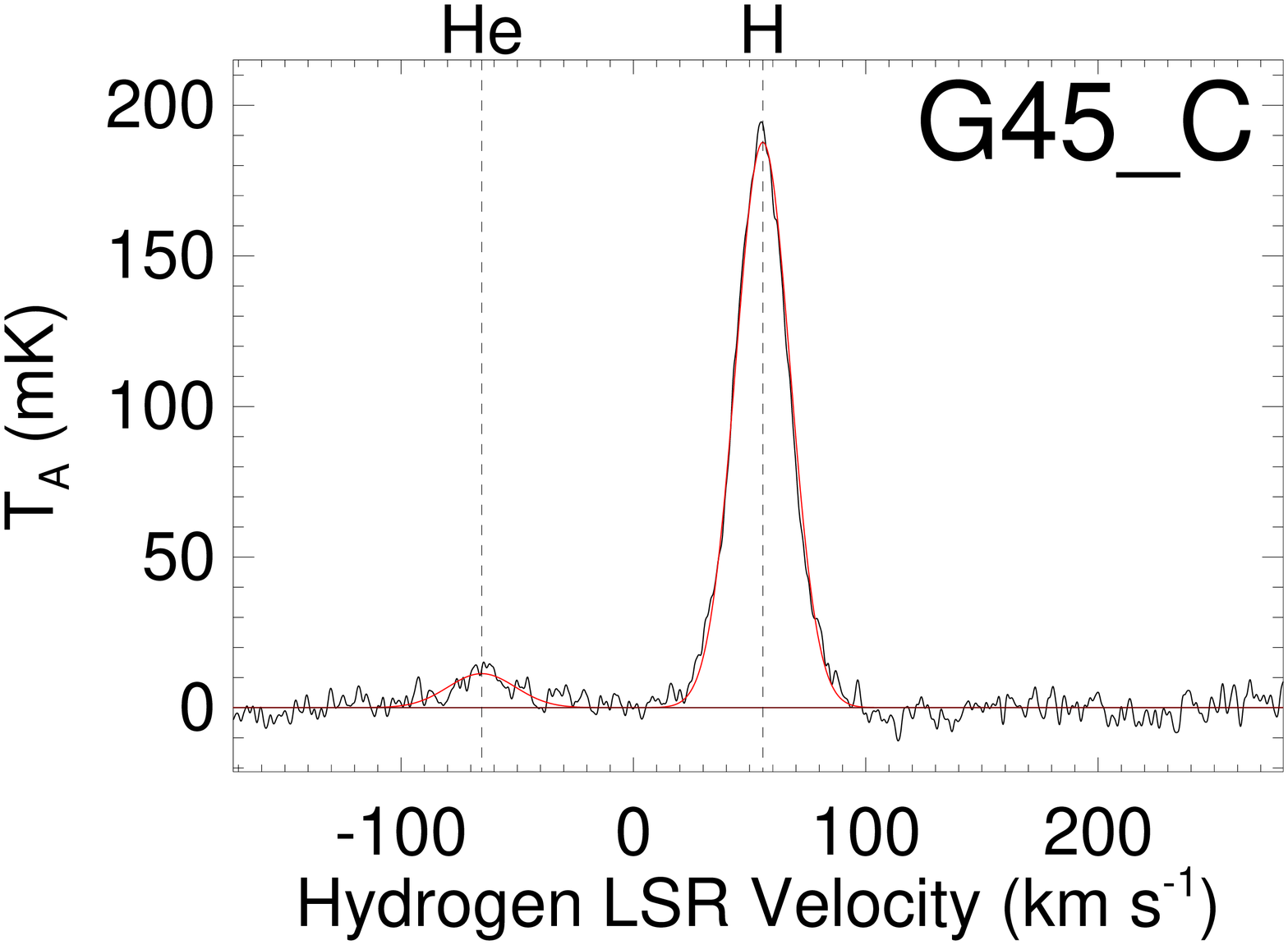} &
\includegraphics[width=.23\textwidth]{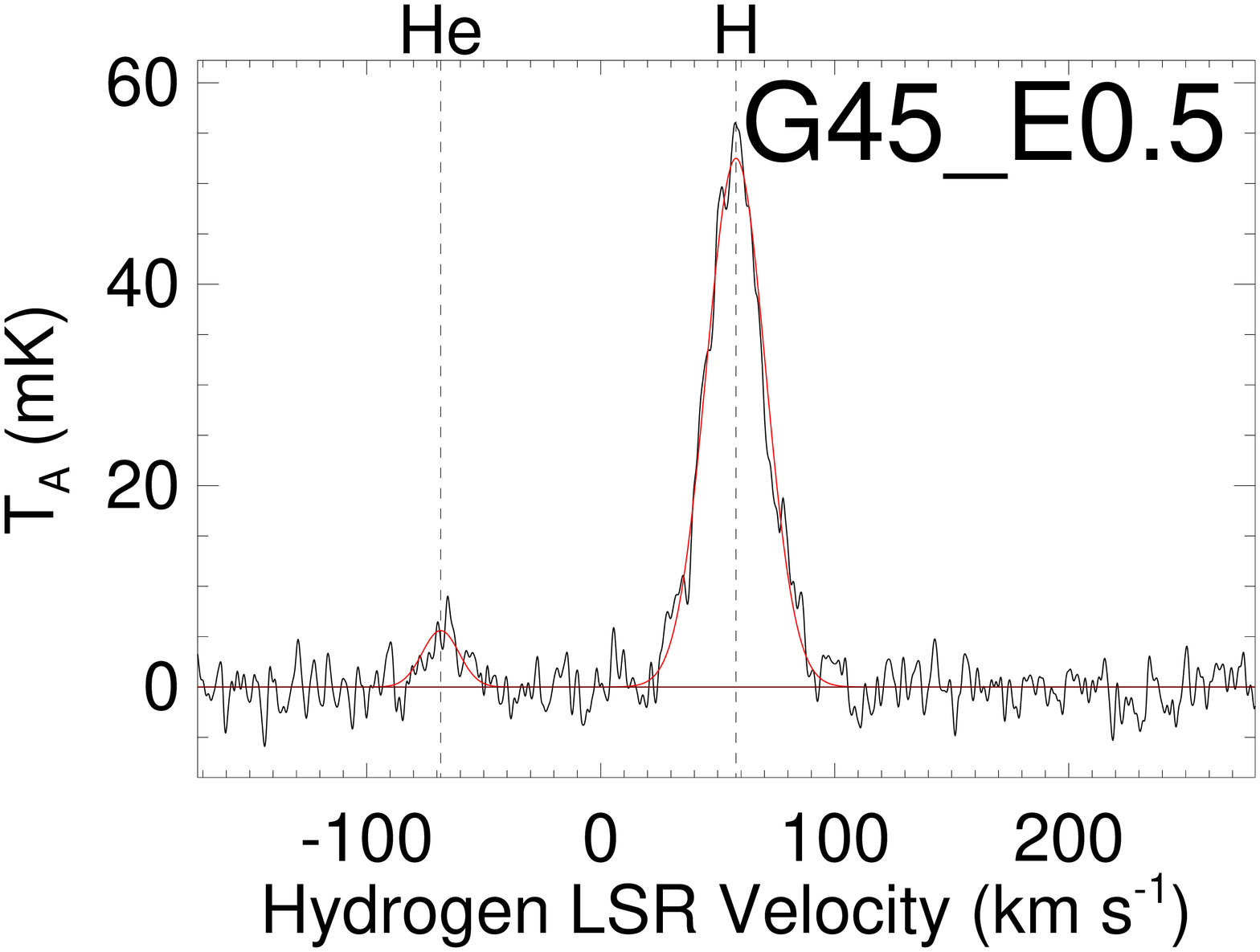} \\
\includegraphics[width=.23\textwidth]{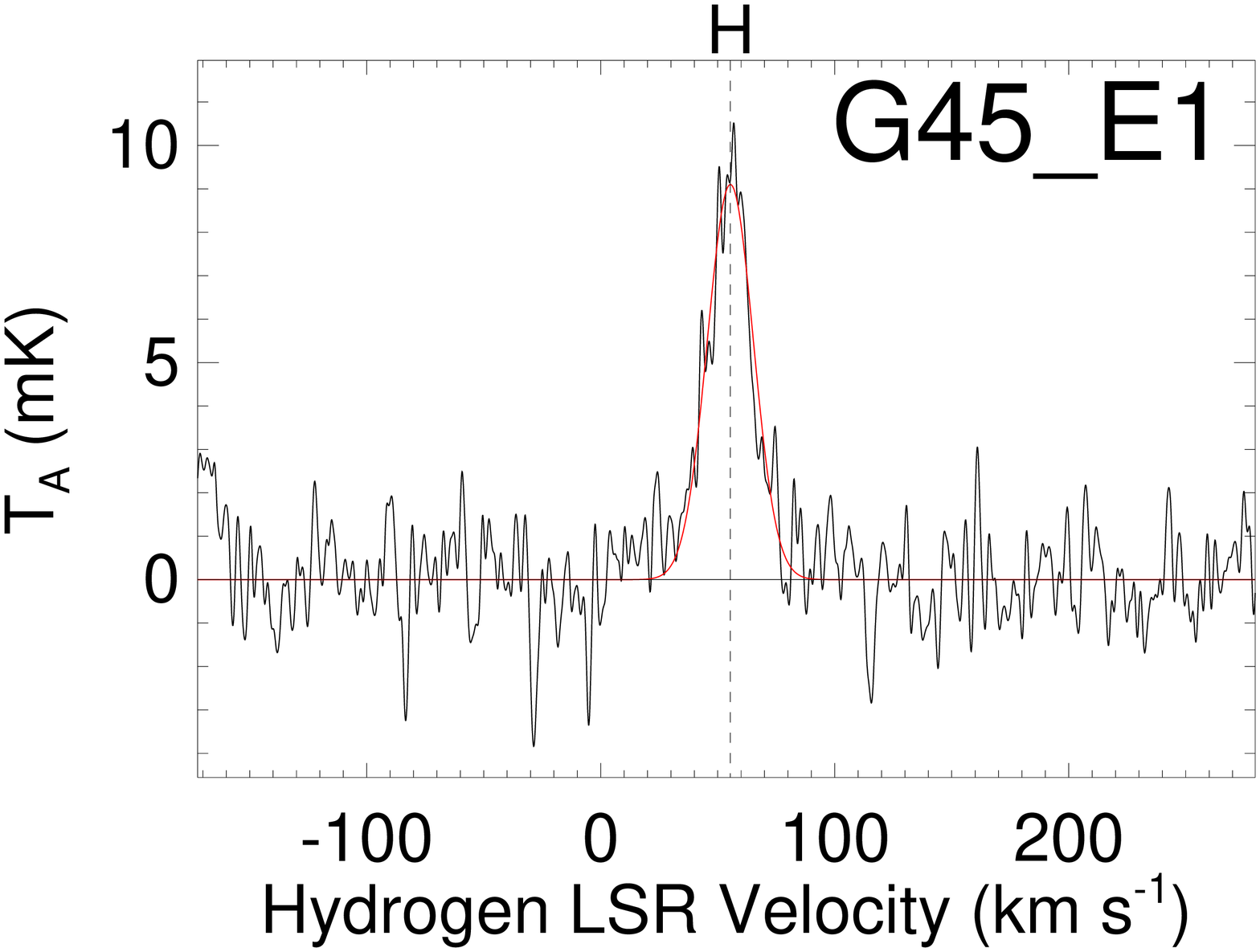} &
\includegraphics[width=.23\textwidth]{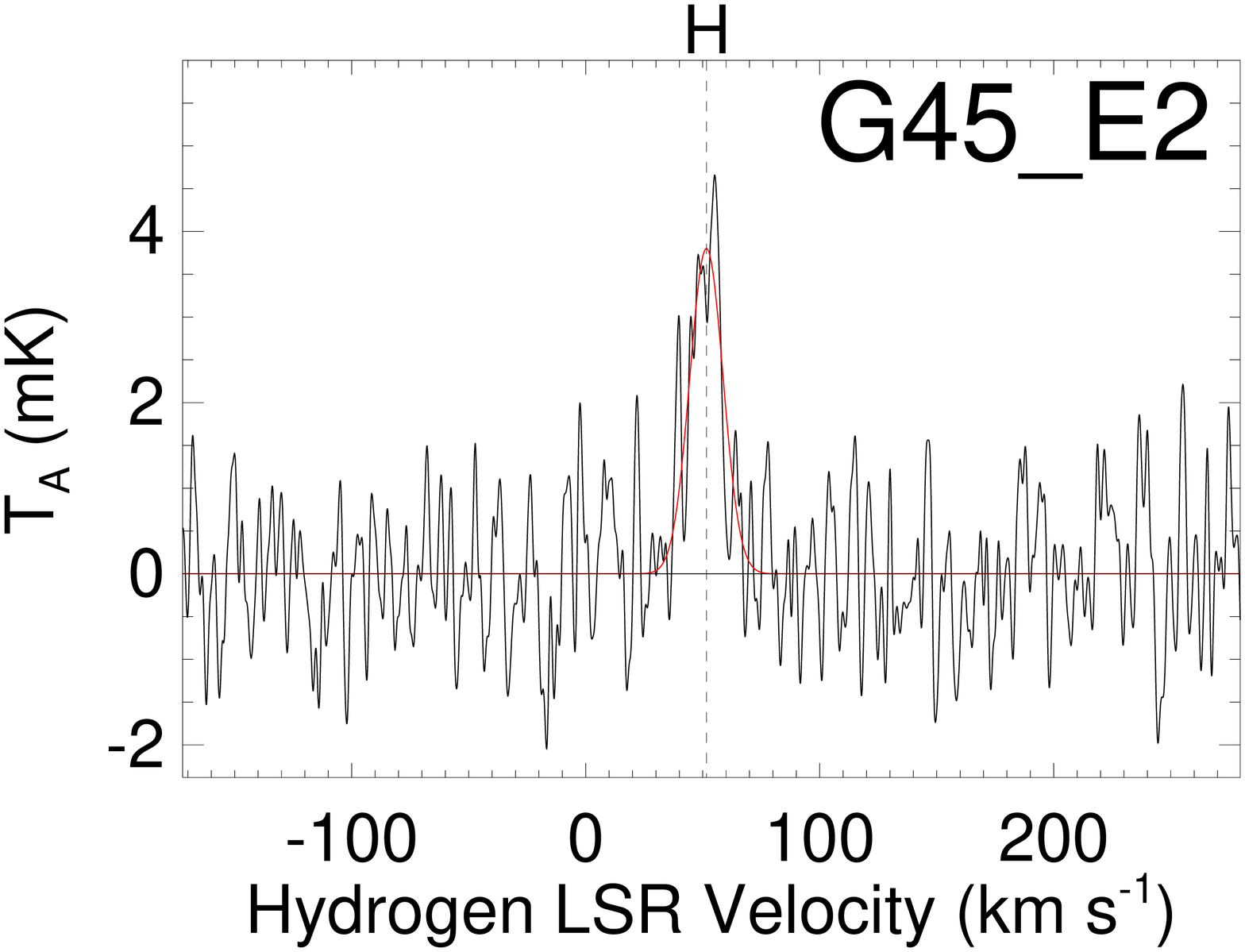} &
\includegraphics[width=.23\textwidth]{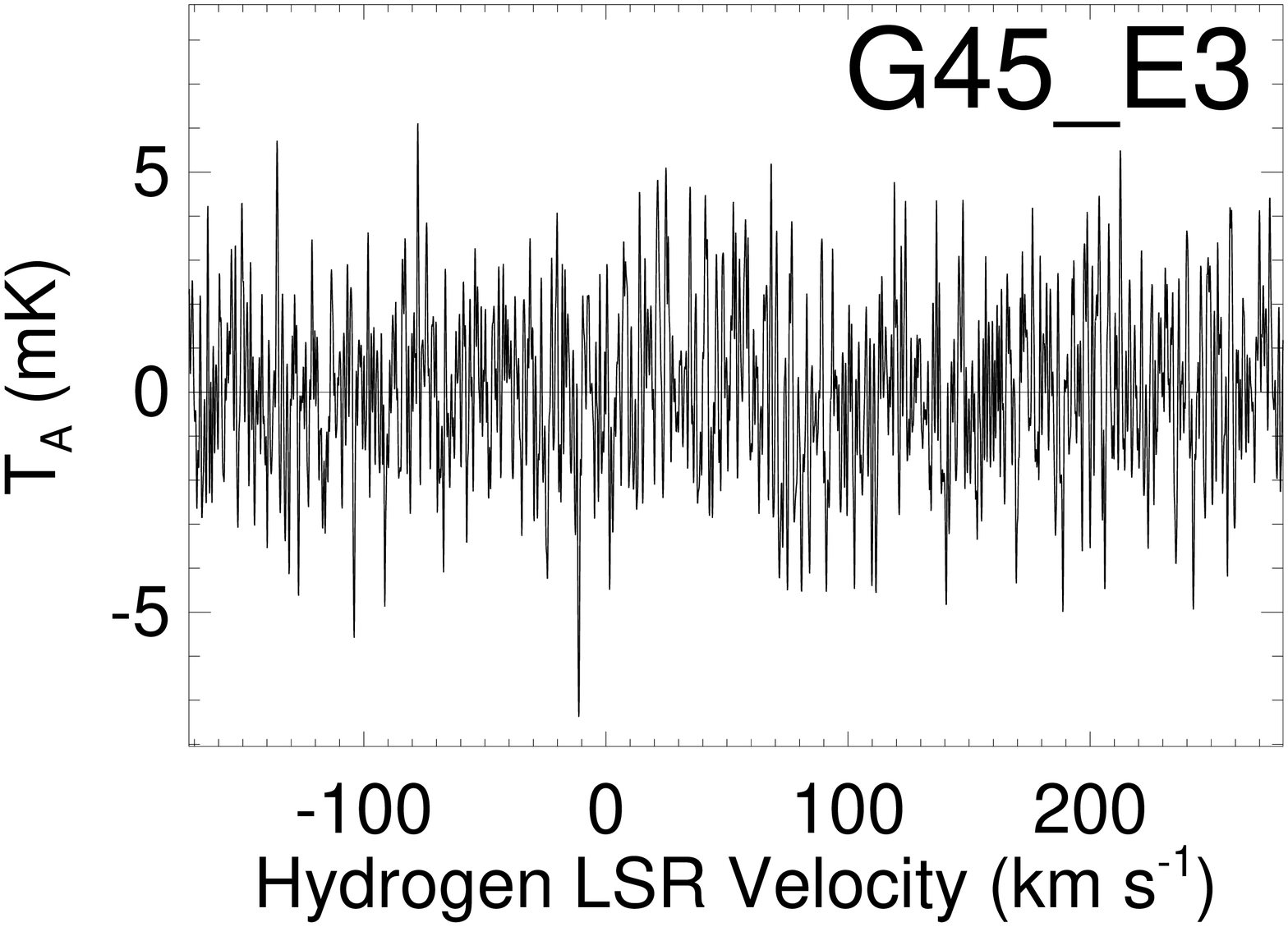} &
\includegraphics[width=.23\textwidth]{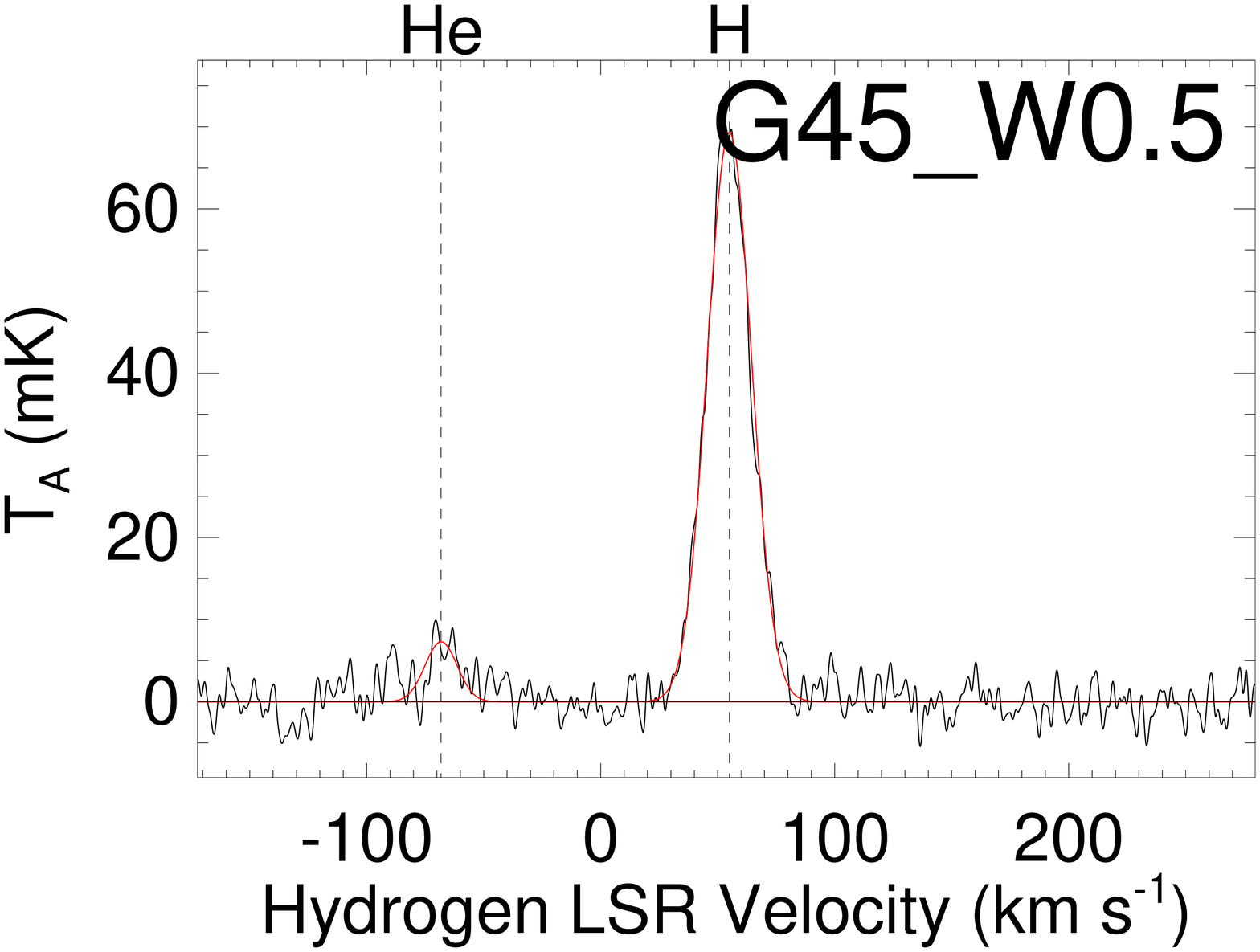} \\
\end{tabular}
\caption{$\beta$ RRL spectra of all observed positions, smoothed to a spectral resolution of 1.86\kms. Plotted is the antenna temperature as a function of hydrogen LSR velocity. The helium and carbon lines are offset from hydrogen by $-124$\kms and $-149$\kms, respectively. We approximate hydrogen, helium, and carbon emission above the S/N threshold defined in \S \ref{sec:obs} with the Gaussian model fits shown in red. The centers of the Gaussian peaks are indicated by dashed vertical lines. The name of the observed position is given in the upper right-hand corner of each plot. \label{fig:spectrabeta}}
\end{figure*}
\renewcommand{\thefigure}{\thesection.\arabic{figure}}

\renewcommand\thefigure{\thesection.\arabic{figure} (Cont.)}
\addtocounter{figure}{-1}
\begin{figure*}
\centering
\begin{tabular}{cccc}
\includegraphics[width=.23\textwidth]{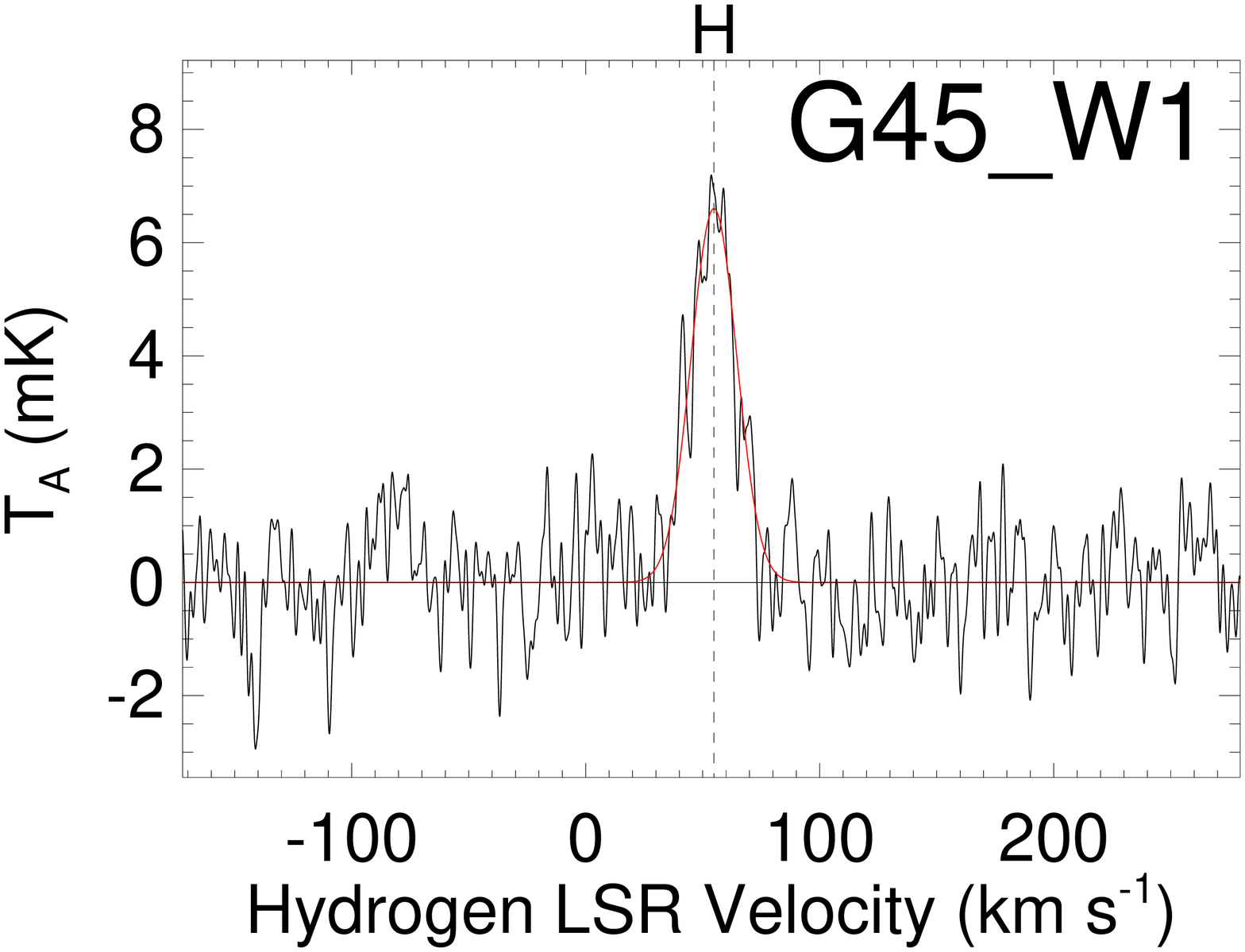} &
\includegraphics[width=.23\textwidth]{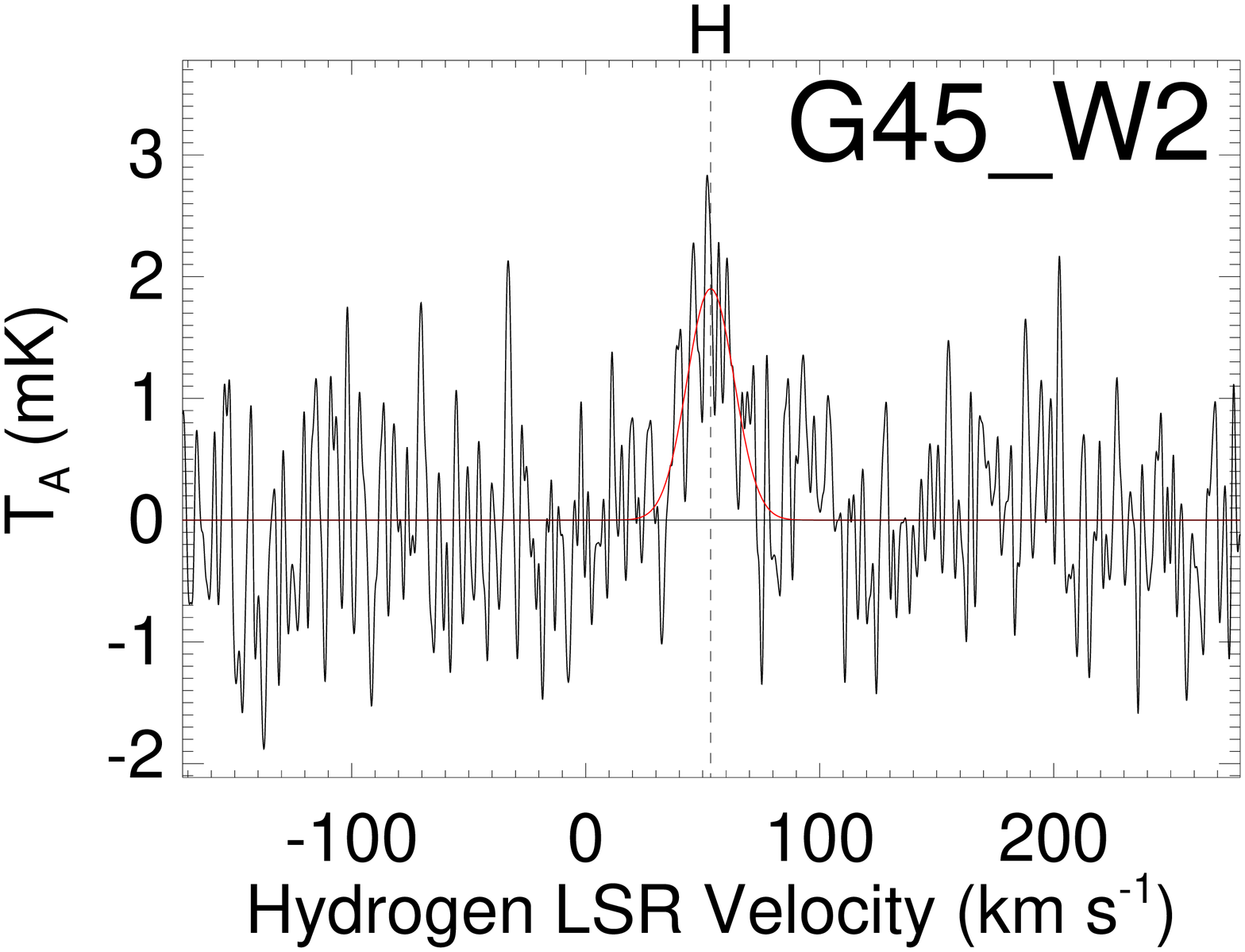} &
\includegraphics[width=.23\textwidth]{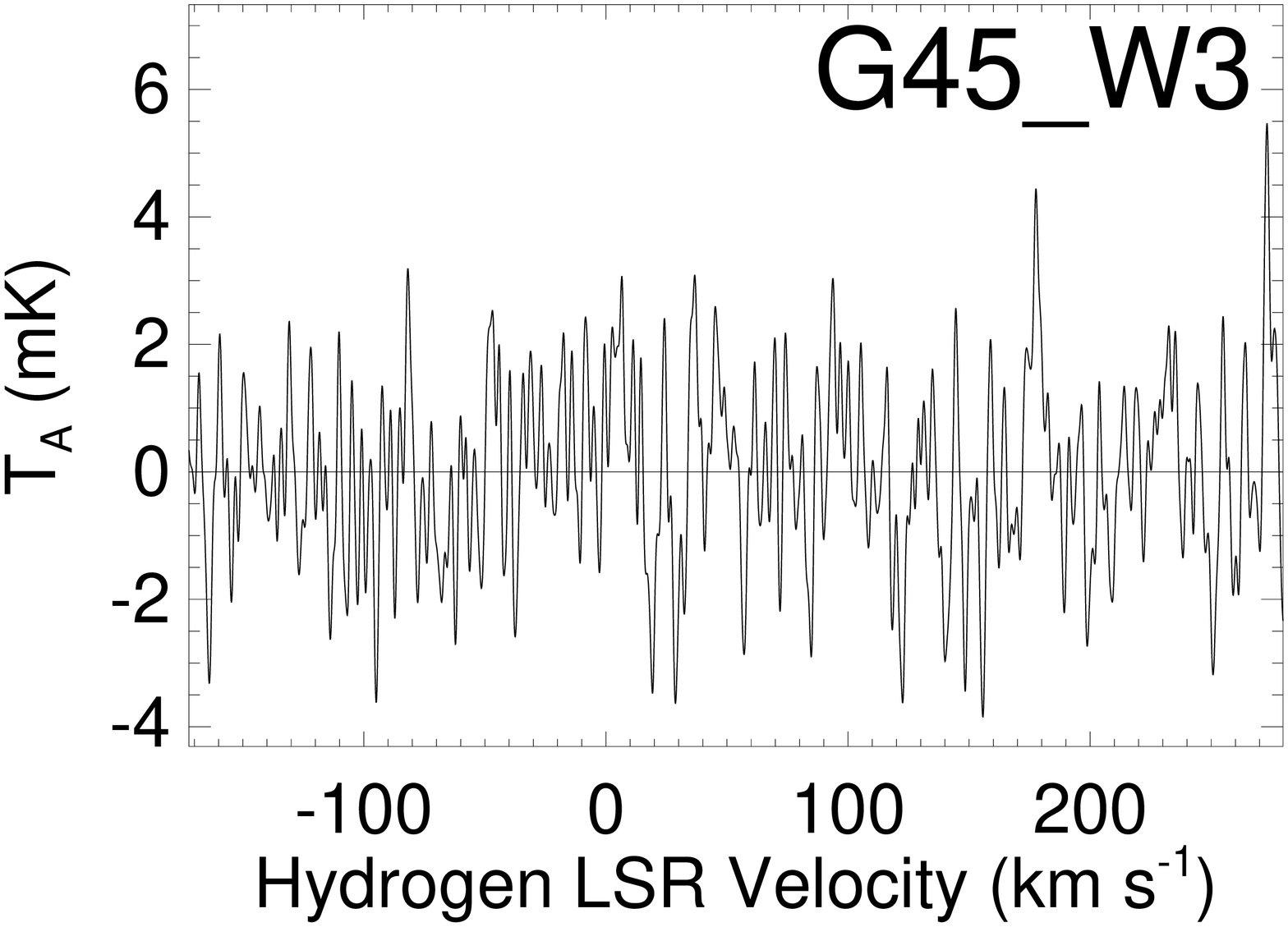} &
\includegraphics[width=.23\textwidth]{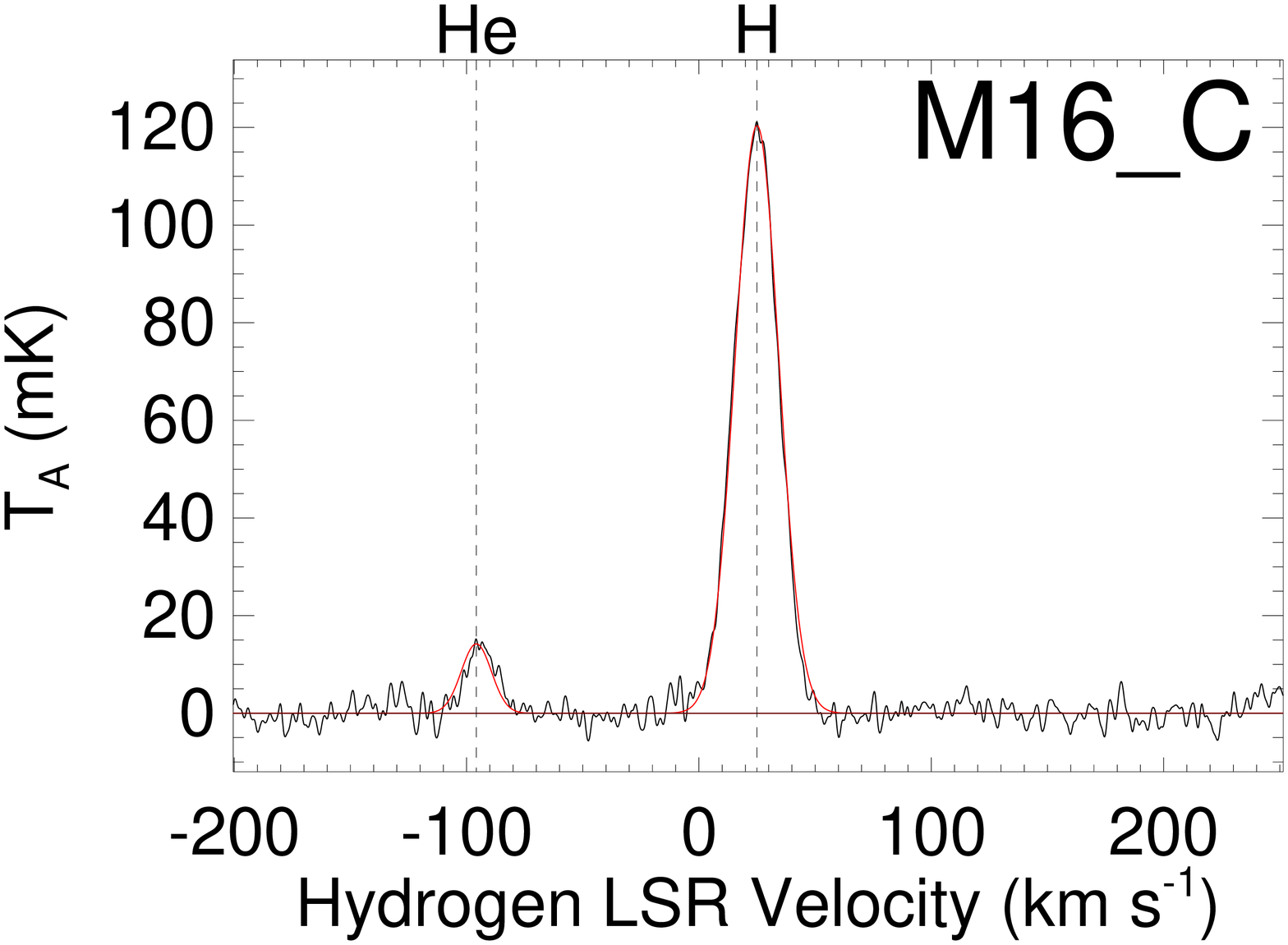} \\
\includegraphics[width=.23\textwidth]{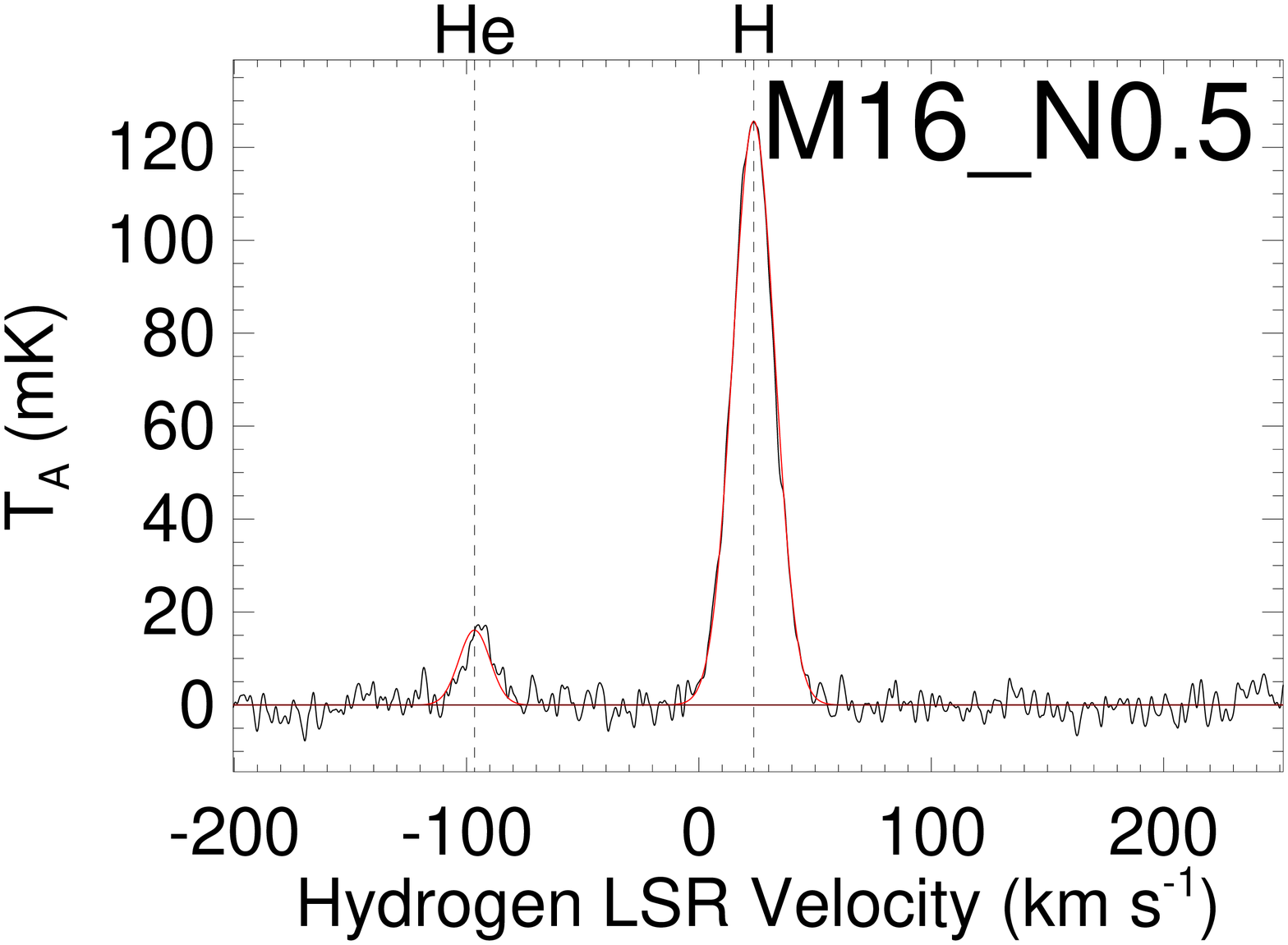} &
\includegraphics[width=.23\textwidth]{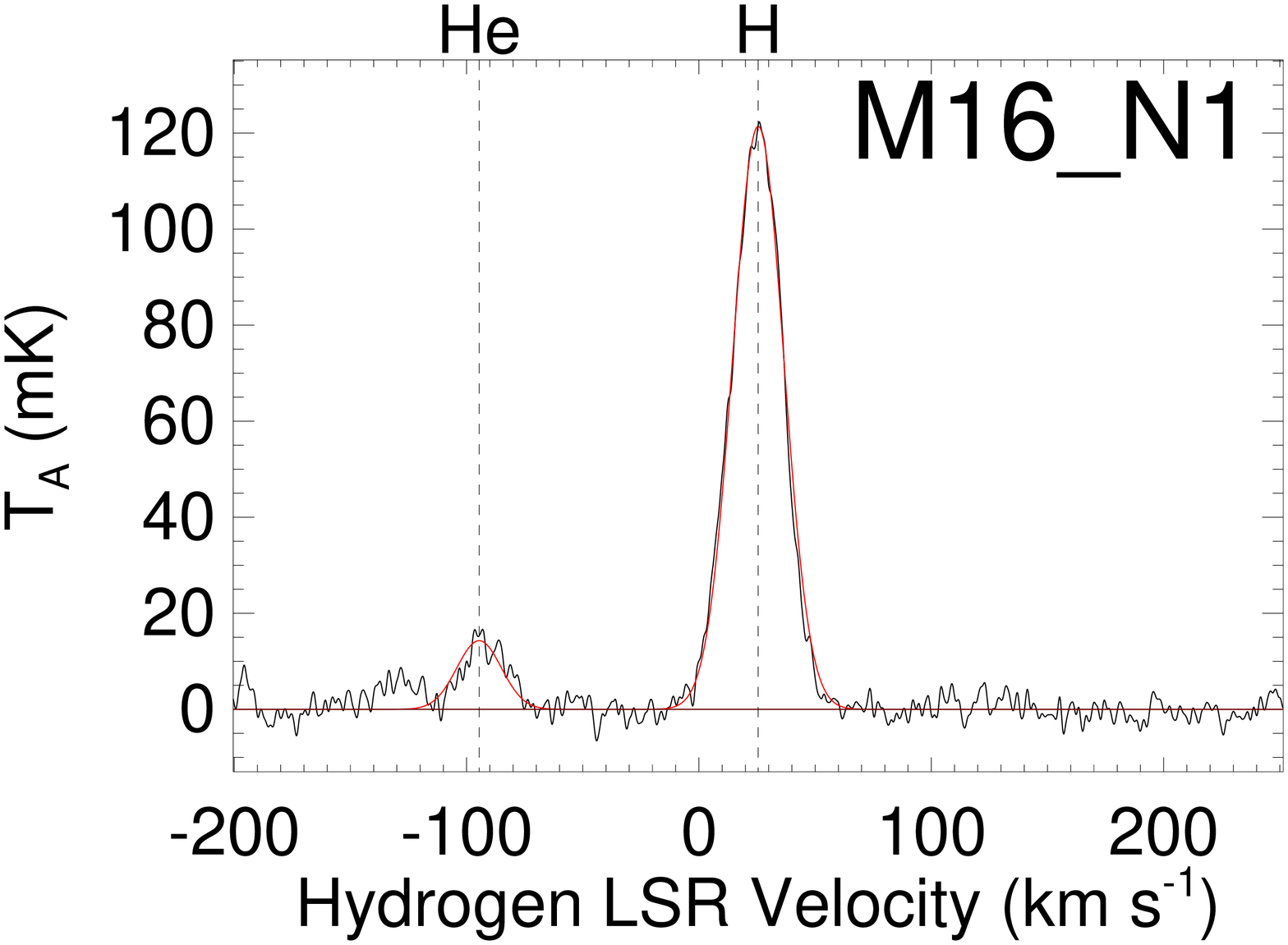} &
\includegraphics[width=.23\textwidth]{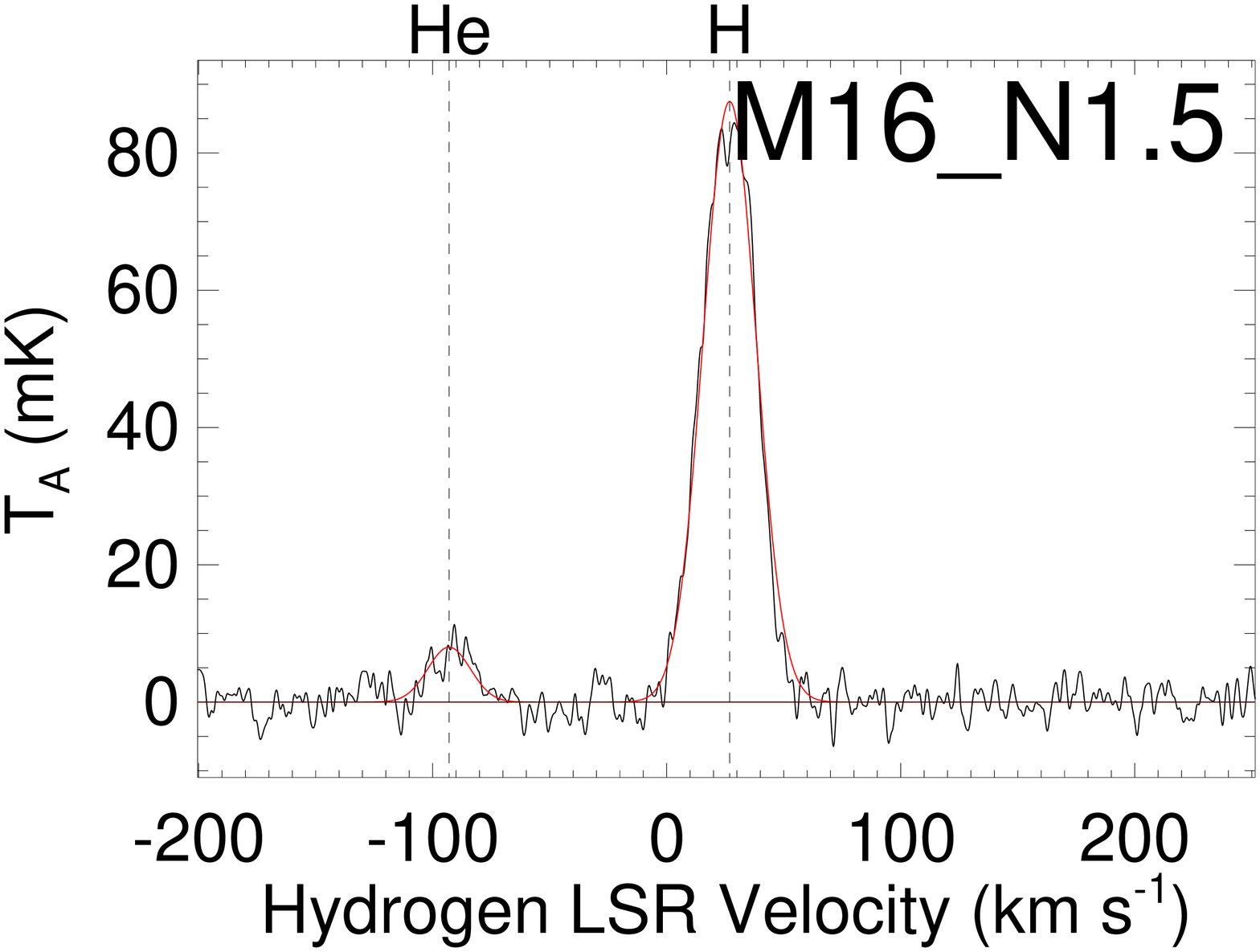} &
\includegraphics[width=.23\textwidth]{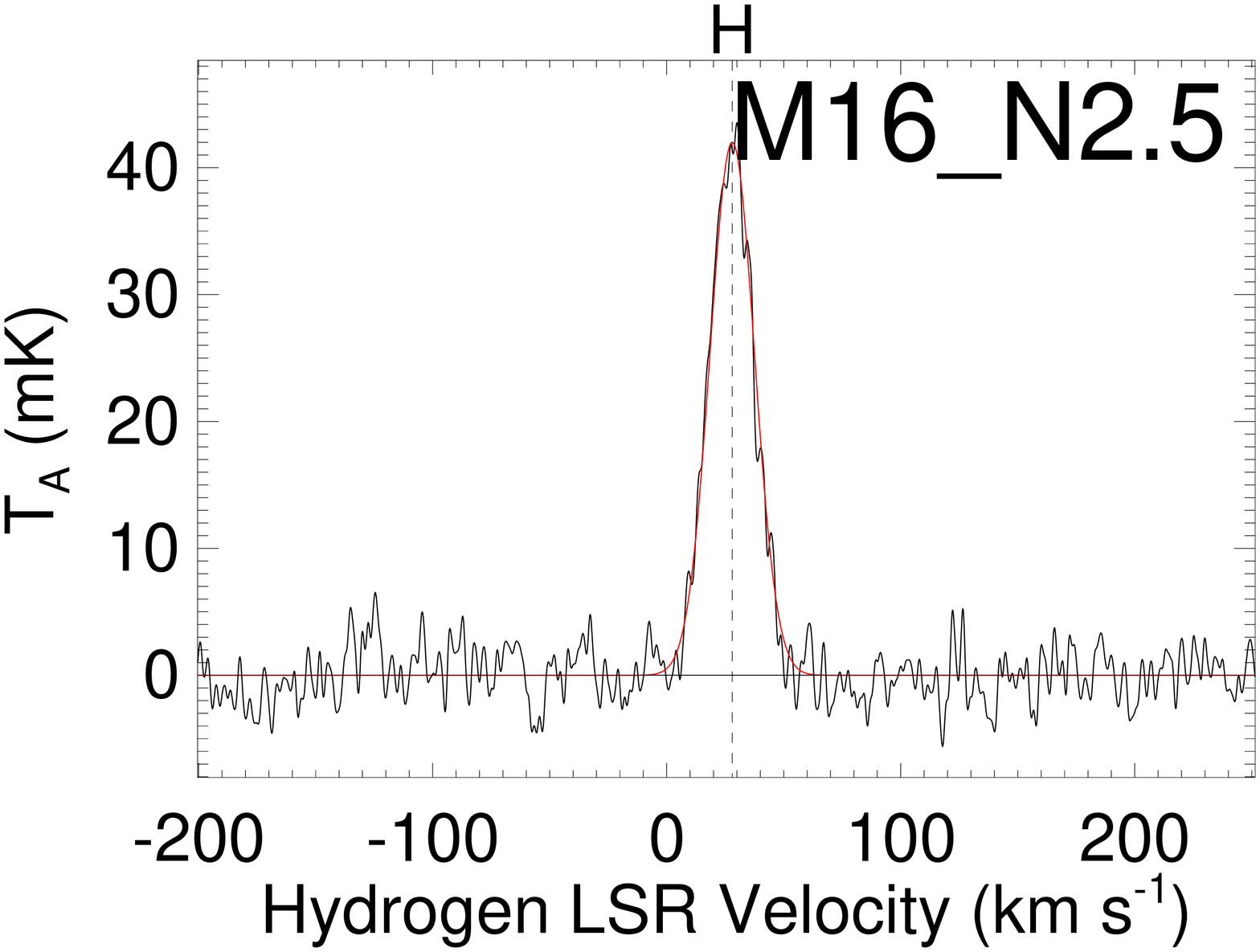} \\
\includegraphics[width=.23\textwidth]{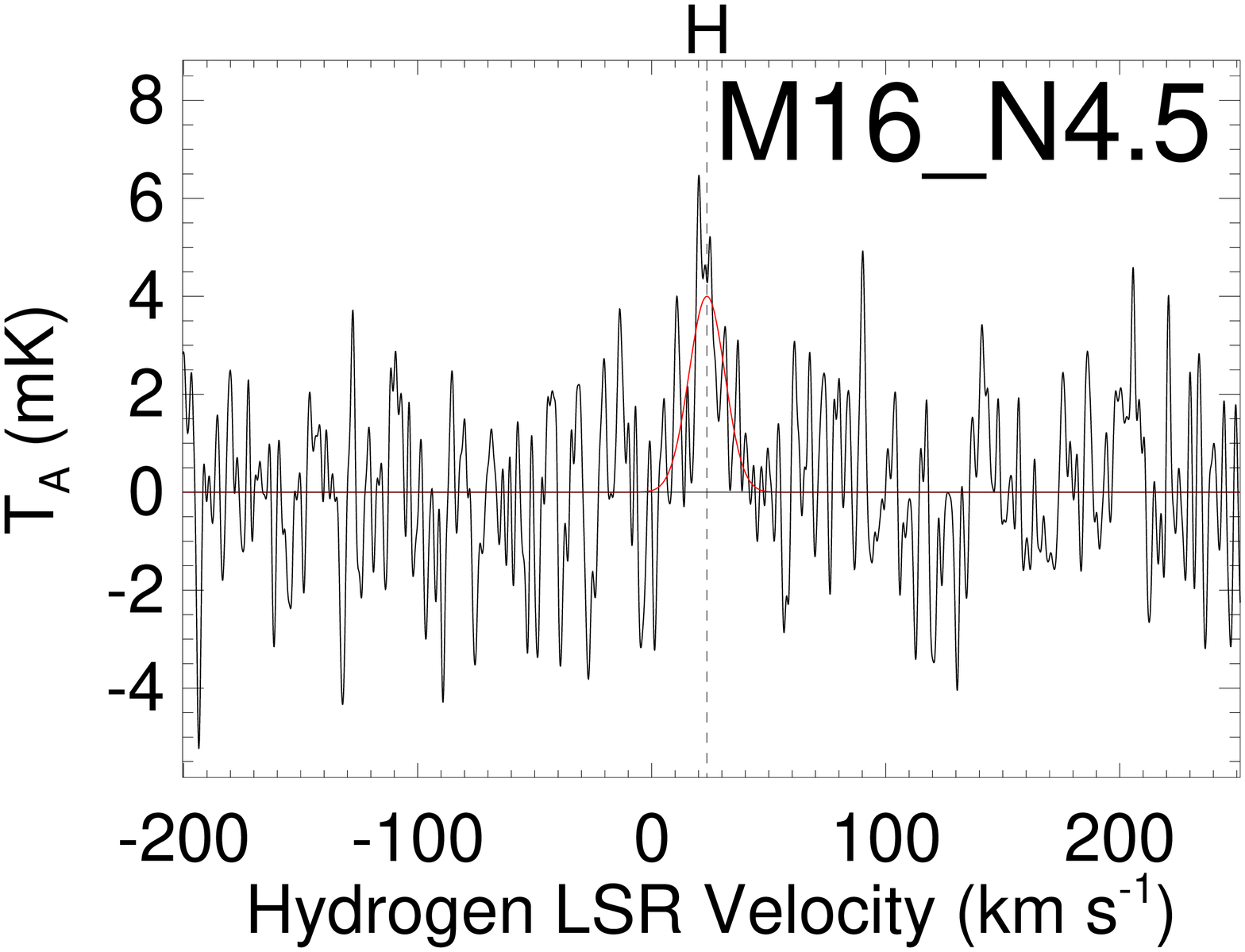} &
\includegraphics[width=.23\textwidth]{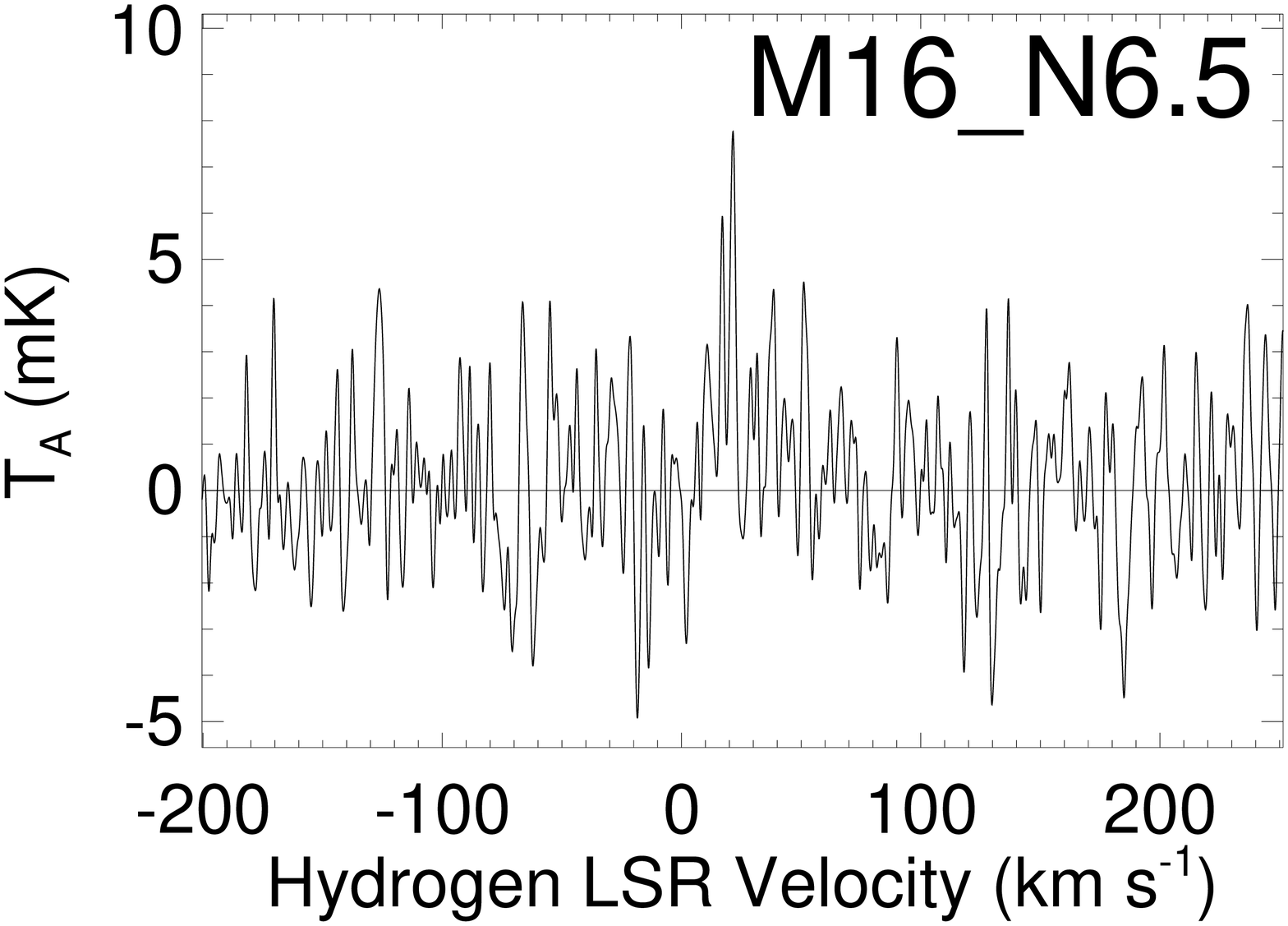} &
\includegraphics[width=.23\textwidth]{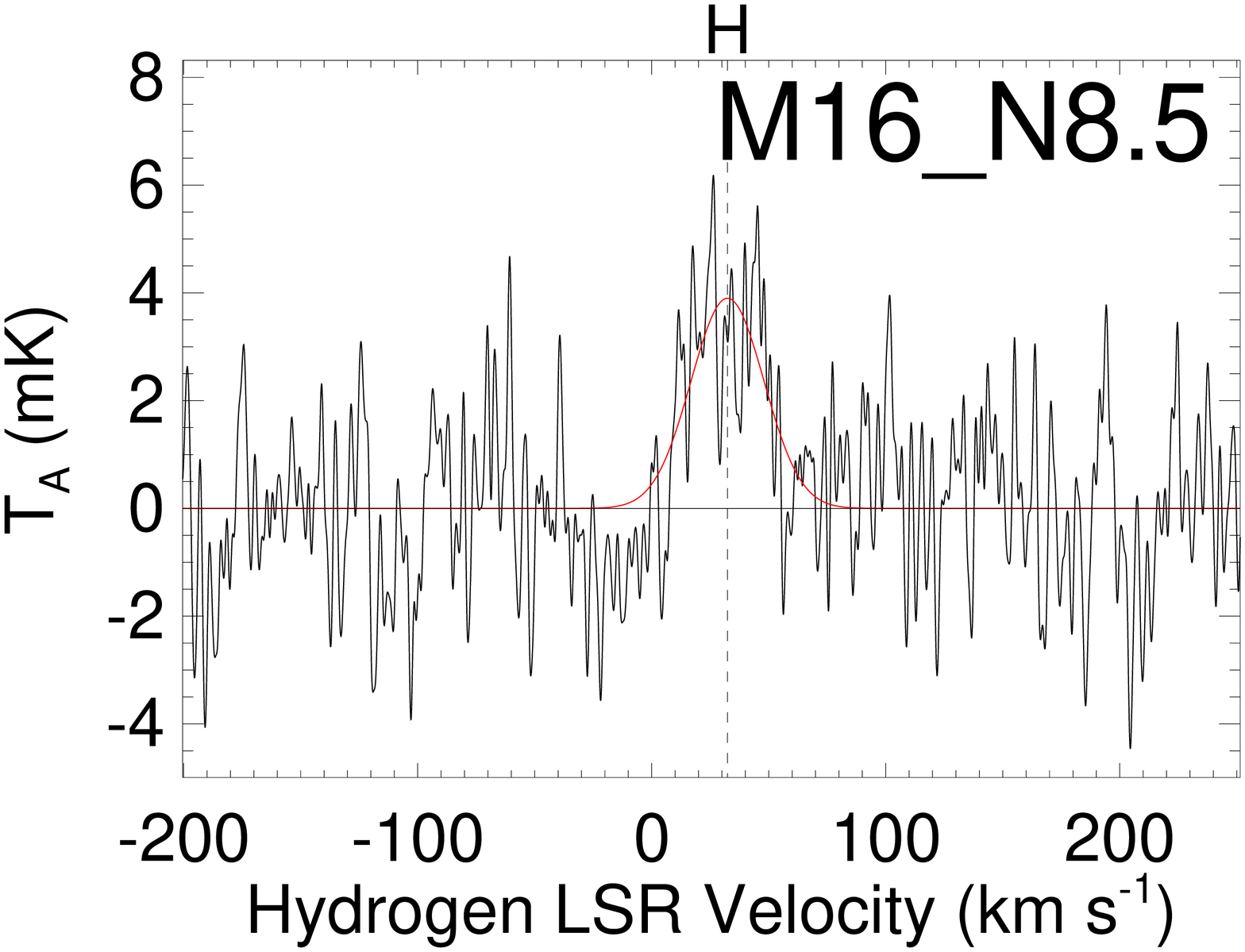} &
\includegraphics[width=.23\textwidth]{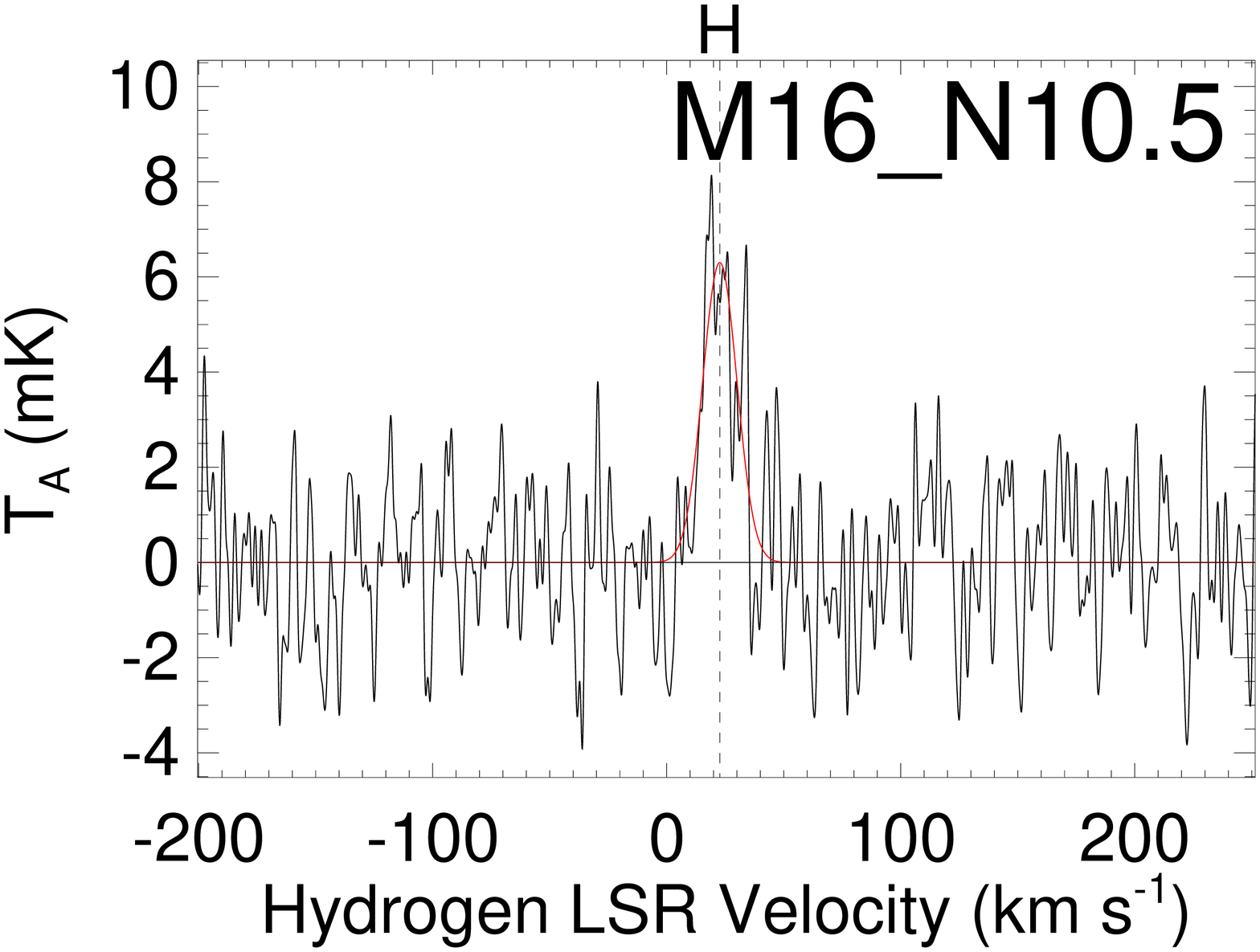} \\
\includegraphics[width=.23\textwidth]{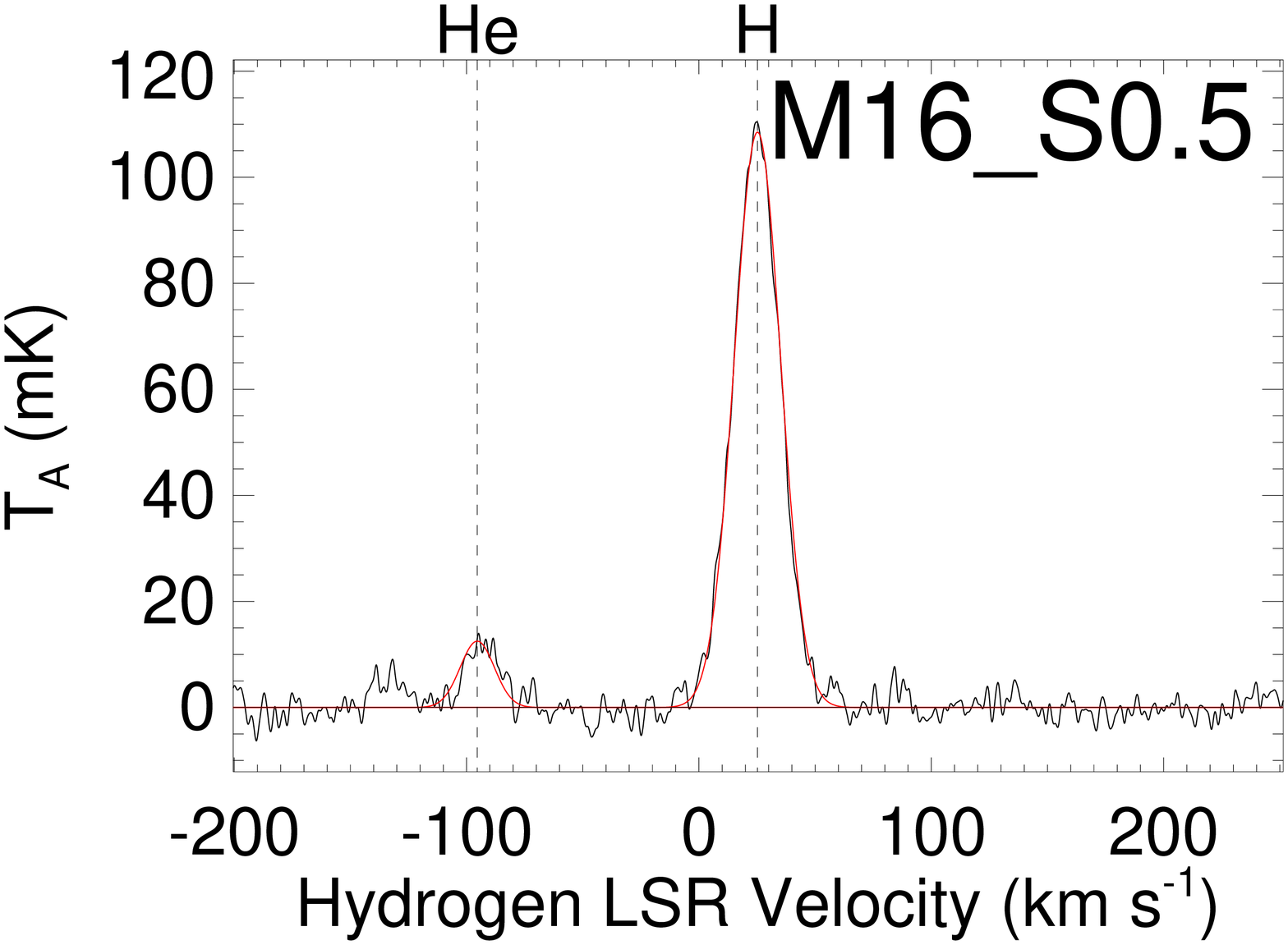} &
\includegraphics[width=.23\textwidth]{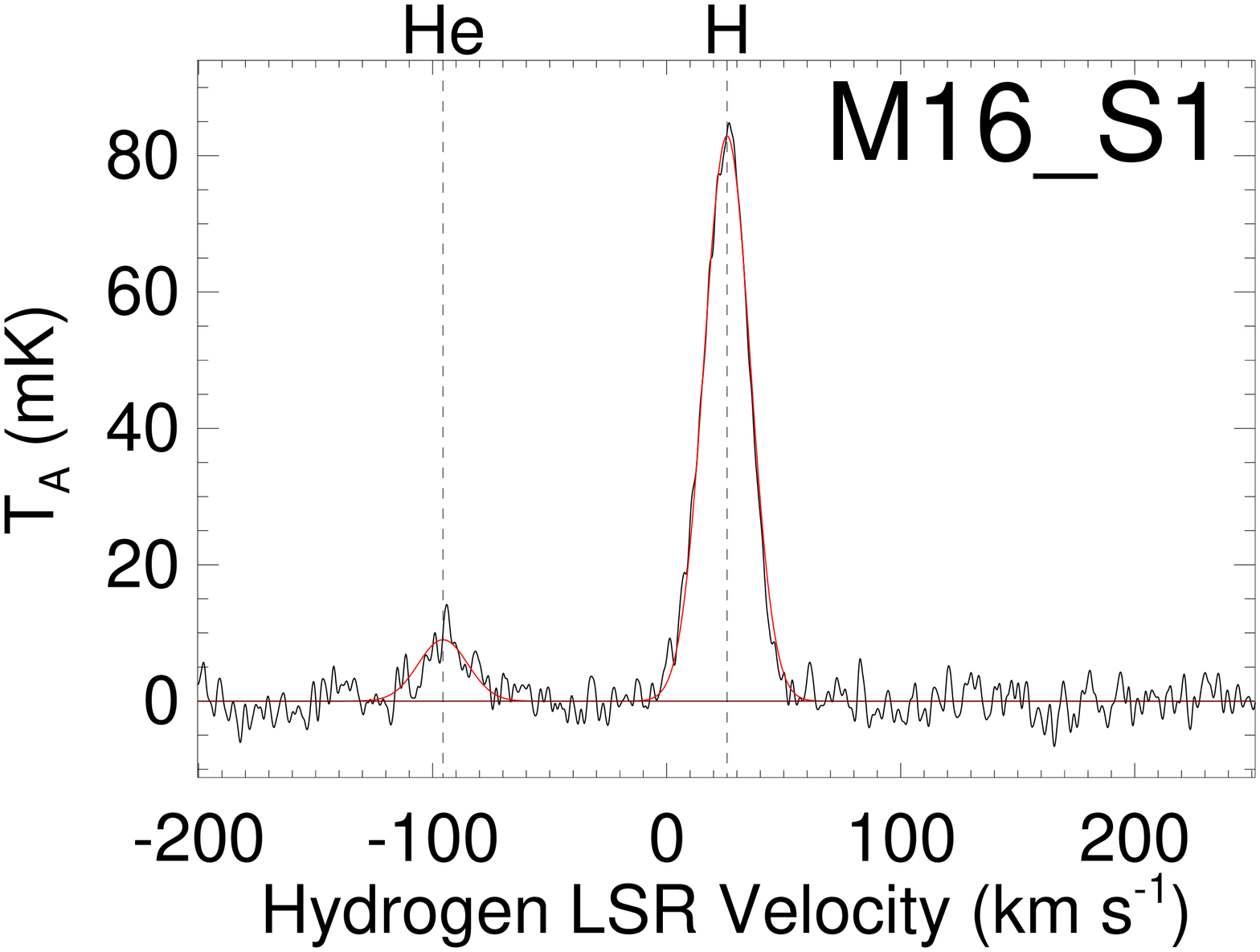} &
\includegraphics[width=.23\textwidth]{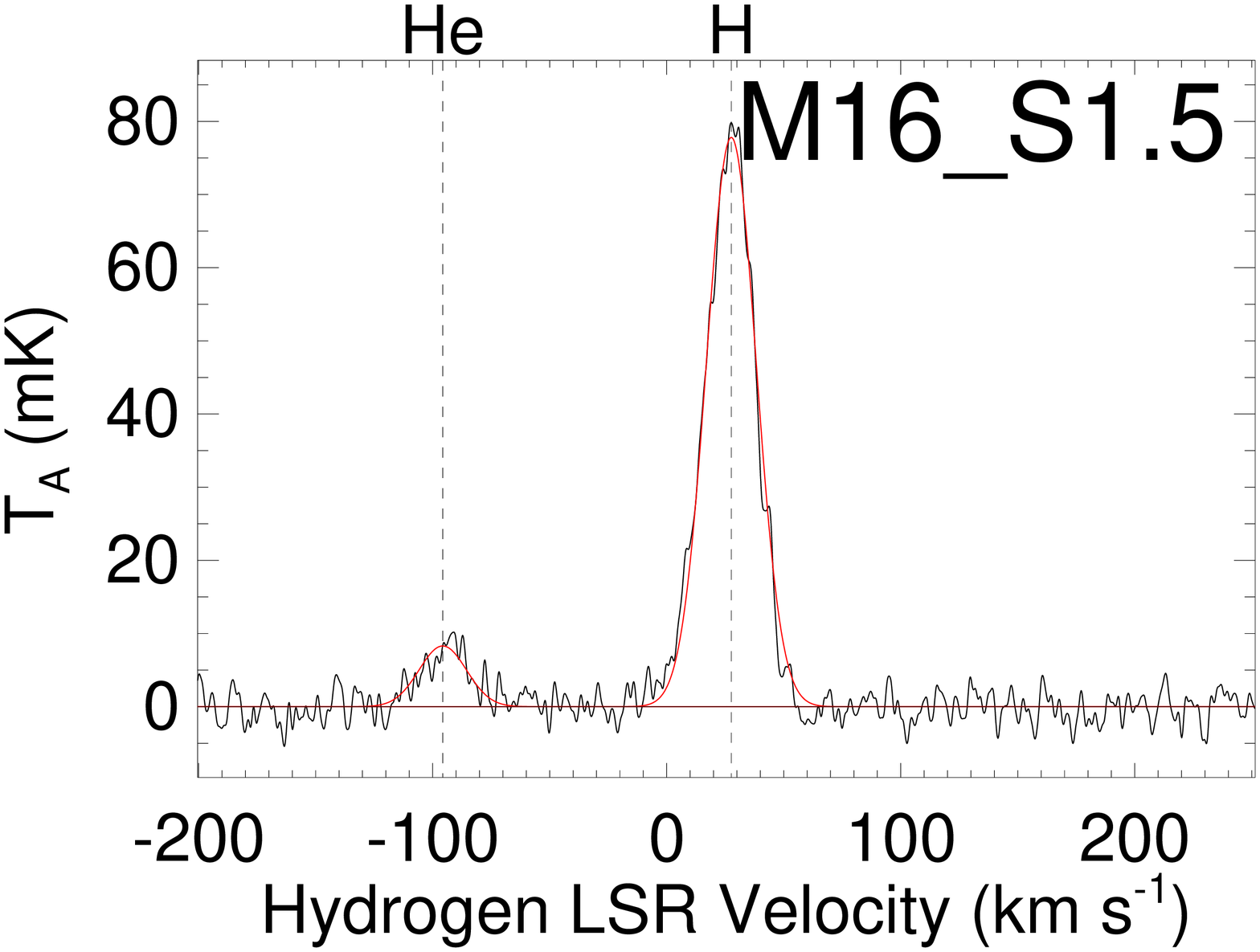} &
\includegraphics[width=.23\textwidth]{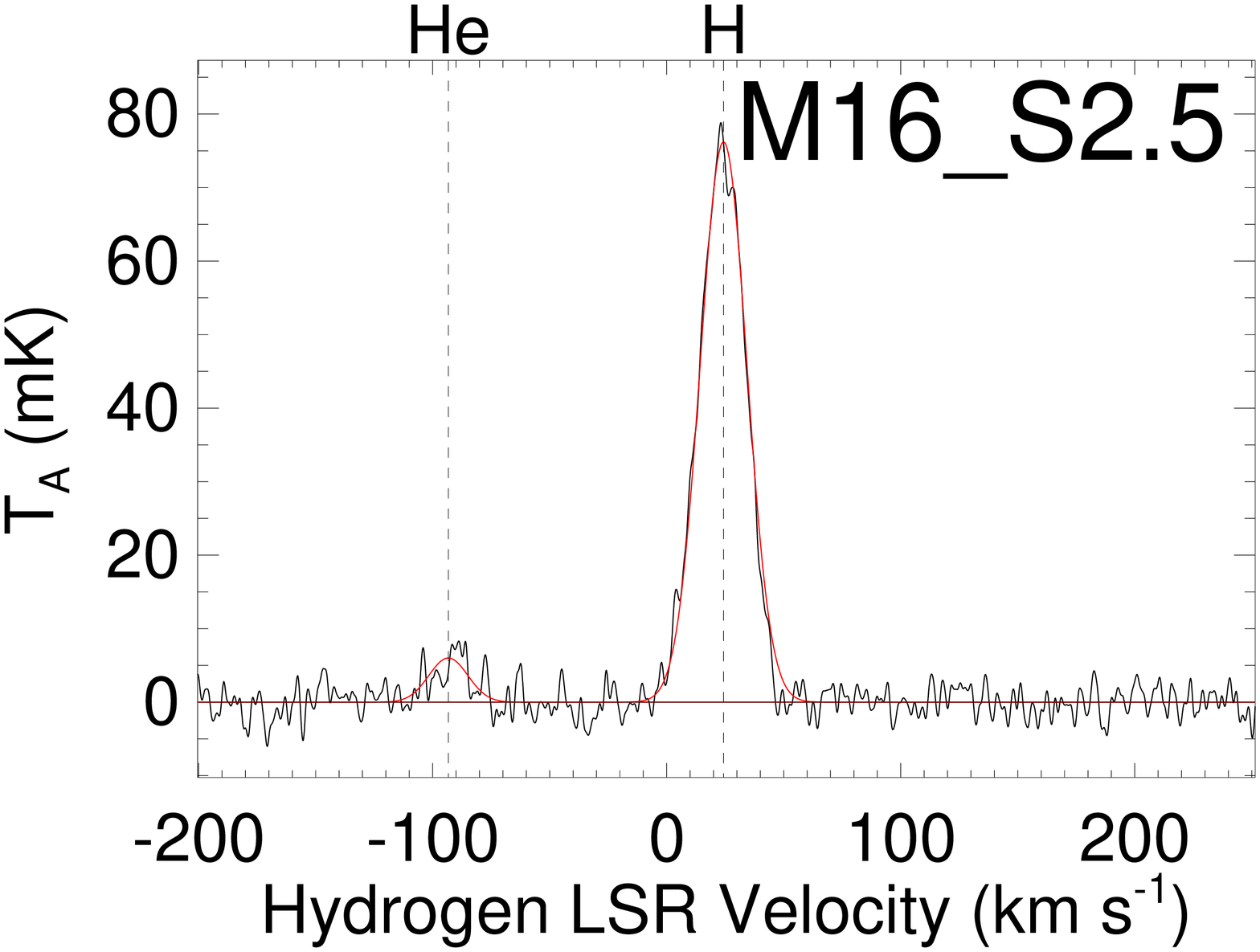} \\
\includegraphics[width=.23\textwidth]{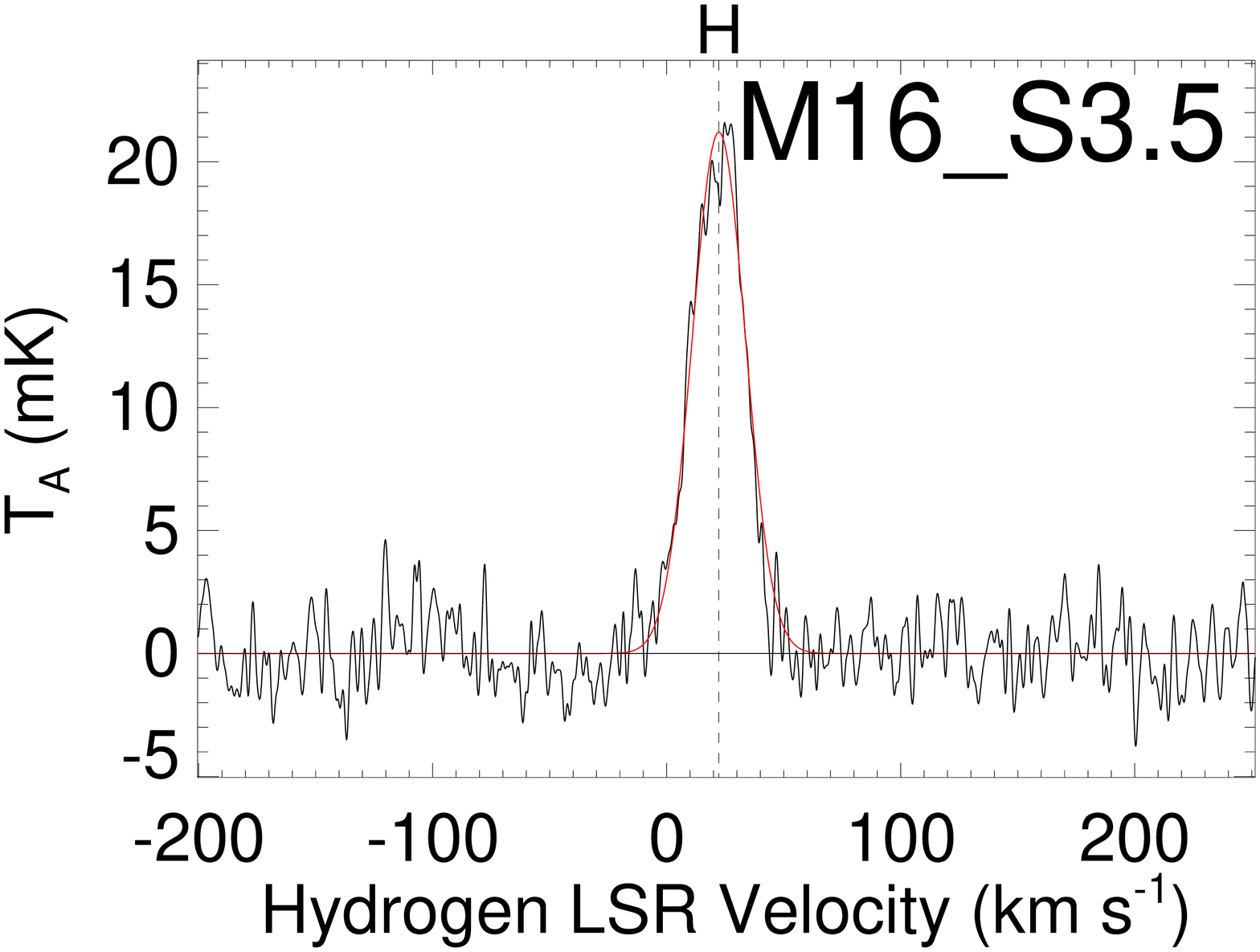} &
\includegraphics[width=.23\textwidth]{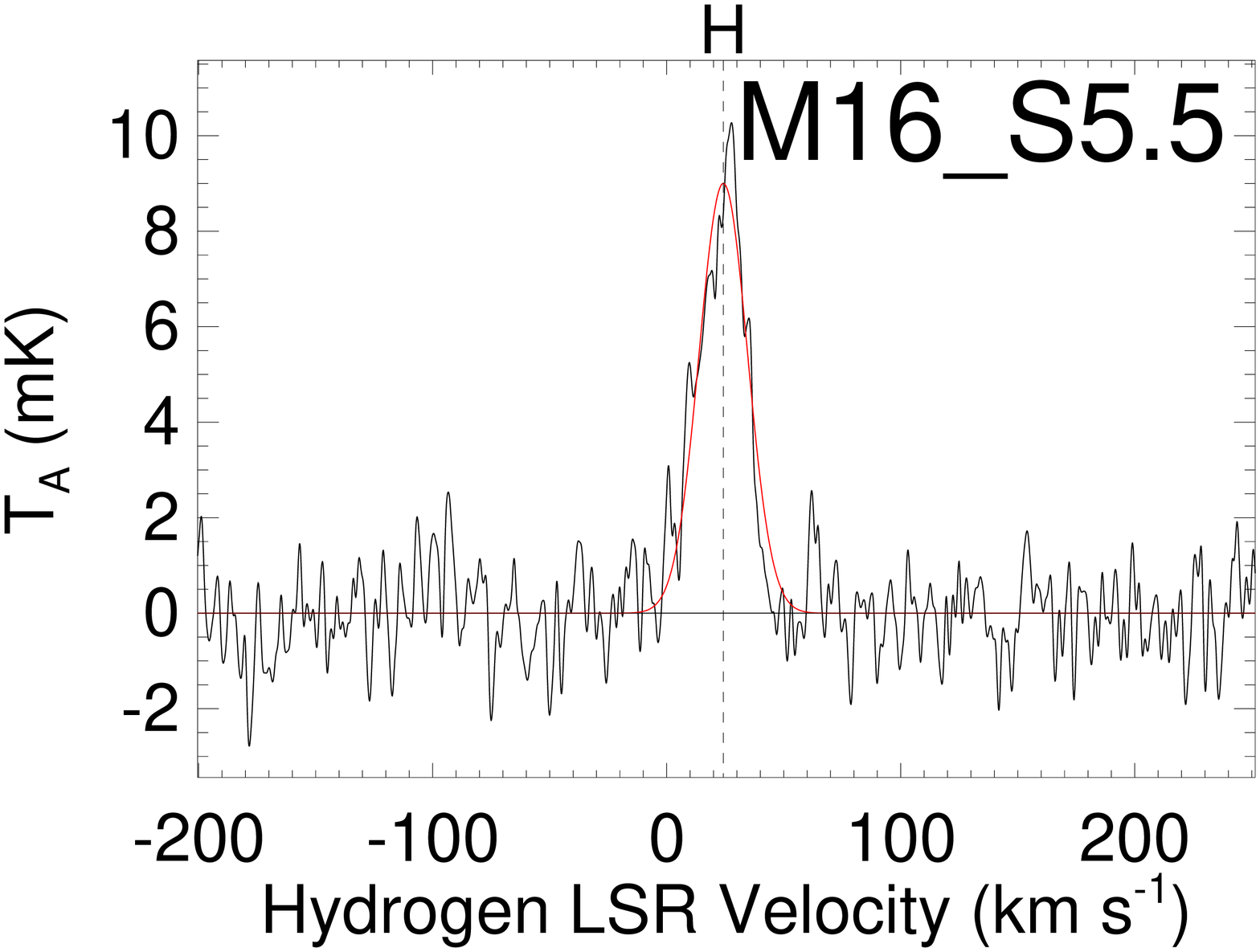} &
\includegraphics[width=.23\textwidth]{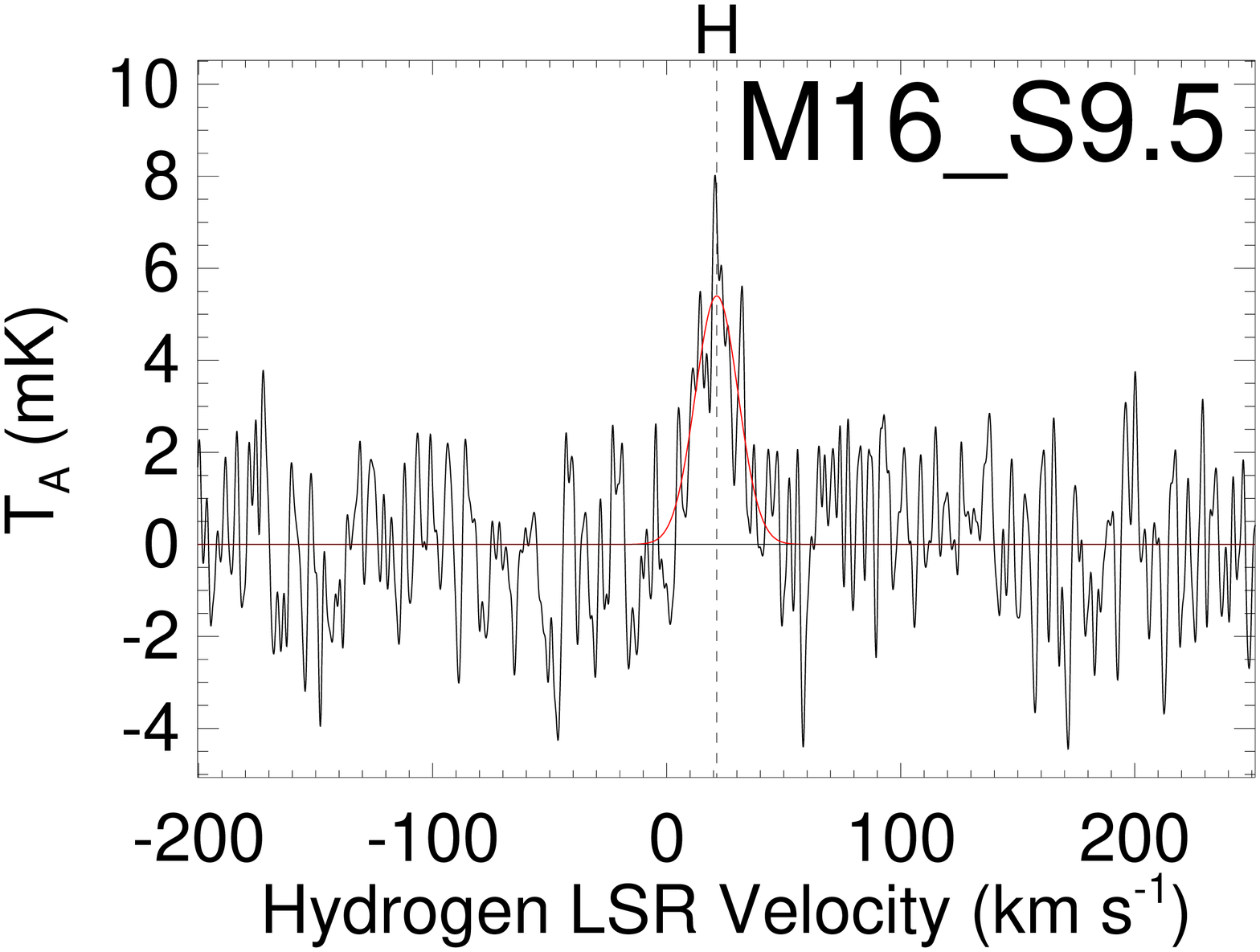} &
\includegraphics[width=.23\textwidth]{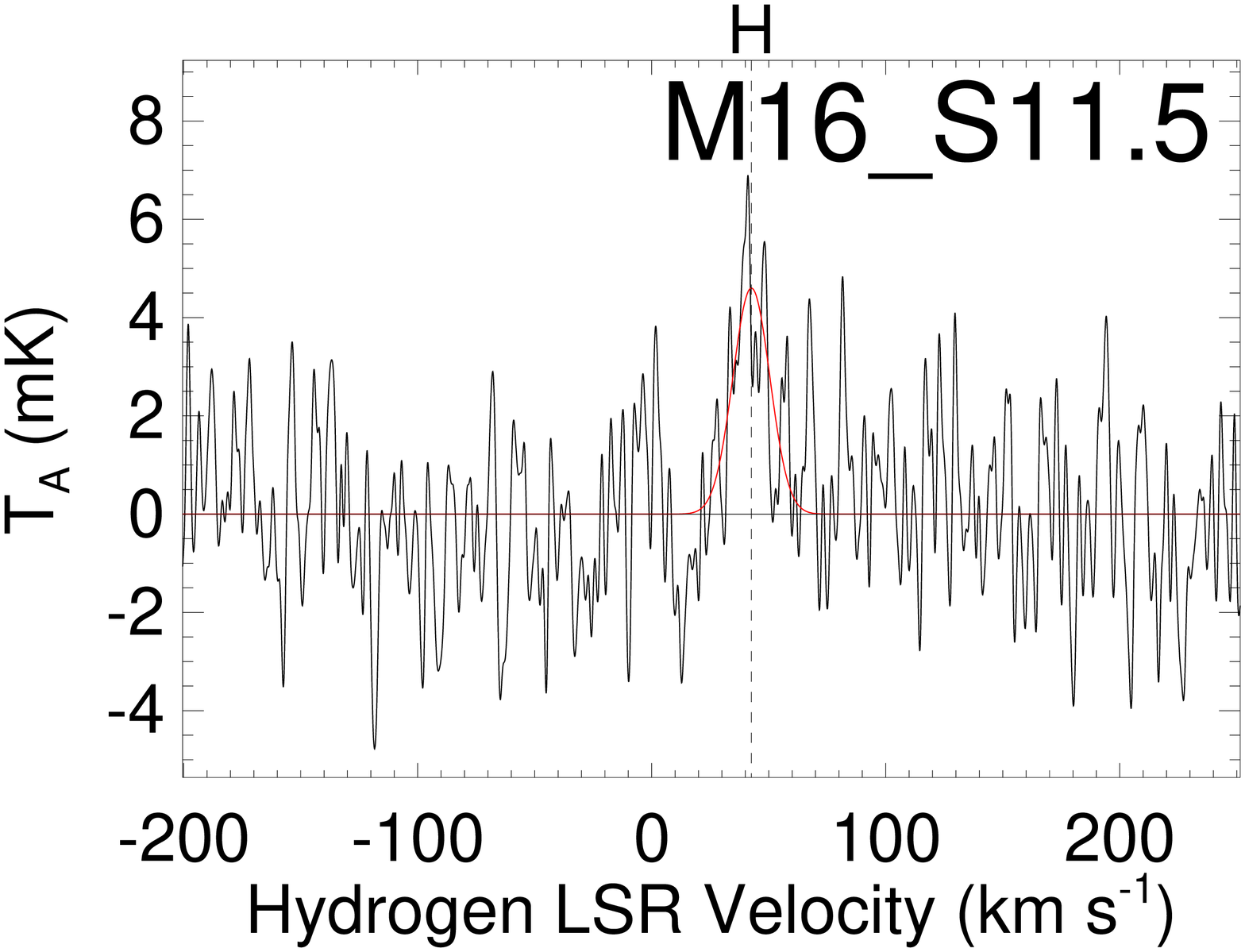} \\
\includegraphics[width=.23\textwidth]{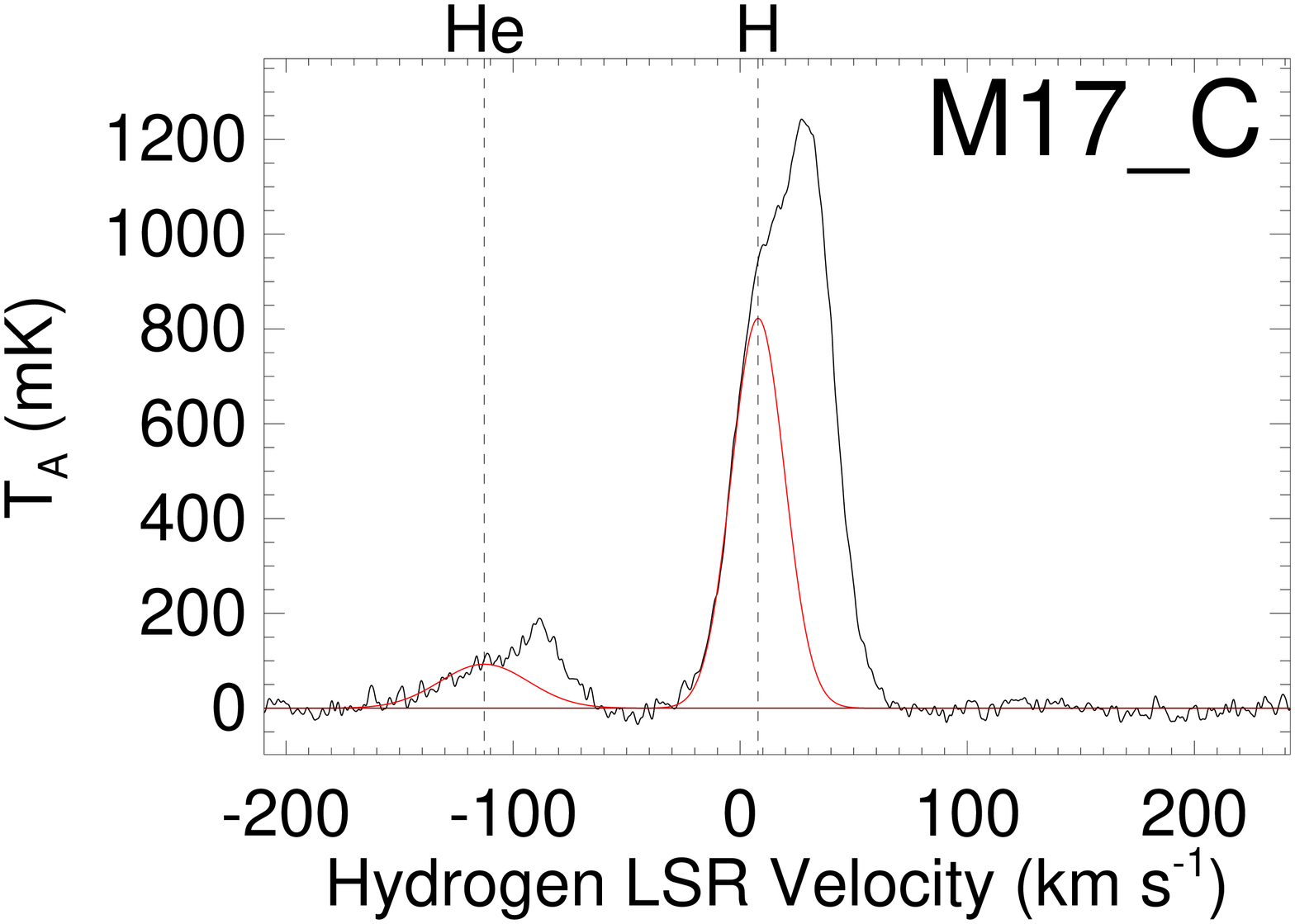} &
\includegraphics[width=.23\textwidth]{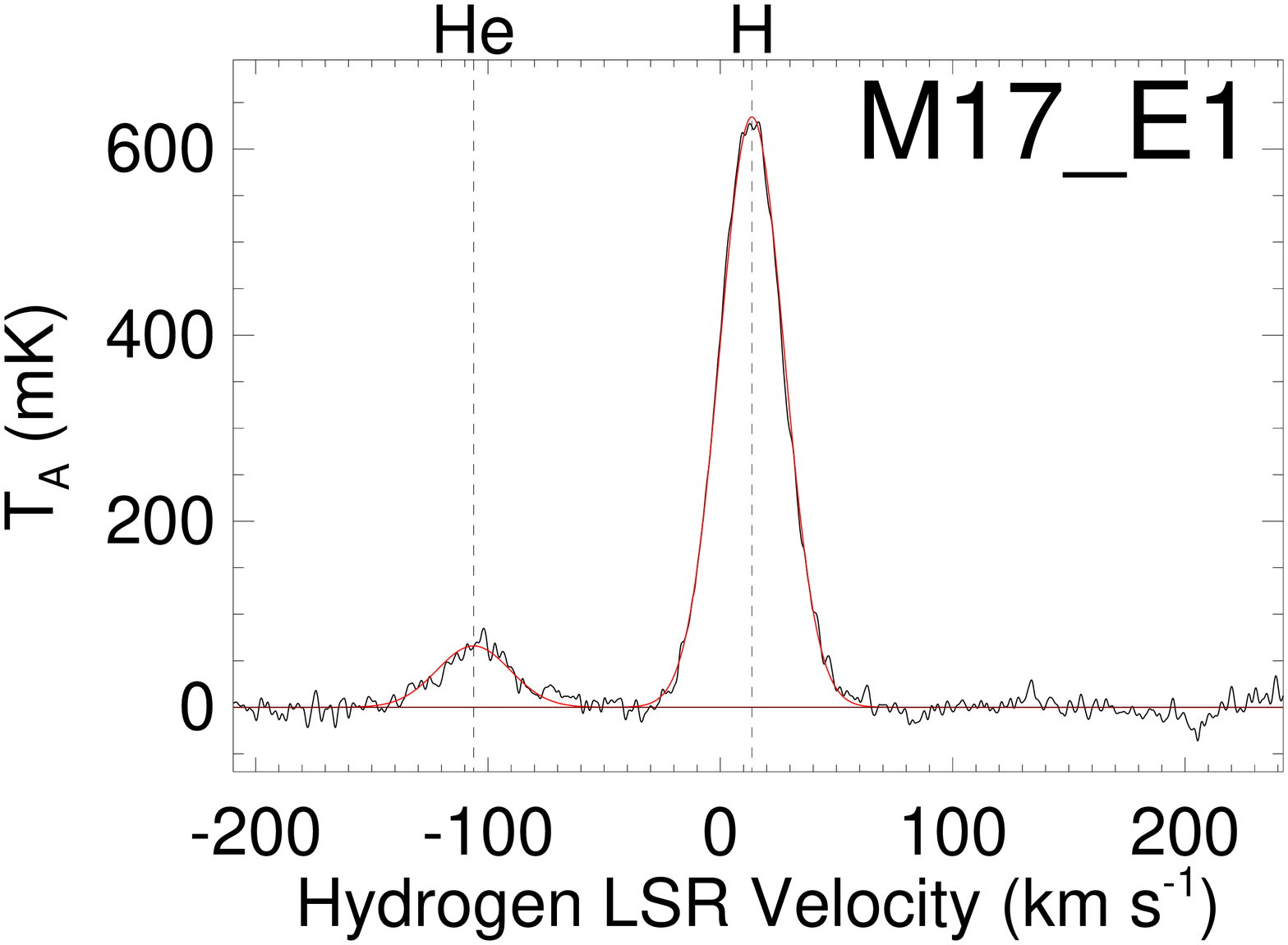} &
\includegraphics[width=.23\textwidth]{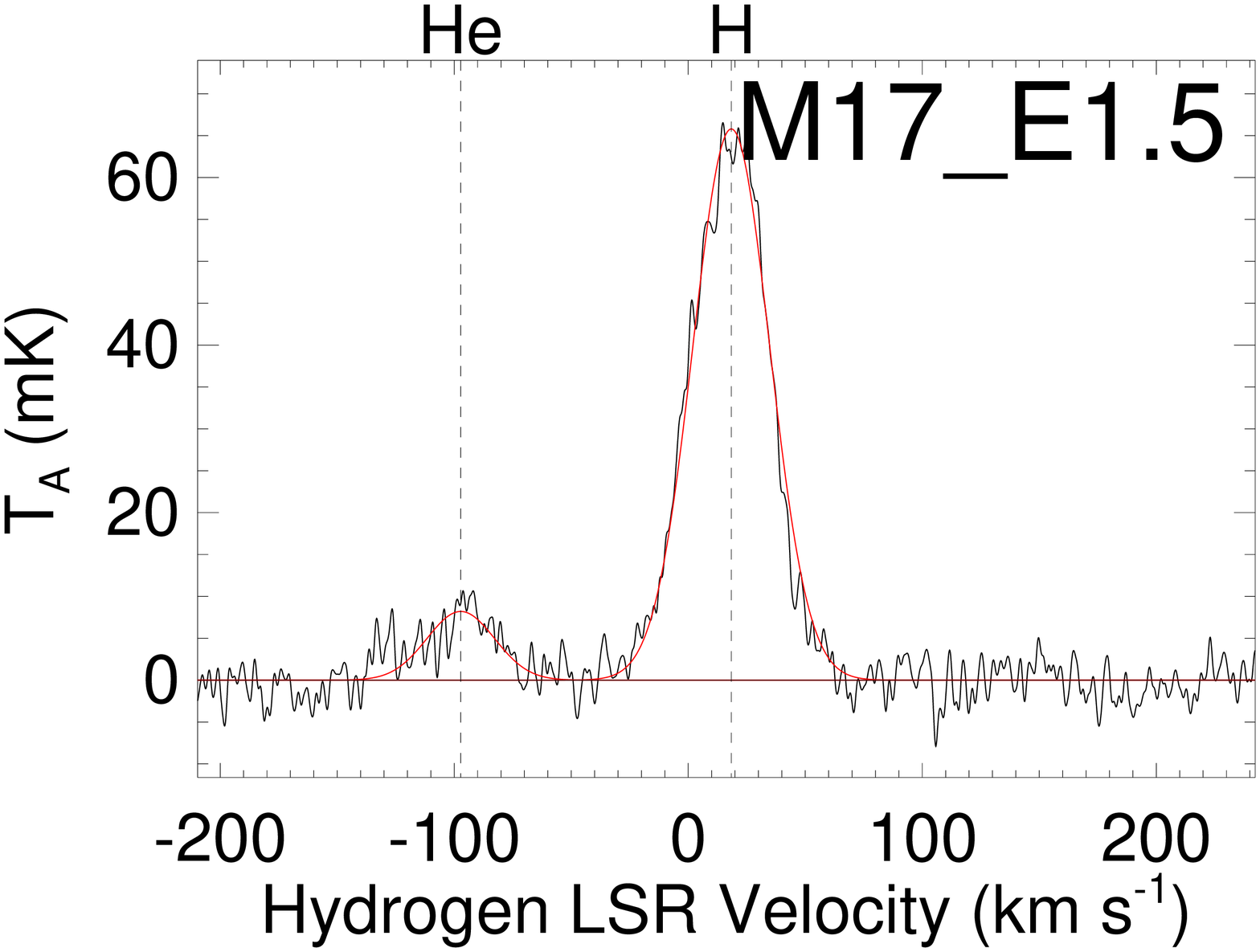} &
\includegraphics[width=.23\textwidth]{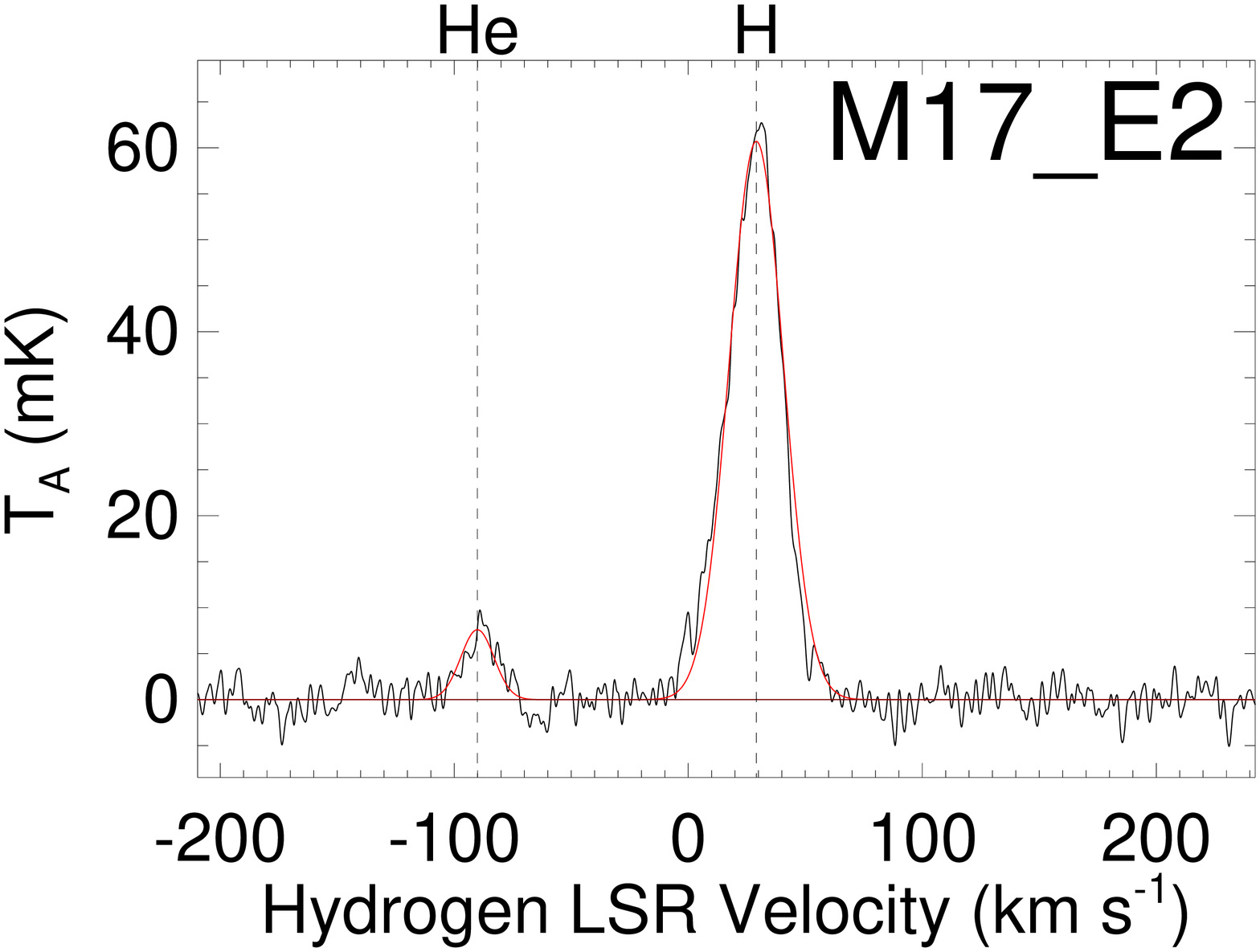} \\
\includegraphics[width=.23\textwidth]{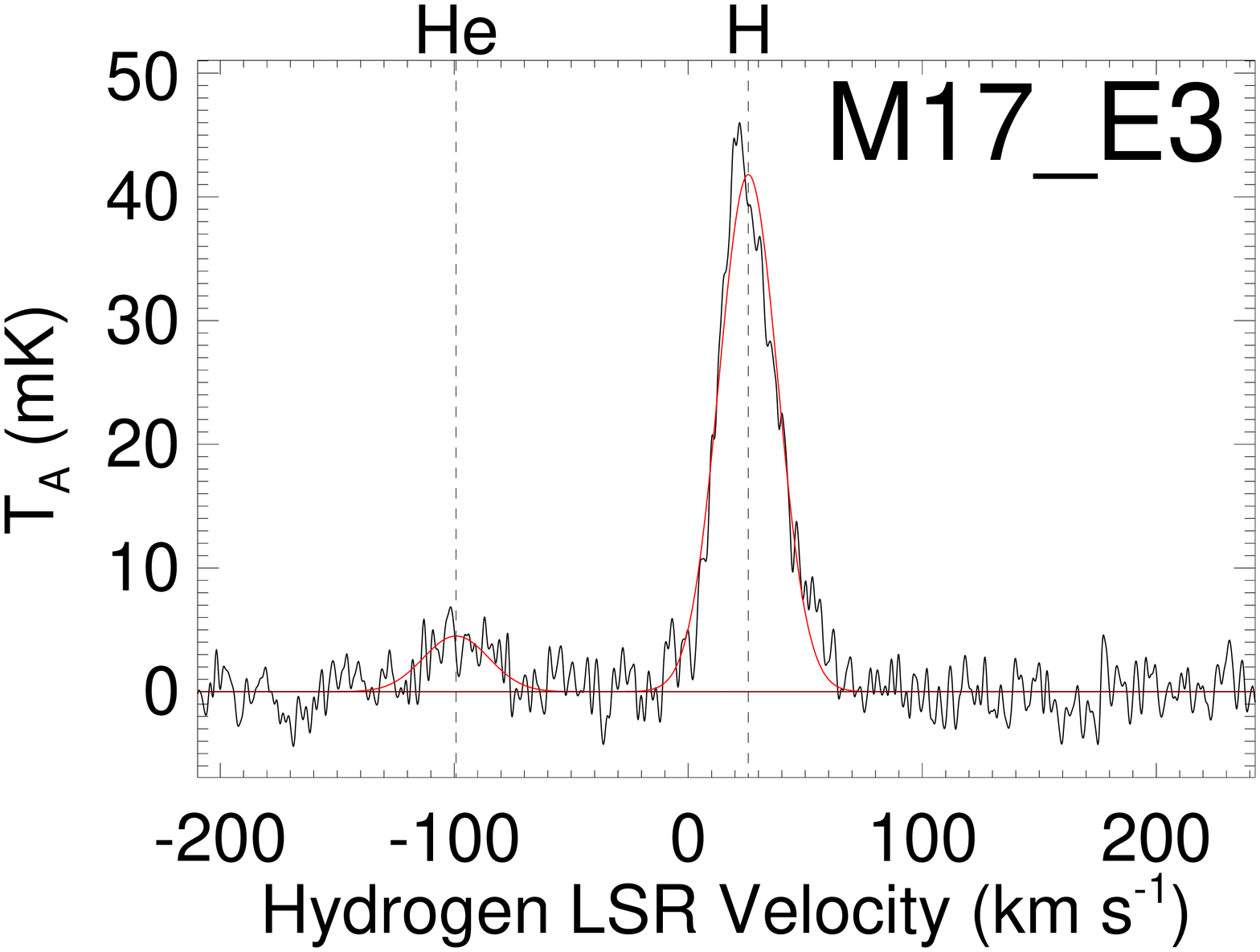} &
\includegraphics[width=.23\textwidth]{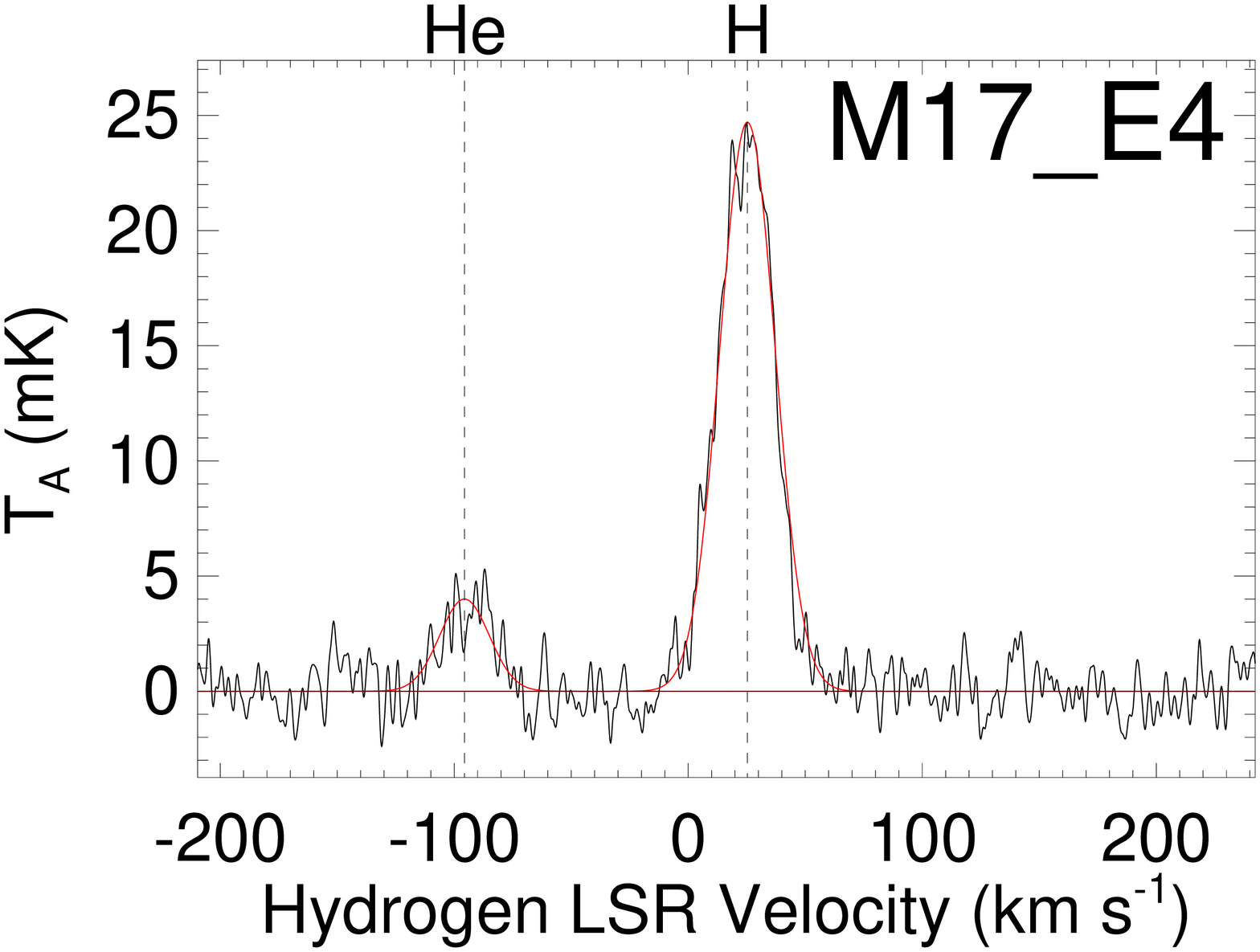} &
\includegraphics[width=.23\textwidth]{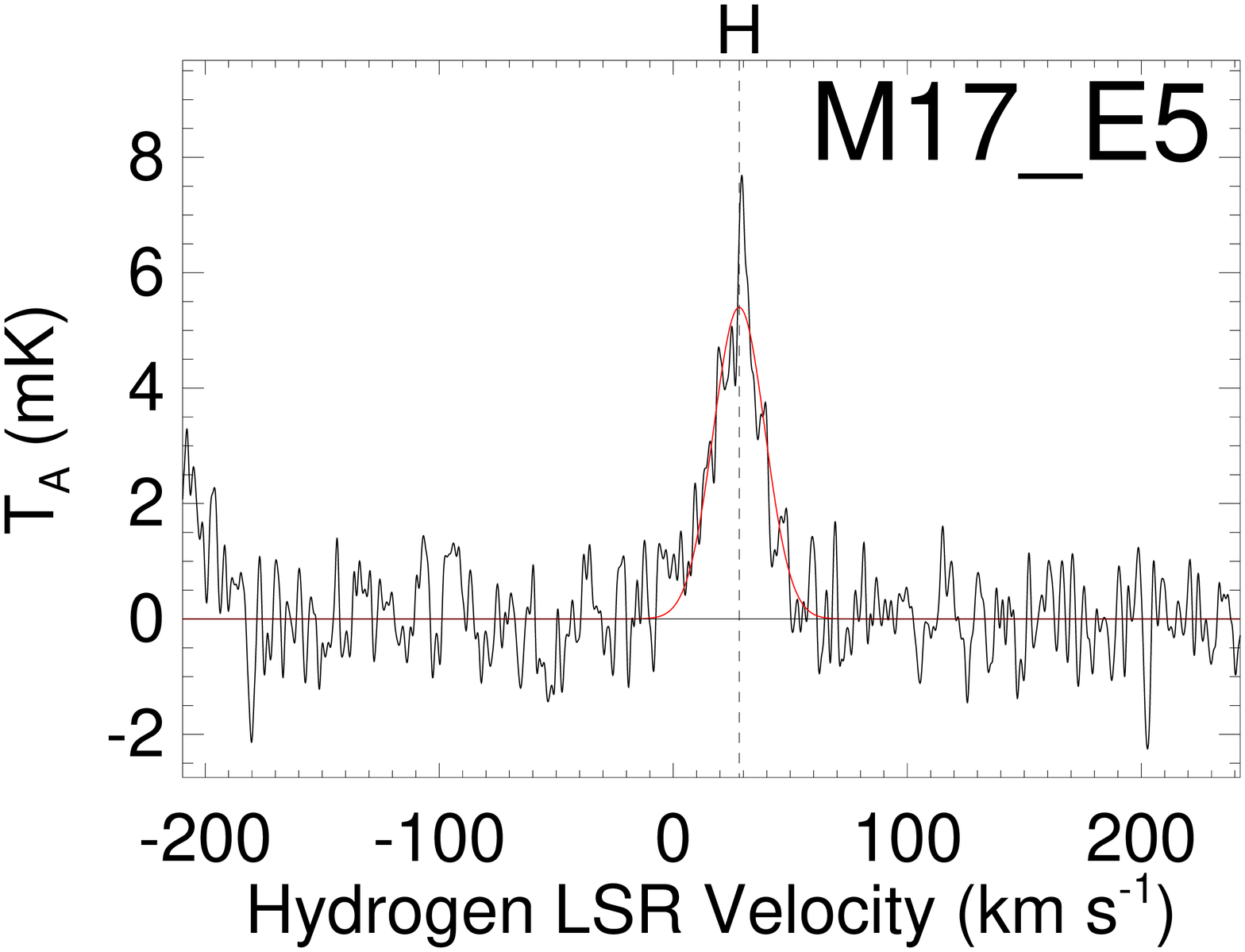} &
\includegraphics[width=.23\textwidth]{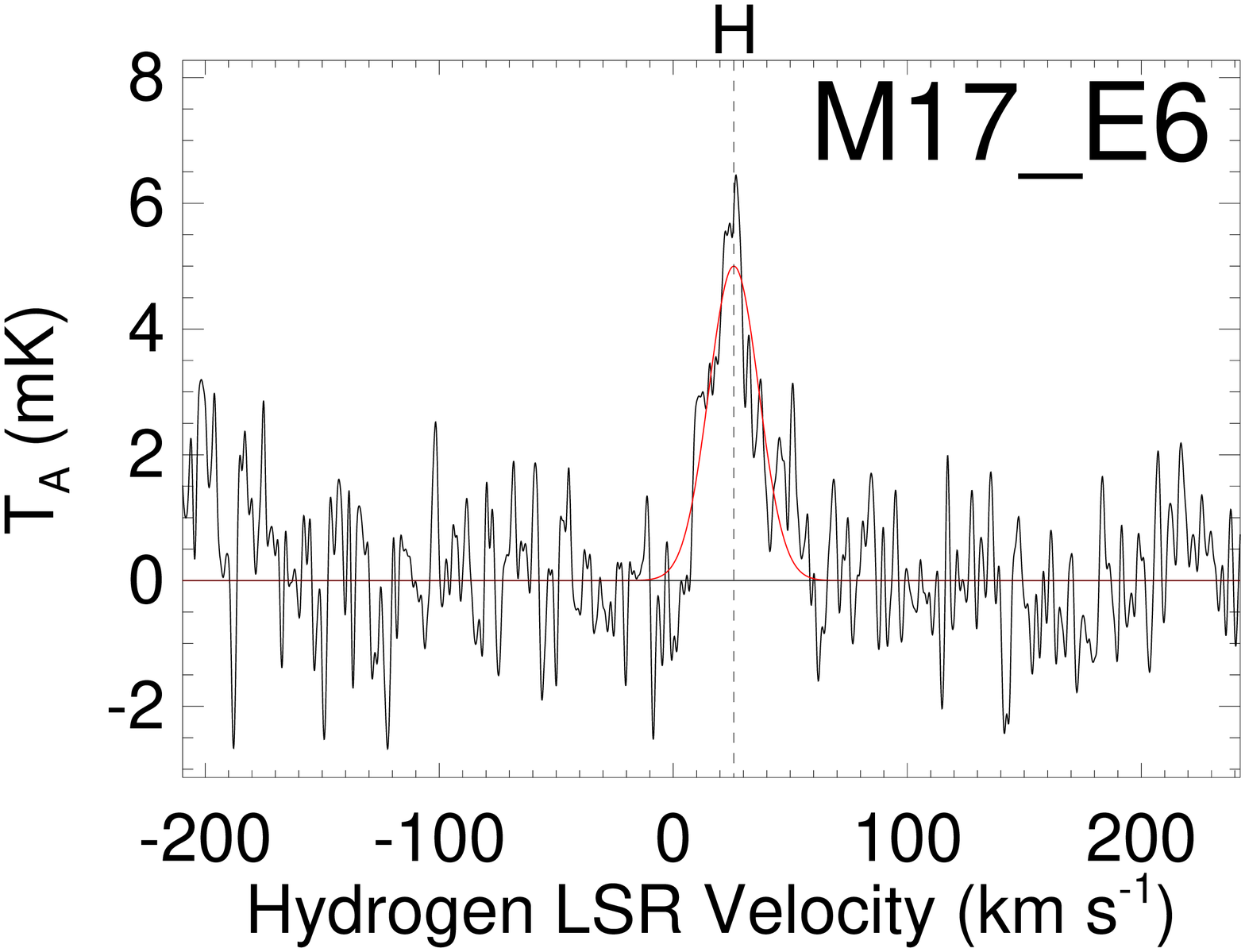} \\
\end{tabular}
\caption{}
\end{figure*}
\renewcommand{\thefigure}{\thesection.\arabic{figure}}

\renewcommand\thefigure{\thesection.\arabic{figure} (Cont.)}
\addtocounter{figure}{-1}
\begin{figure*}
\centering
\begin{tabular}{cccc}
\includegraphics[width=.23\textwidth]{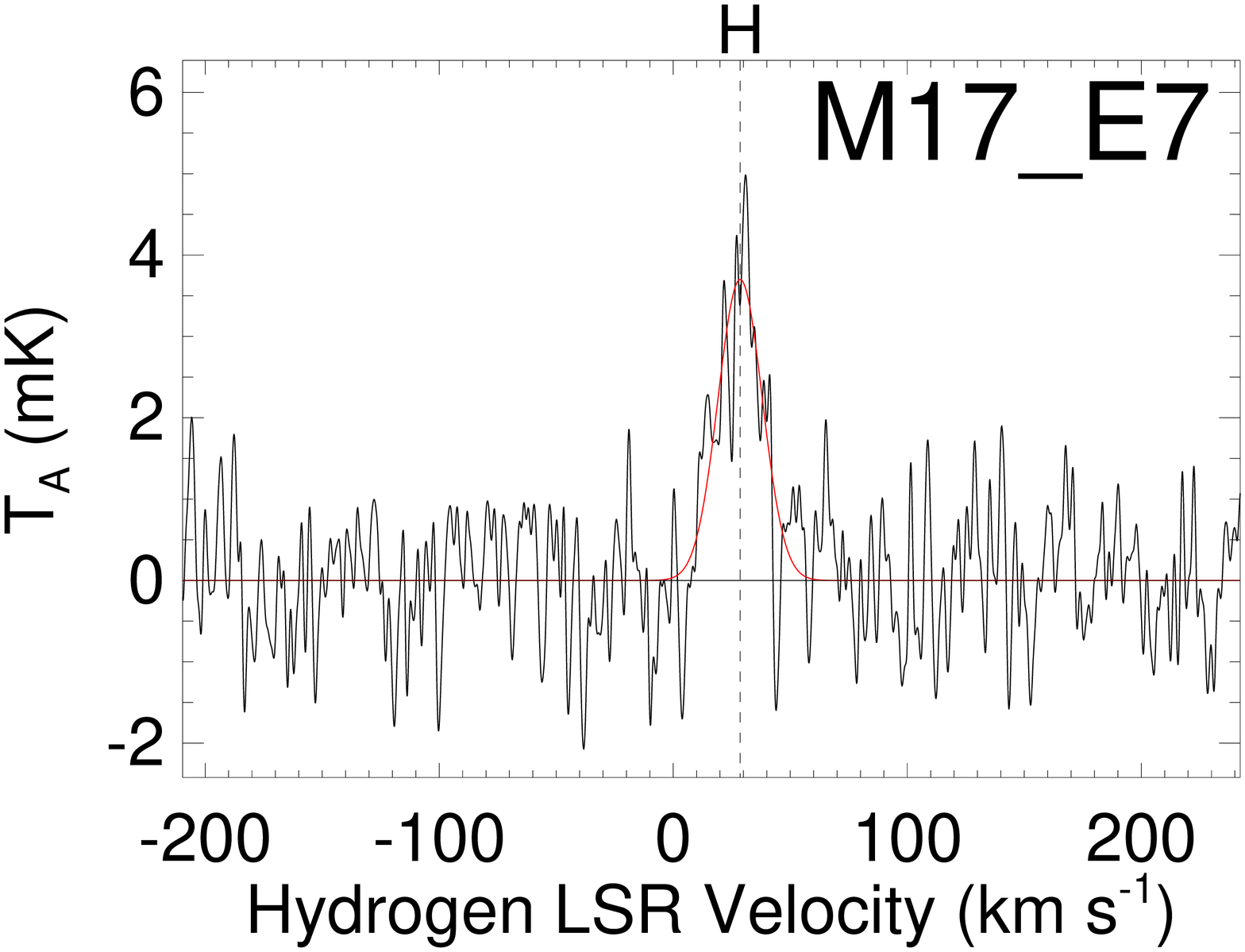} &
\includegraphics[width=.23\textwidth]{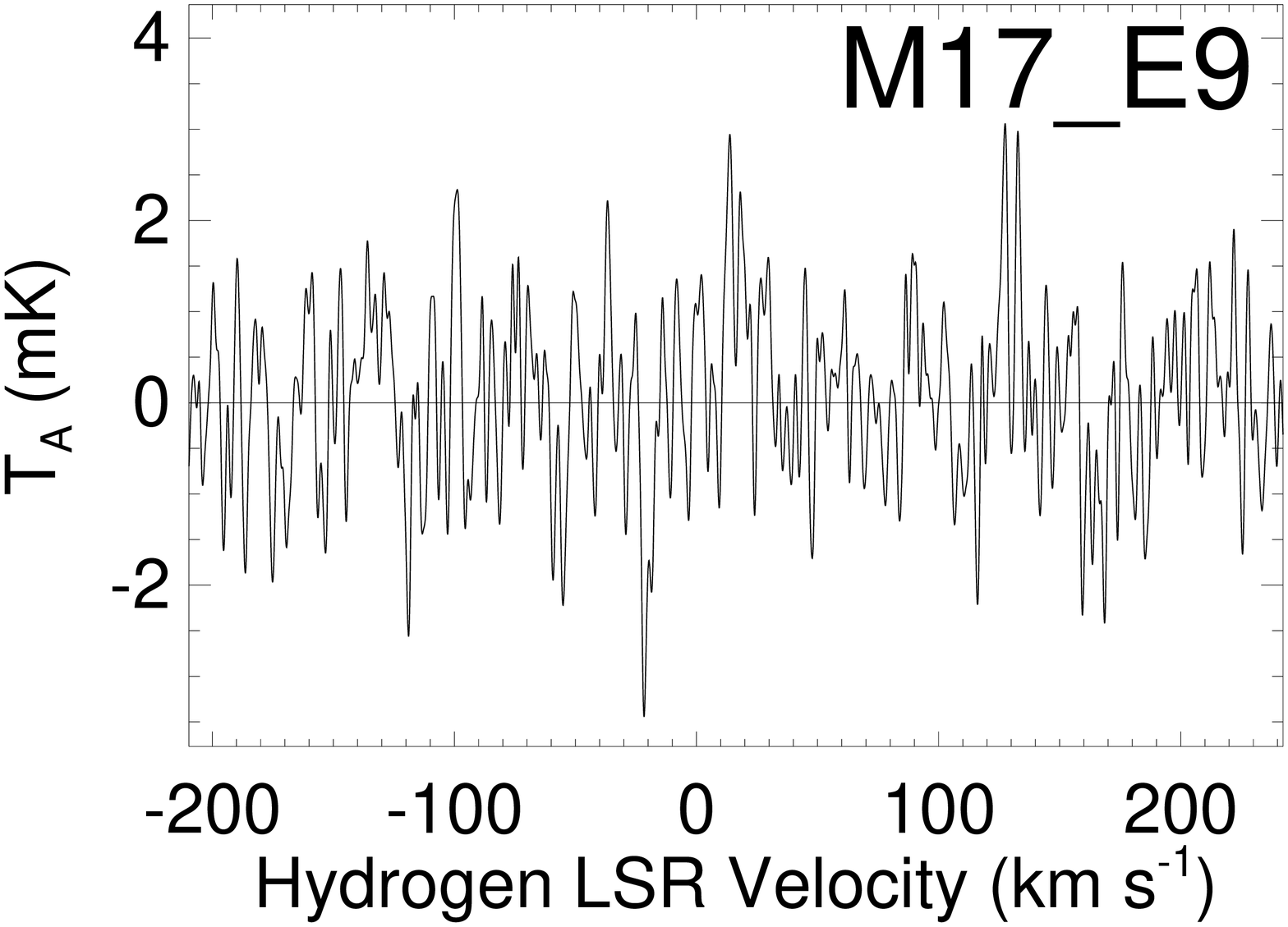} &
\includegraphics[width=.23\textwidth]{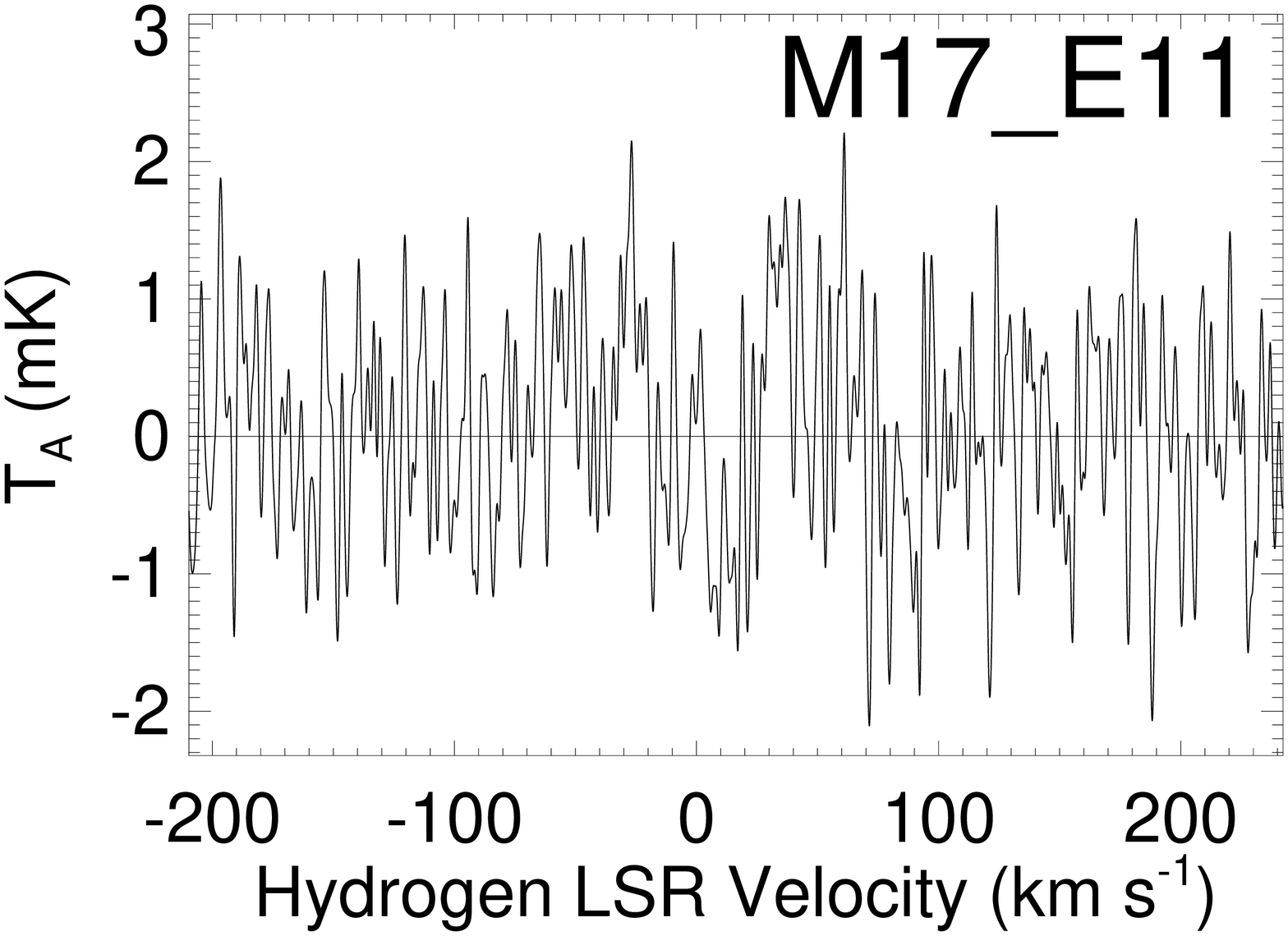} &
\includegraphics[width=.23\textwidth]{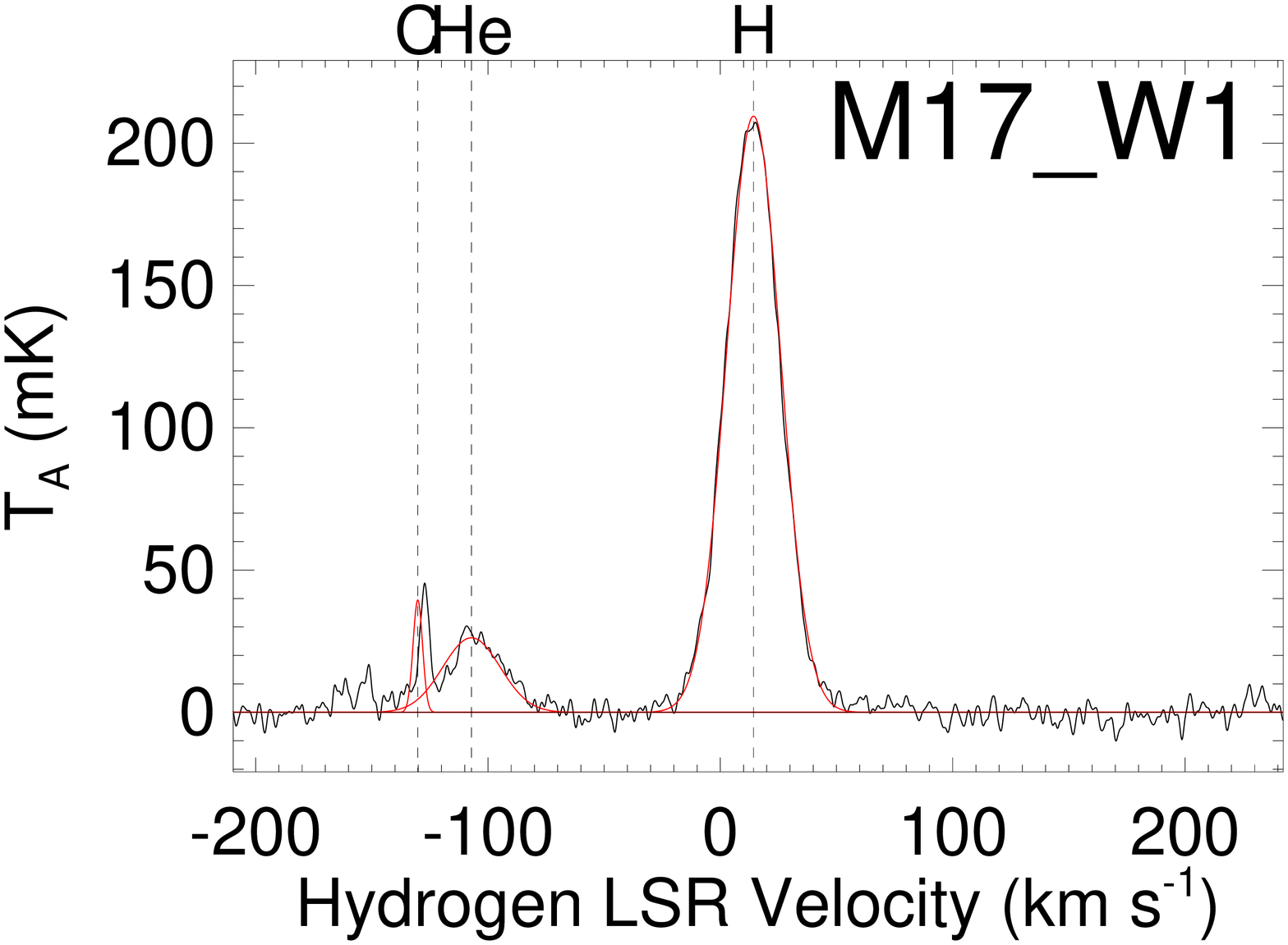} \\
\includegraphics[width=.23\textwidth]{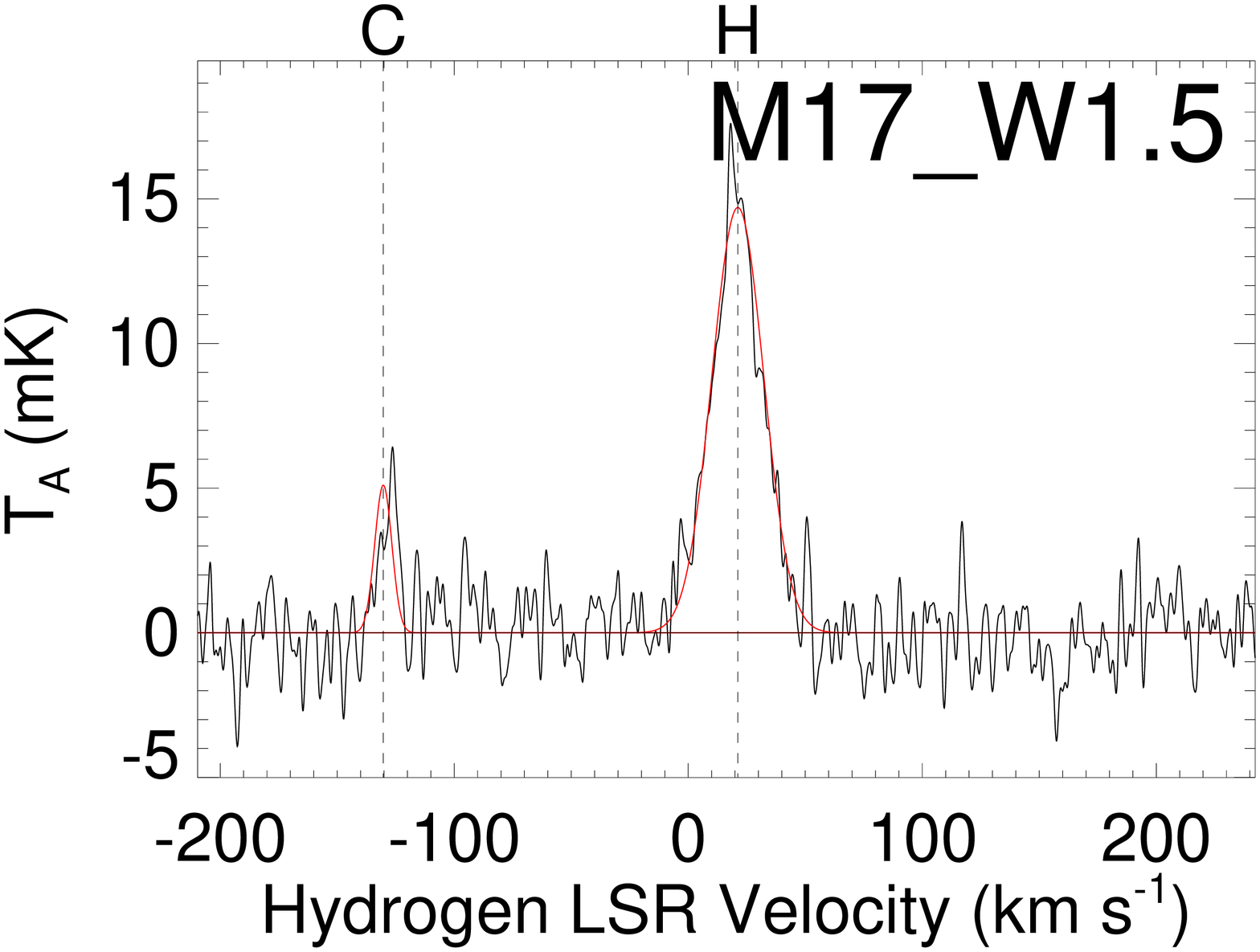} &
\includegraphics[width=.23\textwidth]{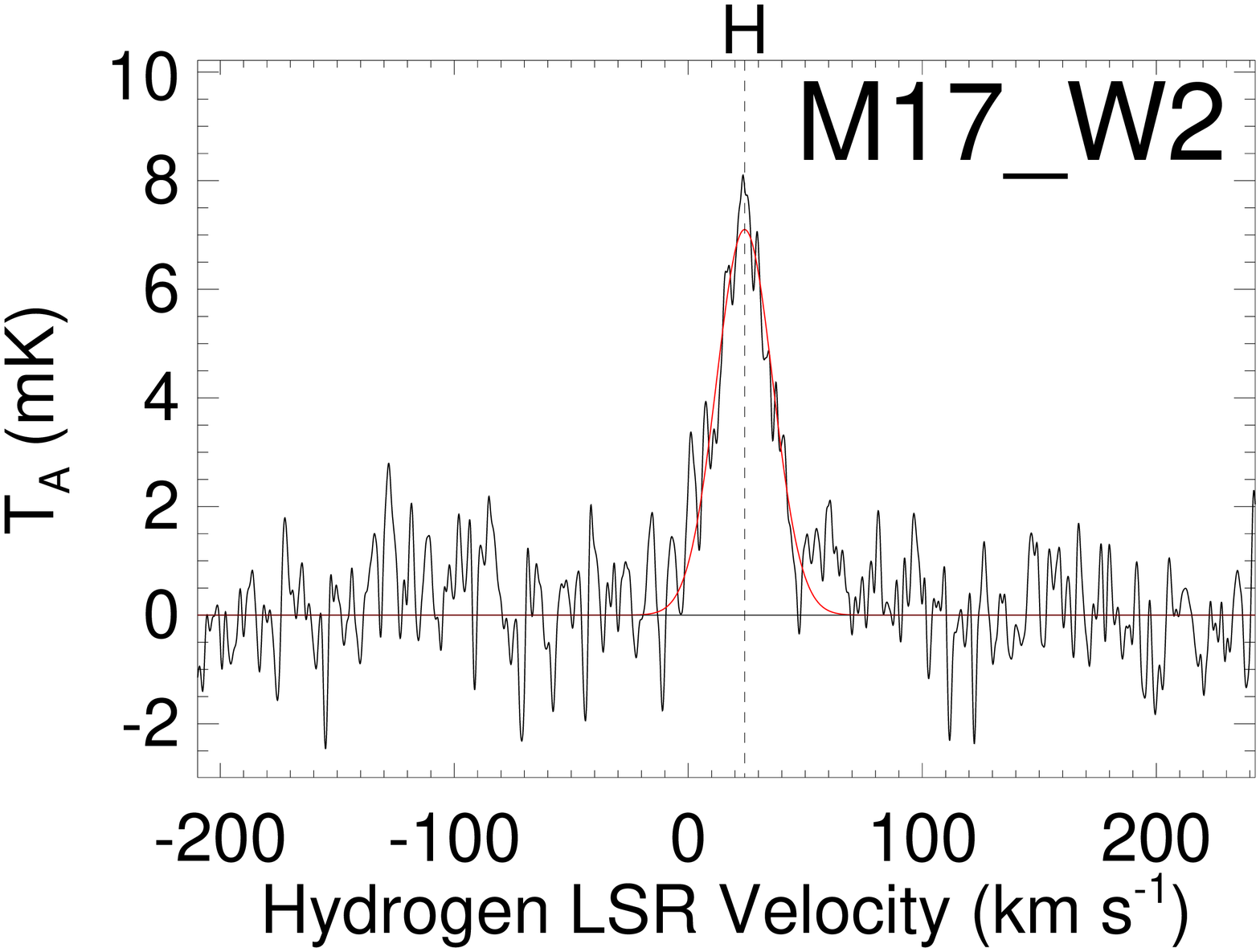} &
\includegraphics[width=.23\textwidth]{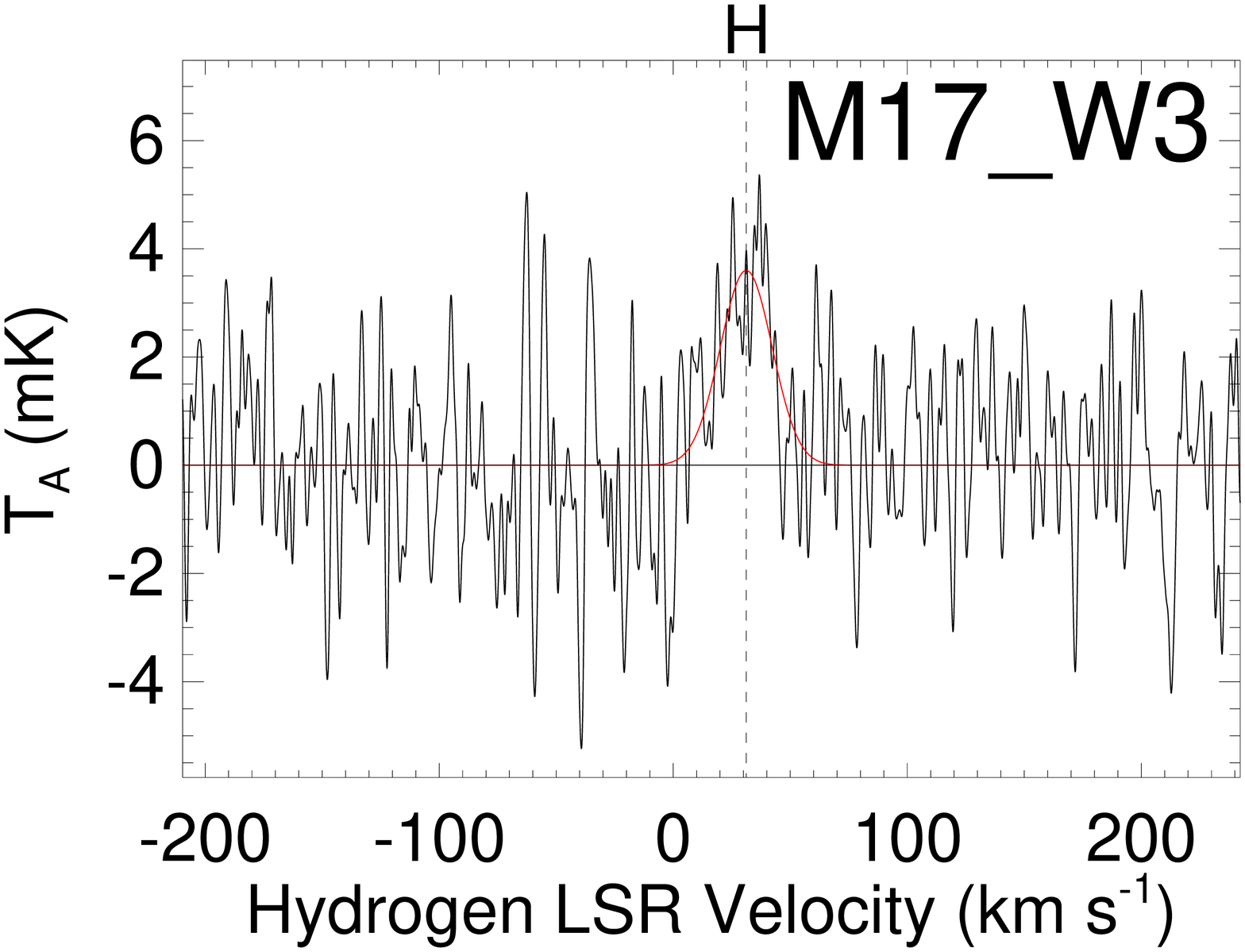} &
\includegraphics[width=.23\textwidth]{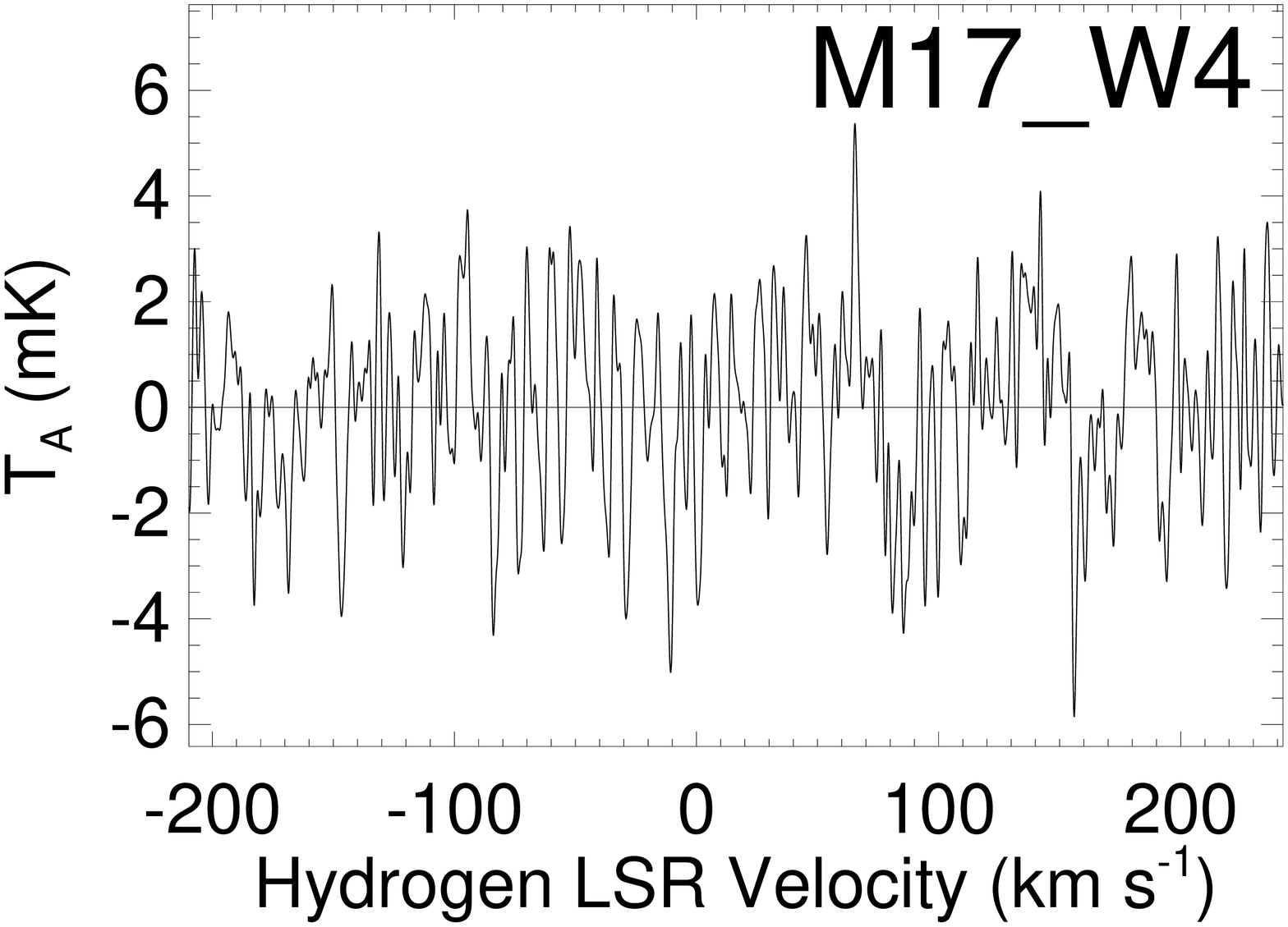} \\
\includegraphics[width=.23\textwidth]{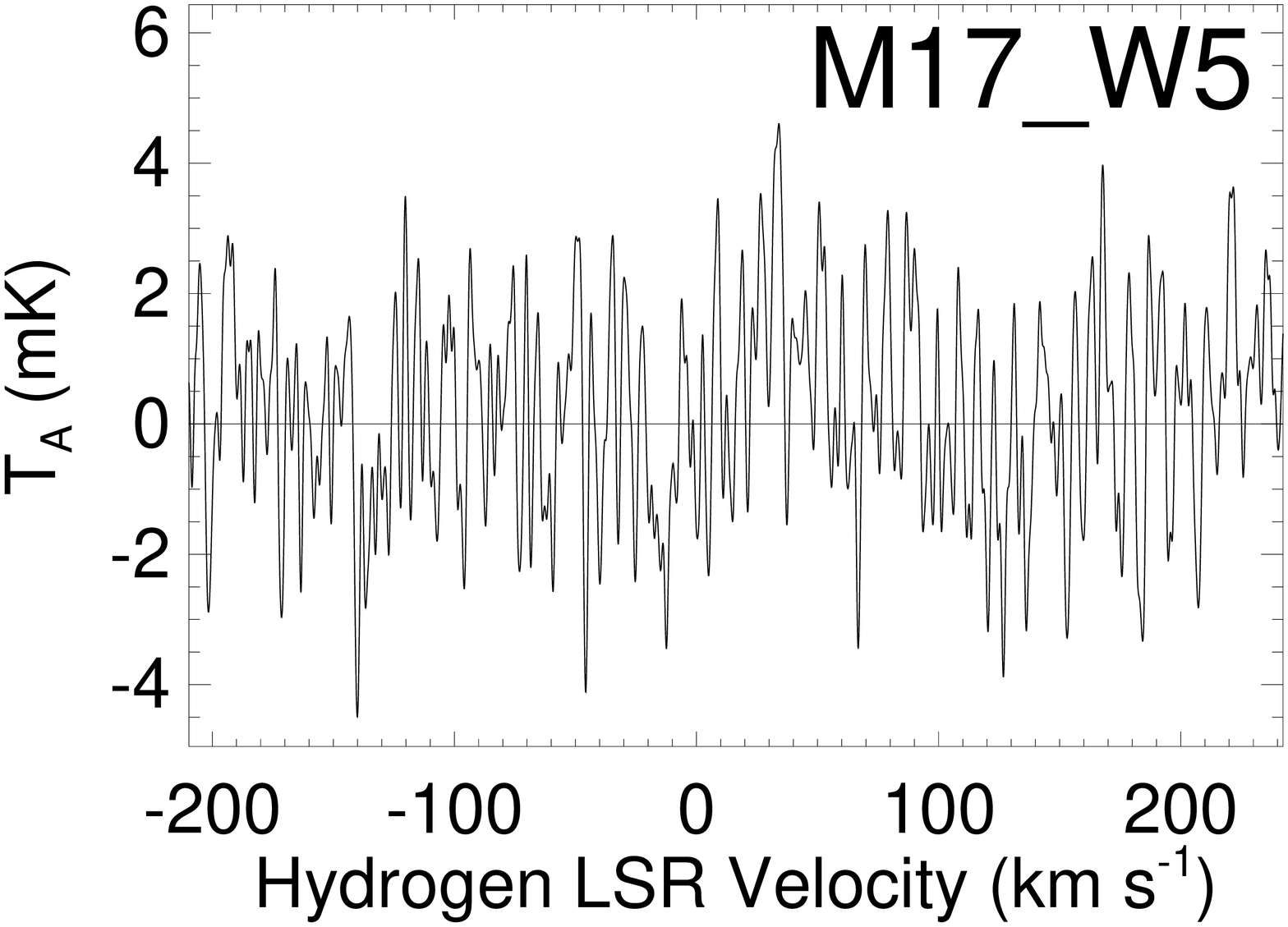} &
\includegraphics[width=.23\textwidth]{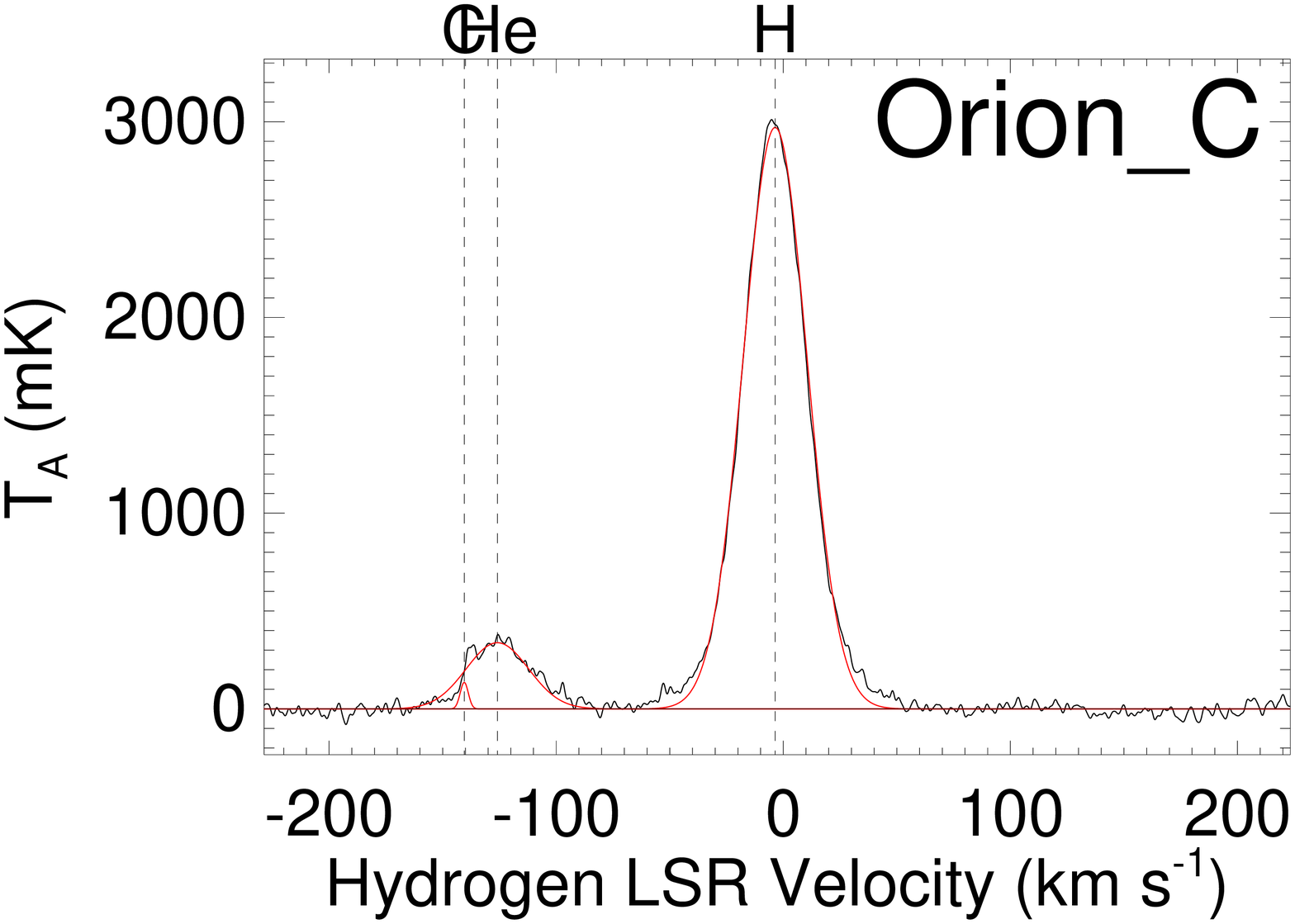} &
\includegraphics[width=.23\textwidth]{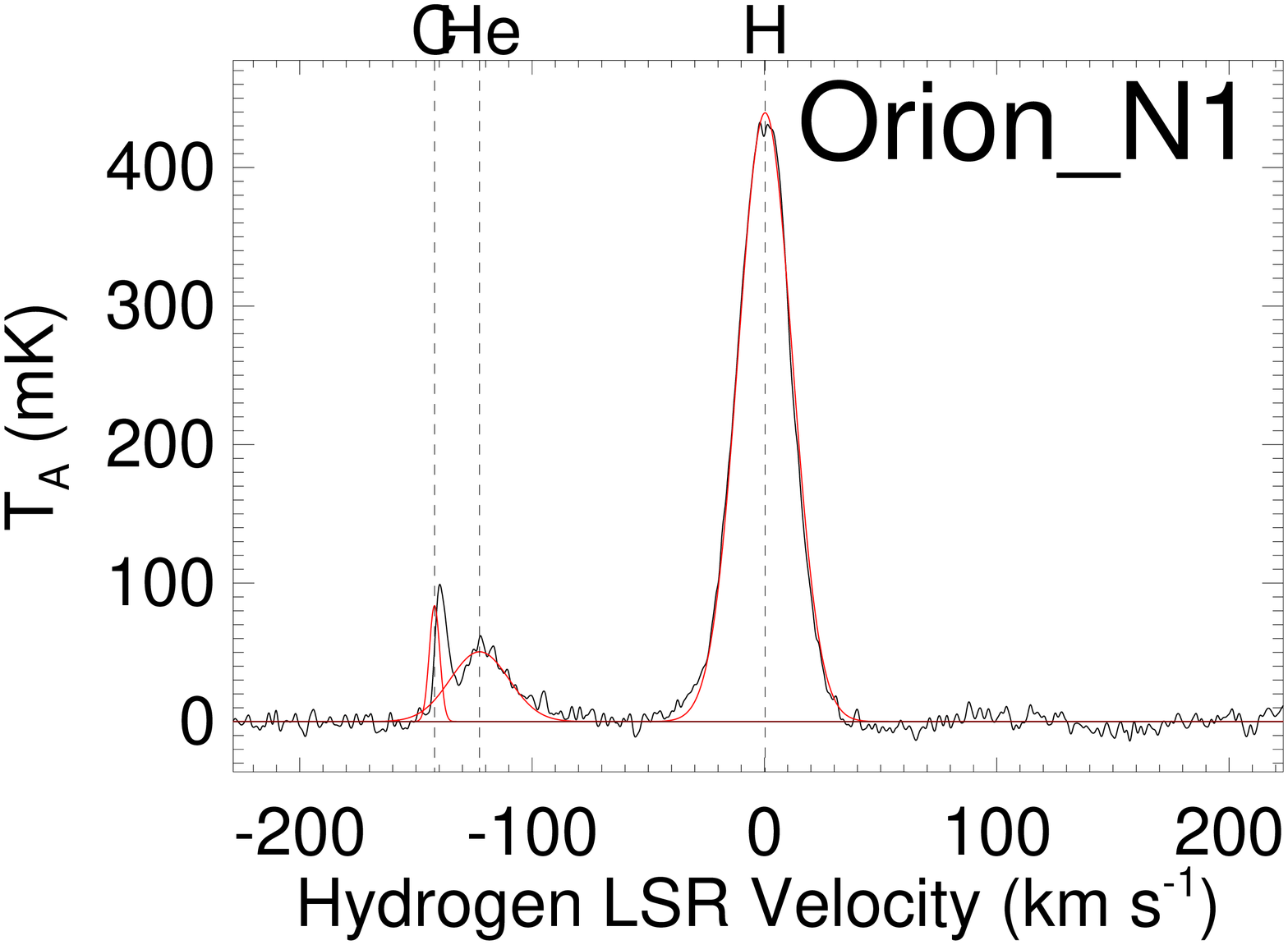} &
\includegraphics[width=.23\textwidth]{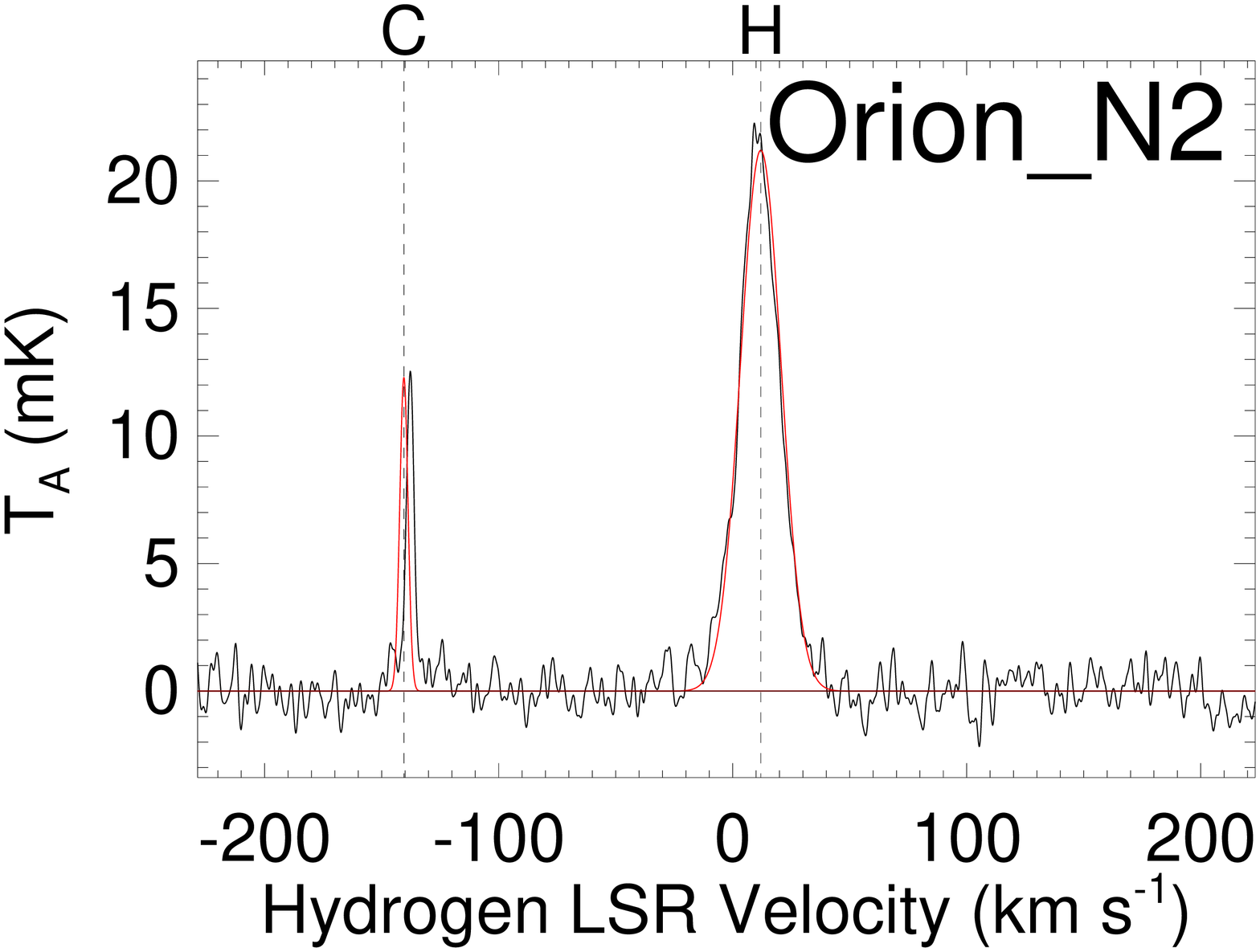} \\
\includegraphics[width=.23\textwidth]{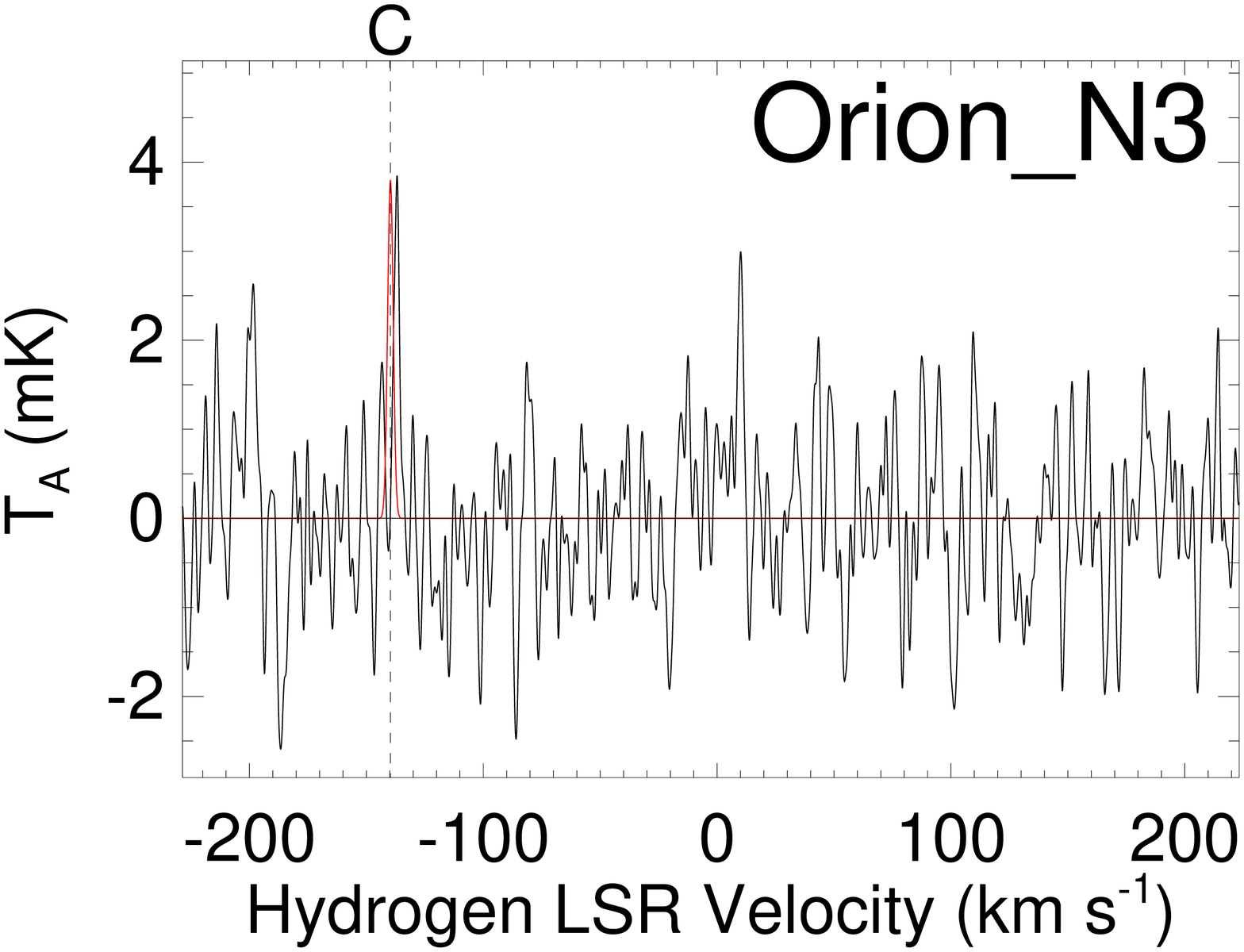} &
\includegraphics[width=.23\textwidth]{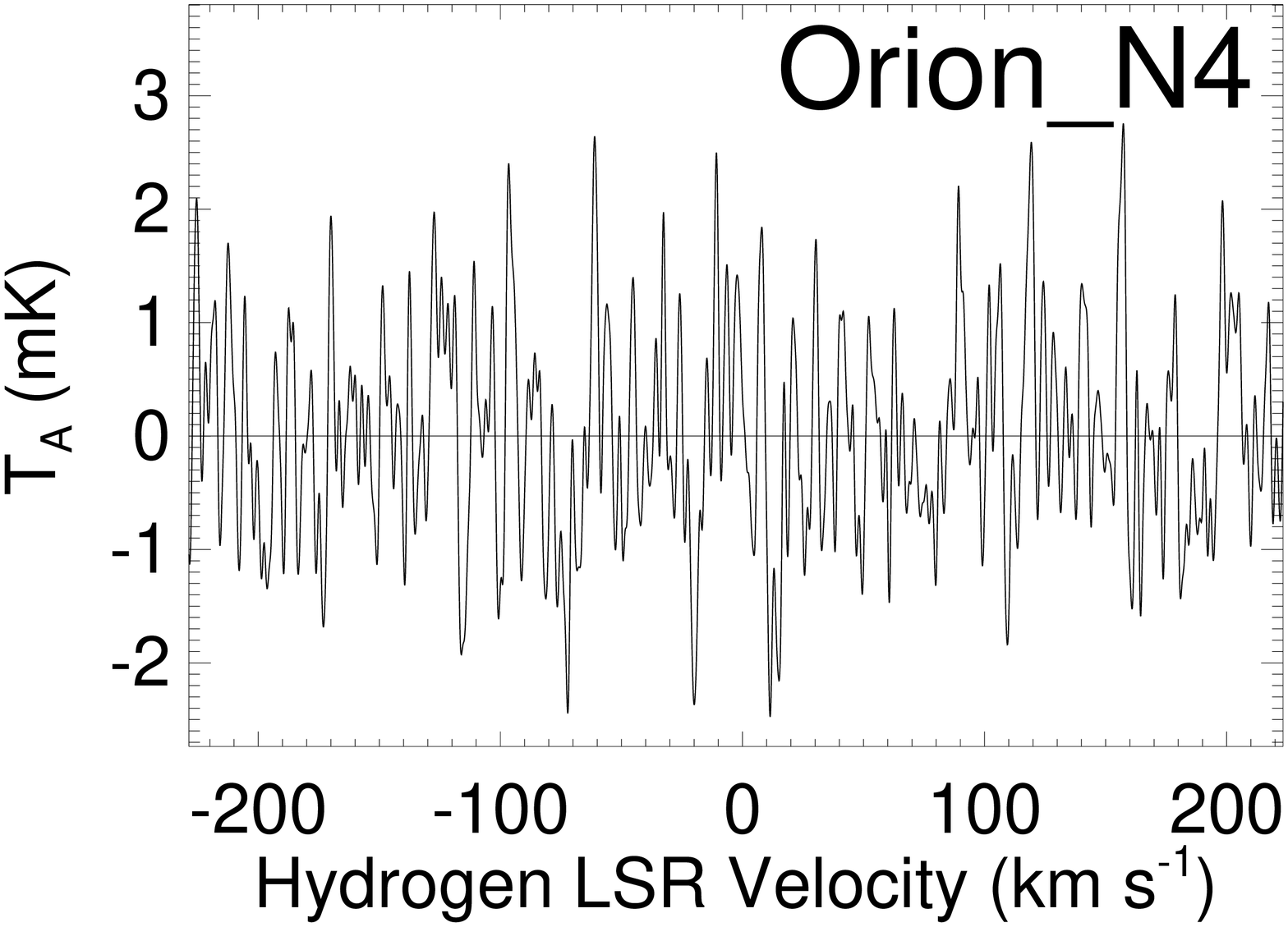} &
\includegraphics[width=.23\textwidth]{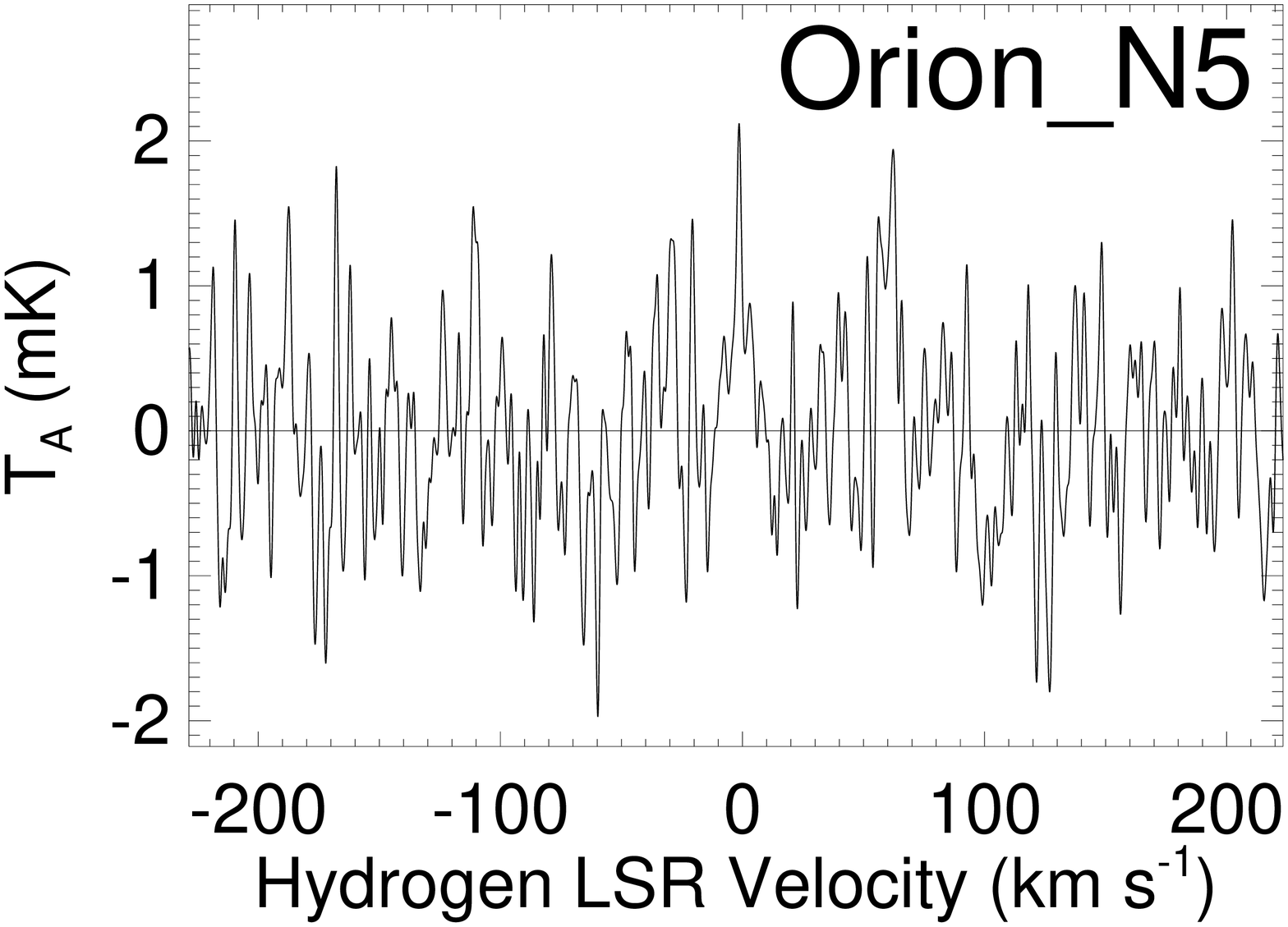} &
\includegraphics[width=.23\textwidth]{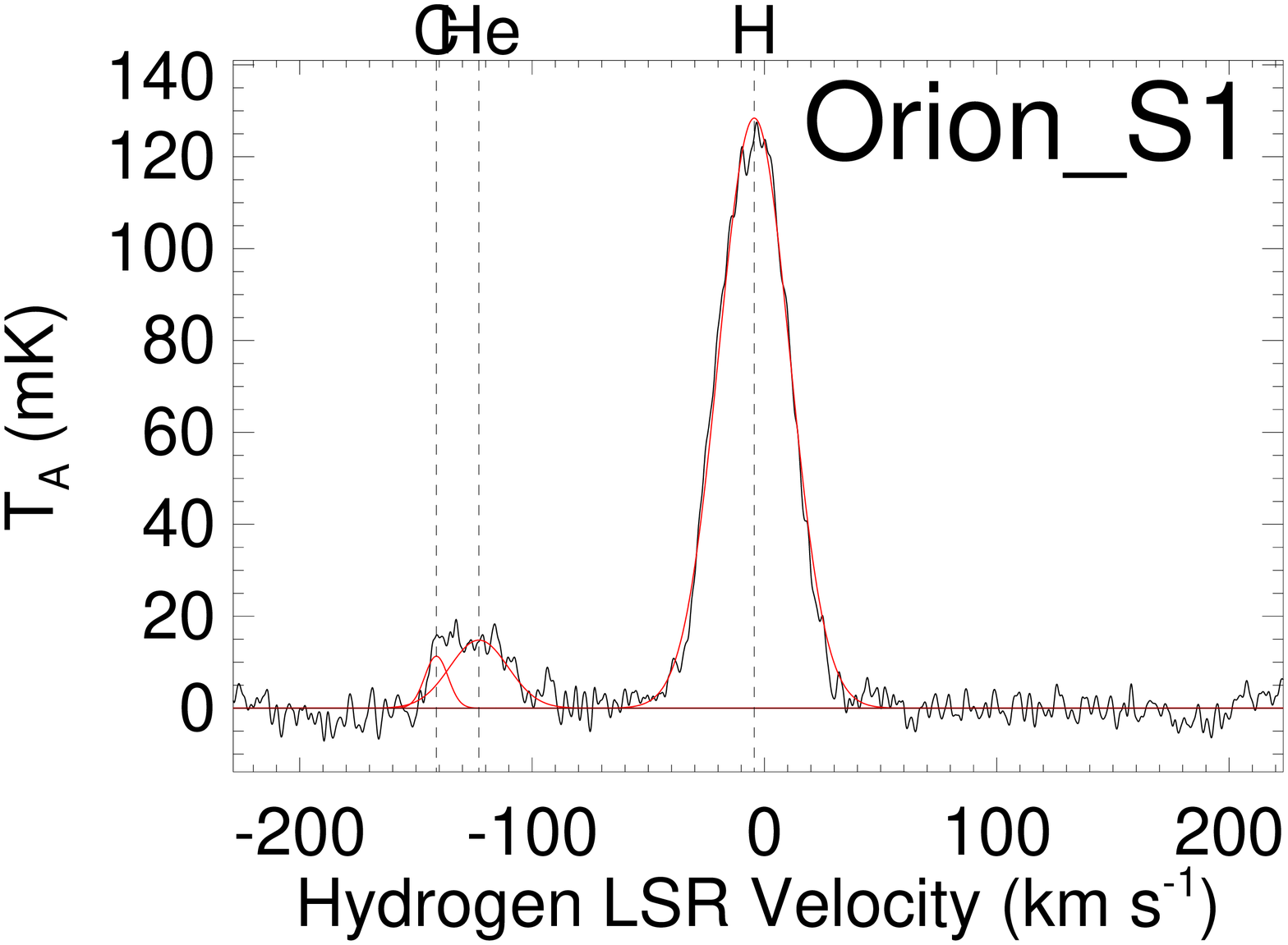} \\
\includegraphics[width=.23\textwidth]{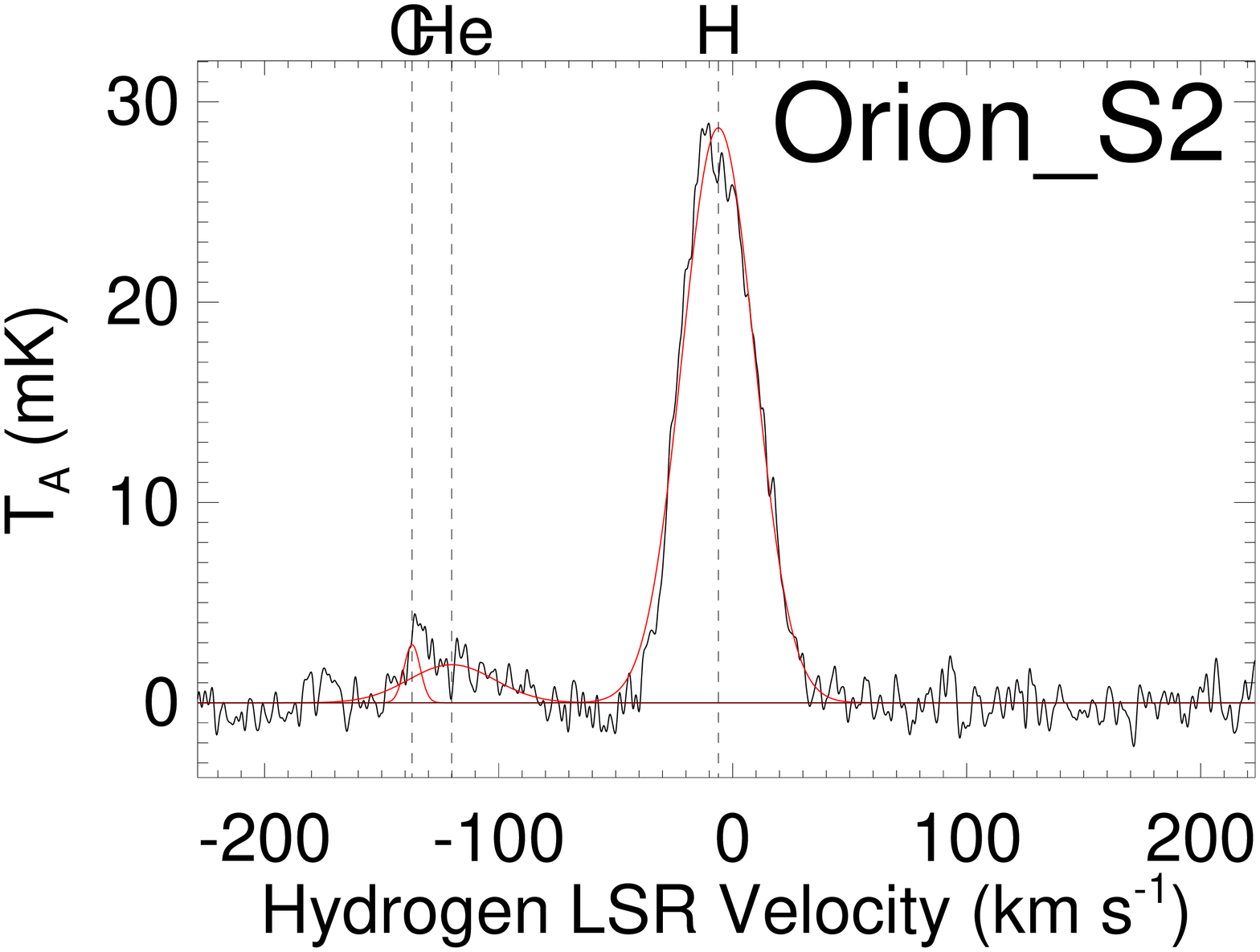} &
\includegraphics[width=.23\textwidth]{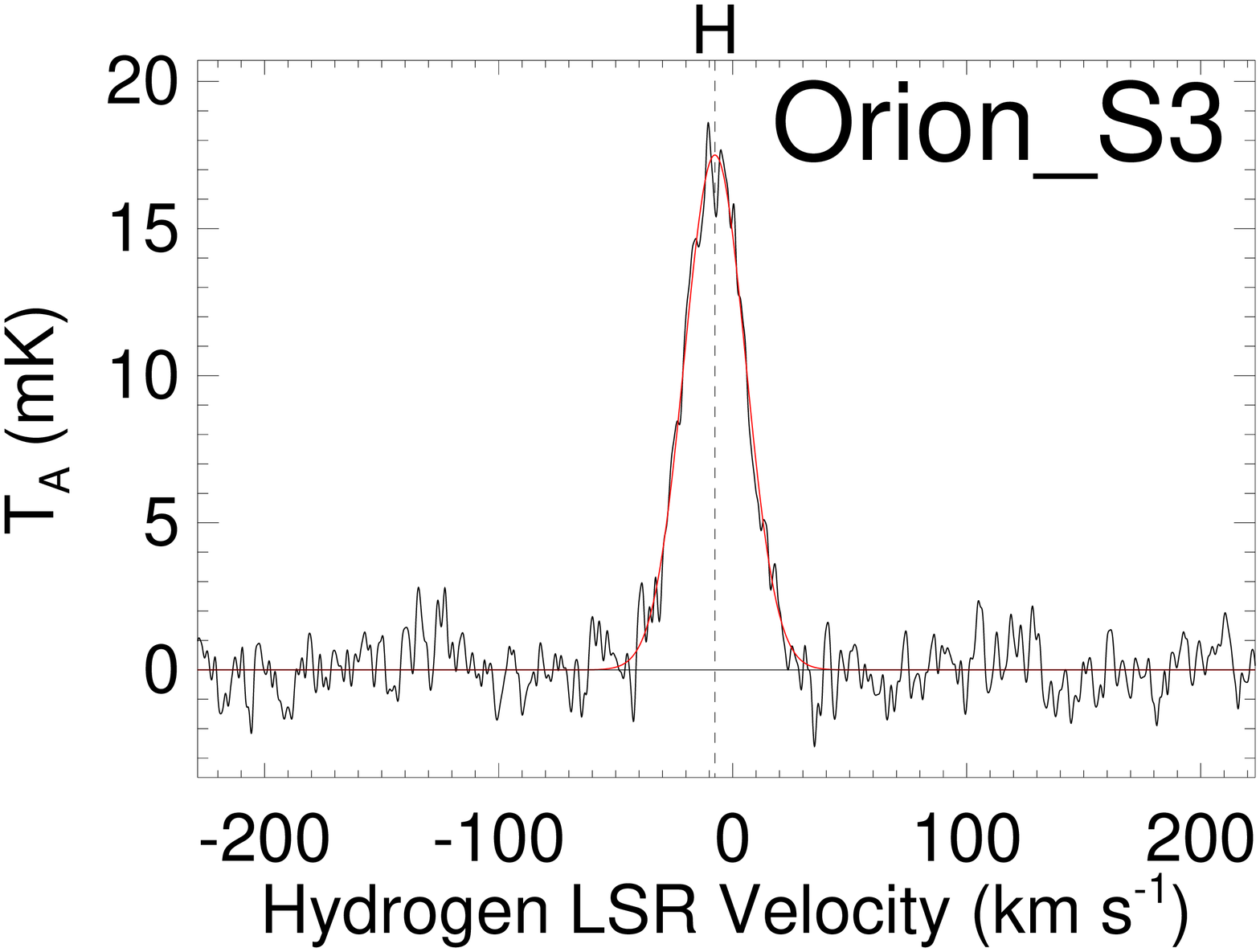} &
\includegraphics[width=.23\textwidth]{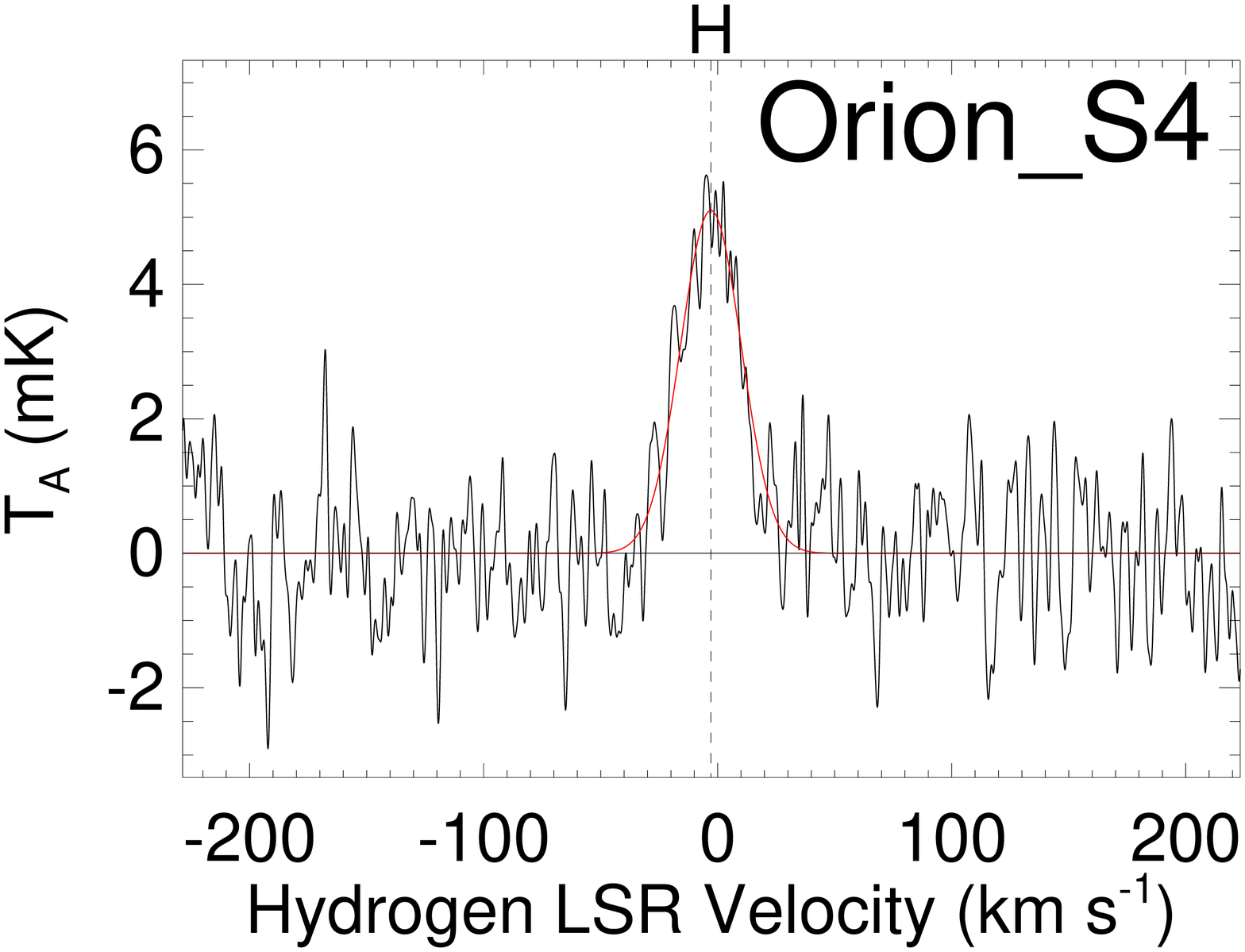} &
\includegraphics[width=.23\textwidth]{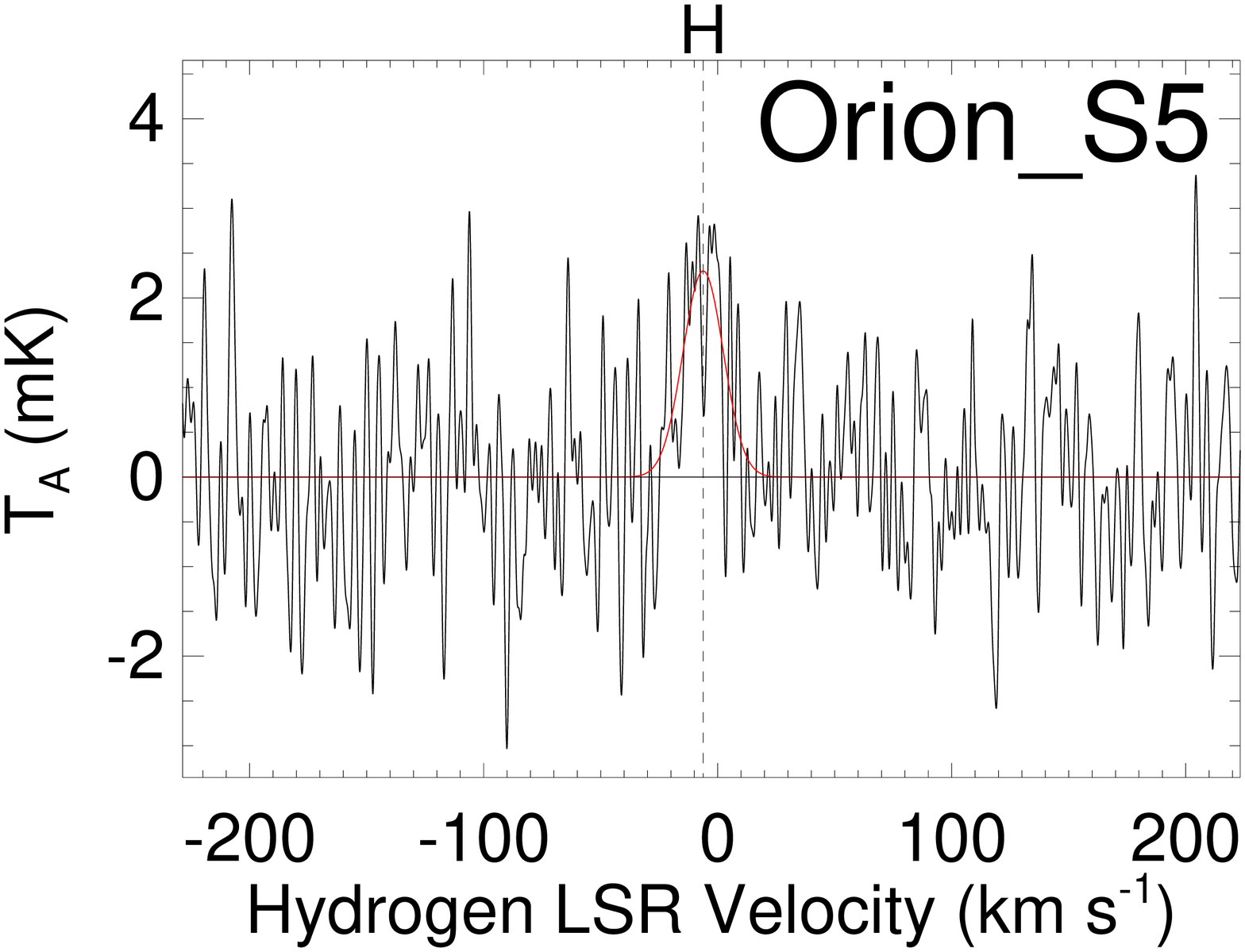}\\
\includegraphics[width=.23\textwidth]{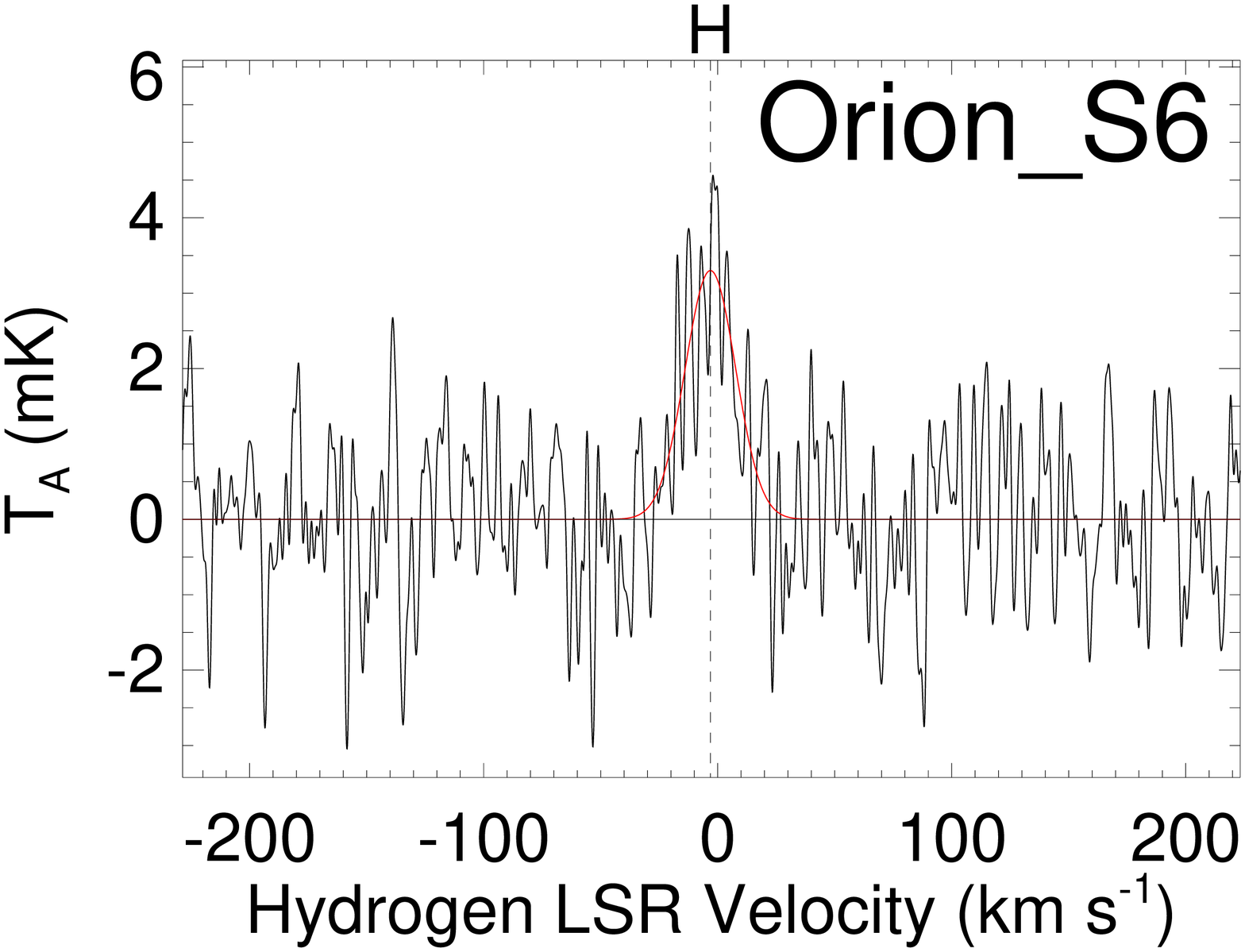} &
\includegraphics[width=.23\textwidth]{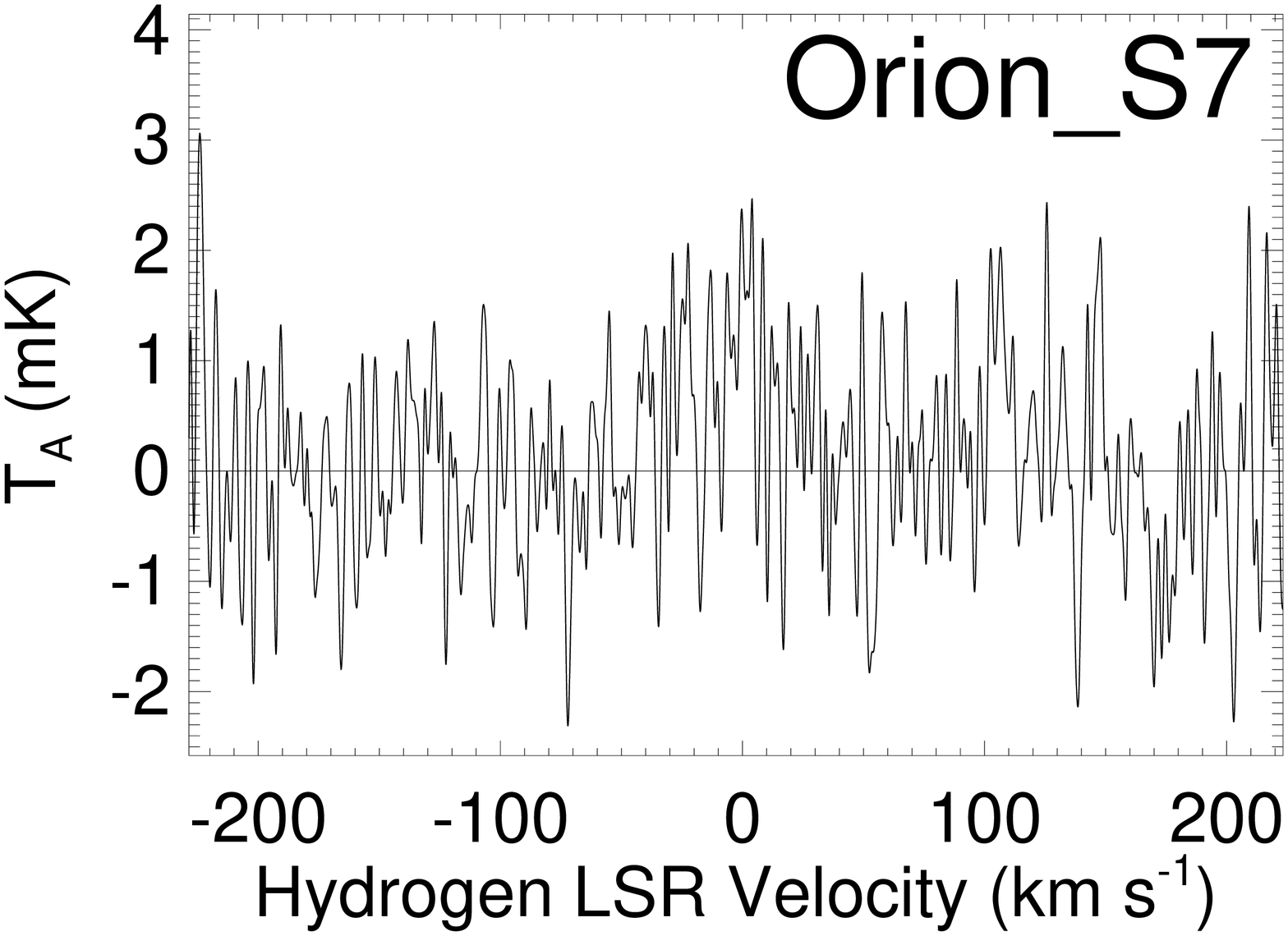} &
\includegraphics[width=.23\textwidth]{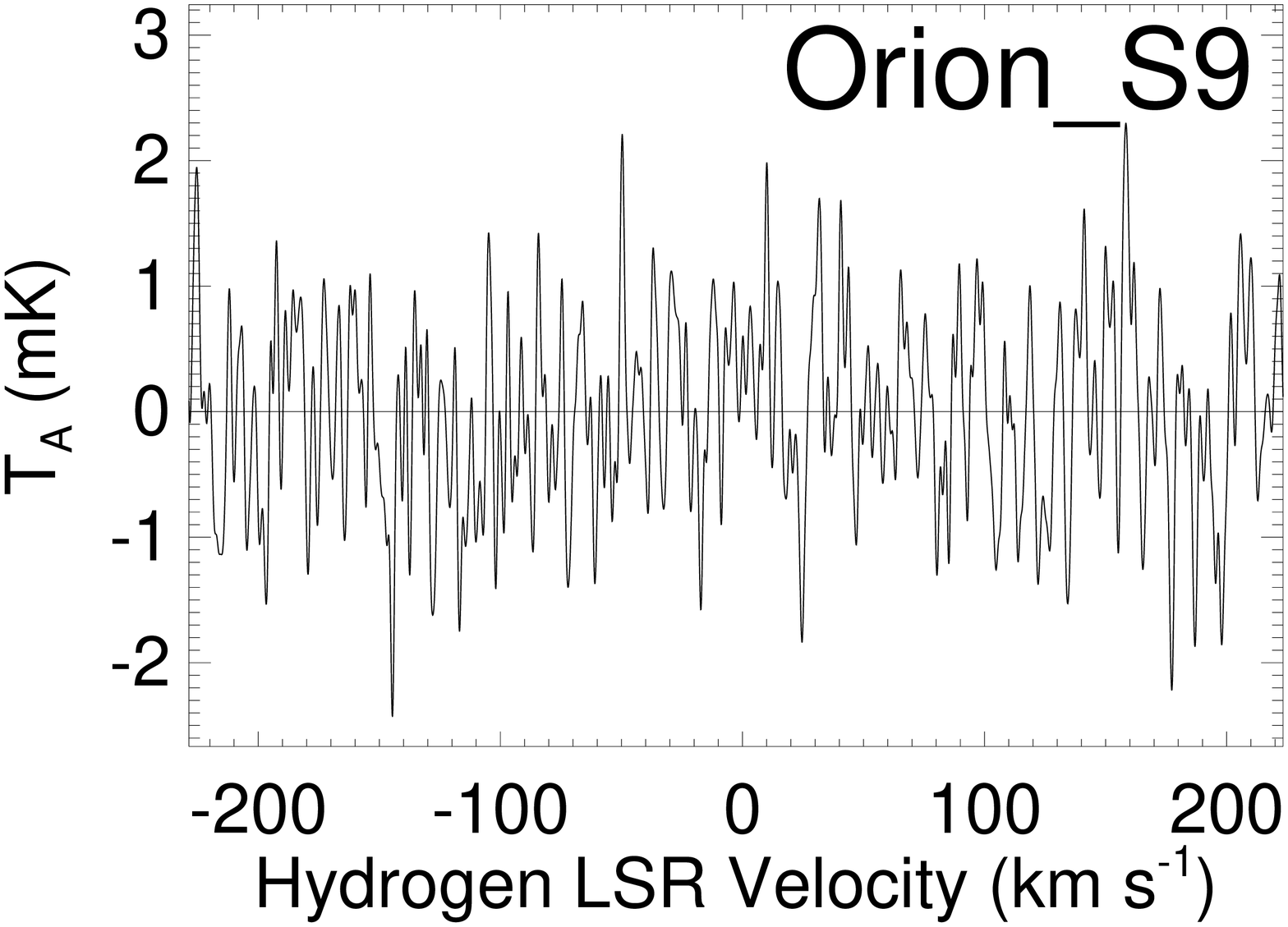} &
\includegraphics[width=.23\textwidth]{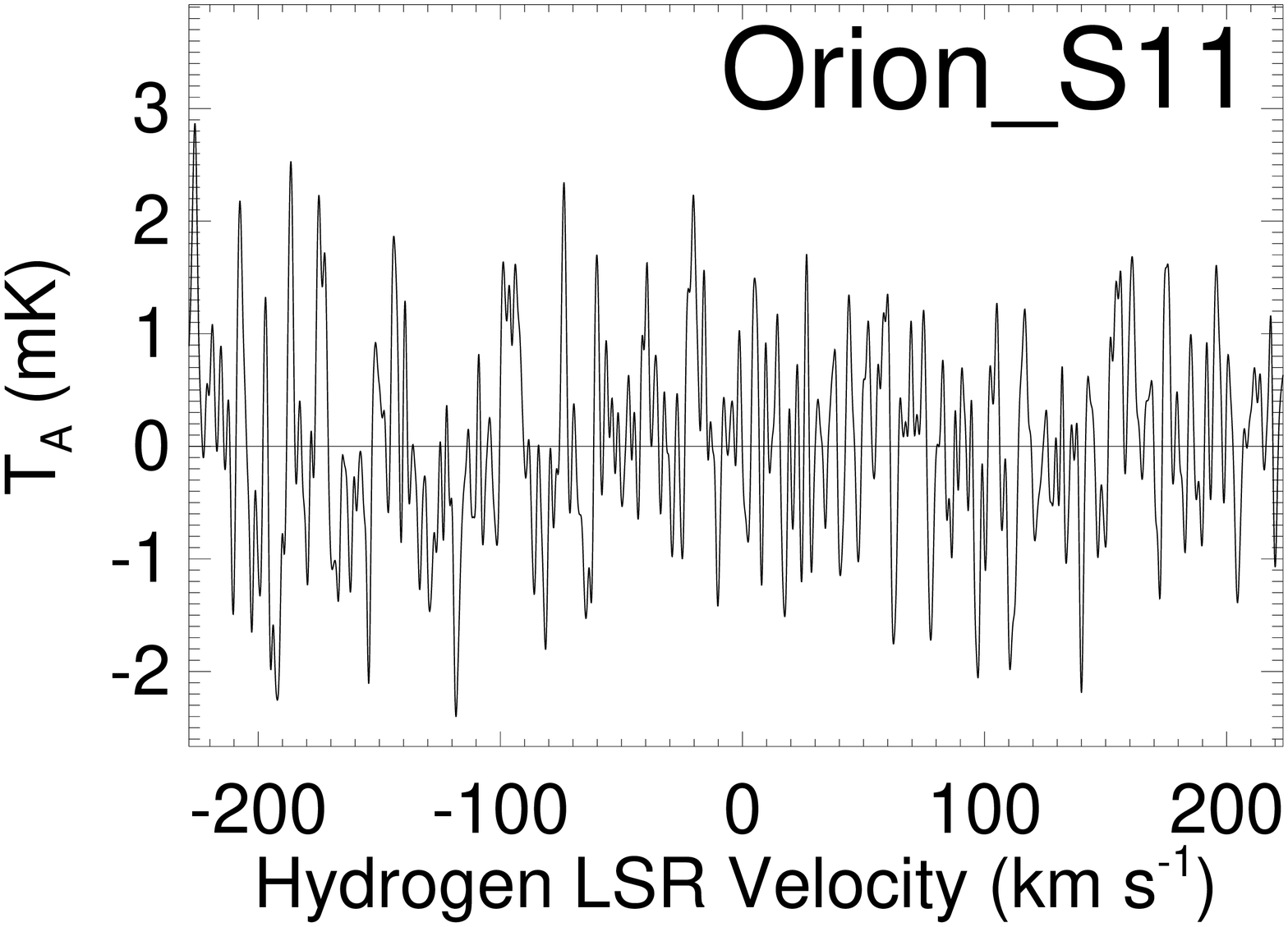} \\
\includegraphics[width=.23\textwidth]{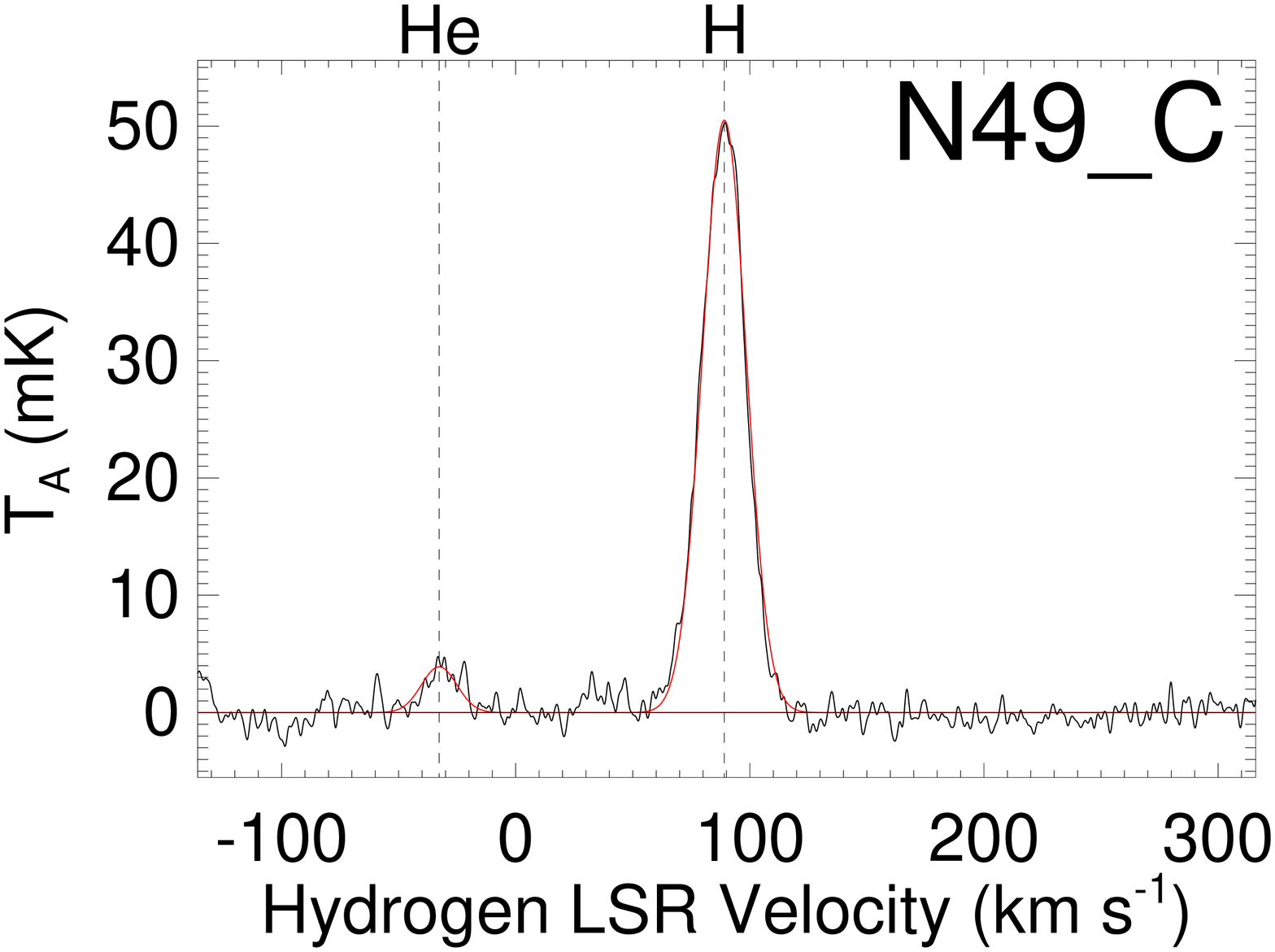} &
\includegraphics[width=.23\textwidth]{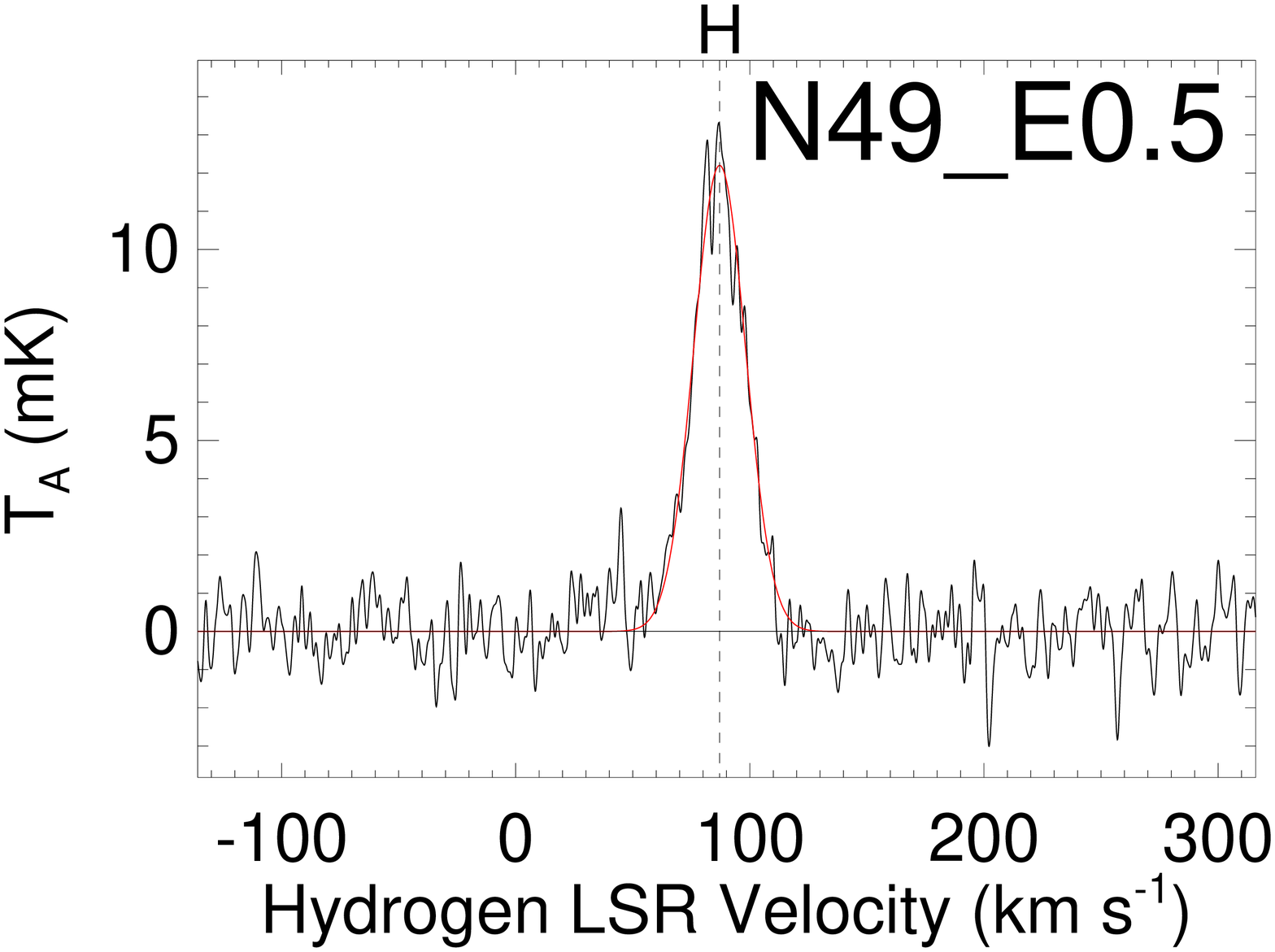} &
\includegraphics[width=.23\textwidth]{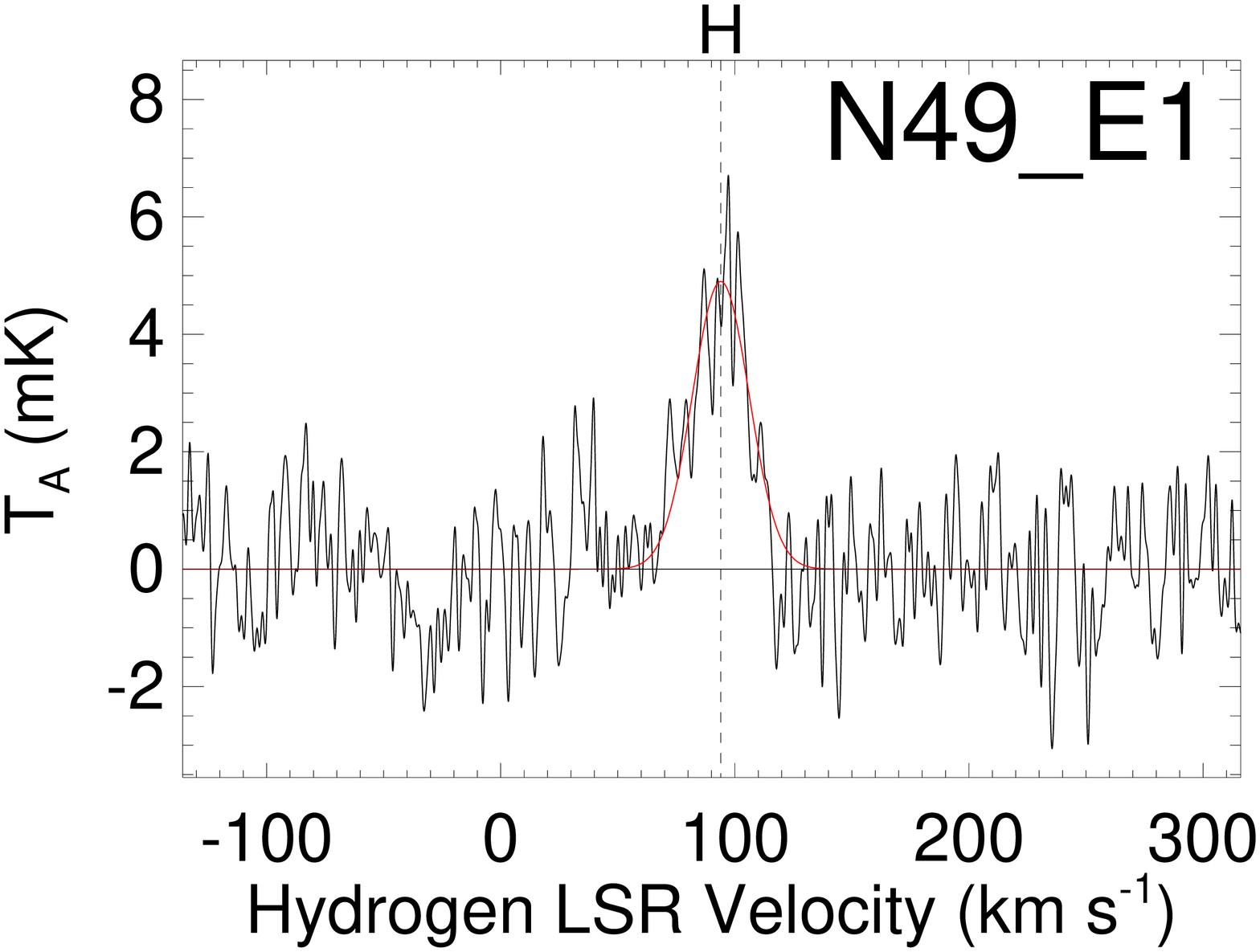} &
\includegraphics[width=.23\textwidth]{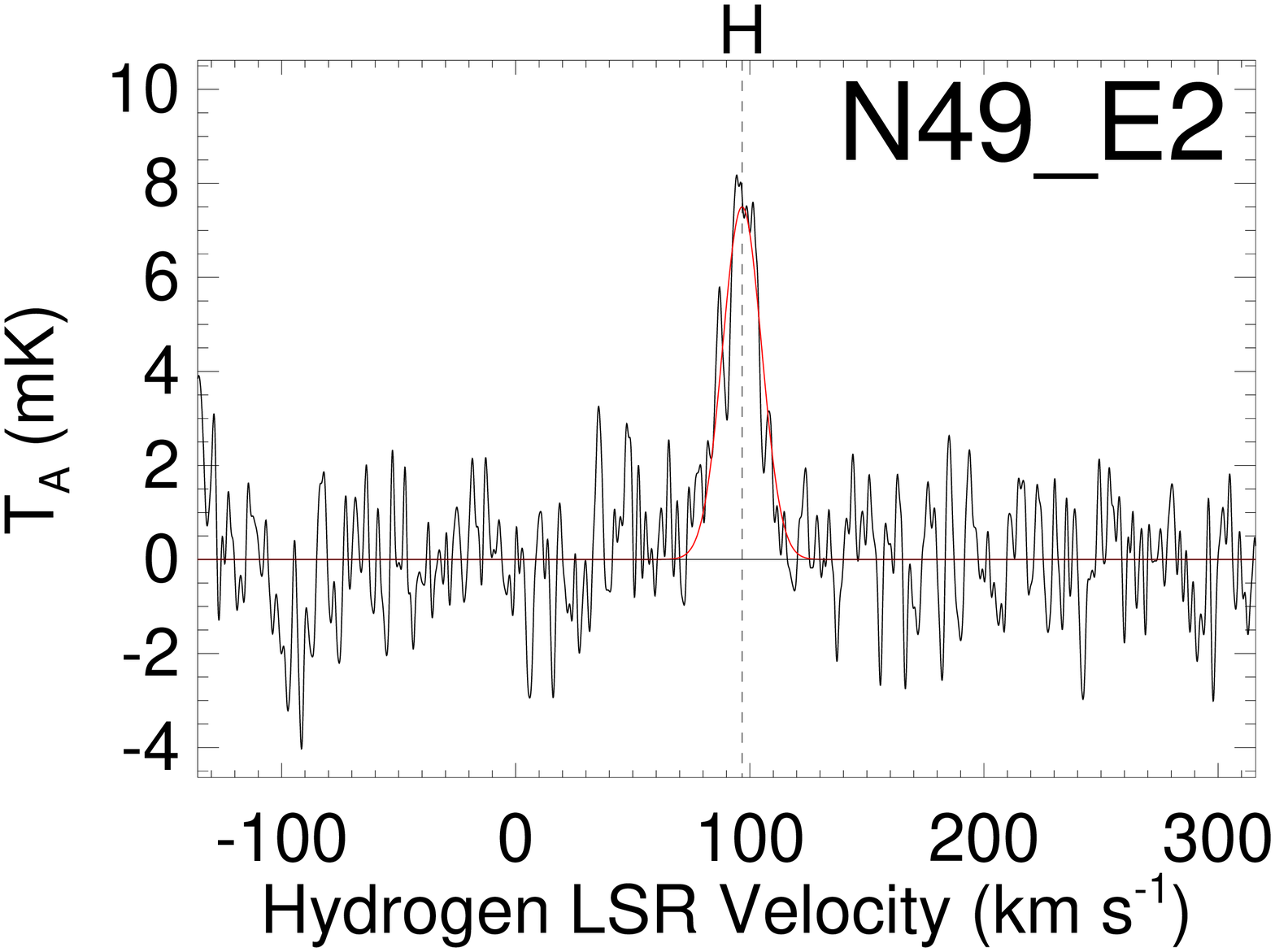} \\
\end{tabular}
\caption{}
\end{figure*}
\renewcommand{\thefigure}{\thesection.\arabic{figure}}

\renewcommand\thefigure{\thesection.\arabic{figure} (Cont.)}
\addtocounter{figure}{-1}
\begin{figure*}
\centering
\begin{tabular}{cccc}
\includegraphics[width=.23\textwidth]{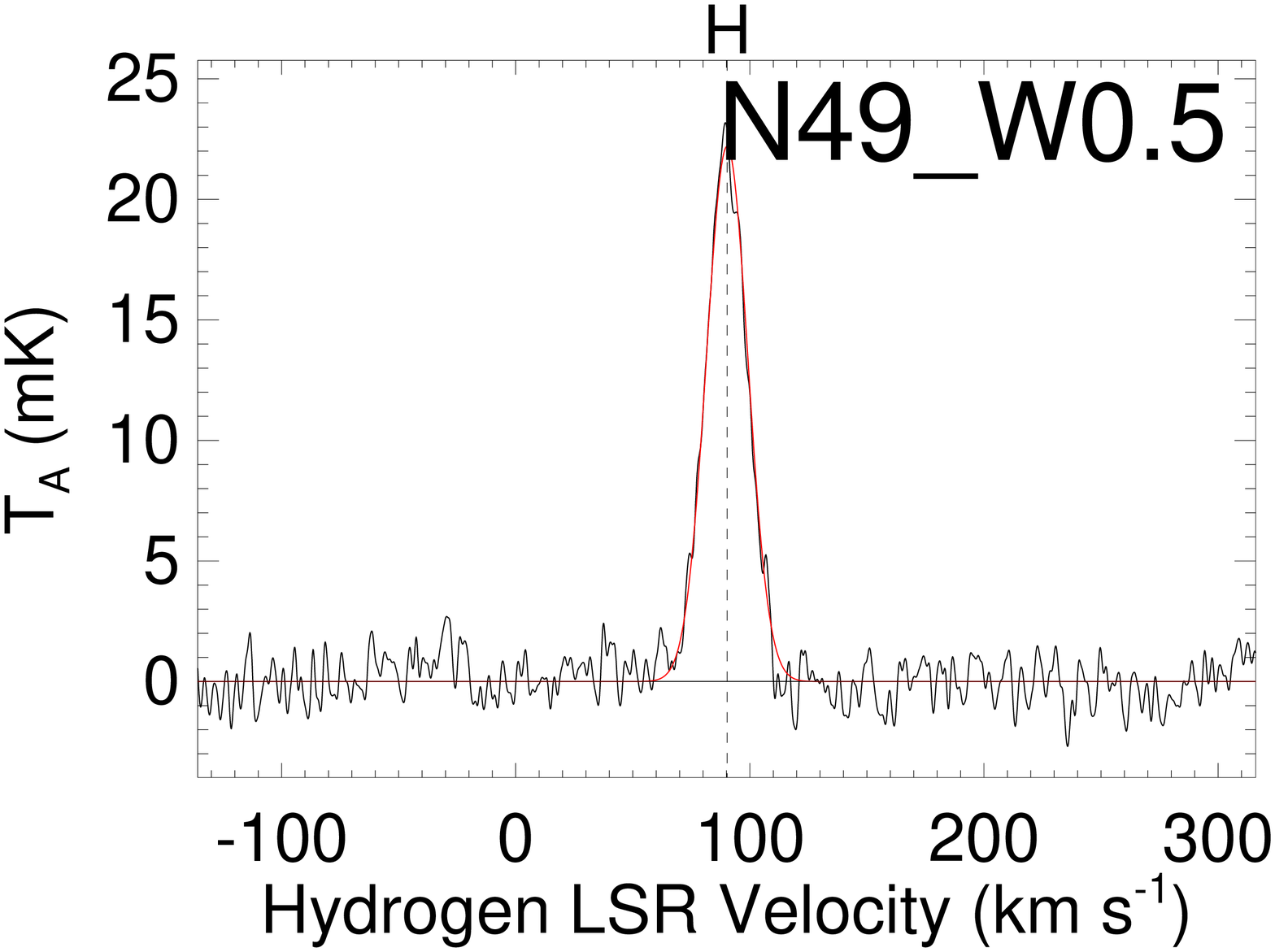} &
\includegraphics[width=.23\textwidth]{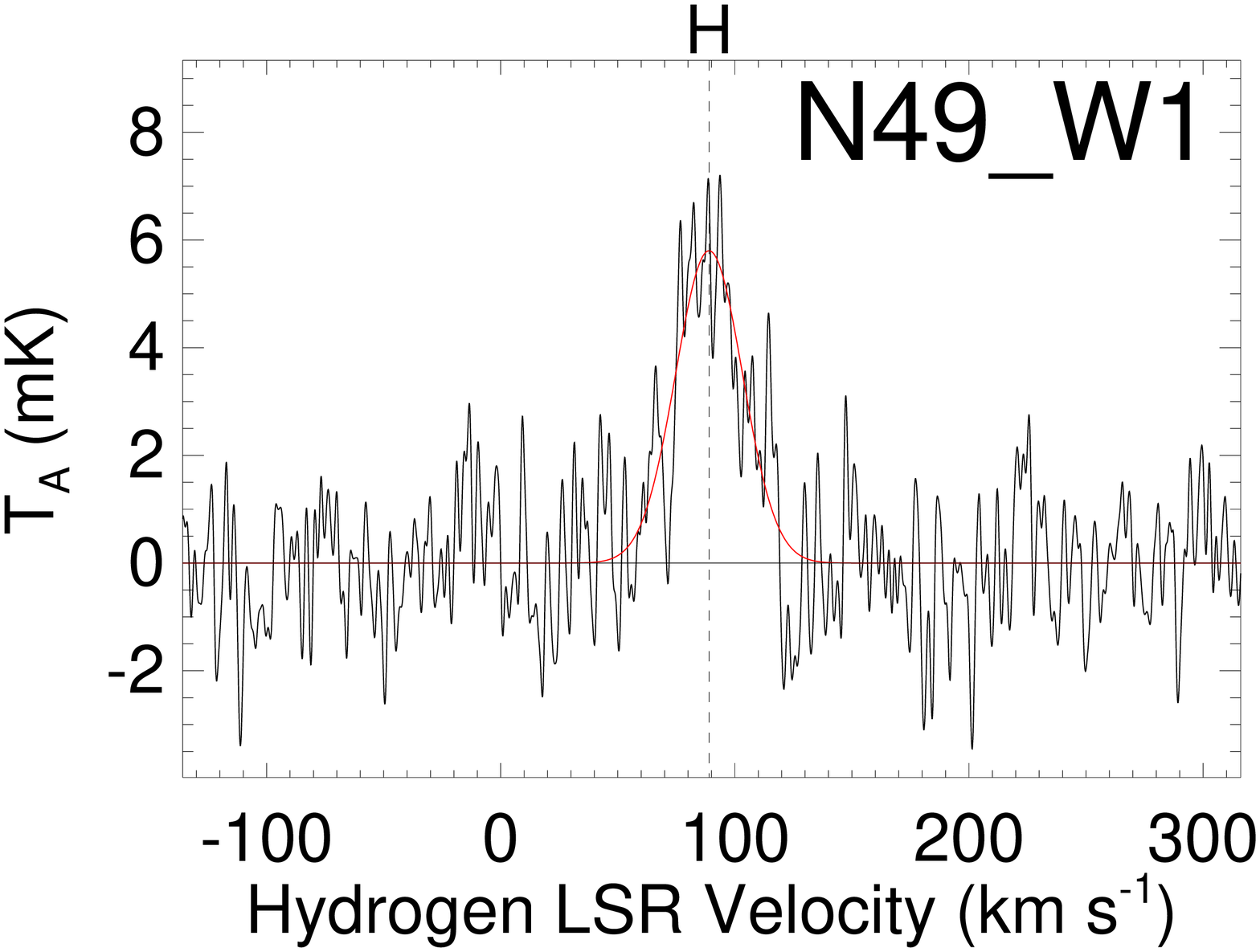} &
\includegraphics[width=.23\textwidth]{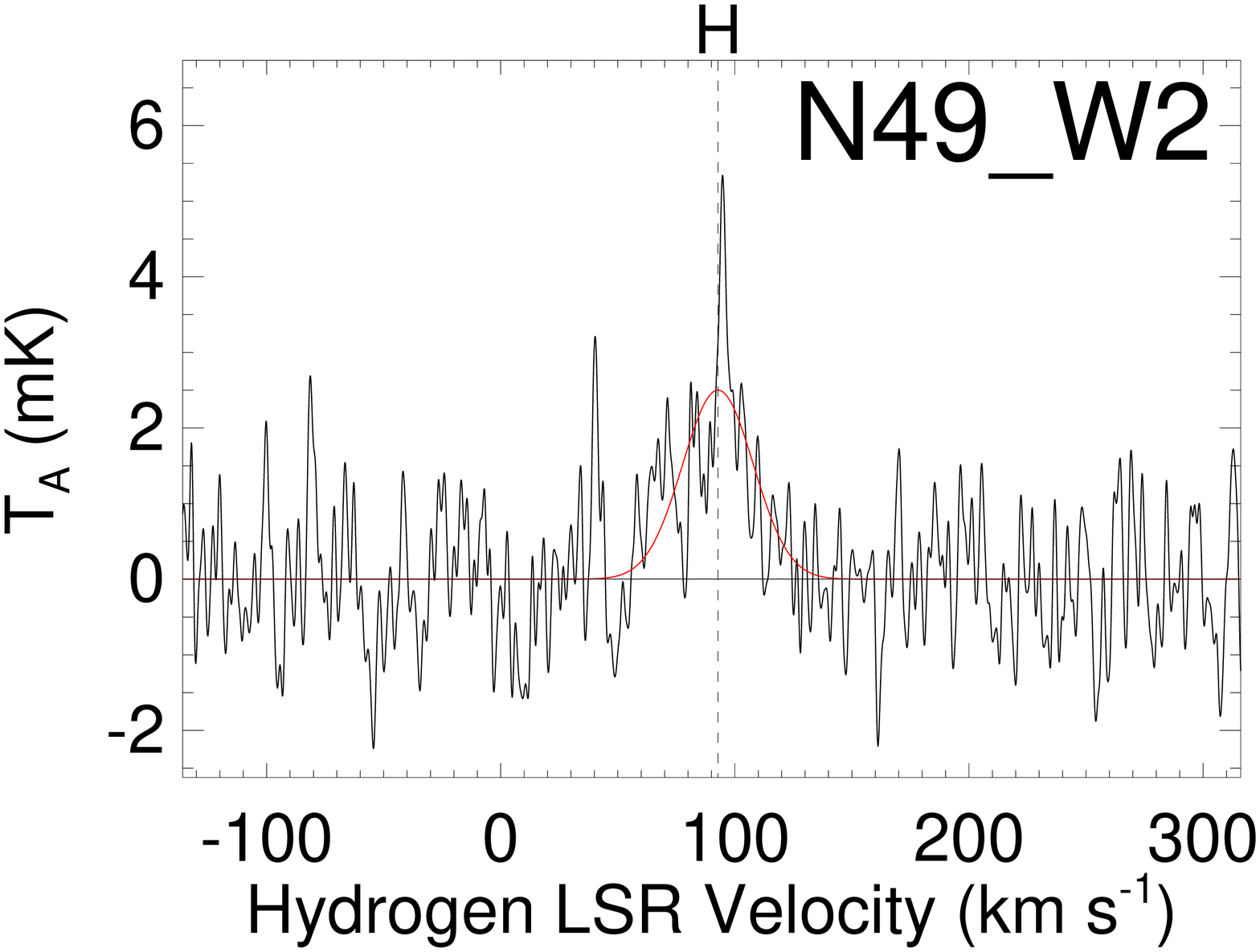} &
\includegraphics[width=.23\textwidth]{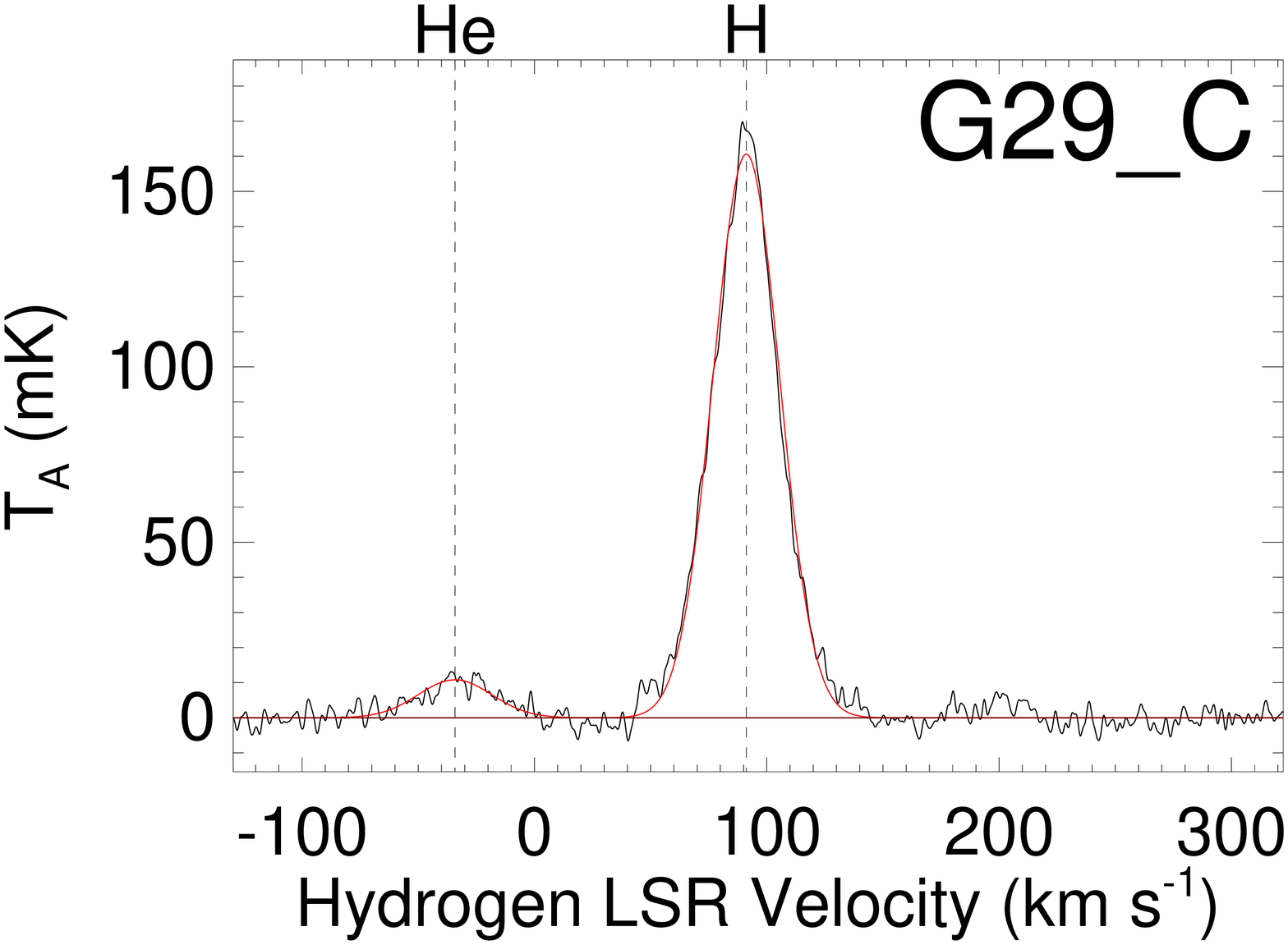} \\
\includegraphics[width=.23\textwidth]{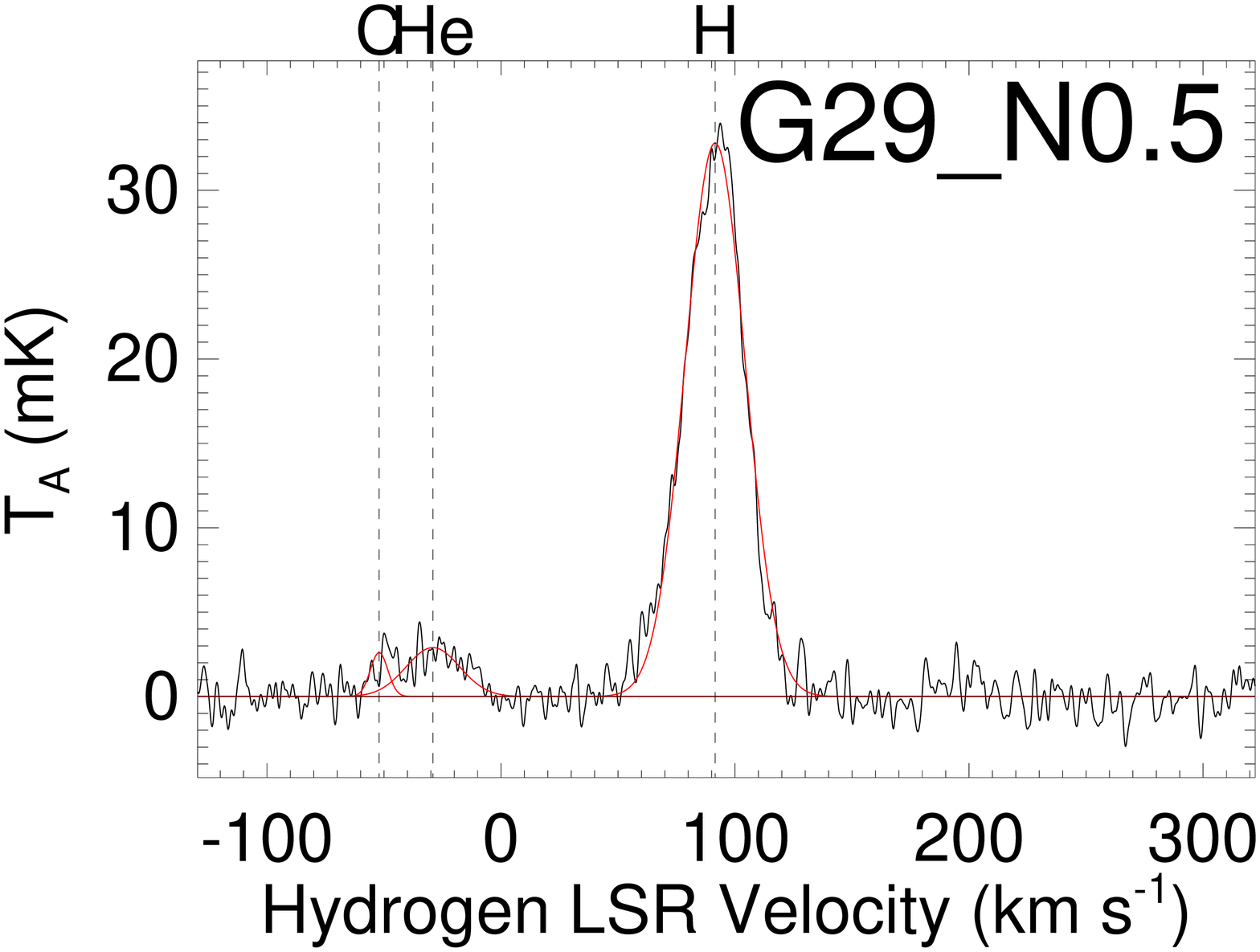} &
\includegraphics[width=.23\textwidth]{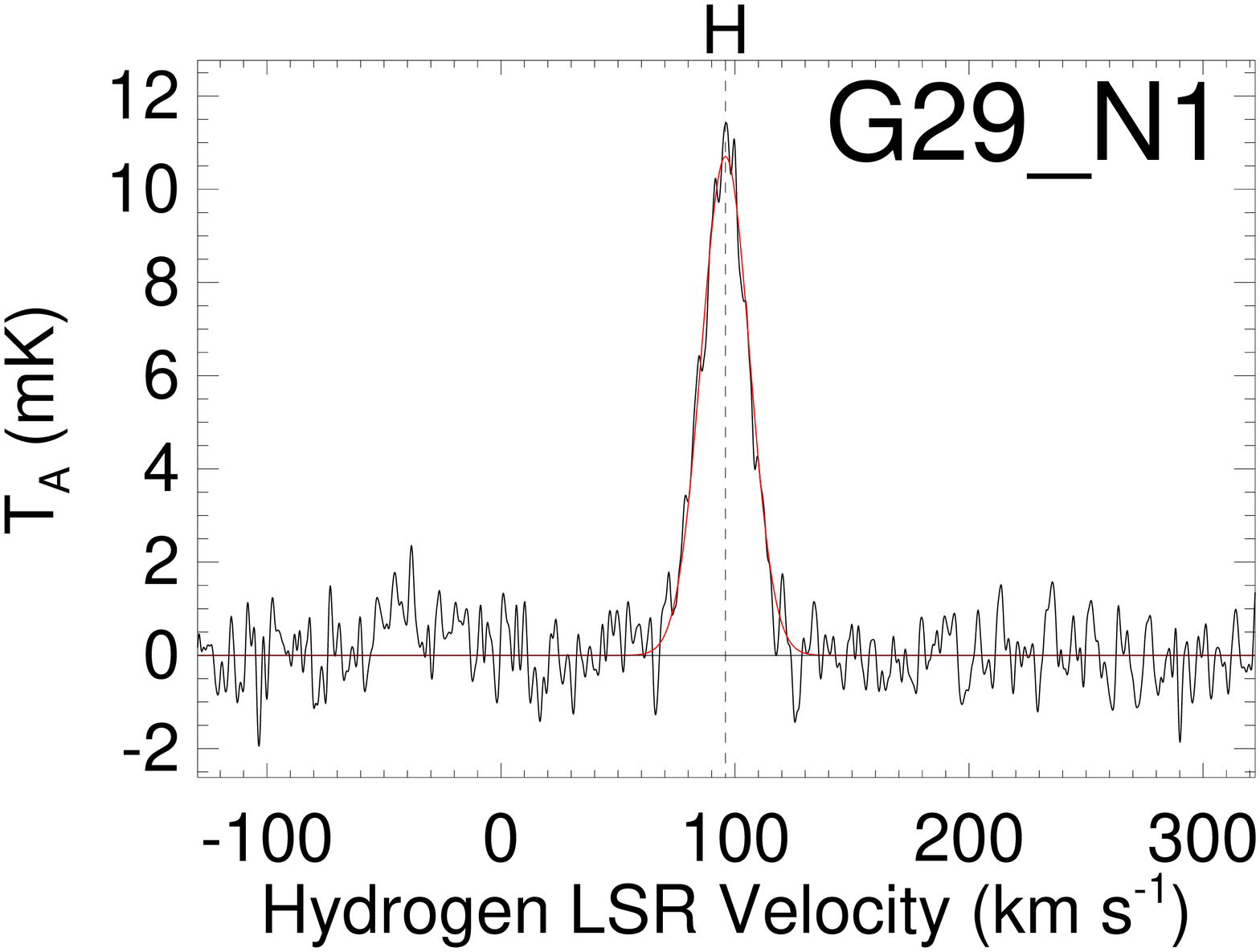} &
\includegraphics[width=.23\textwidth]{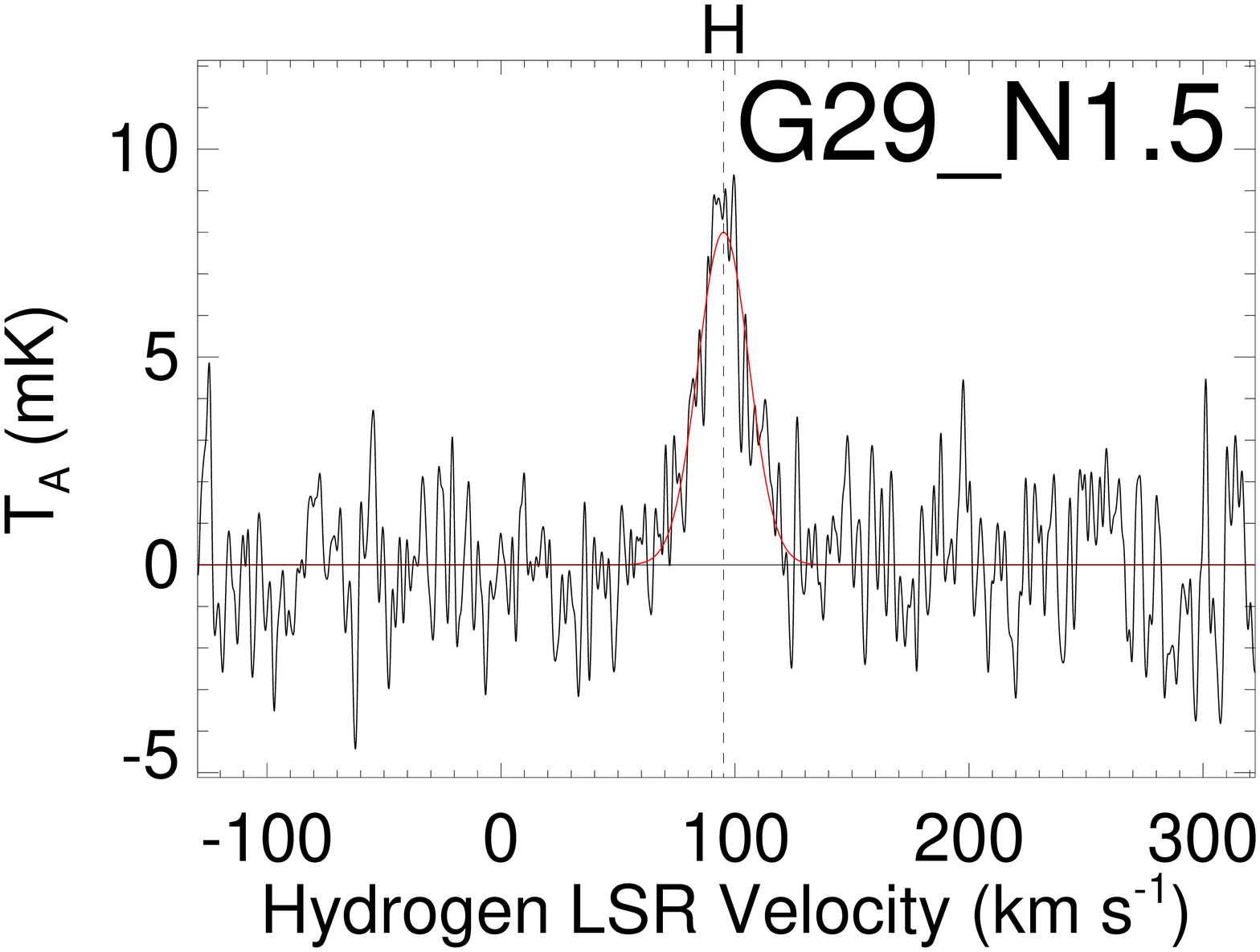} &
\includegraphics[width=.23\textwidth]{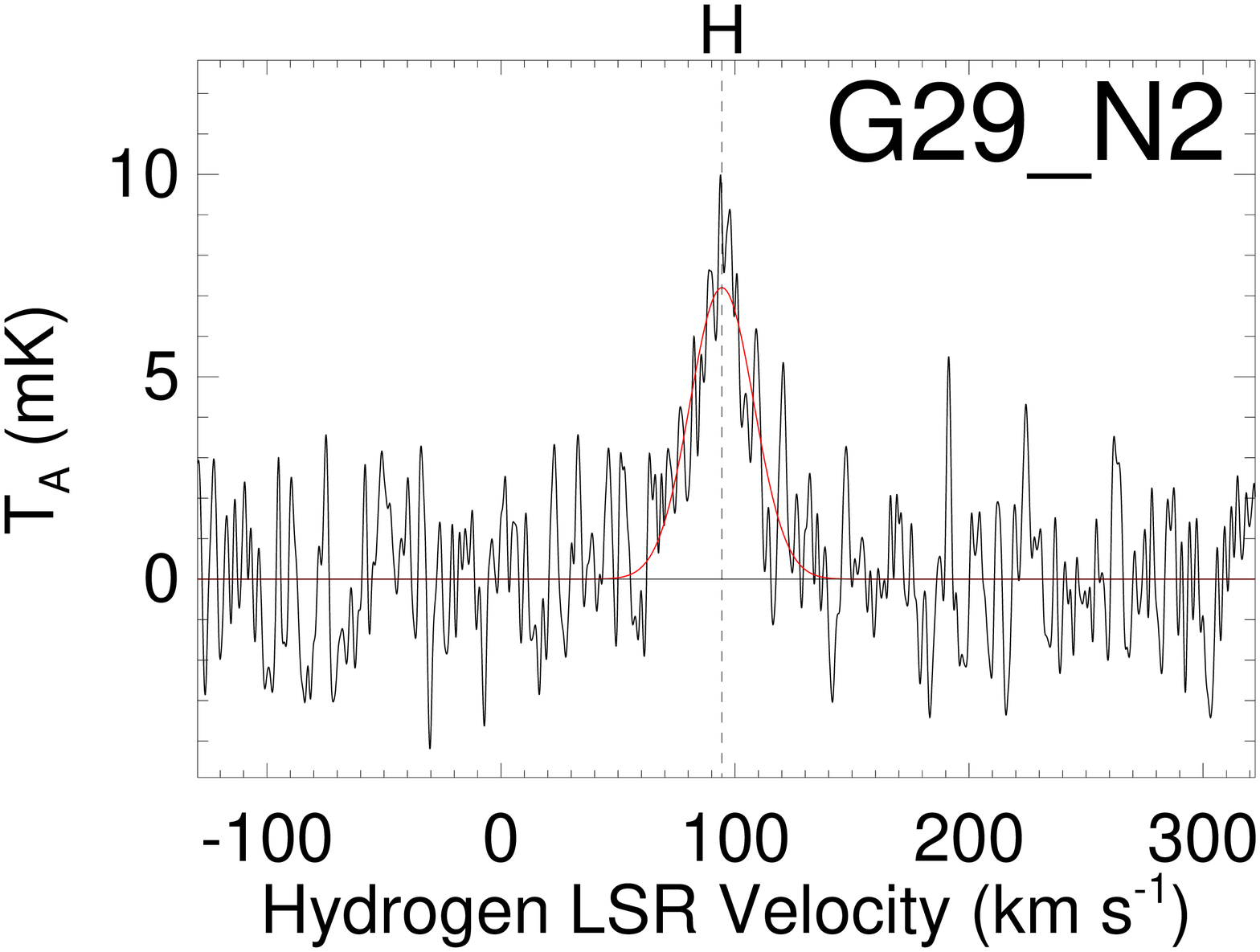} \\
\includegraphics[width=.23\textwidth]{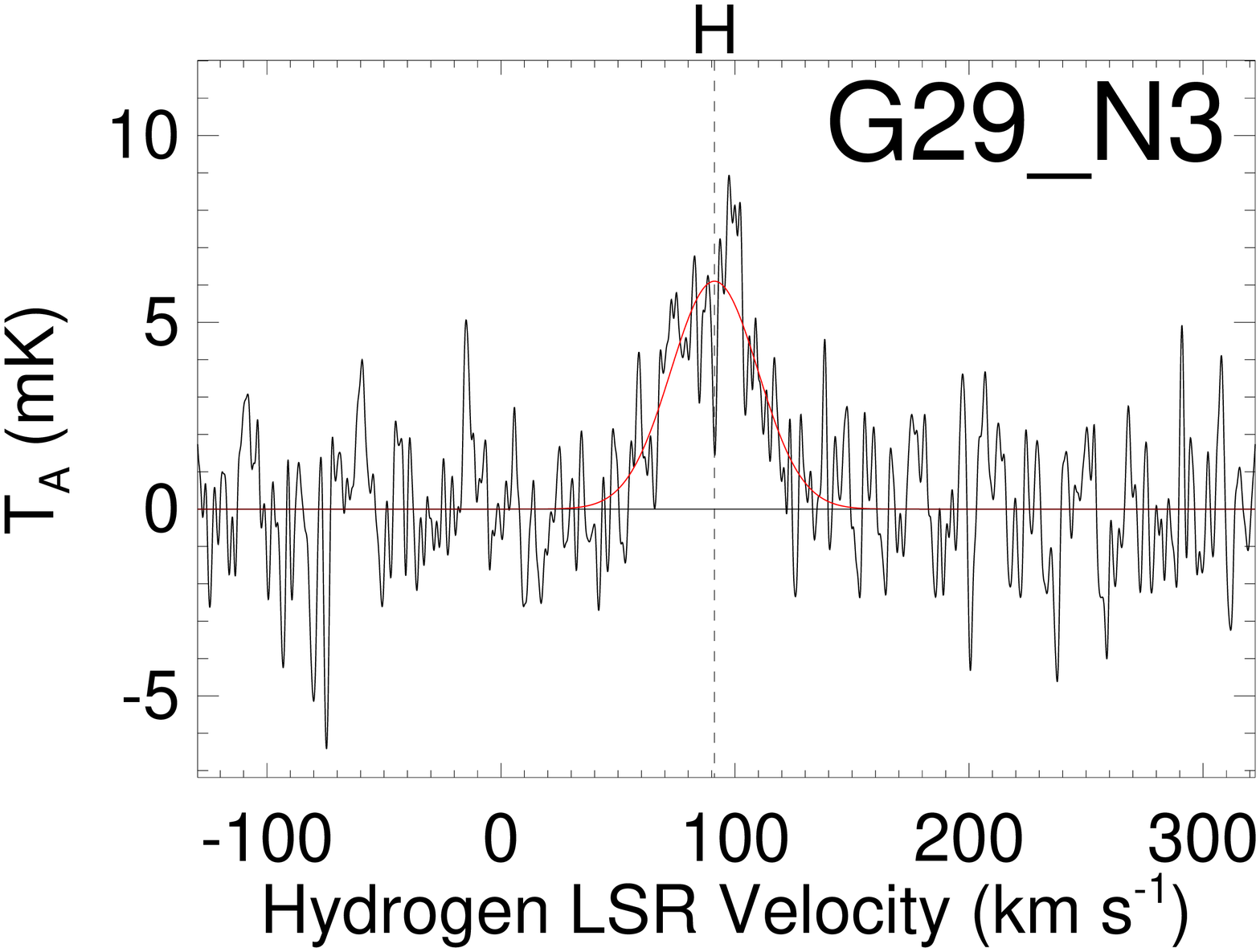} &
\includegraphics[width=.23\textwidth]{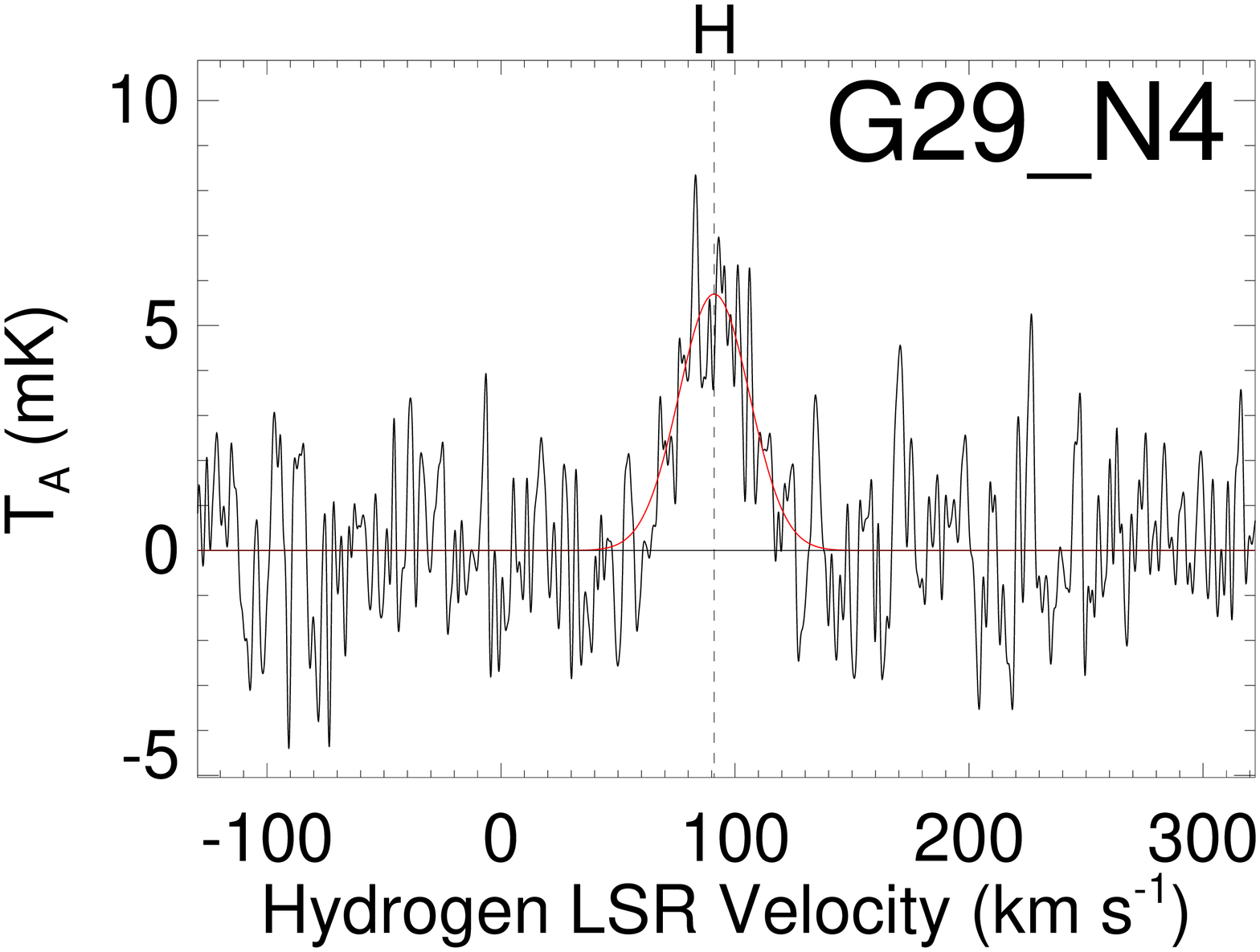} &
\includegraphics[width=.23\textwidth]{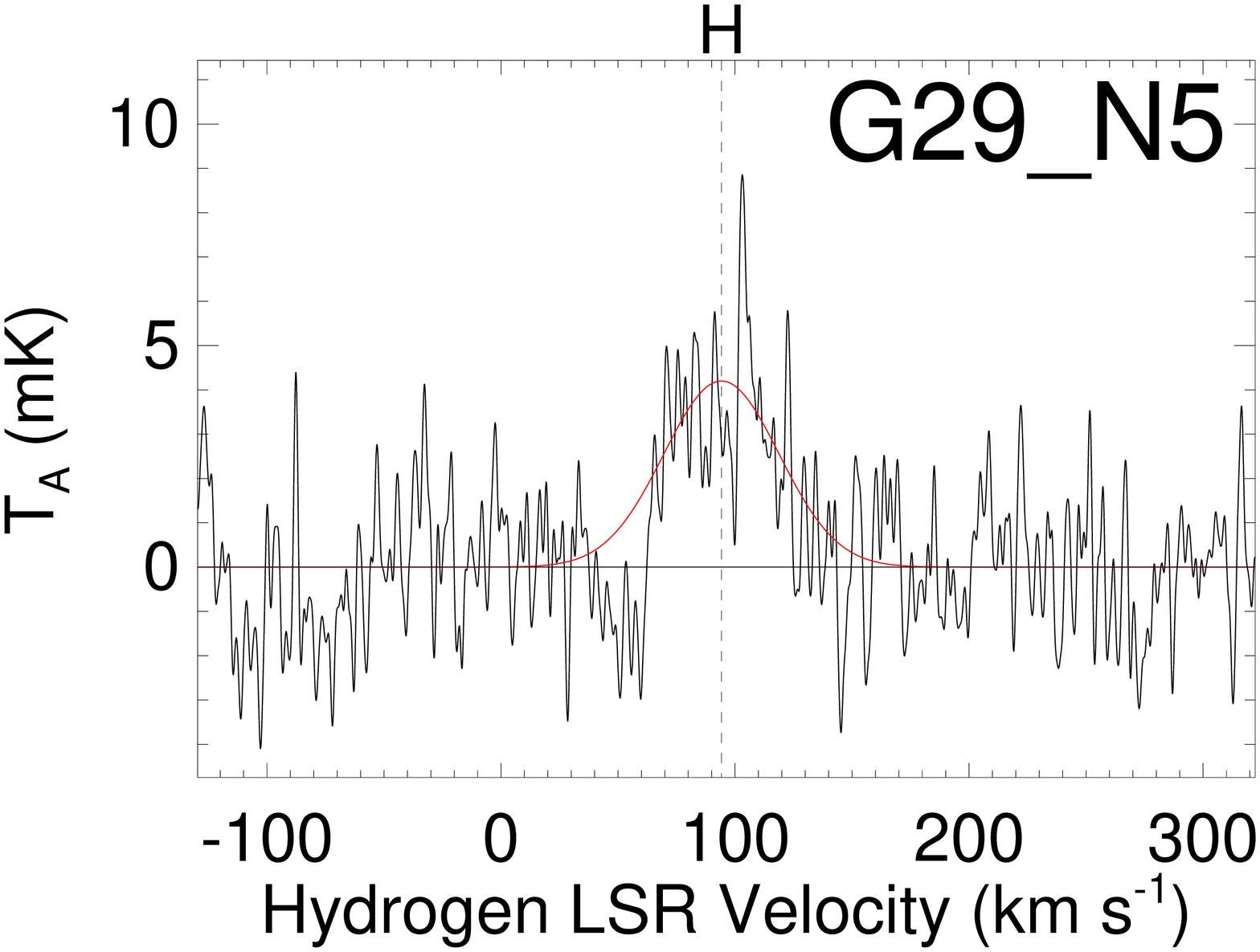} &
\includegraphics[width=.23\textwidth]{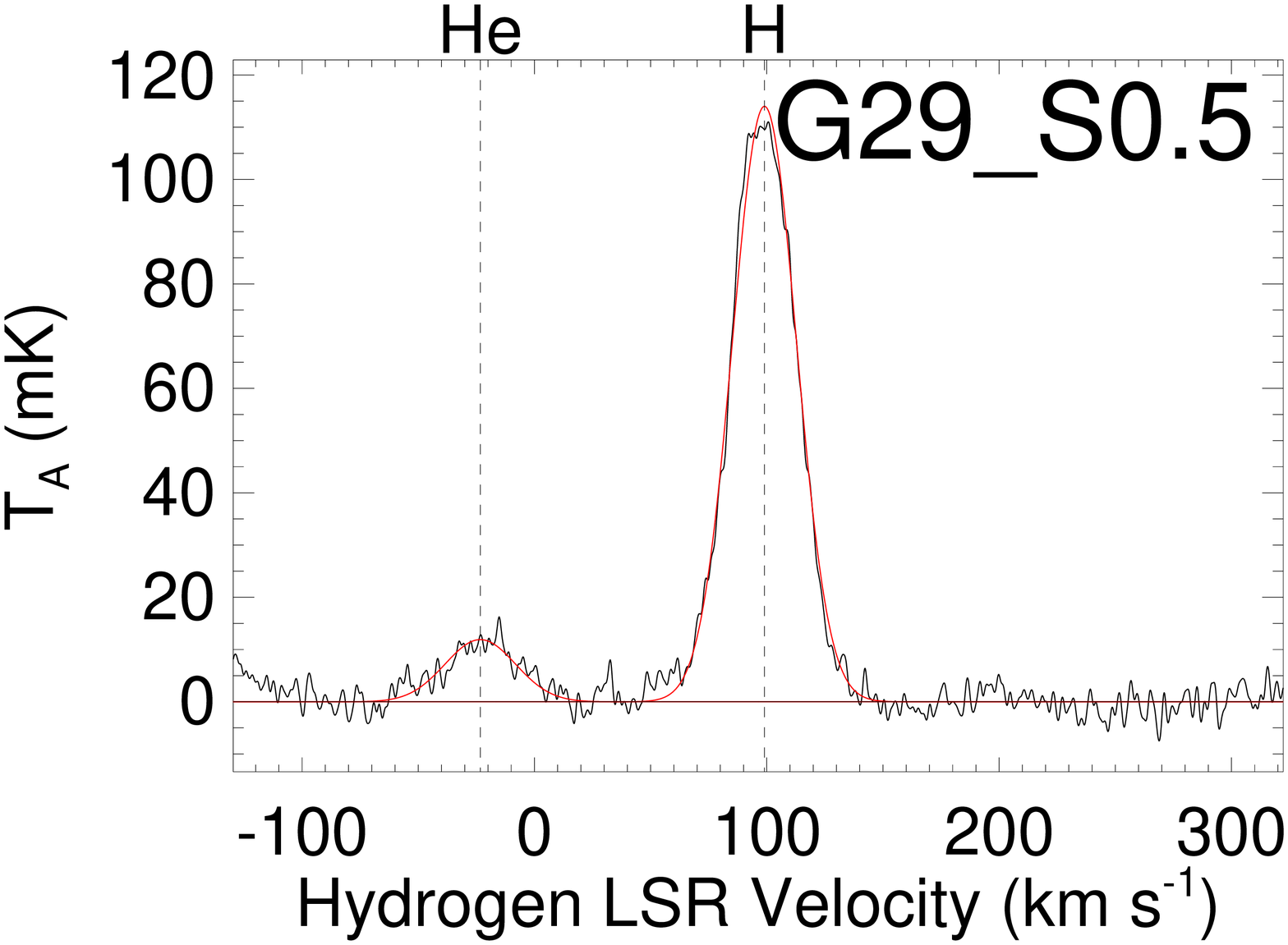} \\
\includegraphics[width=.23\textwidth]{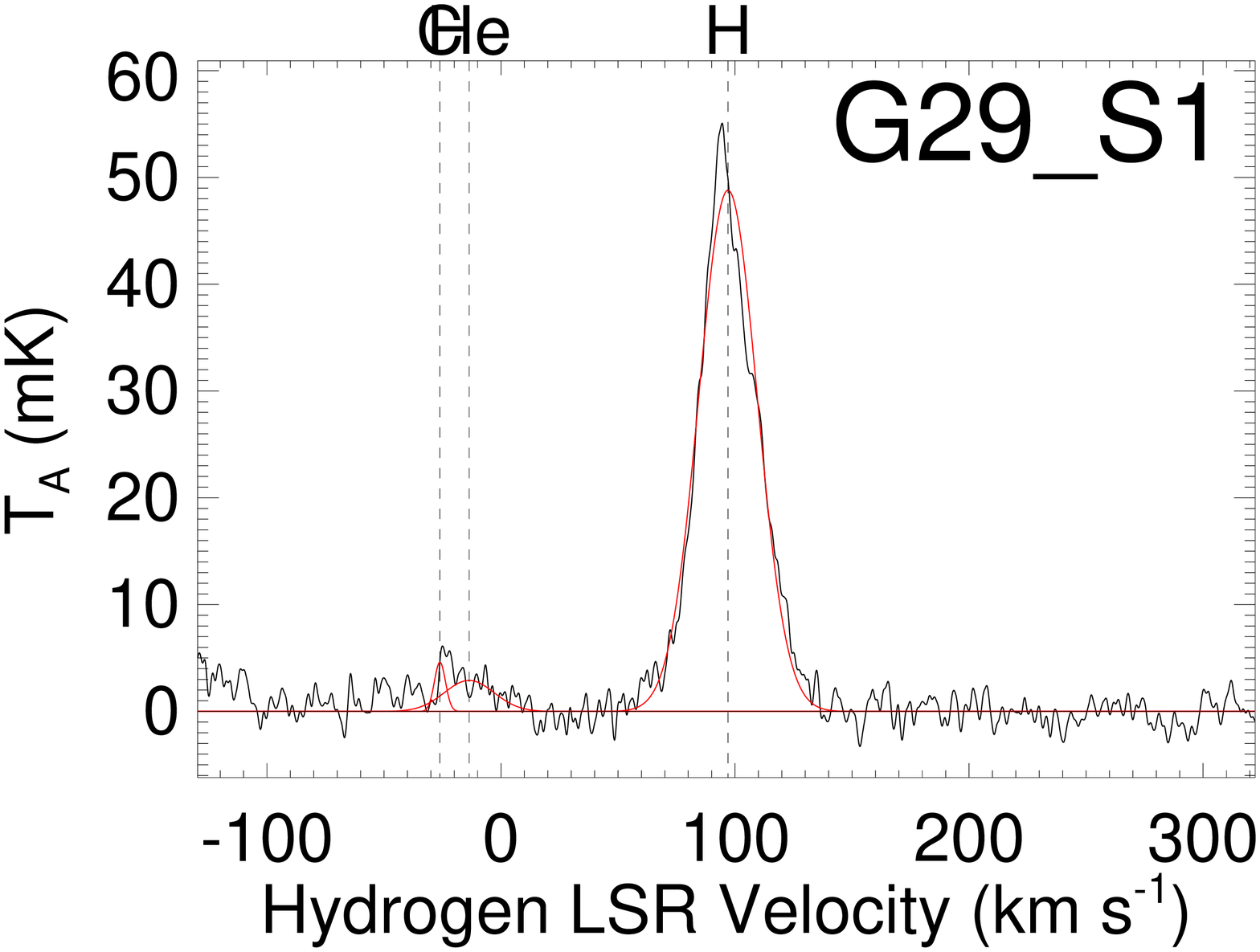} &
\includegraphics[width=.23\textwidth]{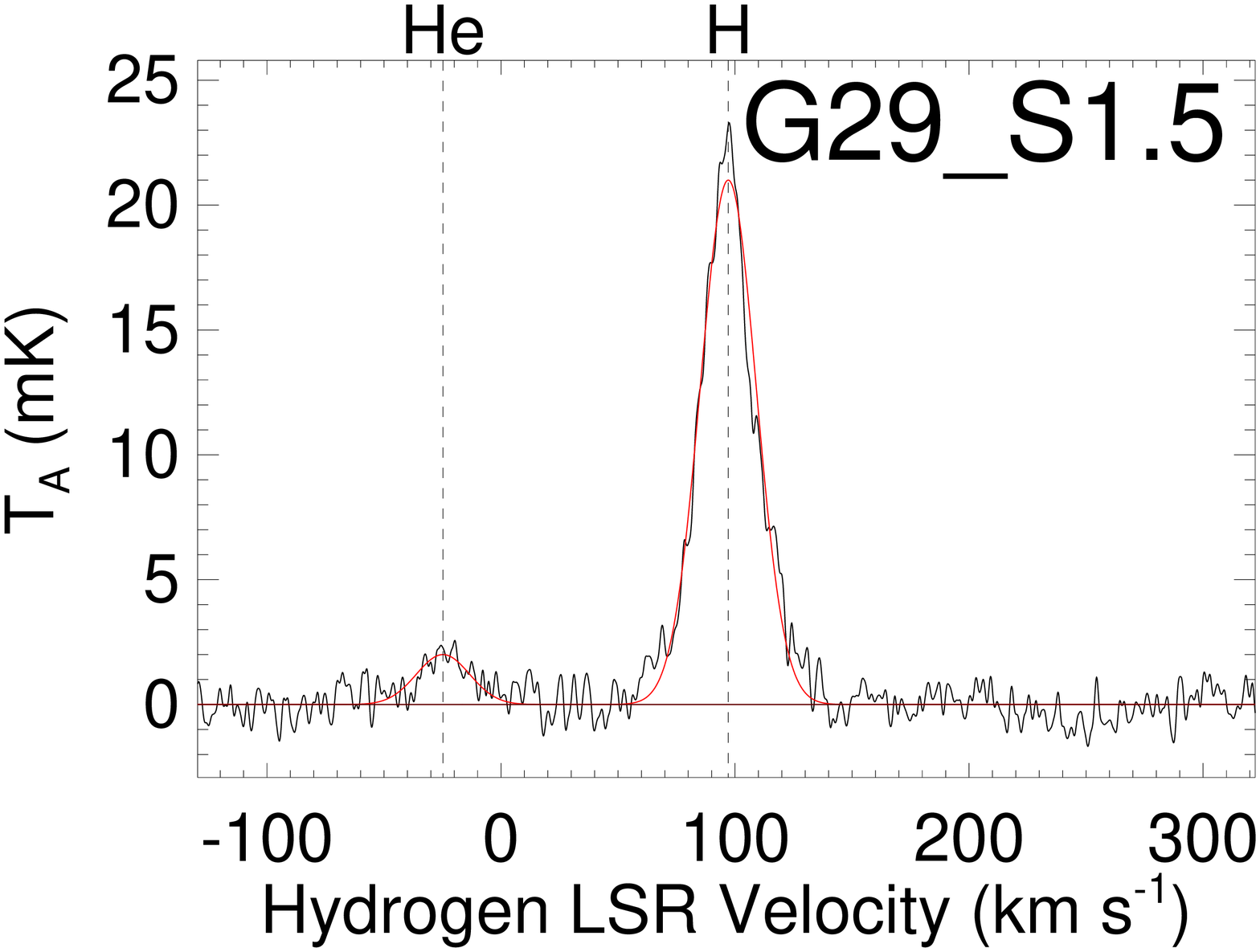} &
\includegraphics[width=.23\textwidth]{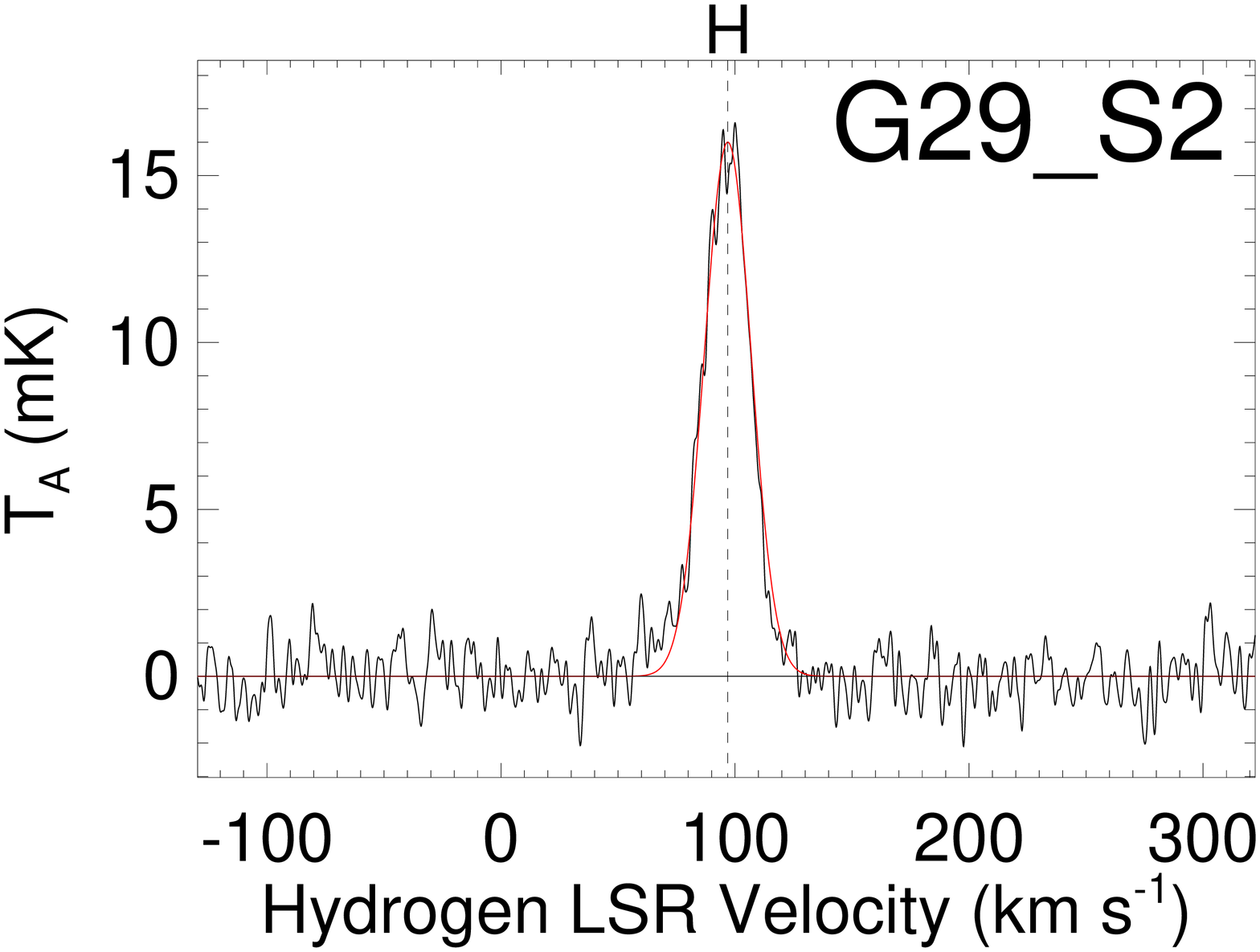} &
\includegraphics[width=.23\textwidth]{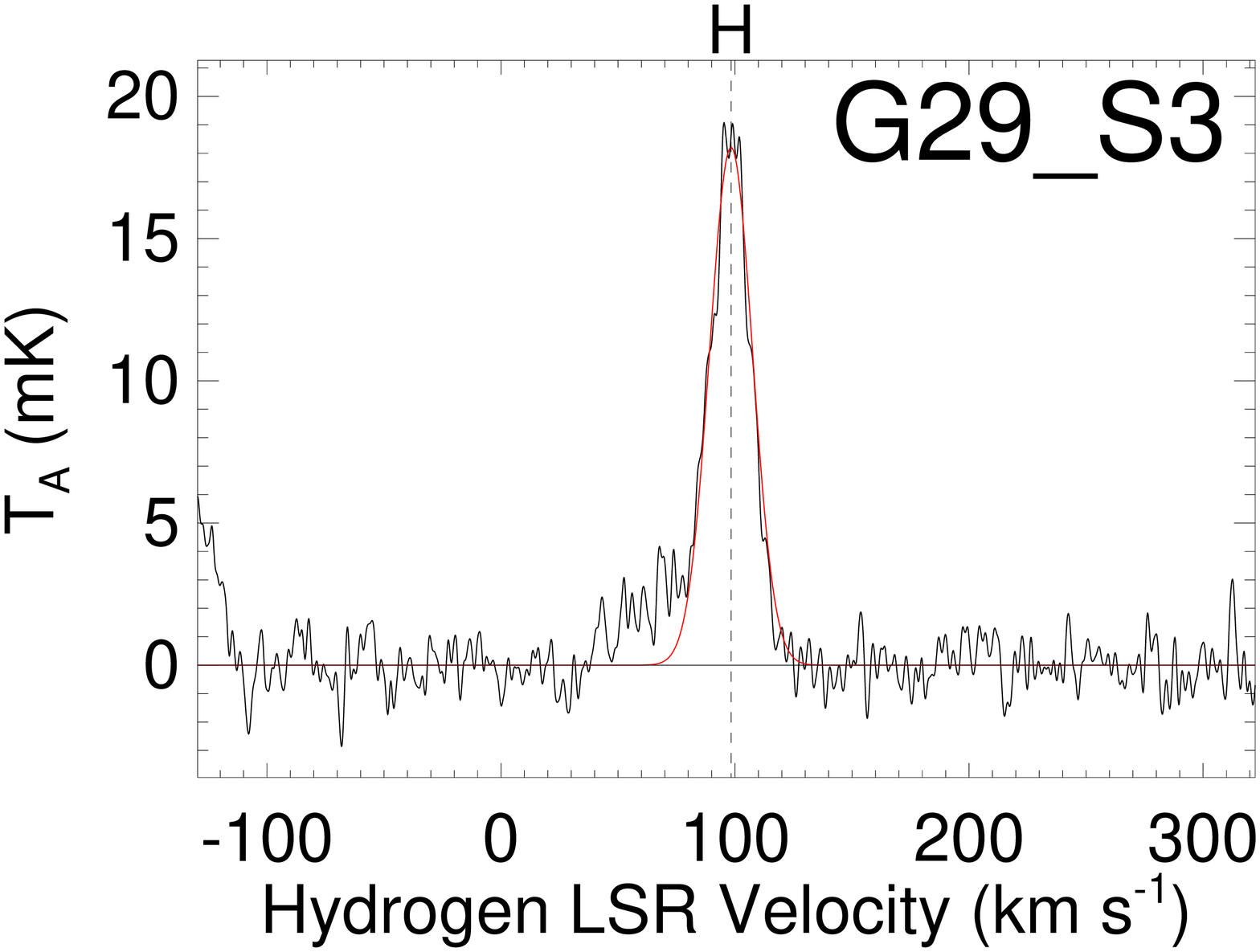} \\
\includegraphics[width=.23\textwidth]{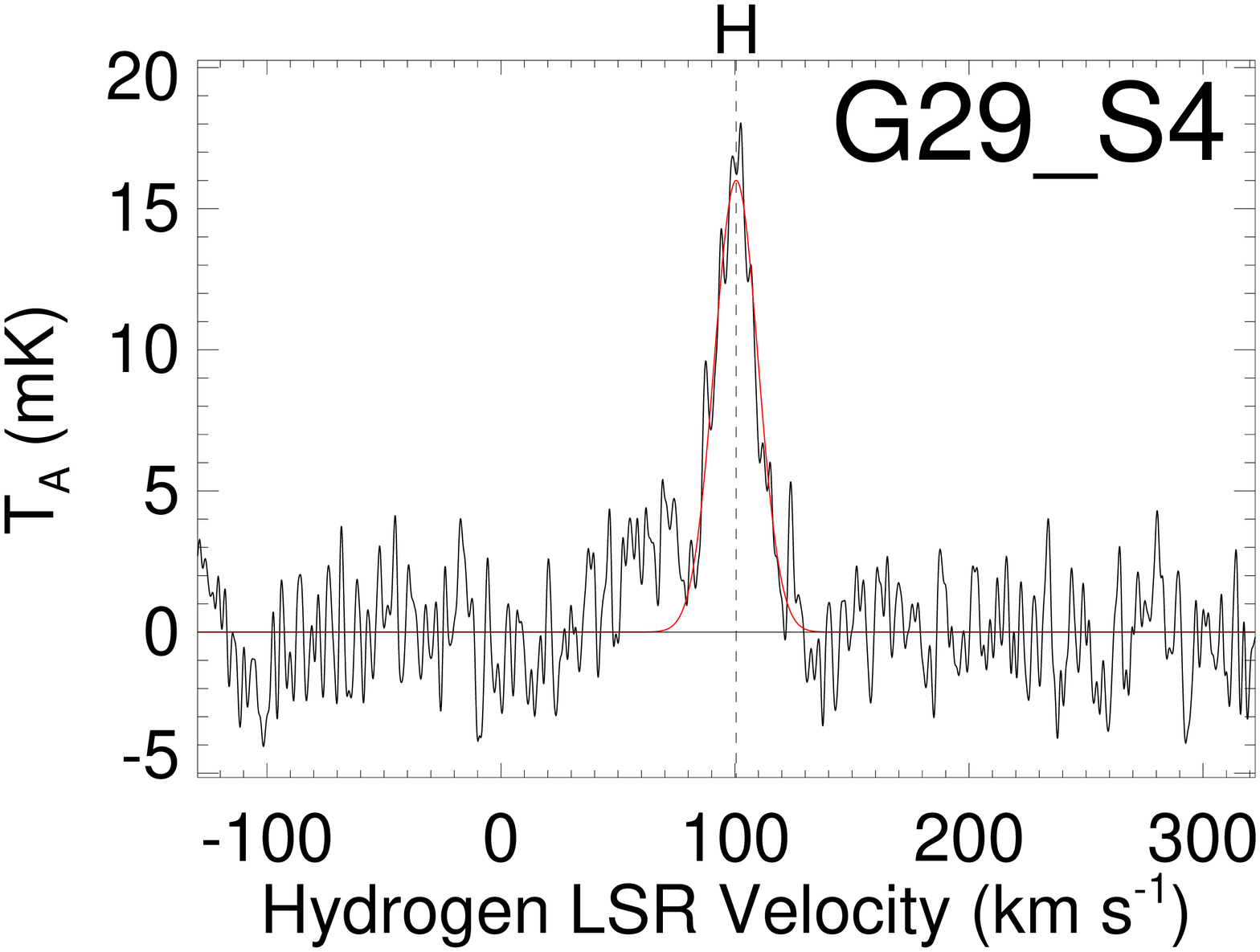} &
\includegraphics[width=.23\textwidth]{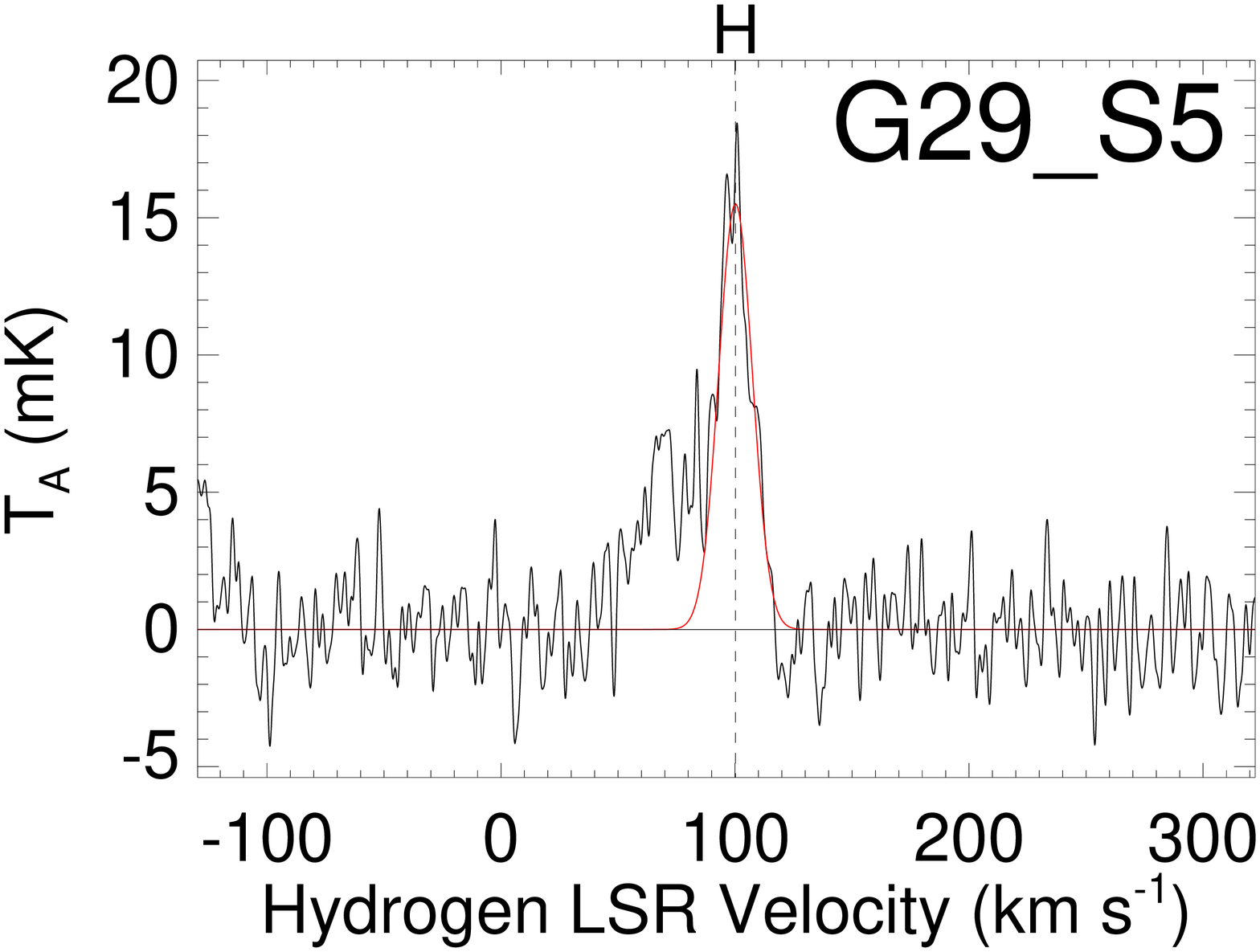} &
&
 \\
\end{tabular}
\caption{}
\end{figure*}
\renewcommand{\thefigure}{\thesection.\arabic{figure}}

\begin{figure*} 
\centering
\begin{tabular}{cccc}
\includegraphics[width=.23\textwidth]{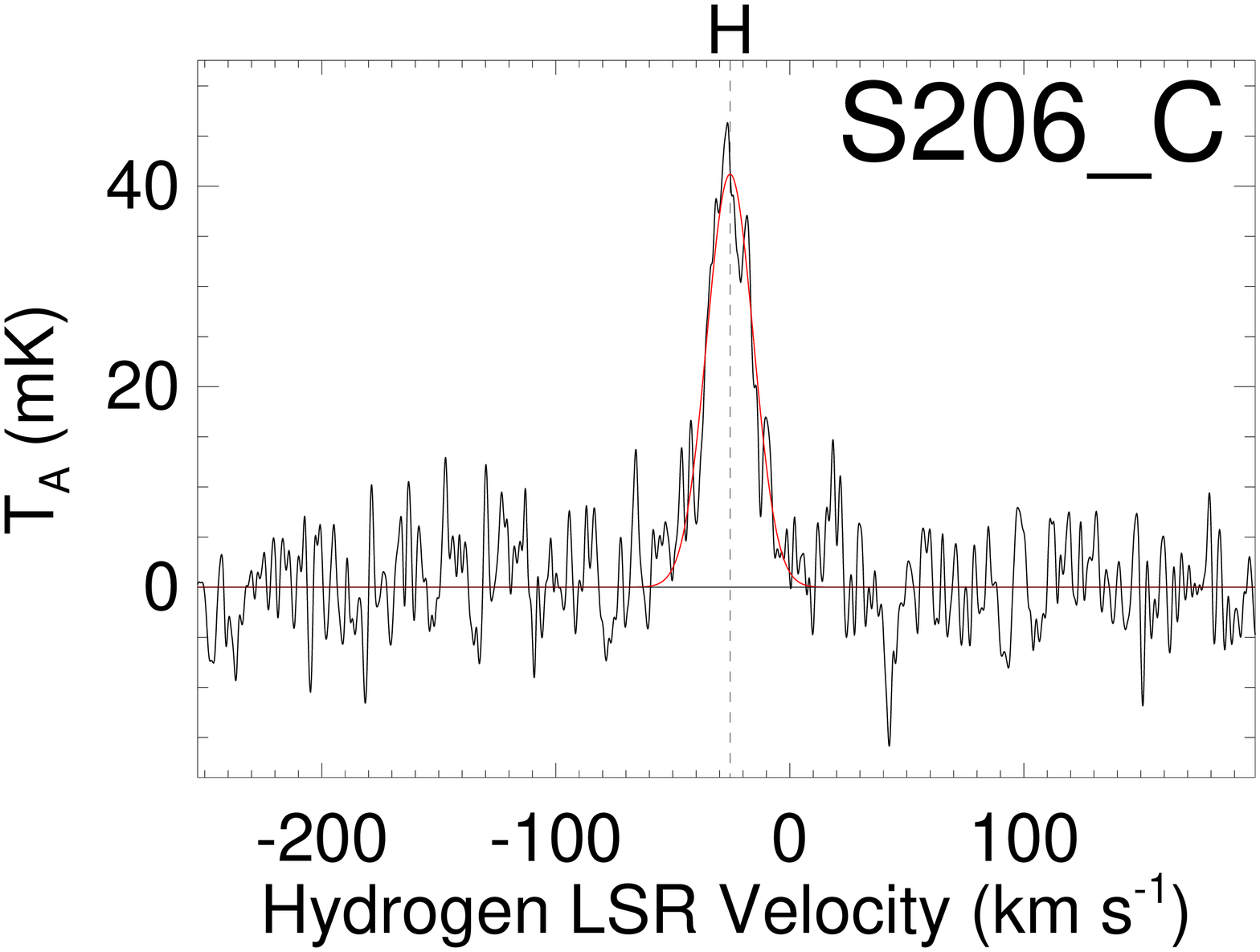} &
\includegraphics[width=.23\textwidth]{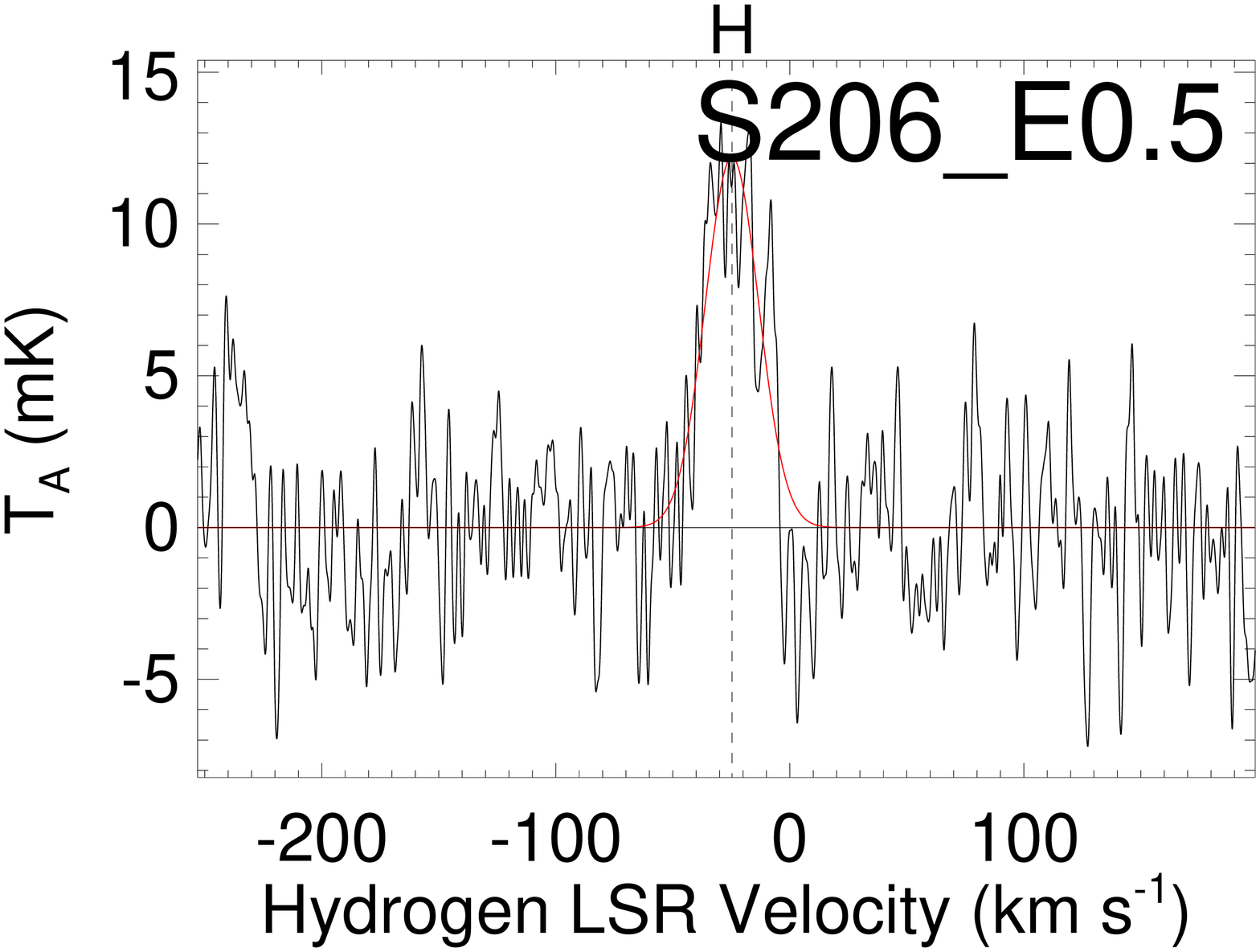} &
\includegraphics[width=.23\textwidth]{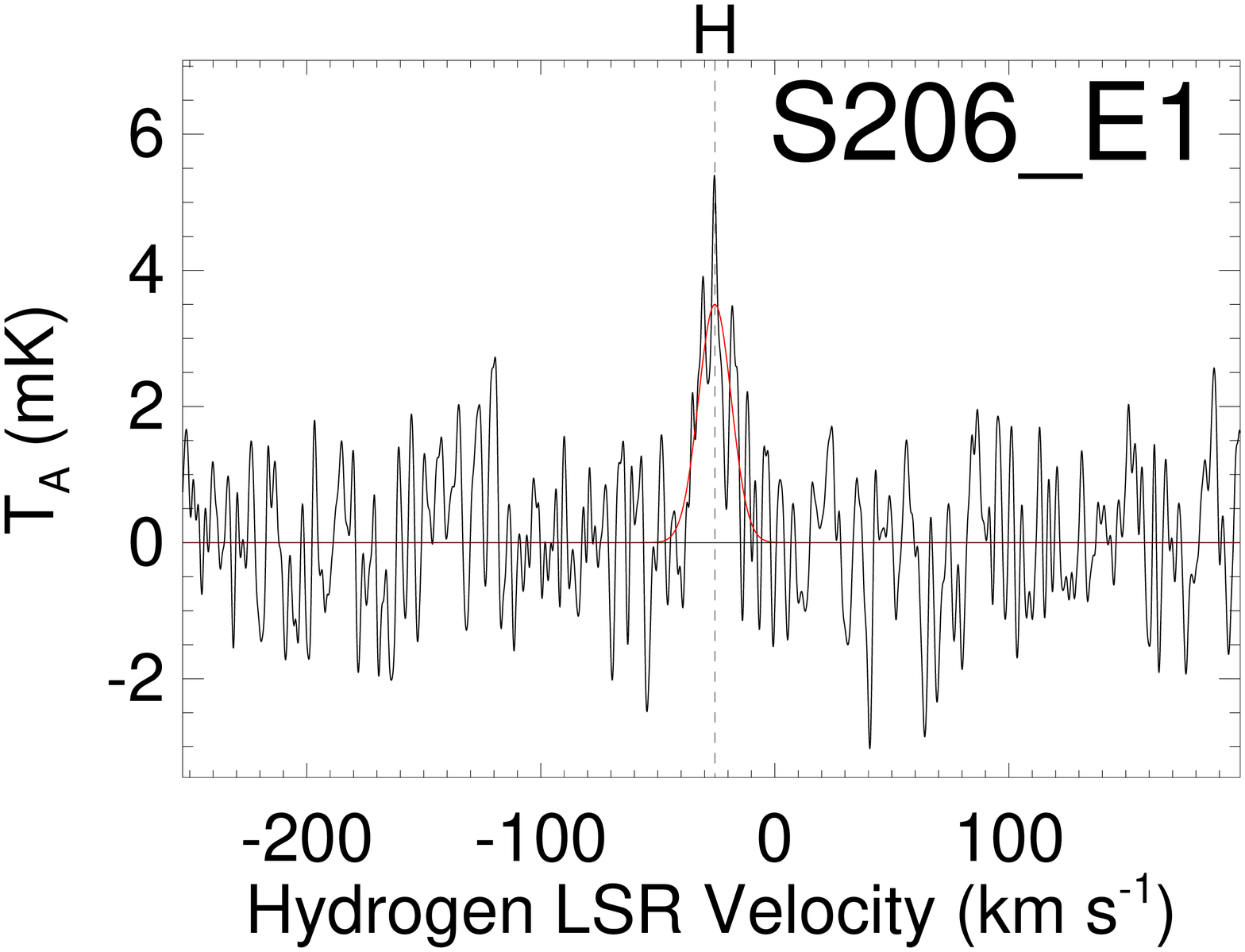} &
\includegraphics[width=.23\textwidth]{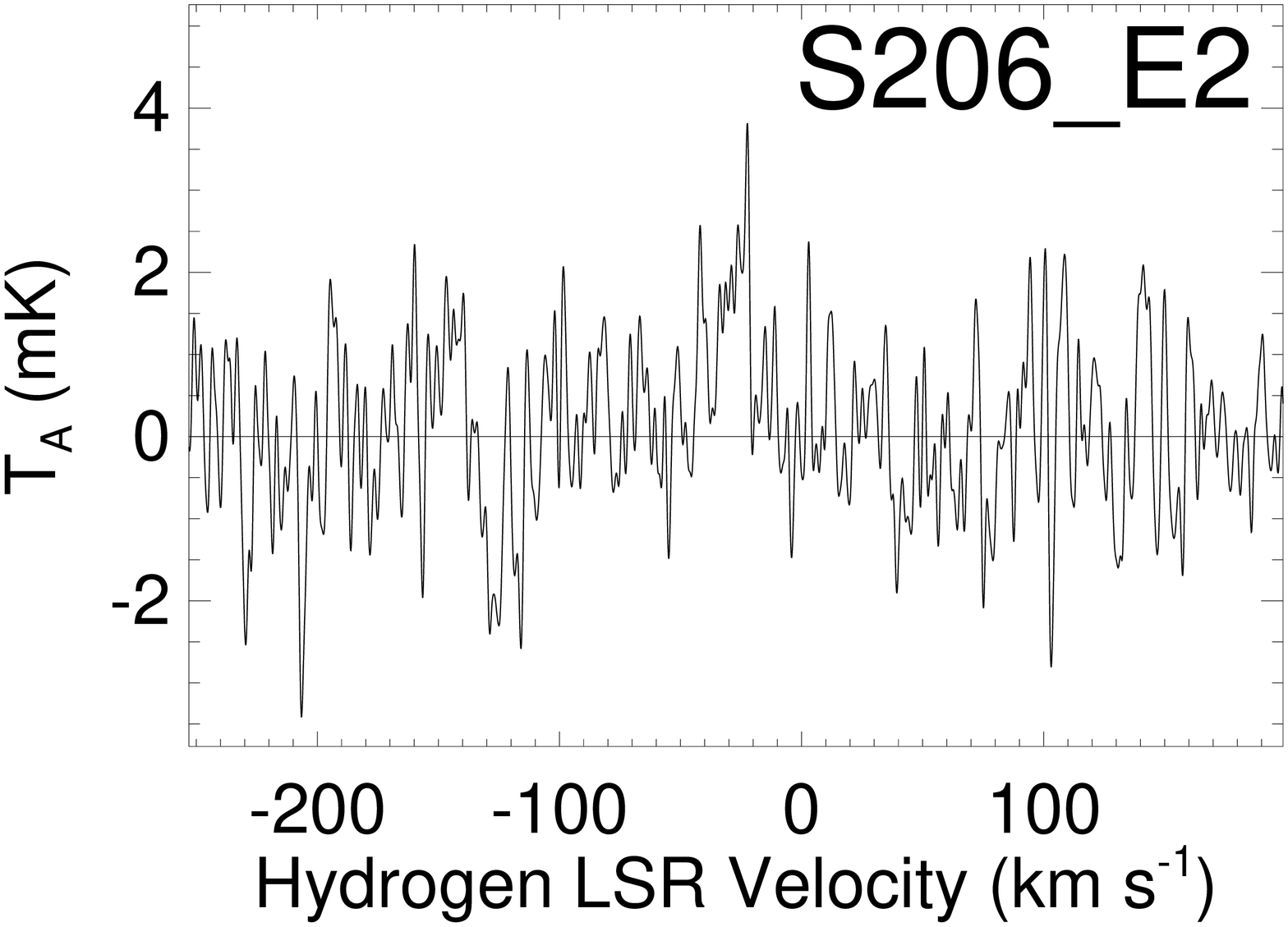} \\
\includegraphics[width=.23\textwidth]{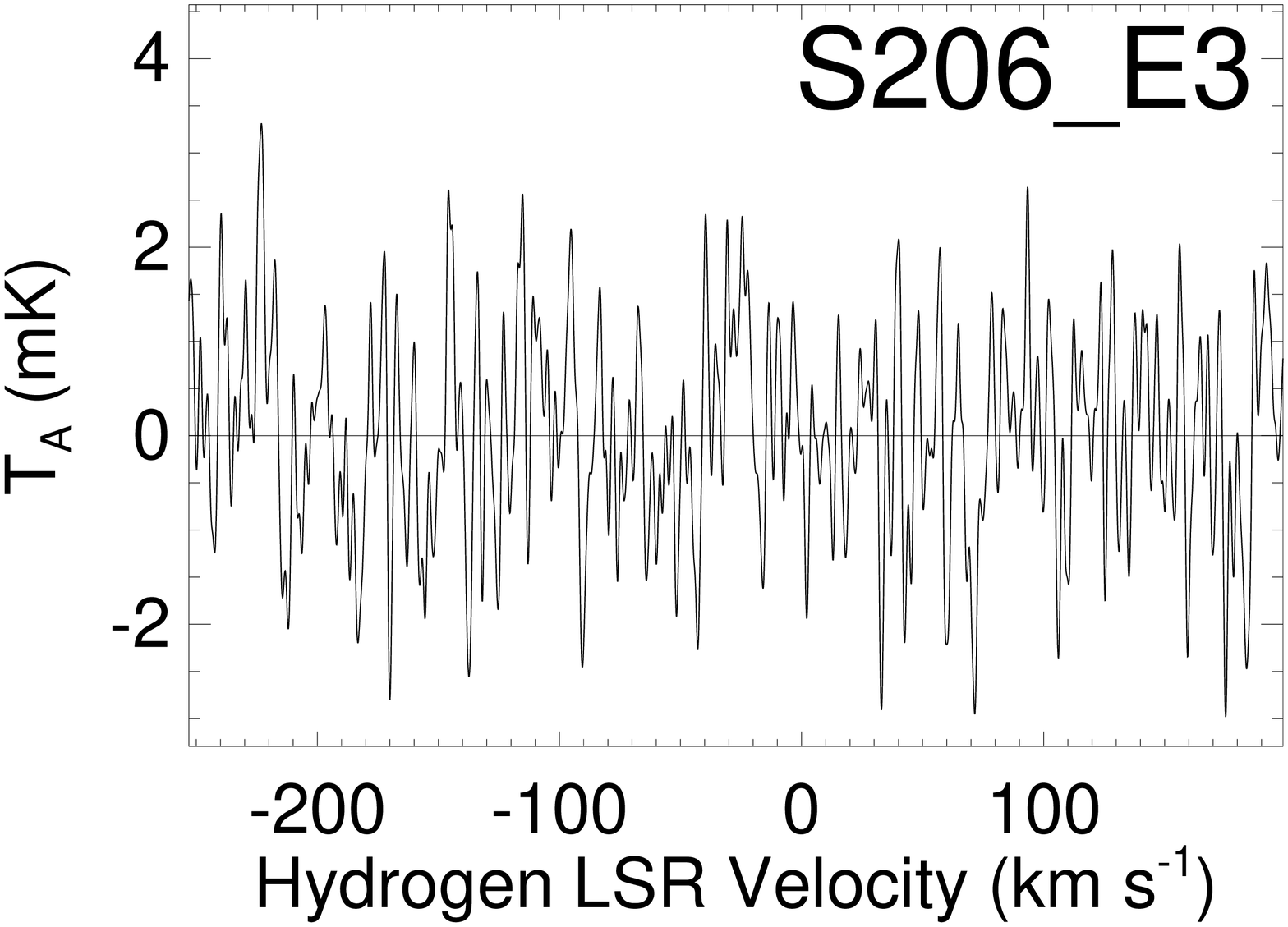} &
\includegraphics[width=.23\textwidth]{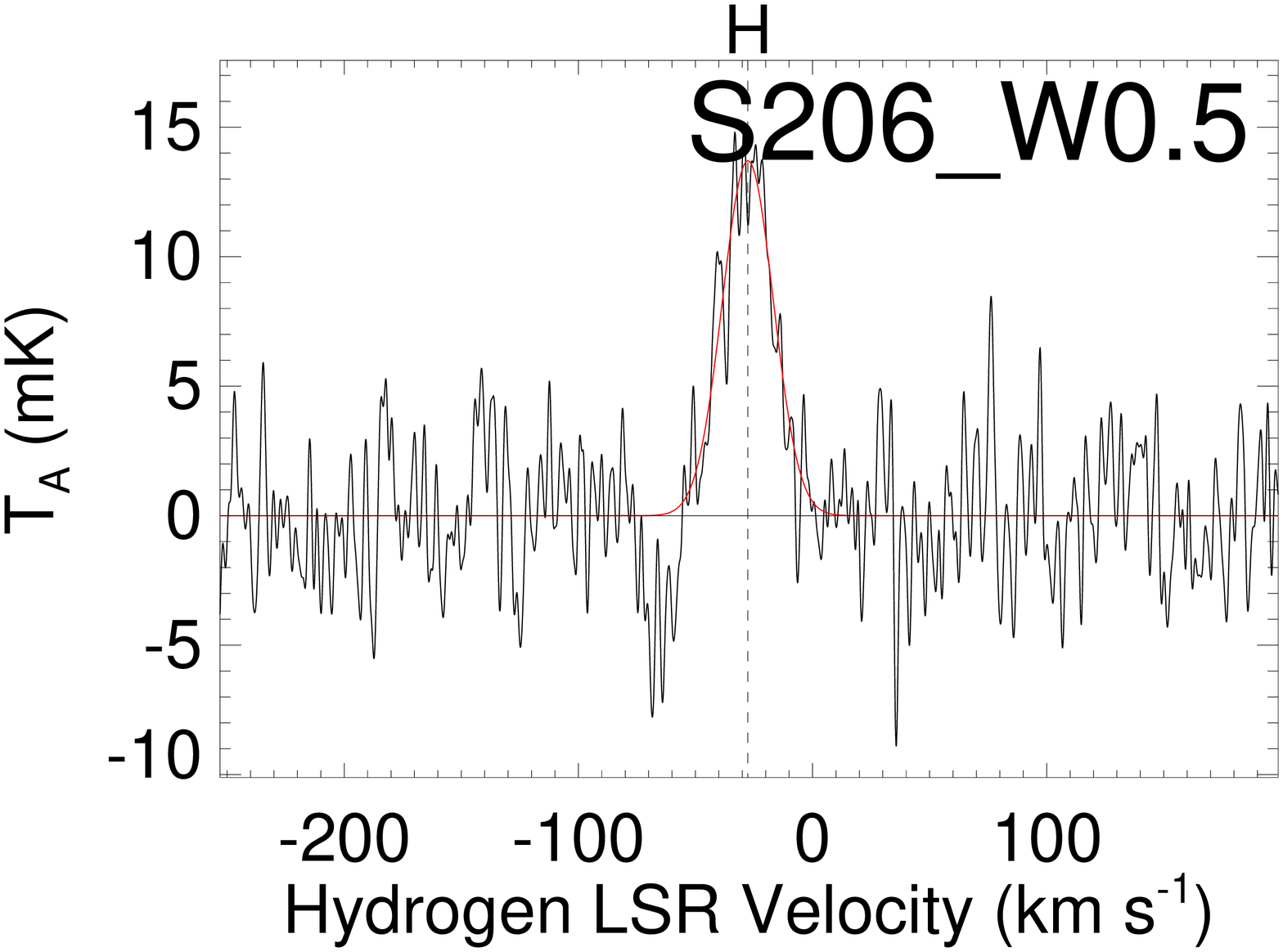} &
\includegraphics[width=.23\textwidth]{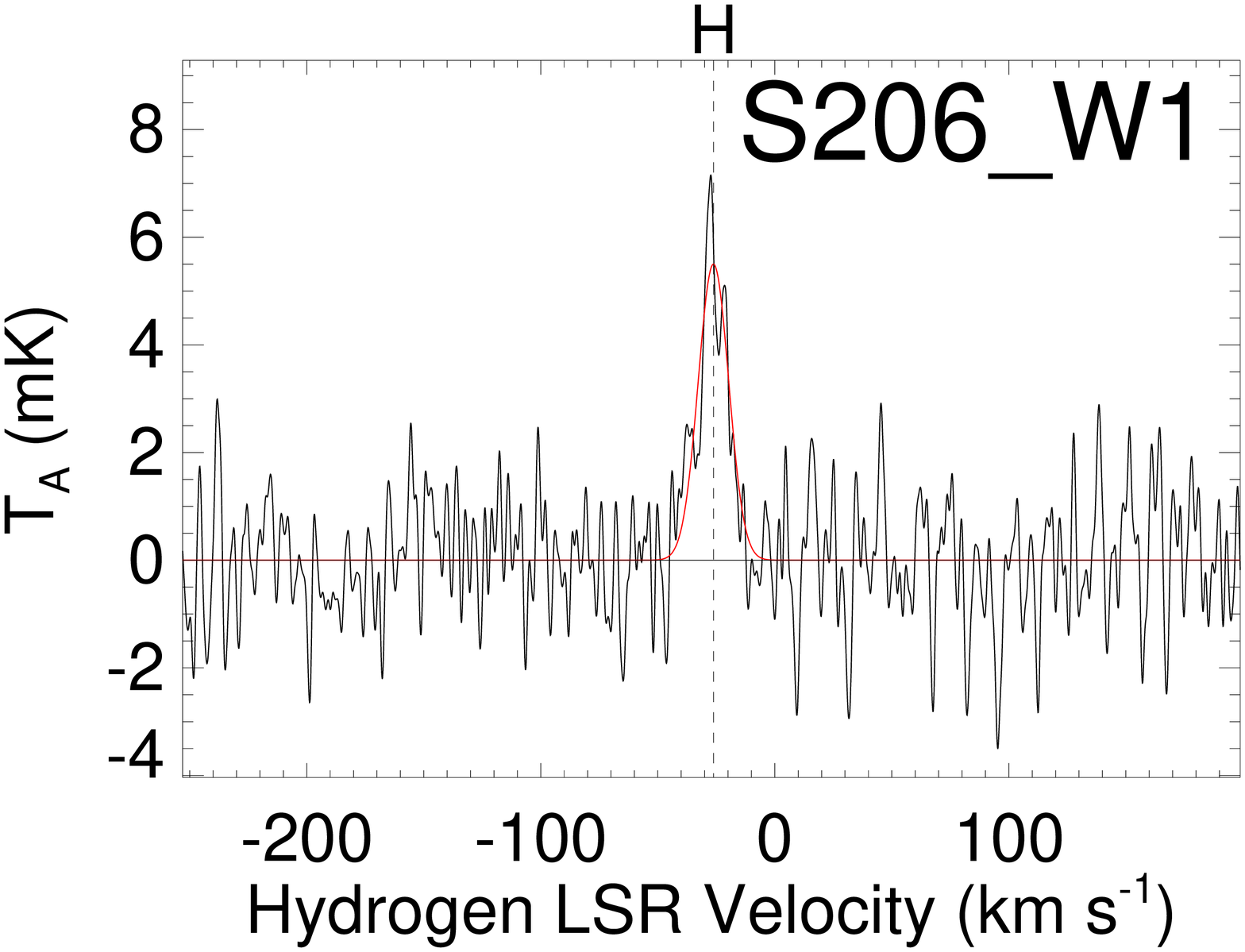} &
\includegraphics[width=.23\textwidth]{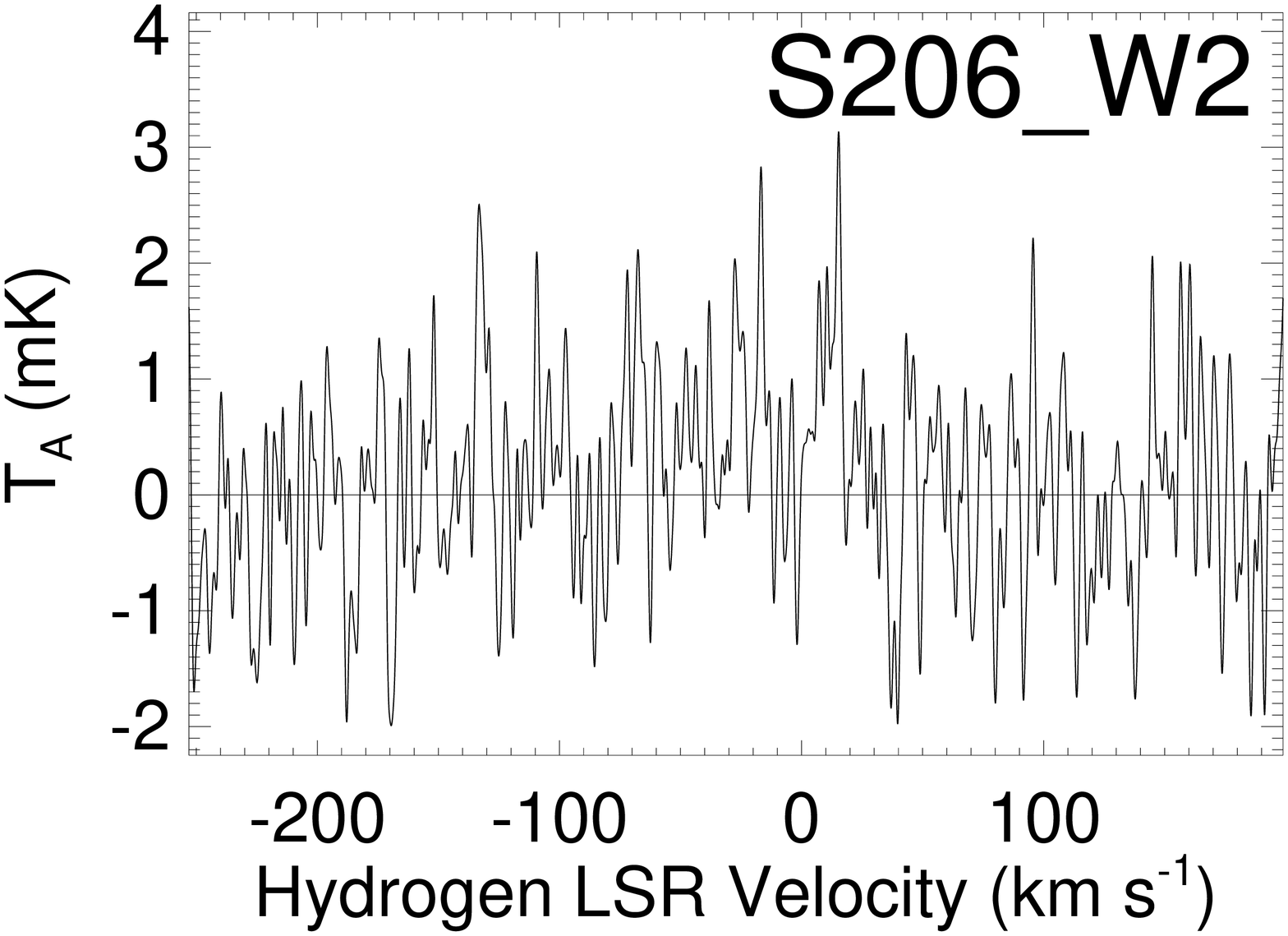} \\
\includegraphics[width=.23\textwidth]{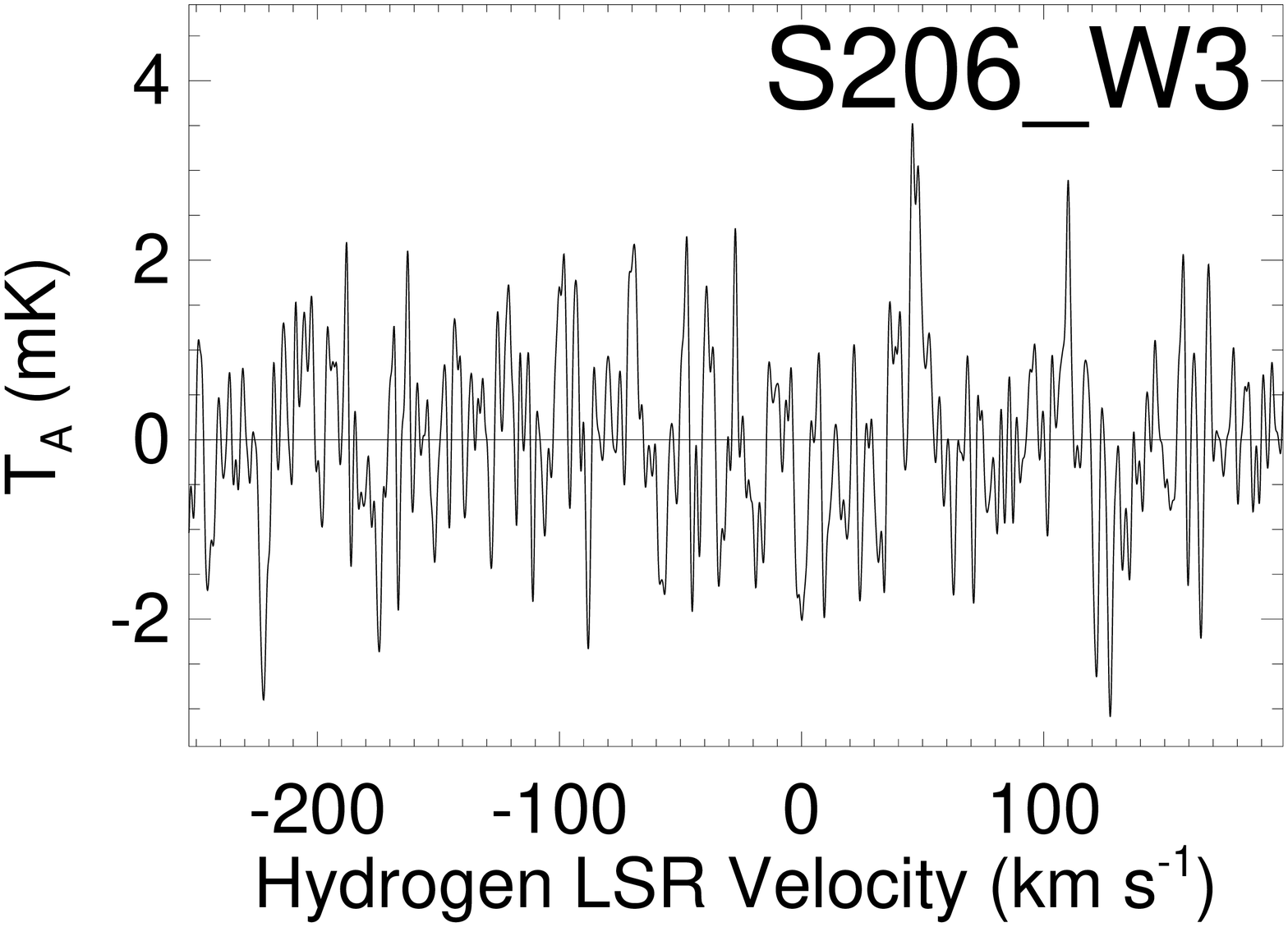} &
\includegraphics[width=.23\textwidth]{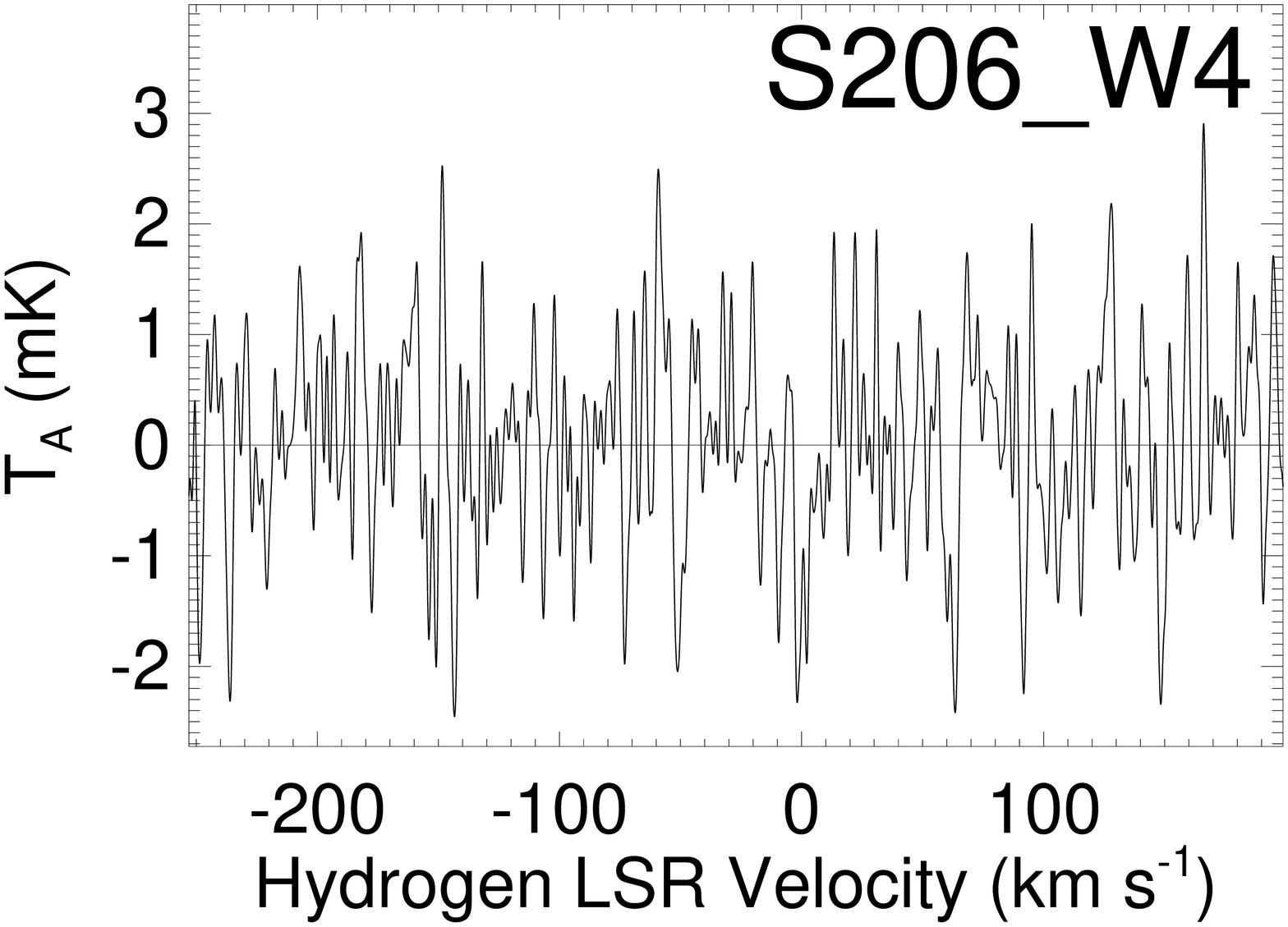} &
\includegraphics[width=.23\textwidth]{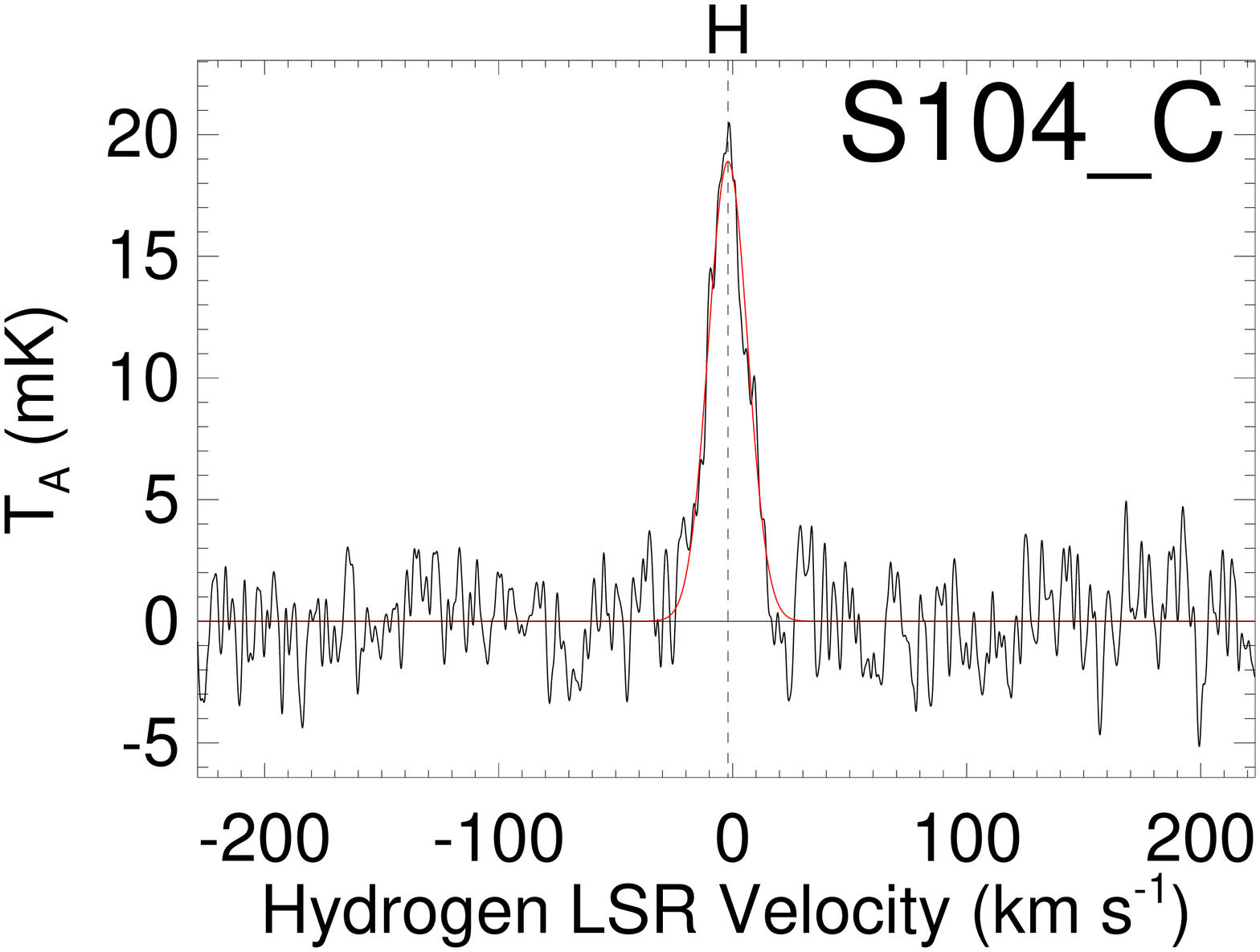} &
\includegraphics[width=.23\textwidth]{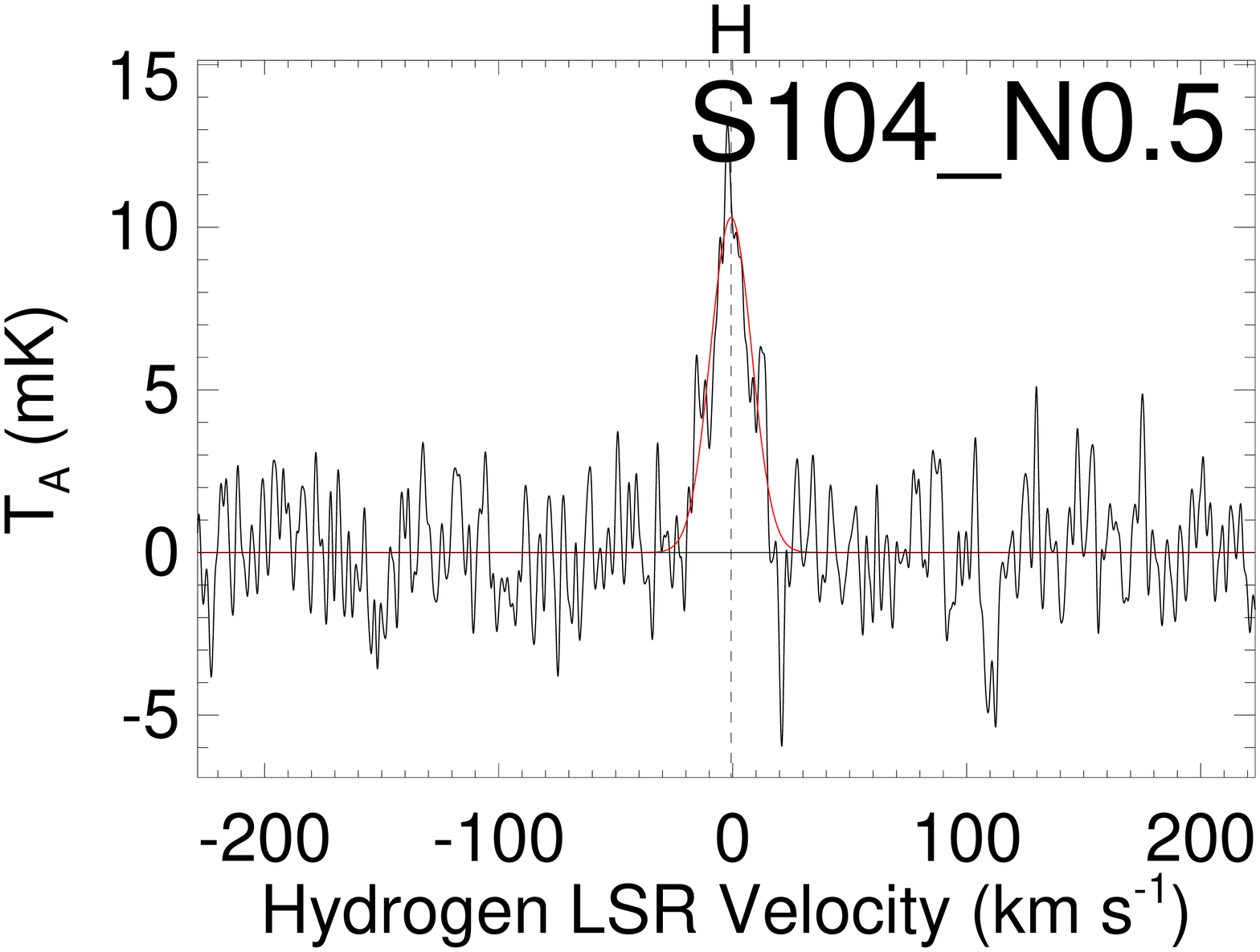} \\
\includegraphics[width=.23\textwidth]{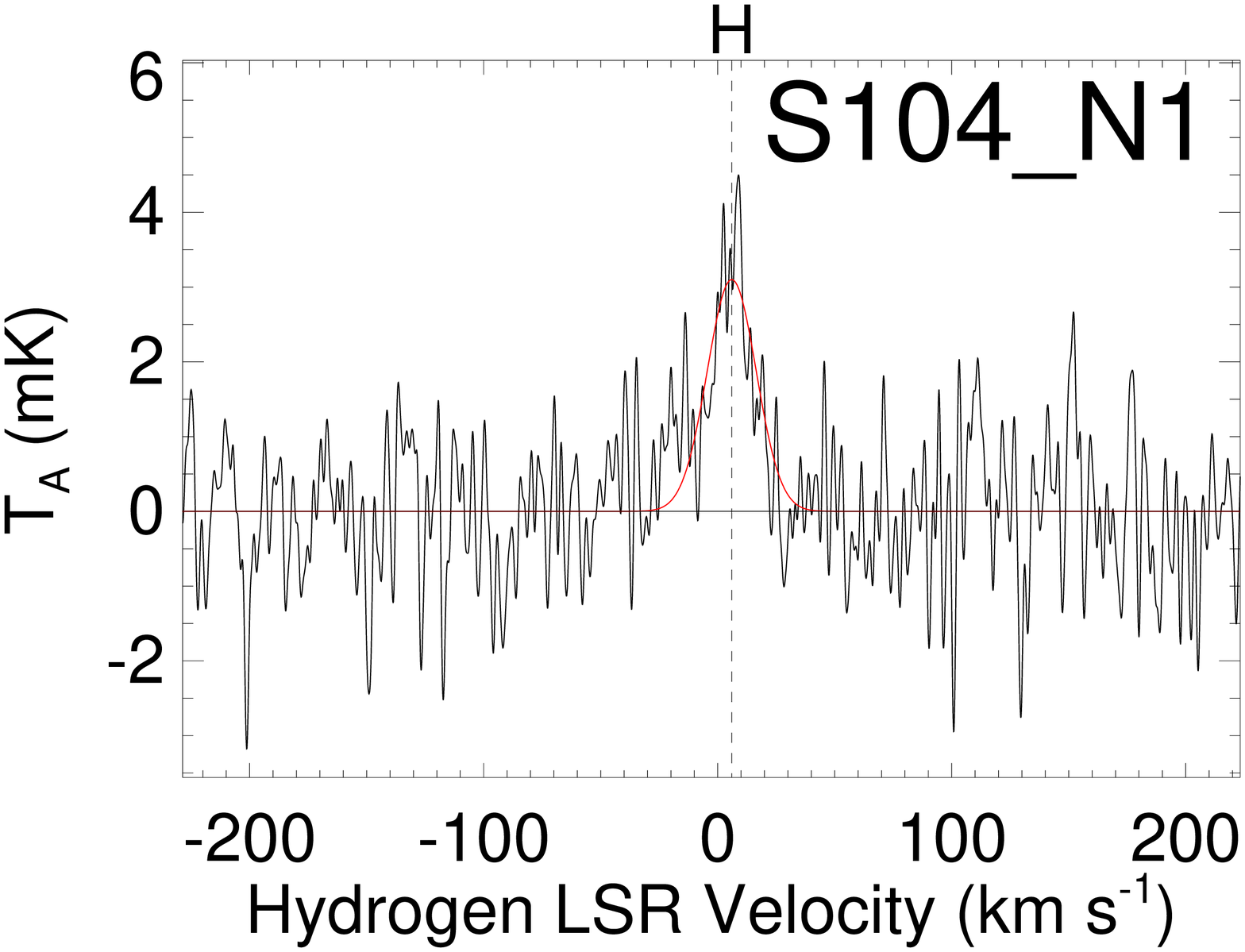} &
\includegraphics[width=.23\textwidth]{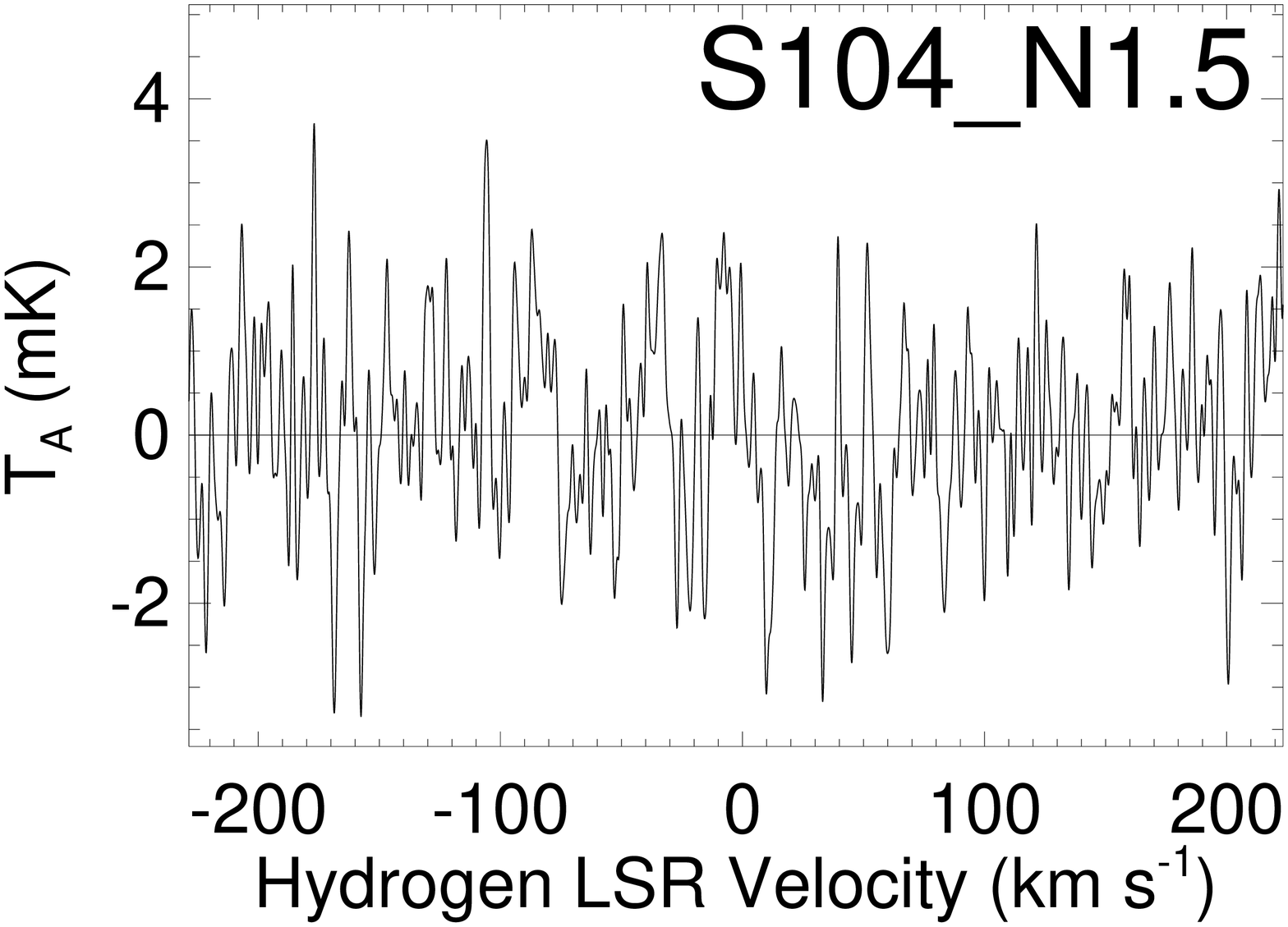} &
\includegraphics[width=.23\textwidth]{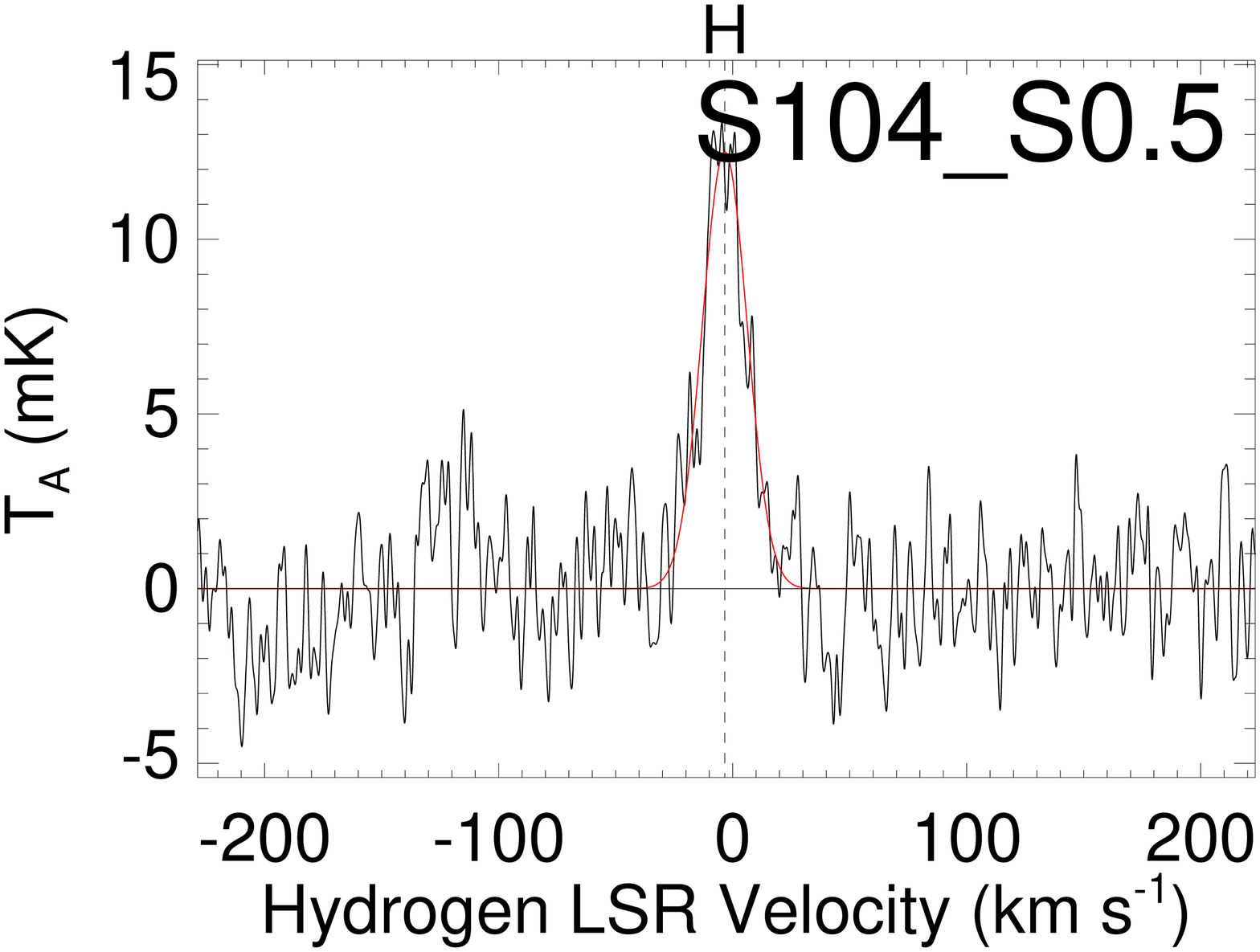} &
\includegraphics[width=.23\textwidth]{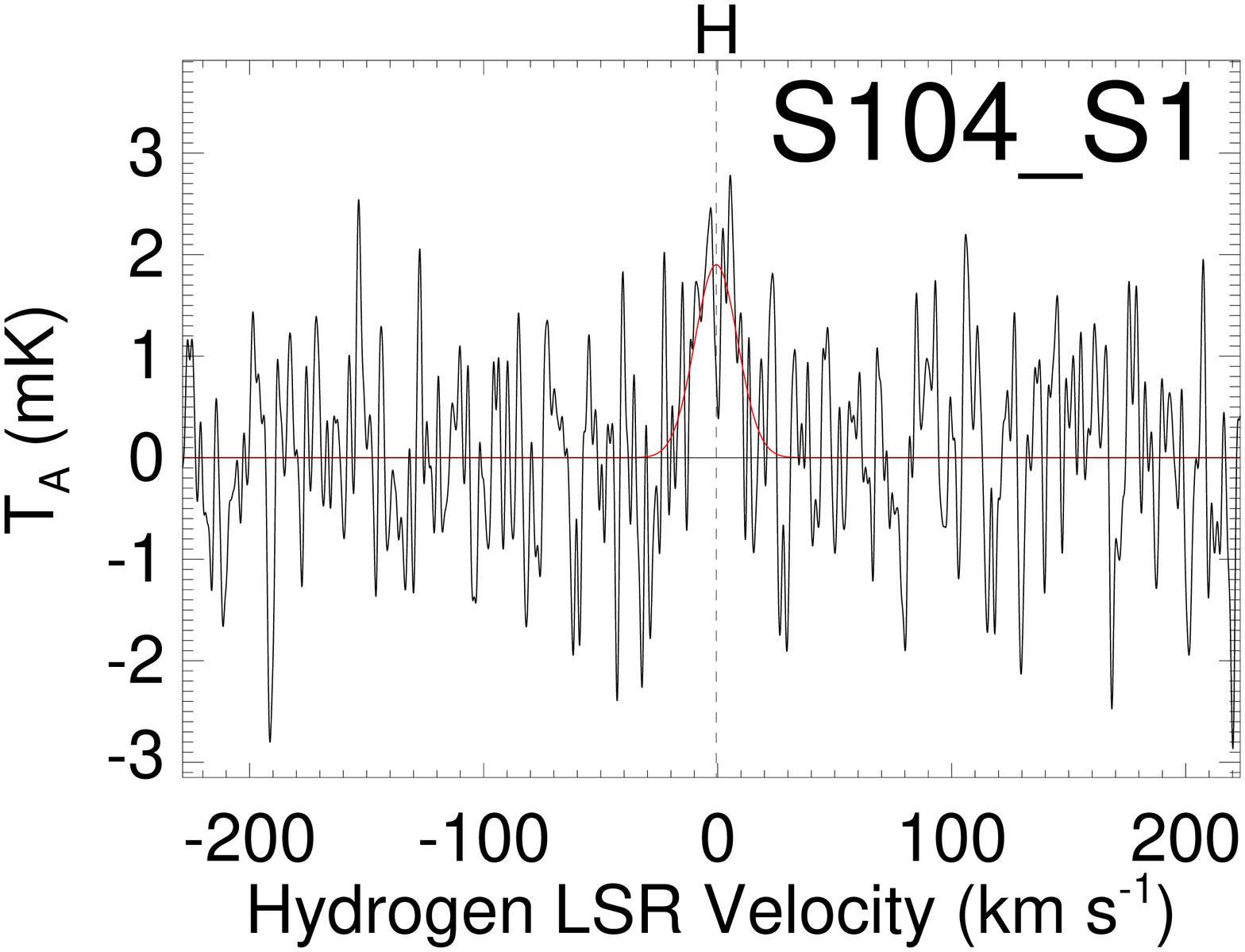} \\
\includegraphics[width=.23\textwidth]{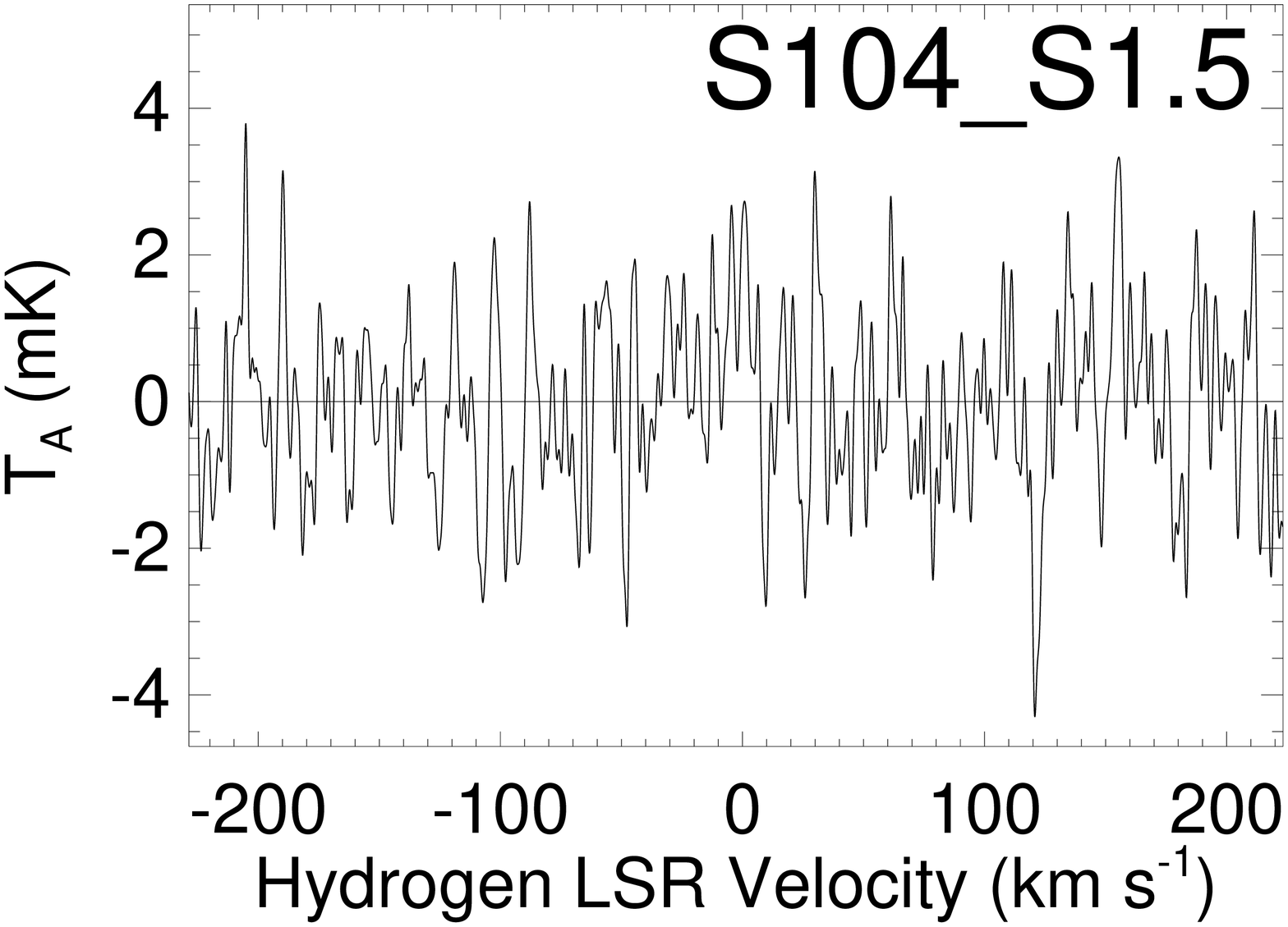} &
\includegraphics[width=.23\textwidth]{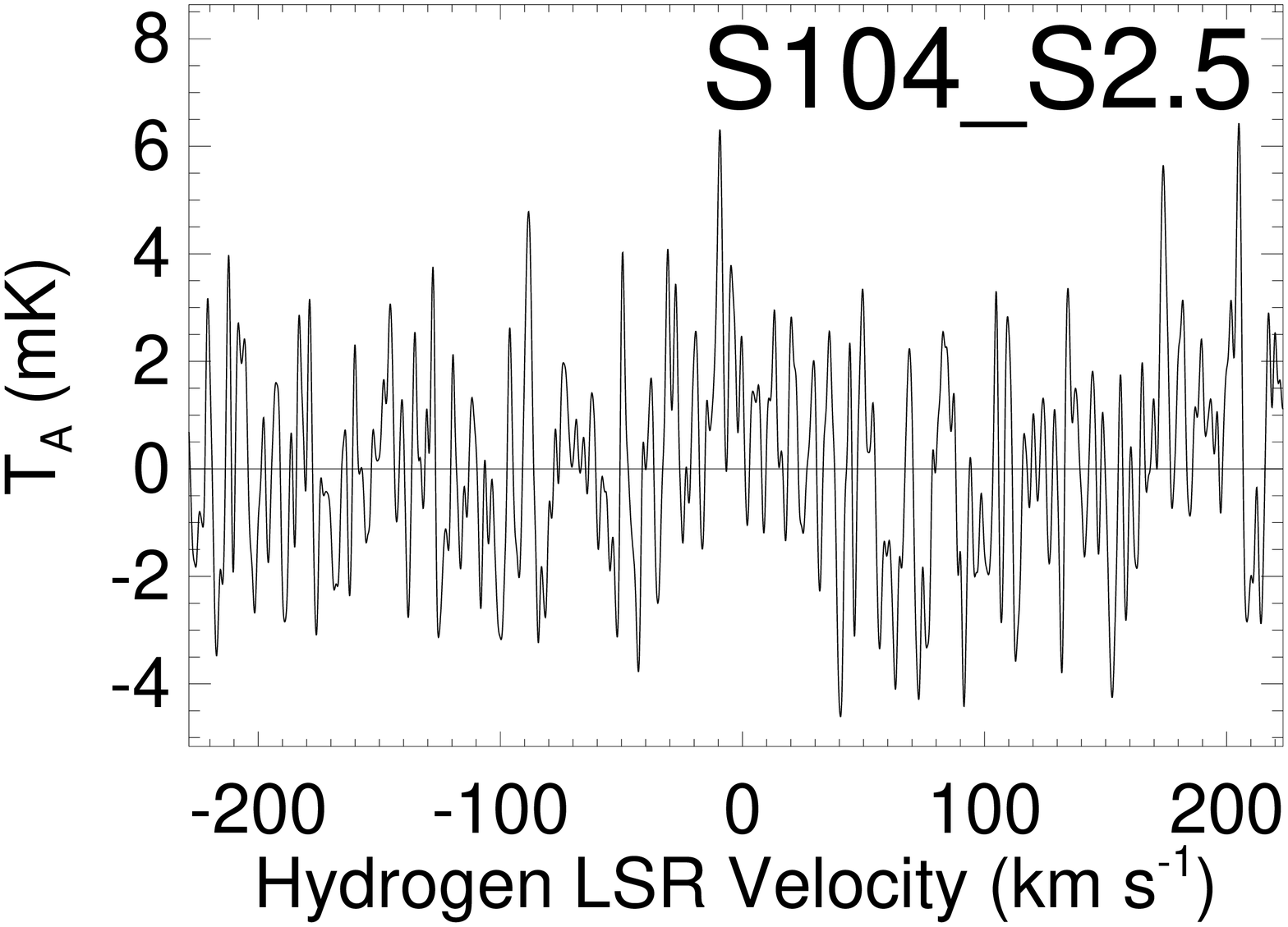} &
\includegraphics[width=.23\textwidth]{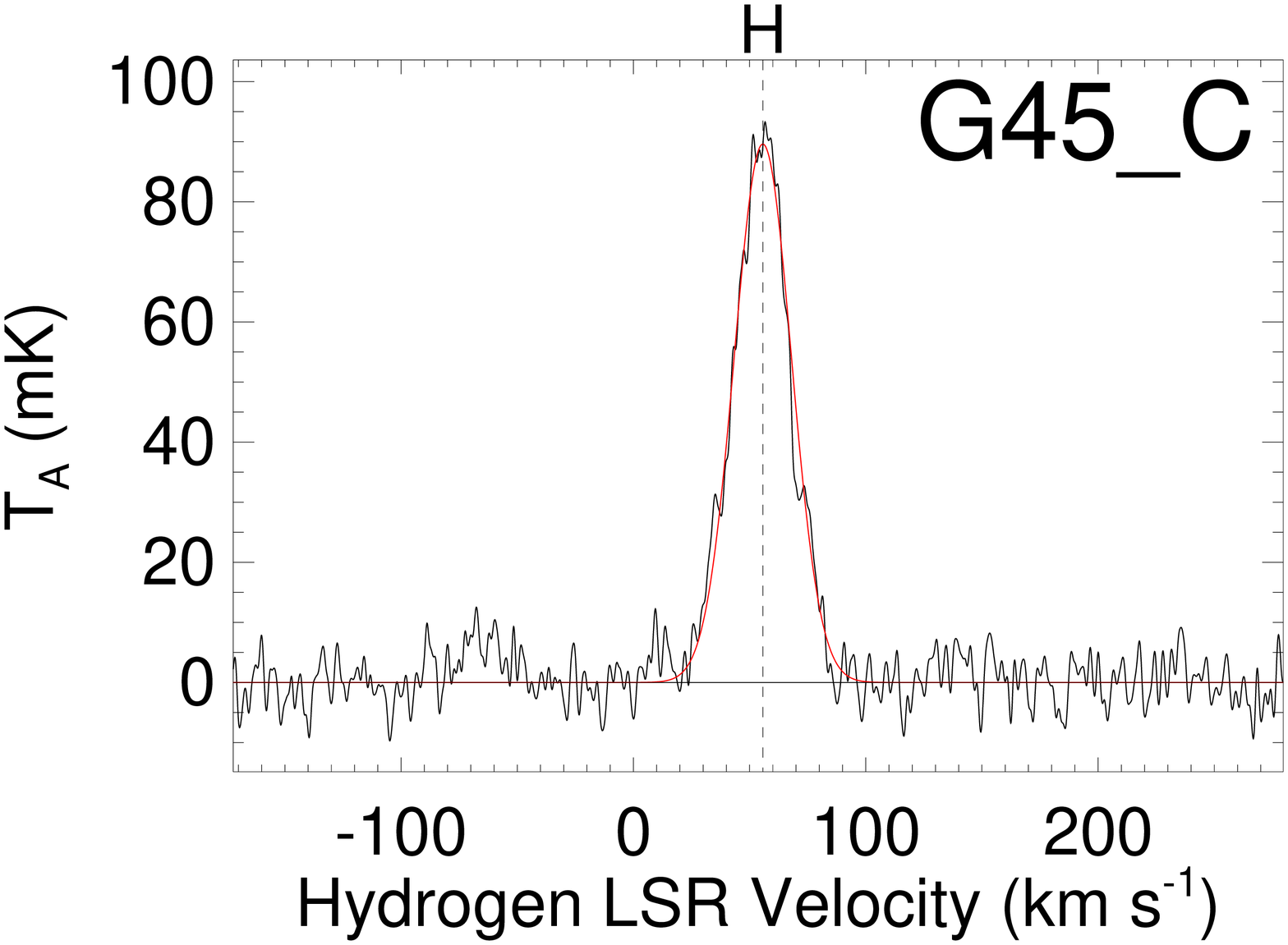} &
\includegraphics[width=.23\textwidth]{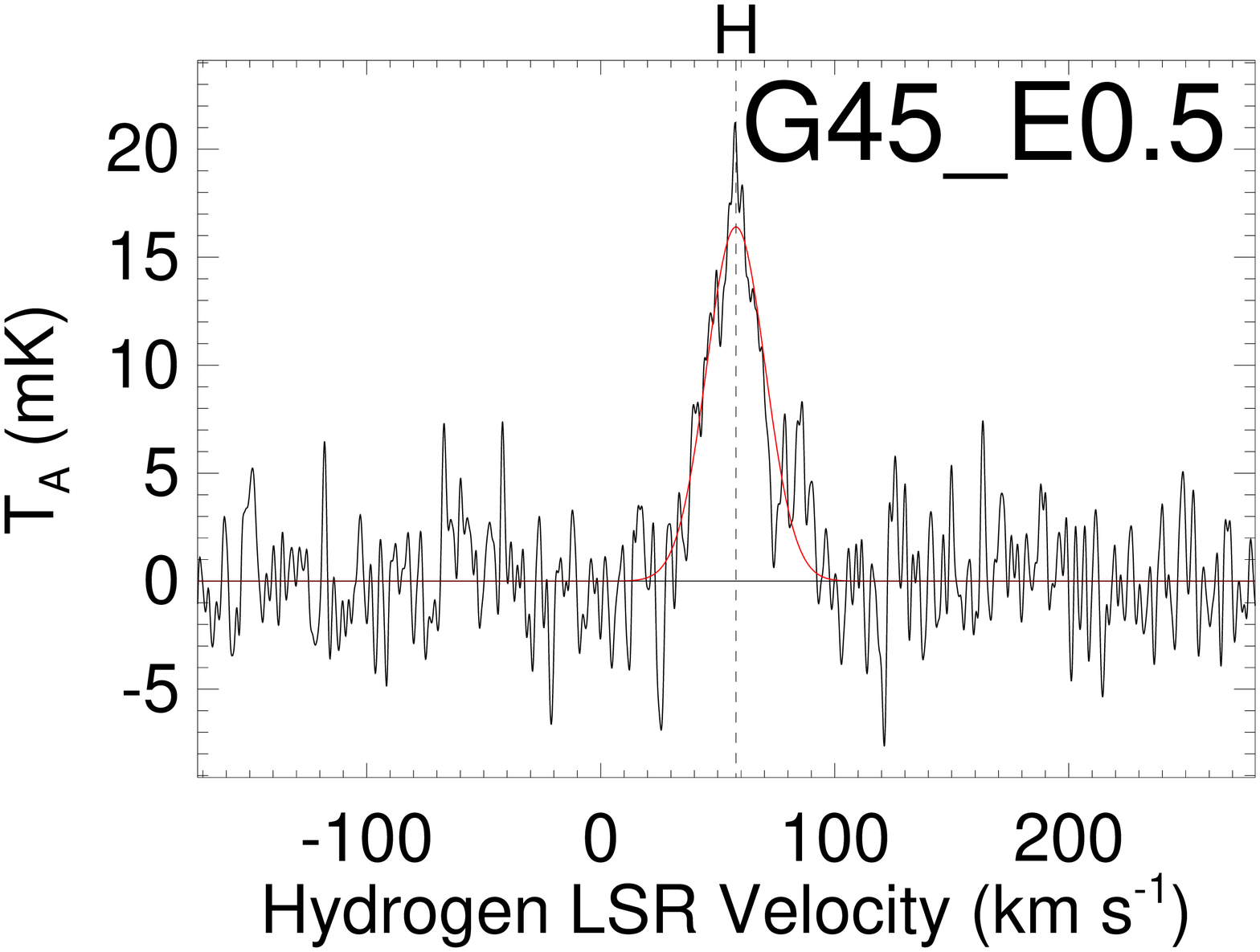} \\
\includegraphics[width=.23\textwidth]{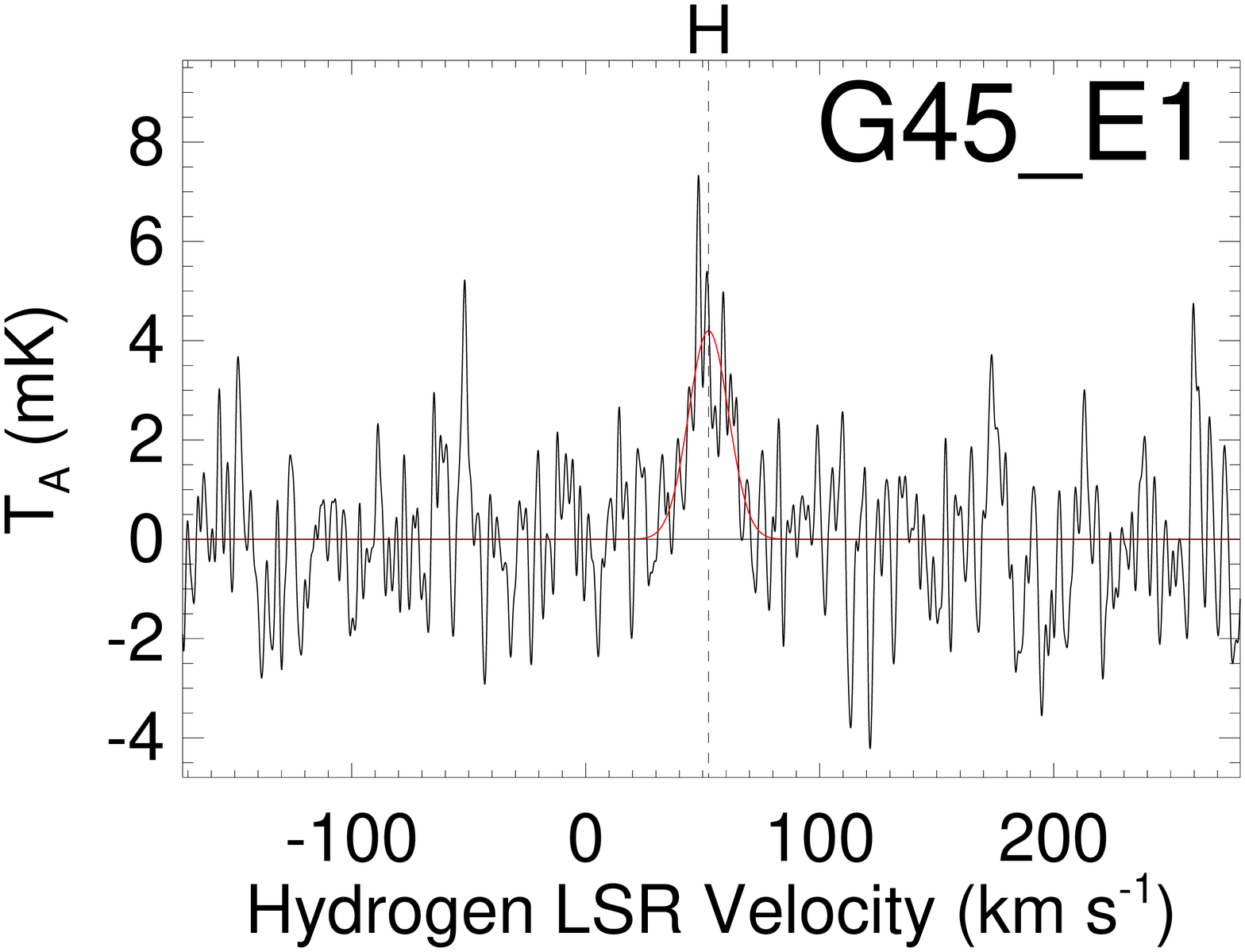} &
\includegraphics[width=.23\textwidth]{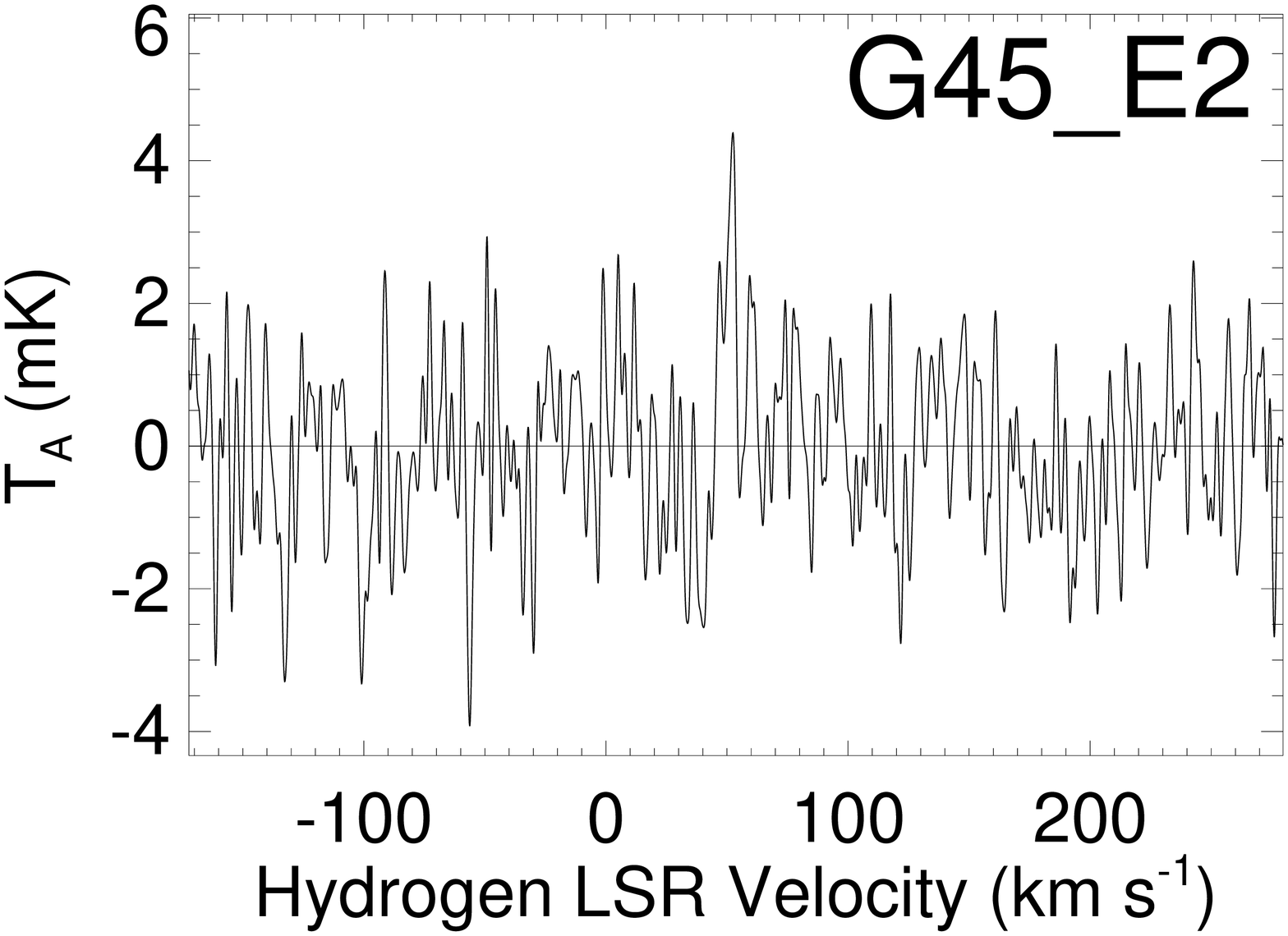} &
\includegraphics[width=.23\textwidth]{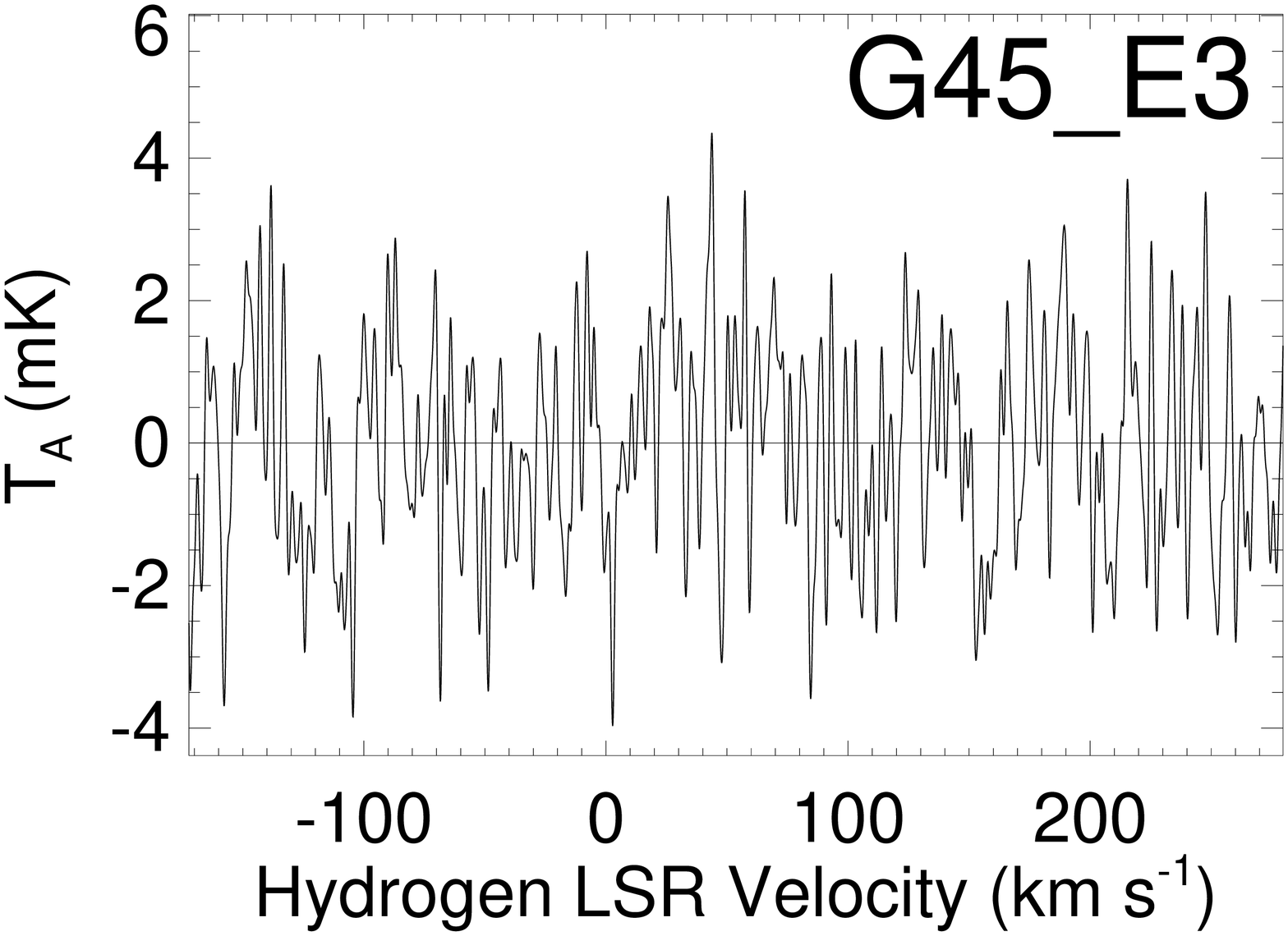} &
\includegraphics[width=.23\textwidth]{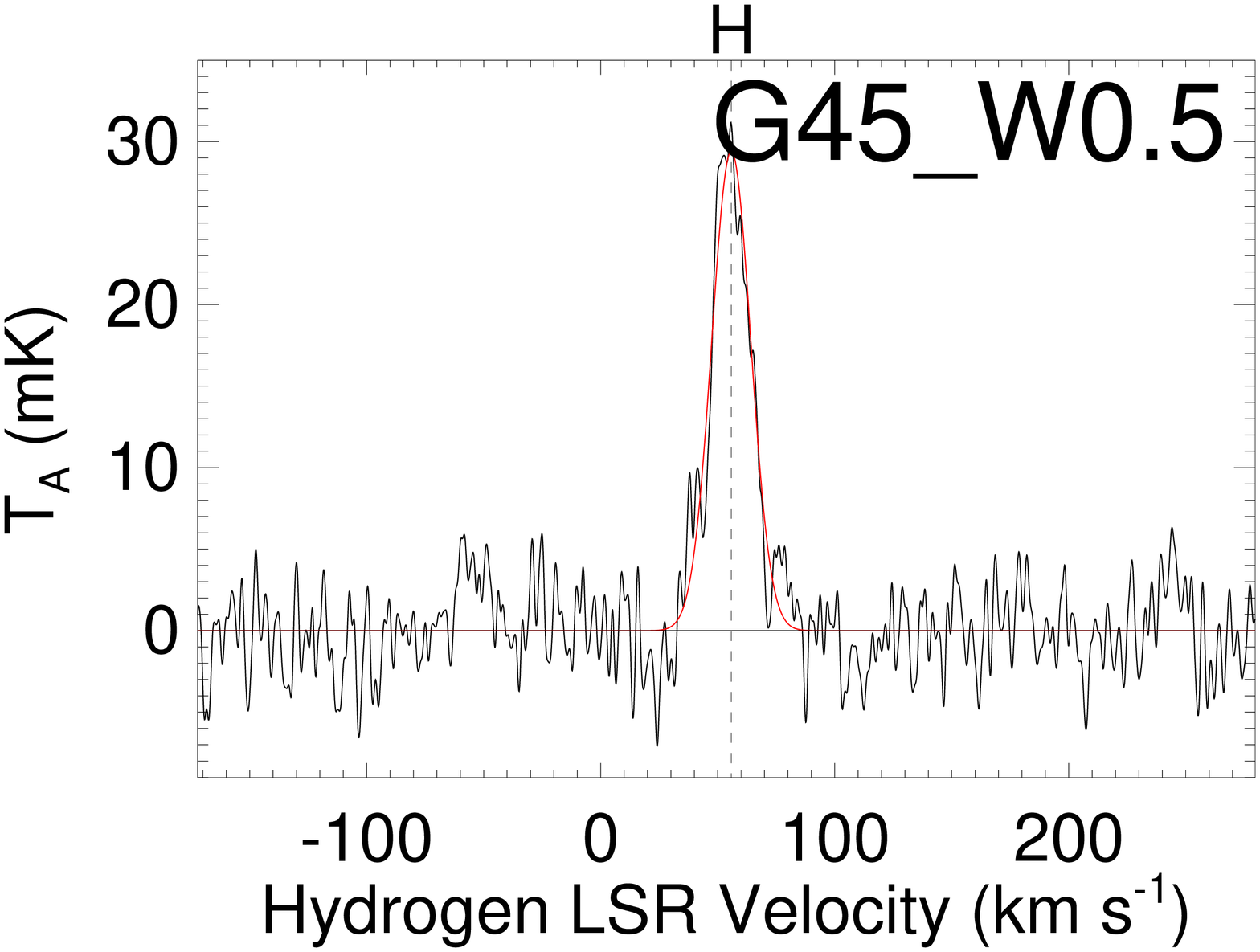} \\
\end{tabular}
\caption{$\gamma$ RRL spectra of all observed positions, smoothed to a spectral resolution of 1.86\kms. Plotted is the antenna temperature as a function of hydrogen LSR velocity. The helium and carbon lines are offset from hydrogen by $-124$\kms and $-149$\kms, respectively. We approximate hydrogen, helium, and carbon emission above the S/N threshold defined in \S \ref{sec:obs} with the Gaussian model fits shown in red. The centers of the Gaussian peaks are indicated by dashed vertical lines. The name of the observed position is given in the upper right-hand corner of each plot. \label{fig:spectragamma}}
\end{figure*}
\renewcommand{\thefigure}{\thesection.\arabic{figure}}

\renewcommand\thefigure{\thesection.\arabic{figure} (Cont.)}
\addtocounter{figure}{-1}
\begin{figure*}
\centering
\begin{tabular}{cccc}
\includegraphics[width=.23\textwidth]{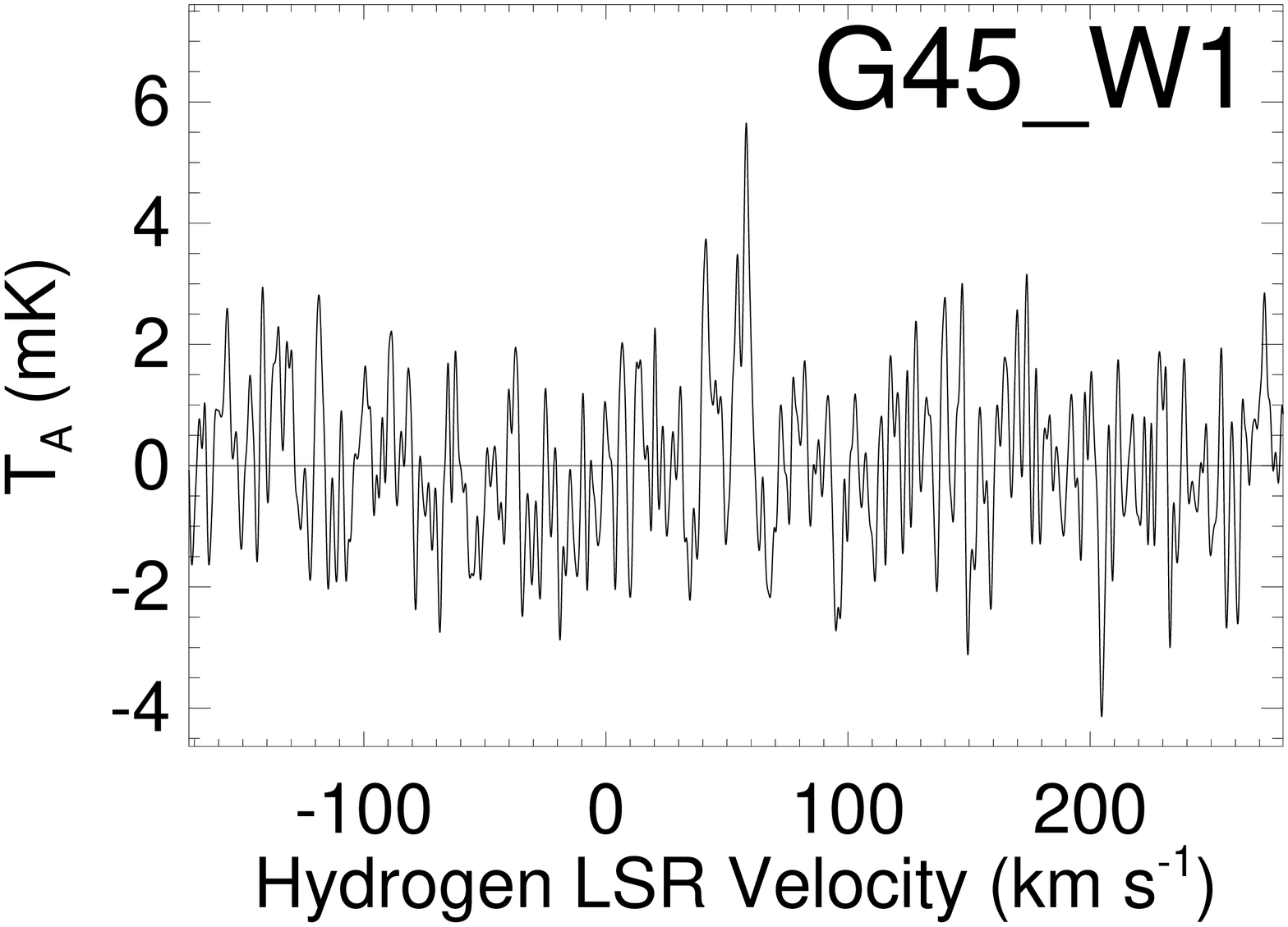} &
\includegraphics[width=.23\textwidth]{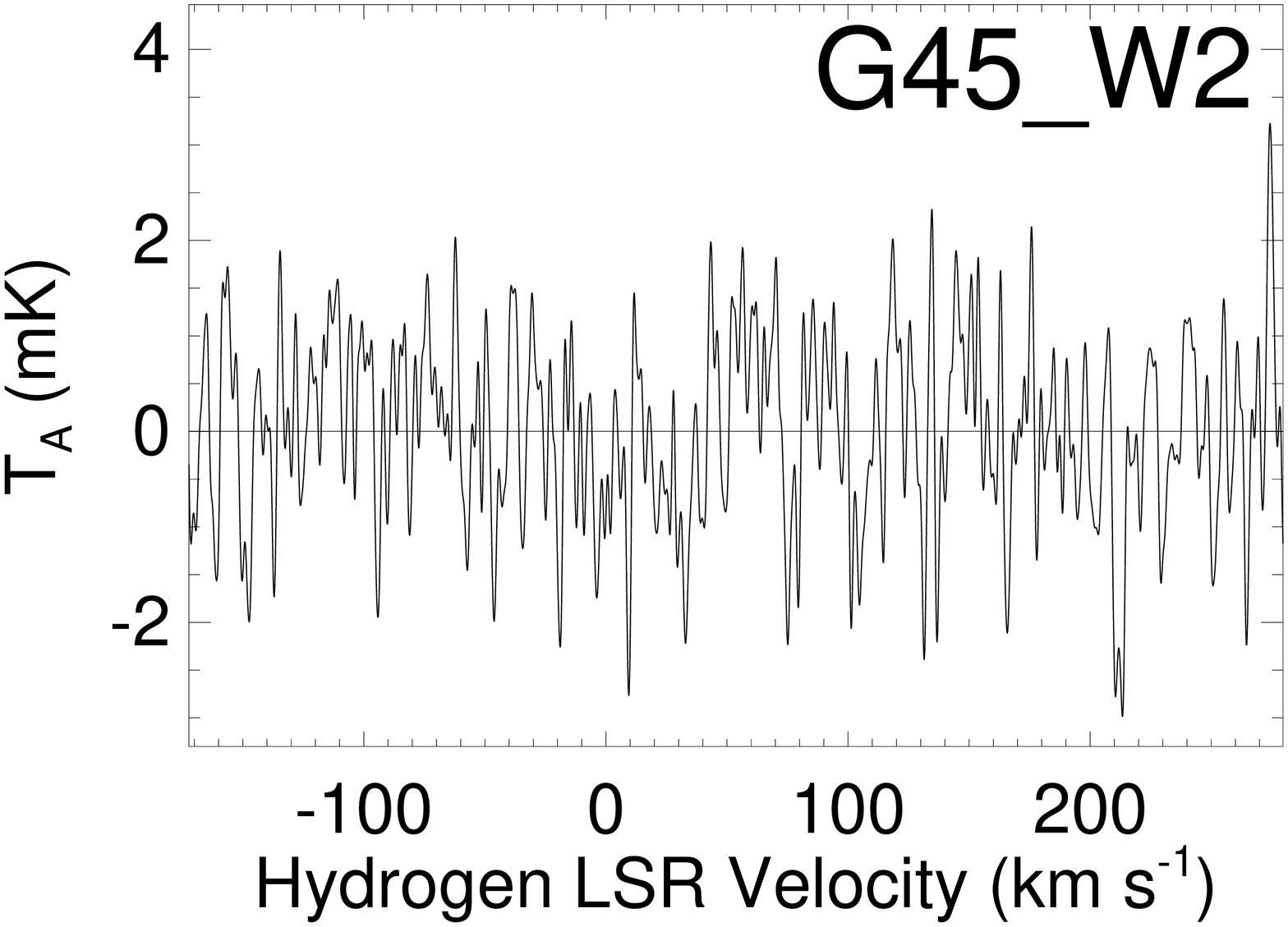} &
\includegraphics[width=.23\textwidth]{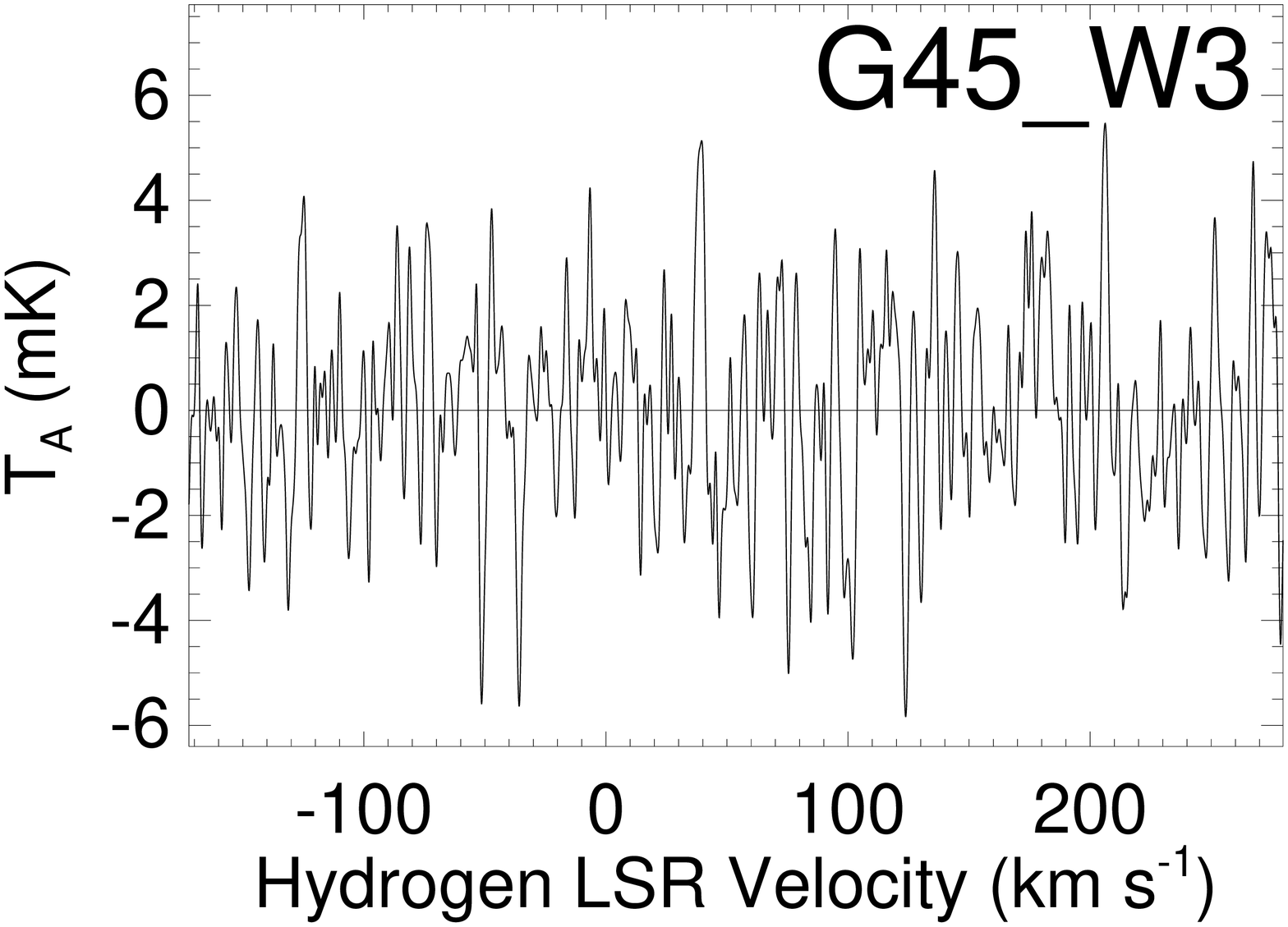} &
\includegraphics[width=.23\textwidth]{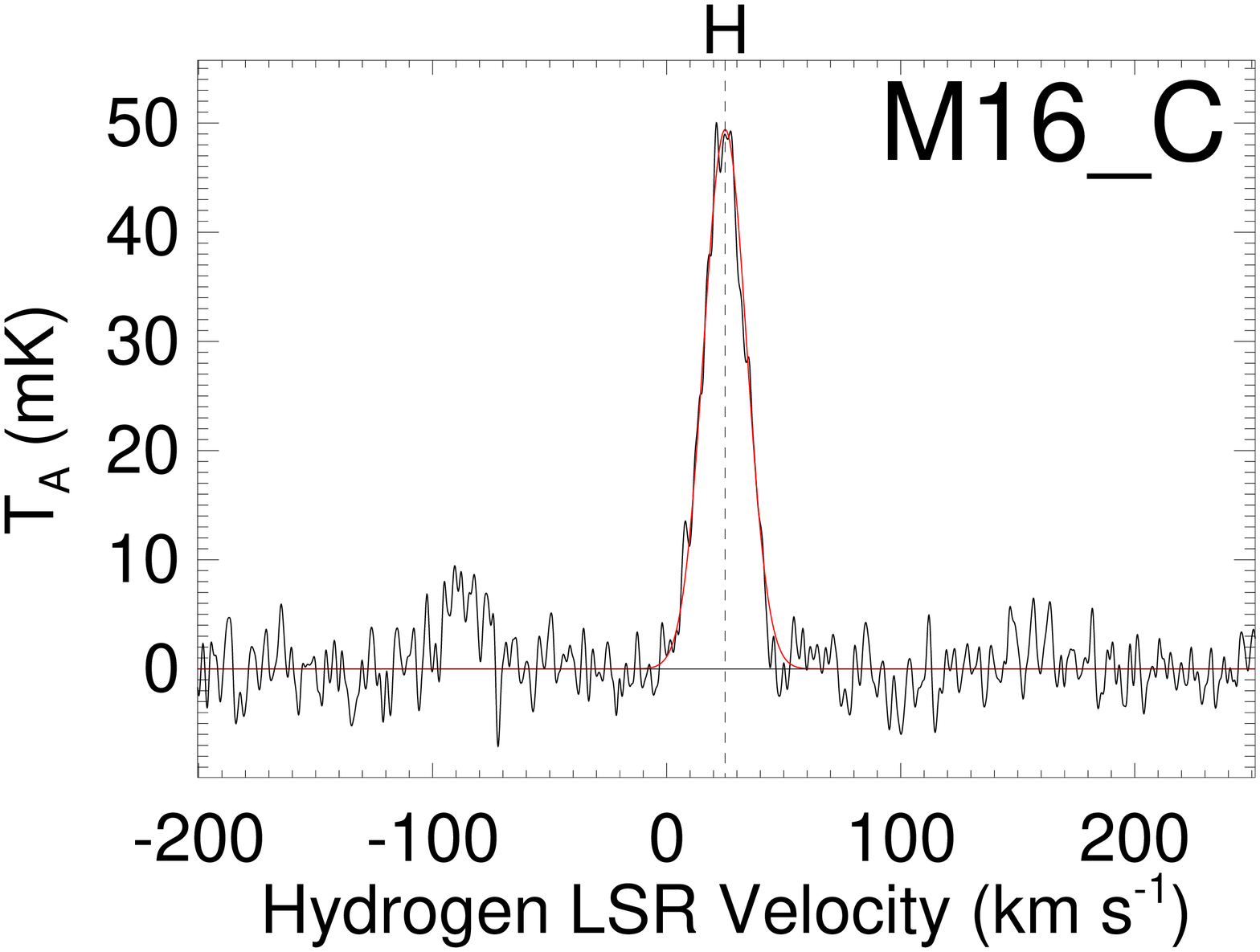} \\
\includegraphics[width=.23\textwidth]{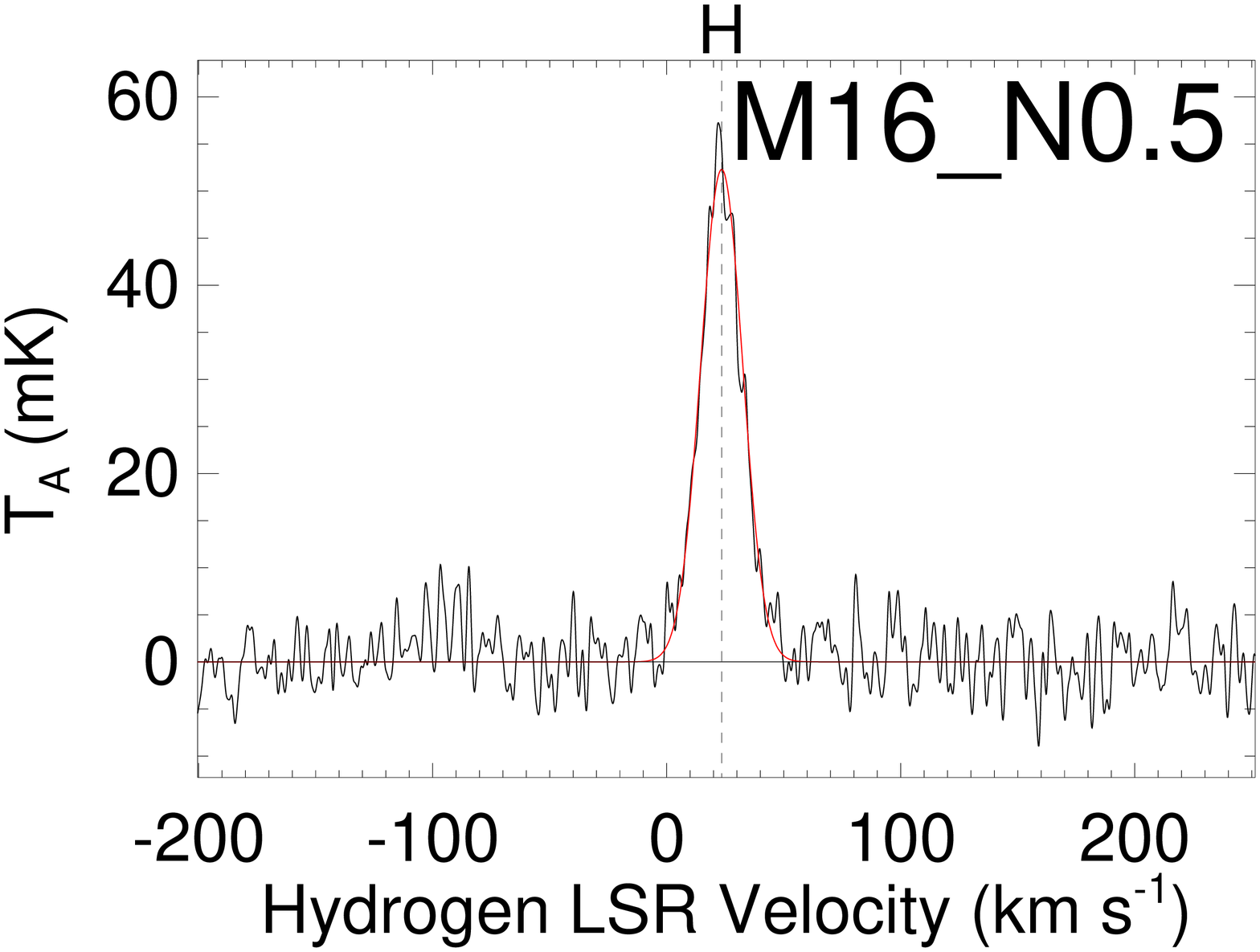} &
\includegraphics[width=.23\textwidth]{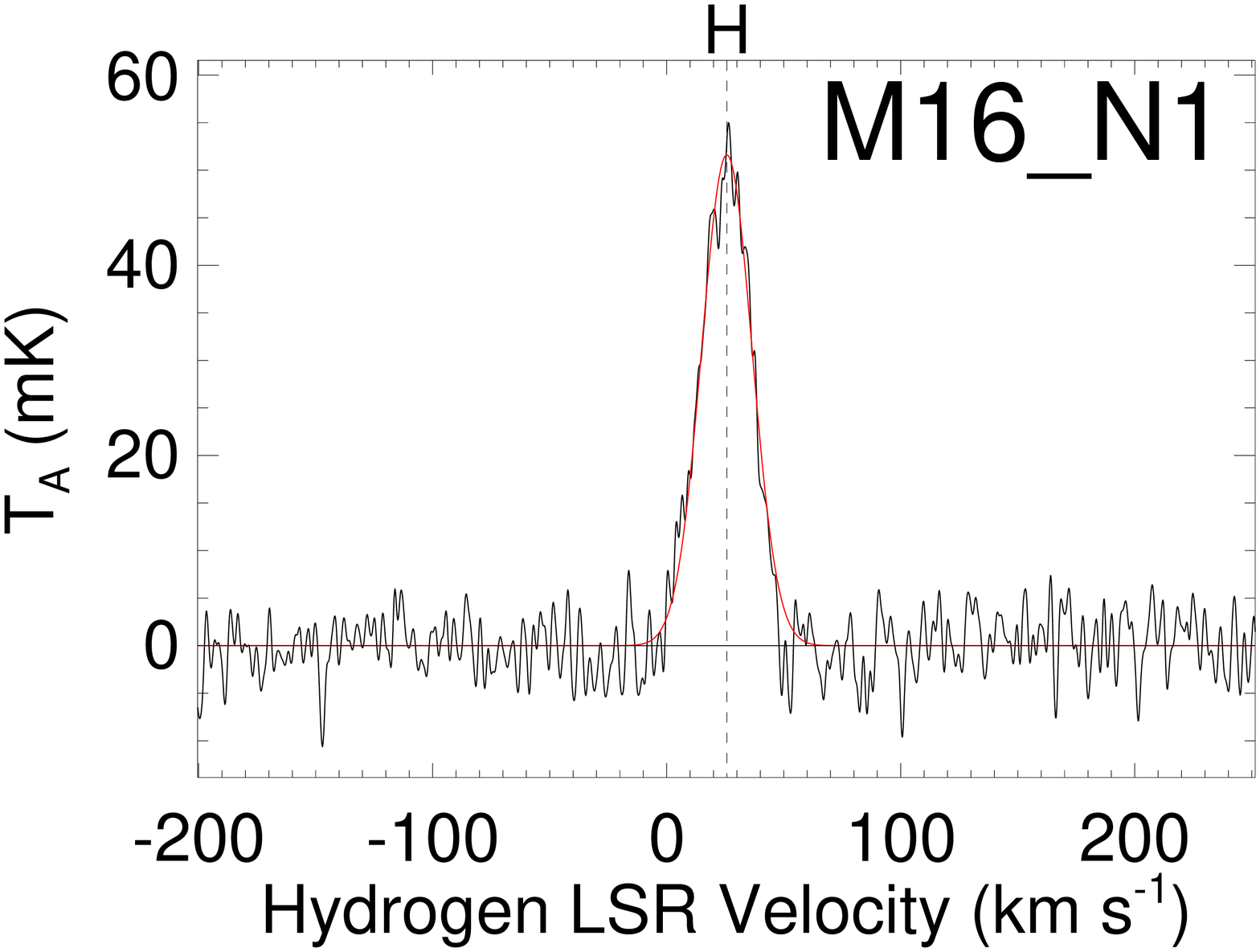} &
\includegraphics[width=.23\textwidth]{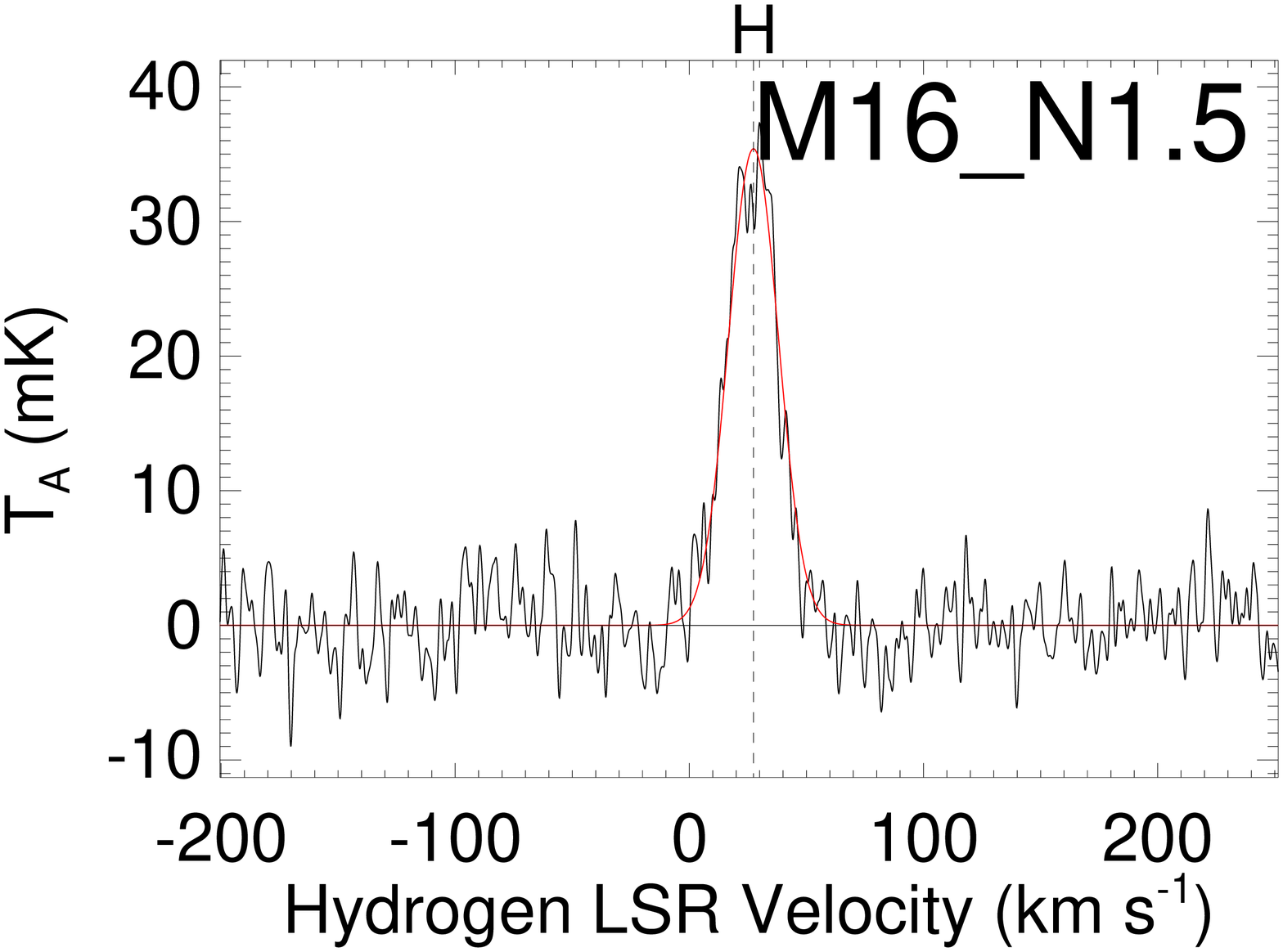} &
\includegraphics[width=.23\textwidth]{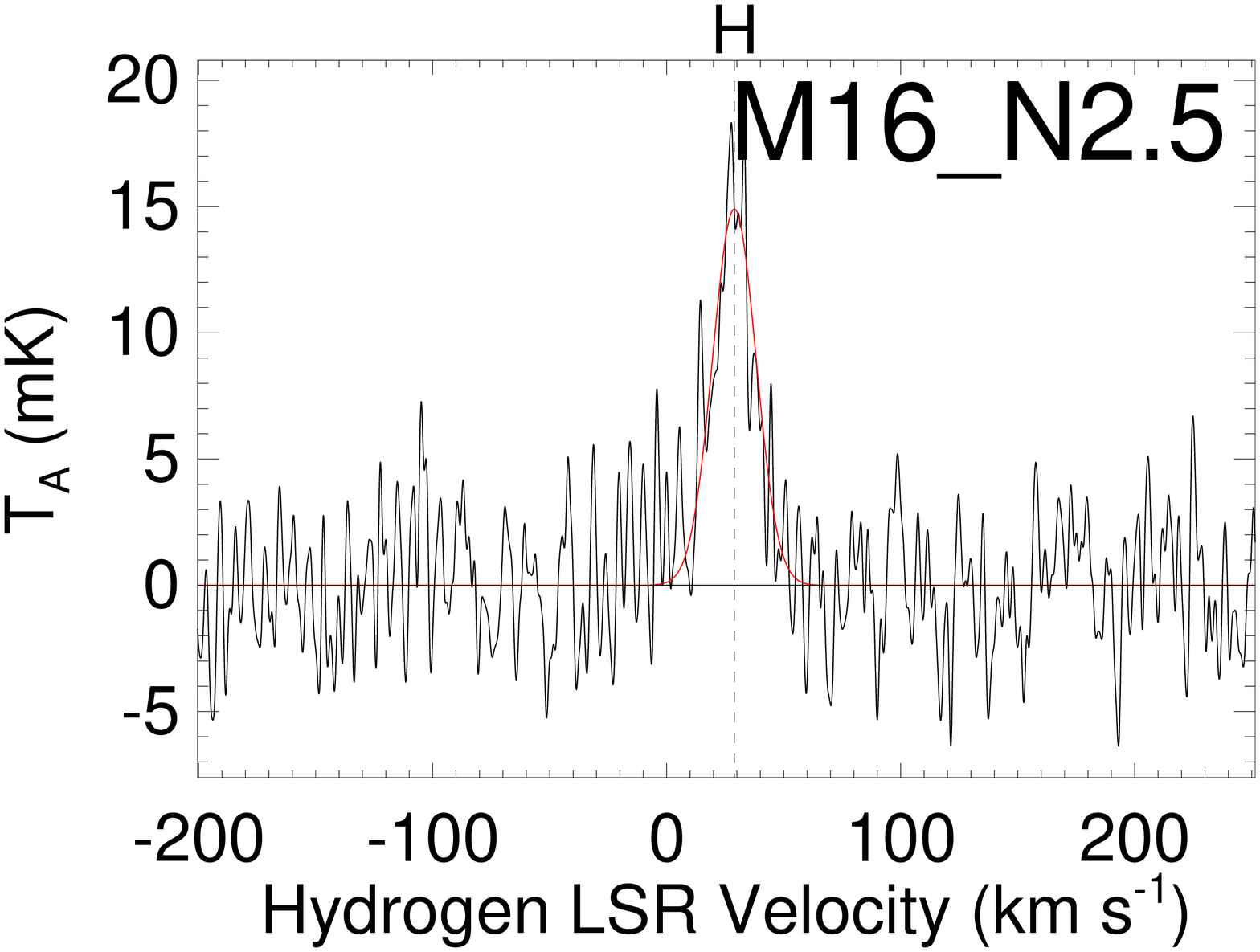} \\
\includegraphics[width=.23\textwidth]{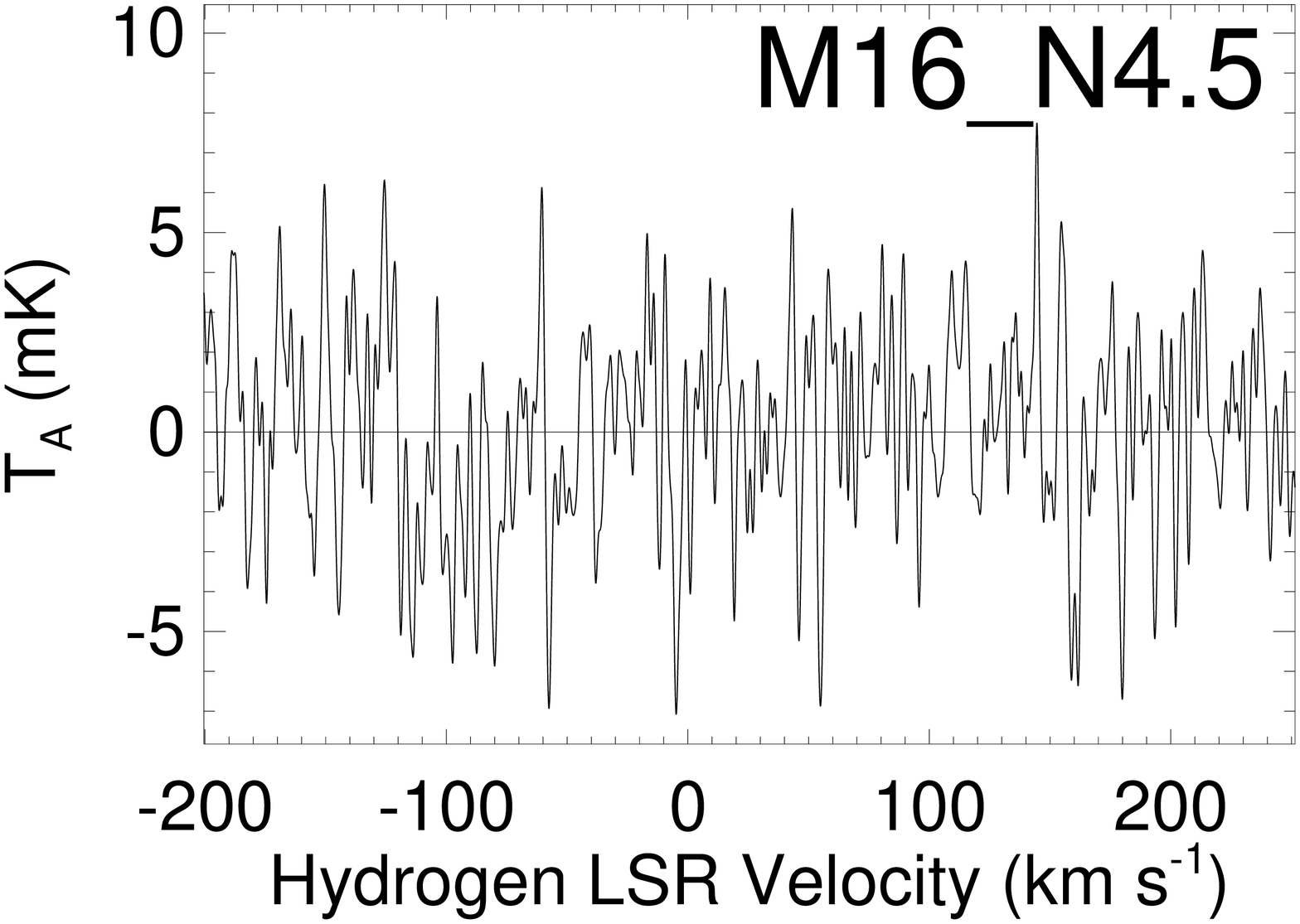} &
\includegraphics[width=.23\textwidth]{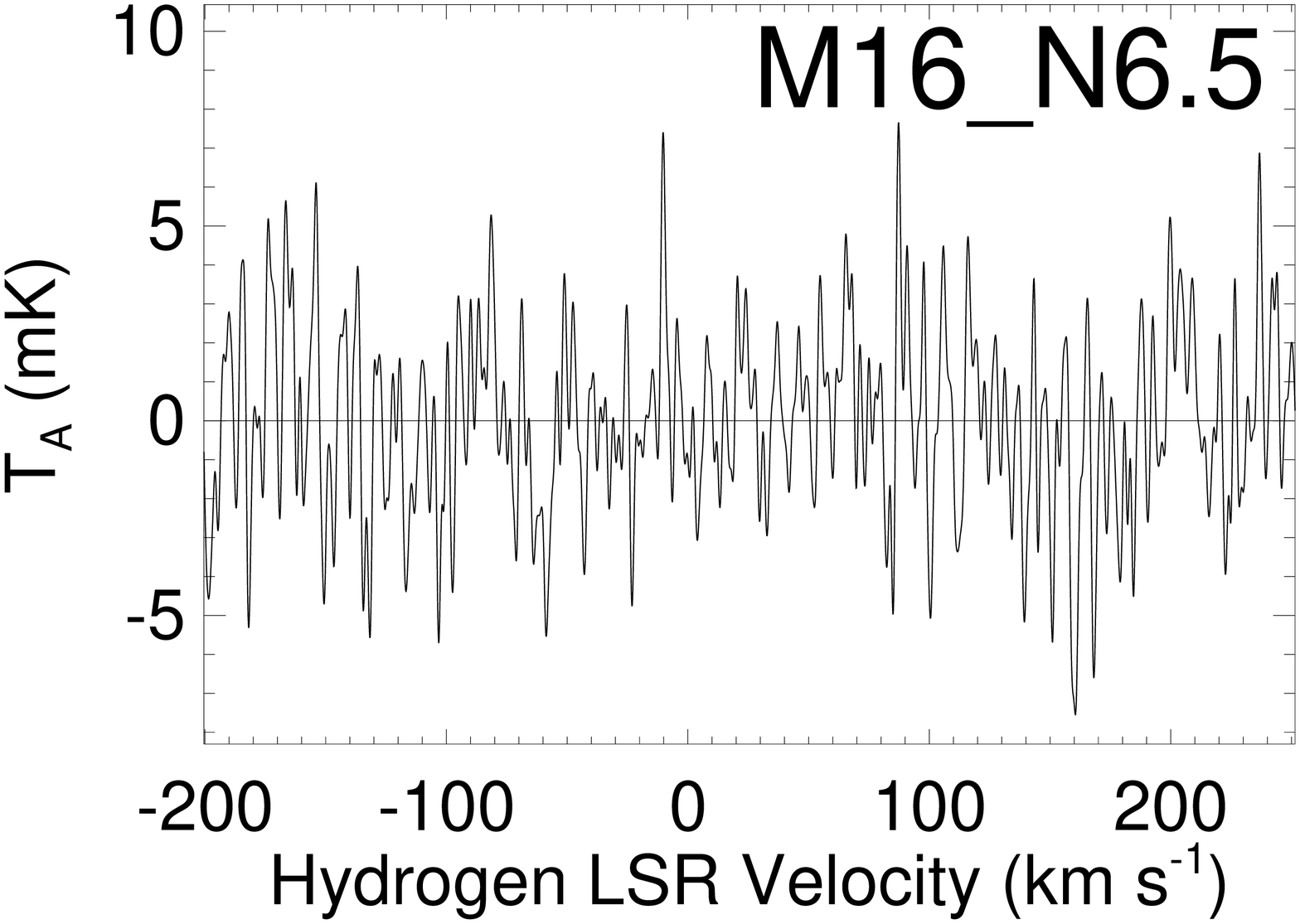} &
\includegraphics[width=.23\textwidth]{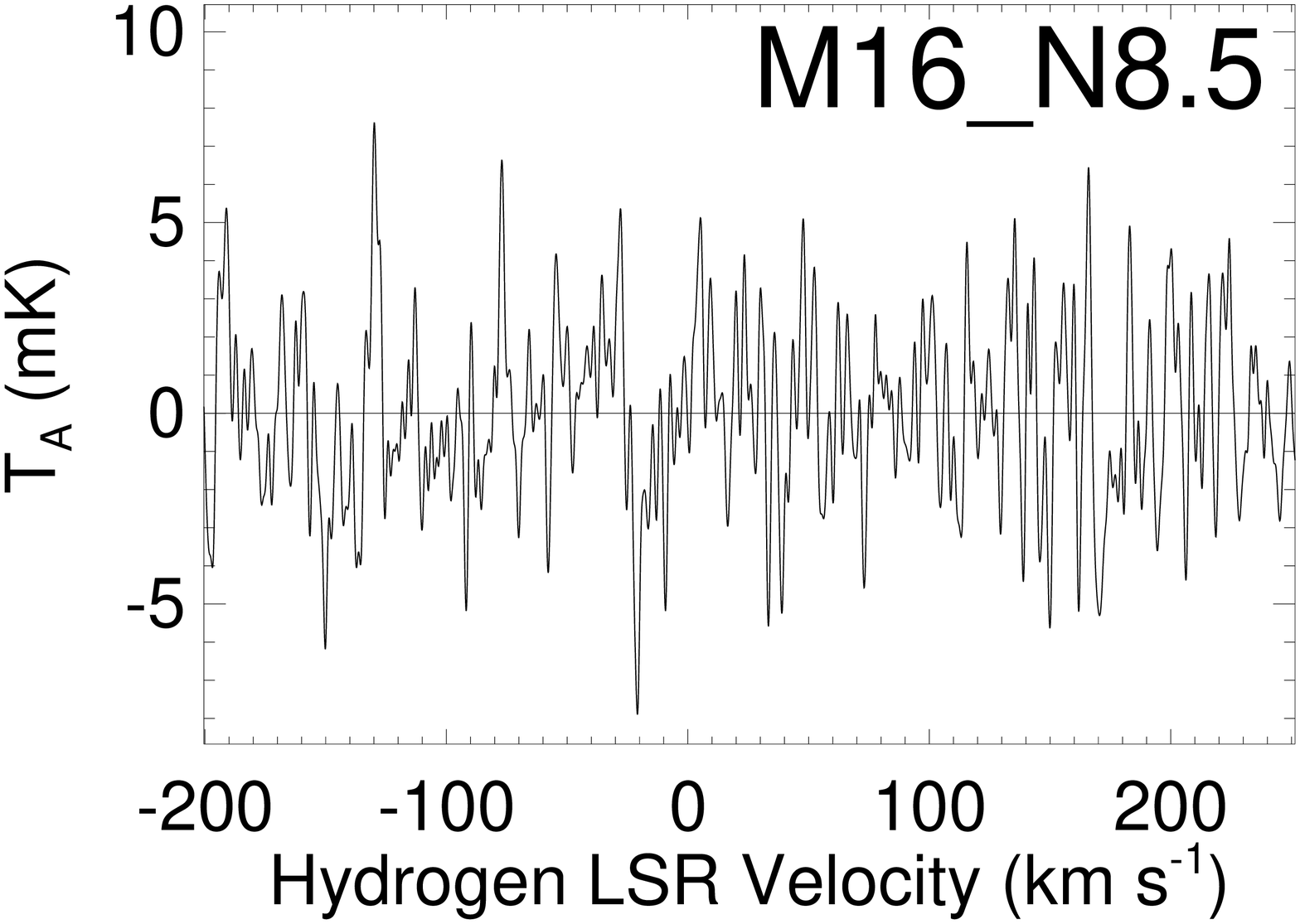} &
\includegraphics[width=.23\textwidth]{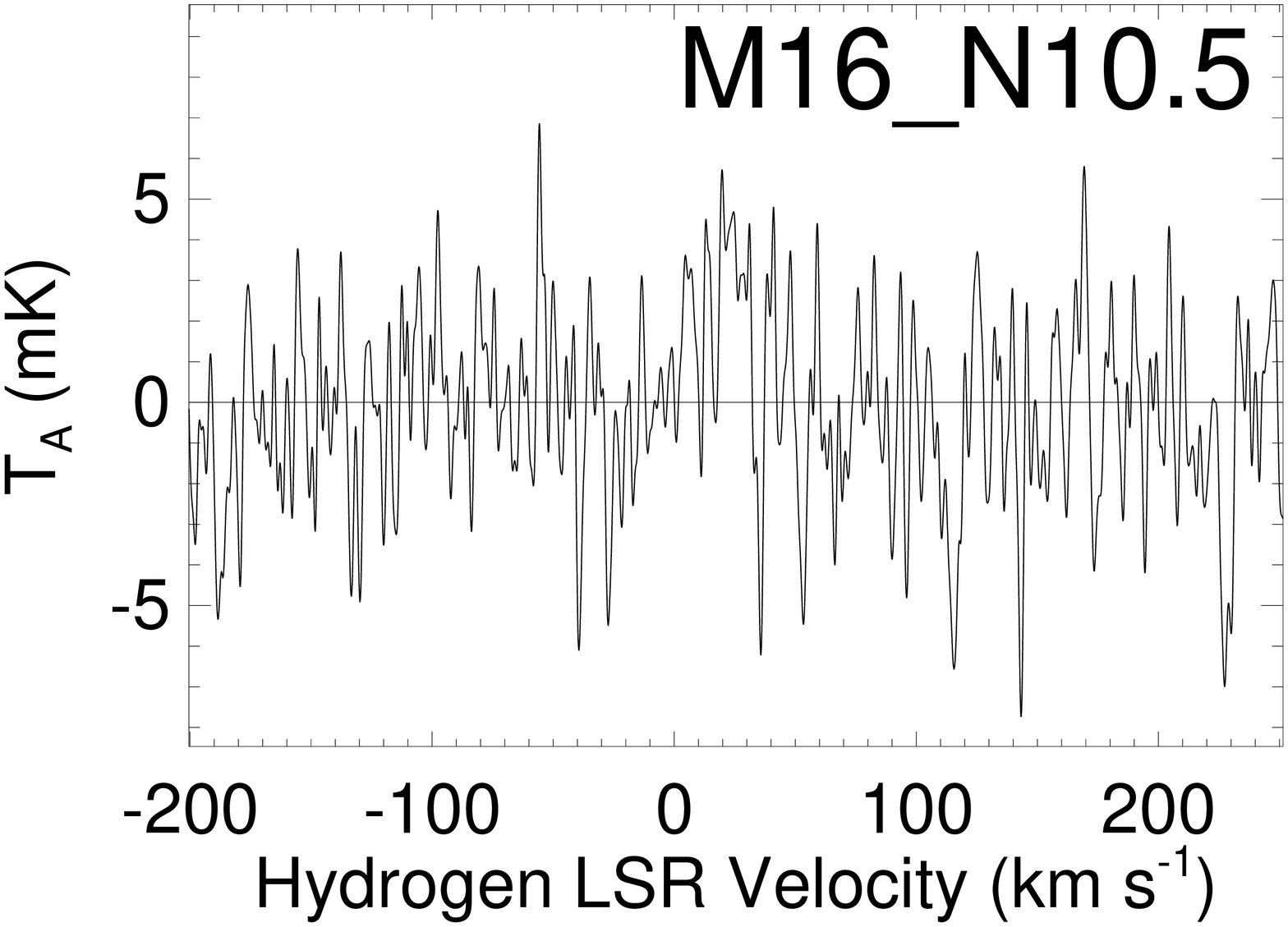} \\
\includegraphics[width=.23\textwidth]{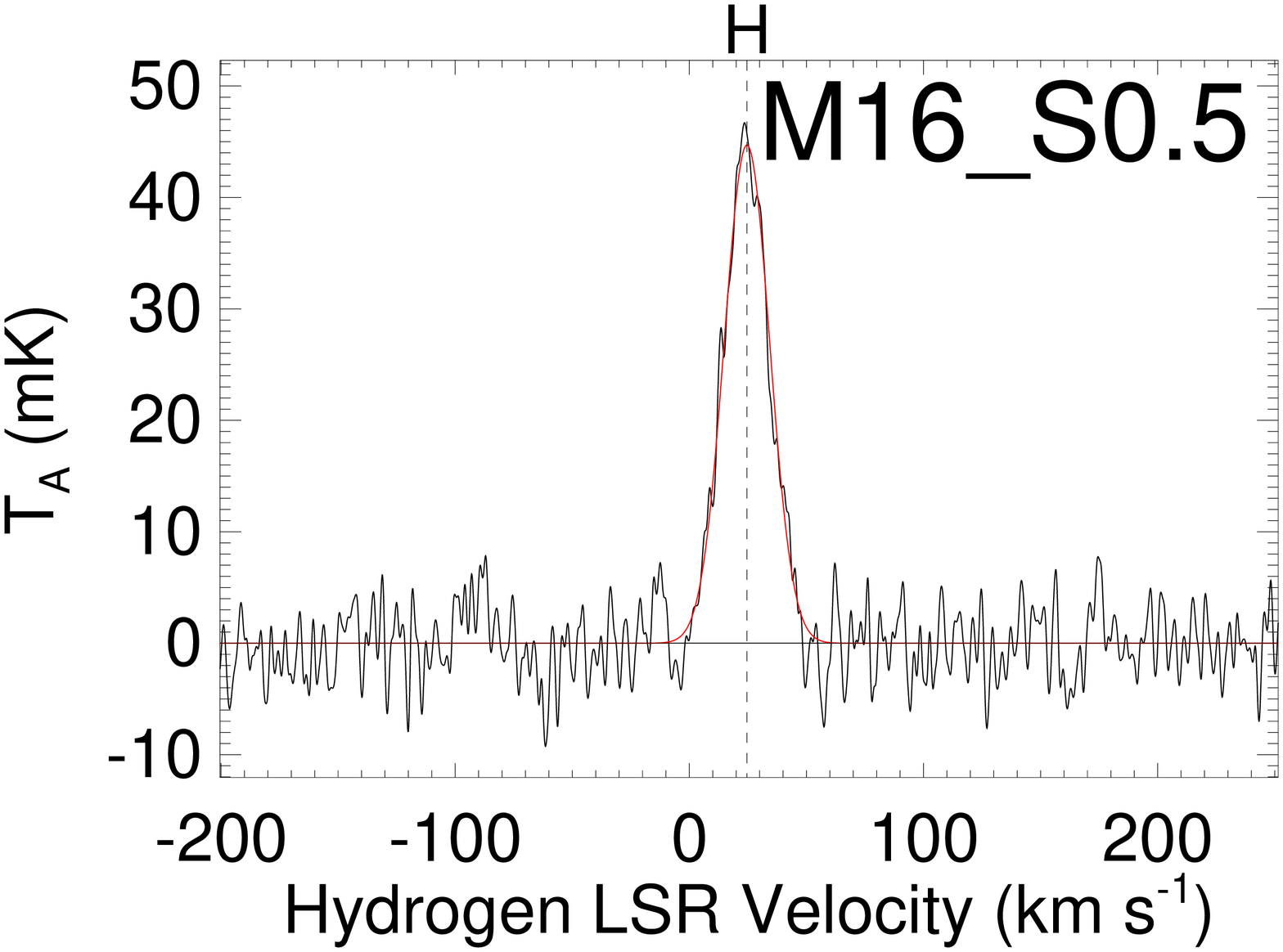} &
\includegraphics[width=.23\textwidth]{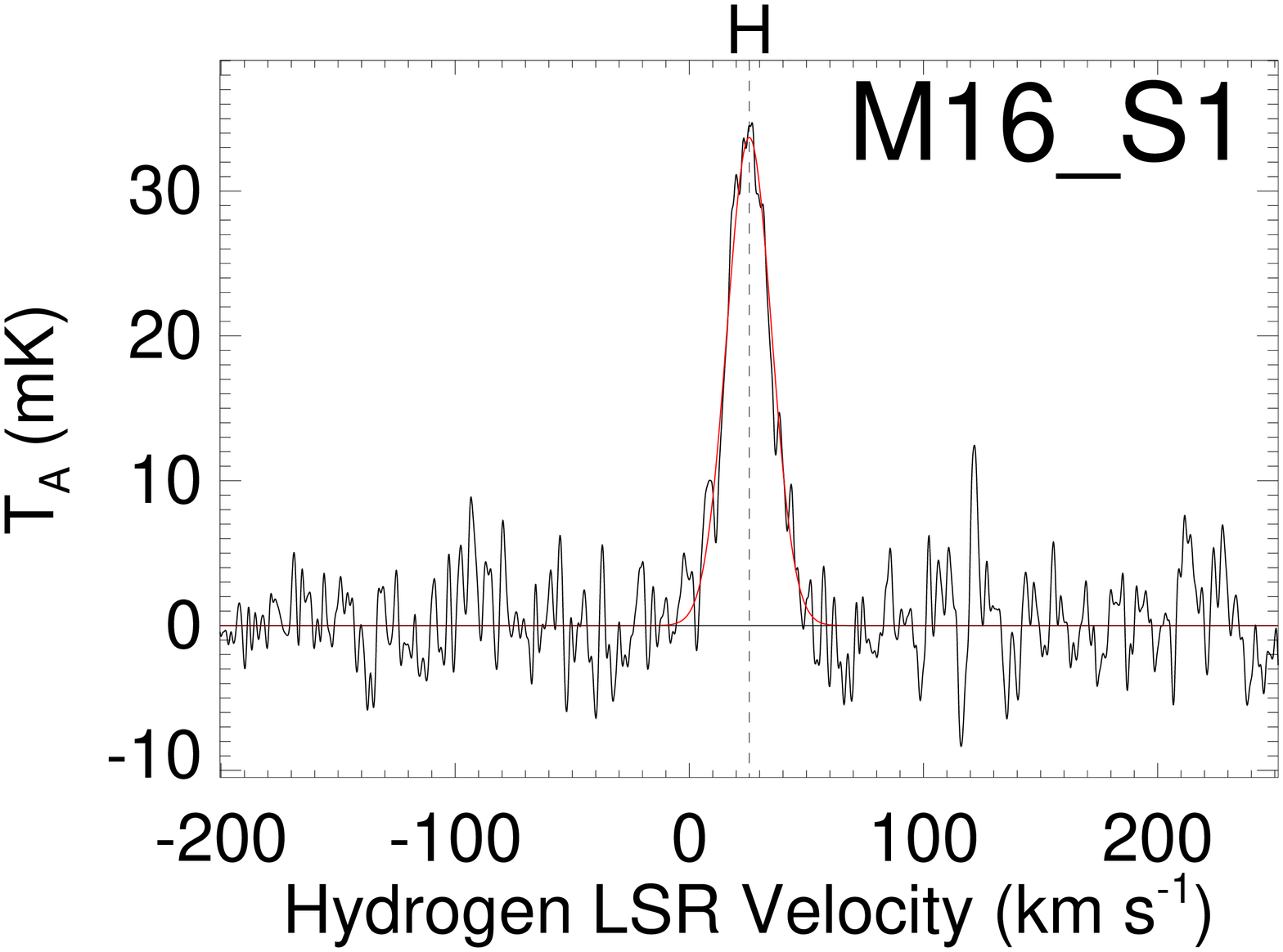} &
\includegraphics[width=.23\textwidth]{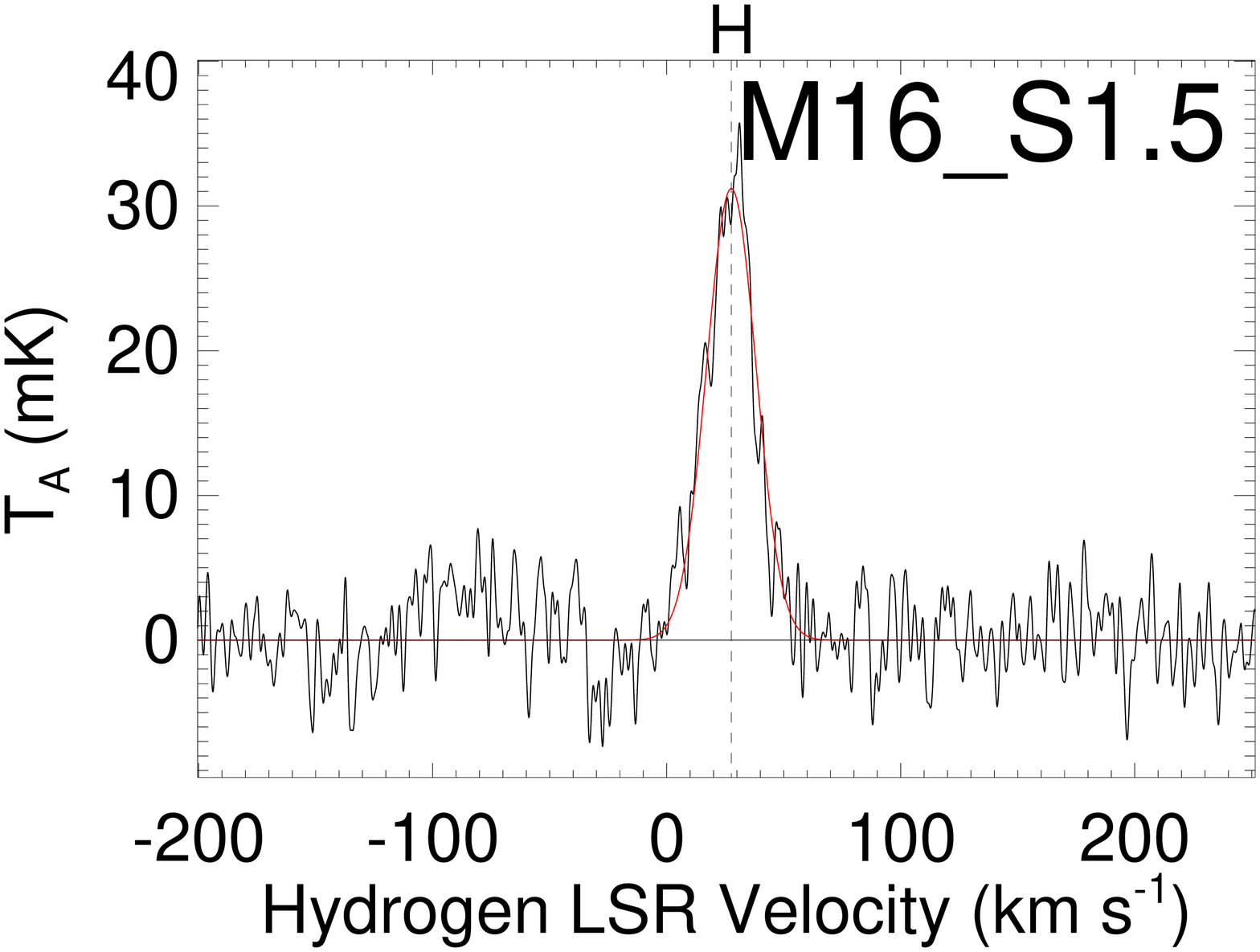} &
\includegraphics[width=.23\textwidth]{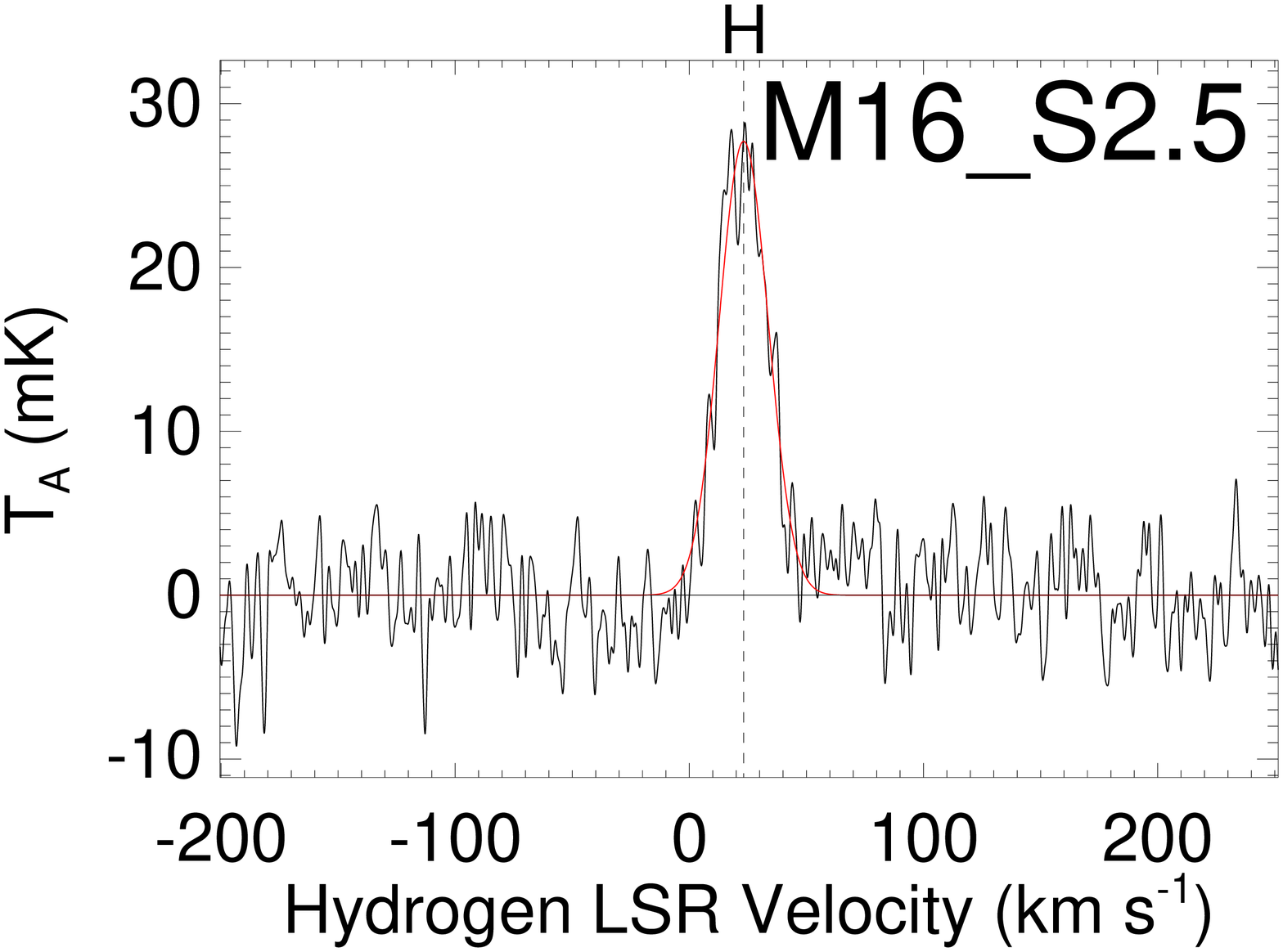} \\
\includegraphics[width=.23\textwidth]{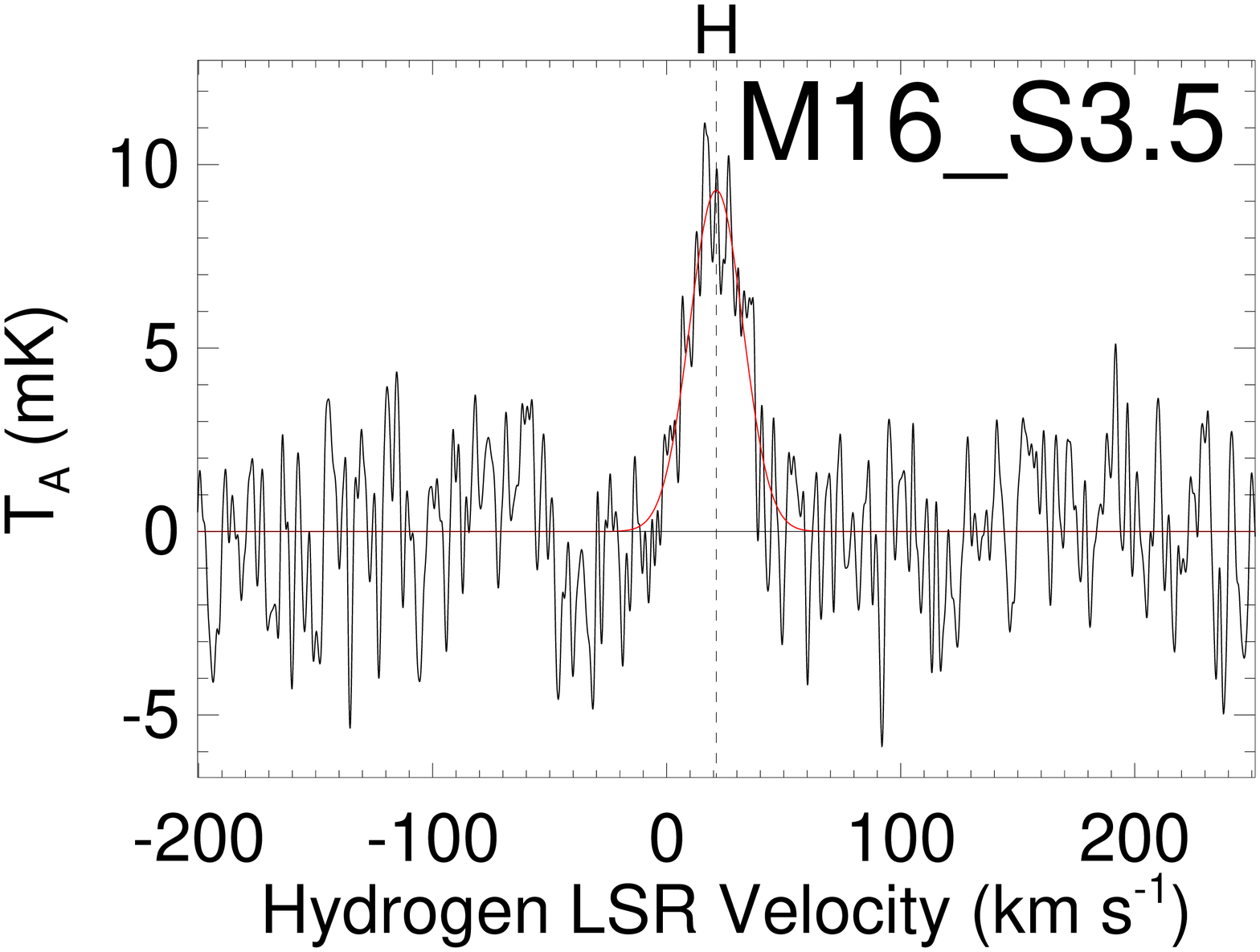} &
\includegraphics[width=.23\textwidth]{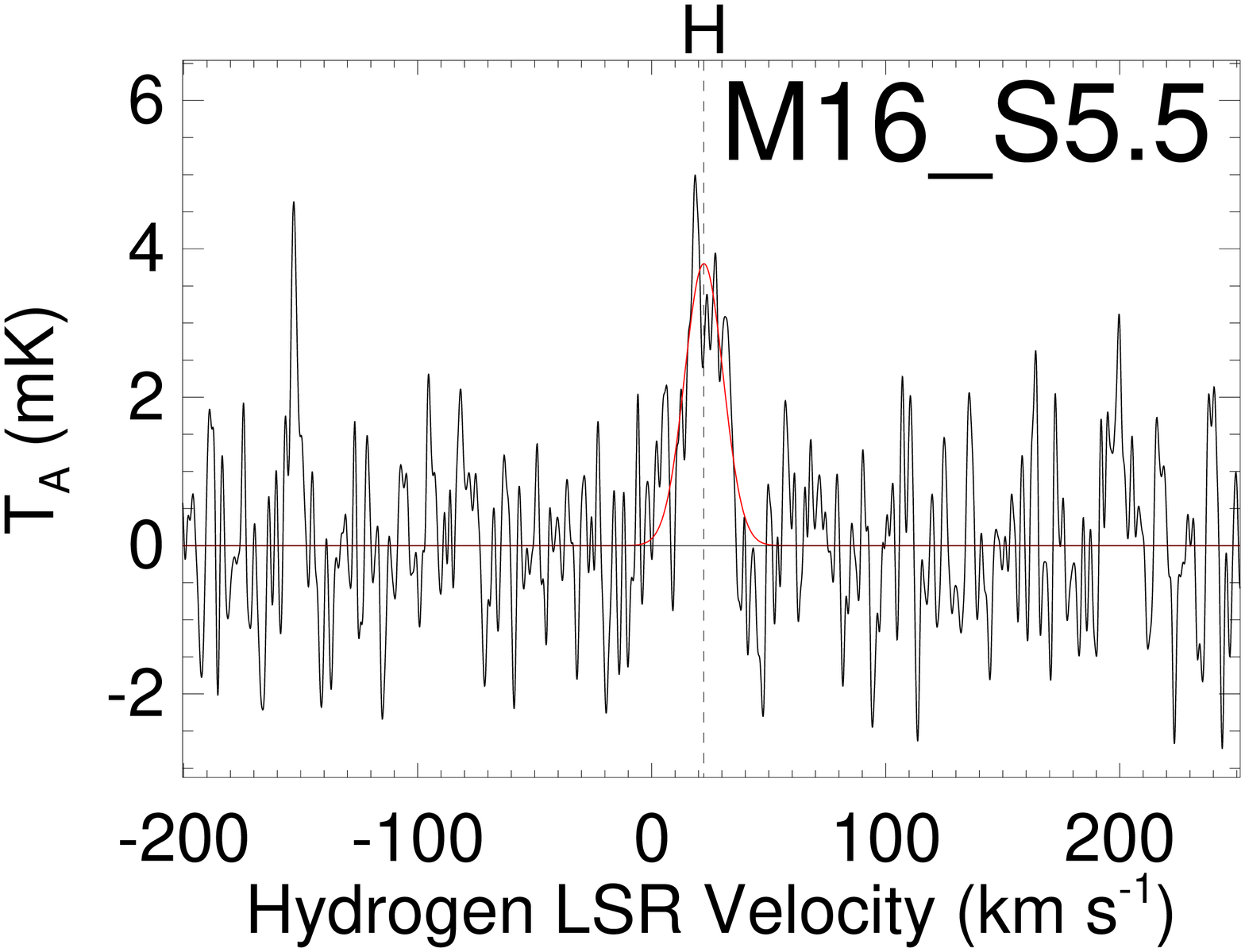} &
\includegraphics[width=.23\textwidth]{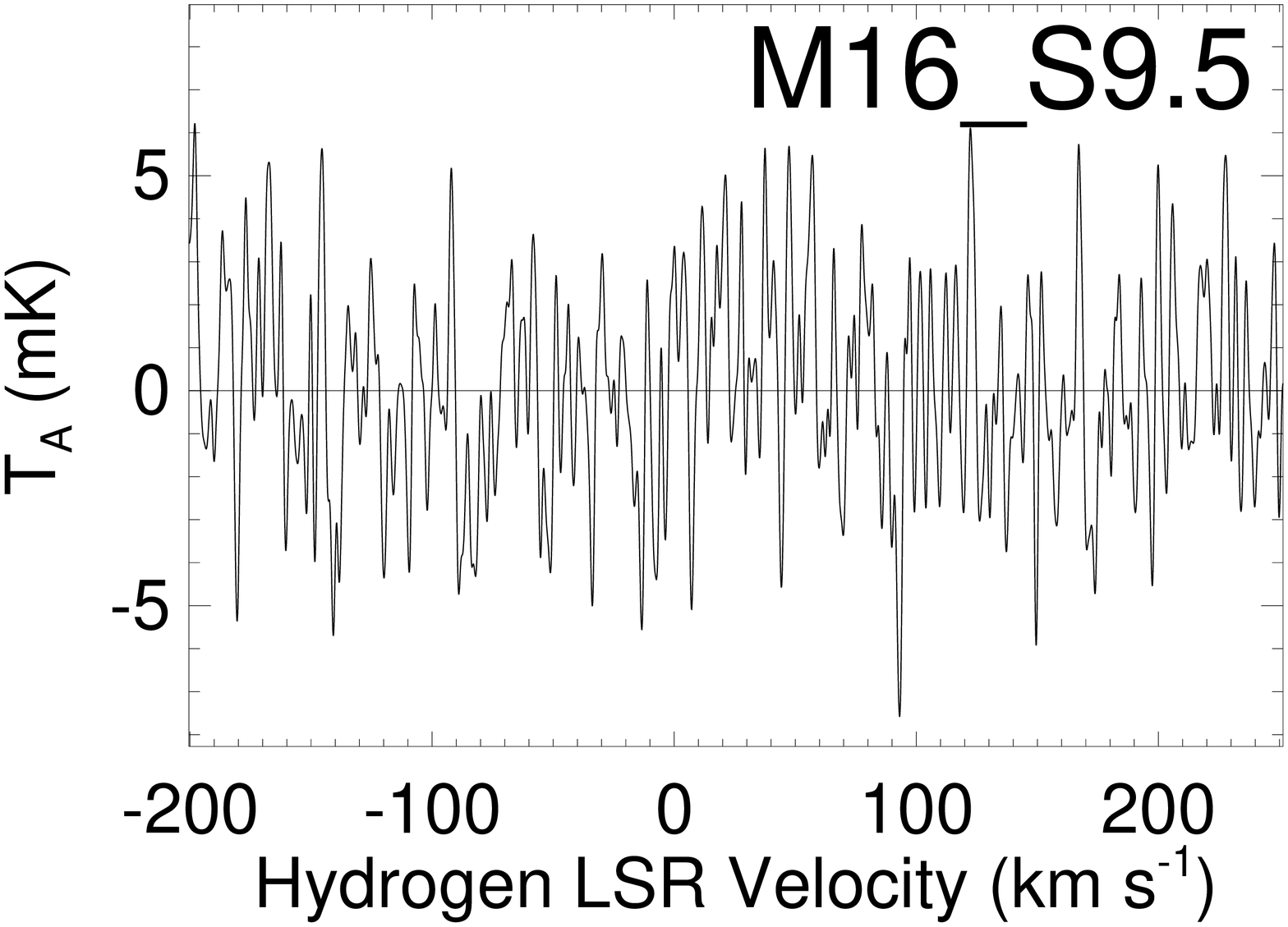} &
\includegraphics[width=.23\textwidth]{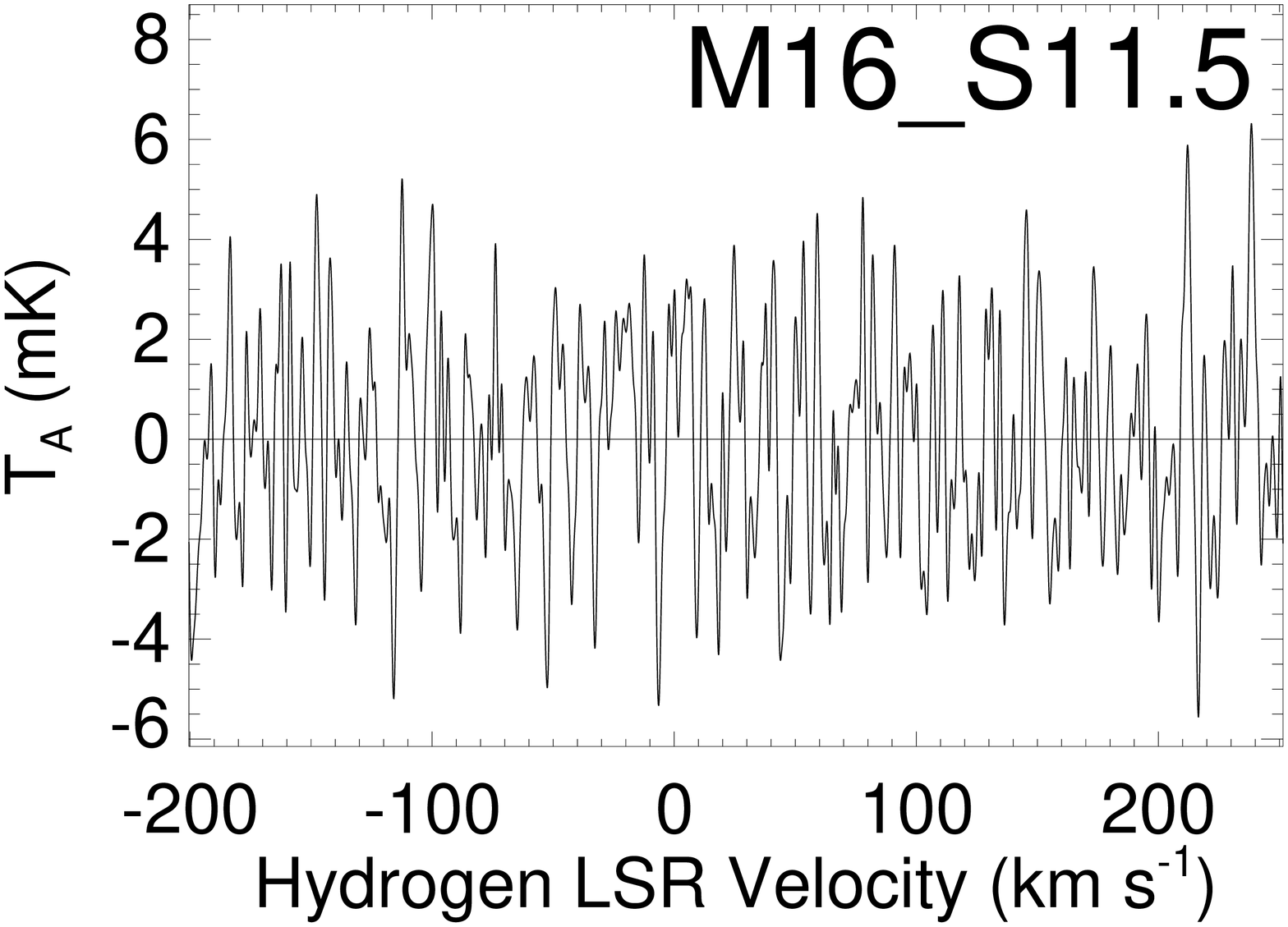} \\
\includegraphics[width=.23\textwidth]{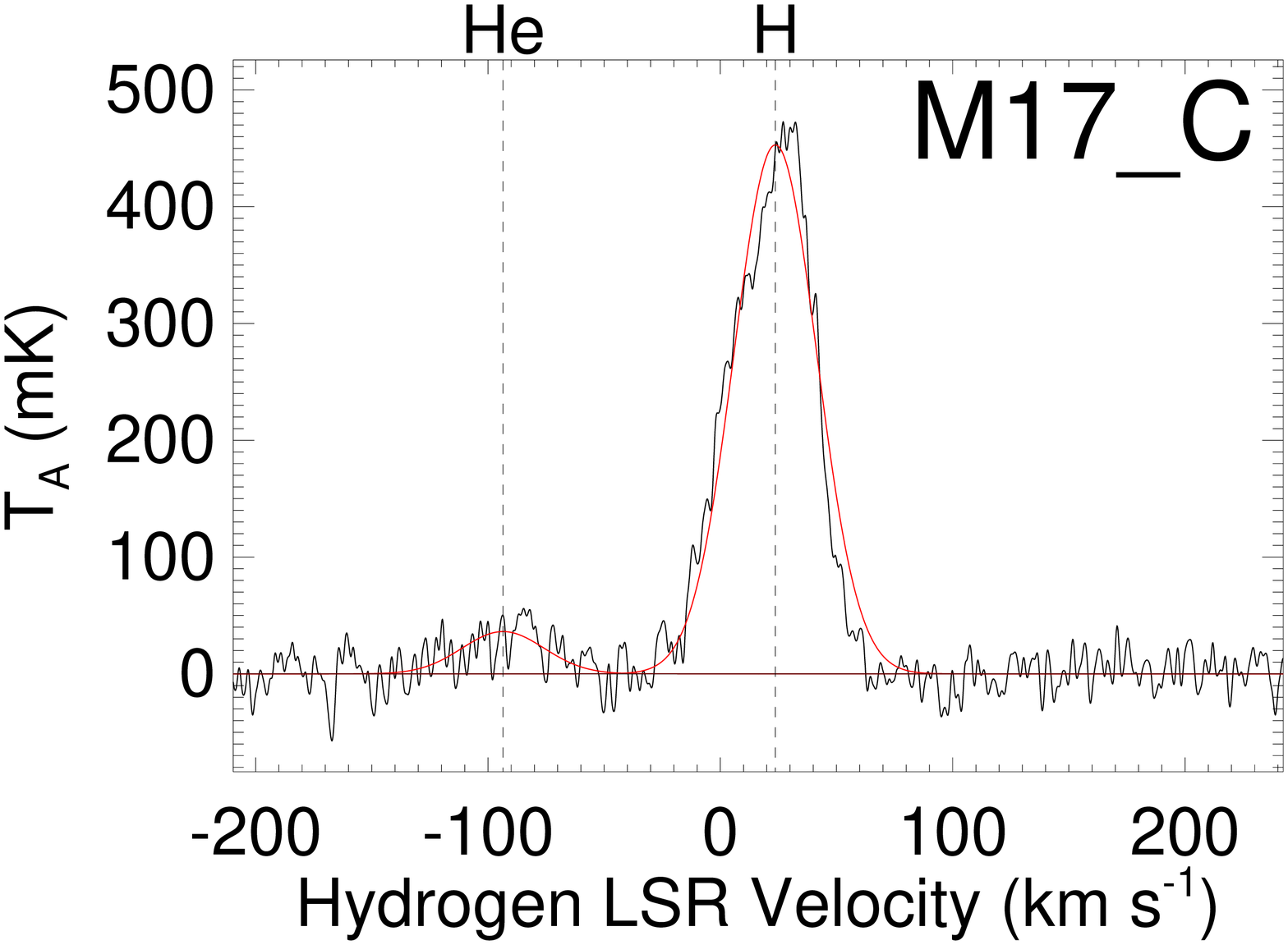} &
\includegraphics[width=.23\textwidth]{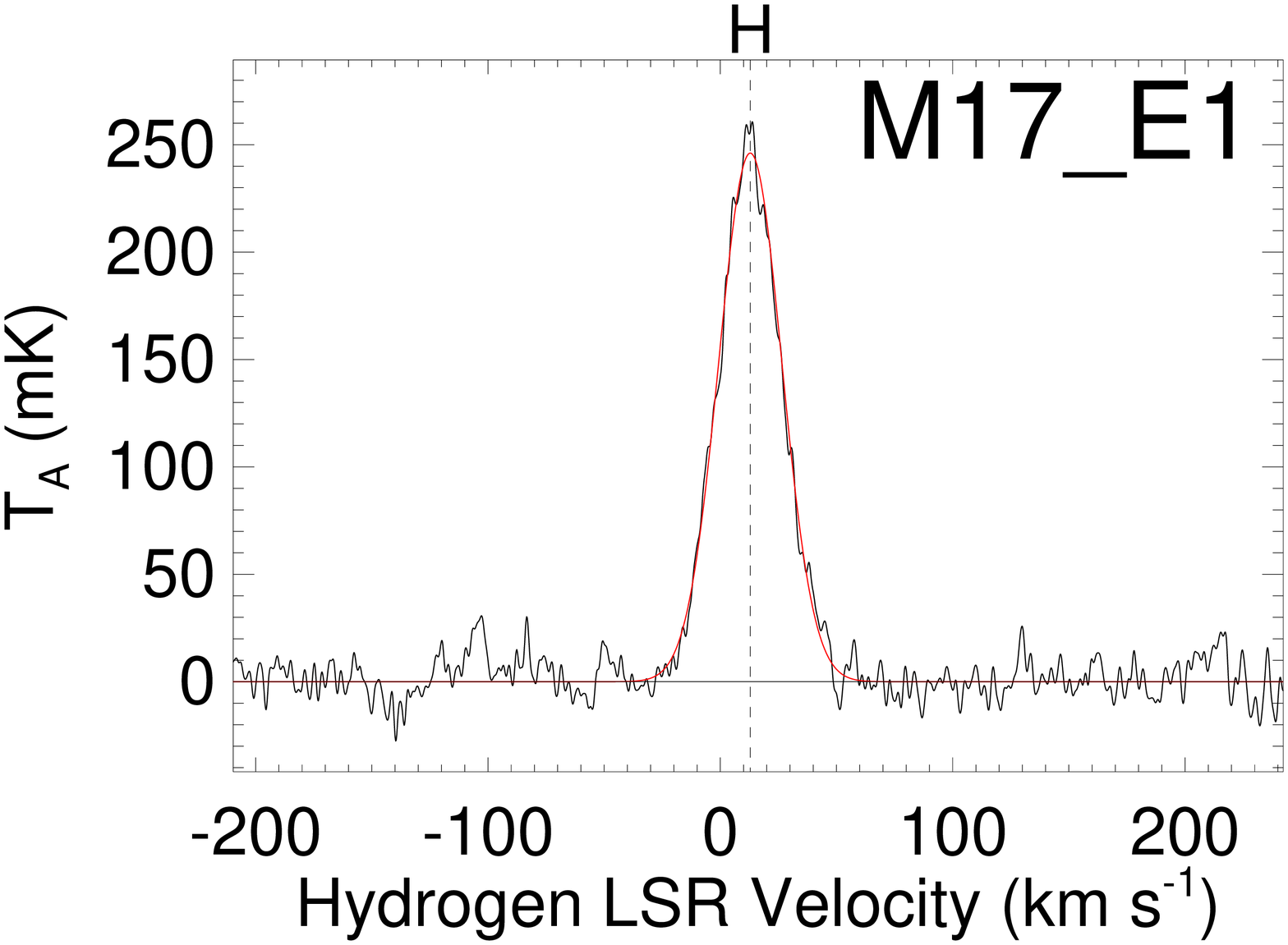} &
\includegraphics[width=.23\textwidth]{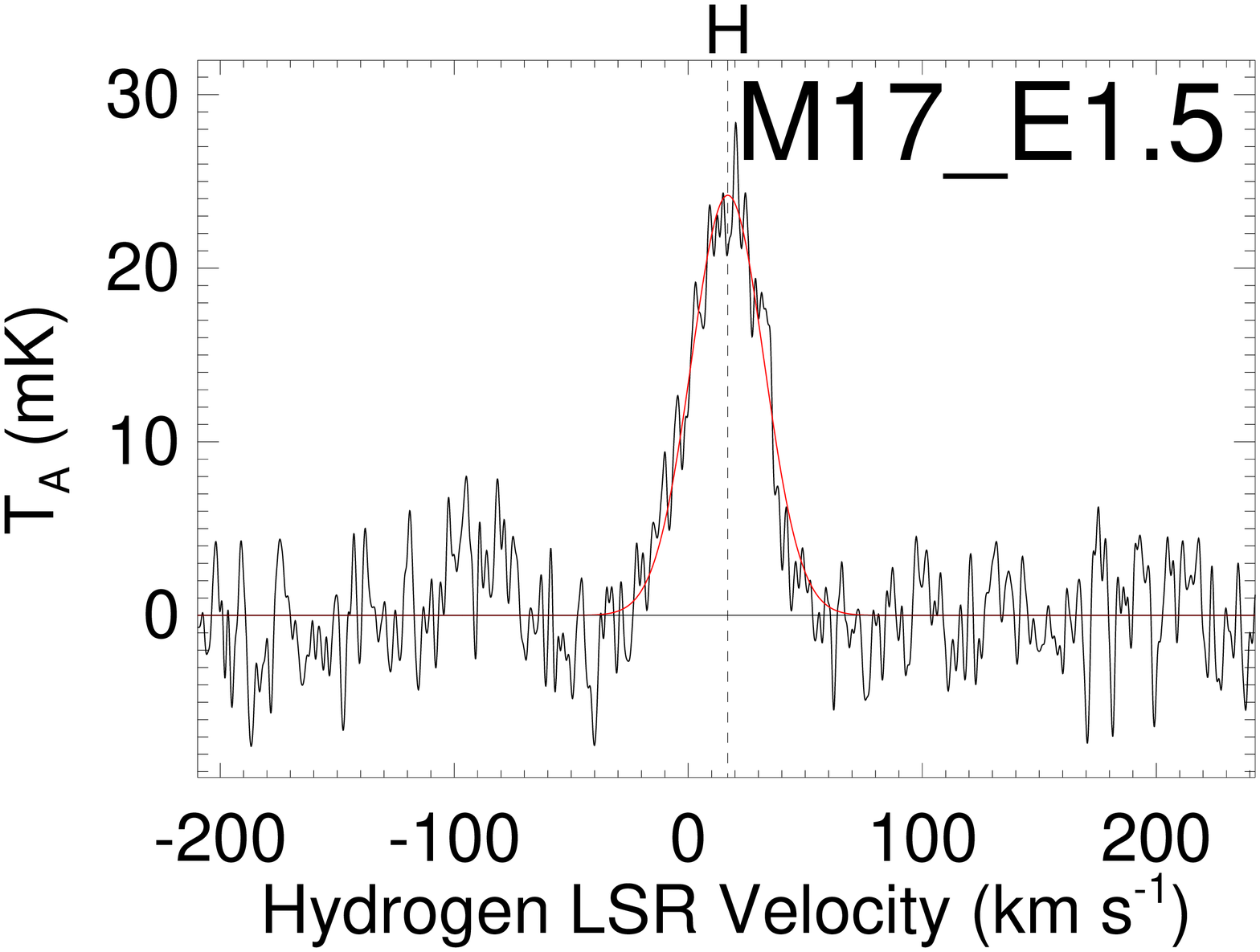} &
\includegraphics[width=.23\textwidth]{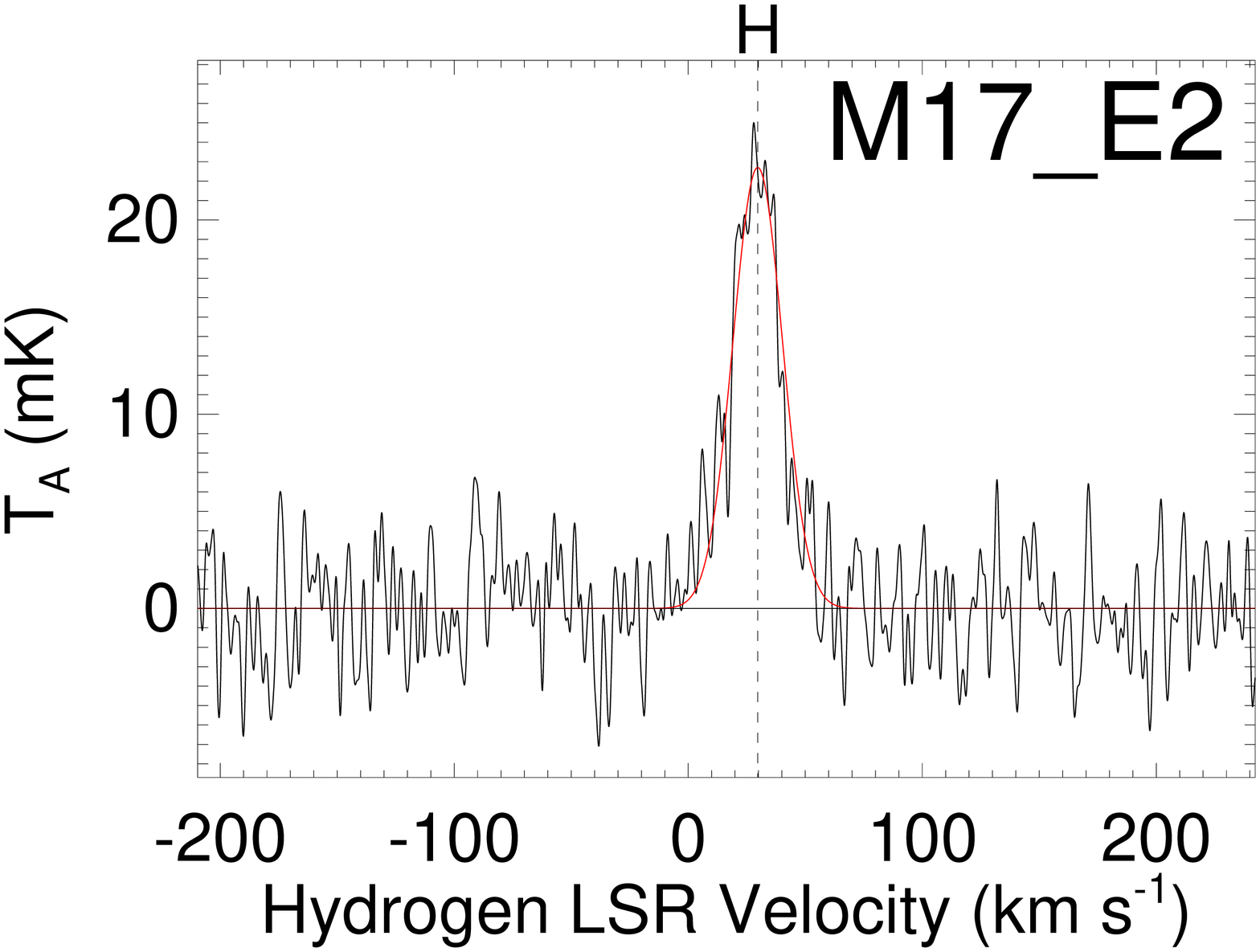} \\
\includegraphics[width=.23\textwidth]{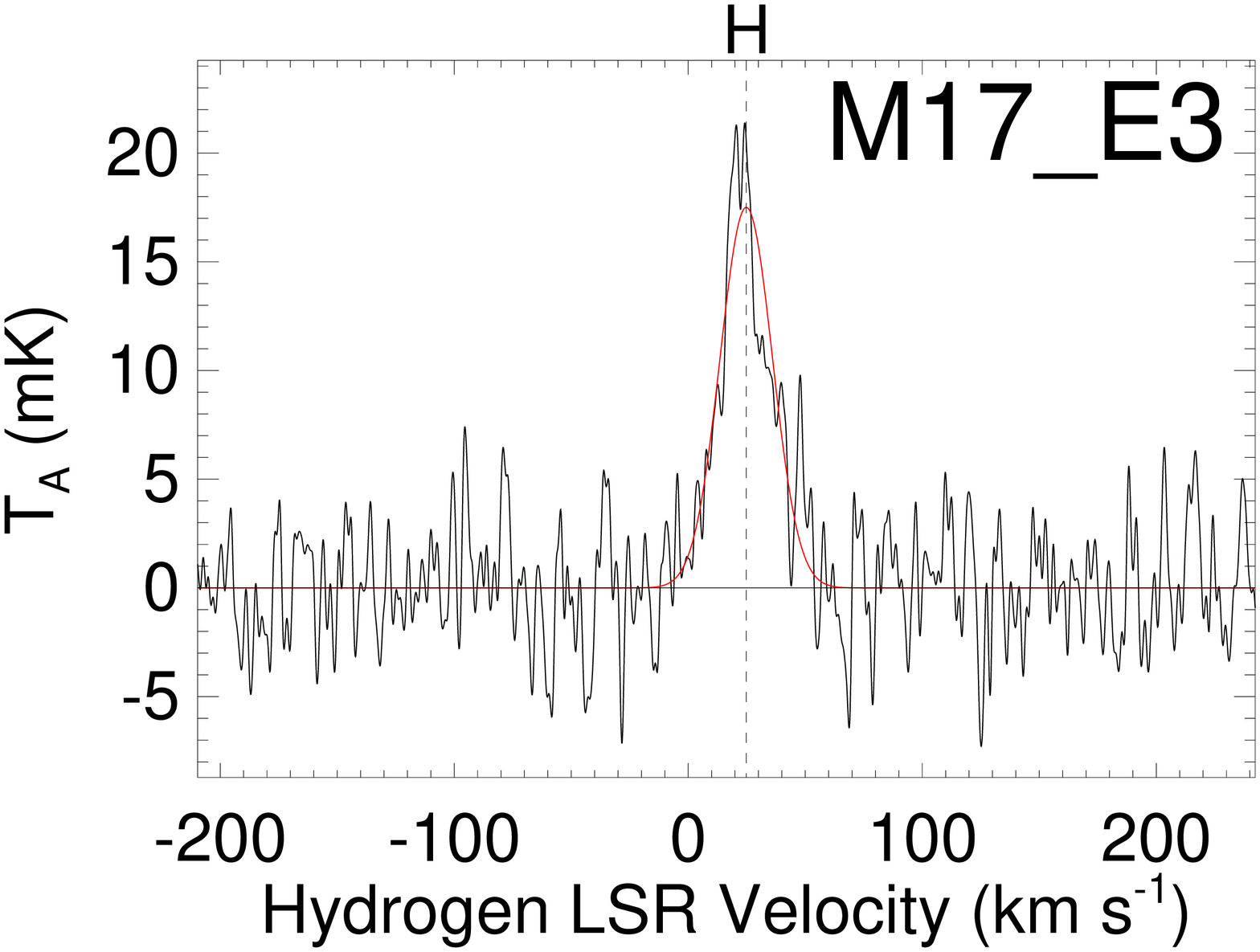} &
\includegraphics[width=.23\textwidth]{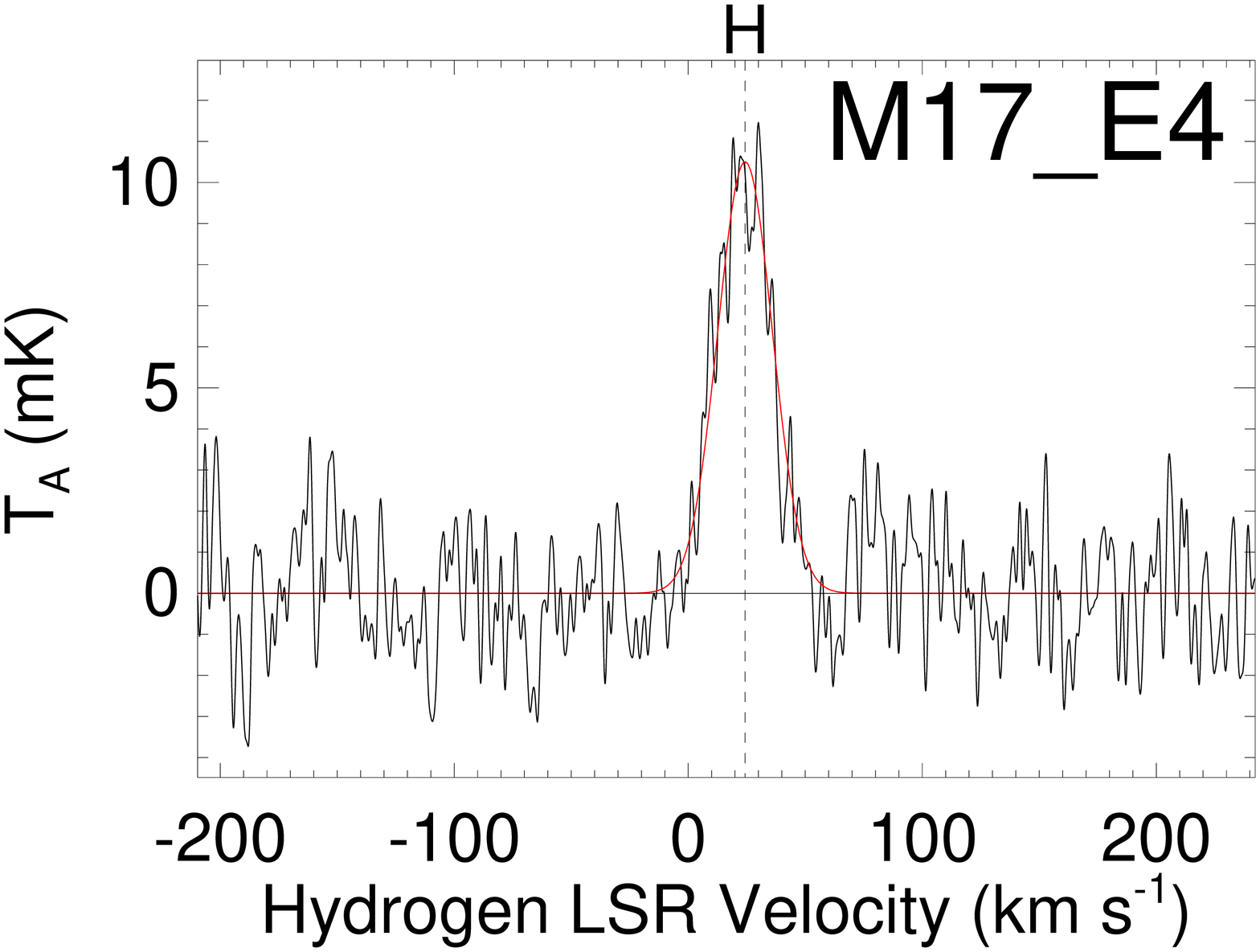} &
\includegraphics[width=.23\textwidth]{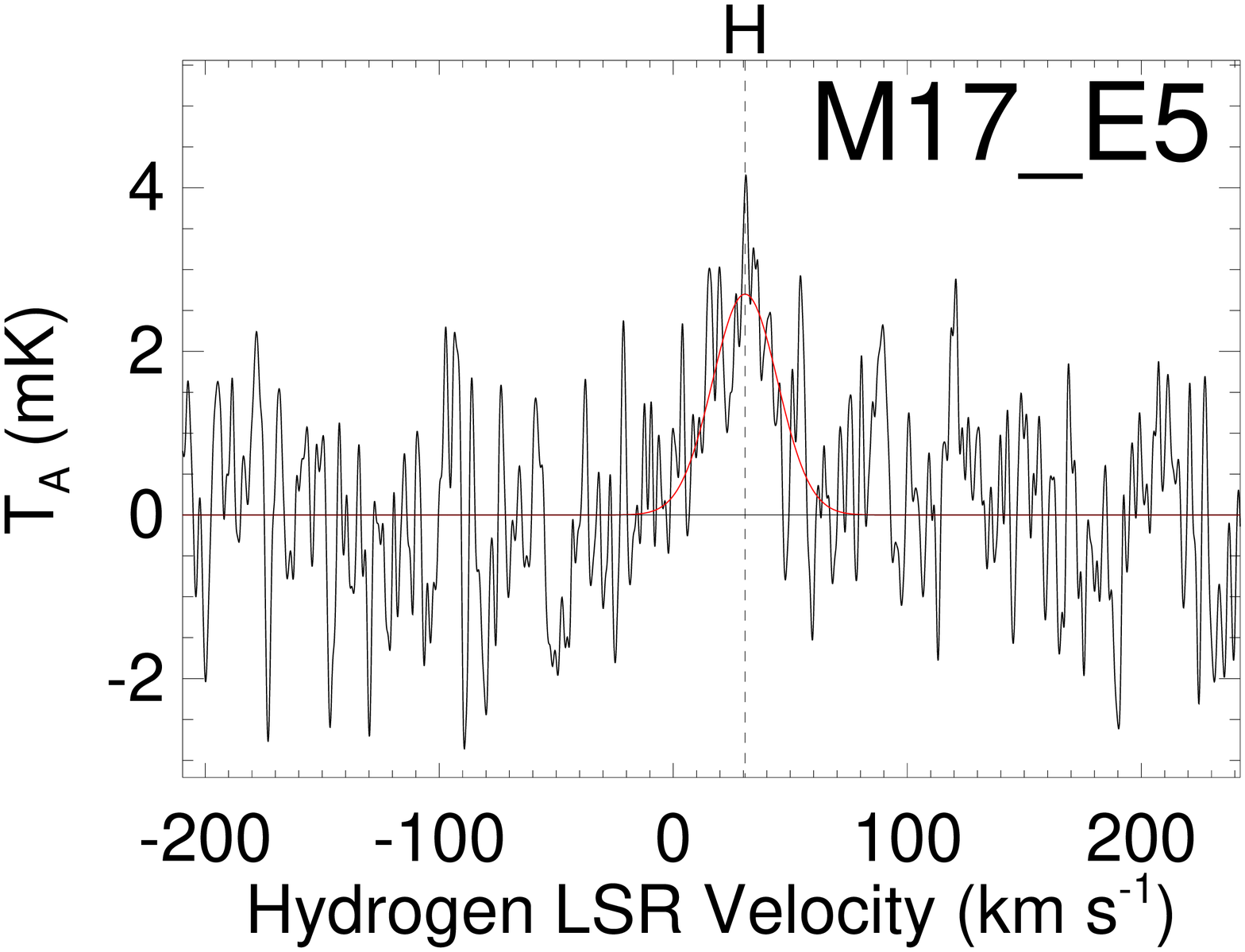} &
\includegraphics[width=.23\textwidth]{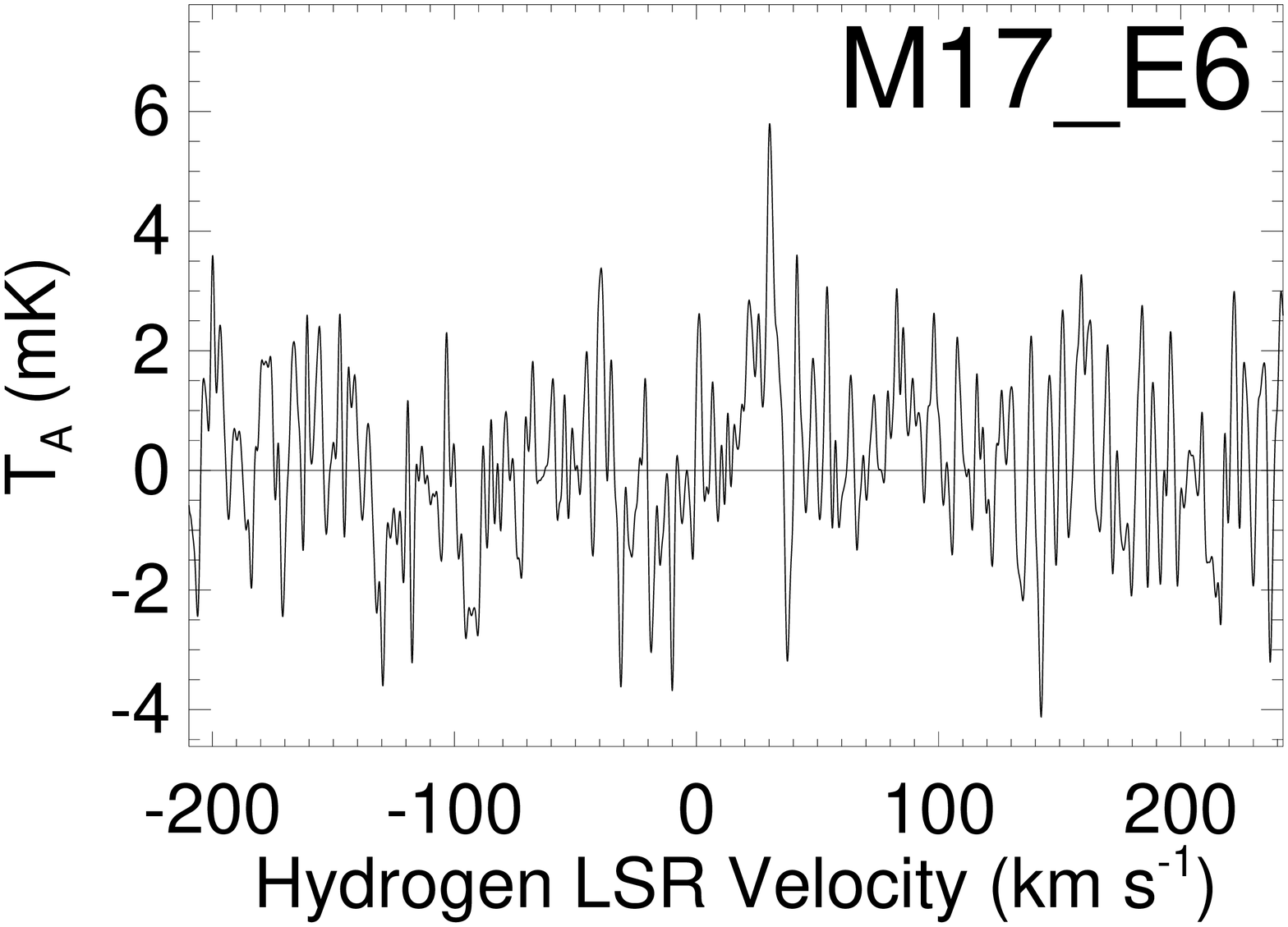} \\
\end{tabular}
\caption{}
\end{figure*}
\renewcommand{\thefigure}{\thesection.\arabic{figure}}

\renewcommand\thefigure{\thesection.\arabic{figure} (Cont.)}
\addtocounter{figure}{-1}
\begin{figure*}
\centering
\begin{tabular}{cccc}
\includegraphics[width=.23\textwidth]{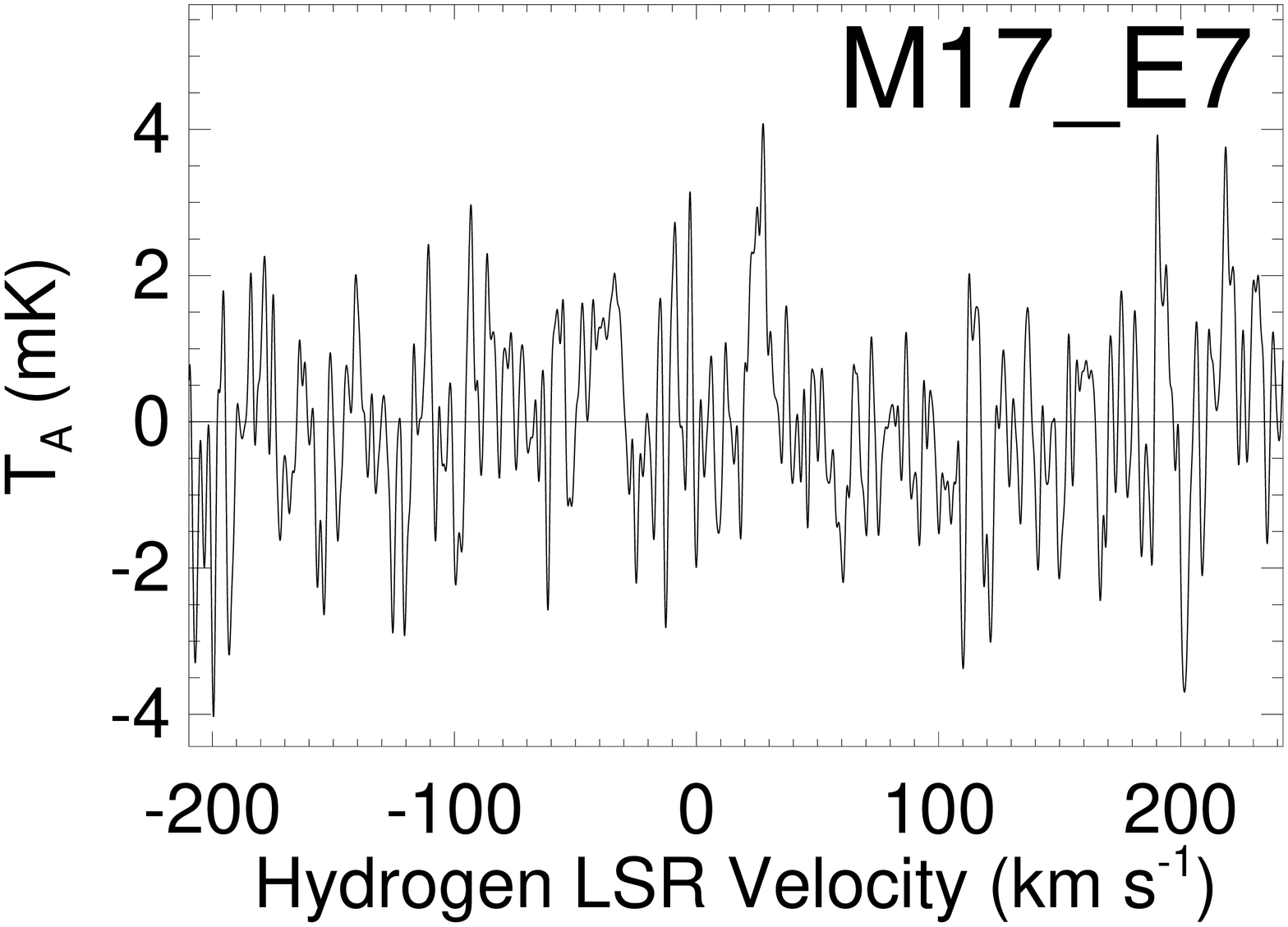} &
\includegraphics[width=.23\textwidth]{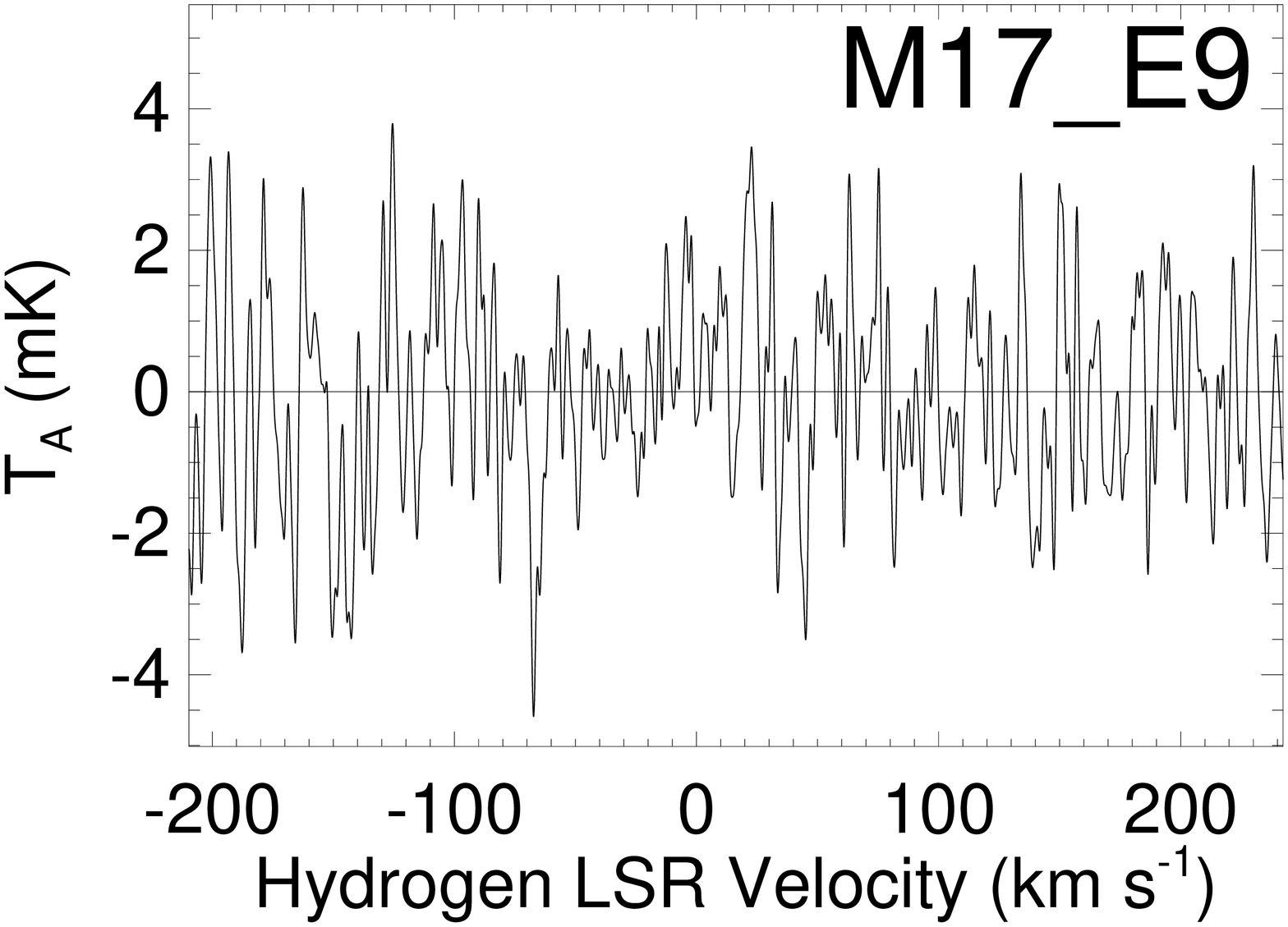} &
\includegraphics[width=.23\textwidth]{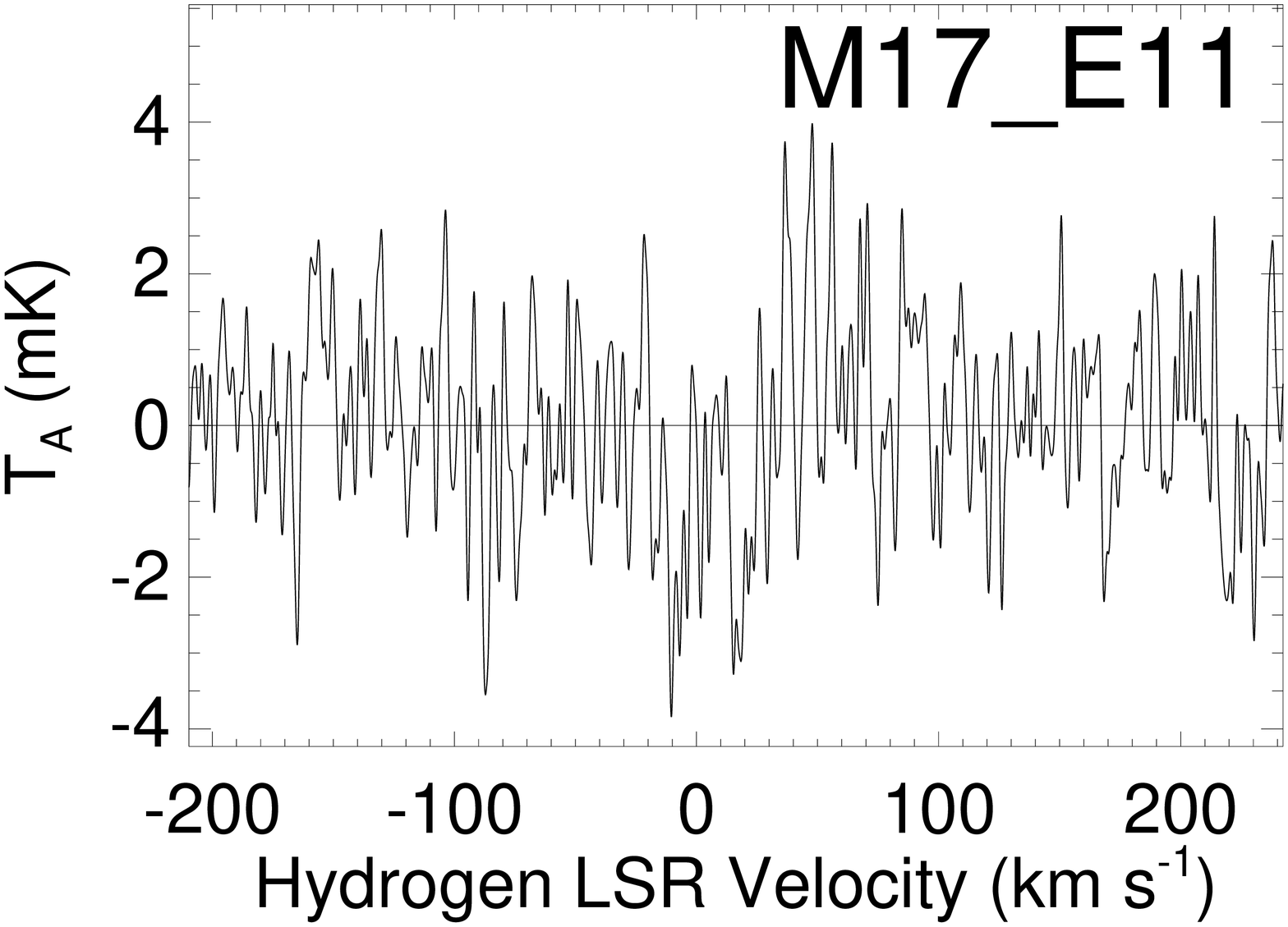} &
\includegraphics[width=.23\textwidth]{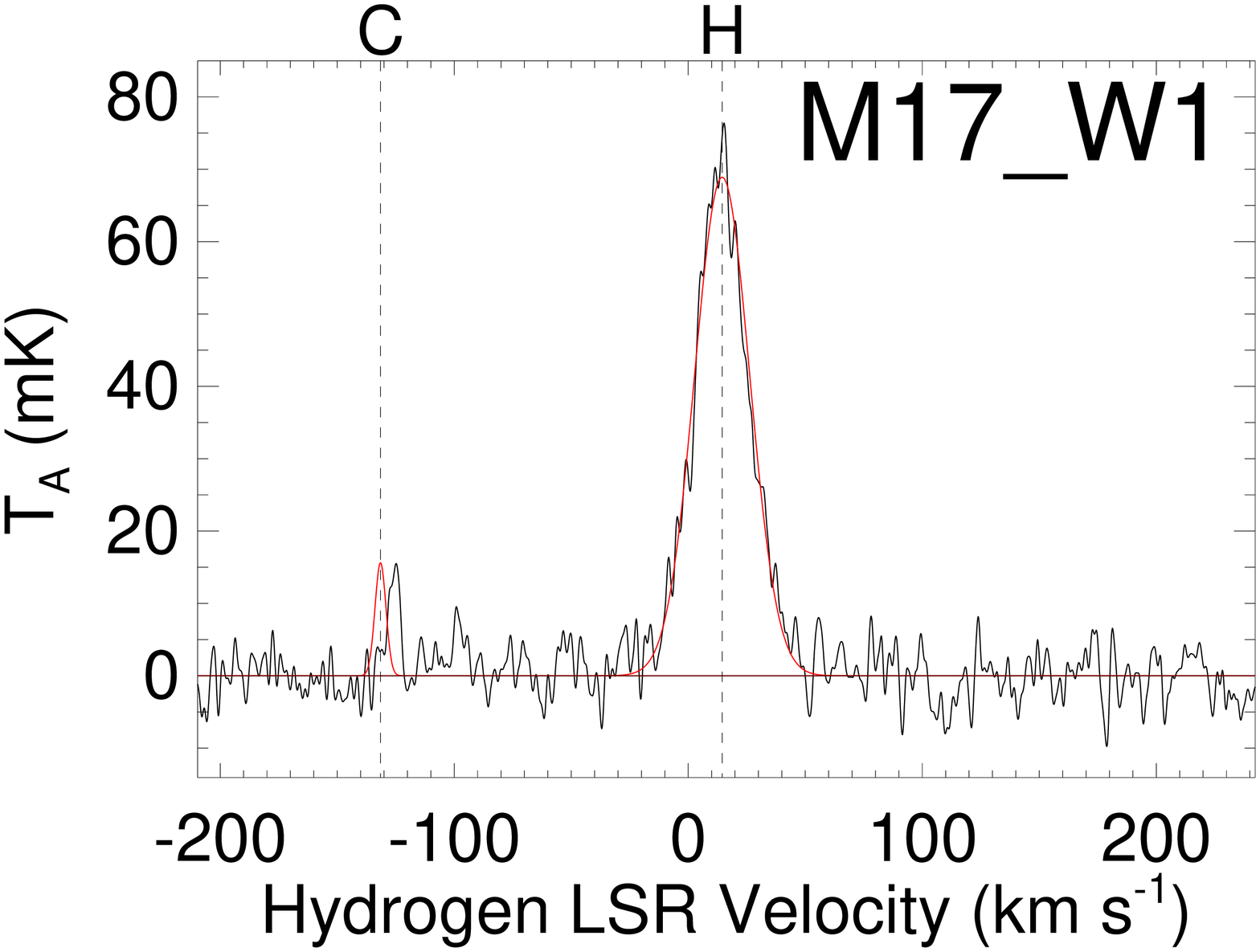} \\
\includegraphics[width=.23\textwidth]{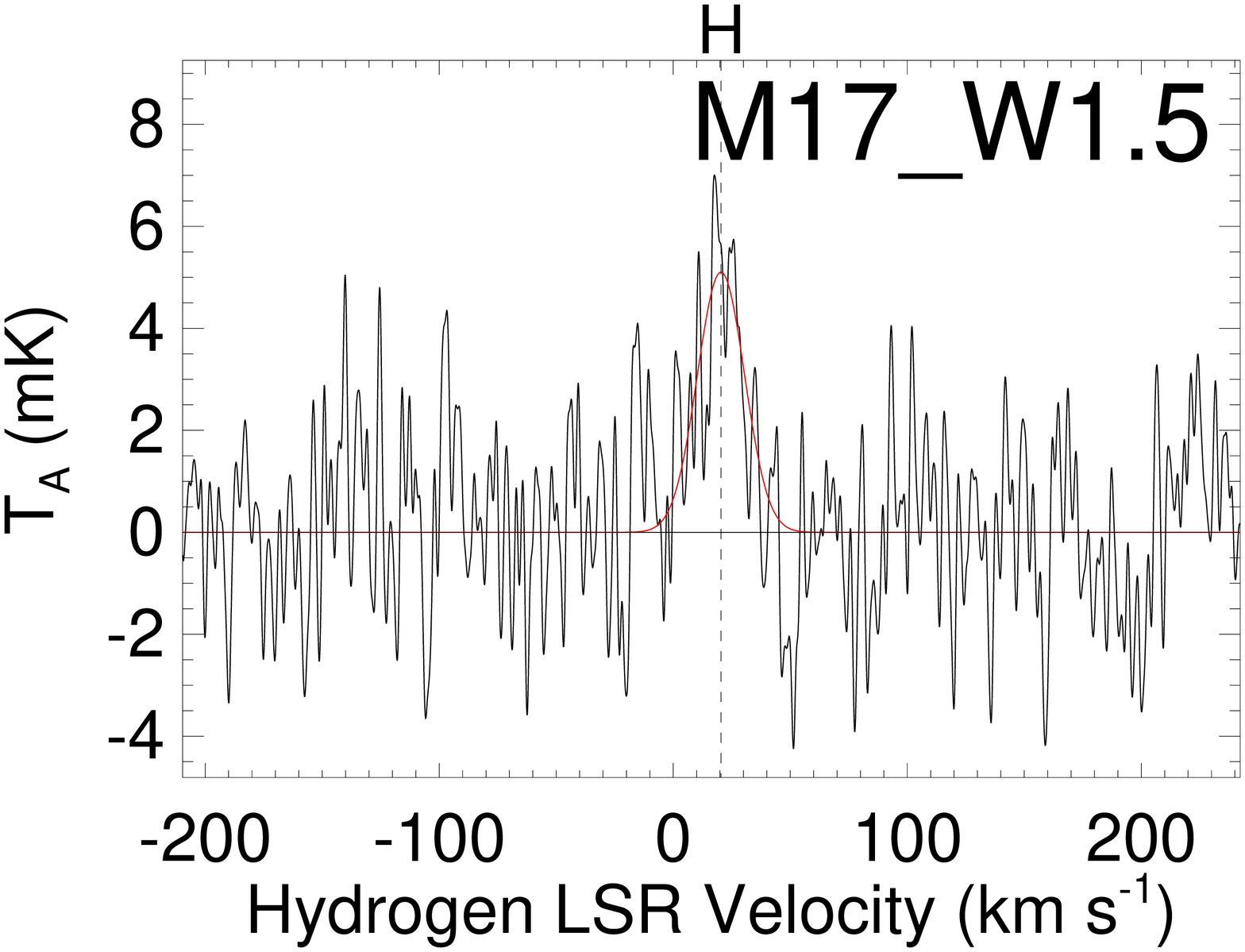} &
\includegraphics[width=.23\textwidth]{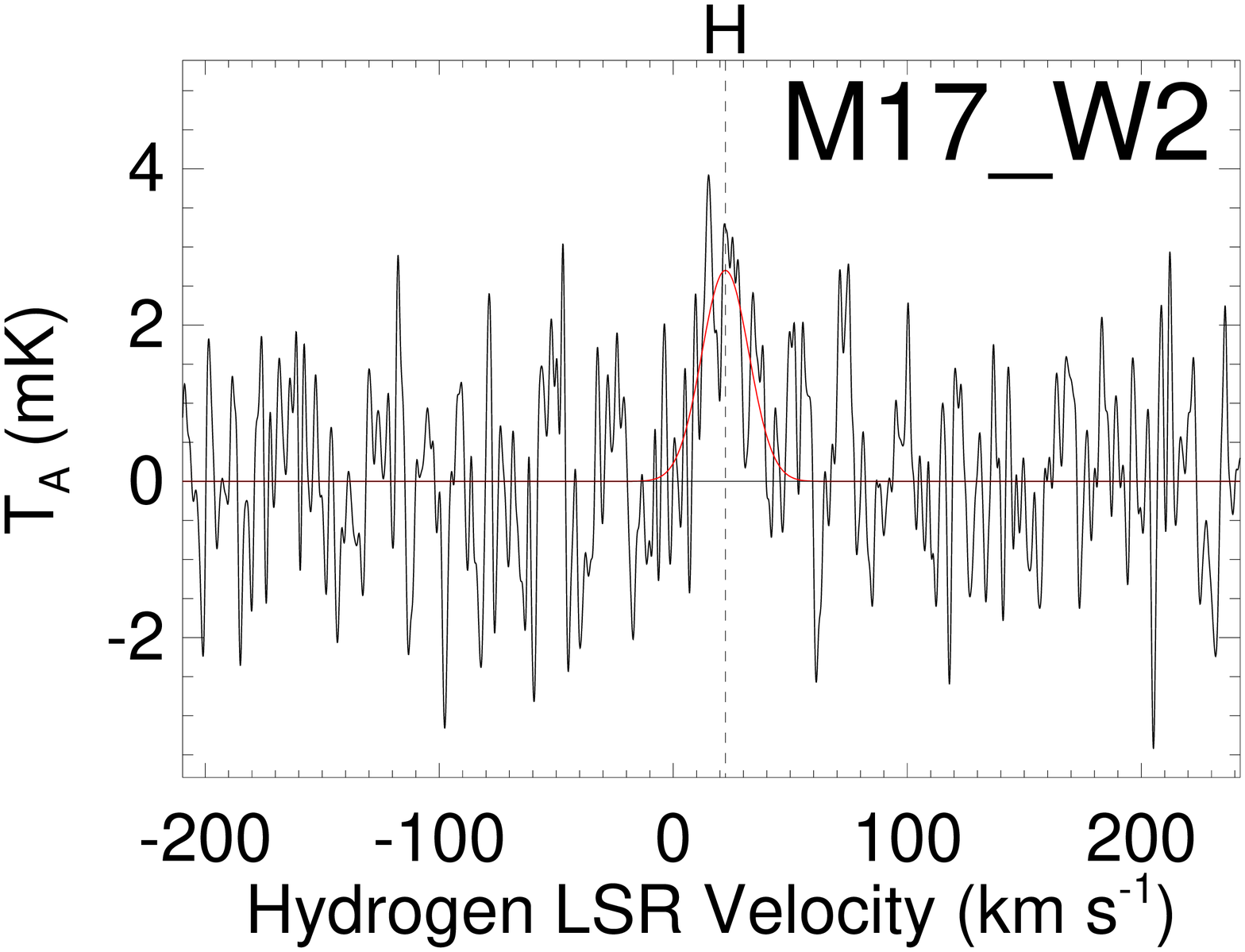} &
\includegraphics[width=.23\textwidth]{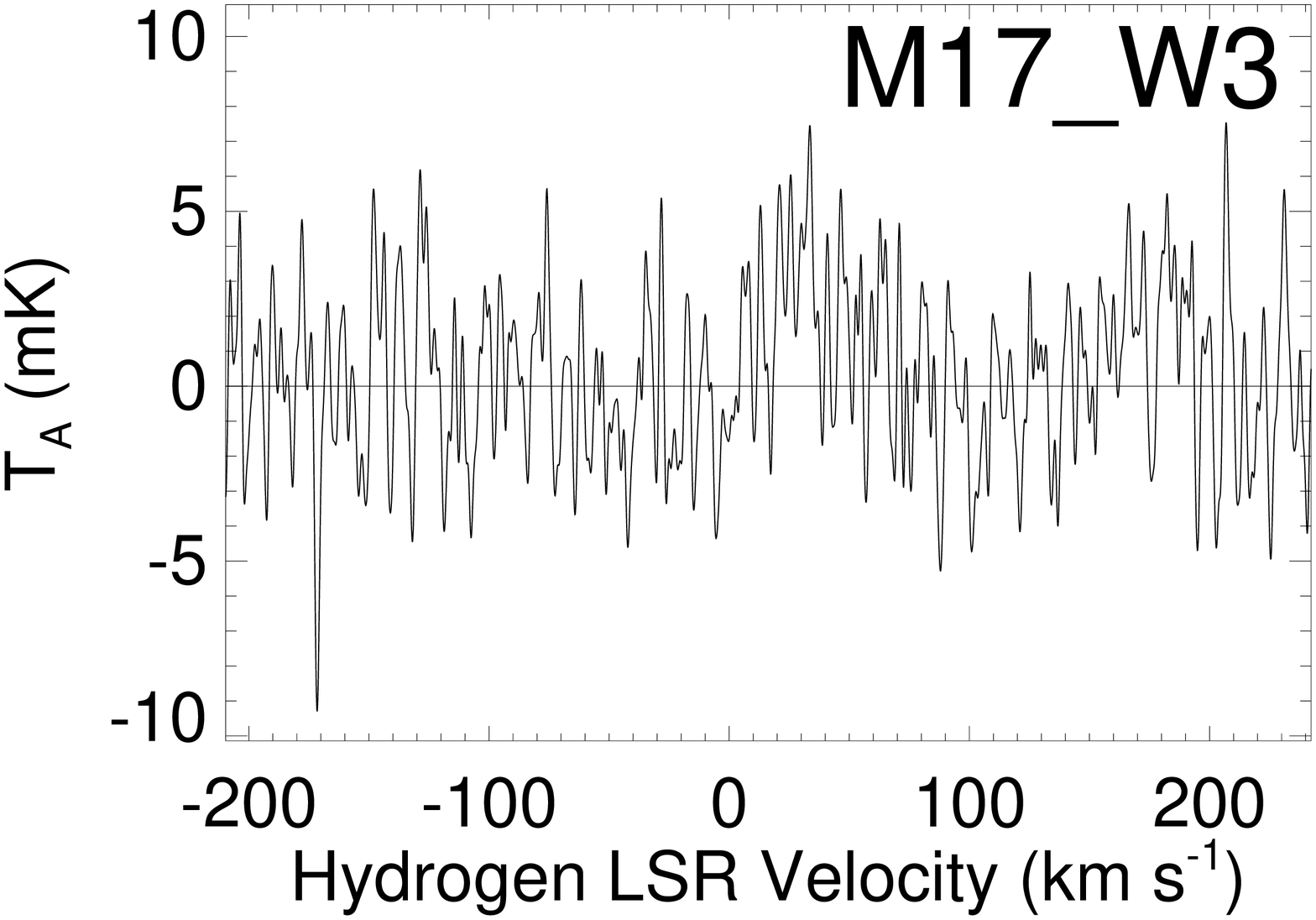} &
\includegraphics[width=.23\textwidth]{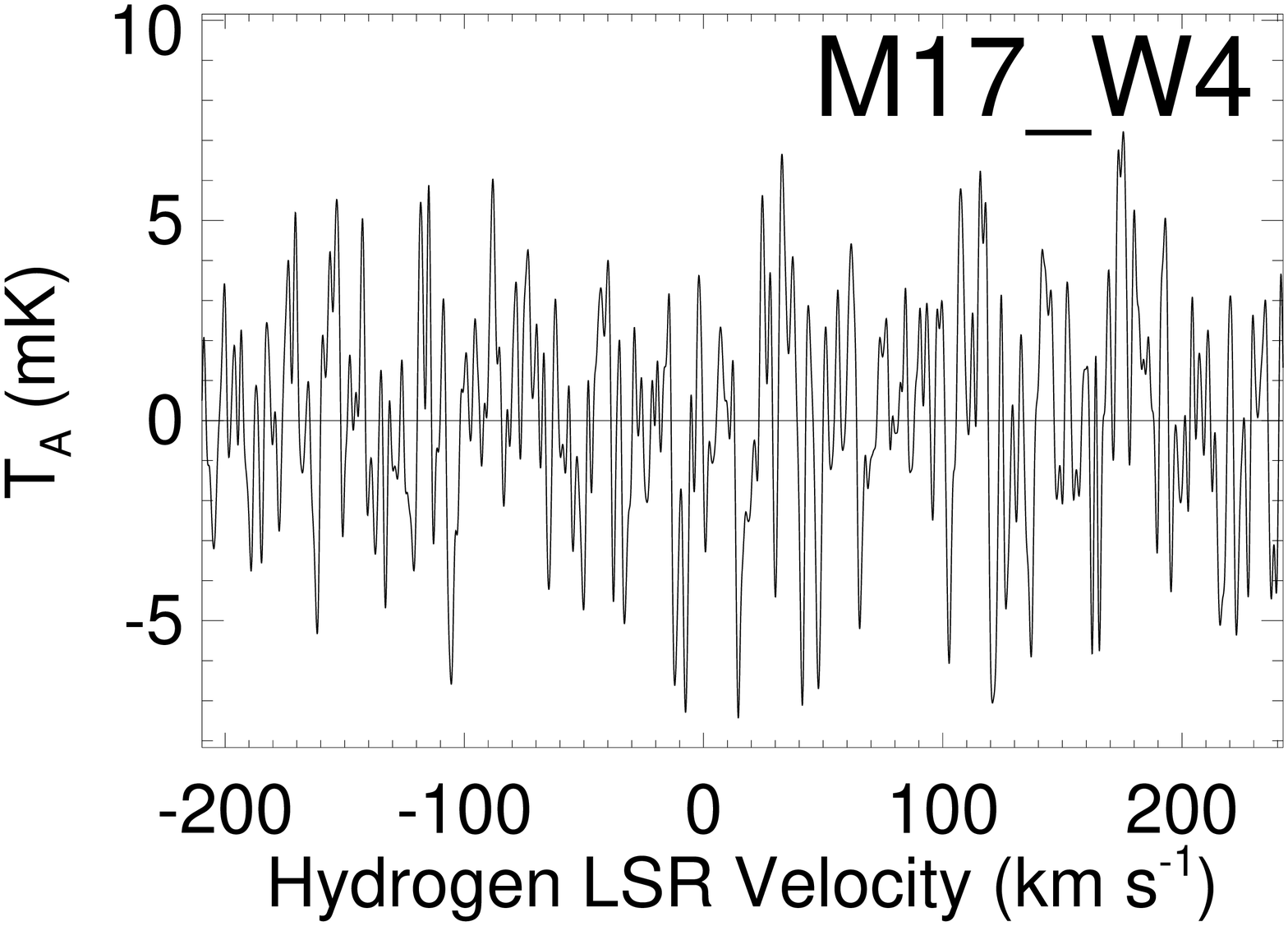} \\
\includegraphics[width=.23\textwidth]{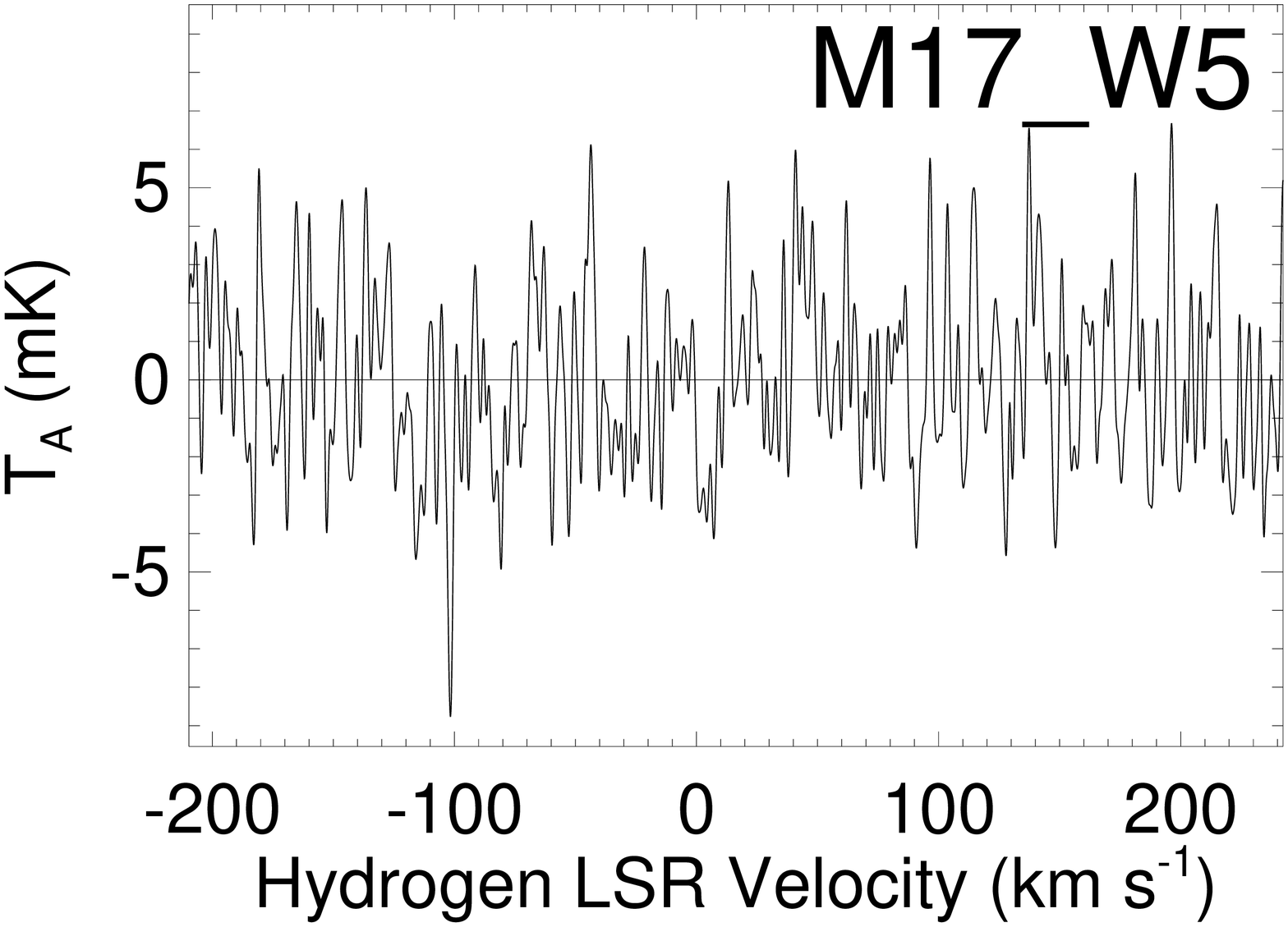} &
\includegraphics[width=.23\textwidth]{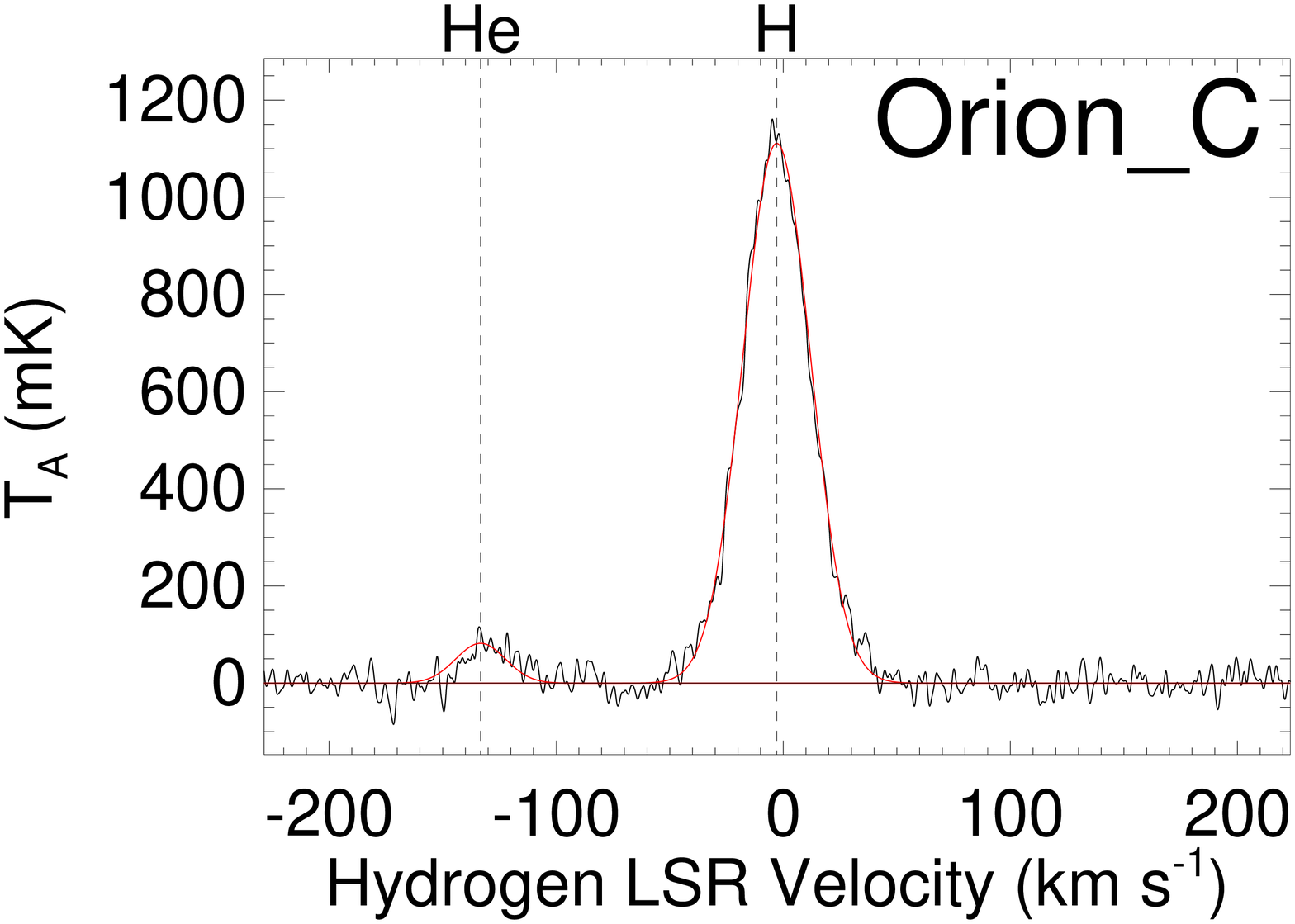} &
\includegraphics[width=.23\textwidth]{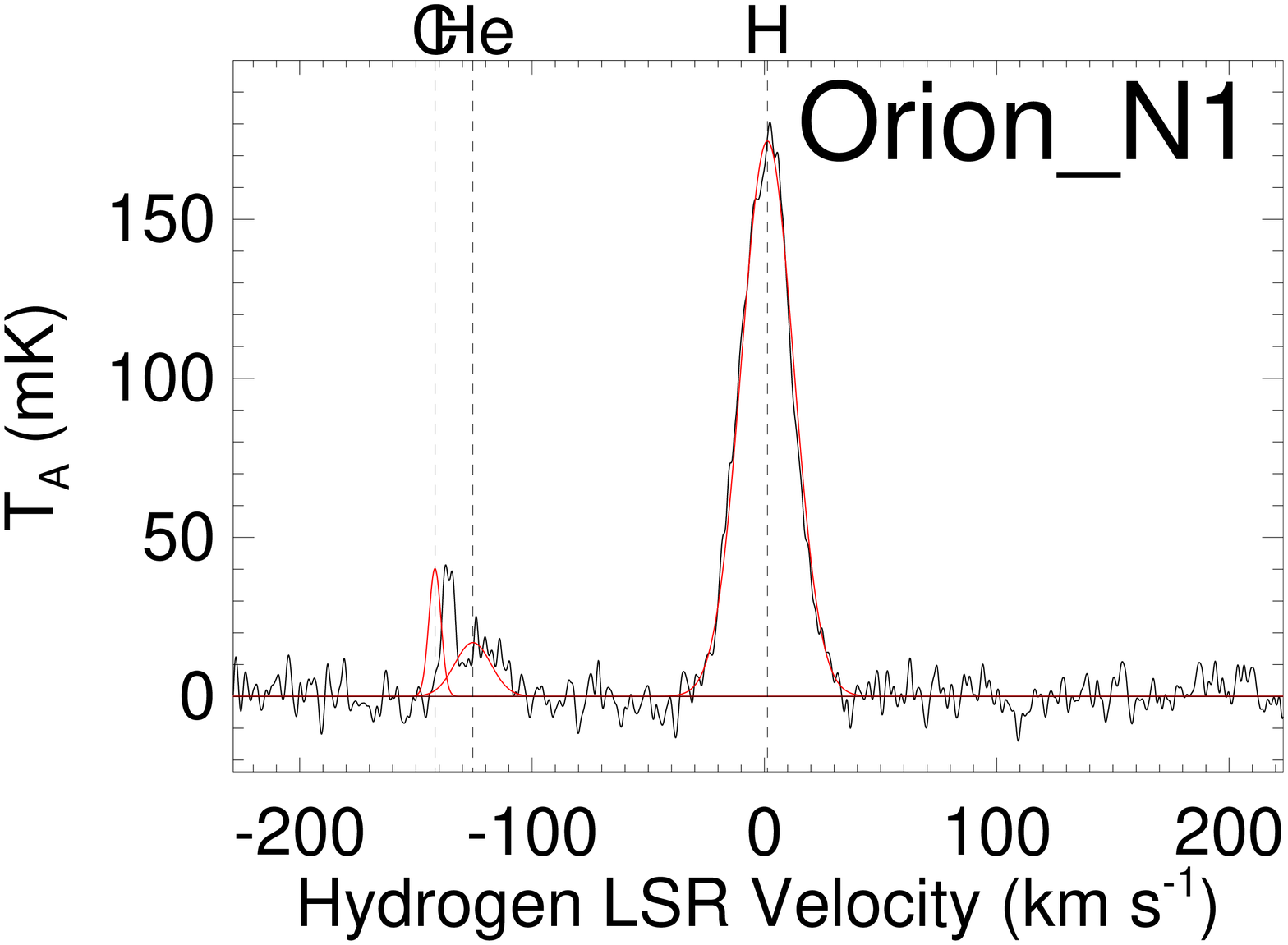} &
\includegraphics[width=.23\textwidth]{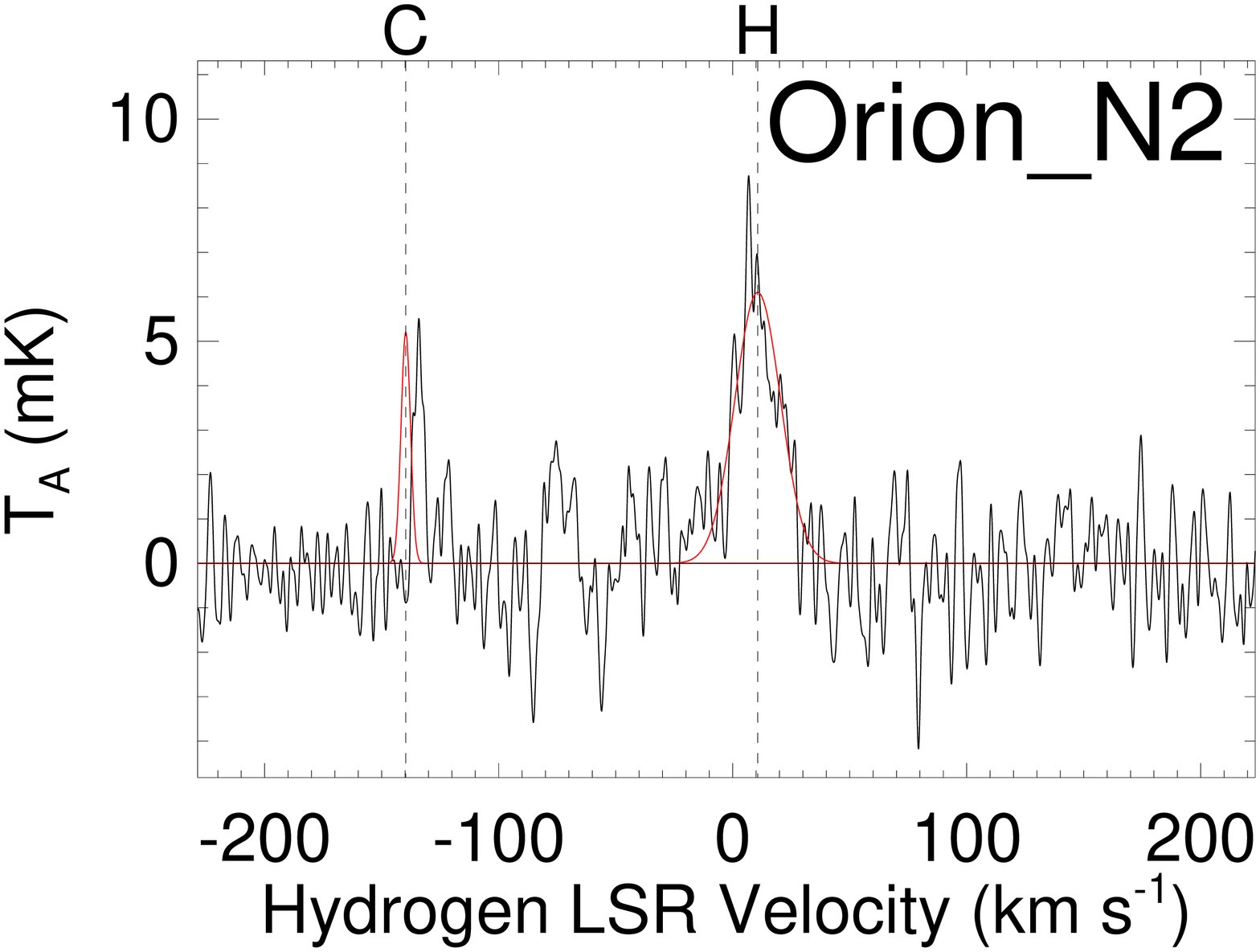} \\
\includegraphics[width=.23\textwidth]{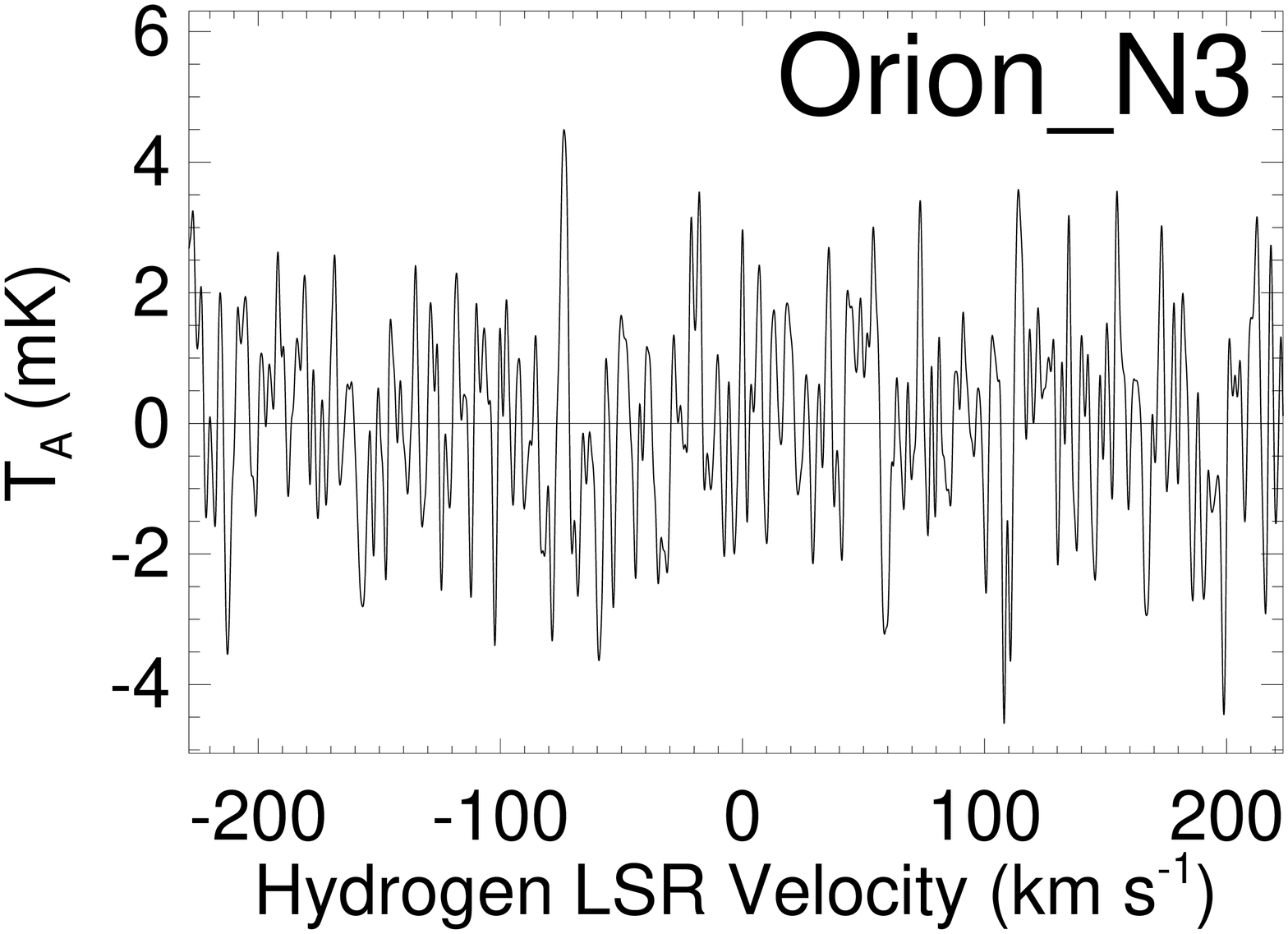} &
\includegraphics[width=.23\textwidth]{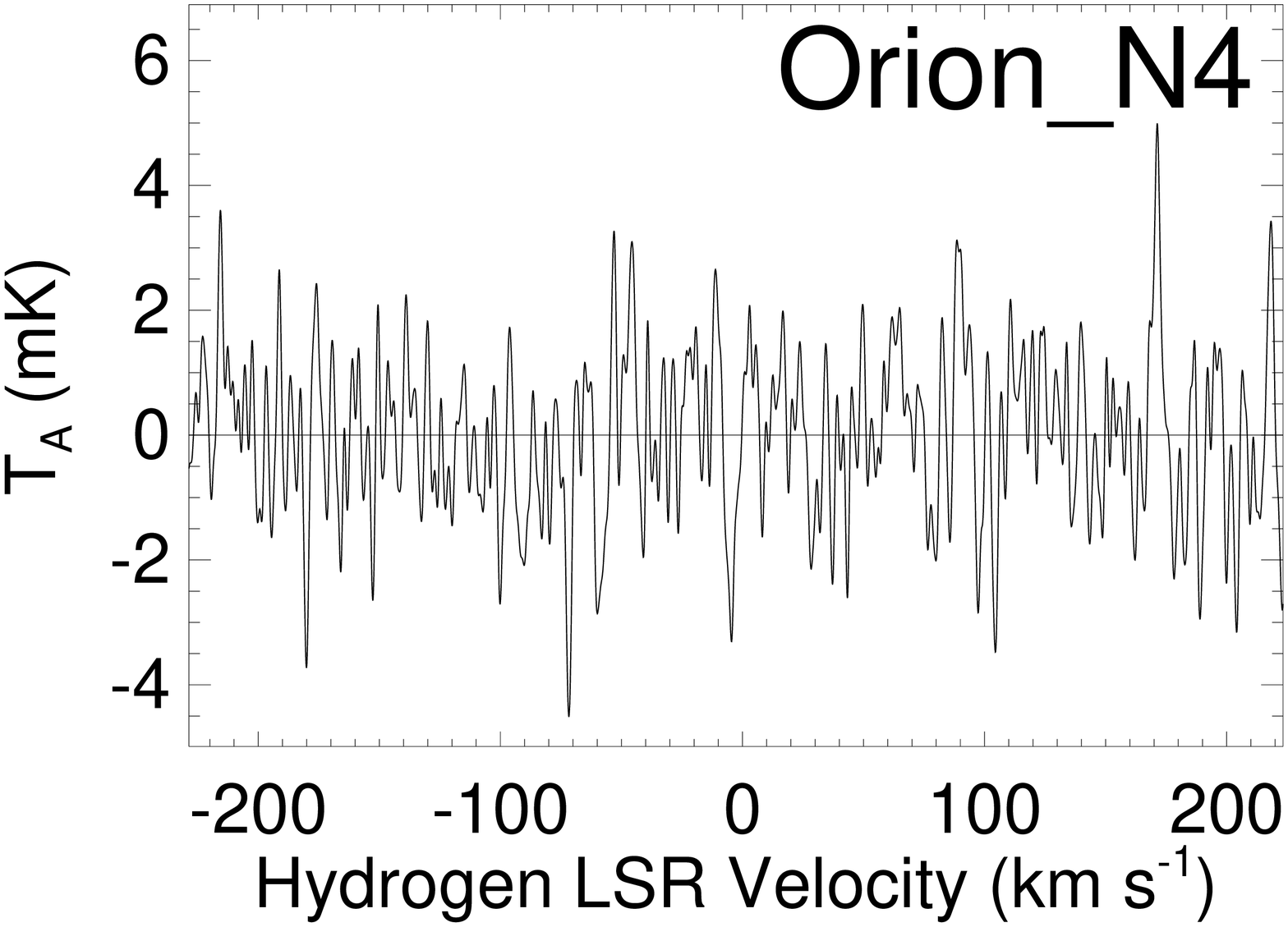} &
\includegraphics[width=.23\textwidth]{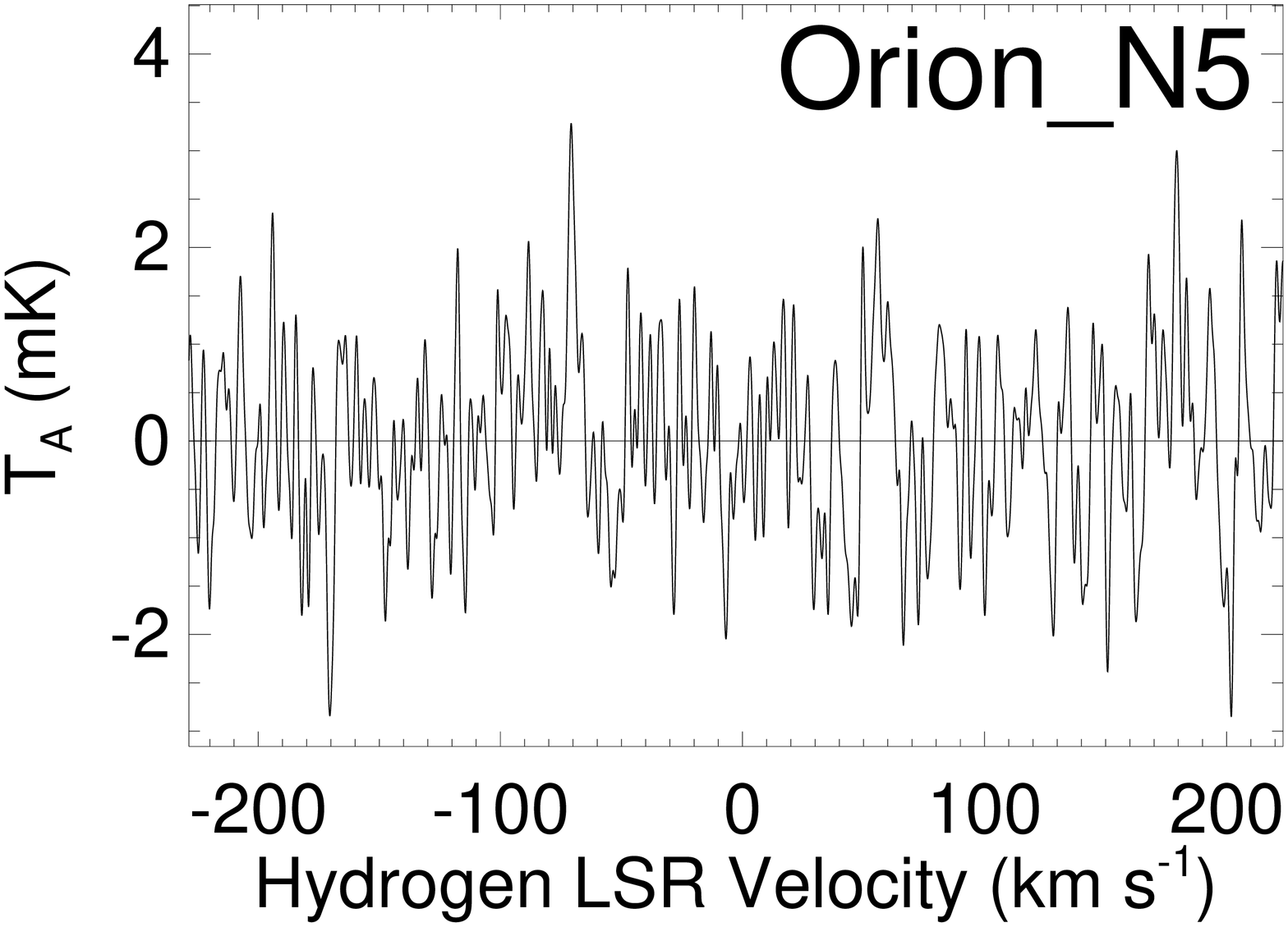} &
\includegraphics[width=.23\textwidth]{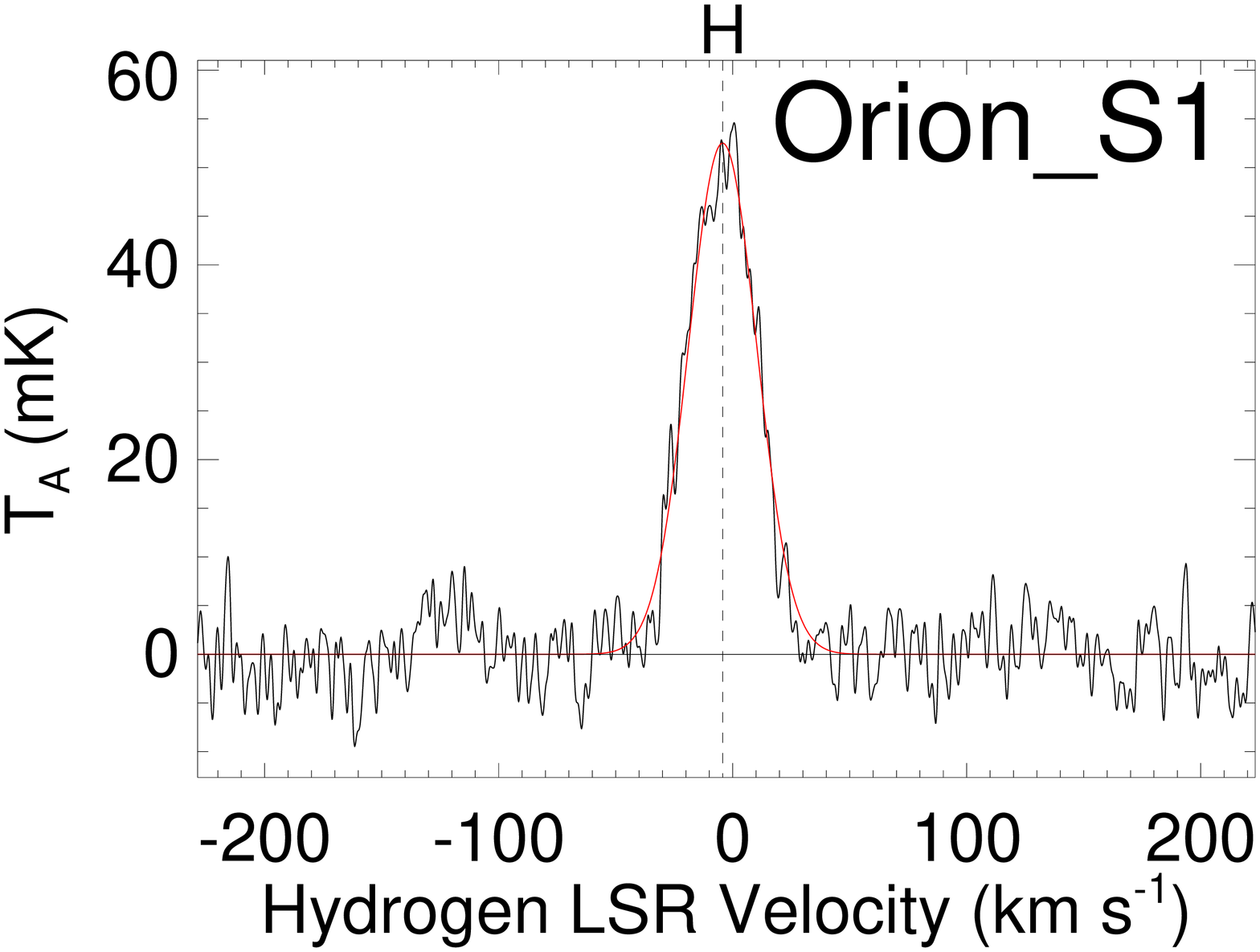} \\
\includegraphics[width=.23\textwidth]{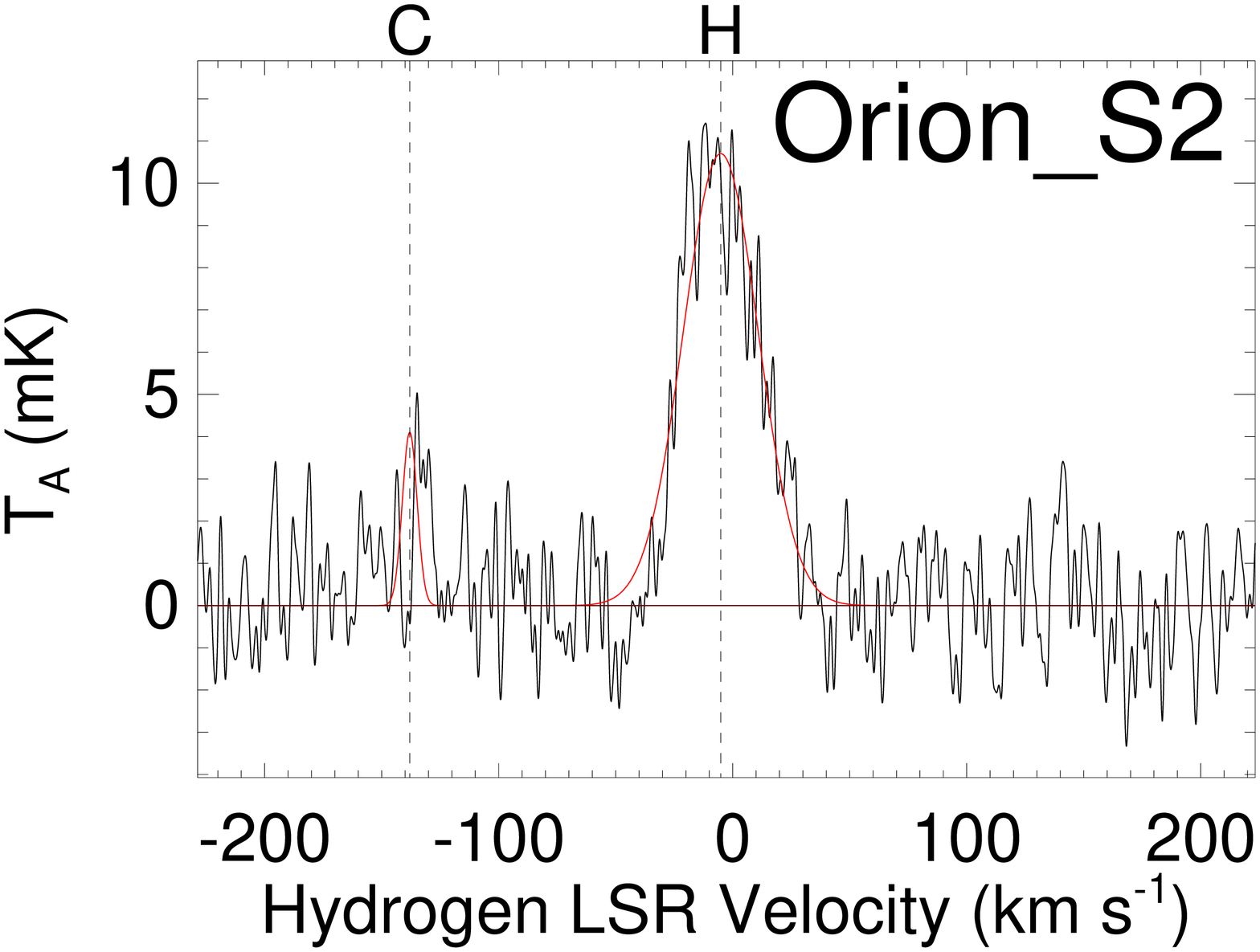} &
\includegraphics[width=.23\textwidth]{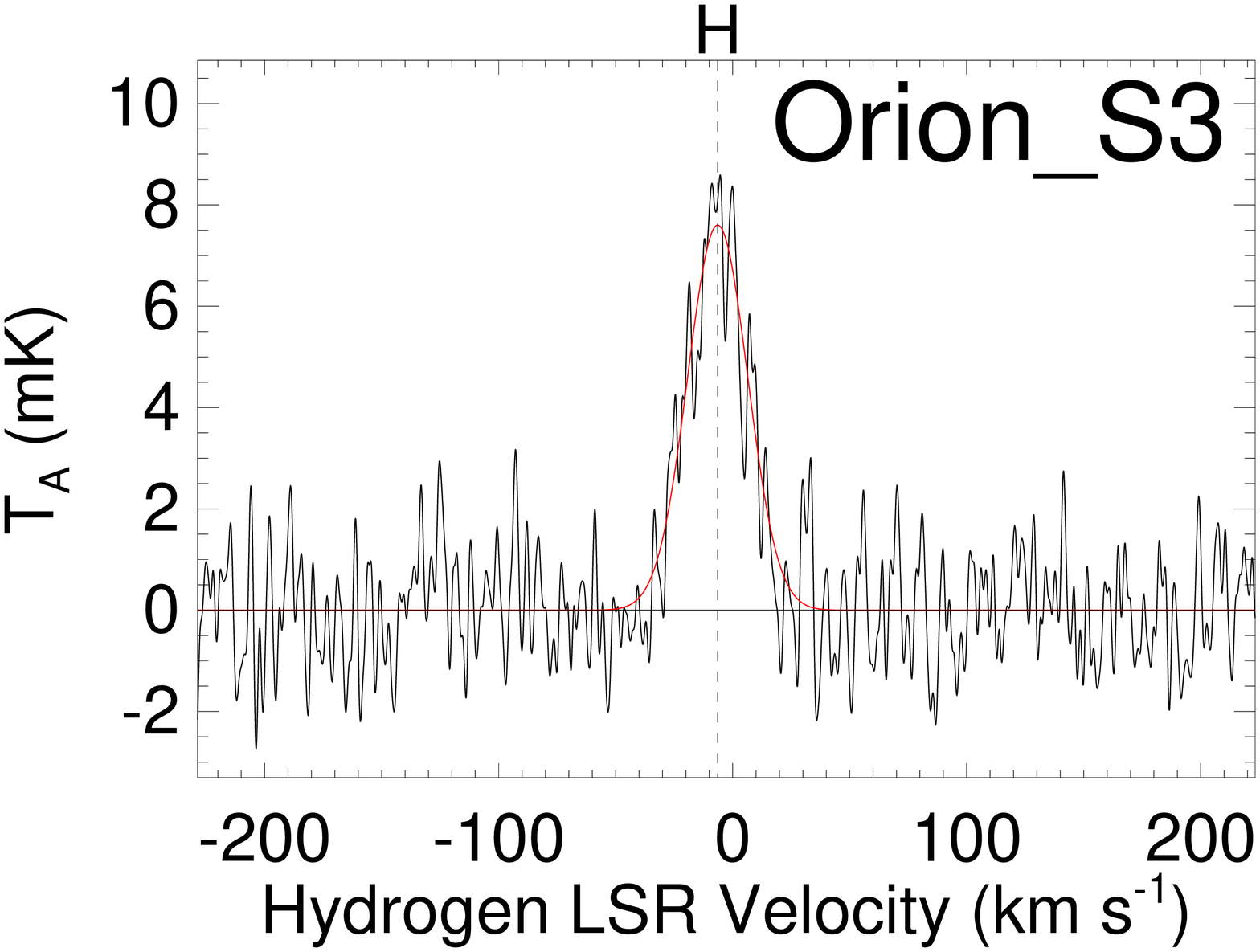} &
\includegraphics[width=.23\textwidth]{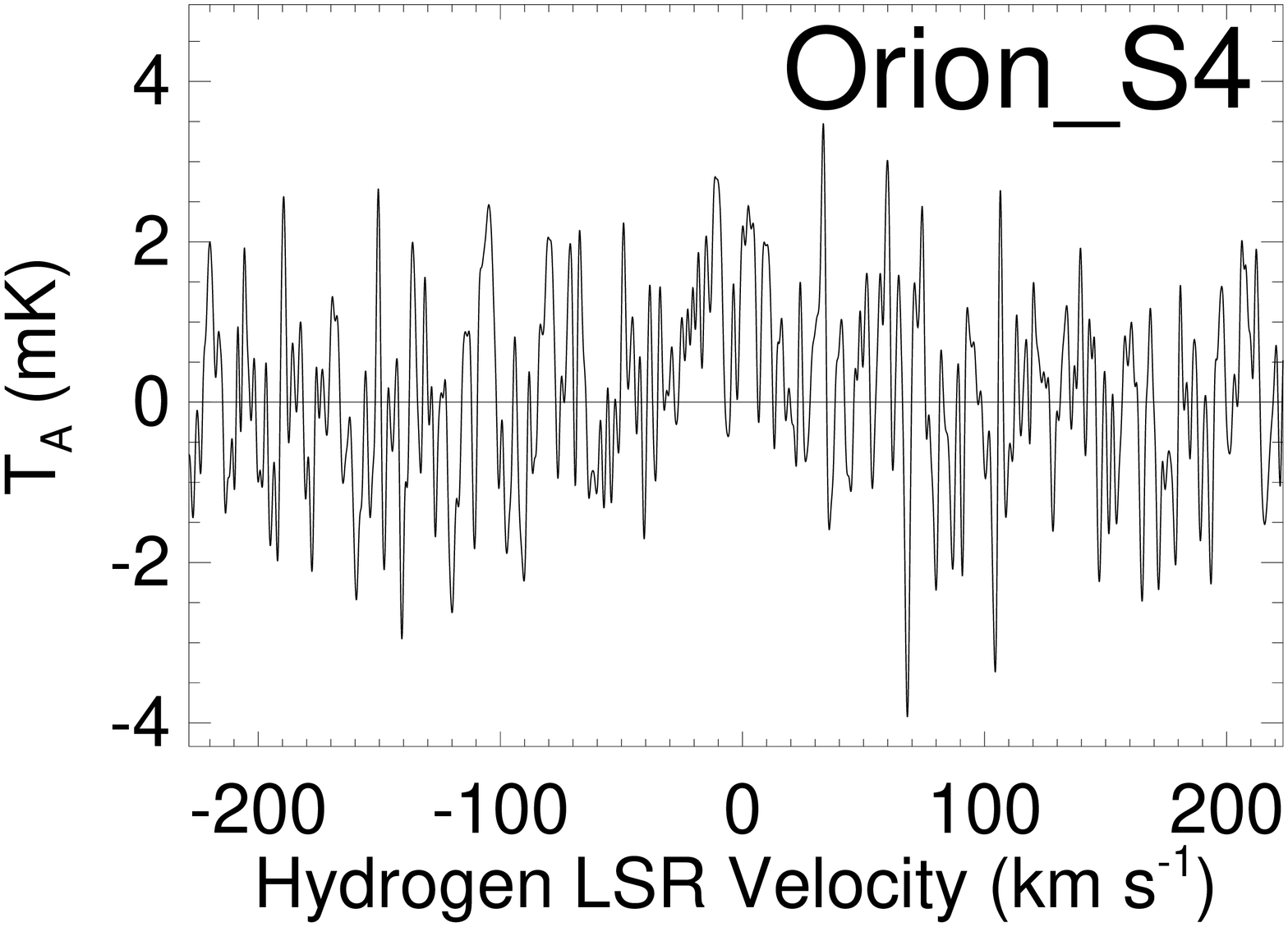} &
\includegraphics[width=.23\textwidth]{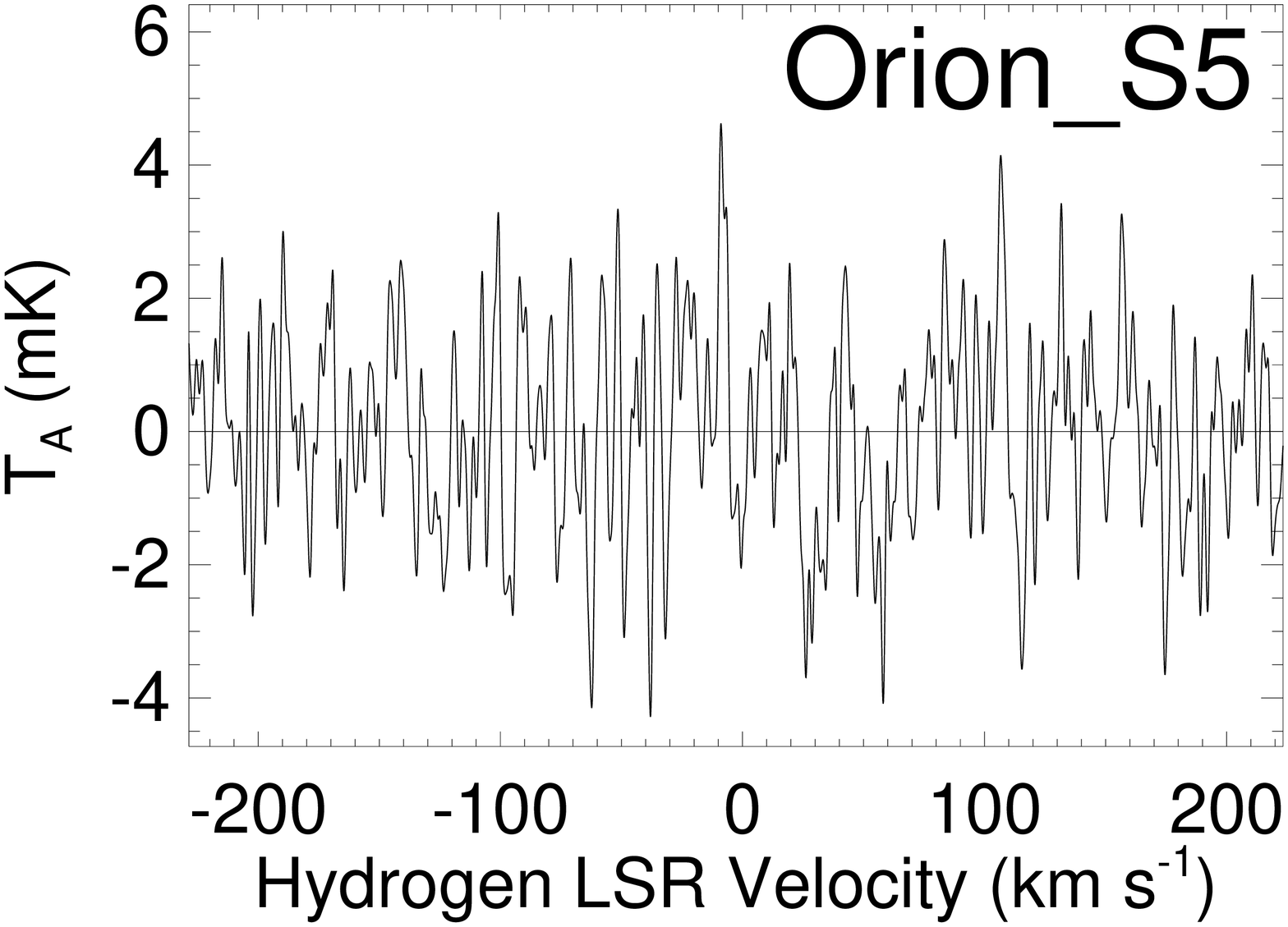}\\
\includegraphics[width=.23\textwidth]{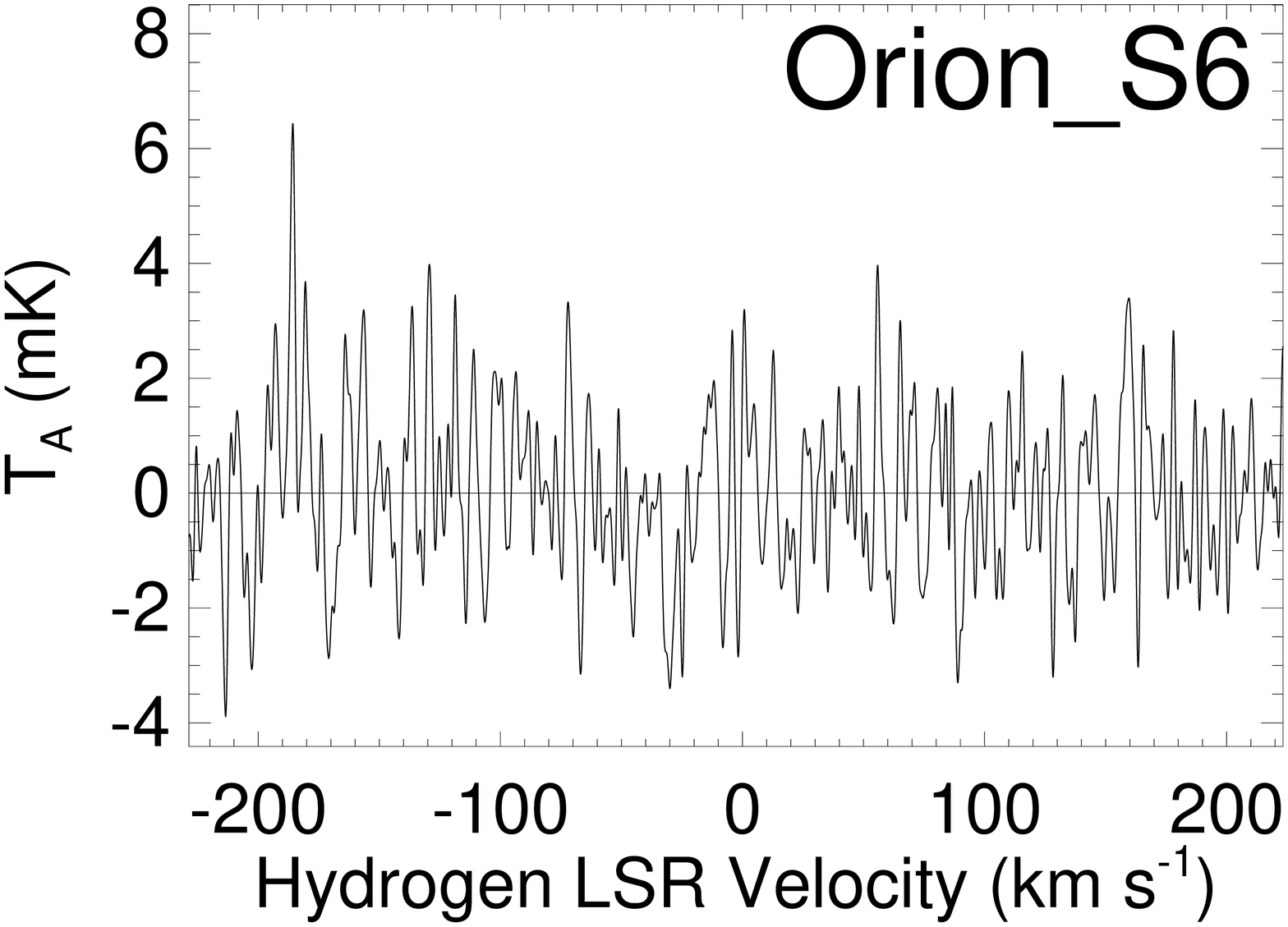} &
\includegraphics[width=.23\textwidth]{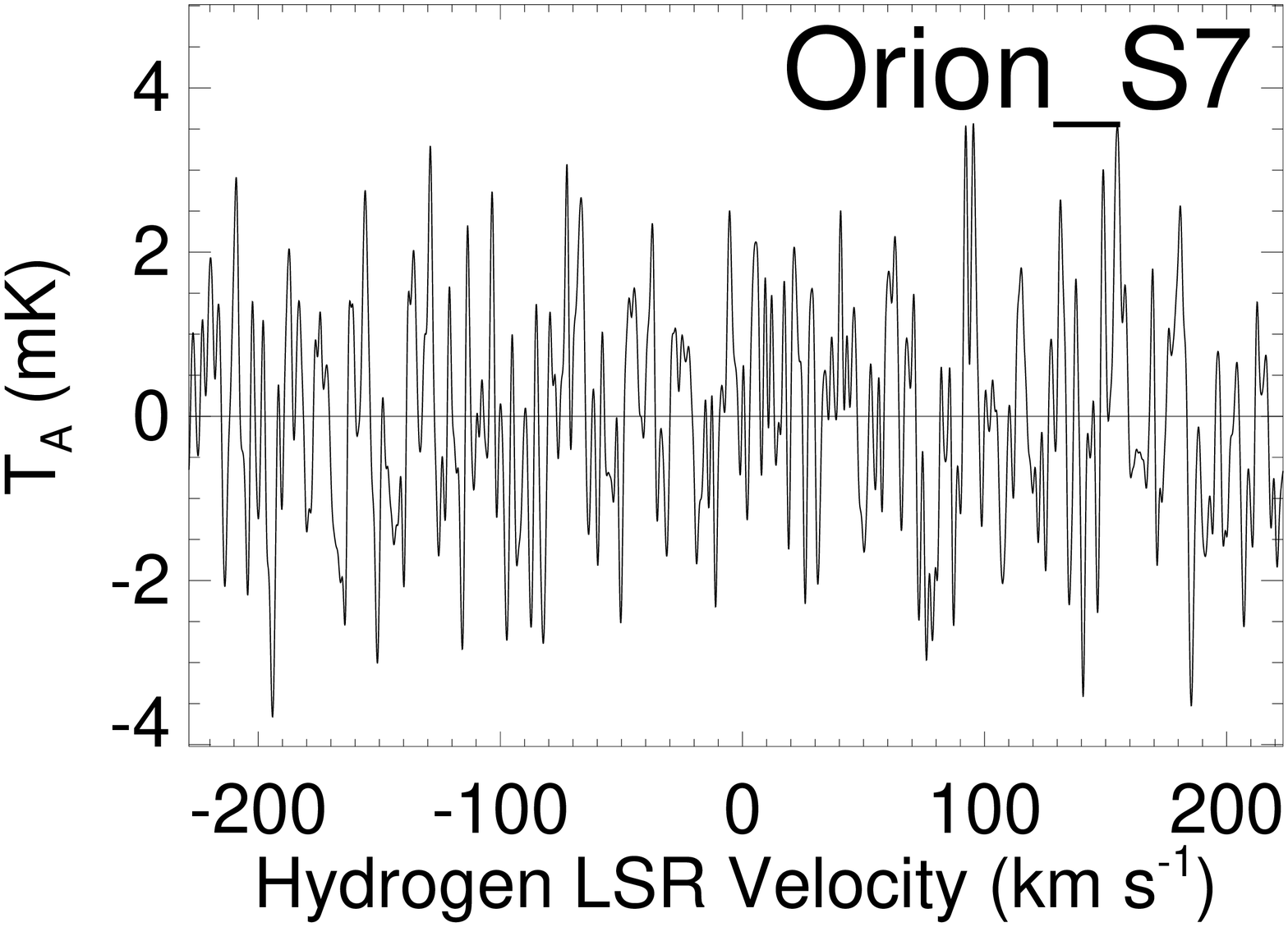} &
\includegraphics[width=.23\textwidth]{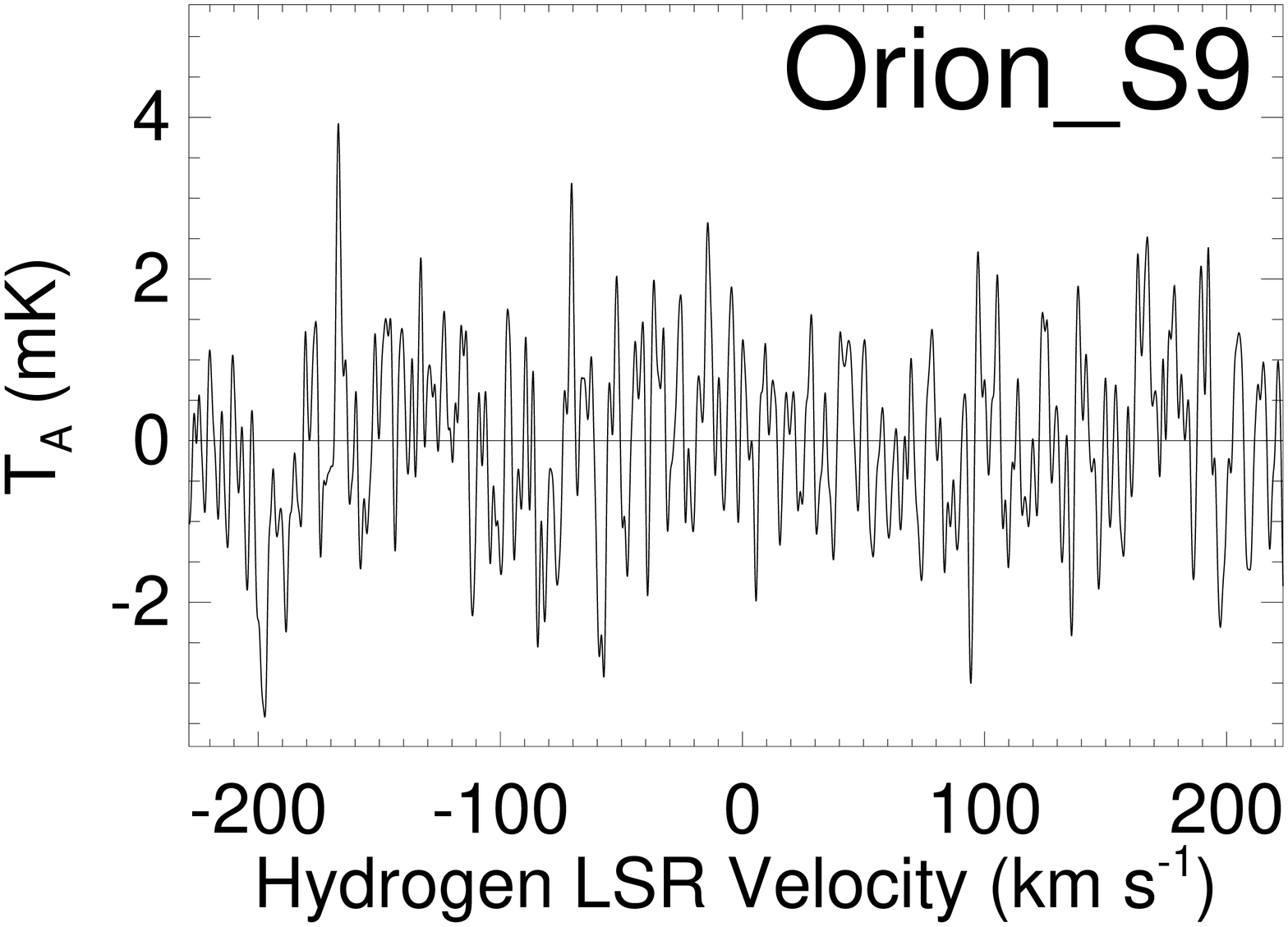} &
\includegraphics[width=.23\textwidth]{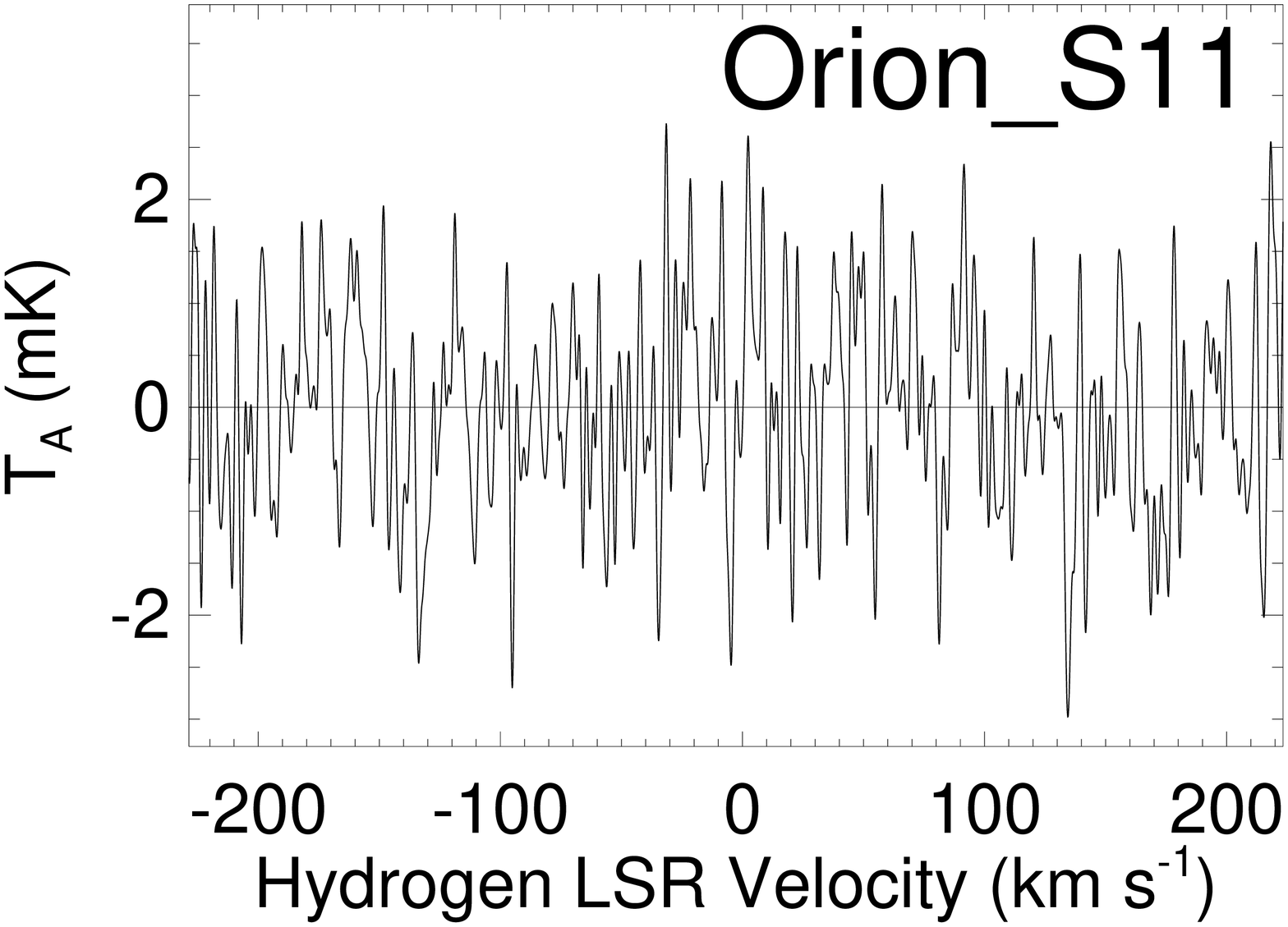} \\
\includegraphics[width=.23\textwidth]{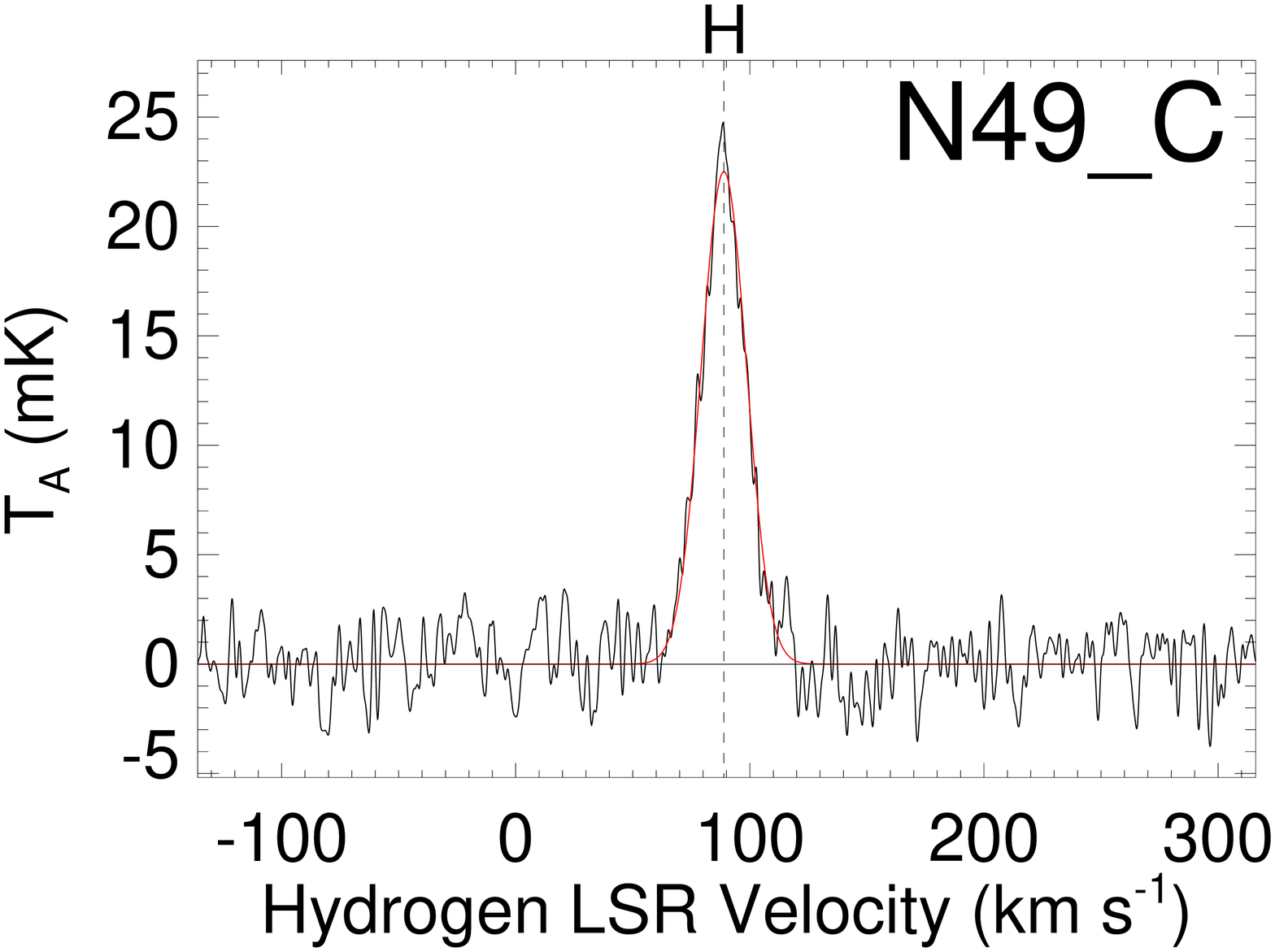} &
\includegraphics[width=.23\textwidth]{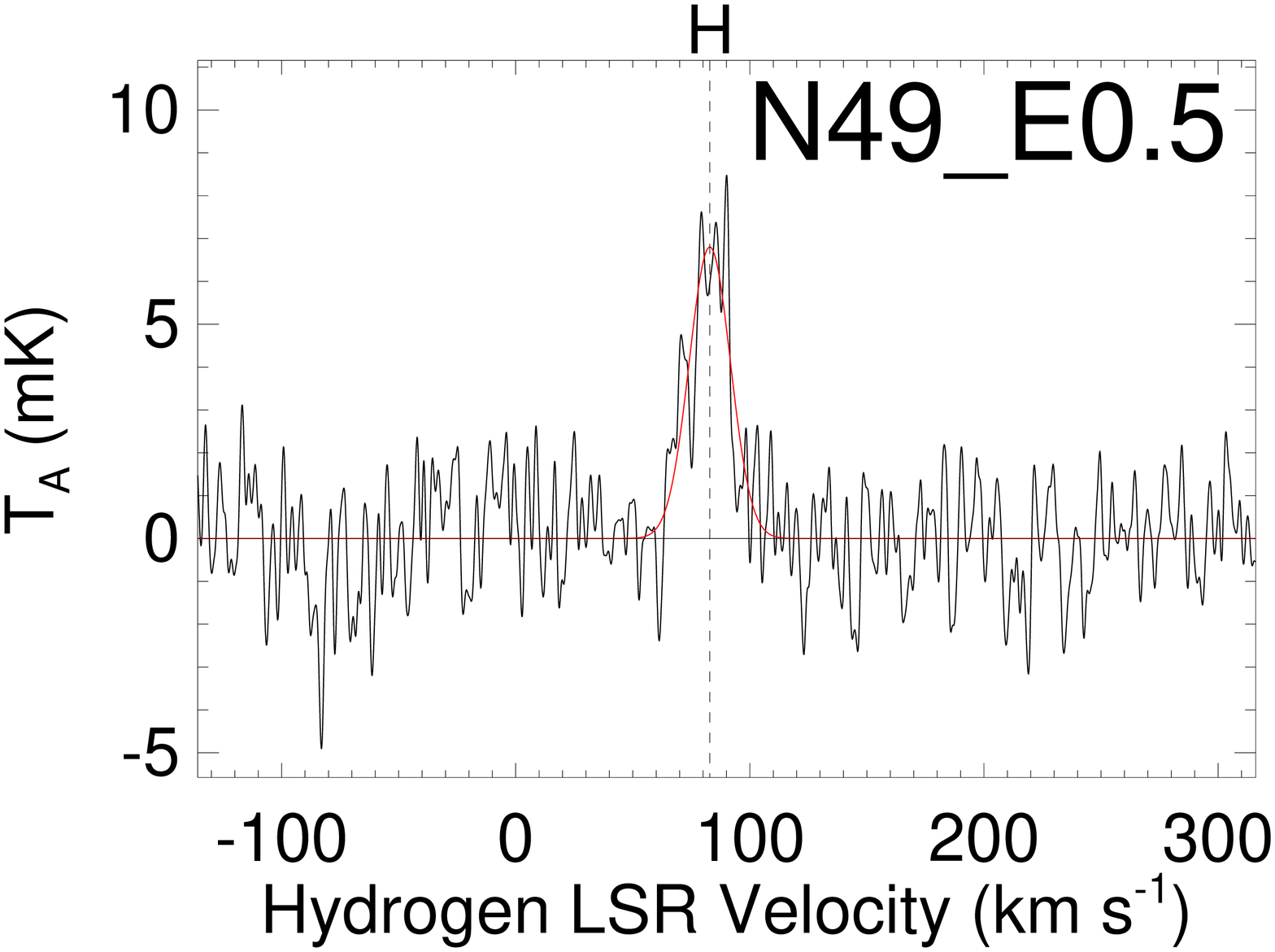} &
\includegraphics[width=.23\textwidth]{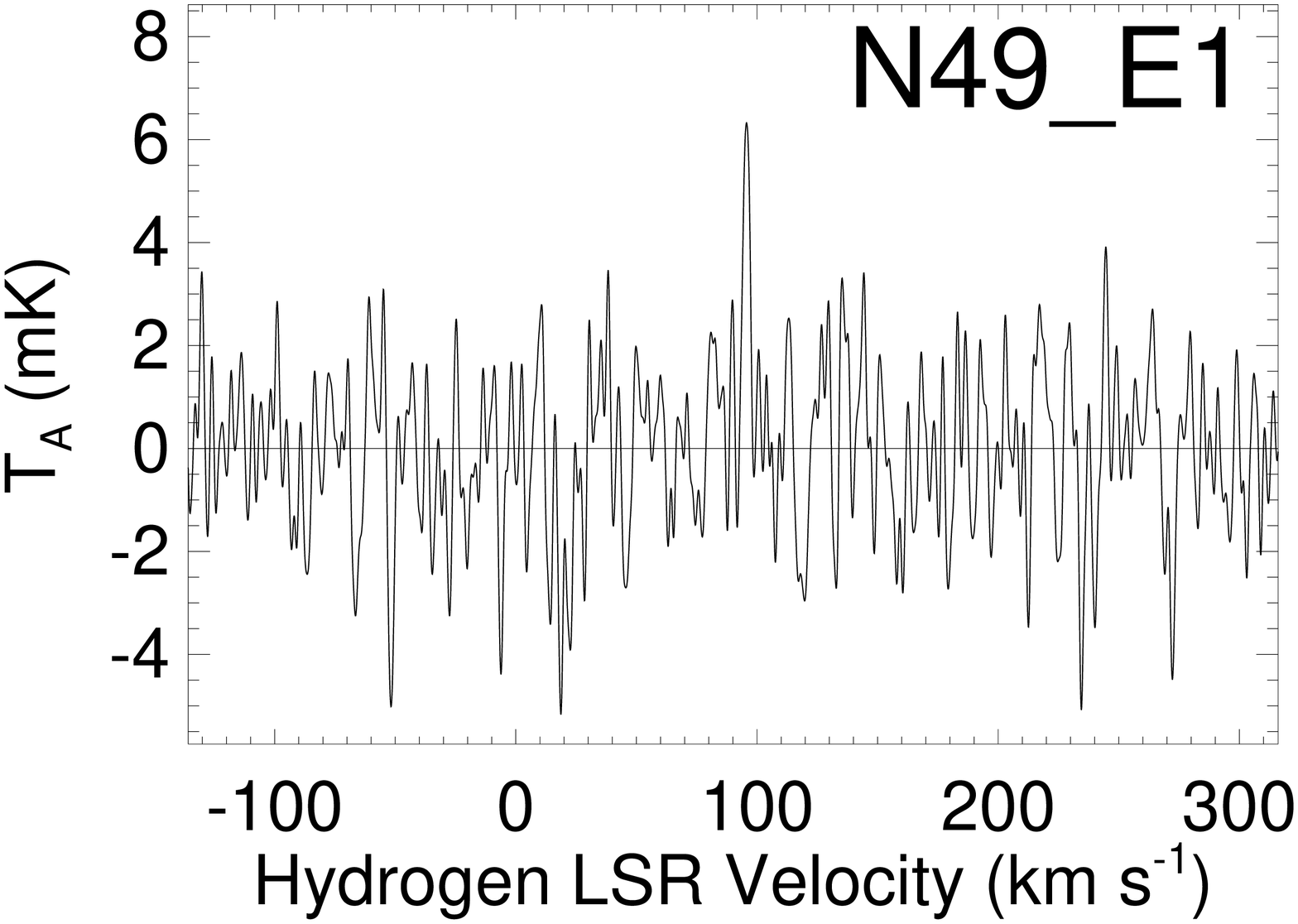} &
\includegraphics[width=.23\textwidth]{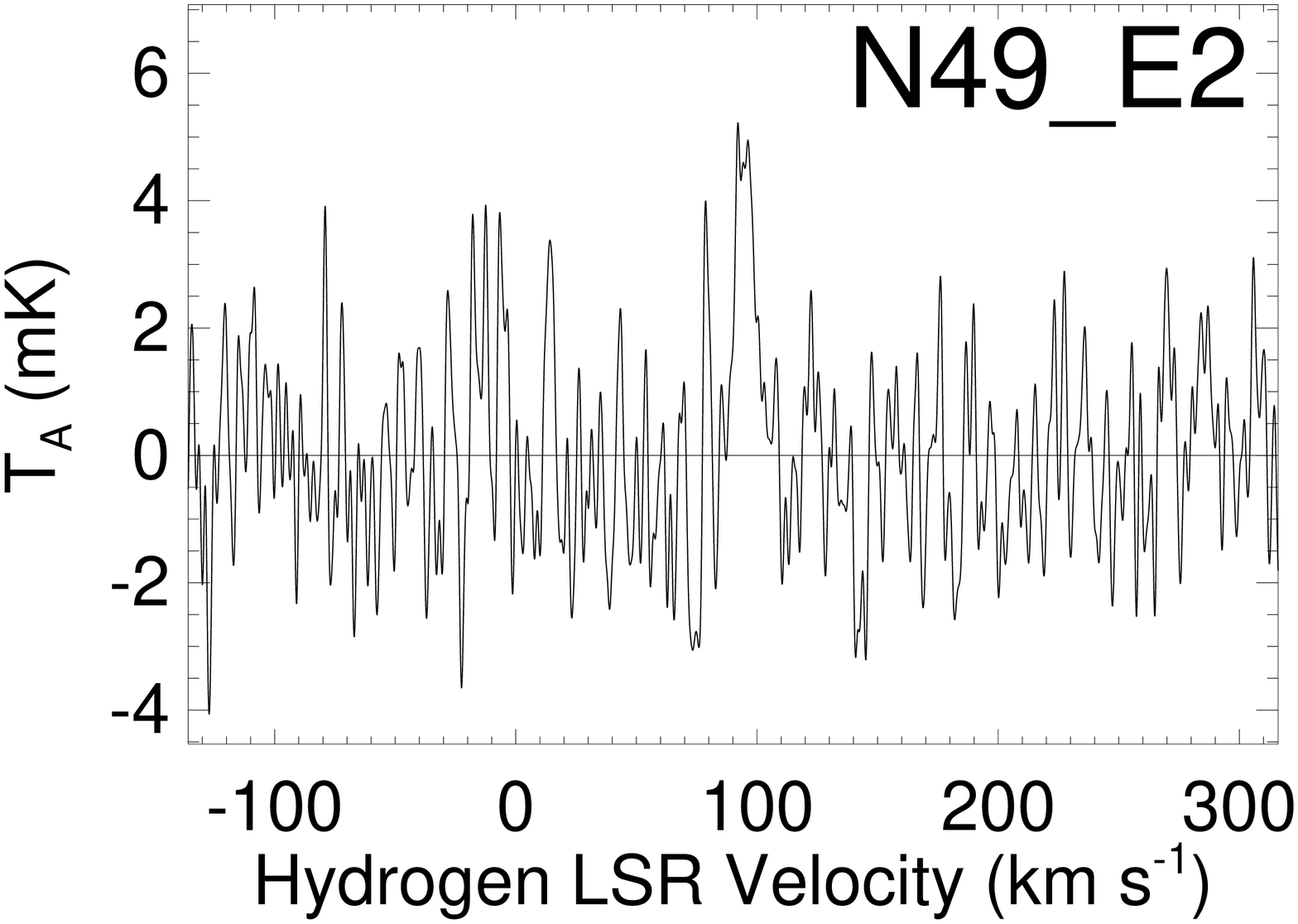} \\
\end{tabular}
\caption{}
\end{figure*}
\renewcommand{\thefigure}{\thesection.\arabic{figure}}

\renewcommand\thefigure{\thesection.\arabic{figure} (Cont.)}
\addtocounter{figure}{-1}
\begin{figure*}
\centering
\begin{tabular}{cccc}
\includegraphics[width=.23\textwidth]{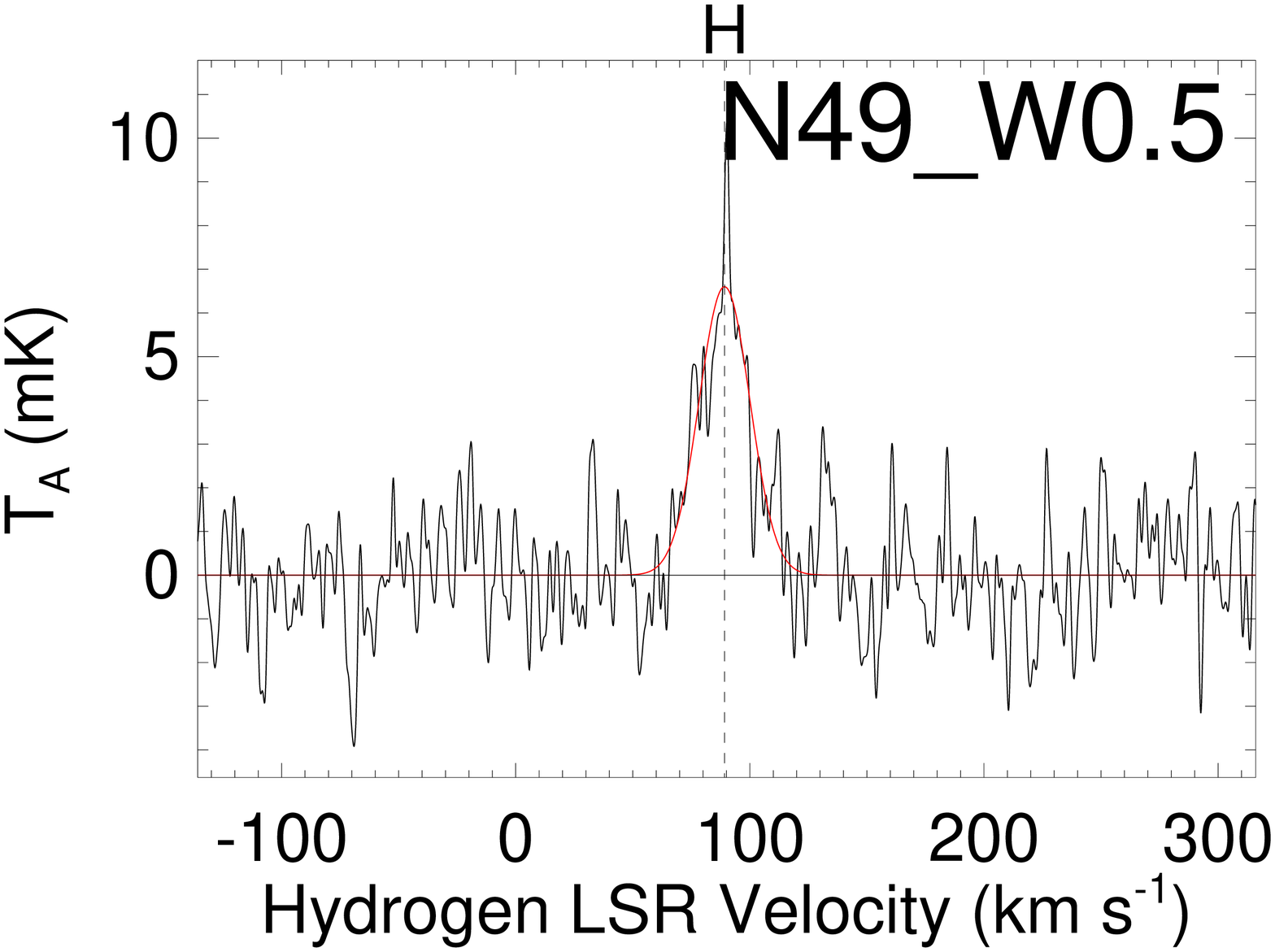} &
\includegraphics[width=.23\textwidth]{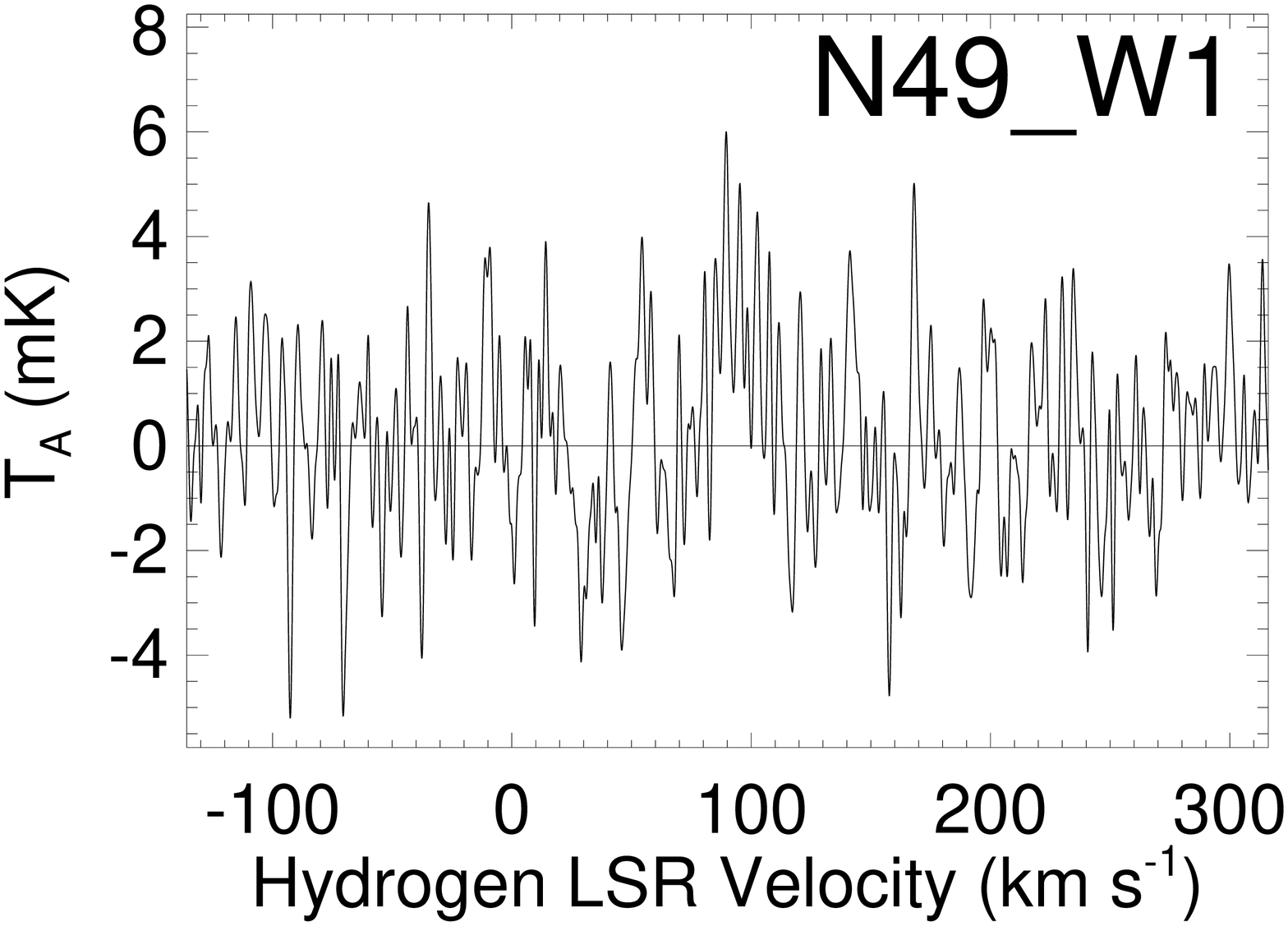} &
\includegraphics[width=.23\textwidth]{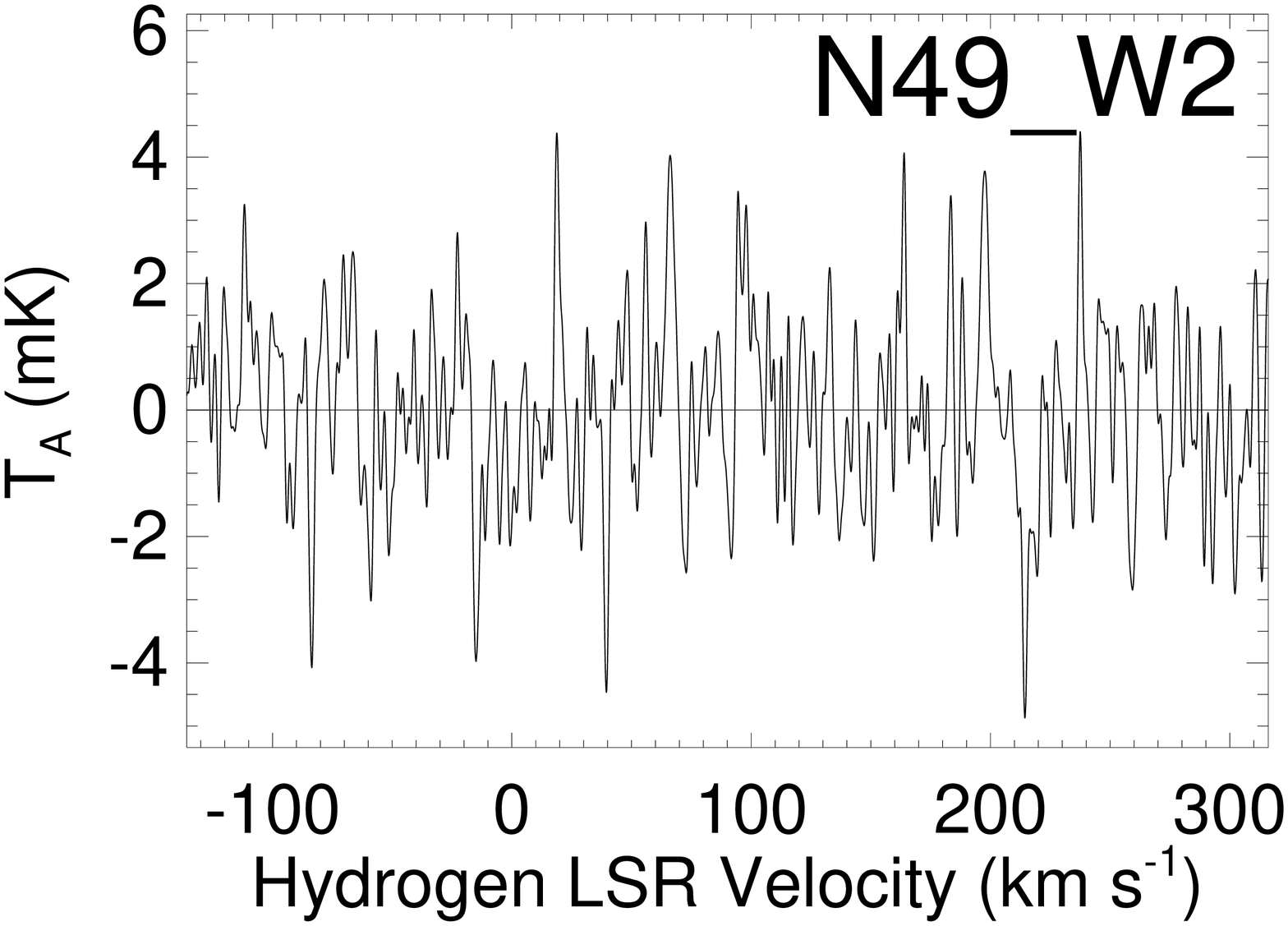} &
\includegraphics[width=.23\textwidth]{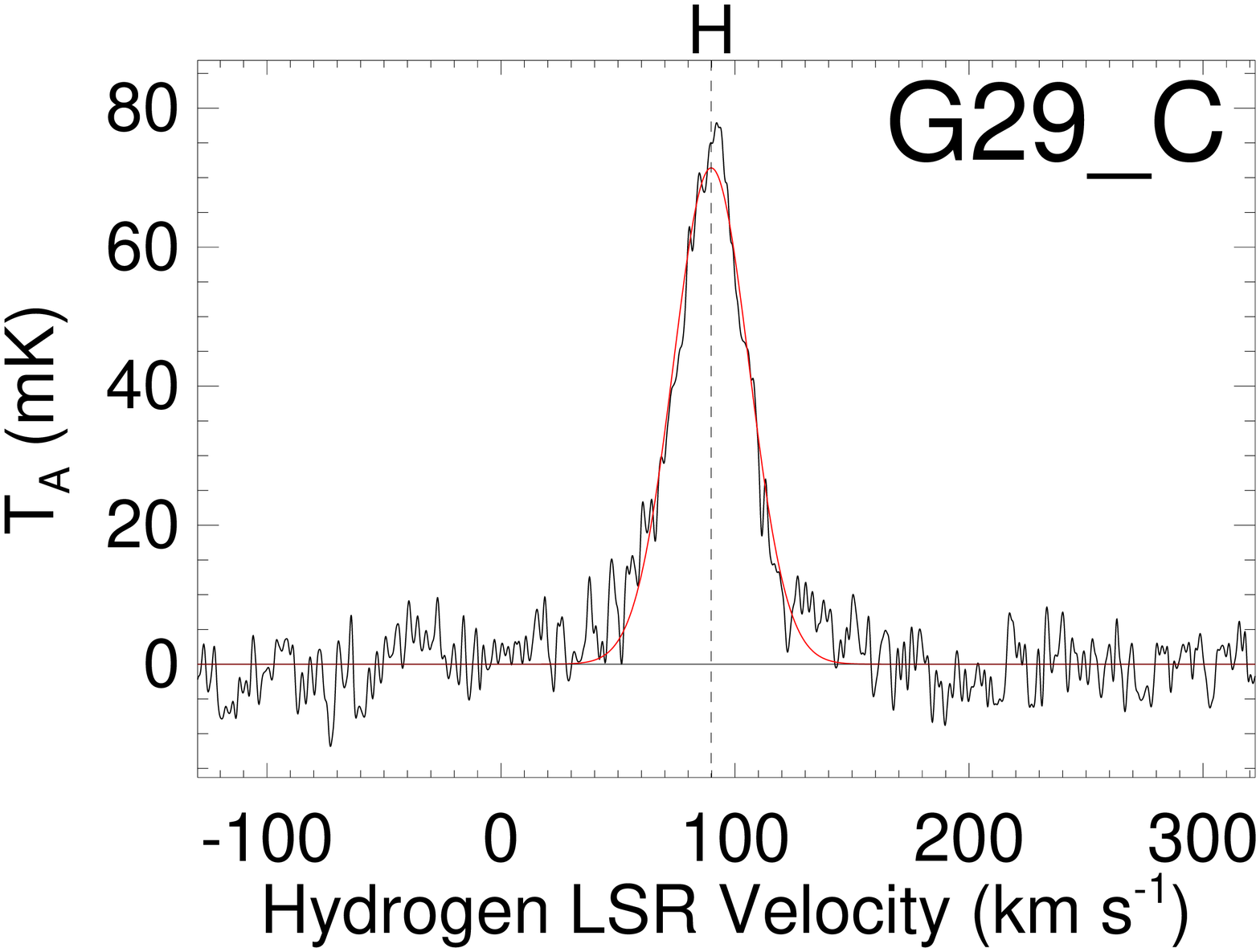} \\
\includegraphics[width=.23\textwidth]{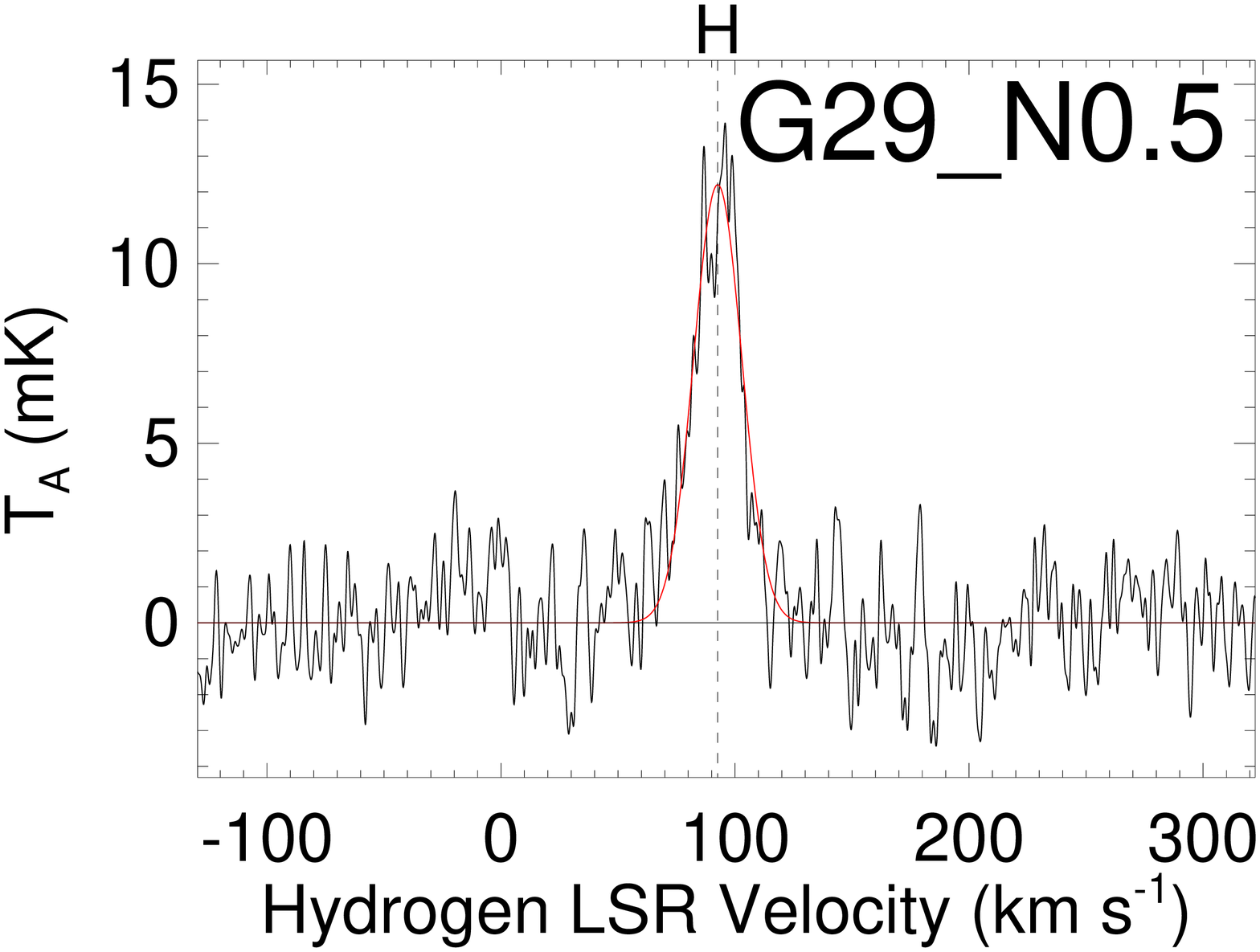} &
\includegraphics[width=.23\textwidth]{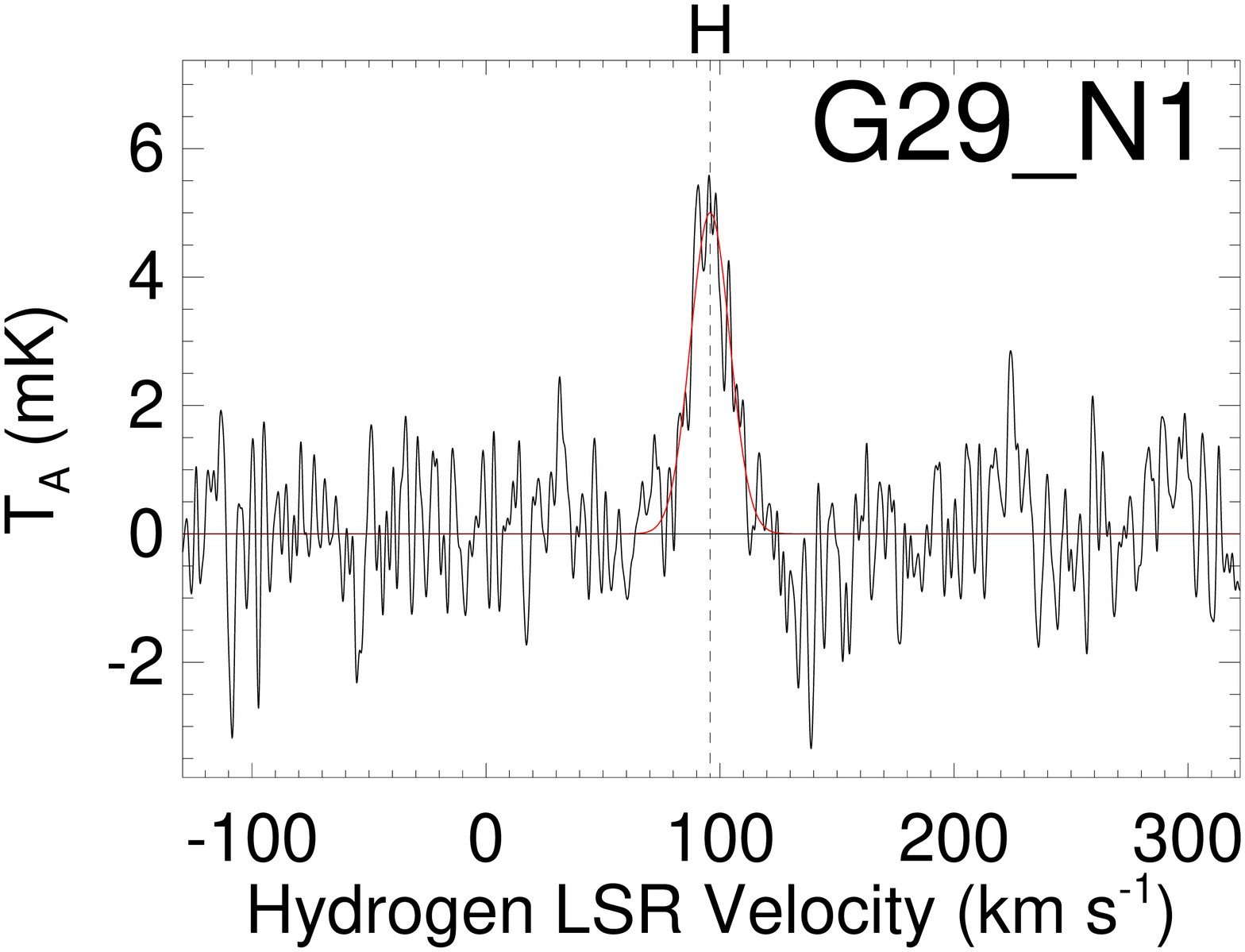} &
\includegraphics[width=.23\textwidth]{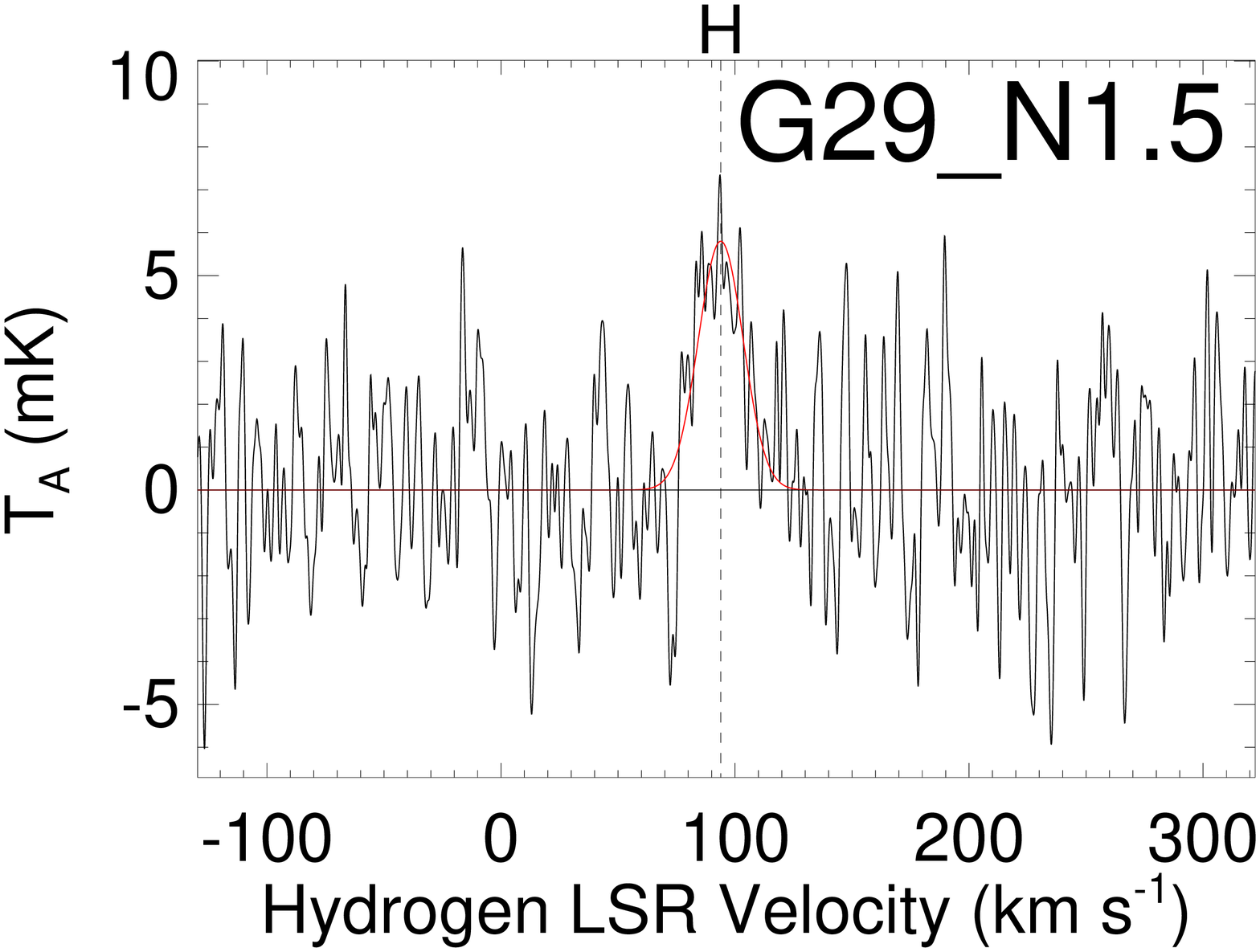} &
\includegraphics[width=.23\textwidth]{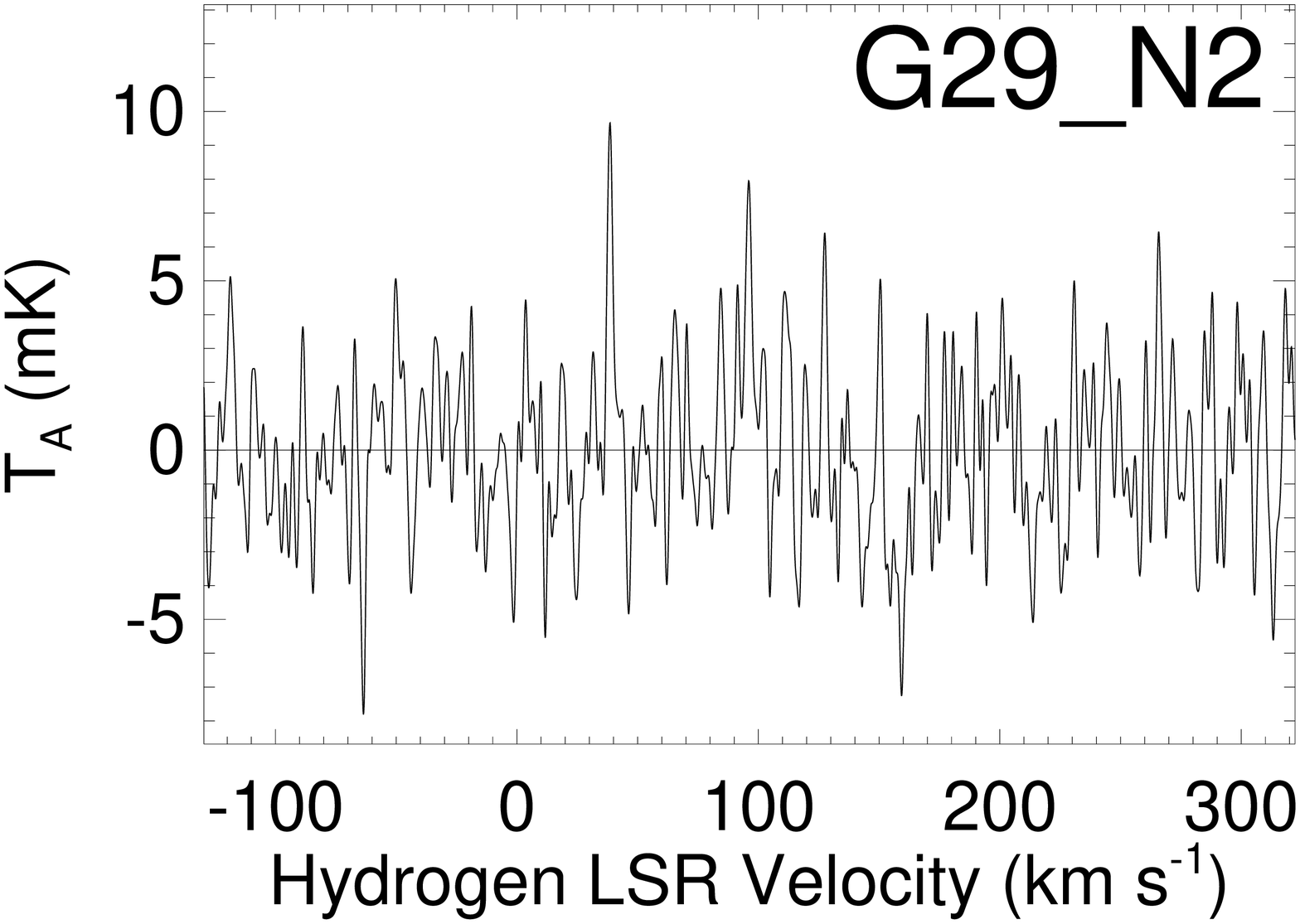} \\
\includegraphics[width=.23\textwidth]{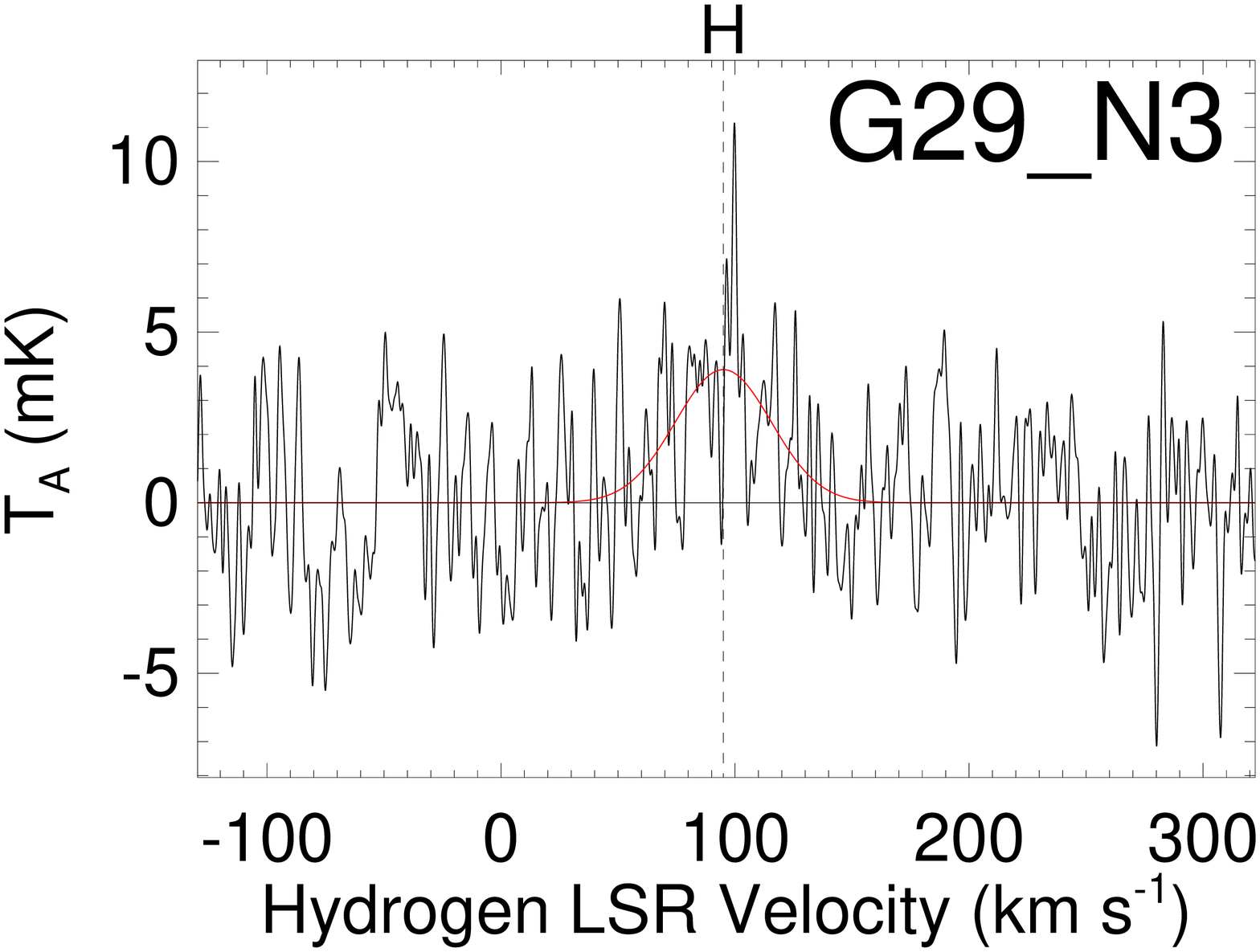} &
\includegraphics[width=.23\textwidth]{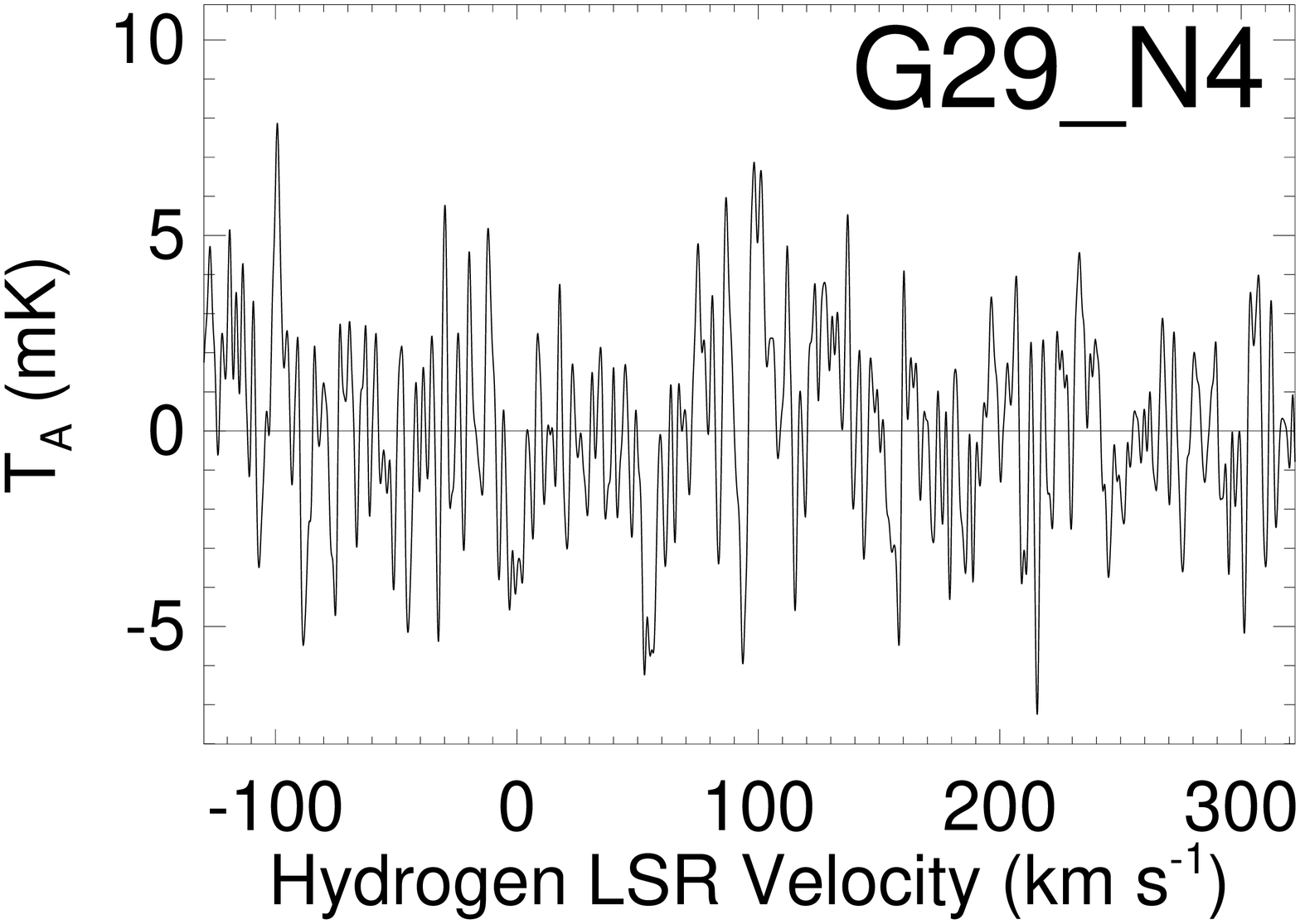} &
\includegraphics[width=.23\textwidth]{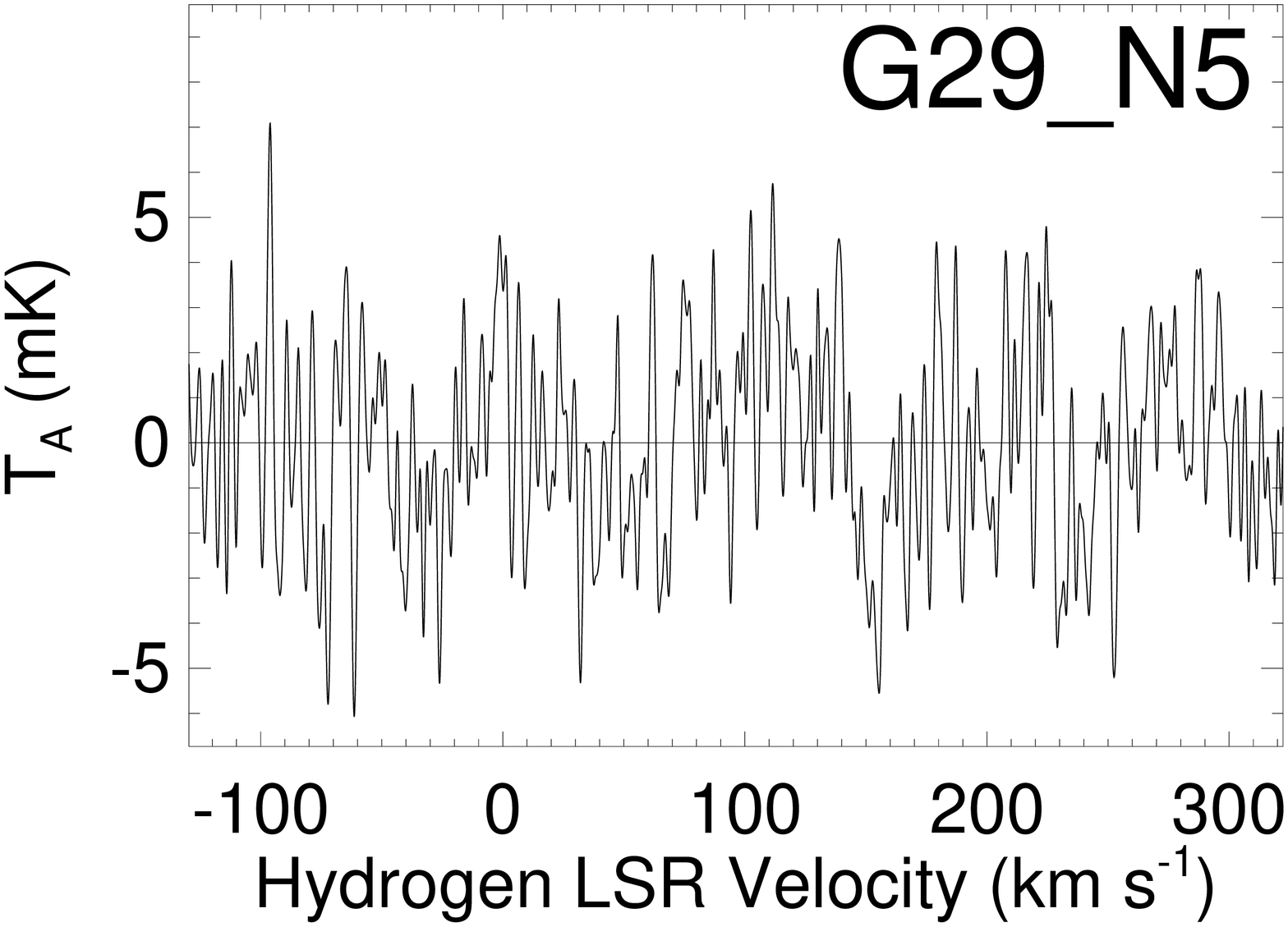} &
\includegraphics[width=.23\textwidth]{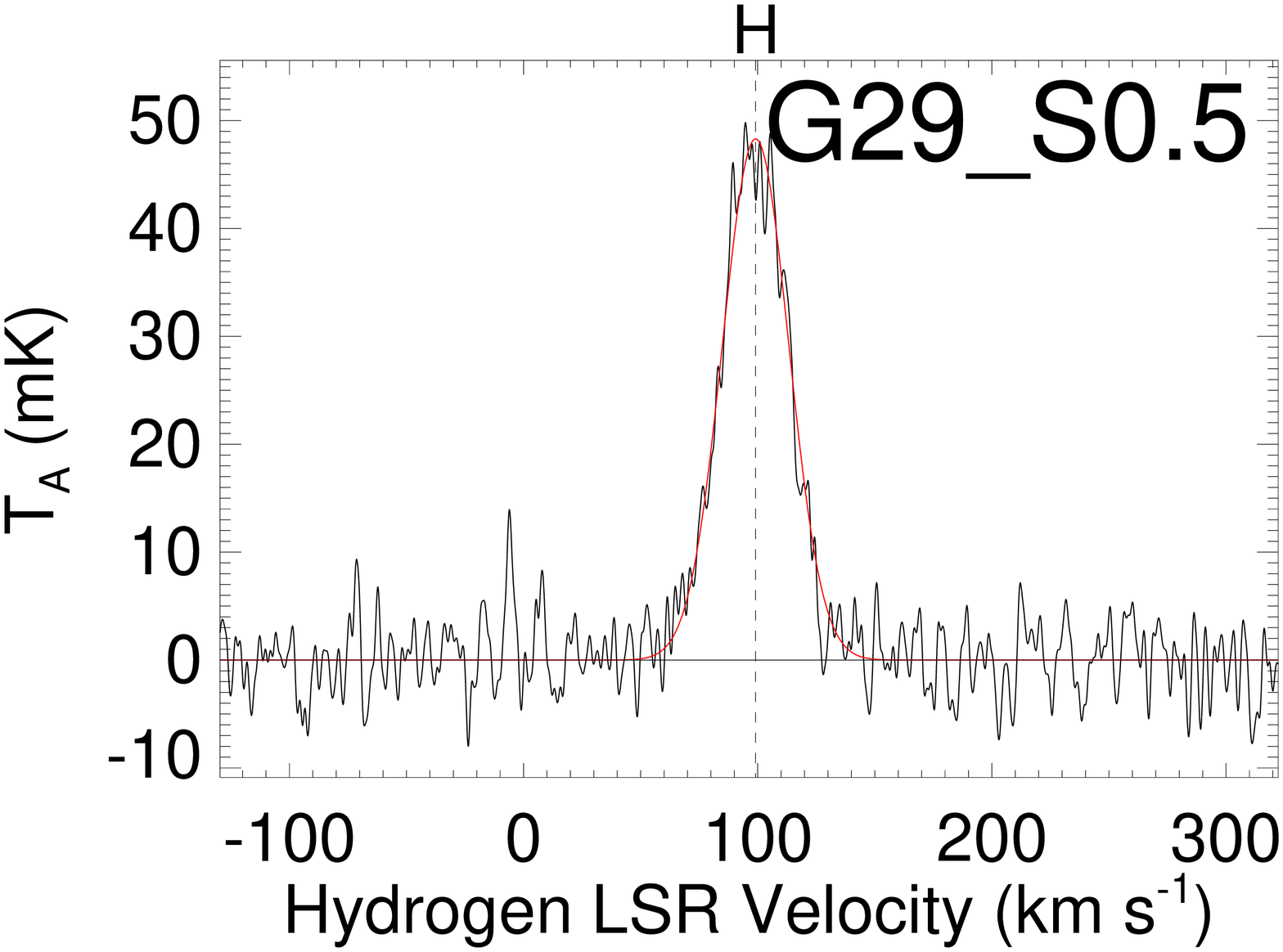} \\
\includegraphics[width=.23\textwidth]{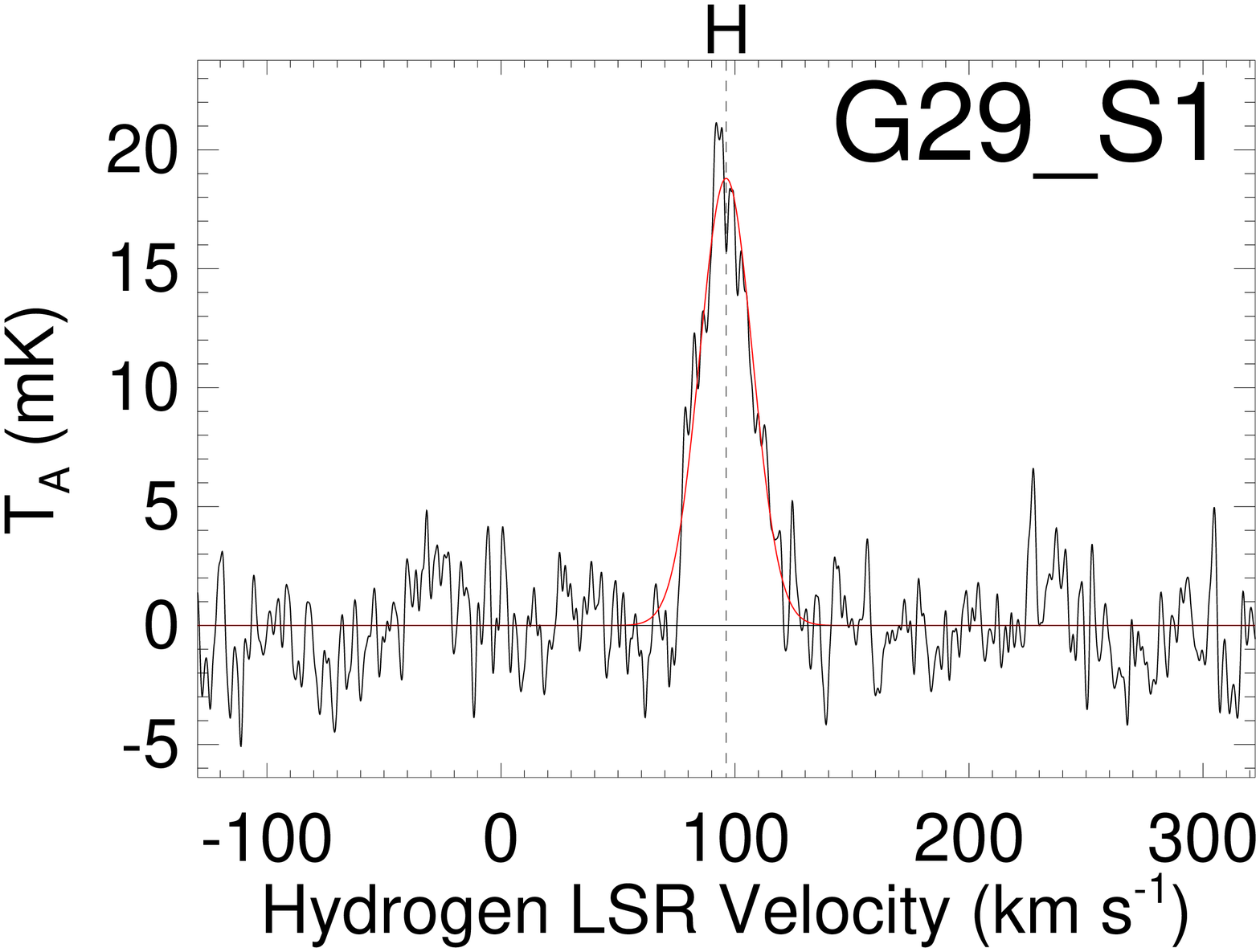} &
\includegraphics[width=.23\textwidth]{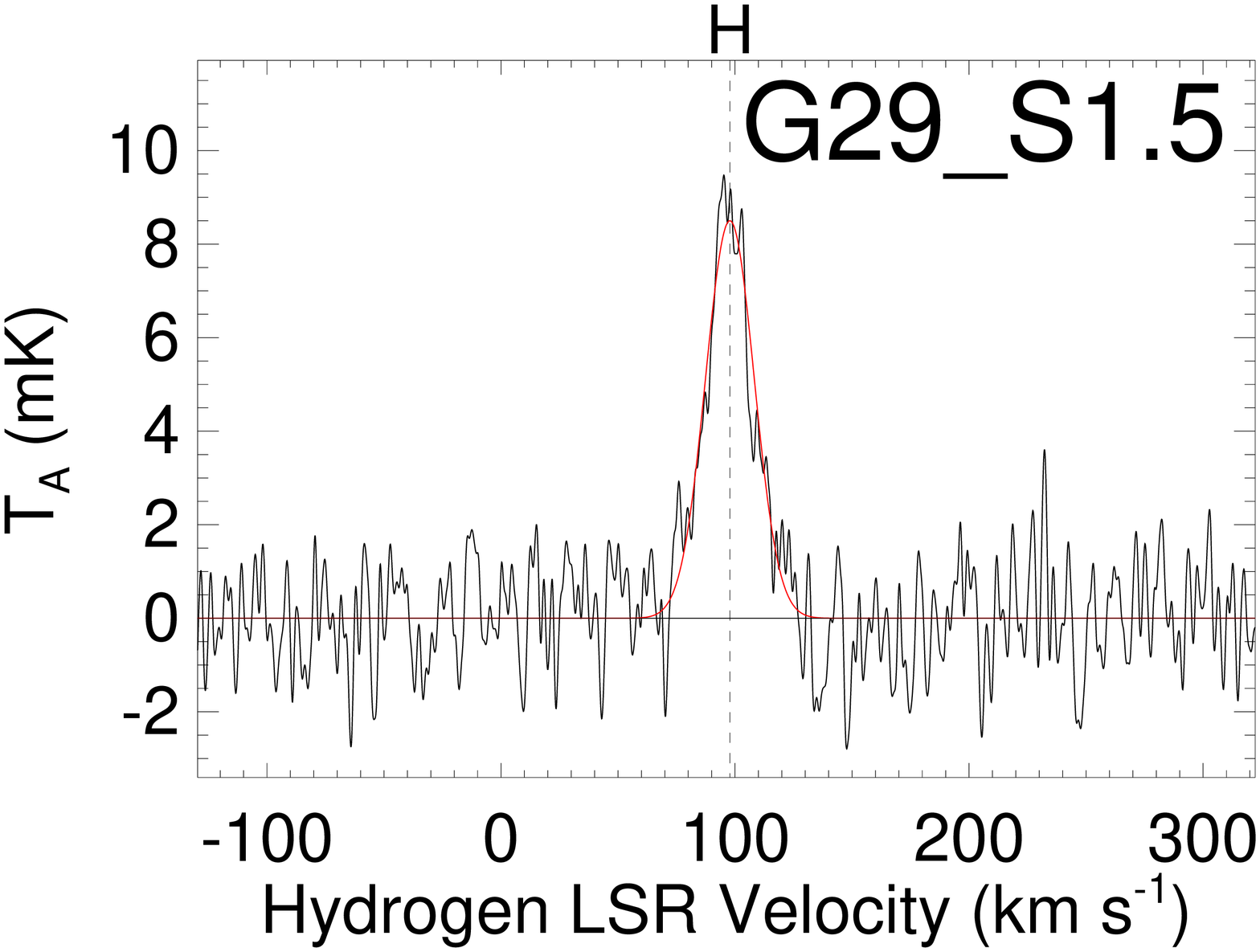} &
\includegraphics[width=.23\textwidth]{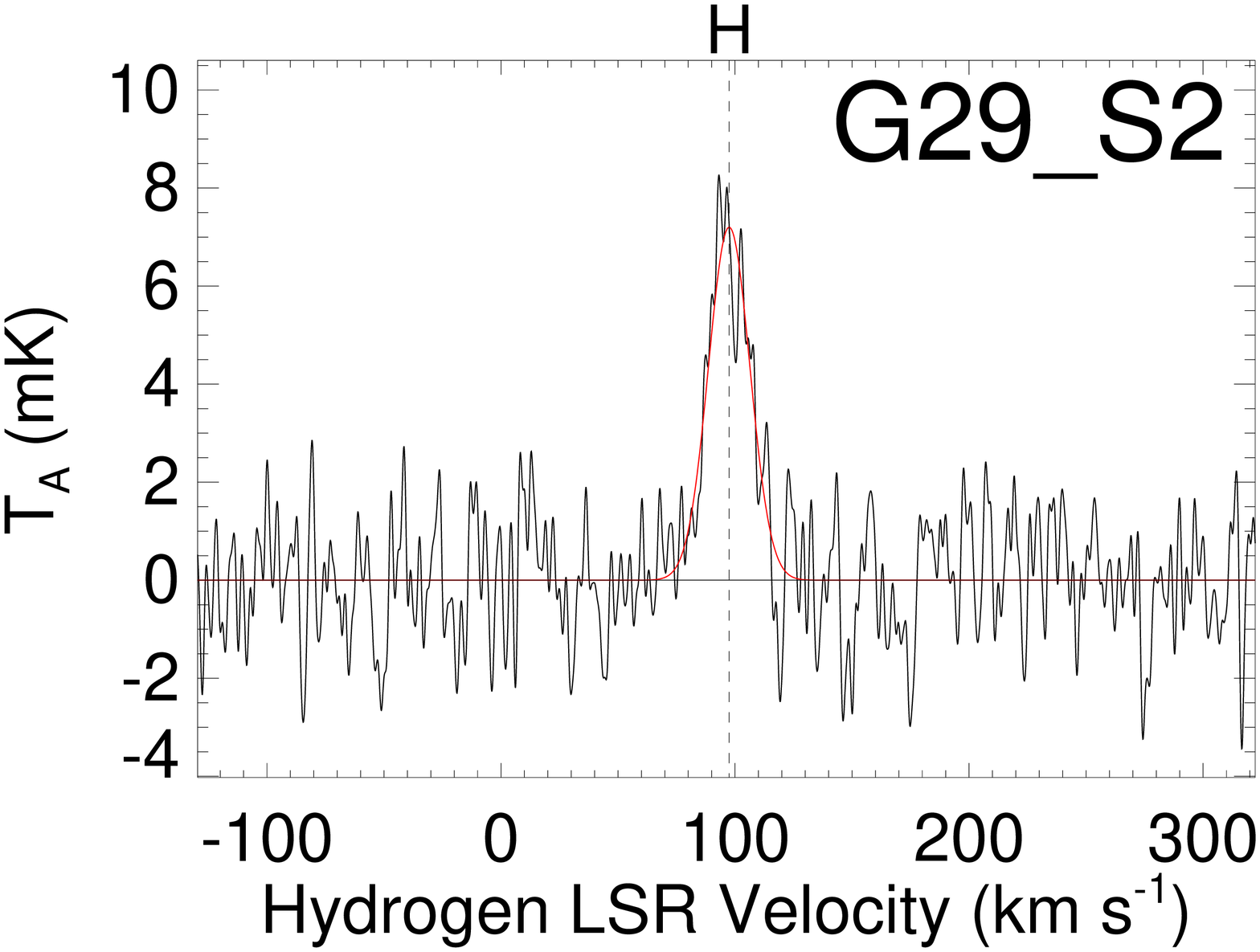} &
\includegraphics[width=.23\textwidth]{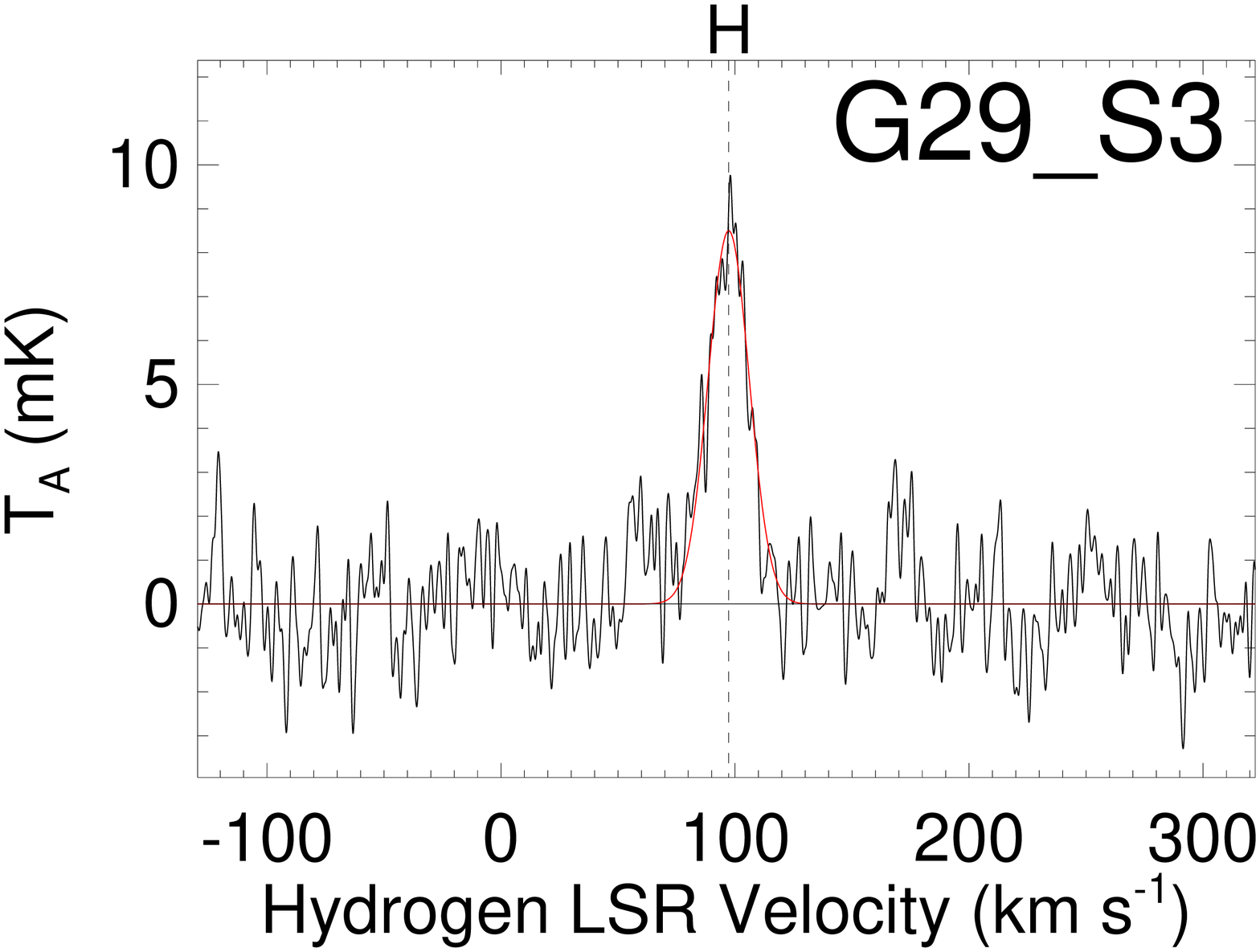} \\
\includegraphics[width=.23\textwidth]{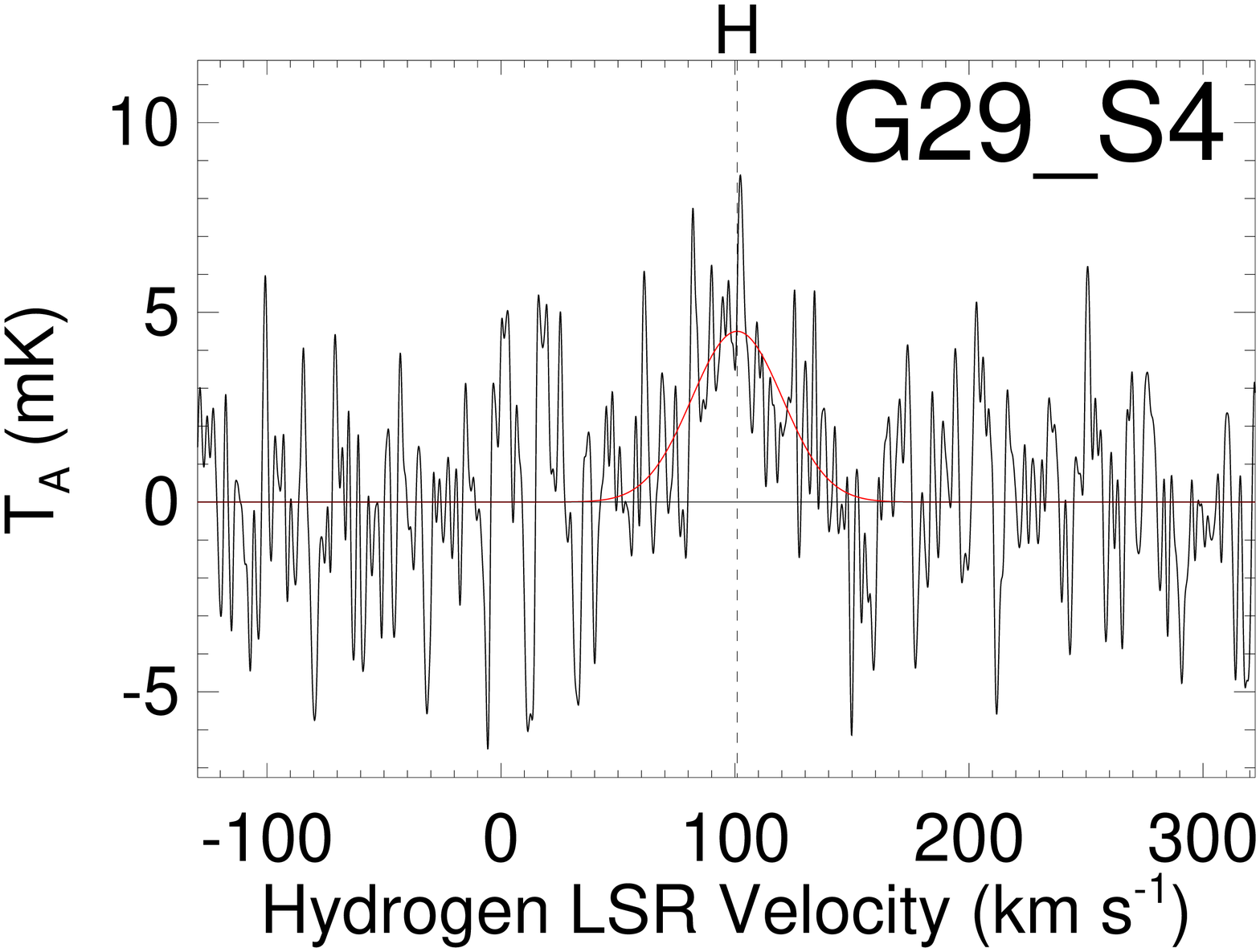} &
\includegraphics[width=.23\textwidth]{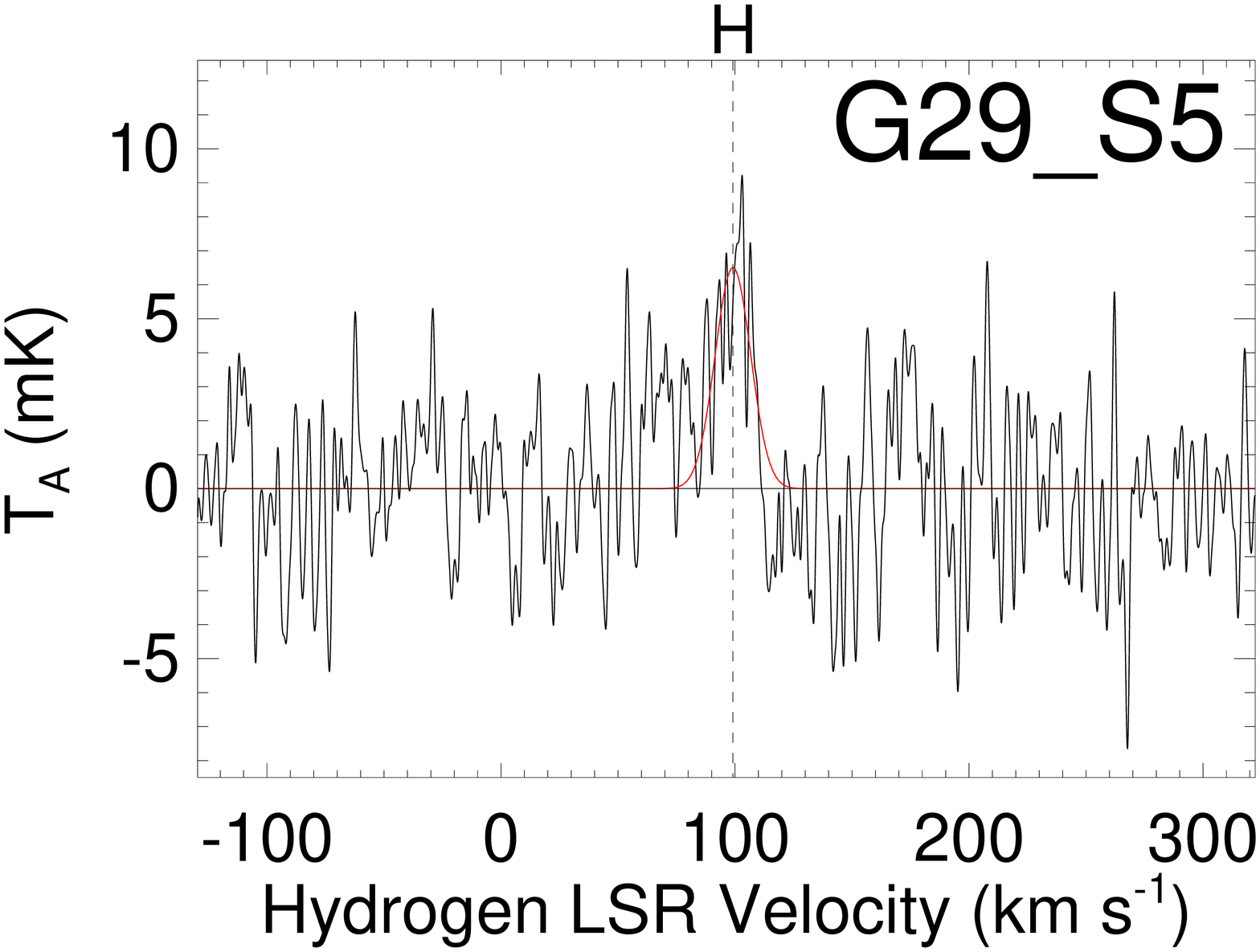} &
&
 \\
\end{tabular}
\caption{}
\end{figure*}
\renewcommand{\thefigure}{\thesection.\arabic{figure}}
\fi

\iftrue
\clearpage
\startlongtable
\begin{deluxetable*}{lcccccc}
\tabletypesize{\footnotesize}
\tablecaption{Radio Continuum Temperatures and Source Properties \label{tab:temperatures}}
\tablehead{Source  & \colhead{$T_{\rm C}$} & \colhead{$\delta T_{\rm C}$} & \colhead{$T_{\rm e}^*$} & \colhead{$\delta T_{\rm e}^*$} & \colhead{EM} & \colhead{$\delta$EM} \\
 &  \colhead{(K)} & \colhead{(K)} & \colhead{(K)} & \colhead{(K)} & \colhead{(pc\,cm$^{-6}$)} &\colhead{(pc\,cm$^{-6}$)} }
\startdata
S206\_C    & 6.90           & 1.00       & \phn 9330 & 1180 & 89030 & 16910  \\
S206\_E0.5 & 2.45           & 0.52       & \phn 9550 & 1770 & 31940 & \phn 8890   \\
S206\_E1   & 0.61           & 0.11       & \phn 9250 & 1540 & \phn 7760 & \phn 1940    \\
S206\_E2   & 0.26           & 0.05     & \phn 9130 & 1690 & \phn 3390 & \phn \phn 940     \\
S206\_E3   & 0.12           & 0.03     & \nodata & \nodata & \nodata & \nodata  \\
S206\_W0.5 & 2.67           & 0.58       & \phn 8860 & 1690 & 33690 & \phn 9650    \\
S206\_W1   & 0.74           & 0.14     & \phn 9130 & 1500 & \phn 9420 & \phn 2320      \\
S206\_W2   & 0.19           & 0.04   & 10190     & 1820 & \phn 2500 & \phn \phn 670    \\
S206\_W3   & 0.05           & 0.02     & \nodata   & \nodata & \nodata & \nodata  \\
S206\_W4   & 0.04           & 0.01     & \nodata   & \nodata & \nodata & \nodata \\
           &                &           &         &        &    & \\   
S104\_C    & 2.06           & 0.41        & 7270      & 1280 & 24400 & 6440    \\
S104\_N0.5 & 1.42           & 0.28        & 8020      & 1380 & 17510 & 4520     \\
S104\_N1   & 0.44           & 0.13       & 7320      & 1950 & \phn 5230 & 2090     \\
S104\_N1.5 & 0.10           & 0.02    & \nodata   & \nodata & \nodata & \nodata  \\
S104\_S0.5 & 1.55           & 0.30      & 8010      & 1350 & 19310 & 4870      \\
S104\_S1   & 0.38           & 0.09     & 9180      & 2020 & \phn 4950 & 1630      \\
S104\_S1.5 & 0.30           & 0.05      & \nodata   & \nodata & \nodata & \nodata   \\
S104\_S2.5 & 0.34           & 0.06     & \nodata   & \nodata & \nodata & \nodata   \\
           &                     & & & & &          \\
G45\_C     & 13.09          & 1.64        & \phn 7500 & \phn 830 & 158190 & 26290   \\
G45\_E0.5  & \phn 3.45      & 0.96        & \phn 7150 & 1750 & \phn 41410 & 15200   \\
G45\_E1    & \phn 0.64      & 0.11       & 10320     & 1740 & \phn \phn 8780 & \phn 2220      \\
G45\_E2    & \phn 0.32      & 0.22      & \nodata   & \nodata & \nodata & \nodata   \\
G45\_E3    & \phn 0.29      & 0.04    & \nodata   & \nodata & \nodata & \nodata    \\
G45\_W0.5  & \phn 3.34      & 0.65      & \phn 7270 & 1240 & \phn 40140 & 10300      \\
G45\_W1    & \phn 0.04      & 0.04         & \phn 1240 & 1080 & \phn \phn \phn 270 & \phn \phn 360   \\
G45\_W2    & \phn 0.15      & 0.05     & \nodata   & \nodata & \nodata & \nodata  \\
G45\_W3    & \phn 0.07      & 0.02      & \nodata & \nodata & \nodata & \nodata  \\
           &                         & & & & &     \\  
M16\_C     & 5.76           & 1.20       & 7420      & 1340 & 69080 & 18750     \\
M16\_N0.5  & 5.59           & 1.11   & 8150      & 1410 & 69420 & 17970      \\
M16\_N1    & 6.49           & 1.24       & 8520      & 1430 & 81790 & 20530    \\
M16\_N1.5  & 4.97           & 0.97    & 8900      & 1570 & 64130 & 17010     \\
M16\_N2.5  & 2.23           & 0.39      & 8690      & 1360 & 28800 & \phn 6770      \\
M16\_N4.5  & 0.19           & 0.06      & \nodata & \nodata & \nodata & \nodata   \\
M16\_N6.5  & 0.03           & 0.05       & \nodata & \nodata & \nodata & \nodata  \\
M16\_N8.5  & 0.26           & 0.05      & \nodata & \nodata & \nodata & \nodata  \\
M16\_N10.5 & 0.31           & 0.06      & \nodata & \nodata & \nodata & \nodata \\
M16\_S0.5  & 5.50           & 1.03        & 8070      & 1310 & 68250 & 16640     \\
M16\_S1    & 4.06           & 0.82        & 9040      & 1590 & 52460 & 13830     \\
M16\_S1.5  & 4.27           & 0.85       & 7790      & 1370 & 51580 & 13600      \\
M16\_S2.5  & 3.73           & 0.72     & 8060      & 1360 & 46360 & 11740      \\
M16\_S3.5  & 1.23           & 0.25      & 7670      & 1470 & 15270 & \phn 4390     \\
M16\_S5.5  & 0.49           & 0.10      & 7810      & 1560 & \phn 6070 & \phn 1820      \\
M16\_S9.5  & 0.39           & 0.08      & \nodata & \nodata & \nodata & \nodata  \\
M16\_S11.5 & 0.29           & 0.04      & \nodata & \nodata & \nodata & \nodata \\
           &    & & &   &      & \\
M17\_C     & 111.11         & 24.07       & 10780     & 2040 & 1500080 & 426320   \\
M17\_E1    & \phn 46.96     & \phn 9.81   & \phn 6870 & 1250 & \phn 543540 & 148400   \\
M17\_E1.5  & \phn \phn 5.73 & \phn 1.39    & \phn 7610 & 1640 & \phn \phn 69350 & \phn 22360 \\
M17\_E2    & \phn \phn 3.72 & \phn 0.82  & \phn 7720 & 1510 & \phn \phn 45360 & \phn 13290    \\
M17\_E3    & \phn \phn 2.59 & \phn 0.58   & \phn 7400 & 1470 & \phn \phn 30580 & \phn \phn 9130    \\
M17\_E4    & \phn \phn 1.57 & \phn 0.30   & \phn 7840 & 1300 & \phn \phn 19160 & \phn \phn 4780   \\
M17\_E5    & \phn \phn 0.32 & \phn 0.07   & \phn 7890 & 1540 & \phn \phn \phn 3760 & \phn \phn 1100  \\
M17\_E6    & \phn \phn 0.28 & \phn 0.08 & \nodata & \nodata & \nodata & \nodata    \\
M17\_E7    & \phn \phn 0.21 & \phn 0.04  & \phn 8720 & 1490 & \phn \phn \phn 2600 & \phn \phn \phn 670  \\
M17\_E9    & \nodata        & \nodata   & \nodata & \nodata & \nodata & \nodata  \\
M17\_E11   & \nodata        & \nodata    & \nodata & \nodata & \nodata & \nodata  \\
M17\_W1    & \phn 14.55     & \phn 5.22  & \phn 6870 & 2140 & \phn 169580 & \phn 79350      \\
M17\_W1.5  & \phn \phn 0.90 & \phn 0.23    & \phn 7410 & 1680 & \phn \phn 10850 & \phn \phn 3690    \\
M17\_W2    & \phn \phn 0.27 & \phn 0.08    & \phn 5440 & 1370 & \phn \phn \phn 2830 & \phn \phn 1070   \\
M17\_W3    & \phn \phn 0.31 & \phn 0.04  & \nodata & \nodata & \nodata & \nodata \\
M17\_W4    & \phn \phn 0.29 & \phn 0.05 & \nodata & \nodata & \nodata & \nodata  \\
M17\_W5    & \phn \phn 0.15 & \phn 0.04  & \nodata & \nodata & \nodata & \nodata  \\
           &              & & & & &         \\
Orion\_C    & 214.56         & 33.39       & 6330      & \phn 860 & 2418330 & 492550    \\
Orion\_N1   & \phn 30.89     & \phn 6.57  & 8040      & 1490 & \phn 384950 & 107070       \\
Orion\_N2   & \phn \phn 0.94 & \phn 0.25  & \nodata & \nodata & \nodata & \nodata  \\
Orion\_N3   & \phn \phn 0.07 & \phn 0.02  & \nodata & \nodata & \nodata & \nodata  \\
Orion\_N4   & \phn \phn 0.05 & \phn 0.01  & \nodata & \nodata & \nodata & \nodata  \\
Orion\_N5   & \phn \phn 0.05 & \phn 0.01 & \nodata & \nodata & \nodata & \nodata  \\
Orion\_S1   & \phn 10.90      & \phn 2.40  & 8010      & 1560 & \phn 136530 & \phn 40000      \\
Orion\_S2   & \phn \phn 2.48 & \phn 0.52    & 8400      & 1660 & \phn \phn 31280 & \phn \phn 9260    \\
Orion\_S3   & \phn \phn 1.32 & \phn 0.25     & 8550      & 1440 & \phn \phn 17030 & \phn \phn 4320   \\
Orion\_S4   & \phn \phn 0.34 & \phn 0.07  & \nodata & \nodata & \nodata & \nodata \\
Orion\_S5   & \phn \phn 0.08 & \phn 0.03  & \nodata & \nodata & \nodata & \nodata   \\
Orion\_S6   & \phn \phn 0.14 & \phn 0.03 & \nodata & \nodata & \nodata & \nodata  \\
Orion\_S7   & \phn \phn 0.09 & \phn 0.02 & \nodata & \nodata & \nodata & \nodata \\
Orion\_S9   & \phn \phn 0.01 & \phn 0.01  & \nodata & \nodata & \nodata & \nodata  \\
Orion\_S11    & \nodata   & \nodata & \nodata & \nodata & \nodata & \nodata \\
               &           & & &        &           &          \\
N49\_C     & 1.90            & 0.32       & 6250      & \phn 920 & 22120 & 4890    \\
N49\_E0.5  & 0.67           & 0.15       & 8100      & 1720 & \phn 8990 & 2870  \\
N49\_E1    & 0.39           & 0.06      & \nodata & \nodata & \nodata & \nodata \\
N49\_E2    & 0.47           & 0.07     & \nodata &\nodata  & \nodata & \nodata  \\
N49\_W0.5  & 0.94           & 0.21     & 7330      & 1580 & 11750 & 3800    \\
N49\_W1    & 0.45           & 0.13      & \nodata & \nodata & \nodata & \nodata   \\
N49\_W2    & 0.35           & 0.10      & \nodata & \nodata & \nodata & \nodata \\
           &                        & & & & &   \\
G29\_C     & 12.94          & 1.46      & 6490      & \phn 680 & 148100 & 23140  \\
G29\_N0.5  & \phn 2.12      & 0.68      & 6050      & 1720 & \phn 24240 & 10340     \\
G29\_N1    & \phn 0.45      & 0.09       & 6890      & 1340 & \phn \phn 5530 & \phn 1620      \\
G29\_N1.5  & \phn 0.40      & 0.08       & \nodata & \nodata & \nodata & \nodata  \\
G29\_N2    & \phn 0.61      & 0.07     & \nodata & \nodata & \nodata & \nodata  \\
G29\_N3    & \phn 0.70      & 0.07     & \nodata & \nodata & \nodata & \nodata \\
G29\_N4    & \phn 0.49      & 0.06      & \nodata & \nodata & \nodata & \nodata \\
G29\_N5    & \phn 0.37      & 0.07     & \nodata & \nodata & \nodata & \nodata  \\
G29\_S0.5  & \phn 6.80      & 1.45      & 5980      & 1120 & \phn 76420 & 21480     \\
G29\_S1    & \phn 2.31      & 0.56     & 5750      & 1270 & \phn 25650 & \phn 8480      \\
G29\_S1.5  & \phn 1.07      & 0.20       & 6850      & 1220 & \phn 12820 & \phn 3420      \\
G29\_S2    & \phn 0.81      & 0.12      & 7780      & 1230 & \phn 10190 & \phn 2410     \\
G29\_S3    & \phn 0.67      & 0.15      & 6980      & 1410 & \phn \phn 8260 & \phn 2510      \\
G29\_S4    & \phn 0.41      & 0.15    & \nodata & \nodata & \nodata & \nodata   \\
G29\_S5    & \phn 0.64      & 0.14     & \nodata & \nodata & \nodata & \nodata 
\enddata
\end{deluxetable*}
\fi

\end{document}